\documentclass[12pt]{report}   
\usepackage{graphicx}  
\usepackage[letterpaper, left=1in, right=1in, top=1in, bottom=1in]{geometry}
\usepackage{setspace}  
\usepackage{times}  
\usepackage[explicit]{titlesec}  
\usepackage[titles]{tocloft}  
\usepackage[utf8]{inputenc} 
\usepackage[backend=biber, style=apa, backref=true, uniquename=false]{biblatex}
\usepackage{appendix}  
\usepackage{rotating}  
\usepackage[normalem]{ulem}  
\usepackage{textcomp} 
\usepackage{indentfirst} 
\usepackage{booktabs,array,arydshln} 
\usepackage{amsmath} 
\usepackage[T1]{fontenc} 
\usepackage{savesym}
\usepackage{amsthm, thmtools, thm-restate}
\savesymbol{openbox}
\usepackage{newtxmath}
\restoresymbol{TX}{openbox}
\usepackage[hidelinks]{hyperref}
\usepackage{csquotes}
\usepackage{tikz, tikz-3dplot}
\usepackage{enumitem}
\usepackage{bbm}
\usepackage{afterpage}
\usepackage{footmisc}
\usepackage{comment}

\setlist[itemize]{noitemsep}
\setlist[enumerate]{noitemsep}

\usepackage[ruled]{algorithm2e} 

\SetCommentSty{mycommfont}

\allowdisplaybreaks

\newcommand{\vect}[1]{\mathbf{#1}}

\newcommand{\RR}{\mathbb{R}}

\newcommand{\FF}{\mathbb{F}}
\newcommand{\PP}{\mathbb{P}}
\newcommand\norm[1]{\left\lVert#1\right\rVert}
\newcommand\abs[1]{\left\lvert#1\right\rvert}
\newcommand\ceil[1]{\left\lceil#1\right\rceil}

\newcommand\parens[1]{\left(#1\right)}
\newcommand\brackets[1]{\left[#1\right]}
\newcommand\angles[1]{\left\langle#1\right\rangle}
\newcommand*{\pr}[2][]{\text{Pr}\ifx\\\left[#1\right]\\\else_{#1}\fi \left[#2\right]}
\newcommand*{\EE}[2][]{\mathbb{E}\ifx\\\left[#1\right]\\\else_{#1}\fi \left[#2\right]}
\newcommand{\partl}[2]{\frac{\partial #1}{\partial #2}}

\newtheorem{theorem}{Theorem}[section]
\newtheorem{lemma}[theorem]{Lemma}
\newtheorem{prop}[theorem]{Proposition}
\newtheorem{corollary}[theorem]{Corollary}
\newtheorem{claim}[theorem]{Claim}
\newtheorem{fact}[theorem]{Fact}
\newtheorem{question}[theorem]{Question}
\newtheorem*{nonothm}{Theorem}
\newtheorem{remark}[theorem]{Remark}
\newtheorem{defin}[theorem]{Definition}

\theoremstyle{definition}
\newtheorem{example}[theorem]{Example}

\makeatletter
\let\c@algorithm\c@theorem
\makeatother

\newcommand{\opt}{\text{Opt}}
\newcommand{\cost}{\text{Score}}
\newcommand{\ind}{\text{Ind}}
\newcommand{\errf}{\text{Error}}
\newcommand{\err}{\text{Err}}
\newcommand{\nstop}{\ensuremath{n_{\mathsf{stop}}}}
\newcommand{\half}{\frac{1}{2}}

\newcommand{\ws}{\text{WS}}
\newcommand{\prb}{\text{pr}}
\newcommand{\wt}{\text{wt}}
\newcommand{\shrink}{\text{Shrink}}

\newcommand{\JB}{\text{JB}}

\newcommand{\proj}{\text{proj}}

\bibliography{references}

\AtEveryBibitem{\clearfield{issn}}
\AtEveryBibitem{\clearlist{issn}}

\AtEveryBibitem{\clearfield{language}}
\AtEveryBibitem{\clearlist{language}}

\AtEveryBibitem{\clearfield{doi}}
\AtEveryBibitem{\clearlist{doi}}

\AtEveryBibitem{\clearfield{url}}
\AtEveryBibitem{\clearlist{url}}

\AtEveryBibitem{%
  \ifentrytype{online}
    {}
    {\clearfield{urlyear}\clearfield{urlmonth}\clearfield{urlday}}}

\begin{document}
\doublespacing  


\begin{titlepage}
\begin{center}

\begin{singlespacing}
\vspace*{6\baselineskip}
Algorithmic Bayesian Epistemology\\
\vspace{3\baselineskip}
Eric Neyman\\
\vspace{18\baselineskip}
Submitted in partial fulfillment of the\\
requirements for the degree of\\
Doctor of Philosophy\\
under the Executive Committee\\
of the Graduate School of Arts and Sciences\\
\vspace{3\baselineskip}
COLUMBIA UNIVERSITY\\
\vspace{3\baselineskip}
\the\year
\vfill

\end{singlespacing}

\end{center}
\end{titlepage}


\begin{titlepage}
\begin{singlespacing}
\begin{center}

\vspace*{35\baselineskip}

\textcopyright  \,  \the\year\\
\vspace{\baselineskip}	
Eric Neyman\\
\vspace{\baselineskip}	
All Rights Reserved
\end{center}
\vfill

\end{singlespacing}
\end{titlepage}

\pagenumbering{gobble}

\begin{titlepage}
\begin{center}

\vspace*{5\baselineskip}
\textbf{\large Abstract}

Algorithmic Bayesian Epistemology

Eric Neyman
\end{center}

One aspect of the algorithmic lens in theoretical computer science is a view on other scientific disciplines that focuses on satisfactory solutions that adhere to real-world constraints, as opposed to solutions that would be optimal ignoring such constraints. The algorithmic lens has provided a unique and important perspective on many academic fields, including molecular biology, ecology, neuroscience, quantum physics, economics, and social science.

This thesis applies the algorithmic lens to Bayesian epistemology. Traditional Bayesian epistemology provides a comprehensive framework for how an individual's beliefs should evolve upon receiving new information. However, these methods typically assume an exhaustive model of such information, including the correlation structure between different pieces of evidence. In reality, individuals might lack such an exhaustive model, while still needing to form beliefs. Beyond such informational constraints, an individual may be bounded by limited computation, or by limited communication with agents that have access to information, or by the strategic behavior of such agents. Even when these restrictions prevent the formation of a \emph{perfectly} accurate belief, arriving at a \emph{reasonably} accurate belief remains crucial. In this thesis, we establish fundamental possibility and impossibility results about belief formation under a variety of restrictions, and lay the groundwork for further exploration.

\vspace*{\fill}
\end{titlepage}

\pagenumbering{roman}
\setcounter{page}{1} 
\renewcommand{\cftchapdotsep}{\cftdotsep}  
\renewcommand{\cftchapfont}{\normalfont}  
\renewcommand{\cftchappagefont}{}  
\renewcommand{\cftchappresnum}{Chapter }
\renewcommand{\cftchapaftersnum}{:}
\renewcommand{\cftchapnumwidth}{5em}
\renewcommand{\cftchapafterpnum}{\vskip\baselineskip} 
\renewcommand{\cftsecafterpnum}{\vskip\baselineskip}  
\renewcommand{\cftsubsecafterpnum}{\vskip\baselineskip} 
\renewcommand{\cftsubsubsecafterpnum}{\vskip\baselineskip} 

\titleformat{\chapter}[display]
{\normalfont\bfseries\filcenter}{\chaptertitlename\ \thechapter}{0pt}{\large{#1}}

\renewcommand\contentsname{Table of Contents}

\begin{singlespace}
\setcounter{tocdepth}{1}
\tableofcontents
\setlength{\cftparskip}{\baselineskip}
\listoffigures
\listoftables
\end{singlespace}

\clearpage

\phantomsection
\addcontentsline{toc}{chapter}{Acknowledgments}

\clearpage
\begin{center}

\vspace*{5\baselineskip}
\textbf{\large Acknowledgements}
\end{center}

First, I would like to thank my Ph.D.\ advisor, Tim Roughgarden. Tim's expansive knowledge of and insight into theoretical computer science and surrounding areas helped me to choose my research directions and understand my work in the context of others' contributions. I benefited greatly from his impeccable advice on communication and presentation. Most of all, Tim was incredibly supportive in my exploration of my interests, both within and outside of grad school.

I would also like to thank my undergraduate mentor, Matt Weinberg. Matt's class on economics and computing inspired me to do research with him. Matt was an incredible mentor, and the research we did together inspired me to go to grad school in theoretical computer science. Moreover, our project on proper scoring rules (Chapter~\ref{chap:precision}) was instrumental in my decision to study algorithmic Bayesian epistemology in particular.

Thanks as well to all of my other research collaborators: Paul Christiano, Raf Frongillo, Jacob Hilton, George Noarov, V\'{a}clav Rozho\v{n}, Bo Waggoner, and Mark Xu. Working with them was great, and I have learned so much from them.

Thanks also to the National Science Foundation, whose graduate research fellowship program funded my research throughout my time as a grad student.

I would like to thank Scott Aaronson, whose blog post and paper on Aumann's agreement theorem I found truly inspiring. Scott may have inspired not just Chapter~\ref{chap:agreement} (which directly follows up on his paper), but this whole direction of my research.

I am so grateful to my family: my mom and my dad and my sister and my grandma. Their never-ending and unconditional love and support means so much to me.

And I am grateful to my longtime friends. Thank you to Mike and Cathy and Dylan and Sarah and Jenny and Ben and Yafah and Mia and Drake and Sam. I owe a great deal of my personal growth to the intellectual conversations and unforgettable adventures that they have shared with me.

\clearpage


\phantomsection
\addcontentsline{toc}{chapter}{Dedication}

\begin{center}

\vspace*{5\baselineskip}
\end{center}

\begin{flushleft}
\emph{To Baba Katya}

\hspace{10mm} \emph{--- who gave me my first lessons on Aumann's agreement theorem.}
\end{flushleft}




\clearpage
\pagenumbering{arabic}
\setcounter{page}{1} 

\phantomsection
\addcontentsline{toc}{chapter}{Preface}

\begin{center}
\vspace*{5\baselineskip}
\textbf{\large Preface}
\end{center}

For me as for most students, college was a time of exploration. I took many classes, read many academic and non-academic works, and tried my hand at a few research projects. Early in graduate school, I noticed a strong commonality among the questions that I had found particularly fascinating: most of them involved reasoning about knowledge, information, or uncertainty under constraints. I decided that this cluster of problems would be my primary academic focus. I settled on calling the cluster \emph{algorithmic Bayesian epistemology:} all of the questions I was thinking about involved applying the ``algorithmic lens'' of theoretical computer science to problems of Bayesian epistemology.

\begin{figure}[ht]
    \centering
    \includegraphics[width=\textwidth]{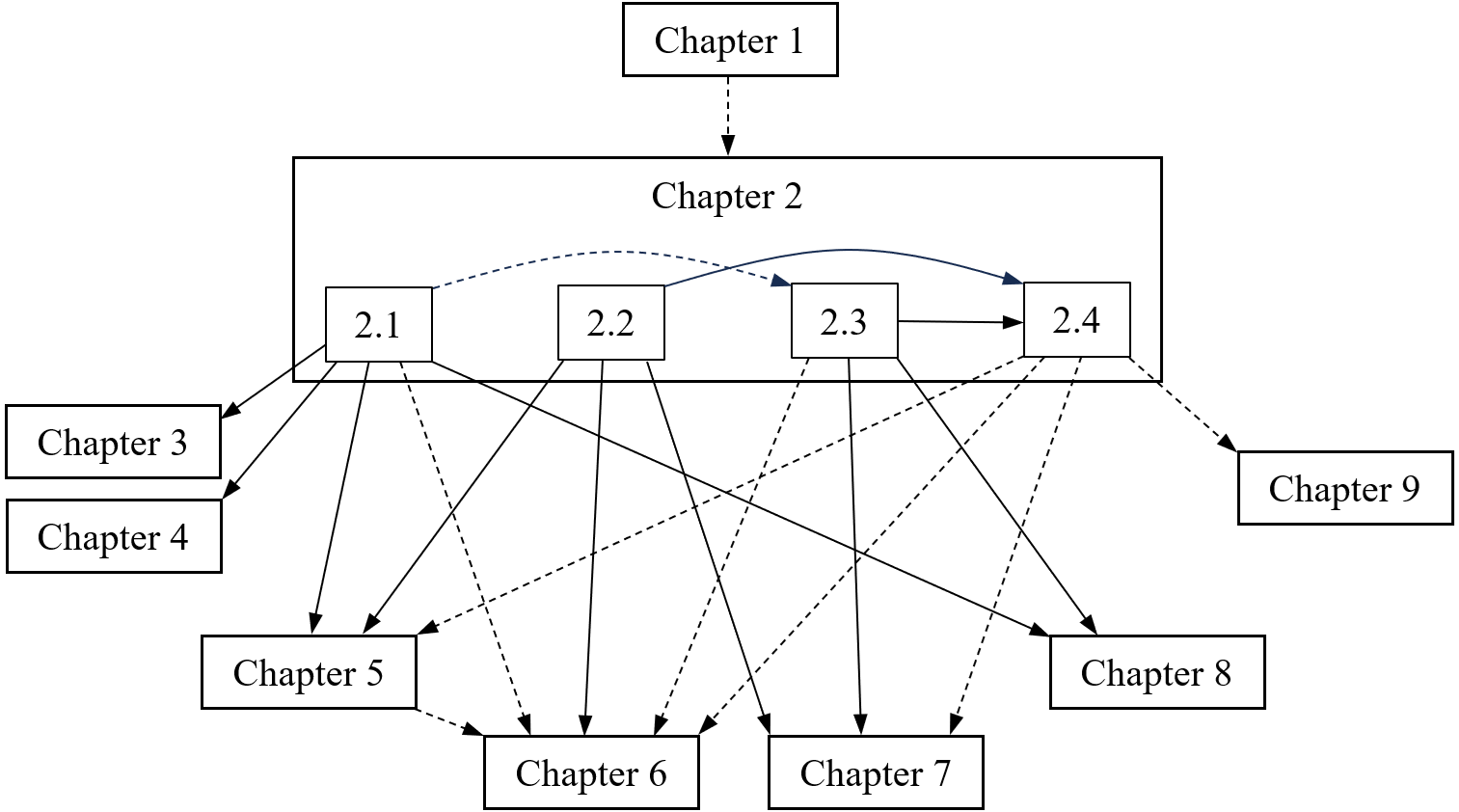}
    \caption[Thesis dependency structure]{The thesis dependency structure. Solid arrows represent required background; dashed arrows represent recommended background.}
    \label{fig:dependencies}
\end{figure}

This thesis showcases my work on these problems. It starts with an introduction (Chapter~\ref{chap:intro}), followed by technical preliminaries (Chapter~\ref{chap:prelims}). Chapters~\ref{chap:precision} through \ref{chap:elk} describe some of my technical contributions. See Figure~\ref{fig:dependencies} for the dependence structure of the thesis: which chapters and sections are necessary for which others.

While not strictly necessary for the technical content, I recommend at least skimming the introduction, where I try to convey what exactly I mean by ``algorithmic Bayesian epistemology'' and why I'm excited about it. I also recommend reading Chapter~\ref{chap:prelims} (Preliminaries), which is intended to be accessible to readers with a general college-level mathematical background.\footnote{Chapter~\ref{chap:prelims} relies most heavily on basic probability theory.} While the nominal purpose of Chapter~\ref{chap:prelims} is to introduce the mathematical tools that we use in later chapters, the topics covered there are interesting in their own right.

Different readers will of course have different opinions about which technical chapters are the most interesting. Each technical chapter begins with a short summary of the content, which may be useful for figuring out whether you want to read it. Naturally, I have my own opinions: I think the most interesting chapters are \ref{chap:qa}, \ref{chap:robust}, and \ref{chap:elk}, so if you are looking for direction, you may want to tiebreak toward reading those.

\vspace{1.5cm}

\titleformat{\chapter}[display]
{\normalfont\bfseries\filcenter}{}{0pt}{\Large\chaptertitlename\ \Large\thechapter : \Large\bfseries\filcenter{#1}}  
\titlespacing*{\chapter}
  {0pt}{0pt}{30pt}	
  



\chapter{Introduction} \label{chap:intro}

The title and subject of this thesis is \emph{algorithmic Bayesian epistemology.} This is an original term, and so in this introduction I will define, explain, and motivate it.

I also hope to convey my excitement about this topic. Problems in algorithmic Bayesian epistemology are theoretically fascinating and practically important. And yet, in many cases, they are surprisingly neglected! This makes the area ripe for exploration. In this introduction and the thesis in general, I will place significant emphasis on exploration: pointing out under-explored areas and asking questions about them. And so, after defining and motivating algorithmic Bayesian epistemology, I will give a whirlwind tour of existing work in the area, with an emphasis on the gaps in our understanding. In the later chapters, I will fill just a few of those many gaps. My primary hope for this thesis is that it will spur new research into directions that I leave uncovered: topics that I acknowledge but don't explore, as well as questions that I didn't even think to ask.\\

But before all that: what is algorithmic Bayesian epistemology?

\section{What is algorithmic Bayesian epistemology?}
Epistemology is the study of knowledge and uncertainty. Bayesian epistemology -- named for Thomas Bayes -- is a particular framework for studying knowledge that aims to understand uncertainty using the tools of probability. In Bayesian epistemology, an observer assigns probabilities to uncertain events (typically called \emph{prior} probabilities) and updates those probabilities in light of new evidence (to get \emph{posterior} probabilities). For example, suppose that the incidence of some disease in the general population is 0.1\%. If I have no further information about whether I have the disease,\footnote{Perhaps the disease is latent until old age, so my lack of symptoms does not constitute evidence against having the disease.} then in the framework of Bayesian epistemology I might say that there's a 0.1\% chance that I have the disease. Now, suppose that my doctor tests me for the disease, and that the test is ten times more likely to come up positive for people who have the disease than for people who don't. If my test comes up positive, I will update my probability to (approximately) 1\% in light of the new evidence.

Unfortunately, forming beliefs in the framework of Bayesian epistemology is often much more complex. Perhaps there is not one test for the disease, but five different tests. Suppose that I take all of the tests, some of which come up positive and others negative. Updating my probability in light of all those tests is simple enough if the tests constitute \emph{independent} evidence -- but they might not. Maybe Test B is more likely to be a false positive if Test A is a false positive, so treating them as independent would cause me to over-update my probability. Maybe Tests B and C test for different sub-types of the disease, such that they complement each other. If I knew all of the statistical relationships between the tests, then I could correctly infer the probability that I have the disease. But in practice, I might not have access to that data.

For another example, consider the task of predicting the weather. State-of-the-art weather forecasting is based on numerical weather prediction (NWP), a method that takes as input observational data about the current state of the atmosphere and solves differential equations in order to simulate future atmospheric conditions. The input data comes from millions of observations made by weather balloons, Doppler radars, satellites, buoys, aircraft, and human volunteers around the world. The forecasting models that implement NWP are incredibly complex, running on some of the world's fastest supercomputers \parencite{noaa17}. Technical breakthroughs have led to a substantial improvement in NWP models over the last three decades \parencite{btb15}.

Despite all that, NWP models are far from perfect: indeed, NWP forecasts beyond ten days perform worse than simply relying on historical averages \parencite{sbg18}. The accuracy of NWP forecasts is limited by informational and computational constraints. First, NWP models rely on observational data that is not very granular: for example, there are only about 1000 automated surface observing systems (ASOS) and fewer than 200 Doppler radar towers in the United States \parencite{noaa17}. Second, even that data is not perfectly precise, as equipment may be miscalibrated. Third, even if perfectly accurate, extremely granular data were available, NWP would be severely limited by computational constraints.

In recent years, weather prediction methods based on machine learning have shown promise as an alternative to NWP methods. These methods involve training neural networks to achieve high predictive performance on past data. A recently-introduced machine learning-based method called GraphCast was shown to improve upon frontier NWP models at low resolution. GraphCast was also able to forecast severe weather events such as atmospheric rivers and tracks of tropical cyclones more accurately \parencite{lam23}. While still outperformed by NWP models on high-resolution forecasts, machine learning-based models are significantly less computationally intensive and are likely to rapidly improve in the coming years.

Even if machine learning-based approaches come to mitigate the issue of computational constraints, weather forecasting will still be limited by informational constraints (imperfect and coarse-grained observational data). Furthermore, these approaches face new statistical challenges, such as overfitting, because they are based on learning from historical data rather than simulating physical laws. Finally, both NWP and machine learning-based approaches face questions of aggregation. Given several disagreeing weather forecasts, how does it make sense to combine them into a single forecast?\\

In light of all of these challenges, it makes sense to ask: \emph{how can we reason about uncertainty, in light of computational, informational, and other challenges that would-be perfect Bayesian reasoners face in practice?} This is the domain of \emph{algorithmic} Bayesian epistemology.

The relationship between Bayesian epistemology and \emph{algorithmic} Bayesian epistemology is the same as the relationship between game theory and algorithmic game theory, and as the relationship between mechanism design and algorithmic mechanism design.

Mechanism design -- traditionally a sub-discipline of economics -- asks the question: how can we design systems containing strategic agents pursuing their own incentives, in a way that produces good outcomes? For example, how can we auction off multiple items to multiple bidders in a way that produces the optimal social welfare for the bidders? The traditional answer from economic theory is the Vickrey-Clarke-Groves (VCG) auction, which elicits bids, computes the optimal allocation of items, and charges each bidder based on their externality on the remaining bidders.

Computer scientists find this answer dissatisfying, for a simple reason: computing the optimal allocation is not feasible, from the standpoint of both communication and computation. First, the bidders' preferences may not be compactly representable, in which case it is infeasible to communicate them to the auctioneer. Second, even if the bidders' preferences \emph{are} compactly representable, actually computing the optimal allocation may still be intractable.\footnote{See e.g.\ the case of single-minded bidders \parencite[\S11.2]{nrtv07}.} And so \emph{algorithmic} mechanism design asks the question: how can we design a \emph{computationally and communicationally tractable} auction mechanism that attains a large fraction of the optimal social welfare \parencite[\S11]{nrtv07}?\\

Algorithmic mechanism design belongs to a longstanding tradition in theoretical computer science: considering problems from other disciplines through an \emph{algorithmic lens.}\footnote{The term \emph{algorithmic lens} was coined at Berkeley by members of the Theory of Computing research group around the year 2000 \emph{(private communication with Christos Papadimitriou).}} That is, instead of asking for the optimal solution to a problem, computer scientists ask: what is the best solution that \emph{can actually be implemented,} given real-world (or real-world-inspired) constraints?

Sometimes, these constraints are computational: what is the best solution that can be found in polynomial time? Other times, the constraints are communciational: what is the best solution if parties are limited in how much they can communicate? Other kinds of constraints are also common. For example:
\begin{itemize}
    \item Constraints on \emph{information.} For example, the subfield of \emph{online algorithms} studies sequential decision making under uncertainty (incomplete information). Often, the goal of an online algorithm is to guarantee a result that is almost as good as the best possible result in hindsight, e.g.\ the prophet inequality from optimal stopping theory \parencite{sc84, ks87}, or no-regret algorithms in online learning \parencite{orabona23}.
    \item Constraints imposed by the \emph{strategic behavior} of agents in a system. For example, many computer scientists study the \emph{price of anarchy}: how much the welfare of a system degrades because of self-interested actors, as compared to a welfare-optimizing central planner \parencite{kp99, rt00}.
\end{itemize}

The study of real-world problems through the algorithmic lens has significantly impacted a variety of disciplines, including molecular biology, ecology, neuroscience, quantum physics, and various social sciences -- see \textcite[\S20]{wig19} for a detailed discussion.

And so, algorithmic Bayesian epistemology is simply the application of the algorithmic lens to the discipline of Bayesian epistemology. It is perhaps best to define algorithmic Bayesian epistemology by its examples, but to attempt a loose description:

\begin{quote}
    \emph{A question belongs to the field of \emph{algorithmic Bayesian epistemology} (henceforth \emph{ABE}) if it involves reasoning about uncertainty from a Bayesian perspective, but under constraints that prevent complete assimilation of all existing information.}
\end{quote}

\section{Prior work in algorithmic Bayesian epistemology}
To further motivate ABE, this section will give a whirlwind tour of prior work in ABE, with a focus on particularly neglected directions and gaps in our understanding. Since ABE involves reasoning about uncertainty under constraints, this section will be sub-categorized by type of constraint. In particular, we will highlight:
\begin{itemize}
    \item Bayesian epistemology (BE) under computational constraints. This sub-topic is extremely important, but also relatively well-studied. Most of our thesis focuses on more neglected parts of ABE, so our summary of prior work on BE under computational constraints will be brief. Computational constraints will be most relevant for Chapter~\ref{chap:elk}.
    \item BE under informational constraints. This sub-topic is important, relatively neglected, and (in my opinion) extremely interesting. Chapters~\ref{chap:learning}, \ref{chap:robust}, and \ref{chap:elk} will all focus on BE under informational constraints.
    \item BE under communication constraints. This sub-topic is the most neglected of the four, despite having many interesting questions. It will be the focus of Chapter~\ref{chap:agreement}.
    \item BE under constraints imposed by the strategic behavior of experts. This sub-topic is reasonably well-studied, though -- I would argue -- not sufficiently well-studied relative to its importance. Chapters~\ref{chap:precision}, \ref{chap:arbitrage}, and \ref{chap:qa} will all focus on this topic.
    \item BE under a combination of the types of constraints listed above.
\end{itemize}

\subsection{Computational constraints}
\paragraph{Approximating Bayesian inference} Exact Bayesian inference -- that is, incorporating evidence in order to exactly compute a posterior probability -- is computationally intractable because it requires taking a sum or integral over a high-dimensional space (one dimension per category of evidence). As a result, there has been extensive research into developing methods for computationally efficient approximate Bayesian inference. Classical results of the field include Markov Chain Monte Carlo (MCMC) methods, such as the Metropolis-Hastings algorithm, approximate Bayesian computation (ABC) methods, and hierarchical Bayesian methods. Such methods are used extensively in many disciplines, such as physics, biology, machine learning, and public health. There is a vast literature on approximate Bayesian inference -- see \textcite{gelman13} for a thorough exposition -- but it is not a focus of this thesis.

\paragraph{Bounded rationality} While theoretical work in economics generally assumes agents to be rational -- that is, to choose the best action in their situation given their knowledge -- behaving fully rationally may involve solving computationally intractable problems. For example, \textcite{cdt09} showed that computing a Nash equilibrium in a game is PPAD-hard. Additionally, empirical studies show that in a variety of settings, people do not behave rationally \parencite{conlisk96}.

This has led to the study of \emph{bounded rationality}: formal models in which agents do not necessarily take the optimal action. There are many models of bounded rationality, including satisficing agents and evolutionary economics; see \textcite{conlisk96} for an overview. Perhaps the one that fits best within the purview of ABE is a model of \emph{level-$k$ rationality} formulated by \textcite{nagel95}. In this model, a level $0$ agent takes a random action; a level $1$ agent best responds under the assumption that all other agents are level $0$ agents; a level $2$ agent best responds under the assumption that all other agents are level $1$ agents; and so on. In a sense, each agent is computing their optimal strategy, but stopping after $k$ steps.\footnote{Note, however, that the limit as $k$ approaches infinity of the strategy of a level $k$ agent is not necessarily well-defined; the strategy may end up cycling among a list of options.} \textcite{chc04} extended this model to allow agents to best-respond under a hypothesized distribution of the levels of other agents, instead of assuming that all agents are one level lower.

\subsection{Informational constraints} \label{sec:informational_constraints}
\paragraph{Forecast aggregation under incomplete information} Suppose that an aggregator receives precipitation forecasts from several different experts. How should the aggregator combine those forecasts into a single number? If the aggregator has a perfect understanding of the likelihood of various possible states of the world and what information each expert has in every world state, then the aggregator can deduce the correct aggregate forecast (given unlimited computation). In practice, however, the aggregator does not have this knowledge and must aggregate the forecasts under incomplete information. Forecast aggregation under incomplete information will be the focus of Chapters~\ref{chap:learning} and \ref{chap:robust}.

One class of approaches to this problem is called \emph{Bayesian} forecast aggregation: the aggregator makes some modeling assumptions about the experts' information and deduces the correct posterior within that model. If the aggregator is tasked with combining probability distributions over a set of outcomes, then perhaps the simplest model would assume that the experts' forecasts are independent conditional on the outcome. In this model, applying Bayes' rule straightforwardly gives an answer. If instead experts are asked to forecast the expectation of a real number $Y$ (such as the amount of rainfall), then one could choose to model each expert as receiving a noisy estimate of $Y$ from some distribution; see \textcite{winkler81} for an early work with this flavor. More recently, \textcite{fck15} considered a setting in which a principal wishes to learn the distribution of a random variable (which is modeled as belonging to a particular parameterized family of distributions) based on information provided by experts, each of whom sees some number of samples from the distribution. \textcite{lgjw22} consider aggregation in a model in which experts share some information and additionally receive private i.i.d.\ samples from an exponential family. See \textcite{licht11} for a introduction to Bayesian forecast aggregation.

Another class of approaches is called \emph{axiomatic} forecast aggregation: the goal of such approaches is to describe a set of axioms that an aggregation method ought to satisfy, and then to characterize methods that satisfy those axioms. \textcite{dl16} describe several such axioms, including unanimity preservation, eventwise independence, and external Bayesianality. No pooling method satisfies all three of these axioms: only linear pooling (i.e.\ a weighted average of the experts' forecasts) satisfies the first two axioms \parencite{aw80}, but it does not satisfy the third axiom. On the other hand, \textcite{gen84} showed that another natural aggregation method called logarithmic pooling satisfies the third axiom. In addition to satisfying natural axioms, linear and logarithmic pooling are the two most well-studied forecast aggregation methods, and they will come up many times in this thesis.

A third class of approaches -- the one most relevant to this thesis -- is called \emph{robust} forecast aggregation. Because the aggregator lacks complete information about the information structure describing the experts' knowledge, it makes sense to seek an aggregation method that performs well in the worst case over a wide class of information structures (see Section~\ref{sec:prelim_info_struct} for an introduction to information structures). For example, \textcite{abs18} find the aggregation method that performs best for the class of information structures of two Blackwell-ordered experts and (separately) for the class of information structures with two conditionally independent experts. \textcite{lr20} take a different approach, analyzing robust aggregation methods under bounds on the correlation between experts' signals. \textcite{dil21} take a similar robust approach to decision problems more generally. Overall, there has been relatively little prior work in this direction, a gap that this thesis aims to fill.

\paragraph{Online learning from expert advice} In \emph{online learning}, a decision-maker must choose from a fixed range of options for each of $T$ time steps. On each time step, after choosing, the decision-maker receives a reward that depends on their chosen option, and also learns the reward they would have received under every other choice they could have made. The decision-maker's goal is to get a total reward that is almost as high as if they had chosen the fixed option that had the highest total reward over the $T$ time steps. This is an example of an algorithmic problem under informational constraints: the constraint here is that the decision-maker knows nothing about the quality of each option beforehand, yet must do nearly as well as if they had known the best fixed option (on average over the time steps). Online learning is a well-studied field; see \textcite{orabona23} for an overview.

We are interested in the subproblem of \emph{online prediction from expert advice}. At each time step, $m$ experts forecast probability distributions over $n$ possible outcomes. The decision-maker then chooses their own probability distribution over the outcomes. Afterward, an outcome is realized and the decision-maker is given a reward based on their distribution and the outcome (generally using a proper scoring rule -- see Section~\ref{sec:strategic_constraints}). The decision-maker aims to compete with the best expert in hindsight, or perhaps even with the best possible mixture (e.g.\ weighted average) of experts in hindsight. See \textcite{cl06} for a survey of this topic. We will explore online prediction from expert advice in Chapter~\ref{chap:learning}.

\paragraph{Estimation theory} In statistics, estimation theory is the study of estimating the parameters of a distribution given samples from the distribution. A simple example is estimating the mean of a normal distribution with variance $1$ based on samples. Given a prior over the true mean, one can apply Bayes' rule to exactly compute the expected value of the mean, using the sample mean as evidence. However, estimation theory typically concerns itself with estimating parameters in settings in which the prior is not known. One common approach is the \emph{minimax estimator}, which is the estimator with the best performance (e.g.\ as measured by expected squared error) in the \emph{worst case} over possible priors (or equivalently, in the worst case over parameters of the distribution). An alternative framing on minimax estimation is one of adversarial robustness: the minimax estimator is the optimal estimator if the prior over parameters is chosen by an adversary. Minimax estimators are thus a classic example of Bayesian epistemology under informational constraints: while not Bayesian in the traditional sense (they do not assume a prior), they can be thought of as the correct Bayesian estimate in the context of an adversarially chosen prior.

The minimax estimator is one example of a natural approach to estimation when lacking a prior over distribution parameters. Other approaches include unbiased minimum-variance estimators and maximum likelihood estimators. See \textcite{kay93} for a thorough treatment of the topic.

\paragraph{Reasoning about Uncertainty} \emph{Reasoning about Uncertainty} \parencite{halpern03} is a book on uncertainty in the context of different notions of probability that depart from the standard one. For example, how might one reason about uncertainty if one is only able to assign probabilities to some subsets of possible events but not other subsets? How should we deal with probabilities that are defined under axioms that are weaker than the standard ones? These perspectives are more distantly related to the topics of this thesis than most topics covered in this introduction.

\subsection{Communication constraints} \label{sec:comm_constraints}
\paragraph{Agreement protocols} There has been little work on Bayesian epistemology under communication constraints, but one stand-out example is \textcite{aar05}. \textcite{aumann76} famously showed that two individuals with a common prior (but possibly different private information) cannot ``agree to disagree''. To be more precise, suppose that there is ``common knowledge'' that Alice's estimate of the chance of rain is $p$, meaning that Alice's estimate is $p$, Bob's estimate of Alice's estimate is $p$, Alice's estimate of Bob's estimate of Alice's estimate is $p$, and so on. And suppose that there is common knowledge that Bob's estimate for the chance of rain is $q$. Then it must be that $p = q$. This result suggests that if Alice and Bob disagree, then they ought to be able to exchange information to reach agreement.

In this model, Alice and Bob can reach agreement by exchanging \emph{all} of their information; however, this might require a prohibitively large amount of communication. \textcite{aar05} showed that simply by repeatedly exchanging their estimates (Alice shares her estimate; Bob shares his estimate after updating on Alice's estimate; and so on), Alice and Bob will quickly reach near-agreement. This leaves many questions open: for example, under what circumstances can Alice and Bob reach agreement in a \emph{computationally} efficient way? Or, under what circumstances is the agreed-upon estimate approximately correct? We address this second question in Chapter~\ref{chap:agreement}.

\subsection{Strategic constraints} \label{sec:strategic_constraints}
\paragraph{Information elicitation} So far we have focused on how an individual with access to information can form beliefs. However, in some settings, the individual must first \emph{learn} the information, e.g.\ by eliciting it from experts. If the experts are self-interested, this can raise strategic questions. The most basic question is one of \emph{truthful elicitation}: how can you pay an expert for information in a way that incentivizes the expert to tell the truth?

For example, suppose that you wish to elicit the probability of rain from a meteorologist. To incentivize the meteorologist, you decide to pay them as a function of the forecast they give you and whether or not it ends up raining. Such a function is called a \emph{scoring rule}. Scoring rules must be chosen carefully. For instance, here is one natural (but poor) choice of scoring rule: if the meteorologist reports a probability $p$ of rain, you will pay them $p$ if it rains and $1 - p$ if it doesn't. (That is, you reward the meteorologist proportionally to the probability that they assigned to the eventual outcome.) This scoring rule incentivizes the meteorologist to report a probability of 100\% if their true belief is anywhere greater than 50\% and to report 0\% if their true belief is anywhere less than 50\%.

A scoring rule is called \emph{proper} if the optimal strategy of an expert who wishes to maximize their expected score is to report their true belief. One commonly used proper scoring rules is the \emph{quadratic score}, which penalizes an expert according to the squared difference between their report and the correct answer (either $0$ or $1$ depending on the outcome). Another is the \emph{logarithmic score}, which is rewards the expert with the logarithm of the probability that the expert assigns to the eventual outcome. We provide an exposition to proper scoring rules in Section~\ref{sec:prelim_proper}; see \textcite{gr07} for a more thorough survey.

Similarly, one can aim to truthfully elicit properties of probability distributions. That is, suppose that you wish to know how much it will rain tomorrow. The meteorologist has a probability distribution over tomorrow's rainfall, and you wish to know the mean of the distribution. If you ask the meteorologist for this mean, how do you pay them (as a function of their report and the eventual rainfall amount -- i.e.\ a draw from the distribution) to truthfully elicit their belief? What if you want to know the median of the distribution, rather than the mean? What about the second moment? The variance? It turns out that the variance is not straightforwardly elicitable -- no reward function will incentivize a truthful report -- though you can compute the variance by separately eliciting the mean and the second moment (both of which \emph{are} straightforwardly elicitable). The foundations of property elicitation were laid by \textcite{savage71, osband85}. See \textcite{lps08, gneiting11, af12, fk20} for more recent work in this area. Overall, most research on property elicitation is fairly recent and a lot of low-hanging fruit remains.

Recently, there has been interest in \emph{contract functions,} which are scoring rules for multiple experts. \textcite{cs11} showed that simply using a proper scoring rule for each expert invites collusion between experts. They asked whether contract functions that do not allow for collusion are possible. We resolve this question in Chapter~\ref{chap:arbitrage} by exhibiting such a contract function. However, interesting questions remain: there are natural definitions of collusion that are broader than the one given by \textcite{cs11}, and we leave open the question of whether any contract function disallows collusion under these broader definitions.

There has also been recent work on \emph{wagering mechanisms,} introduced by \textcite{llwcrsp08}: contract functions that additionally elicit a wager from each expert and redistribute the wagers according to expert performance. The authors defined several desiderata for wagering mechanisms and identified the unique wagering mechanism that satisfied those desiderata. Subsequently, \textcite{cdpv14} explored wagering mechanisms that do not allow collusion (by analogy to the same question for contract functions). More recently, \textcite{fpv17} noted that wagering mechanisms can be thought of as experts trading securities; recent work on wagering mechanisms has often taken advantage of this framing. While this area is not a focus of the thesis, I see wagering mechanisms as a particularly fruitful direction for future work. We further discuss wagering mechanisms in the epilogue.

\paragraph{Prediction markets} Prediction markets are a solution concept to the dual problems of forecast elicitation and forecast aggregation. In a prediction market, experts express their beliefs about the probability of some event (or the value of some unknown quantity) by making profit-maximizing trades. When an expert has private information, they can trade in the market, thus integrating their information into the consensus view. The most well-studied prediction markets in the computer science literature (but not the most common markets in practice) are \emph{market scoring rules} (MSRs), introduced by \textcite{hanson03}. An MSR is based on a proper scoring rule: the experts report their beliefs in sequence, and -- once it is known whether the event happened -- are their score minus the previous expert's score. \textcite{cp07} showed that MSRs can be thought of in terms of experts trading Arrow-Debreu securities (contracts that are worth $1$ if the event happens and $0$ otherwise). See \textcite{acv13} for follow-up work in this vein.

One downside of MSRs, which perhaps accounts for the lack of use of MSRs in practice, is that market liquidity does not increase with total trading volume. \textcite{aflv14} define \emph{volume-parameterized markets,} an MSR-inspired mechanism that aims to fix this issue.

Much work on MSRs and related mechanisms assumes that experts are myopic: that they maximize their expected reward from their current report or trade, without regard to potential future trades. Several recent papers explore conditions under which MSRs incentivize experts to reveal information even when they are not myopic \parencite{cdsrphfg10, cw16, ks18, acwx19}.

An additional line of work explores prediction markets in the context of experts with mutable beliefs, i.e.\ beliefs that may change in light of information revealed by the market \parencite{ost12, crv12}. This is, of course, how experts' beliefs behave in practice. Exploring the dynamics of prediction markets in which participants have mutable beliefs seems like a particularly neglected direction relative to its importance.

\paragraph{Information design} So far we have taken the perspective of an individual who wishes to elicit information from strategic agents. We can also take the perspective of a strategic agent who is interested in \emph{giving away} partial information so as to influence decision-makers. This topic is called information design. In a sense, information design is the mechanism design of ABE: while economic theory focuses on the behavior of strategic agents, mechanism design asks how to design a strategic landscape that is favorable to the designer. Similarly, information design asks how a favorable \emph{informational} landscape can be created.

\emph{Bayesian persuasion} is the sub-case of information design that involves a single information sender and a single receiver \parencite{kg11}. The authors an example of Bayesian persuasion in the context of a prosecutor (the sender) and a judge (the receiver). The judge is tasked with determining the innocence or guilt of 100 defendants, and knows that exactly 30 of the defendants are guilty (but doesn't know which ones). The judge's utility function is taken to be $+1$ for every correctly-classified defendant. The prosecutor knows which defendant are guilty, and their utility is $+1$ for every defendant whom the judge convicts. If the prosecutor reveals all of their information, then the judge will convict the 30 guilty defendants. However, suppose that the prosecutor randomly chooses 29 innocent defendants in addition to the 30 guilty ones and tells the judge that 30 of the 59 selected defendants are guilty. Then the judge will convict all 59 of them, which is a better outcome for the prosecutor. See \textcite{dx21} for a survey of Bayesian persuasion.

Information design more generally concerns itself with a sender giving partial information to multiple receivers, where the sender's utility function depends on the receivers' actions. Additionally, the sender may not know the receivers' utility functions (although the sender has a probability distribution over the utility functions). See \textcite{bm19} for a survey of recent work in information design.

\subsection{Multiple kinds of constraints}
\paragraph{Robust mechanism design} While much of auction theory focuses on buyers with independent valuations, in practice a buyer's valuation may be informed by other buyers' valuations. For example, consider an auction for drilling rights in an oil field, where different companies (buyers) each have private information about the amount of oil in the field. Or consider a used car auction in which different participants have different impressions of the reliability of each car. The study of \emph{interdependent value} auctions goes back to \textcite{wilson69}, who studied common value auctions: auctions in which the item's true value is the same for every participant (but is not known for sure by the participants). \textcite{myerson81} studied a ``weighted sum'' setting, in which each buyer's value is equal to their private signal plus some constant $\beta \in [0, 1]$ times the sum of all other buyers' signals.

A seller may wish to create an auction that maximizes the buyers' welfare or the seller's revenue. The optimal auction depends heavily on the (potentially extremely complex) joint probability distribution over all buyers' signals and values. Studying auctions in such a complex setting has often led to theoretical results that work poorly in practice or are too complex to implement \parencite{wilson87}. This has led to work in \emph{robust} mechanism design, which seeks to relax the assumption that the seller knows the probability distribution over signals and values, instead aiming for mechanisms that have strong guarantees for both incentive compatibility and social welfare (or revenue) under weaker assumptions \parencite{bm13b}. See \textcite{rt19} for a survey of this area. Such work is an example of Bayesian epistemology under simultaneous computational, informational, and strategic constraints.

\paragraph{Incentivizing prediction without ground truth access} Our discussion of information elicitation in Section~\ref{sec:strategic_constraints} (e.g.\ asking a meteorologist how much it will rain tomorrow) crucially relied on access to the eventual outcome (knowing how much it ended up raining). However, sometimes we wish to elicit forecasts for far-future events (``How much will the Earth's average temperature increase over the 21st century?''). \textcite{prelec04} introduced a mechanism called \emph{Bayesian truth serum} for eliciting forecasts in the absence of a knowable ground truth. The basic idea is to elicit from each expert both a forecast and a prediction of other experts' forecasts, and to reward each forecaster both for accurately predicting others' forecasts and for having a ``surprisingly popular'' forecast: a forecast whose frequency among the population of experts was underestimated by other experts. The key idea is that a Bayesian expert should predict that others will underestimate the frequency of that expert's view. \textcite{prelec04} showed that truth-telling is a Nash equilibrium in the Bayesian truth serum mechanism.

The \emph{peer prediction} mechanism \parencite{mrz05} is designed for a similar setting, in which e.g.\ an instructor does not have time to grade all students' homework, and so assigns students to grade each other's homework. Truth-telling is a Nash equilibrium of peer prediction, though under somewhat stricter modeling assumptions. Both of these mechanisms are designed for the elicitation of information under both strategic constraints and computational/informational constraints.

\section{Key takeaways}
Now that I have introduced the concept of algorithmic Bayesian epistemology and given a variety of examples, I will conclude with some overall takeaways.

\paragraph{ABE is important} The ability to form accurate beliefs is self-evidently important. Often, there are obstacles to doing so: maybe you don't have the computational resources to do so perfectly. Or maybe your data is biased but you don't know exactly how. Or maybe you're trying to aggregate estimates from different sources, but you don't know whether your sources' estimates are based on disjoint or overlapping information. Or maybe your information comes from market prices determined by strategic agents. Or maybe you want to incorporate the beliefs of someone who has approached the same question from an unfamiliar perspective. Coming up with solutions in the face of these challenges is the domain of ABE.

\paragraph{ABE is often neglected} Consider the problem of how best to aggregate different forecasts or estimates for a quantity. This question is ubiquitous: it comes up in essentially every branch of science. Yet until quite recently, there was very little empirical work on this question and even less theoretical work. Or consider the problem of forming beliefs under communication constraints: a well-motivated question with very little theoretical work. Proper scoring rules have been relatively well-studied, yet the natural question of \emph{which} proper scoring rule should be used in a given situation has not received much attention. Of the large number of important and well-motivated questions in ABE, relatively few have received a lot of attention. As a consequence, work in ABE is often relatively tractable: much of the low-hanging fruit is left to be picked.

\paragraph{ABE is really interesting} My opinion on this is, of course, subjective. But the question of how to form beliefs is fundamental, and the question of how to form beliefs under constraints seems like a fundamental sub-problem. Also, the mathematical notions that arise from the formal study of this problem are -- at least in my opinion -- very elegant.\\

In the coming chapters, we will look at just a few of the many interesting and well-motivated questions in ABE. Let's begin!
\chapter{Preliminaries} \label{chap:prelims}
While each chapter of this thesis will have its own introduction, some preliminary concepts in ABE will find use throughout the thesis. This chapter introduces those concepts.

\section{Proper scoring rules and the Bregman divergence} \label{sec:prelim_proper}
\emph{This section is necessary for Chapters~\ref{chap:precision}, \ref{chap:arbitrage}, \ref{chap:qa}, and \ref{chap:agreement}, and is also somewhat useful for Chapter~\ref{chap:learning}.}

\subsection{Introduction to proper scoring rules}
Suppose that I want to know whether it will be sunny, cloudy, or rainy tomorrow. I don't know much but weather forecasting, but my friend Skylar does. It's natural for me to ask her for a forecast probability distribution over the options \{sunny, cloudy, rainy\}. If I want to give Skylar an incentive to give me a well-reasoned probability distribution that reflects her true beliefs, then I might promise to pay her according to the accuracy of her forecast. Put another way, I could ask for her forecast, wait until tomorrow, and pay her according to ``how right she was'' -- that is, according to how close her forecast was to the ideal forecast (putting 100\% on whichever outcome actually ends up happening).

Any payment scheme that's a function of Skylar's forecast and the eventual outcome is called a \emph{scoring rule.} For example, one scoring rule gives Skylar a payment (score) equal to the probability that she assigned to the eventual outcome. For example, if Skylar's forecast is (70\% sunny, 20\% cloudy, 10\% rainy), and it ends up being sunny, she would receive a score of 0.7.

Does this scoring rule ``make sense''? Imagine that Skylar's actual belief is (70\% sunny, 20\% cloudy, 10\% rainy). If Skylar is interested in maximizing the expected value of her score, does it make sense for Skylar to tell the truth?

If Skylar tells the truth, then with probability 70\%, it will be sunny and she will receive a score of 0.7; with probability 20\%, it will be cloudy and she will receive a score of 0.2; and with probability 10\%, it will be rainy and she will receive a score of 0.1. Thus, the expected value of her score is
\[70\%(0.7) + 20\%(0.2) + 10\%(0.1) = 0.49 + 0.04 + 0.01 = 0.54.\]
Now, by contrast, suppose that Skylar lies and reports that there's a 100\% chance that it will be sunny tomorrow. If she does this, then her score will be $1$ if it's sunny and $0$ if it's not. Overall, this makes Skylar come out ahead: her expected score is now $70\%(1) = 0.7 > 0.54$. Thus, this scoring rule \emph{gives Skylar an incentive to lie,} which is clearly an undesirable property.

Are there any scoring rules that incentivize Skylar to report her true belief? As we will see, the answer is yes. Any such scoring rule is called a \emph{proper} scoring rule. To talk precisely about proper scoring rules, let's introduce some notation. We will be using this notation through the thesis.

\begin{itemize}
    \item We will use the term \emph{principal} to refer to the person eliciting (asking for) the forecast, and we use the term \emph{expert} to refer to the person reporting their forecast. In the above example, I am the principal and Skylar is the expert.
    \item We will let $n$ be the number of possible outcomes (above, $n = 3$). We will number the outcomes $1$ through $n$ and use the letter $j$ to denote any particular outcome. Thus, $j \in \{1, \dots, n\}$. (In the future, we will write $[n]$ in place of $\{1, \dots, n\}$ for brevity.)
    \item As is standard notation, we will let $\Delta_n$ be the space of all probability distributions over $n$ outcomes. We can think of a probability distribution over $n$ outcomes as a vector of $n$ probabilities: non-negative numbers that add to $1$. In other words, $\Delta_n = \{(x_1, \dots, x_n) \in \RR^n: x_j \ge 0 \enskip \forall j, \sum_j x_j = 1\}$.
    \item Typically, we will use $\vect{x} = (x_1, \dots, x_n)$ to denote the expert's \emph{report} (i.e.\ the probability distribution that the expert reports to the principal) and will use $\vect{p} = (p_1, \dots, p_n)$ to denote the expert's true belief. Both $\vect{x}$ and $\vect{p}$ belong to $\Delta$.
    \item We will use the notation $s(\vect{x}; j)$ for the expert's score, a function of the expert's report $\vect{x}$ and the eventual outcome $j$. Scores are allowed to be negative (or even negative infinity).
\end{itemize}

What does it mean for a scoring rule $s$ to truthfully elicit an expert's belief? If an expert's true belief is $\vect{p}$ and the expert reports $\vect{x}$, then their expected score is $\EE[j \sim \vect{p}]{s(\vect{x}; j)}$. (Here, $j \sim \vect{p}$ means that $j \in [n]$ is selected at random according to the distribution $\vect{p}$.) The expert is incentivized to report the $\vect{x}$ that maximizes this expected value. Therefore:

\begin{defin}
    A scoring rule $s$ for $n$ outcomes is \emph{proper} if for all $\vect{p} \in \Delta_n$, the expression $\EE[j \sim \vect{p}]{s(\vect{x}; j)}$ has a unique maximum at $\vect{x} = \vect{p}$.
\end{defin}

(Note that many sources would call a scoring rule ``proper'' even if the maximum at $\vect{x} = \vect{p}$ is not unique, and would use the term ``strictly proper'' where we say ``proper.'' Under this usage, the scoring rule $s(\vect{x}; j) = 0$ (which gives the expert reward $0$ no matter what) would be considered proper. Generally, we will only be interested in scoring rules that satisfy our stronger definition. When necessary, we will use the term \emph{weakly proper} for the weaker notion.)

There are many proper scoring rules (we characterize them all below), two of which are very well-known. The most well-known is the \emph{quadratic} scoring rule:

\begin{defin}
    The \emph{quadratic scoring rule}, also known as the \emph{Brier score}, is defined by
    \[s_{\text{quad}}(\vect{x}; j) := -(1 - x_j)^2 - \sum_{j' \neq j} x_{j'}^2.\]
    Put otherwise, if $\pmb{\delta}_j$ is the vector with a $1$ in position $j$ and $0$ elsewhere, then $s_{\text{quad}}(\vect{x}; j)$ is equal to $-\norm{\vect{x} - \pmb{\delta}_j}_2^2$.
\end{defin}

The quadratic scoring rule is best thought of as a penalty on the expert equal to the squared distance between their forecast ($\vect{x}$) and the ``perfect'' forecast $\pmb{\delta}_j$. This scoring rule is proper. (Note that the quadratic score can be made nonnegative by adding $2$ to it, if it's desirable for the expert's score to be guaranteed to be nonnegative.)

The second most well-known proper scoring rule is the \emph{logarithmic} scoring rule:

\begin{defin}
    The \emph{logarithmic scoring rule} (often, colloquially, the ``log score'') is defined by
    \[s_{\text{log}}(\vect{x}; j) := \ln(x_j).\]
\end{defin}

Instead of rewarding the expert with the probability assigned to the eventual outcome (which we saw earlier to be improper), the log score rewards the expert with the log of that probability. Note that, unlike the quadratic score, the log score only depends on the probability that the expert assigns to the eventual outcome. It turns out the the log score is the only proper scoring rule with this property (up to adding and multiplying by constants) \parencite{pdl12}.

Unlike the quadratic score, the log score harshly penalizes an expert for assigning an extremely low probability to the eventual outcome (and gives a score of $-\infty$ to an expert who assigns probability zero to the eventual outcome). It makes sense to use the log scoring rule if the principal cares about differentiating between very low and extremely low probabilities. This is because the log scoring rule incentivizes the expert to think carefully about assigning very low probabilities to possible outcomes, and to make sure that their forecast is not overconfident.

\begin{remark} \label{remark:pos_aff}
    Any positive affine transformation of a proper scoring rule is proper. That is, for any $a > 0$ and $b \in \RR$, if $s$ is proper then so is $as + b$.
\end{remark}

\subsection{The Savage representation}
Perhaps the best way to think of proper scoring rules is in terms of their \emph{Savage representation,} named after mathematician Leonard Savage. To introduce the Savage representation, we will first define the \emph{expected score function} of a proper scoring rule:
\begin{defin}
    The \emph{expected score function} $G_s: \Delta_n \to \RR$ of a proper scoring rule $s$ over $n$ outcomes is defined by
    \[G_s(\vect{x}) := \sum_{j \in [n]} x_j s(\vect{x}; j).\]
\end{defin}

$G_s$ is thus the expected score of an honest expert, as a function of their belief. For example:
\begin{itemize}
    \item If $s$ is $s_{\text{quad}}$, then $G_s(\vect{x}) = \sum_j x_j^2 - 1 = \norm{\vect{x}}^2 - 1$. Note that $\norm{\vect{x}}^2$ is large for probability distributions that are more concentrated on particular outcomes, so an expert's expected score is higher if they are more certain about which outcome will happen. This should be intuitive: more informed experts should have higher expected scores.
    \item If $s$ is $s_{\text{log}}$, then $G_s(\vect{x}) = \sum_j x_j \ln(x_j)$. Note that this is exactly the negative of the Shannon entropy of the probability distribution $\vect{x}$. This means that an expert's expected score is higher for lower-entropy distributions, i.e.\ when the expert is more certain about the outcome.
\end{itemize}

Both of these expected score functions are convex. This is not a coincidence:

\begin{theorem}[\cite{savage71}] \label{thm:savage}
    The expected score function of a proper scoring rule is strictly convex. Given a strictly convex function $G: \Delta_n \to \RR$, consider a function $s(\vect{x}; j)$ as follows: at each $\vect{x} \in \Delta_n$, draw a tangent plane to $G$ at $\vect{x}$, and let $s(\vect{x}; j)$ be the value of this tangent plane at $\pmb{\delta}_j$. (Algebraically, $s(\vect{x}; j) = G(\vect{x}) + \angles{\pmb{\delta}_j - \vect{x}, \nabla G(\vect{x})}$, where $\nabla G$ is the gradient\footnote{Or a subgradient, if $G$ is not differentiable.} of $G$ and $\angles{\cdot, \cdot}$ is the dot product.) Then $s$ is a proper scoring rule with expected score function $G$ -- and in fact is the \emph{only}\footnote{Unless $G$ is not differentiable, in which case multiple tangent planes can exist at a point.} proper scoring rule with expected score function $G$.
\end{theorem}

It follows from Theorem~\ref{thm:savage} that for any proper scoring rule $s$, we can rewrite $s$ in terms of its expected score function $G_s$:
\begin{equation} \label{eq:chap2_savage}
    s(\vect{x}; j) = G_s(\vect{x}) + \angles{\pmb{\delta}_j - \vect{x}, \nabla G_s(\vect{x})}.
\end{equation}
This form is known as the \emph{Savage representation} of $s$.

\begin{figure}[ht]
    \centering
    \includegraphics[scale=0.8]{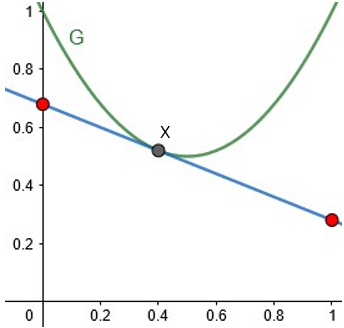}
    \caption[Savage representation of a proper scoring rule]{For the proper scoring rule derived from the function $G$ shown here, suppose that an expert reports a probability $x = 0.4$ on the ``Yes'' outcome. The expert will receive a score equal to the $y$-value of the red point on the right (roughly $0.3$) if the ``Yes'' outcome happens, and a score equal to the $y$-value of the red point on the left (roughly $0.7$) if the ``No'' outcome happens.}
    \label{fig:savage}
\end{figure}

In the case of $n = 2$ -- i.e.\ the setting in which the principal wishes to elicit the probability of a single yes/no outcome -- we can easily visualize the Savage representation. In this setting, we can identify any report with the probability assigned to the ``Yes'' outcome. In Figure~\ref{fig:savage}, $G$ is the expected score function of some proper scoring rule $s$, and the proper scoring rule can be recovered from $G$ by drawing a tangent line at the report $x$ and noticing where it intersects $1$ (that's the score if the ``Yes'' outcome happens) and where it intersects $0$ (that's the score if the ``No'' outcome happens).

It is helpful to have some intuition for Theorem~\ref{thm:savage}. First, why is $G_s$ strictly convex? Suppose for contradiction that there are two forecasts $\vect{p}$ and $\vect{q}$, and a constant $0 < \alpha < 1$, such that $G_s(\alpha \vect{p} + (1 - \alpha) \vect{q}) \ge \alpha G_s(\vect{p}) + (1 - \alpha) G_s(\vect{q})$. Then it would follow that either an expert with belief $\vect{p}$ would be at least as well off reporting $\alpha \vect{p} + (1 - \alpha) \vect{q}$, or that an expert with belief $\vect{q}$ would be at least as well off reporting $\alpha \vect{p} + (1 - \alpha) \vect{q}$. This can be verified algebraically by expanding out the definition of $G_s$.

Second, given a strictly convex $G$, why is the scoring rule $s(\vect{x}; j) = G(\vect{x}) + \angles{\pmb{\delta}_j - \vect{x}, \nabla G(\vect{x})}$ proper? While this fact can be verified algebraically, there is also an intuitive geometric proof. We give this proof below, but to do so we will first introduce concept of a \emph{Bregman divergence.}

\subsection{The Bregman divergence} \label{sec:prelim_bregman}
The \emph{Bregman divergence} is a notion of distance that is defined with respect to a convex function.

\begin{restatable}{defin}{bregmandef}
    For some $n \ge 1$, let $\mathcal{D}$ be a convex subset of $\RR^n$, and let $G: \mathcal{D} \to \RR$ be a differentiable, convex function. Given $\vect{x}, \vect{y} \in \mathcal{D}$, the \emph{Bregman divergence from $\vect{y}$ to $\vect{x}$ with respect to $G$} is defined as
    \begin{equation} \label{eq:chap2_bregman}
        D_G(\vect{y} \parallel \vect{x}) := G(\vect{y}) - G(\vect{x}) - \angles{\vect{y} - \vect{x}, \nabla G(\vect{x})}.
    \end{equation}
\end{restatable}

\begin{figure}[ht]
    \centering
    \includegraphics[scale=0.7]{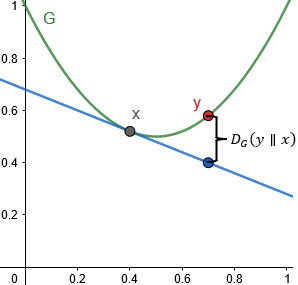}
    \caption[Bregman divergence]{The vertical distance shown is the \emph{Bregman divergence} from $y$ to $x$ with respect to $G$, and is denoted $D_G(y \parallel x)$.}
    \label{fig:bregman}
\end{figure}

While this formal definition is algebraic, the Bregman divergence is fundamentally a geometric object that can be understood without parsing the formal definition. The Bregman divergence from $\vect{y}$ to $\vect{x}$ has the following interpretation. Draw the tangent plane to $G$ at $\vect{x}$; the Bregman divergence is the vertical distance between the plane and the the function $G$ at $\vect{y}$. We show an example (in one dimension) in Figure~\ref{fig:bregman}.

Note that the Bregman divergence is \emph{not} symmetric: $D_G(\vect{y} \parallel \vect{x})$ does not in general equal $D_G(\vect{x} \parallel \vect{y})$. (We discuss the order of arguments more below.) The one exception (up to positive affine transformations) is the function $G(\vect{x}) = \norm{\vect{x}}^2$ (or, if in one dimension, $G(x) = x^2$). In that case, it can be verified that $D_G(\vect{y} \parallel \vect{x}) = \norm{\vect{y} - \vect{x}}^2$, the squared distance between $\vect{x}$ and $\vect{y}$.

The Bregman divergence with respect to $G(\vect{x}) = \norm{\vect{x}}^2$ is thus a very natural notion of distance. A different natural notion of distance is the Bregman divergence with respect to the negative of Shannon entropy (i.e.\ $\sum_j x_j \ln(x_j)$), which came up earlier as the expected score function of the log scoring rule. For this $G$, $D_G(\vect{y} \parallel \vect{x})$ is the KL divergence $D_{\text{KL}}(\vect{y} \parallel \vect{x})$, which is ubiquitous in information theory and statistics.

Now let's return to Theorem~\ref{thm:savage}, which claimed that if a function $G$ is strictly convex, then the scoring rule $s(\vect{x}; j)$ defined as the value of the tangent plane to $G$ at $\vect{x}$, evaluated at $\pmb{\delta}_j$, is proper. Why is this true?

Suppose that an expert's true belief is $\vect{y}$. Then by reporting $\vect{y}$, they will have expected score $G(\vect{y})$. Now, what if, instead, the expert reports some $\vect{x} \neq \vect{y}$? In that case, the expected value of their score is equal to the weighted average of the expert's possible scores depending on the outcome, with weights given by $\vect{y}$. This is just the value of the tangent plane to $G$ at $\vect{x}$, evaluated at $\vect{y}$. (For example, if $\vect{x}$ and $\vect{y}$ are as in Figure~\ref{fig:bregman}, then the expected score of an expert who believes $\vect{y}$ but reports $\vect{x}$ is the height of the point on the blue line below the point labeled $y$.)

Because $G$ is strictly convex, $G(\vect{y})$ is strictly larger than this value. In fact, it is larger by exactly $D_G(\vect{y} \parallel \vect{x})$. Therefore, the expert is best off reporting their true belief $\vect{y}$.

Note that this gives us some interpretations of the Bregman divergence in the context of proper scoring rules. For the points below, let $s$ be a proper scoring rule and $G$ be its expected score function, and consider an expert who is rewarded according to $s$.
\begin{itemize}
    \item The Bregman divergence $D_G(\vect{y} \parallel \vect{x})$ is the expected amount that the expert loses by lying and reporting $\vect{x}$, if their true belief is $\vect{y}$.
    \item Alternatively, if there is a ``true'' probability distribution $\vect{y}$, but the expert mistakenly has belief $\vect{x}$, then $D_G(\vect{y} \parallel \vect{x})$ is the expected amount that the expert loses by reporting $\vect{x}$, as compared to reporting $\vect{y}$.
    \begin{itemize}
        \item Note that Equations~\ref{eq:chap2_savage} and \ref{eq:chap2_bregman} together tell us that for a proper scoring rule $s$, we have $s(\vect{x}; j) = G_s(\pmb{\delta}_j) - D_{G_s}(\pmb{\delta}_j \parallel \vect{x})$. This is a specialization of the above point to the case where we take the ``true'' probability distribution to be $\pmb{\delta}_j$, i.e.\ the distribution that puts 100\% probability on the outcome that actually happens.
    \end{itemize}
    \item Or alternatively, if the expert initially believes $\vect{x}$, and then receives new information that causes them to update their belief to $\vect{y}$, then $D_G(\vect{y} \parallel \vect{x})$ is the value of this new information to the expert (in terms of their expected increase in score).
\end{itemize}

These interpretations give an important intuition about the order of arguments to a Bregman divergence. It generally makes sense to think of the second argument as a less refined (i.e.\ less informed) estimate, and the first argument as a more refined estimate.

Bregman divergences allow us to generalize proper scoring rules beyond the setting in which the realized outcome is simply one of $n$ options. Suppose that a principal wants to elicit some real-valued (or perhaps even vector-valued) quantity from an expert -- for example, the amount that it will rain tomorrow in each of five cities. More precisely, the expert has a probability distribution over the outcome, and the principal wants to elicit the expected value of the distribution. For a strictly convex function $G$, consider the scoring rule
\[s(\vect{x}; \vect{y}) = -D_G(\vect{y} \parallel \vect{x}),\]
where $\vect{x}$ is the expert's reported expected value and $\vect{y}$ is the realized outcome (e.g.\ the amount of rainfall in the five cities). This scoring rule is proper!

\begin{prop}[\cite{banerjee2005clustering}] \label{prop:bregman_max_ev}
    Given a (vector-valued) random variable $\vect{y}$, the quantity $\EE{D_G(\vect{y} \parallel \vect{x})}$ is minimized by $\vect{x} = \EE{\vect{y}}$.
\end{prop}

(This is true by a straightforward generalization of our argument above for the case where $\vect{y} = \pmb{\delta}_j$ for some $j$. See e.g.\ \textcite{banerjee2005clustering} for a full proof.)

We can also add an arbitrary function $f(\vect{y})$ to the scoring rule while maintaining properness, since the expert has no control over the realized outcome $\vect{y}$. Therefore, scoring rules of the form
\[s(\vect{x}; \vect{y}) = f(\vect{y}) - D_G(\vect{y} \parallel \vect{x}),\]
for strictly convex functions $G$ are proper. Indeed, this is an exhaustive characterization: \emph{all} proper scoring rules take this form \parencite[Theorem 12]{af12}.

Proper scoring rules, Bregman divergences, and the intuition behind them are ubiquitous in ABE, and in this thesis in particular.

\section{Forecast aggregation methods} \label{sec:prelim_agg}
\emph{This section (and particularly Subsection~\ref{sec:lin_log}) is necessary for Chapters~\ref{chap:qa}, \ref{chap:learning}, and \ref{chap:robust}. It is also likely to be interesting in its own right.}\\

The question of forecast aggregation -- how to aggregate forecasts from two or more experts into a single forecast -- is so natural that it hardly needs motivating. But to be concrete, here are three examples. First, from \textcite{nr23_qa}:

\begin{example} \label{ex:agg1}
    You are a meteorologist tasked with advising the governor of Florida on hurricane preparations. A hurricane is threatening to make landfall in Miami, and the governor needs to decide whether to order a mass evacuation. The governor asks you for the likelihood of a direct hit, so you decide to consult several weather models. These models all give you different answers: 10\%, 25\%, 70\%. You trust the models equally, but your job is to come up with one number for the governor: your best guess, all things considered. What is the most sensible way for you to aggregate these numbers?
\end{example}

Second, from \textcite{nr22_smarter}:

\begin{example} \label{ex:agg2}
    Suppose that you wish to estimate how much the GDP of the United States will grow next year: perhaps you are making financial decisions and want to know whether to expect a downturn. You don't personally know much about the question -- just that the historical average rate of GDP growth has been 3\% -- but you look online and find several forecasts made by machine learning models. One model predicts 3.5\% growth next year; another predicts 1.5\%; a third predicts a downturn: -1\% growth. How might you take this information into account and turn it into one number: your best guess for next year's growth rate, all things considered?
\end{example}

Third, a new example:

\begin{example} \label{ex:agg3}
    Three different neural networks -- perhaps with different architectures -- are trained to classify pictures of a thousand different animals. Each neural net takes an image as input and outputs a probability distribution over the thousand classes. You want to ensemble (combine) these neural nets into a single classifier by aggregating the distributions that they output. What is the best way to do this?
\end{example}

These examples are diverse in two ways. First, the application domain: the problem of forecast aggregation is very broadly applicable. Beyond these examples, forecast aggregation finds uses in almost every natural and social science. And second, the mathematical domain of the forecasts themselves: the first aggregates probabilities; the second, arbitrary real numbers; and the third, probability distributions over many outcomes.

The ``correct'' way to aggregate forecasts very much depends on the setting. In this section, we will discuss some of the most common aggregation methods.

\subsection{Linear and logarithmic pooling} \label{sec:lin_log}
Let's introduce some basic notation:
\begin{itemize}
    \item We will refer to the forecasters as \emph{experts}. We will let $m$ be the number of experts, and will number them $1$ through $m$.
    \item The forecasts will live inside some space $\mathcal{D}$, which itself lies in $\RR^n$ for some $n$. In Example~\ref{ex:agg2}, $\mathcal{D}$ can be taken to be $\RR$; in Example~\ref{ex:agg3} $\mathcal{D} = \Delta_{1000} \subseteq \RR^{1000}$. Example~\ref{ex:agg1} can be thought of as either $\mathcal{D} = [0, 1] \subseteq \RR$, or as $\mathcal{D} = \Delta_2 \subseteq \RR^2$.
    \item In this section, we will call the forecasts $\vect{x}_1, \dots, \vect{x}_m$ (these are elements of $\mathcal{D}$, and thus are vectors in $\RR^n$). We will let $x_i(j)$ denote the $j$-th component of $\vect{x}_i$. (Our notation will vary slightly throughout the thesis.)
    \item The experts will have \emph{weights} $w_1, \dots, w_m$. The intuition for weights is that some experts may be more informed or reliable than others, so it is natural to weigh their forecasts more heavily in the aggregate.
\end{itemize}

The most straightforward (and most common) way to aggregate forecasts is to average them. In the context of aggregation, averaging forecasts is often called \emph{linear pooling.}

\begin{defin}
    The \emph{linear pool} of forecasts $\vect{x}_1, \dots, \vect{x}_m$ with weights $w_1, \dots, w_m$ is their weighted arithmetic mean: $\sum_{i = 1}^m w_i \vect{x}_i$.
\end{defin}

An important virtue of linear pooling is its simplicity. Linear pooling frequently outperforms attempts at more sophisticated pooling methods, which often lead to overfitting. Linear pooling is also \emph{eventwise-independent,} meaning that the aggregate probability of outcome $j$ only depends on the probabilities that the various experts assign to outcome $j$, and not on the probabilities that they assign to other outcomes. (But, as we will see, this is not always desirable.)

In the context of probabilistic forecasts, a common alternative to linear pooling is called \emph{logarithmic pooling.} Logarithmic pooling involves taking the geometric mean of forecasts, instead of the arithmetic mean. As a simple example, suppose that two experts forecast distributions over three possible outcomes (such as whether tomorrow will be sunny, cloudy, or rainy). Expert 1 forecasts $\vect{x}_1 = (60\%, 36\%, 4\%)$, while Expert 2 forecasts $\vect{x}_2 = (75\%, 5\%, 20\%)$. The logarithmic pool of these two forecasts (with equal weights) first takes the geometric mean of the two forecasts outcome by outcome (in this case, $\parens{\sqrt{0.6 \cdot 0.75}, \sqrt{0.36 \cdot 0.05}, \sqrt{0.04 \cdot 0.2}}$), and then rescales the resulting forecast so that the probabilities add to $1$ (which in this case will give $(75\%, 15\%, 10\%)$).

\begin{defin} \label{def:log_pool}
    Let $\vect{x}_1, \dots, \vect{x}_m$ be probability distributions over $n$ outcomes that assign nonzero probability to every outcome. The \emph{logarithmic pool} of $\vect{x}_1, \dots, \vect{x}_m$ with weights $w_1, \dots, w_m$ is obtained by taking the weighted geometric mean of the forecasts (componentwise) and rescaling the result to add to $1$. That is, the logarithmic pool is the distribution $\vect{x}^*$ defined as
    \[x^*(j) = c \prod_{i = 1}^m (x_i(j))^{w_i},\]
    for all outcomes $j$, where $c$ is the appropriate normalizing constant.\footnote{Specifically, $c = \parens{\sum_{j = 1}^n \prod_{i = 1}^m (x_i(j))^{w_i}}^{-1}$.}
\end{defin}

Why take the geometric mean instead of the arithmetic mean? To illustrate, suppose that a monster is hiding under one of three beds, and two experts are forecasting a probability distribution over which bed the monster is hiding under. Expert 1 checks under Bed 1, doesn't see a monster, and so reports $\vect{x}_1 = (0.04\%, 49.98\%, 49.98\%)$. Meanwhile, Expert 2 checks under Bed 2, doesn't see a monster, and so reports $\vect{x}_2 = (49.98\%, 0.04\%, 49.98\%)$. Then ideally, the aggregate forecast would put almost all probability mass on the monster being under Bed 3. This is not achievable with a linear pool, which will only assign a 49.98\% probability to the monster being under Bed 3. By contrast, a logarithmic pool with equal weights gives an aggregate of roughly $(2.7\%, 2.7\%, 94.6\%)$, which is much more reasonable in this situation.

In other words, logarithmic pooling takes experts seriously when they assign low probabilities to outcomes -- unlike linear pooling. Perhaps for this reason, logarithmic pooling has been found to perform very well on real-world data, typically better than linear pooling \parencite{sbfmtu14}.


Another, perhaps more natural, perspective on the logarithmic pool is that it is the geometric mean of forecasts -- without any normalization -- when the forecasts are viewed as \emph{odds}. Consider our earlier example: $\vect{x}_1 = (60\%, 36\%, 4\%)$, $\vect{x}_2 = (75\%, 5\%, 20\%)$. To take the logarithmic pool, we write each forecast in terms of odds: $(60\%, 36\%, 4\%)$ becomes 60:36:4 odds, or (to simplify) 15:9:1, while $(75\%, 5\%, 20\%)$ becomes 15:1:4 odds. Then we take the geometric mean, which gives 15:3:2. We can then reinterpret these odds as a vector of probabilities: $(75\%, 15\%, 10\%)$.

But perhaps the \emph{most} natural perspective on logarithmic pooling is as an \emph{arithmetic} mean of the experts' \emph{log}-odds (or ``logits,'' in machine learning terminology). That is, while linear pooling takes the arithmetic mean of forecasts in probability space, logarithmic pooling takes the arithmetic mean of forecasts in log-odds space. In other words: the logarithmic pool of $\vect{x}_1, \dots, \vect{x}_m$ with weights $w_1, \dots, w_m$ can also be written as
\[(\ln x^*(1), \dots, \ln x^*(n)) \equiv \sum_{i = 1}^m w_i (\ln x_i(1), \dots, \ln x_i(n)),\]
where $\equiv$ denotes equality up to translation by a multiple of the all-ones vector.\footnote{Log-odds vectors are most naturally thought of belonging to $\RR^n$ modulo the all-ones vector. Translation plays the same role that rescaling plays in Definition~\ref{def:log_pool}.}

To see why this is a natural perspective, let us for simplicity consider the case of a binary event, such that every expert reports the probability $x_i$ of the ``Yes'' outcome. Then the logarithmic pool of $x_1, \dots, x_m$ with weights $w_1, \dots, w_m$ can be defined as the probability $x^*$ that satisfies
\begin{equation} \label{eq:log_pool_odds}
    \ln \frac{x^*}{1 - x^*} = \sum_{i = 1}^m w_i \ln \frac{x_i}{1 - x_i}.
\end{equation}

Thinking of logarithmic pooling in terms of log-odds is natural because log-odds are \emph{units of the strength of Bayesian evidence.} For example, suppose that our $m$ forecasters are estimating the probability of some event $X$, and start with some common prior. Each expert $i$ receives Bayesian evidence $E_i$ and performs a Bayesian update to obtain their posterior probability:
\[\frac{\pr{X \mid E_i}}{\pr{\neg X \mid E_i}} = \frac{\pr{X}}{\pr{\neg X}} \cdot \frac{\pr{E_i \mid X}}{\pr{E_i \mid \neg X}}.\]
Or, taking the log of both sides:
\[\ln \frac{\pr{X \mid E_i}}{\pr{\neg X \mid E_i}} = \ln \frac{\pr{X}}{\pr{\neg X}} + \ln \frac{\pr{E_i \mid X}}{\pr{E_i \mid \neg X}}.\]
That is, in log-odds space, every expert's posterior is equal to their prior, plus a term that represents the strength of the evidence $E_i$ in favor of (or against) $X$.

And so, what does taking the \emph{average} of the experts' posterior log-odds represent? This average is equal to
\[\ln \frac{\pr{X}}{\pr{\neg X}} + \frac{1}{m} \sum_{i = 1}^m \ln \frac{\pr{E_i \mid X}}{\pr{E_i \mid \neg X}}.\]
In other words, the logarithmic pool represents the posterior probability of a hypothetical expert who received evidence whose strength in favor of (or against) $X$ was the average of the strengths of all $m$ experts' evidence.

However, this framing raises an important question: why \emph{average} the experts' Bayesian evidence? Would it not be more appropriate to add them, so as to fully incorporate all experts' evidence? This brings us to the concept of \emph{extremization.}

\subsection{Extremization}
Suppose that two experts are forecasting the probability of an event $X$, and that the experts have a common prior of $\frac{1}{2}$. Then, each expert receives a piece of evidence that causes them to update to $\frac{2}{3}$, and so both experts report a probability of $\frac{2}{3}$ to the aggregator. Both linear and logarithmic pooling will output an aggregate probability of $\frac{2}{3}$. But is that the correct way to aggregate the experts' forecasts?

The answer is: it depends! Suppose, for example, that a coin with unknown bias (i.e.\ probability of heads) -- uniformly selected from $[0, 1]$ -- will be flipped tomorrow, and the experts are forecasting the probability that the coin will come up heads. The experts' prior is, of course, $\frac{1}{2}$.

Now, first, suppose that the evidence that the experts see is one flip of the coin -- the same flip. If the coin comes up heads, then each expert's posterior probability will be $\frac{2}{3}$.\footnote{This can be verified by performing a Bayesian update on this new evidence. It also follows from a theorem known as Laplace's rule of succession.} And because both experts saw the same coin flip, the correct aggregate is $\frac{2}{3}$ as well. The linear and logarithmic pools get this one right.

But now, suppose that the experts see \emph{different, independent} flips of the coin, both of which come up heads. Each expert's posterior will still be $\frac{2}{3}$, but this time the correct aggregate will be \emph{more} than $\frac{2}{3}$ ($\frac{3}{4}$, as it happens). That's because the experts observed different pieces of evidence, both of which caused them to update in the same direction. Accounting for both experts' evidence -- the fact that \emph{both} flips came up heads -- results in an even higher probability.

The key insight from these two contrasting examples is this: if the experts' evidence is identical or heavily overlapping, then taking some sort of average -- perhaps a linear pool (average of probabilities) or logarithmic pool (average of log-odds) -- is sensible. But if the experts' evidence is heavily non-overlapping or disjoint, then these pooling methods are \emph{insufficiently extremizing:} they take a sort of \emph{average} of the experts' evidence, when it would make more sense to do something more like \emph{adding} the evidence. The result is an estimate that is too close to the prior.

A natural solution to this problem is to pool the experts' forecasts and then to \emph{extremize} the pooled forecast -- meaning, to push it away from the prior. The smaller the overlap between the experts' information, the larger the appropriate amount of extremization.\footnote{If the experts' pieces of evidence are independent conditional on the outcome, then \emph{adding} the sizes of their updates from the prior in log-odds space -- or, in other words, extremizing the logarithmic pool by a factor of $m$ -- gives exactly the right answer.} The benefits of extremization are not just theoretical: extremization has been found to improve the quality of aggregate forecasts in practice \parencite{sbfmtu14, sevilla21b}. Indeed, taking the logarithmic pool of forecasts and then extremizing is a state-of-the-art forecast aggregation method.

\subsection{The generalized linear and logarithmic pools} \label{sec:generalized_pool}
One natural way to extremize linear and logarithmic pools is to allow weights to vary freely, rather than requiring them to add to one. For example, we might attempt to define the ``linear pool'' of $\vect{x}_1, \dots, \vect{x}_m$ with weights $w_1, \dots, w_m$ (not necessarily adding to $1$) as before: $\sum_{i = 1}^m w_i \vect{x}_i$.

This does not quite make sense, as it is not invariant to adding a constant. That is, suppose that $\vect{x}_1, \dots, \vect{x}_m$ are all forecasts of some quantity $\vect{x}$. Define $\vect{y} := \vect{x} + \vect{c}$ for some known vector $\vect{c}$. Then the experts' forecasts for $\vect{y}$ (call them $\vect{y}_i$) will be $\vect{y}_i = \vect{x}_i + \vect{c}$. The linear pool of $\vect{y}_1, \dots, \vect{y}_m$ with weights $w_1, \dots, w_m$ ought to be equal to the linear pool of $\vect{x}_1, \dots, \vect{x}_m$ with the same weights, plus $\vect{c}$. This is the case if $\sum_i w_i = 1$, but not in general.

We can fix this problem if we have access to a prior $\vect{x}_0$. (In Example~\ref{ex:agg2} above, for example, the prior might be the historical average growth rate, 3\%.) In that case, we can treat the experts' forecasts in terms of their \emph{updates from the prior,} and aggregate those updates with arbitrary weights. We will call this the \emph{generalized linear pool.}

\begin{defin}
    The \emph{generalized linear pool} of forecasts $\vect{x}_1, \dots, \vect{x}_m$ with weights $w_1, \dots, w_m$, given a prior $\vect{x}_0$, is defined as
    \[\vect{x}_0 + \sum_{i = 1}^m w_i(\vect{x}_i - \vect{x}_0).\]
\end{defin}

If $\sum_i w_i > 1$, we can think of this aggregation strategy as linearly pooling the forecasts and then extremizing the result by pushing it away from the prior by a constant factor. \textcite{su15} called this technique \emph{linear extremization.}

We can define the generalized logarithmic pool similarly. Recall that logarithmic pooling can be thought of as taking an average (i.e.\ linear pool) of the experts' log-odds. We can instead take a \emph{generalized} linear pool of the log-odds.

\begin{defin}
    Given a forecast $\vect{x}$, let $\pmb{\ell}(\vect{x}) := (\ln x(1), \dots, \ln x(n))$. The \emph{generalized logarithmic pool} of forecasts $\vect{x}_1, \dots, \vect{x}_m$ with weights $w_1, \dots, w_m$, given a prior $\vect{x}_0$, is the forecast $\vect{x}^*$ satisfying
    \[\pmb{\ell}(\vect{x}^*) \equiv \pmb{\ell}(\vect{x}_0) + \sum_{i = 1}^m w_i (\pmb{\ell}(\vect{x}_i) - \pmb{\ell}(\vect{x}_0)),\]
    where $\equiv$ denotes equality up to translation by a multiple of the all-ones vector.
\end{defin}

If each expert's evidence is independent conditioned on the eventual outcome, then the generalized logarithmic pool with weights $w_1 = \dots = w_m = 1$ gives exactly the correct aggregate. Since experts' evidence typically has substantial overlap, it usually makes sense to choose much smaller weights. \textcite{sbfmtu14} found that setting all weights to $2/m$ (so that the sum of all of the weights is $2$) results in good performance on real-world data.

\section{Information structures} \label{sec:prelim_info_struct}
\emph{This section is necessary for Chapters~\ref{chap:robust} and \ref{chap:agreement}, and is also useful for Chapter~\ref{chap:learning}.}

\subsection{Introduction to information structures}
Perhaps the most fundamental concept of ABE is the \emph{information structure.} Informally speaking, an information structure is a full description of all information possessed by a set of experts in all possible states of the world.

To be concrete, suppose that we are interested in the value of some (possibly real-valued, possibly vector-valued) random variable $Y$: perhaps the amount that it will rain tomorrow. There are various pieces of information in the world that are relevant to $Y$ (e.g.\ the current dew point, or the air pressure forecast by the GFS weather model). We call these pieces of information \emph{signals}. From a Bayesian standpoint, there is some joint prior probability distribution over the signals and the value of $Y$. We call such a prior an \emph{information structure}.

\begin{restatable}{defin}{infostruct} \label{def:info_struct}
    An \emph{information structure} $\mathcal{I} = (\Omega, \PP, \pmb{\sigma}, Y)$ consists of:
    \begin{itemize}
        \item A set of states of the world $\Omega$, together with a probability distribution $\PP$ over $\Omega$.
        \item A collection of $m$ \emph{signals} $\pmb{\sigma} = (\sigma_1, \dots, \sigma_m)$, which are random variables defined on $\Omega$. That is, for $i = 1 ... m$, we have a signal $\sigma_i: \Omega \to S_i$, where $S_i$ -- the set of possible values that $\sigma_i$ can take on -- is called the $i$-th \emph{signal set}. (We will often think of each signal as belonging to a different expert.)
        \item A random variable $Y: \Omega \to \RR^n$ for some $n$.
    \end{itemize}
\end{restatable}

Let's unpack this definition through a series of examples.
\begin{example} \label{ex:is_1}
    There is a coin that has bias (i.e.\ probability of heads) $Y$, which is either $\frac{1}{3}$ or $\frac{2}{3}$; these two possibilities are equally likely. Two experts each see a different flip of the coin.

    Here, we can think of $S_1 = \{H, T\}$, with $\sigma_1 = H$ if Expert 1 sees heads and $\sigma_1 = T$ if Expert 1 sees tails. $S_2$ and $\sigma_2$ are defined analogously for Expert 2. We can define $\Omega$ as having eight states that together describe $Y$, $\sigma_1$, and $\sigma_2$. Each of these states has a certain probability. For example, the probability of the state $\left\{Y = \frac{2}{3}, \sigma_1 = H, \sigma_2 = T\right\}$ is $\frac{1}{2} \cdot \frac{2}{3} \cdot \frac{1}{3} = \frac{1}{9}$. That's because there's a 50\% chance that $Y = \frac{2}{3}$; and conditional on that, there's a $\frac{2}{3}$ chance that $\sigma_1 = H$ and a $\frac{1}{3}$ chance that $\sigma_2 = T$.

    Often we are interested the expected value of $Y$ conditioned on the experts' information. For this reason, it is often useful to summarize information structures like this one using two tables:

    \singlespacing
    \[\left\{ \EE{Y}: \begin{tabular}{c|cc}&$\sigma_2 = H$&$\sigma_2 = T$\\ \hline $\sigma_1 = H$ & $3/5$ & $1/2$ \\ $\sigma_1 = T$ & $1/2$ & $2/5$\end{tabular} \qquad \PP[\sigma_1, \sigma_2]: \begin{tabular}{c|cc}&$\sigma_2 = H$&$\sigma_2 = T$\\ \hline $\sigma_1 = H$ & $5/18$ & $2/9$ \\ $\sigma_1 = T$ & $2/9$ & $5/18$\end{tabular} \right\}\]
    \doublespacing
    
    For example, the expected value of $Y$ conditioned on $\sigma_1 = H$ and $\sigma_2 = H$ is $\frac{3}{5}$, and the probability that $\sigma_1 = H$ and $\sigma_2 = H$ is $\frac{5}{18}$.

    This table also allows us to compute the expected value of $Y$ conditioned on some individual expert's signal. For example, $\EE{Y \mid \sigma_1 = H} = \frac{3/5 \cdot 5/18 + 1/2 \cdot 2/9}{5/18 + 2/9} = \frac{5}{9}$.
\end{example}

\begin{example} \label{ex:is_2}
    Just like in the previous example, there is a coin with bias $Y$ that is either $\frac{1}{3}$ or $\frac{2}{3}$. But this time, the two experts see the \emph{same} flip of the coin.

    In this case, we can think of $\Omega$, $S_1$, and $S_2$ the same way, but now $\PP$ is different. In particular, the probability of $\left\{Y = \frac{2}{3}, \sigma_1 = H, \sigma_2 = T\right\}$ is now $0$, because $\sigma_1$ and $\sigma_2$ are guaranteed to be either both $H$ or both $T$ in this information structure. The following table summarizes this information structure:

    \singlespacing
    \[\left\{ \EE{Y}: \begin{tabular}{c|cc}&$\sigma_2 = H$&$\sigma_2 = T$\\ \hline $\sigma_1 = H$ & $5/9$ & $-$ \\ $\sigma_1 = T$ & $-$ & $5/9$\end{tabular} \qquad \PP[\sigma_1, \sigma_2]: \begin{tabular}{c|cc}&$\sigma_2 = H$&$\sigma_2 = T$\\ \hline $\sigma_1 = H$ & $1/2$ & $0$ \\ $\sigma_1 = T$ & $0$ & $1/2$\end{tabular} \right\}\]
    \doublespacing
\end{example}

These contrasting examples illustrate that an information structure captures not just the probability distribution of each expert's information individually, but also the interaction between the experts' information, e.g.\ how their signals are correlated.

Next we'll introduce a particular type of information structure that we will find useful.

\begin{restatable}{defin}{pifdef} \label{def:pif} \parencite{spu16}
    An information structure in the \emph{partial information framework} (henceforth, a \emph{PIF information structure}) is an information structure that takes the following form: for each subset $S \subseteq [m]$, there is a random variable $X_S$ (all of these random variables are independent), and $\sigma_i$ is the tuple of random variables $X_S$ for all $S$ containing $i$. The value of $Y$ is equal to $\sum_S X_S$.
\end{restatable}

If we think of the signals as belonging to experts, each $X_S$ as a piece of evidence (a real number, or perhaps a vector of reals), and expert $i$ has access to $X_S$ if $i \in S$. These pieces of evidence behave additively, in the sense that $Y$ is the sum of all of random variables $X_S$.

PIF information structures are interesting to study in the context of aggregation. Expert $i$'s estimate of $Y$ -- that is, the expected value of $Y$ conditioned on $\sigma_i$ -- is equal to $\sum_{S \ni i} X_S + \sum_{S \not \ni i} \EE{X_S}$. For convenience, we typically assume that each $X_S$ has mean zero; this is the case without loss of generality in all of our applications. Under this assumption, expert $i$'s estimate of $Y$ reduces to the sum of the $X_S$'s that Expert $i$ sees: $\sum_{S \ni i} X_S$.

Now consider an aggregator who sees each expert's estimates. If the aggregator knew not just each expert's estimate, but also their signal in full (i.e.\ all of the tuples of $X_S$-values), then their job would be straightforward: just add all the $X_S$-values. But if (as is typical) the aggregator cannot access this information, then aggregation becomes nontrivial. In Section~\ref{sec:bayesian_justifications}, we will discuss the optimal aggregation strategy in the specific case that each $X_S$ is normally distributed.

\subsection{Informational substitutes} \label{sec:prelim_subs}
The space of information structures is vast: there's a huge number of ways in which different experts can have overlapping information about the value of a quantity. Relatively few nontrivial facts are known about \emph{all} information structures. So a typical theorem about information structures imposes conditions: \emph{if an information structure $\mathcal{I}$ satisfies [condition], then...}.

One natural constraint is that, for the purposes of accurately estimating $Y$, there are diminishing marginal returns to learning new signals. An example of diminishing marginal returns would be that learning Signal 7 is more useful if you only know Signal 2, than if you know both Signal 2 and Signal 4. This condition is particularly intuitive in the context of experts with overlapping information: the greater the information overlap between different experts, the fewer returns there are to learning additional experts' information.

The general name for diminishing marginal returns to learning new signals is \emph{informational substitutes.} The notion of informational substitutes was first explored by \textcite{bhk13}, though we will mostly be interested in building on definitions introduced by \textcite{cw16}.\footnote{I recommend the ArXiv version of \textcite{cw16} for the most up-to-date introduction to informational substitutes.}

The concept of diminishing marginal returns to estimating $Y$ only makes sense in the context of some sort of ``value function'' -- that is, a function that describes the quality of an estimate of $Y$. For example, we could judge an estimate of $Y$ based on its squared error. That is, we could say that an information structure satisfies informational substitutes if there are diminishing marginal returns to learning new signals, as measured by the squared error when estimating $Y$. Formalizing this idea results in a concept called \emph{weak informational substitutes} (with respect to squared error).

In the following definition -- and more generally, throughout the thesis -- we will use the following notation: given a subset $A \subseteq [m]$ of signals, $Y_A$ is the expected value of $Y$ conditioned on the signals in $A$. For example, $Y_{\emptyset}$ is the prior $\EE{Y}$; for $i \in [m]$, $Y_i$ is the expected value of $Y$ conditioned on $\sigma_i$; and $Y_{[m]}$ is the expectation of $Y$ conditioned on all information that is present.

\begin{restatable}{defin}{weaksubsquad}(Weak substitutes w.r.t.\ squared error \parencite{cw16}) \label{def:weak_subs_quad}
     Let $\mathcal{I} = (\Omega, \PP, \pmb{\sigma}, Y)$ be an information structure. $\mathcal{I}$ satisfies \emph{weak informational substitutes} (or simply \emph{weak substitutes}) with respect to squared error if, for all $B \subseteq A \subseteq [m]$ and $i \not \in A$, we have\footnote{More abstractly, we could say that $\mathcal{I}$ satisfies weak substitutes with respect to squared error if $-(Y - Y_A)^2$ is a submodular set function on the subsets $A$ of $[m]$.}
     \begin{equation} \label{eq:weak_subs_quad}
         \EE{(Y - Y_A)^2} - \EE{(Y - Y_{A \cup \{i\}})^2} \le \EE{(Y - Y_B)^2} - \EE{(Y - Y_{B \cup \{i\}})^2}.
     \end{equation}
\end{restatable}

That is, the reduction in squared error gained by learning $\sigma_i$ if you already know all the signals in $A$ (that's the left-hand side) is smaller than the reduction in squared error gained by learning $\sigma_i$ if you only know the signals in $B$, which is a subset of $A$ (that's the right-hand side).

\begin{example}
    It is straightforward to see that the information structure in Example~\ref{ex:is_2} satisfies weak substitutes with respect to squared error. That's because $\sigma_1$ and $\sigma_2$ are identical. The value of $\sigma_2$, in terms of reduction in squared error, is positive when no signal is known, but is exactly zero when $\sigma_1$ is already known.
\end{example}

\begin{example}
    Every PIF information structure satisfies weak substitutes with respect to squared error. This is left as an exercise for now, though we will prove a stronger statement in Chapter~\ref{chap:robust}.
\end{example}

The following \emph{non-}example of informational substitutes, which we call the XOR information structure, will be useful in future discussions.

\begin{restatable}{defin}{xoris}
    The \emph{XOR information structure} consists of two signals, $\sigma_1$ and $\sigma_2$, that are independent, random bits (i.e.\ either $0$ or $1$, with equal probability). The value of $Y$ is equal to the binary XOR of $\sigma_1$ and $\sigma_2$, i.e.\ $0$ if $\sigma_1 = \sigma_2$ and $1$ if $\sigma_1 \neq \sigma_2$.
\end{restatable}

To see that the XOR information structure does not satisfy informational substitutes, we observe that the prior is $\frac{1}{2}$ and also the expected value of $Y$ conditioned on any one signal is always $\frac{1}{2}$. That is, having just one signal is completely uninformative! On the other hand, knowing both signals gives away the value of $Y$ exactly. Informally speaking, the opposite of informational substitutes is known as \emph{informational complements} (see \textcite{cw16}), and XOR is a prototypical example of informational complements.

We can also define weak substitutes more generally, for other notions of error. Really, for any differentiable convex function $G$, the Bregman divergence with respect to $G$ from $Y$ to $Y_A$ makes sense in place of $(Y - Y_A)^2$. Bregman divergences are sensible in this context because they \emph{elicit the mean}: for every $Y$, $\EE{D_G(Y \parallel x)}$ is minimized at $x = \EE{Y}$ (see Proposition~\ref{prop:bregman_max_ev} above).

Thus, more generally, we say:
\begin{defin}[Weak substitutes \parencite{cw16}] \label{def:weak_subs}
     Let $\mathcal{I} = (\Omega, \PP, \pmb{\sigma}, Y)$ be an information structure. $\mathcal{I}$ satisfies \emph{weak substitutes} with respect to a differentiable convex function $G$ if, for all $B \subseteq A \subseteq [m]$ and $i \not \in A$, we have
     \[\EE{D_G(Y \parallel Y_A)} - \EE{D_G(Y \parallel Y_{A \cup \{i\}})} \le \EE{D_G(Y \parallel Y_B)} - \EE{D_G(Y \parallel Y_{B \cup \{i\}})}.\]
\end{defin}

Why \emph{weak} substitutes? \textcite{cw16} define stronger notions of substitutes as well. These notions consider partial revelation of signals, thus requiring submodularity over a finer space. We won't go into further detail here: while these definitions are interesting, the work presented in this thesis builds specifically on the notion of weak substitutes.

\subsection{Random variables as vectors and the Pythagorean theorem} \label{sec:prelim_pythag}
It is often useful to think of random variables as vectors with the inner product $\angles{X, Y} := \EE{XY}$. (More formally: given a probability space $(\Omega, \mathcal{F}, \PP)$, the set of random variables with finite variance on $(\Omega, \mathcal{F}, \PP)$, endowed with the inner product $\angles{X, Y} := \EE{XY}$, is a Hilbert space over $\RR$.)

If the state space $\Omega$ is finite, it often makes sense to think of a random variable $X$ geometrically, as a vector with one coordinate per state $\omega \in \Omega$, where the value of the coordinate is the value of $X$ on $\omega$. This perspective on random variables is particularly useful because it gives a very natural notion of an orthogonal projection. As we will soon show, when random variables are thought of as vectors in this way, \emph{orthogonal projections correspond to conditional expectations.}

Consider an information structure $\mathcal{I} = (\Omega, \PP, \pmb{\sigma}, Y)$. Recall our notation $Y_A$ from the previous section, which means the expectation of $Y$ conditioned on all signals in $A$ (a subset of $[m]$). Each $Y_A$ is a random variable that only depends on the values of the signals in $A$. For example, $Y_{\emptyset}$ does not depend on the value of any signals: it is the same across all of $\Omega$. For any $i \in [m]$, $Y_{\{i\}}$ depends on the value of $\sigma_i$, but on no other signals: if $\sigma_i(\omega_1) = \sigma_i(\omega_2)$, then $Y_{\{i\}}(\omega_1) = Y_{\{i\}}(\omega_2)$.

Now, consider two subsets of signals $A$ and $B$, such that $B$ is a subset of $A$. This means that $Y_B$ is a coarser estimate for $Y$ than $Y_A$ is: it is an estimate (conditional expectation) based on a smaller set of signals. In the aforementioned inner product space, $Y_B$ is the orthogonal projection of $Y_A$ onto the subspace of all random variables whose values only depend on the signals in $B$. (Figure~\ref{fig:projection} illustrates this relationship.) Formally:

\begin{figure}[ht]
    \centering
    \includegraphics[scale=0.5]{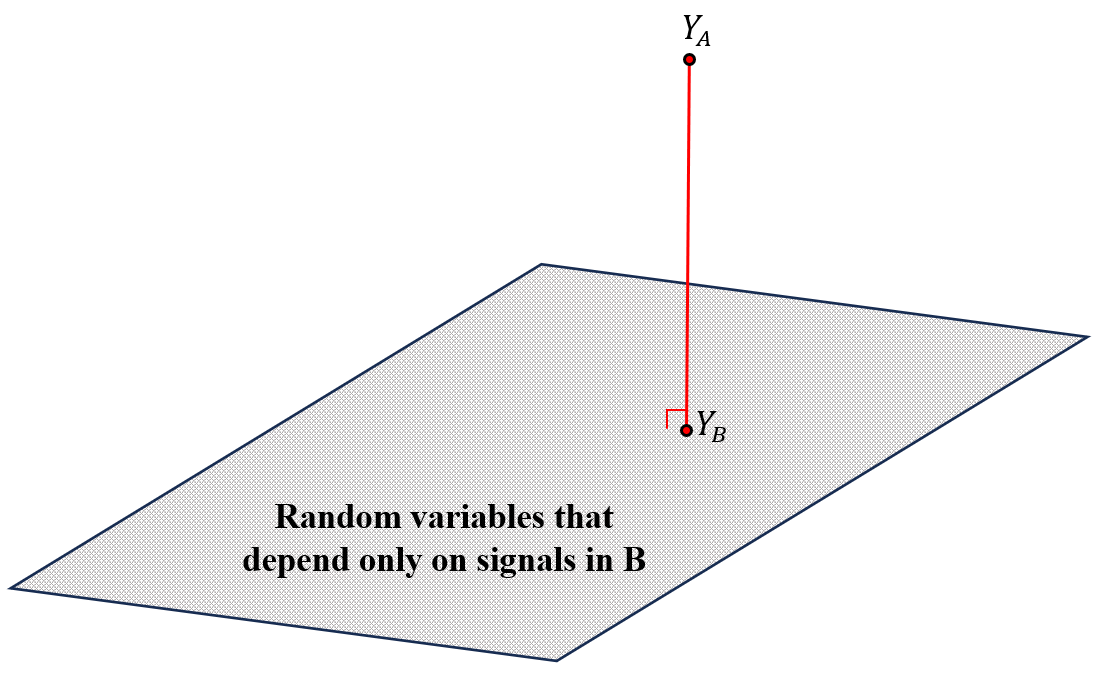}
    \caption[Coarser estimates as orthogonal projections]{For a random variable $Y$, let $B \subseteq A$ be sets of signals. Let $Y_A$ be the expected value of $Y$ conditioned on the signals in $A$, and define $Y_B$ analogously. In the space of random variables with inner product $\angles{X_1, X_2} := \EE{X_1 X_2}$, $Y_B$ is the orthogonal projection of $Y_A$ onto the subspace of random variables whose values only depend on the signals in $B$.}
    \label{fig:projection}
\end{figure}

\begin{prop} \label{prop:projection}
    Let $\mathcal{I} = (\Omega, \PP, \pmb{\sigma}, Y)$ be an information structure, $B \subseteq A \subseteq [m]$, $Y_A = \EE{Y \mid \{\sigma_i\}_{i \in A}}$, and $Y_B = \EE{Y \mid \{\sigma_i\}_{i \in B}}$. In the inner product space\footnote{Technically a Hilbert space, rather than an inner product space.} of random variables on $\Omega$ defined by $\angles{X_1, X_2} = \EE{X_1 X_2}$, $Y_B$ is the orthogonal projection of $Y_A$ onto the subspace of all random variables whose values only depend on the signals in $B$.
\end{prop}

\begin{proof}
    We need to show that $Y_A - Y_B$ has inner product zero with every random variable whose value depends on the signals in $B$. First, note that
    \[Y_B = \EE{Y \mid \{\sigma_i\}_{i \in B}} = \EE{\EE{Y \mid \{\sigma_i\}_{i \in A}} \mid \{\sigma_i\}_{i \in B}} = \EE{Y_A \mid \{\sigma_i\}_{i \in B}}.\]
    In other words, $Y_B$ is the expected value of $Y_A$ conditioned on the signals in $B$.\footnote{Formally, the second equality follows by the \emph{law of iterated expectations,} which says that if $\mathcal{H}_1 \subseteq \mathcal{H}_2 \subseteq \mathcal{F}$ are sub-sigma-algebras, then $\EE{\EE{X \mid \mathcal{H}_2} \mid \mathcal{H}_1} = \EE{X \mid \mathcal{H}_1}$.} This means that
    \[\EE{Y_A - Y_B \mid \{\sigma_i\}_{i \in B}} = \EE{Y_A \mid \{\sigma_i\}_{i \in B}} - Y_B = 0.\]
    Thus, for a variable $Z$ that only depends on the signals in $B$, we can write
    \[\EE{(Y_A - Y_B)Z} = \EE{\EE{(Y_A - Y_B)Z \mid \{\sigma_i\}_{i \in B}}} = \EE{Z \EE{Y_A - Y_B \mid \{\sigma_i\}_{i \in B}}} = \EE{Z \cdot 0} = 0,\]
    as desired.
\end{proof}

(Formally, Proposition~\ref{prop:projection} is a special case of the following more general fact: given a random variable $X$ in a probability space $(\Omega, \mathcal{F}, \PP)$, and a sub-sigma-algebra $\mathcal{H} \subseteq \mathcal{F}$, the orthogonal projection of $X$ onto the subspace of $\mathcal{H}$-measurable random variables is exactly the conditional expectation $\EE{X \mid \mathcal{H}}$ \parencite{zidak57}. However, we only need the more specific fact that we just proved.)

Now, this orthogonal projection property means that we can use the Pythagorean theorem! Concretely, if $Z$ is some random variable whose value depends only on the signals in $A$, then $Z$, $Y_B$, and $Y_A$ form a ``right triangle'' with the right angle at $Y_B$ (see Figure~\ref{fig:projection}). In our inner product space, the squared distance between two variables $X$ and $Y$ is $\angles{X - Y, X - Y} = \EE{(X - Y)^2}$. And so, we have
\[\EE{(Y_A - Z)^2} = \EE{(Y_A - Y_B)^2} + \EE{(Y_B - Z)^2}.\]
More formally (and also more generally):

\begin{restatable}{prop}{pythagsquared}\emph{(Pythagorean theorem for squared distance)}\textbf{.} \label{prop:pythag_squared}
    In a probability space $(\Omega, \mathcal{F}, \PP)$, let $A$ be a real-valued random variable, let $B = \EE{A \mid \mathcal{H}}$ where $\mathcal{H}$ is a sub-sigma-algebra,\footnote{Readers unfamiliar with sigma-algebras can think of $\mathcal{H}$ as supplying partial information about the state of the world $\omega$ (much like a signal). A random variable is \emph{defined on $\mathcal{H}$} if its value depends only on the information that $\mathcal{H}$ provides (i.e.\ the value of the signal).} and $C$ be a real-valued random variable defined on $\mathcal{H}$. Then
	\[\EE{(A - C)^2} = \EE{(A - B)^2} + \EE{(B - C)^2}.\]
\end{restatable}

\begin{proof}
	Observe that $\EE{AB} = \EE{\EE{AB \mid \mathcal{H}}} = \EE{B \EE{A \mid \mathcal{H}}} = \EE{B^2}$, so $\EE{(A - B)^2} = \EE{A^2} - \EE{B^2}$. Also, note that $\EE{AC} = \EE{\EE{AC \mid \mathcal{H}}} = \EE{\EE{A \mid \mathcal{H}} C} = \EE{BC}$,
	where in the second step we use the fact that $C$ is defined on $\mathcal{H}$. Therefore, we have
	\begin{align*}
		\EE{(A - C)^2} &= \EE{A^2} + \EE{C^2} - 2\EE{BC} = \EE{A^2} - \EE{B^2} + \EE{B^2} - 2\EE{BC} + \EE{C^2}\\
		&= \EE{(A - B)^2} + \EE{(B - C)^2},
	\end{align*}
	as desired.
\end{proof}

In fact, the Pythagorean theorem turns out to be true for any Bregman divergence, not just for squared distance! Further, the result extends to vector-valued random variables.

\begin{restatable}{prop}{pythagbregman}\emph{(Pythagorean theorem for Bregman divergence)}\textbf{.} \label{prop:pythag_bregman}
    In a probability space $(\Omega, \mathcal{F}, \PP)$, let $A: \Omega \to \RR^n$ be a random variable, let $B = \EE{A \mid \mathcal{H}}$ where $\mathcal{H}$ is a sub-sigma-algebra, and let $C: \Omega \to \RR^n$ be a random variable that is defined on $\mathcal{H}$. Let $G$ be a differentiable, convex function defined on a convex set that includes all values taken on by $A$, $B$, and $C$. Then
    \[\EE{D_G(A \parallel C)} = \EE{D_G(A \parallel B)} + \EE{D_G(B \parallel C)}.\]
\end{restatable}

\begin{proof}
We have
\begin{align*}
&\EE{D_G(A \parallel B)} + \EE{D_G(B \parallel C)} - \EE{D_G(A \parallel C)}\\
&= \mathbb{E}[G(A) - G(B) - (A - B) \cdot \nabla G(B) + G(B) - G(C) - (B - C) \cdot \nabla G(C)\\
&\qquad - G(A) + G(C) + (A - C) \cdot \nabla G(C)]\\
&= \EE{(A - B) \cdot (\nabla G(C) - \nabla G(B))} = \EE{\EE{(A - B) \cdot (\nabla G(C) - \nabla G(B)) \mid \mathcal{H}}}\\
&= \EE{(\nabla G(C) - \nabla G(B)) \cdot \EE{A - B \mid \mathcal{H}}} = \EE{(\nabla G(C) - \nabla G(B)) \cdot (\EE{A \mid \mathcal{H}} - B)} = 0.
\end{align*}
The third-to-last step follows from the fact that $g(C) - g(B)$ is $\mathcal{H}$-measurable (we are using the ``pulling out known factors'' property of conditional expectations). The last step follows from the fact that $\EE{A \mid \mathcal{H}} = B$.
\end{proof}

While Proposition~\ref{prop:pythag_bregman} is a natural extension of the well-known Proposition~\ref{prop:pythag_squared}, and its proof is straightforward, to my knowledge it first appeared in \textcite{fnw23} (which is presented in Chapter~\ref{chap:agreement} of this thesis).

Note that these Pythagorean theorems allow us to rewrite the definitions of weak substitutes from the previous section:

\begin{prop}
     Let $\mathcal{I} = (\Omega, \PP, \pmb{\sigma}, Y)$ be an information structure. $\mathcal{I}$ satisfies weak substitutes with respect to squared error if, for all $B \subseteq A \subseteq [m]$ and $i \not \in A$, we have
     \[\EE{(Y_{A \cup \{i\}} - Y_A)^2} \le \EE{(Y_{B \cup \{i\}} - Y_B)^2}.\]
     More generally, given a differentiable convex function $G$, $\mathcal{I}$ satisfies weak substitutes with respect to $G$ if, for all $B \subseteq A \subseteq [m]$ and $i \not \in A$, we have
     \[\EE{D_G(Y_{A \cup \{i\}} \parallel Y_A)} \le \EE{D_G(Y_{B \cup \{i\}} \parallel Y_B)}.\]
\end{prop}

\begin{proof}
    We prove the first statement; the proof of the second statement is exactly analogous. We can rewrite the left-hand side of Equation~\ref{eq:weak_subs_quad} as $\EE{(Y_{A \cup \{i\}} - Y_A)^2}$ by using the Pythagorean theorem. (Specifically, we set $A := Y$, $B := Y_{A \cup \{i\}}$, and $C := Y_A$ in the statement of Proposition~\ref{prop:pythag_squared}.) Similarly, we can rewrite the right-hand side of Equation~\ref{eq:weak_subs_quad} as $\EE{(Y_{B \cup \{i\}} - Y_B)^2}$.
\end{proof}

\section{Bayesian justifications for generalized linear and logarithmic pooling} \label{sec:bayesian_justifications}
\emph{This section offers further justification for generalized linear and logarithmic pooling. While not required for any chapter, it provides useful context for Chapters~\ref{chap:qa}, \ref{chap:learning}, \ref{chap:robust}, and \ref{chap:elk}. This section should also be of independent interest, and it contains some original work.}

\subsection{A Bayesian justification for generalized linear pooling}
We are often interested in aggregating forecasts in the context of an information structure. That is, we are interested in a (real-valued) random variable $Y$, and each expert $i$ receives a signal $\sigma_i$ and reports $\EE{Y \mid \sigma_i}$ to the aggregator. (Our notation for this quantity has been $Y_{\{i\}}$, but in this section we will instead use $Y_i$.)

For simplicity, let us suppose that the prior on the value of $Y$ is zero, i.e.\ $Y_\emptyset := \EE{Y} = 0$. In this case, a generalized linear pool of the experts' forecasts (as introduced in Section~\ref{sec:generalized_pool}) is simply some weighted sum of the $Y_i$'s.

Which generalized linear pool is best for a given information structure? This depends on the notion of error, but suppose we care about expected squared error. Then the optimal linear combination of the $Y_i$'s is the orthogonal projection of $Y$ onto the vector subspace of random variables spanned by $Y_1, \dots, Y_m$. Or in other words, it is the linear regression of $Y$ onto the features $Y_1, \dots, Y_m$. The formula for this linear combination is
\[\text{diag}(\Sigma)^\top \Sigma^{-1} \vect{Y},\]
where $\Sigma$ is the covariance matrix of the $Y_i$'s and $\vect{Y}$ is the vector of $Y_i$'s.\footnote{If $\Sigma$ is not invertible, we take the Moore-Penrose pseudoinverse of $\Sigma$ instead of the inverse.}

Is there a natural class of information structures for which the linear regression estimate for $Y$ is not just the best \emph{linear} estimate, but indeed the \emph{best} estimate? Such a class of information structures would offer a \emph{Bayesian justification} for generalized linear pooling, in the sense that a perfect Bayesian with knowledge of the information structure would choose to estimate $Y$ with a particular generalized linear pool of the $Y_i$'s.

Now, the best estimate of $Y$ given $Y_1, \dots, Y_m$ is just its conditional expectation: $\EE{Y \mid Y_1, \dots, Y_m}$. Thus, we are looking for an information structure in which $\EE{Y \mid Y_1, \dots, Y_m}$ is exactly equal to the linear regression estimate for $Y$.

Recall partial information framework (PIF) information structures from Definition~\ref{def:pif}.

\begin{claim}[\cite{sjpu17}]
    Let $\mathcal{I}$ be a PIF information structure in which every $X_S$ is Gaussian with mean zero. Then the expected value of $Y$ conditioned on $Y_1, \dots, Y_m$ is equal to the linear regression estimate for $Y$:
    \[\EE{Y \mid Y_1, \dots, Y_m} = \text{diag}(\Sigma)^\top \Sigma^{-1} \vect{Y}.\]
\end{claim}

\begin{proof}
    Let us write $Y$ as its projection onto the subspace of $Y_i$'s plus the orthogonal remainder, which we will call $Z$:
    \begin{equation} \label{eq:y_in_terms_of_z}
        Y = \text{diag}(\Sigma)^\top \Sigma^{-1} \vect{Y} + Z.
    \end{equation}
    Note that $Y = \sum_S X_S$ and $Y_i = \sum_{S \ni i} X_S$. Thus, $Z$ is also a linear combination of the $X_S$'s, and so $Y_1, \dots, Y_m, Z$ are jointly Gaussian. Further, every $Y_i$ is uncorrelated with $Z$, i.e.\ $\EE{Y_i Z} = 0$, since $Z$ is orthogonal to the subspace of $Y_i$'s. Note also that $\EE{Z} = 0$, as $Z$ is a linear combination of $Y$ and the $Y_i$'s.

    Now, jointly Gaussian and uncorrelated vectors of random variables are independent \parencite{mkb79}. This means that $\EE{Z \mid Y_1, \dots, Y_m} = 0$. Taking the expectation of both sides of Equation~\ref{eq:y_in_terms_of_z} conditioned on $Y_1, \dots, Y_m$ gives us the desired result.
\end{proof}

\subsection{A Bayesian justification for generalized logarithmic pooling}
\emph{To my knowledge, the main result of this section (Claim~\ref{claim:bayes_log_pool}) is original to this thesis.}\\

Suppose that $m$ experts are forecasting a yes/no outcome, and assume for simplicity that they have a common prior of $\frac{1}{2}$. In this case, the generalized logarithmic pool is defined as
\[\ln \frac{x^*}{1 - x^*} = \sum_{i = 1}^m w_i \ln \frac{x_i}{1 - x_i}\]
for some weights $w_1, \dots, w_m$. Can we come up with a Bayesian justification for generalized logarithmic pooling, much as Gaussian PIF information structures gave us a Bayesian justification for generalized linear pooling? The answer is yes!

\begin{claim} \label{claim:bayes_log_pool}
    Consider the following information structure: $Y$ is either $0$ or $1$, uniformly. If $Y = 1$, then the tuple of signals, $(\tau_1, \dots, \tau_m)$, is drawn from a multivariate normal distribution with some mean $\pmb{\mu}$ and covariance matrix $\Sigma$. If $Y = 0$, then $(\tau_1, \dots, \tau_m)$ is drawn from the multivariate normal distribution with mean $-\pmb{\mu}$ and some invertible covariance matrix $\Sigma$. Then the optimal aggregate of the experts' estimates $Y_1, \dots, Y_m$ is a generalized logarithmic pool:
    \[\ln \frac{\pr{Y = 1 \mid \pmb{\tau}}}{\pr{Y = 0 \mid \pmb{\tau}}} = \parens{\frac{\sigma_1^2}{\mu_1} \ln \frac{Y_1}{1 - Y_1}, \dots, \frac{\sigma_m^2}{\mu_m} \ln \frac{Y_m}{1 - Y_m}}^{\top} \Sigma^{-1} \pmb{\mu}.\]
\end{claim}

(We use $\tau_i$ in place of $\sigma_i$ to avoid collision with our notation for signal variances.) Note that this information structure is a kind of \emph{Gaussian mixture model} (see e.g.\ \textcite[\S9.2]{bishop06}).

\begin{proof}
    First, note that $Y_i := \EE{Y \mid \tau_i}$ is simply the probability that $Y = 1$ conditioned on expert $i$'s signal $\tau_i$. Let $\sigma_i^2$ be the variance of $\tau_i$ (so $\sigma_i^2$ is the $i$-th entry of the diagonal of $\Sigma$). Then $\tau_i$ is distributed with mean $\mu_i$ and variance $\sigma_i^2$ if $Y = 1$, and with mean $-\mu_i$ and variance $\sigma_i^2$ if $Y = 0$. Using the formula for a Gaussian PDF, we have that
    \[\frac{\pr{Y = 1 \mid \tau_i}}{\pr{Y = 0 \mid \tau_i}} = \frac{\exp \parens{-\frac{(\tau_i - \mu_i)^2}{2\sigma_i^2}}}{\exp \parens{-\frac{(\tau_i + \mu_i)^2}{2\sigma_i^2}}} = \exp \parens{\frac{2 \mu_i}{\sigma_i^2} \tau_i}.\]
    Thus, we can write $Y_i$ in log-odds space as
    \begin{equation} \label{eq:y_i_in_terms_of_tau_i}
        \ln \frac{Y_i}{1 - Y_i} = \frac{2\mu_i}{\sigma_i^2} \tau_i.
    \end{equation}
    What about the optimal aggregate, $\EE{Y \mid Y_1, \dots, Y_m}$? Note that we can recover the value of $\tau_i$ from the value of $Y_i$ (using the equation we just wrote down\footnote{Unless $\mu_i = 0$, but in that case $\tau_i$ provides no information about $Y$.}), so $\EE{Y \mid Y_1, \dots, Y_m} = \EE{Y \mid \pmb{\tau}}$, where $\pmb{\tau} = (\tau_1, \dots, \tau_m)$. Using the formula for the PDF of a multivariate Gaussian, we have:
    \[\frac{\pr{Y = 1 \mid \pmb{\tau}}}{\pr{Y = 0 \mid \pmb{\tau}}} = \frac{\exp \parens{\frac{-1}{2}(\pmb{\tau} - \pmb{\mu})^\top \Sigma^{-1} (\pmb{\tau} - \pmb{\mu})}}{\exp \parens{\frac{-1}{2}(\pmb{\tau} + \pmb{\mu})^\top \Sigma^{-1} (\pmb{\tau} + \pmb{\mu})}} = \exp(2 \pmb{\tau}^\top \Sigma^{-1} \pmb{\mu}),\]
    where in the last step we used the expanded out $(\pmb{\tau} - \pmb{\mu})^\top \Sigma^{-1} (\pmb{\tau} - \pmb{\mu})$ and $(\pmb{\tau} + \pmb{\mu})^\top \Sigma^{-1} (\pmb{\tau} + \pmb{\mu})$ as sums of four terms and noticed cancellations. (We also used the fact that $\Sigma^{-1}$ is symmetric, so $\pmb{\mu}^\top \Sigma^{-1} \pmb{\tau} = \pmb{\tau}^\top \Sigma^{-1} \pmb{\mu}$.) Therefore, we have
    \[\ln \frac{\pr{Y = 1 \mid \pmb{\tau}}}{\pr{Y = 0 \mid \pmb{\tau}}} = 2 \pmb{\tau}^\top \Sigma^{-1} \pmb{\mu},\]
    which is a particular linear combination of the $\tau_i$'s. Combining this equation with Equation~\ref{eq:y_i_in_terms_of_tau_i}, we have
    \[\ln \frac{\pr{Y = 1 \mid \pmb{\tau}}}{\pr{Y = 0 \mid \pmb{\tau}}} = \parens{\frac{\sigma_1^2}{\mu_1} \ln \frac{Y_1}{1 - Y_1}, \dots, \frac{\sigma_m^2}{\mu_m} \ln \frac{Y_m}{1 - Y_m}}^{\top} \Sigma^{-1} \pmb{\mu}.\]
    This is indeed a generalized logarithmic pool, as it is a particular linear combination of the values of $\ln \frac{Y_i}{1 - Y_i}$.
\end{proof}

We note that a different Bayesian justification of generalized logarithmic pooling was given by \textcite{bg21}. Suppose that $Y$ is uniformly either $0$ or $1$ and that there are $k$ signals that are independent conditioned on $Y$, each known by a subset of experts. If the posterior probability that $Y = 1$ conditioned on each signal individually can be recovered from the experts' forecasts, then the optimal aggregate is a generalized linear pool.\footnote{More formally, let $A$ be the $m \times k$ matrix whose $(i, j)$-entry is $1$ if expert $i$ knows signal $j$. If there is a vector $\vect{h}$ such that $A^\top \vect{h} = \vect{1}_k$, then taking a generalized logarithmic pool of the experts' forecasts with weight vector $\vect{h}$ is optimal.} By contrast, the justification given by Claim~\ref{claim:bayes_log_pool} does not rely on such a ``recovery'' assumption, instead relying on properties of normal distributions.
\chapter{Incentivizing precise forecasts} \label{chap:precision}
\emph{This chapter presents ``Binary Scoring Rules that Incentivize Precision'' \parencite{nnw21}. It assumes background on proper scoring rules presented in Section~\ref{sec:prelim_proper}.}\\

\emph{Summary:} Proper scoring rules -- by definition -- incentivize an expert to predict accurately (report their true belief). However, not all proper scoring rules equally incentivize \emph{precision.} In this chapter, we will consider a model in which a rational expert can refine their belief by repeatedly paying a fixed cost, and is incentivized to do so by a proper scoring rule.

Specifically, our expert aims to predict the probability that a biased coin flipped tomorrow will land heads, and can flip the coin any number of times today at a cost of $c$ per flip. Our first main result defines an \emph{incentivization index} for proper scoring rules,\footnote{We focus specifically on symmetric proper scoring rules for binary outcomes -- more details below.} and proves that this index measures the expected error of the expert's reported probability (where the number of flips today is chosen to maximize the expert's expected payoff, i.e.\ score minus cost). Our second main result finds proper scoring rule that has the lowest (i.e.\ optimal) incentivization index among all proper scoring rules.

\section{Introduction} \label{sec:chap3_intro}
The space of proper scoring rules is vast: as we saw in Section~\ref{sec:prelim_proper}, there is (roughly speaking) one proper scoring rule for every strictly convex function. A principal who wishes to elicit a forecast from an expert must select one such scoring rule to use -- but how? The quadratic and logarithmic scores are common choices because of their simplicity, but is there a more principled way to make this choice?

In many settings, the principal may care about the \emph{precision} of the experts' forecast. As a motivating example, consider the problem of guessing the probability that one of two competing advertisements will be clicked by a user. With zero effort, an expert could blindly guess that each is equally likely. But the expert can expend some cost in order to refine their forecast, i.e.\ make it more precise. For example, the expert could run a crowdsourcing experiment, paying users to see which link they would click. Any proper scoring rule will equally incentivize the expert to accurately report their resulting belief, but not all proper scoring rules equally incentivize the costly gathering of information.

And so, the motivating question of this work is: \emph{Which proper scoring rule most incentivizes the costly gathering of information?}

We propose a simple model to formally measure the extent to which a proper scoring rule incentivizes costly refinement of the expert's beliefs. Specifically, we consider a two-sided coin that comes up heads with probability $p$, and $p$ is drawn uniformly from $(0,1)$ (we refer to $p$ as the \emph{bias} of the coin). Tomorrow the coin will be flipped, and we ask the expert to guess the probability that it lands heads. Today, the expert can flip the coin (with bias $p$) any number of times, at cost $c$ per flip. While we choose this model for its mathematical simplicity, it captures examples like the one above surprisingly well: tomorrow, a user will be shown the two advertisements (clicking one). Today, the expert can run a crowdsourcing experiment and pay any number of workers $c$ to choose between the two ads. This simple model also captures weather forecasting using ensemble methods surprisingly well, and we expand on this connection in Section~\ref{sec:weather}.

With this model in mind, consider the following two extreme forecasts: on the one hand, the expert could never flip the coin, and always output a guess of $1/2$. On the other, the expert could flip the coin infinitely many times to learn $p$ exactly, and output a guess of $p$. Note that both forecasts are accurate: the expert is truthfully reporting their belief, and that belief is correct given the observed flips. However, the latter forecast is more precise. All proper scoring rules incentivize the expert to accurately report their true forecast in both cases, but different scoring rules incentivize the expert to flip the coin a different number of times. More specifically, every proper scoring rule induces a different optimization problem for the expert, thereby leading them to produce forecasts of different quality. In this model, the key question we answer is the following: \emph{which scoring rules best incentivize the expert to produce a precise forecast?} 

As our setting indicates, we will be considering \emph{binary} proper scoring rules -- meaning that the there are two possible outcomes, which we will label ``Yes'' (heads) and ``No'' (tails). Further, we will restrict attention \emph{symmetric} scoring rules -- meaning that the scoring rule treats ``Yes'' and ``No'' symmetrically. That is, we are interested in proper scoring rules $s$ with the property that for all $x \in [0, 1]$, the score of an expert who assigns probability $x$ to Yes, if Yes happens, is equal to the score of an expert who assigns probability $x$ to No, if No happens:
\[s((x \text{ Yes}, 1 - x \text{ No}); \text{Yes}) = s((1 - x \text{ Yes}, x \text{ No}); \text{No}).\]
This condition allows us to simplify notation: we will write $s(x)$ to mean the score of an expert who assigns probability $x$ to whichever outcome is realized. That is, our notation $s(x)$ refers to both $s((x \text{ Yes}, 1 - x \text{ No}); \text{Yes})$ and $s((1 - x \text{ Yes}, x \text{ No}); \text{No})$.

\subsection{Motivation: Relationship to ensemble weather forecasts} \label{sec:weather}
A major shift occurred in the field of weather forecasting around the turn of the 21st century. In the previous century, weather forecasting was viewed as inherently deterministic: a forecasting model would take as input some initial conditions and use differential equations to simulate future states of the atmosphere. However, atmospheric conditions are never perfectly known: our observational equipment only gives us data about bits and pieces of the Earth's atmosphere, while the rest of the picture must be completed with educated guesswork (see our discussion in Chapter~\ref{chap:intro}). Additionally, the chaotic nature of atmospheric phenomena meant that even small inaccuracies in initial conditions would produce substantial forecast inaccuracies even a few days into the future.

Starting in the early 1990s and continuing into the early 2000s, there was a paradigm shift away from deterministic forecasts and toward \emph{ensemble forecasts.} An ensemble is a collection of simulations based on different perturbations of a best guess about the initial conditions. Generally, ensembles consist of five to 100 simulations. The results of these simulations are then used to generate a overall forecast~\parencite{GR05}.

The initial conditions used in ensemble models are typically chosen by ``ensemble prediction systems,'' which attempt to sample initial conditions from a probability distribution based on real-world uncertainty. Each simulation can be thought of as a sample from the probability distribution over the future weather. For instance, if 60\% of simulations predict rain in New York seven days from now, then the ensemble model might estimate the chance of rain in New York seven days from now at 60\%, perhaps slightly adjusted based on a prior inferred from historical climate data~\parencite{GBR07}.

Each simulation can be thought of as a coin flip whose cost is measured in time, energy, or computational resources. Each additional simulation has a constant cost. The final forecast for a weather event is (roughly speaking) the fraction of simulations in which the event occurred. In this way, ensemble forecasting strongly parallels our coin flip-based model of expert learning.

\subsection{Our results}
For a real number $\ell \ge 1$, let $\errf_c^\ell(s)$ be the expected value of the $\ell$-th power of the absolute error that a rational expert makes when incentivized with scoring rule $s$ with cost $c$ per flip. For example, $\errf_c^2(s)$ is the expected squared error of the expert (i.e.\ the squared difference between the true bias of the coin and the expert's report).\footnote{In a sense, $\ell = 2$ is the most ``internally consistent'' choice: the value minimizing the expected squared distance to the true bias is exactly the mean of the expert's probability distribution over the bias, which is the number that $s$ elicits. However, it is reasonable to ask about other values of $\ell$ as well.}

Our first main result is the existence of an \emph{incentivization index}. Specifically, for every $\ell \ge 1$, we give a closed-form index $\ind^\ell(s)$ with the following remarkable property: for all respectful (see Definition~\ref{def:respectful}) proper scoring rules $s_1$ and $s_2$, the inequality $\ind^\ell(s_1) < \ind^\ell(s_2)$ implies the existence of a sufficiently small $c_0 > 0$ such that $\errf_c^\ell(s_1) < \errf_c^\ell(s_2)$ for all $c \leq c_0$ (Theorem~\ref{thm:global}). We formally introduce this index in Definition~\ref{def:index}, but remark here that it is not a priori clear that such an index should exist at all, let alone that it should have a closed form.\footnote{Indeed, a priori it is possible that $\errf_{0.1}^\ell(s_1) < \errf_{0.1}^\ell(s_2)$, but $\errf_{0.01}^\ell(s_1) > \errf_{0.01}^\ell(s_2)$, and $\errf_{0.001}^\ell(s_1) < \errf_{0.001}^\ell(s_2)$, but $\errf_{0.0001}^\ell(s_1) > \errf_{0.0001}^\ell(s_2)$, and so on. The existence of an incentivization index rules out this possibility.}

With an index in hand, we can now pose a well-defined optimization problem for any given $\ell$: which proper scoring rule minimizes the incentivization index? Our second main result nails down this scoring rule precisely; we call it $s_{\ell,\opt}$ (see Theorem~\ref{thm:optimal}).

Some optimal rules $s_{\ell,\opt}$ have a particularly nice closed form (for example, as $\ell \rightarrow \infty$, the optimal rule pointwise converges to a polynomial), but many do not. We also prove, using techniques similar to the Weierstrass approximation theorem \parencite{Weierstrass1885}, that each of these rules can be approximated by polynomial proper scoring rules whose incentivization indices approach the optimum. 

Finally, beyond characterizing the optimal rules, the incentivization indices themselves allow for comparison among popular scoring rules, such as logarithmic, quadratic, and spherical ($s_{\text{sph}}(x):=x/\sqrt{x^2 + (1-x)^2}$). We plot the predictions made by our incentivization index (which provably binds only as $c \rightarrow 0$) for various values of $c$, and also confirm via simulation that the index is predictive for reasonable choices of $c$.

To summarize these results, we find that for all values of $\ell$, some relatively well-known proper scoring rule is very close to optimal. For small values of $\ell = 1, 2$, we find that the scoring rule $s_{\text{hs}}(x) := -\sqrt{\frac{1 - x}{x}}$ -- prominently used by \textcite{bb20} to prove a minimax theorem for randomized algorithms -- is very close to optimal. For larger values of $\ell$ (e.g.\ $\ell = 4$), the log score is near-optimal. For every large values of $\ell$ (e.g.\ $\ell = 16$), the quadratic score is near-optimal. And for much larger values of $\ell$ (e.g.\ $\ell = 128$), the spherical scoring rule is near-optimal. See Table~\ref{tab:first} and Figure~\ref{fig:excel} for more details.

Generally, these scoring rules are in decreasing order of how harshly they penalize assigning a very low probability to the eventual outcome (compare e.g.\ the log and quadratic scoring rules): the smaller the value of $\ell$, the more preferable it is to have a scoring rule that penalizes incorrect forecasts near the extremes.

\subsection{Related work}\label{sec:chap3_related}
To the best of our knowledge, \textcite{Osband89} was the first to consider scoring rules as motivating the predictor to seek additional information about the distribution before reporting their belief. This direction is revisited in \textcite{Clemen}, and has gained more attention recently \parencite{Tsakas, RoughgardenS17, HartlineLSW20}. While these works (and ours) each study the same phenomenon, there is little technical overlap and the models are distinct: each explores a different aspect of this broad agenda. For example, \textcite{RoughgardenS17} consider the predictor's incentive to outperform competing predictors (but there is no costly effort: the predictors' beliefs are still exogenous). \textcite{HartlineLSW20} (which is contemporaneous and independent of our work) is the most similar in motivation, but still has significant technical differences (beyond the two subsequent examples). On one hand, their model is more general than ours in that they consider multi-dimensional state spaces (rather than binary ones, in our model). On another hand, it is more restrictive in that they consider only two levels of effort (versus infinitely many, in our model). 

Our work also fits into the broad category of principal-agent problems. For example, works such as \textcite{CaiDP15, LiuC16, ChenILSZ18, ChenZ19} consider a learning principal who incentivizes agents to make costly effort and produce an accurate data point. Again, the models are fairly distinct, as these works focus on more sophisticated learning problems (e.g.\ regression), whereas we perform a more comprehensive dive into the problem of simply eliciting the (incentivized-to-be-precise) belief.

In summary, there is a sparse, but growing, body of work addressing the study of incentivizing effort in forming predictions, rather than just accuracy in reporting them. The above-referenced works pose various models to tackle different aspects of this agenda. In comparison, our model is arguably the simplest, and we develop a deep understanding of optimal scoring rules in this setting.

\section{Our model and preliminaries}\label{sec:chap3_prelims}
\subsection{Modeling the expert's behavior}\label{sec:chap3_incentives}
We model the expert as Bayesian. Specifically, the expert initially believes the coin bias is uniformly distributed in $(0,1)$. Today, the expert may flip the coin any number of times in order to gauge its true bias, and pays $c$ per flip.

\begin{fact}[Laplace's rule of succession] \label{fact:flipping}
    After having flipped the coin $n$ times, and seen $k$ heads, the expert believes\footnote{By this, we mean the expert believes the coin would land heads with probability $\frac{k+1}{n+2}$, if it were flipped again.} that the coin's bias is $\frac{k+1}{n+2}$.
\end{fact}

Once done flipping, the expert reports their belief about the coin's bias. Tomorrow, the coin is flipped once, and the expert is scored with a proper scoring rule $s$ (known to the expert in advance). 

It remains to define when the expert should stop flipping. Below, an \emph{adaptive strategy} simply refers to a (possibly randomized) stopping rule for the expert, i.e.\ a rule that, given any number of past flips and the proper scoring rule $s$, tells the expert whether to stop or to flip the coin again. The payoff of an adaptive strategy is simply the expected score of an expert who follows that strategy, minus $c$ times the expected number of coin flips.

\begin{defin}
A \emph{globally-adaptive expert} uses the payoff-maximizing adaptive strategy.
\end{defin}

Nailing down the expert's optimal behavior as a function of $c$ is quite unwieldy. Thus, we derive our characterizations up to $o(1)$ terms (as $c \rightarrow 0$). When $c$ is large, one may reasonably worry that these $o(1)$ terms render our theoretical results irrelevant. In Appendix~\ref{app:simulation} we simulate the expert's optimal behavior for large $c$, and confirm that our results hold qualitatively in this regime.

Finally, we define a natural measure of precision for the expert's prediction.

\begin{defin}
The \emph{expected $\ell$-th power error} associated with a proper scoring rule $s$ and cost $c$ is $\errf_c^\ell(s):= \EE{\abs{p-q}^\ell}$. The expectation is taken over the true bias $p$ of the coin, drawn uniformly from $(0,1)$, and $q$, the prediction of a globally-adaptive expert after flipping the coin as many times as they choose.
\end{defin}

\subsection{Scoring rule preliminaries}
Our proofs will make use of fairly heavy single-variable analysis, and therefore will require making some assumptions on $s$ such as differentiability, but also more technical ones. We will clearly state them when necessary, and confirm that all scoring rules of interest satisfy them. For these preliminaries, we need only assume that $s$ is continuously differentiable so that everything which follows is well-defined.

Lemma~\ref{lem:weaklyproper} provides a characterization of proper (and weakly proper) scoring rules in our (binary, symmetric) setting.

\begin{restatable}{lemma}{weaklyproper} \label{lem:weaklyproper}
A continuously differentiable scoring rule $s$ is weakly proper if and only if for all $p \in (0, 1)$, $ps'(p) = (1 - p)s'(1 - p)$ and $s'(p) \ge 0$. It is (strictly) proper if and only if additionally $s'(p) > 0$ almost everywhere\footnote{Almost everywhere on $(0,1)$ refers to the interval $(0,1)$ except a set of measure zero.} in $(0,1)$.
\end{restatable}

\begin{proof}[Proof sketch]
    If the expert believes that the true probability is $p$ and reports $x$, then their expected score is $ps(x) + (1 - p)s(1 - x)$. For every $p$, we want this expression to reach a maximum at $x = p$. The derivative of the expression with respect to $x$ is $ps'(x) - (1 - p)s'(1 - x)$. Thus, for $s$ to be proper, we should have that $ps'(p) = (1 - p)s'(1 - p)$ for all $p$.
\end{proof}

We defer the full proof of Lemma~\ref{lem:weaklyproper} to Appendix~\ref{app:prelim}.

\begin{restatable}{corollary}{strictlyproper} \label{cor:strictlyproper}
Let $s$ be strictly increasing almost everywhere (resp., nondecreasing everywhere) and continuously differentiable on $(0,\frac{1}{2}]$. Then $s$ can be extended to a continuously differentiable proper (resp., weakly proper) scoring rule on $(0,1)$ by defining $s'(p) = \frac{1-p}{p}s'(1-p)$ for $p \in \parens{\frac{1}{2}, 1}$.
\end{restatable}

Put another way: every continuously differentiable proper scoring rule can be defined by first providing a strictly increasing function on $(0,\frac{1}{2}]$, and then extending it as in Corollary~\ref{cor:strictlyproper}. For example, consider the function $s(x)=x$, which is strictly increasing on $(0,\half]$. Defining $s'(x) = \frac{1 - x}{x}\cdot 1 = \frac{1}{x} - 1$ for $x \in [\half, 1)$ results in $s(x) = \ln x - x + 1 + \ln 2$ (where $1 + \ln 2$ is the necessary constant to make $s$ continuous at $x = \half$). Clearly $s'(x) > 0$ on $(0, 1)$ (as promised by Corollary~\ref{cor:strictlyproper}), so we have just constructed a proper scoring rule:
\[s(x) = \begin{cases}x & \text{for } x \le \half \\ \ln x - x + 1 + \ln 2 & \text{for } x \ge \half\end{cases}.\]

\subsection{First steps towards understanding incentivization}\label{sec:chap3_proper}
We will be working with both $s$ and its expected score function $G_s(x) = x s(x) + (1 - x) s(1 - x)$ (see Section~\ref{sec:prelim_proper}). Note that $G_s$ is necessarily symmetric about $\frac{1}{2}$ (this is a consequence of our assumption that $s$ is symmetric). Additionally, the fact that $s$ is continuously differentiable implies that $G_s$ is as well.

\begin{remark}\label{remark:basic} For a weakly proper scoring rule $s$, we have $G_s'(x) = s(x) - s(1-x)$ and $G_s''(x) = s'(x) + s'(1-x) = \frac{s'(x)}{1 - x} \geq 0$ on $(0,1)$.
\end{remark}

(The last equality can be inferred by using Lemma~\ref{lem:weaklyproper}.) Lemma~\ref{lem:increward} observes how an expert's expected score evolves with an additional flip of the coin.

\begin{lemma}\label{lem:increward} For a proper scoring rule $s$, if the expert has already flipped the coin $n$ times, seeing $k$ heads, then their expected increase in score for exactly one additional flip is
\[\frac{k+1}{n+2}G_s\parens{\frac{k+2}{n+3}} + \frac{n-k+1}{n+2}G_s\parens{\frac{k+1}{n+3}} - G_s\parens{\frac{k+1}{n+2}}.\]
This quantity is positive.
\end{lemma}

\begin{proof}
This is a direct application of Laplace's rule of succession (Fact~\ref{fact:flipping}). Currently, the expert believes the probability of heads to be $\frac{k+1}{n+2}$. So their expected score if they stop flipping now is exactly $G_s \parens{\frac{k+1}{n+2}}$. If they flip once more and stop, then with probability $\frac{k+1}{n+2}$ they will get a heads, updating their belief to $\frac{k+2}{n+3}$, and yielding expected score $G_s \parens{\frac{k+2}{n+3}}$. With probability $\frac{n-k+1}{n+2}$ they will get a tails, updating their belief to $\frac{k+1}{n+3}$ and yielding expected score $G_s \parens{\frac{k+1}{n+3}}$.

The fact that the quantity in Lemma~\ref{lem:increward} is positive follows directly from the fact that $G_s$ is strictly convex.
\end{proof}

Proper scoring rules remain proper when scaled by a positive affine transformation. This presents an issue for comparing scoring rules based on their incentivization properties. Because we are interested in incentivizing the expert to take costly actions, so the scale of a proper scoring rule will be relevant. For example, if $s$ is proper, then so is $2s$, and $2s$ clearly does a better job of incentivizing the expert (since the quantity in Lemma~\ref{lem:increward} is larger by a factor of $2$). As such, we will want to first \emph{normalize} any scoring rules under consideration to be on the same scale.

A natural normalization is to consider two scoring rules to be on the same scale if they provide the same expected score to a perfect expert (one who knows the bias exactly). This is a natural choice because the expected score of a perfect expert is an upper bound on the expected payment that the principal must make. Further, as $c \to 0$, the expected score of a globally adaptive expert in fact approaches the expected score of a perfect expert (see Proposition~\ref{prop:cost}). Intuitively, this is because the number of flips approaches infinity as $c$ approaches $0$, so the expert's forecast becomes perfectly precise in this limit.

The expected score of a perfect expert is $\int_0^1 G_s(x) dx$, since a perfect expert has expected payoff $G_s(x)$ if the coin has bias $x$, and the coin's bias is chosen uniformly from $[0, 1]$. For this reason, when evaluating a proper scoring rule in terms of its incentivization properties, we will scale it so that $\int_0^1 G_s(x)dx = 1$.

Scaling proper scoring rules in this way addresses one potential issue, but there is another as well: consider a proper scoring rule $s$ satisfying $\int_0^1 G_s(x)dx = 1$. Then $\tilde{s}(x) := 2s(x) - 1$ is also proper and satisfies $\int_0^1 G_{\tilde{s}}(x)dx = 1$. However, $\tilde{s}$ clearly does a better job incentivizing the expert (again, by the positivity of the quantity in Lemma~\ref{lem:increward}). As such, we will also normalize $s$ so that the score of a completely uninformed expert -- one who flips the coin zero times and says $\frac{1}{2}$ -- is zero.

\begin{defin}\label{def:normalized} A weakly proper scoring rule $s$ is \emph{normalized} if $\int_0^1 G_s(x)dx = 1$, and $s(1/2) = 0$.
\end{defin}

We often use the following equivalent condition (the proof of equivalence is given in Appendix~\ref{app:prelim}).

\begin{restatable}{claim}{normalizedequiv} \label{claim:normalized_equiv}
    For a weakly proper scoring rule $s$, we have
    \[\int_0^1 G_s(x) dx = s \parens{\half} + \int_\half^1 (1 - x) s'(x) dx.\]
    Thus, $s$ is normalized if and only if $s(1/2) = 0$ and $\int_\half^1 (1 - x) s'(x) dx = 1$.
\end{restatable}

\section{An incentivization index}\label{sec:index}
This section presents our first main contribution: an incentivization index that characterizes the expert's expected error. The main result of this section, Theorem~\ref{thm:global}, requires scoring rules to be analytically nice in a specific way. We term such scoring rules \emph{respectful}.

\begin{defin} \label{def:respectful}
A proper scoring rule $s$ is \emph{respectful} if:
\begin{enumerate}[label=(\arabic*)]
    \item $G_s$ is strongly convex on $(0, 1)$. That is, $G_s''(x) \ge a$ on $(0, 1)$ for some $a > 0$.
    \item $G_s'''$ is Riemann integrable on any closed sub-interval of $(0, 1)$.\footnote{Note this does not necessarily require $G_s'''$ be defined on the entire $(0,1)$, just that it is defined almost everywhere.}
    \item There exists $t > \frac{1}{4}$, and $c_0 > 0$ such that for all $c \in (0,c_0)$: $\abs{G_s'''(x)} \le \frac{1}{c^{0.16} \sqrt{x(1 - x)}}G_s''(x)$ on $[c^t, 1 - c^t]$.\footnote{Except in places where $G_s'''$ is undefined.} 
\end{enumerate}
\end{defin}


Recall that $G_s$ is strictly convex for any (strictly) proper scoring rule, so strong convexity is a minor additional assumption. Likewise, the second condition is a minor ``niceness'' assumption. We elaborate on the third condition in detail in Appendix~\ref{app:respect}, and confirm that frequently used proper scoring rules are indeed respectful. We briefly note here that intuitively, the third condition asserts that $G_s''$ does not change too quickly (except possibly near zero and one) for small enough coin-flipping costs $c$. The particular choice of $0.16$ is not special, and could be replaced with any constant less than $1/6$. 

\begin{defin}[Incentivization Index] \label{def:index} For $\ell \ge 1$, we define the \emph{$\ell$-th power incentivization index} of a proper scoring rule $s$:

\[\ind^\ell(s):= \int_0^1 \parens{\frac{x(1 - x)}{G_s''(x)}}^{\ell/4} dx.\]
\end{defin}

\begin{restatable}{theorem}{globalthm} \label{thm:global}
If $\mu_\ell := \frac{2^{\ell/2} \Gamma \parens{\frac{\ell + 1}{2}}}{\sqrt{\pi}}$ is the $\ell^{th}$ moment of the standard normal distribution, then
\[\lim_{c \to 0} c^{-\ell/4}\cdot \errf^\ell_c(s) = \mu_\ell\cdot 2^{\ell/4} \cdot \ind^\ell(s).\]
\end{restatable}

Intuitively, the incentivization index captures the expert's error as $c \rightarrow 0$. More formally, for any two respectful proper scoring rules $s_1, s_2$, $\ind^\ell(s_1) < \ind^\ell(s_2)$ implies that there exists a sufficiently small $c_0 > 0$ such that $\errf_c^\ell(s_1) < \errf_c^\ell(s_2)$ for all $c \leq c_0$. As previously mentioned, Theorem~\ref{thm:global} says nothing about how big or small this $c_0$ might be, although simulations in Appendix~\ref{app:simulation} confirm that it does not appear to be too small for typical scoring rules.

Theorem~\ref{thm:global} says that the expert's expected $\ell$-th power error is proportional to $c^{\ell/4}$. What is the intuition for this asymptotic relationship? Speaking informally, the key facts are that the number of times that the expert flips the coin is proportional to $c^{-1/2}$, and that the expected absolute error is proportional to $n^{-1/2}$, where $n$ is the number of flips. The first fact follows from the fact that the expected improvement in the expert's score from an extra flip is proportional to $\frac{1}{n^2}$ (Claim~\ref{claim:rdoubleprime} below); this quantity thus falls below the cost $c$ when $n$ is on the order of $c^{-1/2}$. The second fact follows from the fact that the expert's estimate of the bias $p$ is (roughly) $\frac{1}{n}$ times a binomial random variable with $n$ trials and probability $p$ of success; the standard deviation of this quantity is proportional to $n^{-1/2}$.

The rest of this section is organized as follows. Sections~\ref{sec:global} through~\ref{sec:five} outline our proof of Theorem~\ref{thm:global}. The key steps are given as precisely-stated technical lemmas with mathematical intuition alongside them, to illustrate where precision is needed for the proof to carry through. Complete proofs of these lemmas can be found in Appendix~\ref{app:index}. In Appendix~\ref{app:respect}, we confirm that natural scoring rules are respectful (which is mostly a matter of validating the third condition in Definition~\ref{def:respectful}).

\subsection{Proof outline of Theorem~\ref{thm:global}}\label{sec:global}
Below, we provide an executive overview of our approach. The concrete steps are separated out as formally-stated technical lemmas in the following sections, with proofs deferred to Appendix~\ref{app:index}. Before beginning, we highlight the main challenge: to prove Theorem~\ref{thm:global}, we need to capture the \emph{precise asymptotics} of the expert's expected error. Upper bounds can be easily shown via concentration inequalities; however, traditional lower bounds via anti-concentration results would simply state that the expected error tends to $0$ as $c \rightarrow 0$ (which holds for every proper scoring rule, and doesn't distinguish among them). So not only are we looking for two-sided bounds on the error, but we need to gauge the precise \emph{rate} at which it approaches zero. Moreover, even obtaining the order of magnitude of the error as $c \to 0$, which turns out to be $c^{-\ell/4}$, still does not suffice: we need to compute the exact coefficient of $c^{-\ell/4}$. This difficulty motivates the need for the technical lemmas stated in this section to be very precise. Our outline is as follows:
\begin{itemize}
\item All of our analysis first considers a locally-adaptive expert, who flips the coin one additional time if and only if the expected increase in score \emph{from that single flip} exceeds $c$.
\item Our first key step, Section~\ref{sec:one}, provides a loose asymptotic \emph{lower bound} on the number of times an expert flips the coin, for all respectful $s$. 
\item Our second key step, Section~\ref{sec:two}, provides a coupling of the expert's flips across all possible true biases $p$. This helps prove uniform convergence bounds over all $p$ for the expert's error: we can now define an unlikely ``bad'' event of overly-slow convergence without reference to $p$.
\item Our third key step, Section~\ref{sec:three}, provides tight bounds on the number of flips by a locally-adaptive expert, up to $(1\pm o(1))$ factors. Note that the first three steps have not referenced an error measure at all, and only discuss the expert's behavior.
\item Our fourth key step, Section~\ref{sec:four}, shows how to translate the bounds in Section~\ref{sec:three} to tight bounds on the error of a locally-adaptive expert, again up to $(1\pm o(1))$ factors.
\item Finally our last step, Section~\ref{sec:five}, shows that the globally-adaptive expert behaves nearly-identically to the locally-adaptive expert, up to an additional $o(1)$ factor of flips.
\end{itemize}

We now proceed to formally state the main steps along this outline, recalling that the first several steps consider a locally-adaptive expert, whose definition is restated formally below:

\begin{defin}[Locally-Adaptive Expert]
The \emph{locally-adaptive expert} flips one more time if and only if making a \emph{single} additional coin flip (and then stopping) increases their expected payoff.
\end{defin}

\subsection{Step one: Lower bounding the expert's number of flips}\label{sec:one}
We begin by tying the expert's expected marginal score from one additional flip to the second derivative of the expected score function, $G_s''$. Below, $Q(n)$ denotes the random variable which is the expert's belief after $n$ flips. The important takeaway from Claim~\ref{claim:rdoubleprime} is that for fixed $n$, the expert's expected belief as a function of $Q(n)$ changes (roughly) as $Q(n)\cdot (1-Q(n)) \cdot G_s''(Q(n))$ -- this takeaway will appear in later sections. 

\begin{restatable}{claim}{rdoubleprime} \label{claim:rdoubleprime}
Let $\Delta_{n + 1}(q):=\EE{G_s(Q(n+1)) \mid Q(n) = q} - G_s(q)$ be the expected increase in the expert's score (not counting the paid cost $c$) from the $(n+1)^{th}$ flip of the coin, given current belief $Q(n)=q$. Then there exist $c_1,c_2 \in [q - 1/n,q + 1/n]$ such that:
\[\Delta_{n + 1} = \frac{q\cdot (1 - q)}{2(n + 3)^2}(q\cdot G_s''(c_1) + (1 - q)\cdot G_s''(c_2))\]
\end{restatable}

Recalling that the locally-adaptive expert decides to flip the coin for the $(n+1)^{th}$ time if and only if $\Delta_{n + 1} \ge c$, and assuming that $G_s''$ is bounded away from zero (Condition 1 in Definition~\ref{def:respectful}), we arrive at a simple lower bound on the number of coin flips.

\begin{restatable}{claim}{onethirdbound} \label{claim:one_third_bound}
For all $s$ such that $G_s''$ is bounded away from zero, there exists $\alpha, c_0$ such that the expert is guaranteed to flip the coin at least $\frac{1}{\alpha c^{1/3}}$ times for all $c \leq c_0$ (no matter the true bias).
\end{restatable}

Using basic concentration inequalities, Claim~\ref{claim:one_third_bound} immediately implies an asymptotic \emph{upper bound} on the expert's error. Recall, however, that we need a two-sided bound, and moreover that we need precise asymptotics of the error. Still, Claim~\ref{claim:one_third_bound} is the first step towards this.

\subsection{Step two: Ruling out irregular coin-flipping trajectories}\label{sec:two}
The expert's coin-flipping behavior depends on $Q(n)$, which depends on the fraction of realized coin flips which are heads, which itself depend on the coin's true bias $p$. Note, of course, that $Q(n)\rightarrow p$ as $n \rightarrow \infty$. If instead we had that $Q(n) = p$ \emph{exactly}, we could leverage Claim~\ref{claim:rdoubleprime} to better understand the number of flips as a function of $p$. Unfortunately, $Q(n)$ will not equal $p$ exactly, and it is even possible to have $Q(n)$ far from $p$, albeit with low probability. 

The challenge, then, is then how to handle these low-probability events, and importantly how to do so \emph{uniformly over $p$}. To this end, we consider the following coupling of coin-flipping processes over all possible biases. Specifically, rather than first drawing bias $p$ and then flipping coins with bias $p$, we use the following identically distributed procedure:
\begin{enumerate}[label=(\arabic*)]
\item Generate an infinite sequence $r_1, r_2, \dots$ of uniformly random numbers in $[0, 1]$.
\item Choose $p$ uniformly at random from $[0, 1]$. \label{step:choose_p}
\item For each $n$, coin $n$ comes up heads if and only if $r_n \le p$.
\end{enumerate}
Under this sampling procedure, $Q_p(n) := \frac{h_p(n) + 1}{n + 2}$ is the expert's estimate after flipping $n$ coins, where $h_p(n)$ is the number of heads in the first $n$ flips, if $p$ is the value chosen in step~\ref{step:choose_p}. 

With this procedure, we can now define a single bad event \emph{uniformly over all $p$}. Intuitively, $\Omega_N$ holds when, no matter what $p$ is chosen in step~\ref{step:choose_p}, the expert's Bayesian estimate of $p$ never strays too far from $p$ after $N$ flips. More formally, the complement of $\Omega_N$ is our single bad event:

\[\overline{\Omega_N} := \bigcup_{n = N}^\infty \bigcup_{j = 1}^{n - 1} \left\{\abs{Q_{j/n}(n) - \frac{j}{n}} > \frac{\sqrt{j(n - j)}}{2n^{1.49}} \right\}.\]
The expression on the right-hand side of the inequality can be rewritten as 
$\sqrt{\frac{\frac{j}{n} \parens{1 - \frac{j}{n}}}{n}} \cdot \frac{n^{.01}}{2}$,
where the radical term gives the order of the expected difference between $Q_{j/n}(n)$ and $\frac{j}{n}$. So intuitively, $\Omega_N$ holds unless the actual difference between $Q_{j/n}(n)$ and $\frac{j}{n}$ far exceeds its expected value. 

We have defined $\Omega_N$ so that, on the one hand, our subsequent analysis becomes tractable when $\Omega_N$ holds, and on the other hand, $\Omega_N$ fails to hold with probability small enough that our asymptotic results are not affected. Below, Claim~\ref{claim:omega_p} gives the property we desire from $\Omega_N$, and Claim~\ref{claim:omega_unlikely} shows that $\overline{\Omega_N}$ is unlikely. The key takeaway from Claim~\ref{claim:omega_p} is that when $\Omega_N$ holds, the expert's prediction is close to $p$ \emph{for all $n \geq N$ and $p \in (0,1)$} and this closeness \emph{shrinks with $n$}.

\begin{restatable}{claim}{omegap} \label{claim:omega_p}
The exists a sufficiently large $N_0$ such that for all $N\geq N_0$: if $\Omega_N$ holds, then \[\abs{Q_p(n) - p} \le \frac{\sqrt{p(1 - p)}}{n^{.49}} \quad \text{ for all } n \ge N \text{ and } p \in [1/n,1-1/n].\]
\end{restatable}

\begin{restatable}{claim}{omegaunlikely} \label{claim:omega_unlikely}
\[\pr{\overline{\Omega_N}} = O \parens{e^{-N^{.01}}}.\]
\end{restatable}

While it is trivial to see that $Q_p(n)$ approaches $p$ as $n \rightarrow \infty$, we reiterate that Claims~\ref{claim:omega_p} and~\ref{claim:omega_unlikely} guarantee quantitatively that: (a) when $\Omega_N$ holds, $|Q_p(n)-p|$ \emph{shrinks with $n$}, (b) the probability that $\Omega_N$ fails \emph{shrinks exponentially fast in $N$}, and (c) both previous bounds are \emph{uniform over $p$}. 

\subsection{Step three: Tightly bounding the expert's number of flips}\label{sec:three}

We now nail down the precise asymptotics of the number of the expert's flips as a function of the true bias $p$. This becomes significantly more tractable after assuming $\Omega_N$ holds. Below, the random variable $\nstop$ denotes the number of flips that a locally-adaptive expert chooses to make.

\begin{restatable}{prop}{nstopbound} \label{prop:n_stop_bound}
Assume that $\Omega_N$ holds for some $N$, and let $t$ be as in Definition~\ref{def:respectful}. There exists a constant $\gamma$ and cost $c_0 > 0$ such that for all $c \leq c_0$ and all $p \in [2c^t, 1 - 2c^t]$, we have
\[\sqrt{\frac{p(1 - p)G_s''(p)}{2c}(1 - \gamma c^{1/300})} \le \nstop \le \sqrt{\frac{p(1 - p)G_s''(p)}{2c}(1 + \gamma c^{1/300})}.\]
\end{restatable}
Proposition~\ref{prop:n_stop_bound} has two key aspects. First, the upper and lower bounds on $\nstop$ match up to a $1\pm o(1)$ factor. Second, the $o(1)$ term is independent of $p$. To get intuition for why $\nstop \approx \sqrt{\frac{p(1-p)G_s''(p)}{2c}}$, recall that Claim~\ref{claim:rdoubleprime} shows after $n$ flips, the expected marginal gain is $\Delta_{n + 1} \approx \frac{p(1 - p)}{2n^2} G_s''(p)$. This quantity first falls below $c$, the cost per flip, after $n = \sqrt{\frac{p(1 - p)G_s''(p)}{2c}}$ flips.

\subsection{Step four: Translating number-of-flips bounds to error bounds}\label{sec:four}
Having pinned down $\nstop$ quite precisely, we will now obtain a tight bound on the error of the locally-adaptive expert's reported prediction. By contrast, the previous three steps performed an analysis of the locally-adaptive expert's coin-flipping behavior, which does not depend on the choice of error metric. Lemma~\ref{lem:close} below is a formal statement of the main step of this process, which nails down the asymptotics of the error conditioned on $\Omega_N$. Below, $\err_c(p)$ denotes a random variable equal to the locally-adaptive expert's error (i.e.\ absolute difference between their report and the true bias) when the cost is $c$ and the true bias is $p$ (and the scoring rule $s$ is implicit). 

\begin{restatable}{lemma}{close} \label{lem:close}
Let $\ell\ge 1$ and $\mu_\ell := \frac{2^{\ell/2} \Gamma \parens{\frac{\ell + 1}{2}}}{\sqrt{\pi}}$ be the $\ell^{th}$ moment of a standard Gaussian. Let $N = \frac{1}{\alpha c^{1/3}}$ (so $N$ is implicitly a function of $c$). For all $p \in [2c^t, 1 - 2c^t]$ we have
\[(1 - o(1))\cdot  \mu_\ell \cdot \parens{\frac{2p(1 - p)}{G_s''(p)}}^{\ell/4} \le c^{-\ell/4} \cdot \EE{(\err_c(p))^\ell \mid \Omega_N} \le (1 + o(1)) \cdot\mu_\ell \cdot \parens{\frac{2p(1 - p)}{G_s''(p)}}^{\ell/4}\]
where the $o(1)$ term is a function of $c$ (but not $p$) that approaches zero as $c$ approaches zero.
\end{restatable}

Lemma~\ref{lem:close} is the key, but far from only, step in translating Proposition~\ref{prop:n_stop_bound} to tight bounds on the locally-adaptive expert's error. Intuitively, it states that the value of the expert's error will be, up to a $1\pm o(1)$ factor, consistent with what one would expect from using a quantitative central limit theorem in conjunction with the bound on \nstop\ from Proposition~\ref{prop:n_stop_bound}.

\subsection{Step five: From locally-adaptive to globally-adaptive behavior}\label{sec:five}
Finally, we extend our previous analysis from locally-adaptive to globally-adaptive experts. In particular, for a scoring rule that gives a finite expected score to a perfect expert, we prove that the globally-adaptive expert does not flip significantly more than a locally-adaptive expert would, and therefore their achieved errors are equal up to a $1\pm o(1)$ factor. Below, the random variable $n_g$ denotes the number of flips by the globally-adaptive expert.

\begin{restatable}{lemma}{globalhelper} \label{lem:global_helper}
Assume $s$ is respectful and normalizable (i.e.\ $\int_0^1 G_s(x) dx < \infty$). Let $\gamma$ be as in Proposition~\ref{prop:n_stop_bound}. There exists a $c_0 > 0$, such that for all $c \leq c_0$: If $\Omega_{\nstop}$ holds and $4c^t \le Q(\nstop) \le 1 - 4c^t$, then \[\nstop \leq n_g \le (1 + 6\gamma c^{1/300}) \nstop.\]
\end{restatable}

Lemma~\ref{lem:global_helper} is the key step in this portion of the analysis. The remaining work is to bound the impact of negligible events (such as $\Omega_{\nstop}$ failing, or $ Q(\nstop)$ being extremely close to $0$ or $1$) on our analysis. This completes our outline of the proof of Theorem~\ref{thm:global}.

\section{Finding optimal scoring rules} \label{sec:optimal}
Now that we have shown that the incentivization index characterizes how well any respectful scoring rule incentivizes a globally-adaptive expert to minimize error, we have a well-defined optimization problem: \emph{which normalized proper scoring rule has the lowest incentivization index} (and therefore minimizes the expert's expected error)? Recall the following necessary and sufficient set of conditions for a continuously differentiable and normalized scoring rule $s$ to be weakly proper:\footnote{Including weakly proper scoring rules in our optimization domain makes the analysis simpler. The optimal scoring rules are in fact strictly proper.}
\begin{itemize}
\item (Lemma~\ref{lem:weaklyproper}) For all $x \in (0, 1)$, $xs'(x) = (1 - x)s'(1 - x)$ and $s'(x) \ge 0$.
\item (Definition~\ref{def:normalized}, Claim~\ref{claim:normalized_equiv}) $s(1/2) = 0$ and $\int_\half^1 (1 - x)s'(x) dx = 1$.
\end{itemize}

So our goal is just to find the scoring rule which satisfies these constraints and minimizes the incentivization index:
\[\ind^\ell(s) = \int_0^1 \parens{\frac{x(1 - x)}{G_s''(x)}}^{\ell/4} dx = \int_0^1 \parens{\frac{x(1 - x)^2}{s'(x)}}^{\ell/4} dx.\]

(The last step follows from Remark~\ref{remark:basic}.) The main result of this section is the following theorem, whose proof we defer to Appendix~\ref{app:optimal}.

\begin{restatable}{theorem}{optimal} \label{thm:optimal}
The unique continuously differentiable normalized proper scoring rule which minimizes $\ind^\ell(s)$ is:
\[s_{\ell,\opt}(x) := \begin{cases}\kappa_\ell \int_\half^x (t^{\ell - 8} (1 - t)^{2\ell + 4})^{1/(\ell + 4)} dt & x \le \half \\
\kappa_\ell \int_\half^x (t^\ell (1 - t)^{2\ell - 4})^{1/(\ell + 4)} dt & x \ge \half.\end{cases}\]
\end{restatable}

While $s_{\ell,\opt}$ is certainly challenging to parse, importantly it has a closed form, and can thus be numerically evaluated. Section~\ref{sec:plots} contains several plots of these scoring rules, alongside traditional ones. Below we give an overview of our proof of Theorem~\ref{thm:optimal} (the full details of the proof can be found in Appendix~\ref{app:optimal}).

\subsection{Proof overview of Theorem~\ref{thm:optimal}}
As shown in Corollary~\ref{cor:strictlyproper}, the equation $xs'(x) = (1 - x)s'(1 - x)$ lets us extend $s$ uniquely in a continuous manner to $(0, 1)$ if we know $s$ on $[\half, 1)$. Thus, we can simply consider $s$ on $[\half, 1)$. For $s'$ to be nonnegative everywhere, it suffices for it to be nonnegative on $[\half, 1)$, because of the relation $xs'(x) = (1 - x)s'(1 - x)$. Also, observe that the integrand in the definition of the incentivization index is symmetric about $\half$; this is clear from the fact that $G_s''$ is symmetric about $\half$. This means that
\[\ind^\ell(s) = 2 \int_{\half}^1 \parens{\frac{x(1 - x)^2}{s'(x)}}^{\ell/4} dx.\]
Thus, our question can be phrased as follows: find the continuously differentiable function $s: [\half, 1) \to \RR$ satisfying $s \parens{\half} = 0$, $s'(x) \ge 0$, and $\int_\half^1 (1 - x)s'(x) dx = 1$, that minimizes
\[\int_{\half}^1 \parens{\frac{x(1 - x)^2}{s'(x)}}^{\ell/4} dx.\]

From this point, our problem is simply a continuous mathematical program. It is not obvious that the program should admit a closed-form solution, but it does. We defer all details to Appendix~\ref{app:optimal}, and just briefly note that we can formulate the problem exclusively as a function of $s'$, and then uniquely reconstruct $s$ using $s \parens{\half} = 0$. Once we have done this, we can take a Lagrangian relaxation by putting a multiplier on the constraint $\int_\half^1 (1 - x)s'(x) dx = 1$, and hope that the solution to the relaxation is continuous and satisfies $s'(x) \geq 0$. While this is not guaranteed to succeed, this method does in fact nail down the optimum. Below are the main technical lemmas that yield Theorem~\ref{thm:optimal}. Note that $h$ plays the role of $s'$ in these lemmas.

\begin{restatable}{lemma}{besth} \label{lem:besth}
For any $\ell \ge 1$, a function $h: [\half, 1) \to \RR_{\ge 0}$ satisfying $\int_\half^1 (1 - x)h(x) dx = 1$ that minimizes $\int_{\half}^1 \parens{\frac{x(1 - x)^2}{h(x)}}^{\ell/4} dx$ is $\tilde{h}_\ell(x) = \kappa_\ell (x^\ell (1 - x)^{2\ell - 4})^{1/(\ell + 4)}$, where $\kappa_\ell = \parens{\int_\half^1 (x(1 - x)^3)^{\ell/(\ell + 4)} dx}^{-1}$.
\end{restatable}

(Note that $\kappa_\ell$ is simply a normalization constant, so as to make $\int_\half^1 (1 - x)h(x) dx$ equal $1$.)

\begin{restatable}{corollary}{hunique} \label{cor:hunique}
The unique continuous function $h: [\half, 1) \to \RR_{\ge 0}$ satisfying $\int_\half^1 (1 - x)h(x) dx = 1$ that minimizes $\int_{\half}^1 \parens{\frac{x(1 - x)^2}{h(x)}}^{\ell/4} dx$ is $\tilde{h}_\ell$.
\end{restatable}

Theorem~\ref{thm:optimal} then follows from Corollary~\ref{cor:hunique} by setting $s_{\ell,\opt}$ to the integral of $\tilde{h}_\ell$ on $[1/2,1)$, and extending it to $(0,1/2)$ via Corollary~\ref{cor:strictlyproper}. For some choices of $\ell$, the particular scoring rule $s_{\ell,\opt}$ has an interesting closed form (see Section~\ref{sec:compare}), but this is not true for all $\ell$. Even in cases where the particular closed form is not illuminating, the fact that $s_{\ell,\opt}$ even exists is already interesting, and the fact that Theorem~\ref{thm:optimal} nails down the closed form allows us to compare other scoring rules to the optimum. We conclude with a remark, confirming that our analysis in Section~\ref{sec:index} indeed is meaningful for all derived optimal scoring rules.

\begin{restatable}{remark}{respect} \label{rem:respect}
For every $\ell \in [1, 8]$, $s_{\ell, \opt}$ is respectful. For $\ell > 8$, and all $\varepsilon > 0$, there exists a respectful normalized proper scoring rule $s$ such that $\abs{s(x) - s_{\ell, \opt}(x)} \le \varepsilon$ for all $x \in (0, 1)$, with $\ind^\ell(s) \leq \ind^\ell(s_{\ell,\opt})+\varepsilon$.
\end{restatable}

We give a proof for $\ell \in [1,8]$ in Appendix~\ref{app:index}. Meanwhile, the proof for $\ell > 8$ follows from the proof of Theorem~\ref{thm:approx} in Appendix~\ref{app:weierstrass}.\footnote{More specifically, the scoring rules $s_\varepsilon$ defined in the proof of Theorem~\ref{thm:approx} uniformly converge to $s_{\ell, \opt}$.}

The following corollary follows from Remark~\ref{rem:respect}.
\begin{corollary}
For $\ell \ge 1$, let
\[\errf_\opt^\ell := \inf_s \lim_{c \to 0} c^{-\ell/4} \cdot \errf_c^\ell(s)\]
where $s$ ranges over all normalized, respectful, continuously differentiable proper scoring rules. Let
\[\ind_\opt^\ell := \inf_s \ind^\ell(s)\]
where $s$ ranges over all normalized, continuously differentiable proper scoring rules. Then:
\begin{enumerate}[label=(\arabic*)]
\item \label{item:corollary_1} $\errf_\opt^\ell = \mu_\ell\cdot  2^{\ell/4}\cdot \ind_\opt^\ell$.
\item \label{item:corollary_2} For $1 \le \ell \le 8$, the first infimum is uniquely achieved by $s = s_{\ell, \opt}$.
\item \label{item:corollary_3} For $\ell > 8$, no (respectful) function achieves the first infimum, but the infimum is reached in the limit by uniform approximations of $s_{\ell, \opt}$ (which are normalized, respectful, and continuously differentiable).
\end{enumerate}
\end{corollary}

\section{Comparing scoring rules}\label{sec:compare}
In this section we compare various scoring rules by their incentivization indices, for various values of $\ell$. Of particular interest are the values $\ell = 1$ (expected absolute error), $\ell = 2$ (expected squared error), and the limit as $\ell \to \infty$ (which penalizes bigger errors ``infinitely more'' than smaller ones, so this regime corresponds to minimizing the probability of being very far off).

\subsection{Optimal scoring rules for particular values of $\ell$} \label{sec:compare_opt}
We begin by noting some values of $\ell$ for which the function $s_{\ell, \opt}$ takes a nice closed form. $\ell = 1$ happens to not be one such value. For $\ell = 2,4,8$, the functions $s_{\ell,\opt}$ can be written in terms of elementary functions on the entire interval $(0,1)$. For $\ell = 2$, the closed form on $(1/2,1)$ is a polynomial, although its extension via Corollary~\ref{cor:strictlyproper} to $(0,1/2)$ is not. For $\ell = 8$, the closed form on both $(0,1/2)$ and $(1/2,1)$ is a polynomial, although they are different. Interestingly, as $\ell \rightarrow \infty$, the closed form converges pointwise to a single polynomial. Specifically, for these values of $\ell$:

\textbf{For $\mathbf{\ell = 2}$:} On $[\half, 1)$, we have
\[s_{2, \opt}(x) = \kappa_2 \int_\half^x t^{2/3} dt = \frac{3}{5} \kappa_2\parens{x^{5/3} - \parens{\half}^{5/3}}.\]

\textbf{For $\mathbf{\ell = 8}$:} On $(0, \half]$, we have
\[s_{8, \opt}(x) = \kappa_8 \int_\half^x (1 - t)^{5/3} dt = \frac{3}{8} \kappa_8 \parens{\parens{\half}^{8/3} - (1 - x)^{8/3}}\]
and on $[\half, 1)$, we have
\[s_{8, \opt}(x) = \kappa_8 \int_\half^x (t^{2/3} - t^{5/3}) dt = \kappa_8 \parens{\frac{3}{5}\parens{x^{5/3} - \parens{\half}^{5/3}} - \frac{3}{8} \parens{x^{8/3} - \parens{\half}^{8/3}}}.\]

Finally, \textbf{as $\mathbf{\ell \to \infty}$}: on the entire interval $(0, 1)$, $s_{\ell, \opt}$ pointwise converges to
\[\lim_{\ell \to \infty} \kappa_\ell \cdot \int_\half^x t(1 - t)^2 dt = \frac{320}{3} \parens{\frac{1}{4}x^4 - \frac{2}{3} x^3 + \frac{1}{2}x^2 - \frac{11}{192}} = \frac{5}{9}(48x^4 - 128x^3 + 96x^2 - 11).\]

We refer to this last rule as $s_{\infty, \opt}$. Intuitively, minimizing the expected value of error raised to a power that approaches infinity penalizes any error infinitely more than an even slightly smaller error. Put otherwise, this metric judges a scoring rule by the maximum (over $p \in (0, 1)$) of the spread of the distribution of expert error. The scoring rule $s_{\infty, \opt}$ has a very special property, which is that the quantity $\frac{x(1 - x)}{G_{s_{\infty, \opt}}''(x)} = \frac{x(1 - x)^2}{s_{\infty, \opt}'(x)}$, which appears in the incentivization index, is a constant regardless of $x$. This means that, in the limit as $c \to \infty$, the distribution of the expert's error is the same regardless of $p$. It makes intuitive sense that making the spread of the distribution of expert error uniform over all $p$ also minimizes the maximum of these spreads, which explains why $s_{\infty, \opt}$ has this interesting property.

As some of these rules are not infinitely differentiable, a natural question to ask is: what infinitely differentiable normalized function minimizes $\ind^\ell$? While (as we have shown by virtue of $s_{\ell, \opt}$ being the unique minimizer) achieving an incentivization index equal to $\ind^\ell(s_{\ell, \opt})$ with an infinitely differentiable scoring rule is impossible, it turns out that it is possible to get arbitrarily close -- and in fact it is possible to get arbitrarily close with \emph{polynomial} scoring rules. The main idea of the proof is to use the Weierstrass approximation theorem to approximate $s_{\ell, \opt}$ with polynomials. See Section~\ref{sec:weierstrass} for a full proof.

\subsection{Plots of some relevant scoring rules} \label{sec:plots}
\begin{figure}[ht]
    \centering
    \includegraphics[scale=0.85]{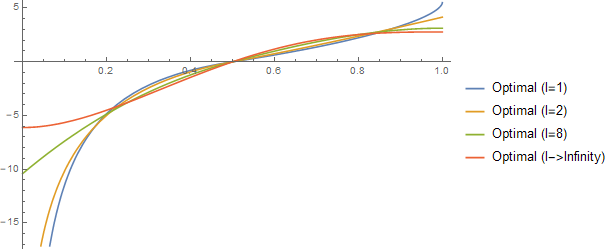}
    \caption[Optimal scoring rules under different error measures (plot 1)]{Plots of $s_{\ell, \opt}$ for $\ell = 1, 2, 8, \infty$.}
    \label{fig:opt_s_comparison}
\end{figure}

Figure~\ref{fig:opt_s_comparison} plots $s_{\ell,\opt}$ for $\ell = 1, 2, 8, \infty$. It demonstrates that optimal scoring rules for larger values of $\ell$ are ``flatter,'' choosing to sacrifice rewarding precision near $0$ and $1$, in favor of rewarding precision closer to $\frac{1}{2}$. An expert who is scored by $s_{\infty, \opt}$ does not particularly care to distinguish between 98\% and 99\% probabilities, since the scoring rule is basically flat near the tails; this is not the case for $s_{1, \opt}$. Conversely, because $s_{\infty, \opt}$ is steeper than $s_{1, \opt}$ near $\frac{1}{2}$, an expert cares more about differentiating between a 50\% and a 51\% chance if scored with $s_{\infty, \opt}$ than with $s_{1, \opt}$.

\begin{figure}[ht]
    \centering
    \includegraphics[scale=0.85]{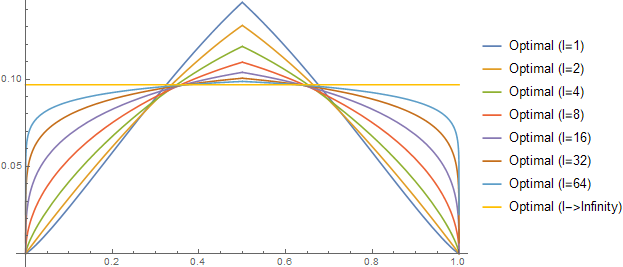}
    \caption[Optimal scoring rules under different error measures (plot 2)]{Plots of $\sqrt{\frac{x(1 - x)}{G_s''(x)}}$ for $s = s_{\ell, \opt}$ for a variety of values of $\ell$. Up to a constant factor, this quantity is the variance of the normal distribution that approximates the distribution of the expert's response if the true bias of the coin is $x$ (for small $c$).}
    \label{fig:opt_s_variance_comparison}
\end{figure}

Another, perhaps more enlightening way to view these scoring rules is through the quantity $\sqrt{\frac{x(1 - x)}{G_s''(x)}}$. Up to a constant factor depending on the cost $c$ of a flip, this is the variance of the normal distribution that approximates the distribution of the expert's response if the true bias of the coin is $x$ (for small $c$) -- or, put otherwise, the expected squared error. Figure~\ref{fig:opt_s_variance_comparison} plots this quantity for a variety of the scoring rules we have discussed.

Figure~\ref{fig:opt_s_variance_comparison} reinforces our previous point: optimal rules for small value of $\ell$ result in very small errors near $0$ and $1$, but relatively large errors in the middle. In Section~\ref{sec:compare_opt}, we discussed in brief why it makes sense that the value of $\frac{x(1 - x)}{G_s''(x)}$ is constant for the scoring rule $s_{\infty, \opt}$. We see this in Figure~\ref{fig:opt_s_variance_comparison}: since our normalization constraints force a trade-off between minimizing expert error for different values of the coin's bias $p$, the scoring rule whose error is independent of $p$ will have the minimum possible value of the maximum error over all $p$.

\begin{figure}[ht]
    \centering
    \includegraphics[scale=0.85]{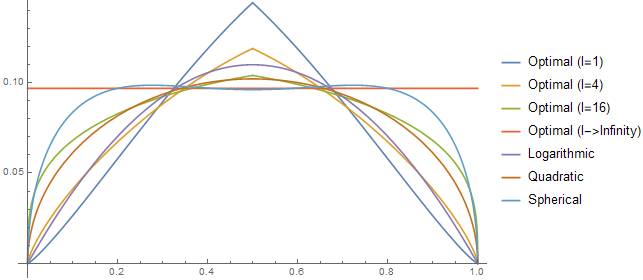}
    \caption[Comparison of common proper scoring rules to optimal ones]{Plots of $\sqrt{\frac{x(1 - x)}{G_s''(x)}}$ for $s = s_{\text{log}}$, $s_{\text{quad}}$, and $s_{\text{sph}}$, as well as some optimal rules. Up to a constant factor, this quantity is the variance of the normal distribution that approximates the distribution of the expert's response if the true bias of the coin is $x$ (for small $c$).}
    \label{fig:s_variance_comparison}
\end{figure}

Finally, Figure~\ref{fig:s_variance_comparison} is similar to Figure~\ref{fig:opt_s_variance_comparison}, except that it also includes the (normalized) logarithmic, quadratic, and spherical scoring rules. In Section~\ref{sec:comparison_ind}, we noted that the logarithmic rule is near-optimal for $\ell \approx 4$, the quadratic rule for $\ell \approx 16$, and the spherical rule for even larger $\ell$. Figure~\ref{fig:s_variance_comparison} helps provide some intuition for this fact: the log scoring rule is similar in shape to $s_{4, \opt}$ and similarly for the quadratic scoring rule and $s_{16, \opt}$, and for the spherical scoring rule and $s_{\infty, \opt}$.

\subsection{Comparison of incentivization indices of scoring rules} \label{sec:comparison_ind}
We compare commonly studied scoring rules such as quadratic, logarithmic, and spherical, and in this section we refer to their normalizations as $s_\text{quad},s_\text{log},s_\text{sph}$, respectively. Additionally we include for comparison $s_\text{hs}$, which is the normalization of the $hs$ scoring rule: $-\sqrt{\frac{1 - x}{x}}$. This scoring rule was prominently used by \textcite{bb20} to prove their minimax theorem for randomized algorithms. 

Table~\ref{tab:first} states $\ind^\ell(s)$ for various scoring rules $s$ (the lower the better). It lets us compare the performance of various scoring rules by our metric for any particular value of $\ell$. However, as one can see, $\ind^\ell$ decreases as $\ell$ increases. This makes sense, since $\ind^\ell$ measures the expected $\ell$-th power of error. For this reason, if we wish to describe how a given scoring rule performs over a range of values of $\ell$, we need to normalize these values. We do so by taking the $\ell$-th root and dividing these values by the $\ell$-th root of the optimal (smallest) index (and take the inverse so that larger numbers are better). This gives us the following measure of scoring rule precision, which makes sense across different values of $\ell$:
\[\parens{\frac{\ind^\ell(s_{\ell, \opt})}{\ind^\ell(g)}}^{1/\ell}.\]

\begin{table}[ht]
\begin{center}
\begin{tabular}{r||c|c|c}
$\ind^\ell(\cdot)$ & $\ell = 1$ & $\ell = 2$ & $\ell = 4$\\
\hline
$s_\text{log}$ & 0.260 & 0.0732 & 0.00644\\
$s_\text{quad}$ & 0.279 & 0.0802 & 0.00694\\
$s_\text{sph}$ & 0.296 & 0.0889 & 0.00819\\
$s_\text{hs}$ & 0.255 & 0.0723 & 0.00658\\
$s_{1,\opt}$ & 0.253 & 0.0728 & 0.00719\\
$s_{2,\opt}$ & 0.255 & 0.0718 & 0.00661\\
$s_{4,\opt}$ & 0.261 & 0.0732 & 0.00639\\
$s_{\infty,\opt}$ & 0.311 & 0.0968 & 0.00974
\end{tabular}
\end{center}\caption[Incentivization indices of various scoring rules and error measures]{A table of incentivization index values for various proper scoring rules under various error measures.}\label{tab:first}
\end{table}

\begin{figure}[ht]
\begin{center}
\includegraphics[scale=.7]{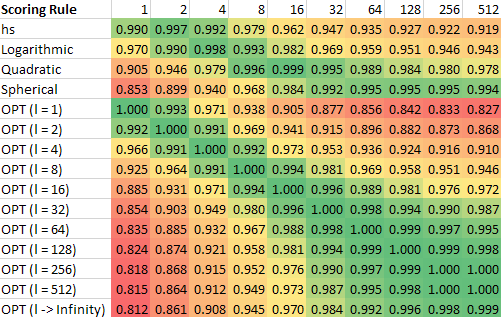}
\end{center}\caption[Incentivization indices of scoring rules relative to optimal values]{Incentivization indices of various proper scoring rules relative to the optimal proper scoring rule, for a variety of error measures. The values in the table are $\parens{\frac{\ind^\ell(s_{\ell, \opt})}{\ind^\ell(s)}}^{1/\ell}$.}\label{fig:excel}
\end{figure}

\begin{figure}[!ht]
\begin{center}
\includegraphics[scale=.45]{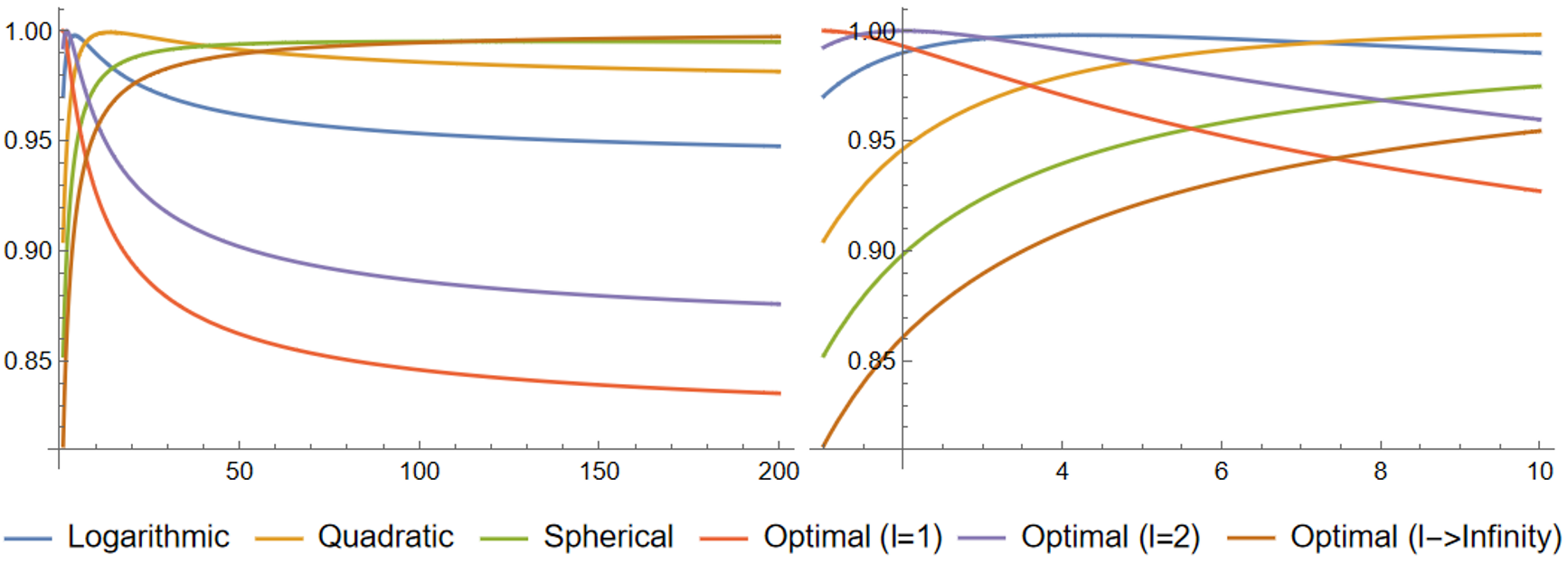}
\end{center}\caption[Incentivization indices relative to optimal values, continuous version]{Incentivization indices of various proper scoring rules relative to the optimal proper scoring rule. On the left, $\ell$ varies between $1$ and $200$. The chart on the right is a zoomed-in version of the same plot, showing values of $\ell$ between $1$ and $10$. The plotted values are $\parens{\frac{\ind^\ell(s_{\ell, \opt})}{\ind^\ell(s)}}^{1/\ell}$.}\label{fig:comparison_200_10}
\end{figure}

Figure~\ref{fig:excel}, which evaluates this expression for a selection of proper scoring rules and values of $\ell$, reveals some interesting patterns. Of the hs, logarithmic, quadratic, and spherical scoring rules, the hs scoring rule is the best one for the smallest values of $\ell$ and is in fact near-optimal for $\ell = 2$. The logarithmic rule is the best one for somewhat larger values of $\ell$ and is in near-optimal for $\ell \approx 4$. For larger values of $\ell$, the quadratic scoring rule is best, and is near-optimal for $\ell \approx 16$. For even larger values of $\ell$, the spherical scoring rule is the best of the four. This pattern suggests that for any given proper scoring rule there is a trade-off between incentivizing precision at low and at high values of $\ell$; it would be interesting to explore this further. Figure~\ref{fig:comparison_200_10} is a continuous version of Figure~\ref{fig:excel}. It shows how the numbers in the table vary over a continuum of values of $\ell$.

\section{Polynomial scoring rules with almost-optimal incentivization indices} \label{sec:weierstrass}
The main result of this section is the following theorem, stating that polynomial,\footnote{When we say a scoring rule $s$ is polynomial, we mean simply that $s$ is a polynomial function.} respectful proper scoring rules suffice to get arbitrarily close to the optimal incentivization index.

\begin{restatable}{theorem}{approxthm} \label{thm:approx}
For $\ell \ge 1$ and $\varepsilon > 0$, there exists a respectful polynomial normalized proper scoring rule $s$ satisfying $\ind^\ell(s) \leq \ind^\ell(s_{\ell,\opt})+\varepsilon$.
\end{restatable}

The proof of Theorem~\ref{thm:approx} uses ideas from the Weierstrass approximation theorem. However, the Weierstrass approximation theorem gives a particular measure of ``distance'' between two functions, which does not translate to these functions having similar incentivization indices. So one challenge of the proof is ensuring convergence of a sequence of polynomials to $s_{\ell,\opt}$ \emph{in a measure related to $\ind^\ell$}. A second challenge is to ensure that all polynomials in this sequence are themselves proper, respectful scoring rules. Like previous technical sections, we include a few concrete lemmas to give a sense of our proof outline.

For example, one step in our proof is to characterize all \emph{analytic} proper scoring rules (that is, proper scoring rules that have a Taylor expansion which converges on their entire domain $(0,1)$). A necessary condition to be analytic is to be infinitely differentiable, which rules of the form $s_{\ell,\opt}$ are not, for any fixed $\ell$. We therefore seek to approximate such scoring rules with polynomial scoring rules (which are analytic), which are also respectful and proper.

\begin{restatable}{theorem}{analytic} \label{thm:analytic}
Let $s: (0, 1) \to \RR$ be analytic (and symmetric, as we have been assuming). Then $s$ is a proper scoring rule if and only if $s$ is nonconstant, $s'(x) \ge 0$ everywhere, and
\[s(x) = c_0 + \sum_{k > 0 \text{\emph{ odd}}} c_k(2k + 1 - 2kx)\parens{x - \half}^k\]
for some $c_0, c_1, c_3, c_5, \dots \in \RR$.
\end{restatable}

As an example to help parse Theorem~\ref{thm:analytic}, the quadratic scoring rule has $c_1 < 0$, and $c_i = 0$ for all other $i$. Using Theorem~\ref{thm:analytic}, we can conclude the following about $G_s''$ for any proper scoring rule $s$:

\begin{restatable}{lemma}{taylorphil} \label{lem:taylorphil}
Let $s: (0, 1) \to \RR$ be analytic and symmetric. Then $s$ is a proper scoring rule if and only if $G_s''$ is not uniformly zero, nonnegative everywhere, and can be written as
\[G_s''(x) = \sum_{k \ge 0 \text{ even}} d_k \parens{x - \half}^k.\]
\end{restatable}

Lemma~\ref{lem:taylorphil} provides clean conditions on what functions $G_s''$ are safe to use in our sequence of approximations, and our proof follows by following a Weierstrass approximation-type argument while keeping track of these conditions. The rest of the details for the proof of Theorem~\ref{thm:approx} can be found in Appendix~\ref{app:weierstrass}.

\section{Conclusion}
We proposed a simple model in which an expert can expend costly effort to refine their prediction, and studied the effectiveness of different proper scoring rules in incentivizing the expert to form a precise belief. Our first main result (Theorem~\ref{thm:global}) identified the existence of a closed-form incentivization index: scoring rules with a lower index incentivize the expert to be more precise. Our second main result (Theorem~\ref{thm:optimal}) identified the unique optimal scoring rule with respect to this index. Section~\ref{sec:compare} then used the incentivization index to compare common proper scoring rules (including our newly-found optimal ones), and Section~\ref{sec:weierstrass} showed that one can get arbitrarily close to the optimal incentivization index with polynomial proper scoring rules.

Our model is mathematically simple to describe, and yet it captures realistic settings surprisingly well (see Section~\ref{sec:chap3_intro}). As such, there are many interesting directions for future work.

First: what about measures of error other than the $\ell$-th power of absolute distance? In Section~\ref{sec:prelim_proper}, we introduced Bregman divergences and explained their relationship to proper scoring rules. In particular, if an expert is scored according to a proper scoring rule $s$, then the expert's expected loss relative to perfectly reporting the coin's true bias is equal to the Bregman divergence from the true bias to their report with respect to $G_s$. It is thus natural to consider the expected Bregman divergence as a notion of the expert's error. One might ask several questions related to this:

\begin{itemize}
    \item For a given $G$, what is the optimal proper scoring rule for Bregman divergence with respect to $G$? (We have already answered this question for $G(x) = x^2$: namely, $s_{2, \opt}$.) The case of $G$ being the negative of binary entropy (in which case the Bregman divergence is KL divergence) seems particularly interesting.
    \item Conversely, is there a Bregman divergence with respect to which the quadratic scoring rule is optimal? What about the logarithmic scoring rule?
    \item As a general rule, will a proper scoring rule $s$ be unusually good at incentivizing precision, if the measure of error is Bregman divergence with respect to $G_s$? We already know that a proper scoring rule $s$ isn't necessarily \emph{optimal} for Bregman divergence with respect to $G_s$: in particular, $s_{\text{quad}}$ isn't optimal for $\ell = 2$. However, perhaps a weaker statement to this effect could be true. Is there a proper scoring rule $s$ that is optimal for Bregman divergence with respect to $G_s$?
\end{itemize}

Other parameters of our model can also be varied to yield interesting questions:

\begin{itemize}
    \item In our model, the expert receives information in discrete chunks (one coin flip at a time). The expert's probability distribution over the bias of the coin is always a beta distribution with integer parameters $\alpha$ and $\beta$. Instead, we could model an expert as continually receiving information, so that their probability distribution over the bias of the coin is modeled as a beta distribution with continuously changing real parameters $\alpha$ and $\beta$. (This is inspired by an approach taken by \textcite{fck15} for a related problem.) In addition to being an interesting variant of our model, a continuous model like this one may simplify the analysis. Other natural priors and processes of information gain may also be interesting to explore.
    \item Our work considers a globally-adaptive expert, and establishes that they behave nearly identically to a locally-adaptive expert. What about a non-adaptive expert, who must decide a priori how many coin flips to make before seeing the flip outcomes?
\end{itemize}
\chapter{Arbitrage-free contract functions} \label{chap:arbitrage}
\emph{This chapter presents ``Strictly Proper Contract Functions Can Be Arbitrage-Free'' \parencite{nr22_contract}, although all contents starting with Definition~\ref{def:redist_arb} are new. It assumes background on proper scoring rules presented in Section~\ref{sec:prelim_proper}.}\\

\emph{Summary:} We consider mechanisms for truthfully eliciting probabilistic predictions from a group of experts. The standard approach -- using a proper scoring rule to separately reward each expert -- is not robust to collusion: experts may collude to misreport their beliefs in a way that guarantees them a larger total reward no matter the eventual outcome. It is a long-standing open question whether there is a truthful elicitation mechanism that makes any such collusion (also called \emph{arbitrage}) impossible. We resolve this question positively, exhibiting a class of proper, arbitrage-free contract functions. These contract functions have two parts: one ensures that the total reward of a coalition of experts depends only on the average of their reports; the other ensures that changing this average report hurts the experts under at least one outcome.

\section{Introduction}
If a principal wishes to elicit a probabilistic forecast from an expert, they may pay the expert using a proper scoring rule. But in many settings, the principal may want to elicit forecasts from multiple experts, so as to get a better sense of expert opinion and the extent to which there is a consensus. The principal could use a proper scoring rule to elicit each expert's forecast. If experts are not allowed to collude, then this strategy is incentive-compatible; however, \textcite{french83} observed that experts can collude in a way that increases the sum total profit of all experts, no matter the final outcome.

For example, recall the quadratic scoring rule from Section~\ref{sec:prelim_proper}:
\[s_{\text{quad}}(\vect{x}; j) := -(1 - x_j)^2 - \sum_{j' \neq j} x_{j'}^2.\]
Suppose that three experts believe that there is a 40\%, 50\%, and 90\% chance of rain, and that they are paid according to the proper scoring rule $s(\vect{x}; j) = s_{\text{quad}}(\vect{x}; j) + 1$. If it ends up raining, then the sum of their scores will be $0.28+0.5+0.98=1.76$; if it doesn't rain, then the sum of their scores will be $0.68+0.5-0.62=0.56$. However, suppose that the experts all collude to report 60\%. Then the sum of their scores is $0.68+0.68+0.68=2.04$ if it rains and $0.28+0.28+0.28=0.84$ if it doesn't: a larger number in both cases! The experts can agree beforehand to a redistribution of their rewards that guarantees every expert a larger profit than if they had not colluded.

\textcite{cs11} called this phenomenon -- in which experts collude to misreport in a way where their total reward is larger no matter the outcome -- \emph{arbitrage}. They showed that every proper scoring rule admits arbitrage -- indeed, that there is an arbitrage opportunity for any group of experts so long as they do not all agree on the probability of the event. Specifically, a coalition of experts can risklessly make a profit by deviating to report an aggregate of their beliefs (in the case of the quadratic scoring rule, this aggregate is the arithmetic mean). In Chapter~\ref{chap:qa}, we will extend this observation to probability distributions over more than two possible outcomes.

For many reasons, the expert may wish to make arbitrage impossible. First, the principal may wish to know whether the experts are in agreement: if they are not, for instance, the principal may want to elicit opinions from more experts. If the experts collude to report an aggregate value (as in our example), the principal does not find out whether they originally agreed. Second, even if the principal only seeks to act based on some aggregate of the experts' opinions, their method of aggregation may be different from the one that experts use to collude. For instance, the principal may have a private opinion on the trustworthiness of each expert and wishes to average the experts' opinions with corresponding weights. Collusion among the experts denies the principal this opportunity. Third, a principal may wish to track the accuracy of each individual expert (to figure out which experts to trust more in the future, for instance), and collusion makes this impossible. Fourth, the space of collusion strategies that constitute arbitrage is large. In our example above, any report in $[0.546, 0.637]$ would guarantee a profit; and this does not even mention strategies in which experts report different probabilities. As such, the principal may not even be able to recover basic information about the experts' beliefs from their reports.

As we have discussed, preventing arbitrage is impossible if the principal chooses a proper scoring rule and uses it to reward all of the experts. However, the principal has more freedom than this: they may choose to make each expert's reward depend not only on that expert's report but also \emph{other} experts' reports. \textcite[\S5]{cs11} ask whether there is any mechanism for rewarding experts that makes arbitrage impossible, concluding that this ``seems unlikely.'' \textcite{fppw20} explore this question further, proposing a mechanism that prevents arbitrage if the experts' reports are guaranteed to be in the range $[\epsilon, 1 - \epsilon]$ for some positive $\epsilon$ (though their mechanism may require very large payments if $\epsilon$ is small). However, they leave open Chun and Shachter's question of whether an incentive-compatible, arbitrage-free reward mechanism exists.

We resolve this question in the affirmative by exhibiting a class of incentive-compatible mechanisms in which arbitrage from collusion is impossible. Our mechanism takes inspiration from Brier's quadratic scoring rule, but modifies it to take into account the aggregate performance of the remaining experts.

\section{Related work}
\textcite{fppw20} explore the question of whether proper arbitrage-free mechanisms exist by proving positive results under different relaxations of these constraints. Their main result is a proper arbitrage-free mechanism under the restriction that the range of allowed reports is restricted to $[\epsilon, 1 - \epsilon]$. However, their mechanism necessitates payments that are exponentially large in $1/\epsilon$. Alternatively, these payments can be scaled down, but at the expense of giving essentially zero reward to each expert on the vast majority of the interval of possible reports, thus providing little incentive for truthful reporting. They also exhibit a positive result if the properness criterion is somewhat relaxed to allow for some contract functions that are weakly proper but not proper.

\textcite{cdpv14} explore the different but related topic of \emph{arbitrage-free wagering mechanisms.} In a wagering mechanism, each expert wagers a certain amount of money along with their report, and the pool of wagers is redistributed among the experts depending on each expert's report and wager and the eventual outcome. In this setting, they define arbitrage as any opportunity for an individual to risklessly make a profit. That is, an arbitrage opportunity is one in which an expert may unilaterally deviate by submitting a report that guarantees a profit no matter the final outcome. This differs from Chun and Shachter's definition of arbitrage, which is concerned with riskless profit opportunities stemming from collusion between experts.

The most well-known wagering mechanism is the \emph{weighted score wagering mechanism}, which rewards each expert based on their performance compared to other experts according to a proper scoring rule. An expert may risklessly profit from a weighted score wagering mechanism by reporting an aggregate of other experts' reports. This is the same aggregate as the one that a coalition of experts who are rewarded with a proper scoring rule may report in order to risklessly make a profit in our setting. \textcite[\S4.1]{cdpv14} define \emph{no-arbitrage wagering mechanisms}, which modify the reallocation rule of weighted score wagering mechanisms to reward each expert based on their performance relative to the performance of the aggregate of all other experts' reports. No-arbitrage wagering mechanisms can be re-interpreted in our setting as contract functions that prevent the \emph{entire group of experts} from colluding. However, as we discuss in the next section, this is easy to accomplish; we are instead faced with the challenge of preventing collusion between any coalition of experts of any size. Thus, our mechanism and theirs share some of the same spirit, but are different mechanisms that solve different problems.

\section{Preliminaries on contract functions}
Contract functions, defined by \textcite{cs11}, generalize scoring rules to multiple experts. We say that there are $m$ experts; for $i \in [m]$, expert $i$ reports a probability distribution $\vect{p}_i \in \Delta_n$ over $n$ outcomes. We denote the $j$-th coordinate of $\vect{p}_i$ as $p_{i, j}$.

A \emph{contract function} is any function that takes as input the $m$ experts' reports and the outcome, and outputs the reward of each expert. Formally, a contract function is any function $\Pi: (\Delta_n)^m \times [n] \to \RR^m$; if the experts report distributions $\vect{p}_1, \dots, \vect{p}_m$ and the outcome is $j$, then the vector of expert rewards is $\Pi(\vect{p}_1, \dots, \vect{p}_m; j)$. We let $\Pi_i(\cdot)$ denote the $i$-th coordinate of $\Pi(\cdot)$, i.e.\ expert $i$'s reward. We will generally use $\vect{P}$ to denote the $m$-tuple of reports $(\vect{p}_1, \dots, \vect{p}_m)$.

A contract function is \emph{weakly proper} if for each $i \in [m]$, expert $i$ maximizes their expected reward by reporting their belief $\vect{b}_i$, no matter the reports $\vect{p}_{-i}$ of the other experts. Formally, $\Pi$ is weakly proper if for all $i \in [m]$, for all $\vect{b}_i$ and all $\vect{p}_{-i}$, $\sum_j b_{i, j} \Pi_i(\vect{x}, \vect{p}_{-i}; j)$ is maximized at $\vect{x} = \vect{b}_i$. We say that $\Pi$ is \emph{proper} if $\vect{x} = \vect{b}_i$ is the unique maximizer, i.e.\ that an expert does strictly worse by misreporting their belief.

Our goal is to exhibit a proper contract function that does not permit arbitrage from collusion. We use the definition of arbitrage given by \textcite{fppw20}, which was adapted from \textcite{cs11}.

A contract function $\Pi$ \emph{admits arbitrage} if there is a coalition (i.e.\ subset) $C \subseteq [m]$ of experts and $m$-tuples of expert reports $\vect{P}$ and $\vect{Q}$, with $\vect{p}_i = \vect{q}_i$ for all $i \not \in C$, such that
\[\sum_{i \in C} \Pi_i(\vect{Q}; j) \ge \sum_{i \in C} \Pi_i(\vect{P}; j)\]
for all $j \in [n]$, and the inequality is strict for some $j$. We say that $\Pi$ is \emph{arbitrage-free} if it does not admit arbitrage. Intuitively, $\Pi$ admits arbitrage if it is possible for a coalition of experts to collude to misreport their values in such a way that the total reward of the experts in the coalition ends up larger, no matter the outcome. (Above, the misreport is $\vect{Q}$; the constraint that $\vect{p}_i = \vect{q}_i$ for $i \not \in C$ means that only experts in $C$ change their reports.) If this is possible, then the experts in $C$ can commit beforehand to a redistribution of the extra reward in a way that makes every expert in the coalition better off no matter the eventual outcome $j$.

\begin{remark}
	Positive affine transformations preserve both properness and arbitrage-freeness. That is, if $\Pi$ is proper then so is $a\Pi + b$ for any $a > 0$ and $b$, and this is likewise true for arbitrage-freeness.
\end{remark}

The question posed by \textcite{cs11} and explored by \textcite{fppw20}, which we answer affirmatively in this work, is: \textbf{Does there exist a proper arbitrage-free contract function?}

In the case of $m = 2$ experts, there is a straightforward solution:
{ \begin{equation} \label{eq:solution_m2}
		\Pi(\vect{p}_1, \vect{p}_2; j) = \parens{s_{\text{quad}}(\vect{p}_1; j) - s_{\text{quad}}(\vect{p}_2; j), s_{\text{quad}}(\vect{p}_2; j) - s_{\text{quad}}(\vect{p}_1; j)}.
\end{equation}}%
This contract function is proper because expert 1's reward is the (proper) quadratic score of their report plus a term that does not depend on their report, and likewise for expert 2. It is arbitrage-free because the total reward of the two experts is $0$ no matter what. Indeed, this contract function is arbitrage-free with any proper scoring rule in place of the quadratic scoring rule.

This idea does not extend to $m > 2$ experts, because an arbitrage-free contract function must not admit arbitrage by a coalition of experts of any size. While it is easy to construct a contract function that does not admit arbitrage by a coalition of size $m$ (by making the total reward always equal to $0$), this does not automatically make the contract function free of arbitrage opportunities for coalitions of sizes between $2$ and $m - 1$. In the next section we address this challenge and exhibit a proper contract function that is arbitrage-free for $m > 2$ experts.

\section{A class of proper arbitrage-free contract functions} \label{sec:mechanism}
Suppose that -- as before -- there are $m \ge 2$ experts who are forecasting an event with $n \ge 2$ outcomes. Given experts with reports $\vect{P} = (\vect{p}_1, \dots, \vect{p}_m)$ and a nonempty subset $S \subseteq [m]$ of the experts, we will let $\overline{\vect{p}}_S := \frac{1}{\abs{S}} \sum_{i \in S} \vect{p}_i$ be the average of the experts' reports. We will use $\overline{\vect{p}}_{-i}$ to denote $\overline{\vect{p}}_{[m] \setminus \{i\}}$.

We now state our main theorem, which exhibits a class of proper, arbitrage-free contract functions.

\begin{theorem} \label{thm:construction}
	Let $\alpha$ be a real number such that $\alpha < 0$ or $\alpha \ge 2(m - 1)^2n$. Let $\Pi$ be the contract function defined by
	\[\Pi_i(\vect{P}; j) = s_{\text{quad}}(\vect{p}_i; j) - (m - 1)^2 s_{\text{quad}}(\overline{\vect{p}}_{-i}; j) + \alpha \overline{\vect{p}}_{-i, j}\]
	for each $i$, $j$. Then $\Pi$ is proper and arbitrage-free.
\end{theorem}

Note that in the case of $m = 2$, setting $\alpha = 0$ yields our aforementioned solution for two experts in Equation~\ref{eq:solution_m2}. Unfortunately, setting $\alpha = 0$ for $m > 2$ experts causes arbitrage-freeness to fail in certain edge cases.

One can think of the contract function in Theorem~\ref{thm:construction} as having two parts. The first part, $s_{\text{quad}}(\vect{p}_i; j) - (m - 1)^2 s_{\text{quad}}(\overline{\vect{p}}_{-i}; j)$, ensures that any coalition's total reward depends only on the average of the coalition's reports. In effect this significantly limits the degrees of freedom that a coalition has when colluding. The second part, $\alpha \overline{\vect{p}}_{-i, j}$, ensures that any deviation in this average report causes a decrease in total reward under at least one outcome.

We first present the proof of Theorem~\ref{thm:construction} for $n = 2$ outcomes, as this allows us to simplify notation while still explaining the core ideas.

\begin{proof}[Proof of Theorem~\ref{thm:construction} for $n = 2$]
	First, note that $\Pi$ is proper, because expert $i$'s reward is their quadratic score plus a term that does not depend on their report. It remains to show that $\Pi$ is arbitrage-free.
	
	Let $C \subseteq [m]$ be a coalition of experts. Properness entails that no expert can unilaterally find an arbitrage opportunity, so we may assume that $\abs{C} \ge 2$.
	
	For an outcome $j$ and a subset $S \subseteq [m]$, let $p_{S, j} := \sum_{i \in S} p_{i, j}$.

	\begin{lemma} \label{lem:pmx}
		Let $d = m - 1 - \frac{\alpha}{4(m - 1)}$. The expression for $\Pi_i(\vect{P}; j)$ is equal to
		\begin{equation} \label{eq:adapted_n2}
			2(p_{[m], j} - d - 1)(p_{[m], j} - 2p_{i, j} - d + 1) + f(m, \alpha),
		\end{equation}
		for some function $f$.
	\end{lemma}

    \begin{proof}[Proof of Lemma~\ref{lem:pmx}]
    	For some $f$ (whose exact form does not concern us), we have
    	{
    		\begin{align*}
    			\Pi_i(\vect{P}; j) &= s_{\text{quad}}(\vect{p}_i; j) - (m - 1)^2 s_{\text{quad}}(\overline{\vect{p}}_{-i}; j) + \alpha \overline{\vect{p}}_{-i,j}\\
    			&= -2(1 - p_{i, j})^2 - (m - 1)^2 \parens{-2\parens{1 - \frac{p_{-i, j}}{m - 1}}^2} + \frac{\alpha}{m - 1} p_{-i, j}\\
    			&= -2(1 - p_{i, j})^2 + 2(m - 1 - p_{-i, j})^2 + \frac{\alpha}{m - 1}p_{-i, j}\\
    			&= -2(1 - p_{i, j})^2 + 2 \parens{d - p_{-i, j}}^2 + f(m, \alpha)\\
    			&= 2 \parens{d - p_{-i, j} + (1 - p_{i, j})} \parens{d - p_{-i, j} - (1 - p_{i, j})} + f(m, \alpha)\\
    			&= 2(p_{[m], j} - d - 1)(p_{[m], j} - 2p_{i, j} - d + 1) + f(m, \alpha),
    	\end{align*}}%
    	as desired.
    \end{proof}
	
	Equation~\ref{eq:adapted_n2} makes it evident that rewards add nicely across experts in a coalition $C$, as the first term of the product is the same for all experts in $C$. We will use the notation $\Pi_C(\vect{P}; j)$ to denote $\sum_{i \in C} \Pi_i(\vect{P}; j)$. The key idea is that, as we are about to show, if the reports of experts not in $C$ are held fixed, $\Pi_C(\vect{P}; j)$ depends \emph{only} on $p_{C, j}$. Thus, the experts in $C$ have only one degree of freedom available for colluding: the sum of their reports.
	
	We write $\overline{C}$ to mean $[m] \setminus C$. We have
	{
		\begin{align} \label{eq:pi_c_cong}
			\Pi_C(\vect{P}; j) &= 2 \sum_{i \in C} (p_{[m], j} - d - 1)(p_{[m], j} - 2p_{i, j} - d + 1) + \abs{C} f(m, \alpha) \nonumber\\
			&= 2(p_{C, j} + p_{\overline{C}, j} - d - 1) ((\abs{C} - 2)p_{C, j} + \abs{C}(p_{\overline{C}, j} - d + 1)) + \abs{C} f(m, \alpha) \nonumber\\
			&= 2((\abs{C} - 2) p_{C, j}^2 + 2((\abs{C} - 1)(p_{\overline{C}, j} - d) + 1)p_{C, j}) + g(m, \alpha, \abs{C}, p_{\overline{C}, j}),
		\end{align}
	}%
	for some function $g$. Now, recall the constraints on $\alpha$ in Theorem~\ref{thm:construction}, and note that $\alpha < 0 \Leftrightarrow d > m - 1$ and $\alpha \ge 4(m - 1)^2 \Leftrightarrow d \le 0$. With this in mind, we now prove the following claim, which is sufficient to complete our proof.
	
	\begin{claim} \label{claim:monotonicity}
		If $d \le 0$, then for each $j$ and for all possible reports of experts not in $C$, $\Pi_C(\vect{P}; j)$ is a strictly increasing function of $p_{C, j}$. If $d > m - 1$, it is a strictly decreasing function of $p_{C, j}$.
	\end{claim}
	
	By virtue of deriving Equation~\ref{eq:pi_c_cong}, we have already proven the most difficult part of Claim~\ref{claim:monotonicity}, which is that $\Pi_C(\vect{P}; j)$ is a function of (i.e.\ determined by) $p_{C, j}$. Why is this function's monotonicity sufficient to complete our proof of Theorem~\ref{thm:construction}? Since $p_{C, 1} + p_{C, 2} = \abs{C}$, it follows from Claim~\ref{claim:monotonicity} that for $d \le 0$ and $d > m - 1$, colluding in a way that increases the total reward in the case of one outcome necessarily decreases it in the case of the other outcome.
	
	\begin{proof}[Proof of Claim~\ref{claim:monotonicity}]
		We first consider the case of $\abs{C} = 2$. In this case we have
		\[\Pi_C(\vect{P}; j) = 4(p_{\overline{C}, j} - d + 1)p_{C, j} + g(m, \alpha, \abs{C}, p_{\overline{C}, j}).\]
		Now, $0 \le p_{\overline{C}, j} \le m - 2$, which means that $1 - d \le p_{\overline{C}, j} - d + 1 \le m - 1 - d$. If $d \le 0$, this quantity is guaranteed to be strictly positive, so $\Pi_C(\vect{P}; j)$ is a strictly increasing function of $p_{C, j}$; if $d > m - 1$, it is guaranteed to be negative, so $\Pi_C(\vect{P}; j)$ is a strictly decreasing function of $p_{C, j}$.\\
		
		Now assume that $\abs{C} > 2$. In this case, it follows from Equation~\ref{eq:pi_c_cong} that $\Pi_C(\vect{P}; j)$ is a parabola with a minimum at
		\[\frac{(\abs{C} - 1)(d - p_{\overline{C}, j}) - 1}{\abs{C} - 2}.\]
		We wish to show that if $d \le 0$ then this quantity is at most $0$, and that if $d > m - 1$ then it is at least $\abs{C}$ (since the range of possible values of $p_{C, j}$ is $[0, \abs{C}]$). If $d \le 0$ then, since $p_{\overline{C}, j} \ge 0$, we have
		\[\frac{(\abs{C} - 1)(d - p_{\overline{C}, j}) - 1}{\abs{C} - 2} \le \frac{-1}{\abs{C} - 2} \le \frac{-1}{m - 2} \le 0.\]
		If $d > m - 1$ then, since $p_{\overline{C}, j} \le m - \abs{C}$, we have
		\[\frac{(\abs{C} - 1)(d - p_{\overline{C}, j}) - 1}{\abs{C} - 2} \ge \frac{(\abs{C} - 1)^2 - 1}{\abs{C} - 2} = \abs{C}.\]
	\end{proof}
	
	\noindent Having proved the claim, we have completed the proof of Theorem~\ref{thm:construction} for $n = 2$.
\end{proof}

We note that setting $\alpha = 0$ results in a contract function that is arbitrage-free except in one edge case: in the event that all but two experts assign a probability of zero to some outcome $j$, the remaining experts can collude to adjust their probabilities -- in particular, lowering the total probability they assign to outcome $j$ -- in a way that increases their total reward under outcome $j$ and leaves the remaining rewards unchanged. If we are willing to put this exception aside (e.g.\ if we only allow reports strictly between $0$ and $1$), then we may regard the resulting contract function $\Pi_i(\vect{P}; j) = s_{\text{quad}}(\vect{p}_i; j) - (m - 1)^2 s_{\text{quad}}(\overline{\vect{p}}_{-i}; j)$ as arbitrage-free. This contract function has a natural interpretation: it rewards an expert for the accuracy of their forecast but penalizes the expert if others are accurate in aggregate. This rule is reminiscent of the no-arbitrage wagering mechanism for the quadratic scoring rule given by \textcite{cdpv14}, except that the penalty is multiplied by a factor of $(m - 1)^2$.

We now present the proof of Theorem~\ref{thm:construction} in full generality.

\begin{proof}[Proof of Theorem~\ref{thm:construction} for general $n$]
	First, note that $\Pi$ is proper, because expert $i$'s reward is their quadratic score plus a term that does not depend on their report. It remains to show that $\Pi$ is arbitrage-free.
	
	Let $C \subseteq [m]$ be a coalition of experts. Properness entails that no expert can unilaterally find an arbitrage opportunity, so we may assume that $\abs{C} \ge 2$.
	
	For an outcome $j$ and a subset $S \subseteq [m]$, let $p_{S, j} := \sum_{i \in S} p_{i, j}$. Let $d = m - 1 - \frac{\alpha}{2(m - 1)}$. For some $f$ whose particular form does not concern us, we have
	\begin{align*}
        &\Pi_i(\vect{P}; j) = s_{\text{quad}}(\vect{p}_i; j) - (m - 1)^2 s_{\text{quad}}(\overline{\vect{p}}_{-i}; j) + \alpha \overline{\vect{p}}_{-i,j}\\
        &= (m - 1)^2 \parens{1 - \frac{1}{m - 1}p_{-i, j}}^2 - (1 - p_{i, j})^2 + \sum_{\ell \neq j} \parens{(m - 1)^2 \parens{\frac{1}{m - 1} p_{-i, \ell}}^2 - p_{i, \ell}^2} + \frac{\alpha}{m - 1} p_{-i, j}\\
        &= (d - p_{-i, j})^2 - (1 - p_{i, j})^2 + \sum_{\ell \neq j} \parens{p_{-i, \ell}^2 - p_{i, \ell}^2} + f(m, n, \alpha)\\
        &= (d - p_{-i, j} + (1 - p_{i, j}))(d - p_{-i, j} - (1 - p_{i, j})) + \sum_{\ell \neq j} (p_{-i, \ell} + p_{i, \ell})(p_{-i, \ell} - p_{i, \ell}) + f(m, n, \alpha)\\
        &= (p_{[m], j} - d - 1)(p_{[m], j} - 2p_{i, j} - d + 1) + \sum_{\ell \neq j} p_{[m], \ell}(p_{[m], \ell} - 2p_{i, \ell}) + f(m, n, \alpha).
	\end{align*}
	
	We will use the notation $\Pi_C(\vect{P}; j)$ to denote $\sum_{i \in C} \Pi_i(\vect{P}; j)$. We also write $\overline{C}$ to mean $[m] \setminus C$ and $\vect{P}_{\overline{C}}$ to mean the collection of reports $\vect{p}_i$ for $i \in \overline{C}$. We have
    \begin{align*}
        &\Pi_C(\vect{P}; j) = \sum_{i \in C} \parens{(p_{[m], j} - d - 1)(p_{[m], j} - 2p_{i, j} - d + 1) + \sum_{\ell \neq j} p_{[m], \ell}(p_{[m], \ell} - 2p_{i, \ell})} + \abs{C} f(m, n, \alpha)\\
        &= (p_{C, j} + p_{\overline{C}, j} - d - 1)((\abs{C} - 2)p_{C, j} + \abs{C}(p_{\overline{C}, j} - d + 1))\\
        &\qquad + \sum_{\ell \neq j} (p_{C, \ell} + p_{\overline{C}, \ell})(\abs{C} p_{\overline{C}, \ell} + (\abs{C} - 2) p_{C, \ell}) + \abs{C} f(m, n, \alpha)\\
        &= (\abs{C} - 2) p_{C, j}^2 + ((2\abs{C} - 2)(p_{\overline{C}, j} - d) + 2)p_{C, j}\\
        &\quad + \sum_{\ell \neq j} \parens{(\abs{C} - 2) p_{C, \ell}^2 + (2\abs{C} - 2) p_{\overline{C}, \ell} p_{C, \ell}} + g(m, n, \alpha, \abs{C}, \vect{P}_{\overline{C}})\\
        &= (2 - (2\abs{C} - 2)d) p_{C, j} + \sum_\ell \parens{(\abs{C} - 2) p_{C, \ell}^2 + (2\abs{C} - 2) p_{\overline{C}, \ell} p_{C, \ell}} + g(m, n, \alpha, \abs{C}, \vect{P}_{\overline{C}})
	\end{align*}
	
	for some function $g$. Consider a different vector $\vect{Q}$ that agrees with $\vect{P}$ on $\overline{C}$.\\
	
	\paragraph{Case 1: $\alpha < 0$.} In this case, $2 - (2\abs{C} - 2)d < 2 - (2\abs{C} - 2)(m - 1)$. Let $\tilde{j} = \arg \max_\ell q_{C, \ell} - p_{C, \ell}$, and let $\epsilon = q_{C, \tilde{j}} - p_{C, \tilde{j}}$. We note that
	\[\sum_\ell q_{C, \ell}^2 - p_{C, \ell}^2 = \sum_\ell (q_{C, \ell} - p_{C, \ell})(q_{C, \ell} + p_{C, \ell}) \le \epsilon \sum_\ell (q_{C, \ell} + p_{C, \ell}) = 2 \epsilon \abs{C}.\]
	Thus, we have
	\begin{align*}
			\Pi_C(\vect{Q}; \tilde{j}) - \Pi_C(\vect{P}; \tilde{j}) &= (2 - (2\abs{C} - 2)d) \epsilon + \sum_\ell (\abs{C} - 2)(q_{C, \ell}^2 - p_{C, \ell}^2) + (2\abs{C} - 2)(q_{C, \ell} - p_{C, \ell}) p_{\overline{C}, \ell}\\
			&\le (2 - (2\abs{C} - 2)d) \epsilon + (\abs{C} - 2) \cdot 2 \epsilon \abs{C} + (2\abs{C} - 2)(m - \abs{C}) \epsilon\\
			&\le (2 - (2\abs{C} - 2)(m - 1)) \epsilon + (\abs{C} - 2) \cdot 2 \epsilon \abs{C} + (2\abs{C} - 2)(m - \abs{C}) \epsilon\\
			&= 2\epsilon(1 + (\abs{C} - 2)\abs{C} + (\abs{C} - 1)(1 - \abs{C})) = 0,
	\end{align*}
	with equality in the second step only when $\epsilon = 0$, i.e.\ $q_{C, \ell} = p_{C, \ell}$ for all $\ell$. Thus, either the total reward of the experts in $C$ is the same under $\vect{Q}$ as under $\vect{P}$ for every outcome, or it is strictly smaller under $\vect{Q}$ in the case of outcome $\tilde{j}$.\\
	
	\paragraph{Case 2: $\alpha \ge 2(m - 1)^2 n$.} In this case, $2 - (2\abs{C} - 2)d \ge 2 + (2\abs{C} - 2)(m - 1)(n - 1)$. $\tilde{j} = \arg \max_\ell p_{C, \ell} - q_{C, \ell}$, and let $\epsilon = p_{C, \tilde{j}} - q_{C, \tilde{j}}$. Since $\sum_\ell (q_{C, \ell} - p_{C, \ell}) = 0$, it follows that $q_{C, \ell} - p_{C, \ell} \le (n - 1)\epsilon$ for all $\ell$. We note that
	\begin{align*}
		\sum_\ell q_{C, \ell}^2 - p_{C, \ell}^2 &\le \sum_\ell (q_{C, \ell} + p_{C, \ell}) \max(q_{C, \ell} - p_{C, \ell}, 0)\\
		&\le 2\abs{C} \sum_\ell \max(q_{C, \ell} - p_{C, \ell}, 0) \le 2 \abs{C}(n - 1) \epsilon.
	\end{align*}
	We also have that
	\[\sum_\ell (q_{C, \ell} - p_{C, \ell}) p_{\overline{C}, \ell} \le (m - \abs{C}) \sum_\ell \max(q_{C, \ell} - p_{C, \ell}, 0)\le (m - \abs{C})(n - 1)\epsilon.\]
	Therefore,
	\begin{align*}
		&\Pi_C(\vect{Q}; \tilde{j}) - \Pi_C(\vect{P}; \tilde{j}) = -(2 - (2\abs{C} - 2)d) \epsilon + \sum_\ell (\abs{C} - 2)(q_{C, \ell}^2 - p_{C, \ell}^2) + (2\abs{C} - 2)(q_{C, \ell} - p_{C, \ell}) p_{\overline{C}, \ell}\\
		&\le -(2 - (2\abs{C} - 2)d) \epsilon + (\abs{C} - 2) \cdot 2\abs{C}(n - 1)\epsilon + (2\abs{C} - 2) \cdot (m - \abs{C})(n - 1)\epsilon\\
		&= 2\epsilon(-1 - (\abs{C} - 1)(m - 1)(n - 1) + (n - 1)(m(\abs{C} - 1) - \abs{C}))\\
		&= 2\epsilon(-1 - (n - 1)) = -2\epsilon n \le 0,
	\end{align*}
	with equality in the last step only when $\epsilon = 0$, i.e.\ $q_{C, \ell} = p_{C, \ell}$ for all $\ell$. As in the previous case, this means that either the total reward of the experts in $C$ is the same under $\vect{Q}$ as under $\vect{P}$ for every outcome, or it is strictly smaller under $\vect{Q}$ in the case of outcome $\tilde{j}$. This completes the proof.
\end{proof}

\section{Future directions in arbitrage-freeness} \label{sec:stronger_defs}
While the contract functions defined in Theorem~\ref{thm:construction} are proper and arbitrage-free, there are other desirable notions of arbitrage-freeness that they do not satisfy. To see this, consider the following concrete example: suppose that there are two outcomes, and all $m > 2$ experts think that Outcome 1 will happen with probability $1$. Suppose that $\alpha$ is chosen to be only slightly negative -- very close to zero. In that case, all experts reporting their true belief is just about the \emph{worst} possible outcome for the experts: they each believe that they will each receive a score of $\alpha$. By comparison, if all experts were to instead lie and report $\vect{p}_i = (0, 1)$ (i.e.\ that Outcome 2 will happen with probability $1$), then each expert believes that they will receive reward $2((m - 1)^2 - 1)$. That is, the experts are much better off if everyone lies than if everyone tells the truth.

Why is this not a counterexample to arbitrage-freeness? Well, if Outcome 2 happens, then the experts will get reward $2((m - 1)^2 - 1)$ if they tell the truth and $\alpha$ if they lie. The experts believe that Outcome 2 will happen with probability $0$, but if Outcome 2 \emph{does} happen, then they will end up better off if they tell the truth. However, since every expert believes that Outcome 1 is guaranteed to happen, it seems that in practice the experts would want to collude in this situation: they believe that they can collude to guarantee themselves a greater profit with probability 100\%.

How might we expand the definition of arbitrage to include the collusion scenario we just described? One way is to say that $\Pi$ admits arbitrage if there is a collusion strategy for the experts in $C$ that, in the opinion of every expert in $C$, increases the expected total reward of the experts in $C$. Formally:

\begin{defin} \label{def:exp_arb}
	A contract function $\Pi$ \emph{admits expected arbitrage} if there is a coalition $C \subseteq [m]$ of experts and vectors of reports $\vect{P} = (\vect{p}_1, \dots, \vect{p}_m)$, $\vect{Q} = (\vect{q}_1, \dots, \vect{q}_m)$, with $p_i = q_i$ if $i \not \in C$, such that for all $i \in C$ we have
	\[\sum_{j \in [n]} p_{i, j} \sum_{k \in C} \Pi_k(\vect{P}; j) \le \sum_{j \in [n]} p_{i, j} \sum_{k \in C} \Pi_k(\vect{Q}; j),\]
	and the inequality is strict for some $i$. We say that $\Pi$ is \emph{free of expected arbitrage} if it does not admit expected arbitrage.
\end{defin}

Up to edge scenarios,\footnote{It is possible for a coalition of experts to collude in a way that increases their total reward under an outcome to which they all assign probability $0$. If their reward in the case of all other outcome is unchanged, such a deviation would constitute arbitrage but not expected arbitrage.} if a contract function admits arbitrage then it also admits expected arbitrage. On the other hand, in the case of $m > 2$ experts, the scoring rules described by Theorem~\ref{thm:construction} (which do not admit arbitrage) do admit expected arbitrage. As an example, consider two outcomes and $m$ experts with beliefs $(\frac{1}{2}, \frac{1}{2})$. If all experts report their beliefs, then each expert's reward is $\frac{\alpha}{2} + \frac{1}{2}((m - 1)^2 - 1)$, no matter the outcome. If all experts instead report $(1, 0)$ then each expert expects a reward of $\frac{\alpha}{2} + (m - 1)^2 - 1$, which is larger. This raises the following question:

\begin{question} \label{q:expected_arb}
	Is there a proper scoring rule that does not admit expected arbitrage?
\end{question}

We can define an even stronger notion of arbitrage-freeness:

\begin{defin} \label{def:redist_arb}
    We say that $\Pi$ \emph{admits redistributional arbitrage} if there is:
    \begin{itemize}
        \item A coalition $C \subseteq [m]$ of experts
        \item A \emph{redistribution agreement} $A$ that takes as input a list of probability distributions $\vect{X} = (\vect{x}_1, \dots, \vect{x}_m)$, another list of probability distributions $\vect{Y} = (\vect{y}_1, \dots, \vect{y}_m)$ such that $\vect{x}_i = \vect{y}_i$ for any $i \not \in C$,\footnote{Here, the $\vect{x}_i$ represent the reports given to $\Pi$, the $\vect{y}_i$ represent the pre-collusion probabilities (these can be thought of as the ``true beliefs''), and $j$ represents the outcome that happens.} and an outcome $j$, and outputs a list of rewards, one for each $i \in C$, such that
        \[\sum_{i \in C} A_i(\vect{X}; \vect{Y}; j) = \sum_{i \in C} \Pi_i(\vect{X}; j).\]
        (We call this the \emph{budget balance} property of $A$.) \dots
    \end{itemize}
    \dots such that there are vectors of reports $\vect{P} = (\vect{p}_1, \dots, \vect{p}_m)$, $\vect{Q} = (\vect{q}_1, \dots, \vect{q}_m)$, with $p_i = q_i$ if $i \not \in C$, so that for all $i \in C$ we have
    \[\sum_{j \in [n]} p_{i, j} A_i(\vect{Q}; \vect{P}; j) \ge \sum_{j \in [n]} p_{i, j} \Pi_i(\vect{P}; j),\]
    and the inequality is strict for some $i$. We say that $\Pi$ is \emph{free of redistribution arbitrage} if it does not admit expected arbitrage.
\end{defin}

We show that every $\Pi$ that admits expected arbitrage also admits redistributional arbitrage -- that is, that if there is a misreport that each experts believes will make the \emph{pool} of colluding experts better off in expectation, then there is a way to redistribute the earnings in a way that makes each expert believe that \emph{they} will be better off in expectation.

\begin{theorem}
    Every contract function $\Pi$ that admits expected arbitrage admits redistributional arbitrage.
\end{theorem}

\begin{proof}
    Let $\Pi$ be a contract function that admits expected arbitrage, and let $C$, $\vect{P}$, and $\vect{Q}$ be as in Definition~\ref{def:exp_arb}. For any $\vect{X} = (\vect{x}_1, \dots, \vect{x}_m)$ and $\vect{Y} = (\vect{y}_1, \dots, \vect{y}_m)$ that coincide outside of $C$, let
    \[W(\vect{X}; \vect{Y}; j) := \sum_{i \in C} \Pi_i(\vect{X}; j) - \sum_{i \in C} \Pi_i(\vect{Y}; j)\]
    be the total gain of experts in $C$ by colluding to report $X$ instead of $Y$. Thus, $\sum_j p_{i, j} W(\vect{Q}; \vect{P}; j) \ge 0$ for all $i$ and the inequality is strict for some $i$. Define
    \[A_i(\vect{X}; \vect{Y}; j) := \Pi_i(\vect{Y}; j) + \frac{1}{\abs{C}} W(\vect{X}; \vect{Y}; j).\]
    (Clearly $A$ is budget-balanced.) Then for any $i \in C$ we have
    \begin{align*}
        \sum_j p_{i, j} A_i(\vect{Q}; \vect{P}; j) &= \sum_j p_{i, j} \parens{\Pi_i(\vect{P}; j) + \frac{1}{\abs{C}} W(\vect{Q}; \vect{P}; j)}\\
        &\ge \sum_j p_{i, j} \Pi_i(\vect{P}; j),
    \end{align*}
    with the inequality strict for some $i$, as desired.
\end{proof}

We close with the question analogous to Question~\ref{q:expected_arb} for this stronger notion:

\begin{question}
    Is there a proper scoring rule that does not admit redistributional arbitrage?
\end{question}
\chapter{Quasi-arithmetic pooling} \label{chap:qa}
\emph{This chapter presents ``From Proper Scoring Rules to Max-Min Optimal Forecast Aggregation'' \parencite{nr23_qa}, although the contents of Section~\ref{sec:overconfidence} are original to this thesis. It assumes background on proper scoring rules presented in Section~\ref{sec:prelim_proper}, as well as background on forecast aggregation presented in Section~\ref{sec:prelim_agg}.}\\

\emph{Summary:} This chapter forges a strong connection between two seemingly unrelated forecasting problems: incentive-compatible forecast elicitation and forecast aggregation.  We have previously discussed proper scoring rules as a solution to the former problem (see Section~\ref{sec:prelim_proper}).  To each proper scoring rule $s$ we associate a corresponding method of aggregation, mapping expert forecasts and expert weights to a ``consensus forecast,'' which we call \emph{quasi-arithmetic (QA) pooling} with respect to $s$.  We justify this correspondence in several ways:

\begin{itemize}
\item QA pooling with respect to the two most well-studied scoring rules (quadratic and logarithmic) corresponds to the two most well-studied forecast aggregation methods (linear and logarithmic). 

\item Given a scoring rule $s$ used for payment, a forecaster agent who sub-contracts several experts, paying them in proportion to their weights, is best off aggregating the experts' reports using QA pooling with respect to $s$, meaning this strategy maximizes its worst-case profit (over the possible outcomes). 

\item The score of an aggregator who uses QA pooling is concave in the experts' weights. As a consequence, online gradient descent can be used to learn appropriate expert weights from repeated experiments with low regret.

\item QA pooling can be used to define overconfidence with respect to a scoring rule. The resulting notion of overconfidence turns out to be equivalent to another natural definition of overconfidence.

\item The class of all QA pooling methods is characterized by a natural set of axioms (generalizing classical work by Kolmogorov on quasi-arithmetic means).
\end{itemize}

\section{Introduction and motivation} \label{sec:chap5_intro}
\subsection{Choice of scoring rule as a value judgment}
There are infinitely many proper scoring rules. How might a principal go about deciding which one to use? To gain some intuition, recall the quadratic and logarithmic scoring rules from Section~\ref{sec:prelim_proper}:
\begin{align*}
    s_{\text{quad}}(\vect{x}; j) &:= -(1 - x_j)^2 - \sum_{j' \neq j} x_{j'}^2\\
    s_{\text{log}}(\vect{x}; j) &:= \ln(x_j)
\end{align*}

\begin{figure}[ht]
	\centering
	\includegraphics[scale=1]{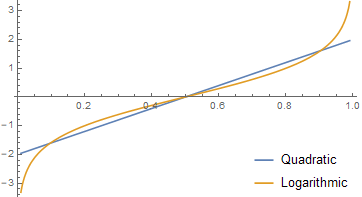}
	\caption[Quadratic score vs.\ log score: expert preference over the outcome]{Difference between expert's score if an outcome happens and if it does not happen, as a function of the expert's report, for the quadratic and logarithmic scoring rules. For example, if the expert reports a 70\% probability of an outcome, then under the quadratic rule they receive a score of $-\norm{(0.7, 0.3) - (1, 0)}_2^2 = - 0.18$ if the outcome happens and $-\norm{(0.7, 0.3) - (0, 1)}_2^2 = - 0.98$ if it does not: a difference of $0.8$. If scored with the logarithmic rule, this difference would be $0.61$.}
	\label{fig:quad_log}
\end{figure}

Let us consider the quadratic and logarithmic scoring rules in the case of a binary Yes/No outcome. In Figure~\ref{fig:quad_log}, for both scoring rules, we show the difference between the expert's score if a given outcome happens and their score if it does not happen, as a function of the probability that they assign to the outcome.\footnote{We scale down the logarithmic rule by a factor of $2 \ln 2$ to make the two rules comparable. The factor $2 \ln 2$ was chosen to make the range of values taken on by the expected score functions of the two scoring rules the same.}

For the quadratic scoring rule, this difference scales linearly with the expert's report. Meanwhile, for the logarithmic rule, the difference changes more slowly than for the quadratic rule for probabilities near 50\%, but much more quickly at the extremes. Informally speaking, this means that the logarithmic rule indicates a preference (of the elicitor) for high precision close to $0$ and $1$, while the quadratic rule indicates a more even preference for precision across $[0, 1]$. Put another way, an elicitor who chooses to use the logarithmic scoring rule renders a judgment that the probabilities $0.01$ and $0.001$ are qualitatively quite different; one who uses the quadratic rule indicates that these probabilities are very similar.\\

Now, recall from Section~\ref{sec:prelim_agg} the two most widely-used forecast aggregation methods: linear and logarithmic pooling. Given probabilistic forecasts $\vect{x}_1, \dots, \vect{x}_m$, the linear pool with weights $w_1, \dots, w_m$ (adding to $1$) is given by the weighted arithmetic mean: $\vect{x}^* = \sum_i w_i \vect{x}_i$. Meanwhile, the logarithmic pool is given by the (renormalized) geometric mean: $x^*(j) = c \prod_{i = 1}^m (x_i(j))^{w_i}$.

On its surface, the \emph{elicitation} of forecasts has seemingly little to do with their \emph{aggregation}. However, given that the choice of scoring rule implies a subjective judgment about how different probabilities compare to one another, it makes sense to apply this judgment to the aggregation of forecasts as well.

As an example, consider the setting of weather prediction, with models playing the role of experts. In such contexts we often care about low-probability extreme events: a 0.1\% chance of an imminent major hurricane may not be worth preparing for; a 1\% chance could mean significant preparations, and a 10\% chance could mean mandatory evacuations. The need to distinguish very unlikely events from somewhat unlikely events has two consequences. First, as discussed above, this is a reason to use the logarithmic scoring rule to assess the quality of weather models. Second, we wish to avoid the failure mode in which an ill-informed forecaster assigns a high probability due to lack of evidence and thereby drowns out a better-informed low-probability forecast. We would expect to encounter this failure mode with linear pooling; for example, if the more informed model predicts a 0.1\% chance and the less informed model predicts a 20\% chance, linear pooling with equal weights\footnote{Assigning equal weights makes sense when there is not enough information to predict in advance which model will be less informed.} would predict roughly a 10\% chance. Logarithmic pooling, by contrast, assigns roughly a 1.6\% chance to the event, avoiding this failure mode. In general, a calibrated model that predicts a very low probability must have good evidence, so it may make sense to give the model more weight.

By contrast, a consultant whose job is to determine the closest races in an election may not much care about the difference between a 0.1\% and a 1\% chance of victory. After all, attention and resources are generally devoted to races with highly uncertain outcomes. As such, it might make sense to assess the qualities of forecast models using the quadratic scoring rule. Similarly, without a compelling reason to pay extra attention to extreme probabilities, it may make more sense to simply take the average of the forecasts' opinions.

In the case of extreme weather prediction, we have argued in favor of using the logarithmic scoring rule to assess the models and the logarithmic pooling method to aggregate them, for similar reasons. In the case of political prediction for targeting close races, we have argued in favor of the quadratic scoring rule and linear pooling, also for similar reasons. Could there be a formal connection between proper scoring rules and forecast aggregation methods that captures this intuition? This brings us to the main focus of this work: namely, we prove a novel correspondence between proper scoring rules and forecast aggregation methods.

\subsection{Our definitions} \label{subsec:definitions}
Recall from Section~\ref{sec:prelim_proper} the \emph{Savage representation} of a proper scoring rule:
\begin{equation} \label{eq:chap5_savage}
s(\vect{x}; j) = G(\vect{x}) + \angles{\pmb{\delta}_j - \vect{x}, \vect{g}(\vect{x})}
\end{equation}
where $G$ is the expected score function of $s$ and $\vect{g}$ is the gradient of $G$. The function $\vect{g}$, which will be central to our work, describes the difference in the expert's score depending on which outcome happens. More precisely, the vector $(s(\vect{x}; j_1), \dots, s(\vect{x}; j_n))$ is exactly the vector $\vect{g}(\vect{x})$, except possibly for a uniform translation in all coordinates. Put otherwise, for any two outcomes $j_1$ and $j_2$, we have $g_1(\vect{x}) - g_2(\vect{x}) = s(\vect{x}; j_1) - s(\vect{x}; j_2)$ (this can be verified from Equation~\ref{eq:chap5_savage}). This is precisely the quantity plotted in Figure~\ref{fig:quad_log} for the quadratic and logarithmic scoring rules, if $j_1$ is ``Yes'' and $j_2$ is ``No.'' This observation about the function $\vect{g}$ motivates the correspondence that we will establish between proper scoring rules and forecast aggregation methods.

\paragraph{\textbf{Quasi-arithmetic pooling}}
We can now define the aforementioned correspondence between proper scoring rules and forecast aggregation methods. Given a proper scoring rule $s$ used for elicitation, and given $m$ probability distributions $\vect{p}_1, \dots, \vect{p}_m$ and expert weights $w_1, \dots, w_m$, the aggregate distribution $\vect{p}^*$ that we suggest is the one satisfying
\[\vect{g}(\vect{p}^*) = \sum_{i = 1}^m w_i \vect{g}(\vect{p}_i).\]
(In Section~\ref{sec:chap5_prelims} we will define this notion more precisely using subgradients of $G$ instead of gradients; this will ensure that $\vect{p}^*$ is well defined, i.e.\ that it exists and is unique.) This definition of $\vect{p}^*$ can be restated as the forecast that minimizes the weighted average Bregman divergence (with respect to $G$) to all experts' forecasts.

We refer to this pooling method as \emph{quasi-arithmetic pooling} with respect to $\vect{g}$ (or the scoring rule $s$), or \emph{QA pooling} for short.\footnote{This term comes from the notion of quasi-arithmetic means: given a continuous, strictly increasing function $f$ and values $x_1, \dots, x_m$ , the \emph{quasi-arithmetic mean with respect to $f$} of these values is $f^{-1}(1/m \sum_i f(x_i))$.} To get a sense of QA pooling, let us determine what this method looks like for the quadratic and logarithmic scoring rules.

\paragraph{\textbf{QA pooling with respect to the quadratic scoring rule}} Recall from Section~\ref{sec:prelim_proper} that the expected score function $G_{\text{quad}}$ of the quadratic scoring rule $s_{\text{quad}}$ is equal to $\sum_j x_j^2 - 1$. We thus have $\vect{g}_{\text{quad}}(\vect{x}) = (2x_1, \dots, 2x_n)$, so we are looking for the $\vect{p}^*$ such that
\[(2p^*(1), \dots, 2p^*(n)) = \sum_{i = 1}^m w_i (2p_i(1), \dots, 2p_i(n)).\]
This is $\vect{p}^* = \sum_{i = 1}^m w_i \vect{p}_i$. Therefore, \emph{QA pooling for the quadratic scoring rule is precisely linear pooling,} which we introduced in Section~\ref{sec:prelim_agg} as the simplest and most widely-used forecast aggregation method.

\paragraph{\textbf{QA pooling with respect to the logarithmic scoring rule}} Recall that the expected score function $G_{\text{log}}$ of the logarithmic scoring rule $s_{\text{log}}$ is equal to $\sum_j x_j \ln(x_j)$. We thus have $\vect{g}_{\log}(\vect{x}) = (\ln x_1 + 1, \dots, \ln x_n + 1)$, so we are looking for the $\vect{p}^*$ such that
\[(\ln p^*(1) + 1, \dots, \ln p^*(n) + 1) = \sum_{i = 1}^m w_i (\ln p_i(1) + 1, \dots, \ln p_i(n) + 1).\]
By exponentiating the components on both sides, we find that $p^*(j) = c\prod_{i = 1}^n (p_i(j))^{w_i}$ for all $j$, for some proportionality constant $c$. \emph{This is precisely the definition of the logarithmic pooling method,} which we introduced in Section~\ref{sec:prelim_agg} as a simple, well-motivated, and effective way of aggregating probabilistic forecasts. (The constant $c$ comes from the fact that values of $\vect{g}(\cdot)$ should be interpreted modulo translation by the all-ones vector; see Remark~\ref{remark:mod_T1n}.)\\

The fact that this pooling scheme maps the two most well-studied scoring rules to the two most well-studied forecast aggregation methods has not been noted previously, to our knowledge. This correspondence suggests that -- beyond just our earlier informal justification -- QA pooling with respect to a given scoring rule may be a fundamental concept. The rest of this work argues that this is indeed the case.

This correspondence may have practical implications for forecasters. While the quadratic and logarithmic scoring rules are both ubiquitous in practice, linear pooling is far more common than logarithmic pooling \parencite{rg10}. This is despite empirical evidence that logarithmic pooling often outperforms linear pooling \parencite{sbfmtu14}. The connection that we establish between the logarithmic scoring rule and logarithmic pooling provides further reason to think that logarithmic pooling has been somewhat overlooked.

\subsection{Our results}
\paragraph{\textbf{(Section~\ref{sec:max_min}) Max-min optimality}} Suppose that a principal asks you to issue a forecast and will pay you according to $s$. You are not knowledgeable on the subject but know some experts whom you trust on the matter (perhaps to varying degrees). You sub-contract the experts, promising to pay each expert $i$ according to $w_i \cdot s$. By using QA pooling according to $s$ on the experts' forecasts, you guarantee yourself a profit; in fact, this strategy maximizes your worst-case profit, and is the unique such report. Furthermore, this profit is the same for all outcomes. This fact can be interpreted to mean that you have, in a sense, pooled the forecasts ``correctly'': you do not care which outcome will come to pass, which means that you have correctly factored the expert opinions into your forecast. We give an additional interpretation of this optimality notion as maximizing an aggregator's guaranteed improvement over choosing an expert at random.

\paragraph{\textbf{(Section~\ref{sec:convex_losses}) Learning expert weights}} Pooling forecasts entails assigning weights to experts. Where do these weights come from? How might one learn them from experience?

Suppose we have a fixed proper scoring rule $s$, and further consider fixing the reports of the $m$ experts as well as the eventual outcome. One can ask: what does the score of the aggregate distribution (per QA pooling with respect to $s$) look like as a function of $\vect{w}$, the vector of expert weights? We prove that this function is concave. This is useful because it allows for online convex optimization over expert weights.

\begin{nonothm}[informal]
Let $s$ be a bounded proper scoring rule.\footnote{For which QA pooling is well defined (we discuss this below).} For time steps $t = 1 \dots T$, $m$ experts report forecasts to an aggregator, who combines them into a forecast $\vect{p}^t$ using QA pooling with respect to $s$ and suffers a loss of $-s(\vect{p}^t; j^t)$, where $j^t$ is the outcome at time step $t$. If the aggregator updates the experts' weights using online gradient descent, then the aggregator's regret compared to the best weights in hindsight is $O(\sqrt{T})$.
\end{nonothm}

The aforementioned concavity property is a nontrivial fact that demonstrates an advantage of QA pooling with respect to the proper scoring rule used for elicitation, as compared with using e.g.\ linear or logarithmic pooling regardless of the method of elicitation: linear and logarithmic pooling satisfy the concavity property for some proper scoring rules $s$ but not others.

\paragraph{\textbf{(Section~\ref{sec:overconfidence}) Overconfidence}} Informally, an expert is \emph{overconfident} if the expert's forecasts would be more accurate if they were less extreme (closer to uniform). Much as it makes sense to aggregate forecasts in a way that depends on the scoring rule, it also makes sense to define overconfidence in a way that depends on the scoring rule. We give two natural definitions of overconfidence with respect to a proper scoring rule. The first definition says that a series of forecasts made by an expert is overconfident if the expert's total score is lower than the score the expert expects. The second definition says that a series of forecasts made by an expert is overconfident if using QA pooling to make the forecasts less extreme would increase the expert's total score. We prove that these two definitions are equivalent.

\paragraph{\textbf{(Section~\ref{sec:axiomatization}) Natural axiomatization for QA pooling methods}} \textcite{kol30} and \textcite{nag30} independently came up with a simple axiomatization of quasi-arithmetic means. We show how to change these axioms to allow for weighted means; the resulting axiomatization is a natural characterization of all quasi-arithmetic pooling methods in the case of $n = 2$ outcomes. Furthermore, although quasi-arithmetic means are typically defined for scalar-valued functions, we demonstrate that these axioms can be extended to describe quasi-arithmetic means with respect to vector-valued functions, as is necessary for our purposes if $n > 2$. This extension is nontrivial but natural, and to our knowledge has not previously been described.\footnote{For $n > 2$, these axioms characterize the class of all QA pooling methods with respect proper scoring rules that satisfy \emph{convex exposure}, a natural condition that we introduce in Section~\ref{sec:chap5_prelims}.}

\subsection{Related work} \label{sec:chap5_related_work}
Since this work's contribution is a connection between two well-studied problems (elicitation and aggregation of forecasts), there is naturally a large volume of related work, some of which we have discussed in previous chapters. In this section, we briefly discuss the most closely-related work to ours.

\paragraph{\textbf{Quasi-arithmetic means}} Our notion of quasi-arithmetic pooling is an adaptation (and extension to higher dimensions) of the existing notion of quasi-arithmetic means. These were originally defined and axiomatized independently by \textcite{kol30} and \textcite{nag30}. \textcite{aczel48} generalized this work to include weighted quasi-arithmetic means, though these means have weights baked in rather than taking them as inputs, which is different from our setting. See \textcite[\S3.1]{gmmp11} for an overview of this topic.

\paragraph{\textbf{Arbitrage from collusion}} In Chapter~\ref{chap:arbitrage} we discussed the problem of colluding experts: suppose that a principal uses a proper scoring rule $s$ to elicit forecasts from multiple experts. If the experts have different opinions, they can collude to report the forecast, thus earning them a larger total score than if they had been honest, no matter the outcome. We did not detail how exactly this collusion works, but the answer turns out to be precisely QA pooling: for the case of $n = 2$ outcomes, \textcite{cs11} showed that if all experts report what we are calling the QA pool of their opinions with respect to $s$, then their total score is guaranteed to increase. Our Theorem~\ref{thm:max_min} recovers this result as a special case.

\paragraph{\textbf{Connections to Bregman divergence}} In part, our work provides an alternative interpretation of prior work on forecast aggregation via minimizing Bregman divergence, see e.g.\ \textcite{ada14, pet19}. Concretely, \textcite[\S4]{pet19} defines a notion of aggregation analogous to ours, though in a different context. The main focus of their line of work is on connecting forecast aggregation to Bregman divergence; our approach connects aggregation with proper scoring rules, and a connection to Bregman divergence falls naturally out of this pursuit.

\paragraph{\textbf{Dual averaging}} One perspective on QA pooling is that, instead of directly averaging experts' forecasts, QA pooling prescribes considering forecasts as elements in the dual space of gradients (of the function $G$) and taking the average in this space before converting the result back to the primal space of probabilities. Gradient methods in online machine learning often take the \emph{sum} of gradients of losses. Taking the average is of gradients is a less ubiquitous technique known as \emph{dual averaging,} which was introduced by \textcite{nesterov09} and generalized further by \textcite{xiao10}. However, the contexts of QA pooling and dual averaging are quite different.

\paragraph{\textbf{Aggregation via prediction markets}} In Chapter~\ref{chap:intro}, we saw market scoring rules (MSRs) as a way of aggregating forecasts. \textcite{cp07} introduced \emph{cost-function markets}, in which a market maker sells $n$ types of shares -- one for each outcome -- where the price of a share depends on the number of shares sold thus far according to some cost function. They established a connection between cost-function markets and MSRs, where a market with a given cost function will behave the same way as a certain MSR. In particular, the cost function $C$ of a cost function market is the convex dual of the expected score function $G$ of the proper scoring rule associated with the corresponding MSR \parencite[\S8.3]{acv13}.

QA pooling has a simple interpretation in terms of cost function markets: for the cost function market corresponding to the scoring rule $s$, let $\vect{q}_i$ be the quantity vector that implies each expert $i$'s probability $\vect{p}_i$ (or in other terms, $i$ would buy a bundle $\vect{q}_i$ of shares as the first participant in the market). Then the QA pool with respect to $s$ is the probability implied by the weighted average quantity vector $\sum_i w_i \vect{q}_i$.

Our work is also superficially similar to \textcite{cv10, acv13}, which tied cost-function market making to online learning of probability distributions. Our results on online learning are in a different context: the goal of their online learning problem is to learn a \emph{probability distribution over outcomes}, whereas our goal in Section~\ref{sec:convex_losses} is to learn \emph{expert weights.}

\section{Preliminaries} \label{sec:chap5_prelims}
Throughout this work, we will let $m$ be the number of experts and use the index $i$ to refer to any particular expert. We will let $n$ be the number of outcomes and use the index $j$ to refer to any particular outcome. We will let $\Delta_n$ be the standard in $\RR^n$ and $\pmb{\delta}_j$ be the basis-aligned unit vector with a $1$ in the $j$-th coordinate.

\subsection{Proper scoring rules}
We have already introduced proper scoring rules, along with their Savage representations, in Section~\ref{sec:prelim_proper}. However, it will serve us well to be more precise in this chapter. First, we will clarify the range of possible scores: we will allow $s(\vect{x}; j)$, for $\vect{x} \in \Delta_n$ and $j \in [n]$, to be any real number, \emph{or} negative infinity. For example, $s_{\text{log}}((1, 0); 2) = -\infty$.

We define the \emph{forecast domain} $\mathcal{D}$ associated with $s$ to be the set of forecasts $\vect{x}$ such that $s(\vect{x}; j)$ is real-valued for all $j$. For example, if $s = s_{\text{log}}$, then $\mathcal{D}$ is the interior of $\Delta_n$. When discussing forecast aggregation, we will assume that all forecasts belong to $\mathcal{D}$.\footnote{This choice removes from consideration cases such as two experts reporting $(1, 0)$ and $(0, 1)$ under the logarithmic scoring rule; aggregating these forecasts using our method is tantamount to adding positive and negative infinity.}

We will let $G$ denote the expected score function of $s$.\footnote{It will typically be clear which scoring rule we are working with, so there is no need to write $G_s$.} We will assume that $s(\cdot; j)$ is continuous on $\Delta_n$ for all $j$, as is $G(\cdot)$; to our knowledge, this is the case for all frequently-used proper scoring rules.\footnote{For $s$, continuity is with respect to the standard topology on $\RR \cup \{-\infty\}$, i.e.\ the one that includes sets of the form $[-\infty, r)$ as open sets. This means that e.g.\ the log scoring rule is continuous on all of $\Delta_n$.}

In Section~\ref{sec:prelim_proper} we introduced the Savage representation of a proper scoring rule. We do so here again, this time more carefully.

\begin{prop}[{\cite[Theorem 2]{gr07}}] \label{prop:G_convex}
	A regular scoring rule $s$ is proper if and only if
	\begin{equation} \label{eq:s_from_g_2}
		s(\vect{x}; j) = G(\vect{x}) + \angles{\vect{g}(\vect{x}), \pmb{\delta}_j - \vect{x}},
	\end{equation}
	for some convex function $G: \Delta_n \to \RR$ and subgradient function\footnote{That is, $\vect{g}$ satisfies $G(\vect{y}) \ge G(\vect{x}) + \angles{\vect{g}(\vect{x}), \vect{y} - \vect{x}}$ for all $\vect{x}, \vect{y}$. Note that $g_i(\vect{x})$ may be $-\infty$ if $x_i = 0$; see \textcite{waggoner21} for an examination of subgradients of functions to the extended reals.} $\vect{g}$ of $G$. The function $G$ is then the expected score function of $s$.
\end{prop}

Since we are assuming that $s$ and $G$ are continuous, it follows from Equation~\ref{eq:s_from_g_2} that $\vect{g}$ must be continuous as well.\footnote{Continuity of each component $g_\ell$ of $\vect{g}$ (for $\ell \in [n]$) is (as with $s$) with respect to the standard topology on $\RR \cup \{-\infty\}$. To see that $g_\ell$ is continuous at a given $\vect{p}$, consider the limit of Equation~\ref{eq:s_from_g_2} as $\vect{x} \to \vect{p}$ with $j = \ell$ if $p_j = 0$ and $j \neq \ell$ if $p_j \neq 0$.} A convex function with a continuous finite subgradient is differentiable \parencite[Proposition 17.41]{bc11}, which means that $G$ is differentiable on the interior of $\Delta_n$, with gradient $\vect{g}$.

As discussed in Section~\ref{sec:prelim_proper}, the key intuition of Equation~\ref{eq:s_from_g_2} is that the score of an expert who reports probability distribution $\vect{p}$ is determined by drawing the tangent plane to $G$ at $\vect{p}$; the value of this plane at $\pmb{\delta}_j$, where $j$ is the outcome that happens, is the expert's score.

We refer to $\vect{g}$ as the \emph{exposure function} of $s$. We borrow this term from finance, where \emph{exposure} refers to how much an agent stands to gain or lose from various possible outcomes -- informally speaking, how much the agent cares about which outcome will happen. If we view $G(\vect{p}) - \angles{\vect{g}(\vect{p}), \vect{p}}$ as the agent's ``baseline profit,'' then the $j$-th component of $\vect{g}(\vect{p})$ is the amount that the agent stands to gain (or lose) on top of the baseline profit if outcome $j$ happens.

\begin{remark} \label{remark:mod_T1n}
Because the domain of $G$ is a subset of $\Delta_n$ (and thus lies in a plane that is orthogonal to the all-ones vector $\vect{1}_n$), it makes the most sense to think of its gradient function $\vect{g}$ as taking on values in $\RR^n$ modulo translation by the all-ones vector $\vect{1}_n$; we will denote this space by $\RR^n/T(\vect{1}_n)$. Sometimes we find it convenient to treat $G$ as a function of $n$ variables rather than $n - 1$ variables out of convenience, thus artificially extending the domain of $G$ outside of the plane containing $\Delta_n$. The component of the gradient of $G$ that is parallel to $\vect{1}_n$ is not relevant.\footnote{Formally, consider the change of coordinates given by $z_j = x_n - x_j$ for $j \le n - 1$ and $z_n = \sum_j x_j$, so that the domain of $G$ lies in the plane $z_n = 1$. Then for $j \le n - 1$, $\partl{G}{z_j}$ at a given point in the domain of $G$ does not change if $1$ is substituted for $z_n$; only $\partl{G}{z_n}$ changes (to zero). Equivalently in terms of our original coordinates, the change that $\vect{g}$ undergoes when we consider $G$ to be a function only defined on $\mathcal{D}$ instead of $\RR^n$ is precisely a projection of $\vect{g}$ onto $\{\vect{x} \in \RR^n: \sum_i x_i = 0\}$.}
\end{remark}

Next, recall the Bregman divergence from Section~\ref{sec:prelim_bregman}:
\bregmandef*

We note the following well-known facts about Bregman divergence:

\begin{prop}[Well-known facts about Bregman divergence] \label{prop:bregman_facts}
Let $G$ be a differentiable, strictly convex function. Then:
\begin{itemize}
\item $D_G(\vect{p} \parallel \vect{q}) \ge 0$, with equality only when $\vect{p} = \vect{q}$.
\item For every $\vect{q}$, $D_G(\vect{x} \parallel \vect{q})$ is a strictly convex function of $\vect{x}$.
\end{itemize}
\end{prop}

Finally, we make a note about interpreting the $n = 2$ outcome case in one dimension.

\begin{remark} \label{remark:n2}
Because $\Delta_n$ is $(n - 1)$-dimensional, we can think of the case of $n = 2$ outcomes in one dimension. All probabilities in are of the form $(p, 1 - p)$; we map $\Delta_2$ to $[0, 1]$ via the first coordinate. Thus, we let $G(p) := G(p, 1 - p)$. We let $g(p) := G'(p) = \angles{\vect{g}(p, 1 - p), (1, -1)}$. The tangent line to $G$ at $p$ will intersect the line $x = 1$ at $s(p; 1)$ (i.e.\ the score if Outcome 1 happens) and intersect the line $x = 0$ at $s(p; 2)$ (i.e.\ the score if Outcome 2 happens). See Figure~\ref{fig:savage} for an illustration.
\end{remark}

The formulation in Remark~\ref{remark:n2} will be helpful when discussing the two-outcome case, e.g.\ in Section~\ref{sec:axiomatization}.

\subsection{Quasi-arithmetic pooling}
We now introduce the central concept of this chapter: quasi-arithmetic pooling.

\begin{defin}[quasi-arithmetic pooling] \label{def:qa_pooling}
	Let $s$ be a proper scoring rule with expected score function $G$ and exposure function $\vect{g}$. Given forecasts $\vect{p}_1, \dots, \vect{p}_m \in \mathcal{D}$ with non-negative weights $w_1, \dots, w_m$ adding to $1$, the \emph{quasi-arithmetic (QA) pool} of these forecasts \emph{with respect to $s$} (or with respect to $\vect{g}$), denoted by $\sideset{}{_\vect{g}}\bigoplus\limits_{i = 1}^m (\vect{p}_i, w_i)$, is the unique $\vect{p}^* \in \Delta_n$ such that $\sum_{i = 1}^m w_i \vect{g}(\vect{p}_i)$ is a subgradient of $G$ at $\vect{p}^*$.
	
	If the forecasts and weights are clear from context, we may simply write $\vect{p}^*$ to refer to their quasi-arithmetic pool; or, if only the forecasts are clear, we may write $\vect{p}^*_{\vect{w}}$, where $\vect{w}$ is the vector of weights.
\end{defin}

\begin{remark}
Equivalently \parencite[Theorem 23.5]{roc70_book}, we can define
\begin{equation} \label{eq:qa_min_1}
	\sideset{}{_\vect{g}}\bigoplus\limits_{i = 1}^m (\vect{p}_i, w_i) := \arg \min_{\vect{x}} G(\vect{x}) - \angles{\vect{x}, \sum_{i = 1}^m w_i \vect{g}(\vect{p}_i)}.
\end{equation}
Also equivalently, we can define
\[\sideset{}{_\vect{g}}\bigoplus\limits_{i = 1}^m (\vect{p}_i, w_i) := \arg \min_{\vect{x}} \sum_{i = 1}^m w_i D_G(\vect{x} \parallel \vect{p}_i).\]
The expression being minimized here differs from the one in Equation~\ref{eq:qa_min_1} by a constant: namely, $\sum_i w_i(\angles{\vect{p}_i, \vect{g}(\vect{p}_i)} - G(\vect{p}_i))$.
\end{remark}

The QA-pool is well-defined, i.e.\ exists and is unique. It is unique because a strictly convex function cannot have the same subgradient at two different points \parencite[Lemma 3.11]{waggoner21}. It exists because $G(\vect{x}) - \angles{\vect{x}, \sum_i w_i \vect{g}(\vect{p}_i)}$ is continuous and thus attains its minimum on the (compact) domain $\Delta_n$.

In light of the fact that $D_G(\vect{p} \parallel \vect{q})$ is the expected loss in score by an expert who, believing $\vect{p}$, instead reports $\vect{q}$ (as discussed in Section~\ref{sec:prelim_bregman}), the Bregman divergence formulation of QA pooling gives another natural interpretation.

\begin{remark} \label{remark:bregman_interpretation}
	Consider a proper scoring rule $s$ with forecasts $\vect{p}_1, \dots, \vect{p}_m$ with weights $w_1, \dots, w_m$. The QA pool of these forecasts is the forecast $\vect{p}^*$ that, if it is the correct answer (i.e.\ if the outcome is drawn according to $\vect{p}^*$), would minimize the expected loss of a randomly chosen (according to $\vect{w}$) expert relative to reporting $\vect{p}^*$.
\end{remark}

In this sense, QA pooling reflects a \emph{compromise} between experts: it is the probability that, if it were correct, would make the experts' forecast least wrong overall.

\begin{remark}
	Since the Bregman divergence is convex in its first argument, computing the QA pool is a matter of convex optimization. In particular, given oracle access to $\vect{g}$, the ellipsoid method (see e.g.\ \textcite{gls93}) can be used to efficiently find the QA pool of a list of forecasts.
\end{remark}

Note that although Definition~\ref{def:qa_pooling} only specifies that $\vect{p}^* \in \Delta_n$, in fact it lies in $\mathcal{D}$:

\begin{claim}
	For any $\vect{p}_1, \dots, \vect{p}_m$ and $w_1, \dots, w_m$, the QA pool $\vect{p^*}$ lies in $\mathcal{D}$.
\end{claim}

\begin{proof}
	Let $\vect{x} \in \Delta_n \setminus \mathcal{D}$. We show that $\sum_i w_i \vect{g}(\vect{p}_i)$ is not a subgradient of $G$ at $\vect{x}$.
	
	We have $s(\vect{x}; j) = -\infty$ for some $j$ (satisfying $x_j = 0$), so $\angles{\vect{g}(\vect{x}), \pmb{\delta}_j - \vect{x}} = -\infty$. Since $\vect{g}$ is continuous, for sufficiently small $\epsilon$, we have
	\[\angles{\vect{g}(\vect{x} + \epsilon'(\pmb{\delta}_j - \vect{x})), \pmb{\delta}_j - \vect{x}} < \angles{\sum_i w_i \vect{g}(\vect{p}_i), \pmb{\delta}_j - \vect{x}}\]
	for all $\epsilon' \le \epsilon$. This means that
	\[G(\vect{x} + \epsilon(\pmb{\delta}_j - \vect{x})) - G(\vect{x}) < \angles{\sum_i w_i \vect{g}(\vect{p}_i), \epsilon(\pmb{\delta}_j - \vect{x})},\]
	so $\sum_i w_i \vect{g}(\vect{p}_i)$ is not a subgradient of $G$ at $\vect{x}$, as desired.
\end{proof}

While our max-min optimality result (Section~\ref{sec:max_min}) holds unconditionally, our results in Section~\ref{sec:convex_losses} and \ref{sec:axiomatization} require that our proper scoring rule $s$ satisfy a property that we term \emph{convex exposure}.

\begin{defin}[convex exposure] \label{def:convex_exposure}
	A proper scoring rule $s$ with forecast domain $\mathcal{D}$ has \emph{convex exposure} if the range of its exposure function $\vect{g}$ on $\mathcal{D}$ is a convex set.
\end{defin}

The key fact about proper scoring rules with convex exposure is that for all $\vect{p}_1, \dots, \vect{p}_m$ and $w_1, \dots, w_m$, $\sum_i w_i \vect{g}(\vect{p}_i) = \vect{g}(\vect{x})$ for some $\vect{x} \in \mathcal{D}$. This means that $\sum_i w_i \vect{g}(\vect{p}_i)$ is a subgradient of $G$ at $\vect{x}$, so $\vect{x}$ is the weighted QA pool of the $p_i$'s. In other words, if $s$ has convex exposure, then we may write
\begin{equation} \label{eq:g_convex_exposure}
	\vect{g}(\vect{p}^*) = \sum_i w_i \vect{g}(\vect{p}_i),
\end{equation}
where $\vect{p}^*$ is the weighted QA pool of the given forecasts. The convex exposure property thus allows us to write down relations between exposures of forecasts that would otherwise not necessarily be true.

The quadratic and logarithmic scoring rules, as well as all proper scoring rules for binary outcomes, have convex exposure. In Appendix~\ref{appx:convex_exposure} we explore in more depth the question of which commonly used proper scoring rules have convex exposure property.

\section{QA pooling as a max-min optimal method} \label{sec:max_min}
Our goal is to give a formal justification for quasi-arithmetic pooling. Remark~\ref{remark:bregman_interpretation} established that the QA pool is optimizes (i.e.\ minimizes) the weighted average Bregman divergence to the experts' forecasts. This section gives another justification for QA pooling in terms of max-min optimality. We will give additional justifications in later sections.

\begin{theorem} \label{thm:max_min}
Let $s$ be a proper scoring rule and let $\vect{g}$ be the exposure function of $s$. Fix any forecasts $\vect{p}_1, \dots, \vect{p}_m \in \mathcal{D}$ with non-negative weights $w_1, \dots, w_m$ adding to $1$. Define
\[u(\vect{p}; j) := s(\vect{p}; j) - \sum_{i = 1}^m w_i s(\vect{p}_i; j).\]
Then the quantity $\min_j u(\vect{p}; j)$ is uniquely maximized by setting $\vect{p}$ to $\vect{p}^* := \sideset{}{_\vect{g}}\bigoplus\limits_{i = 1}^m (\vect{p}_i, w_i)$. Furthermore, this minimum (across outcomes $j$) is achieved simultaneously by all $j$ with $p^*_j > 0$. This quantity is non-negative, and is positive unless all reports $\vect{p}_i$ with positive weights are equal.
\end{theorem}

One interpretation for this theorem statement is as follows. Consider an agent who is tasked with submitting a forecast, and who will be paid according to $s$. The agent decides to sub-contract $m$ experts to get their opinions, paying expert $i$ the amount $w_i s(\vect{p}_i; j)$ if the expert reports $\vect{p}_i$ and outcome $j$ happens. (Perhaps experts whom the agent trusts more have higher $w_i$'s.) Finally, the agent reports some (any) forecast $\vect{p}$. Then $u(\vect{p}; j)$ is precisely the agent's profit (utility).

The quantity $\min_j u(\vect{p}; j)$ is the agent's minimum possible profit over all outcomes. It is natural to ask which report $\vect{p}$ maximizes this quantity. Theorem~\ref{thm:max_min} states that this maximum is achieved by the QA pool of the experts' forecasts with respect to $s$, and that this is the unique maximizer.

A possible geometric intuition to keep in mind for the proof (below): for each expert $i$, draw the plane tangent to $G$ at $\vect{p}_i$. For any $j$, the value of this plane at $\pmb{\delta}_j$ is $s(\vect{p}_i; j)$. Now take the weighted average of all $m$ planes; this is a new plane whose intersection with any $\pmb{\delta}_j$ is the total score received by the experts if $j$ happens. Since $G$ is convex, this plane lies below $G$. To figure out which point maximizes the agent's guaranteed profit, push the plane upward until it hits $G$. It will hit $G$ at $\vect{p}^*$ and the agent's worst-case profit will be the vertical distance that the plane was pushed.

\begin{proof}[Proof of Theorem~\ref{thm:max_min}]
By Equation~\ref{eq:qa_min_1}, computing the QA pool amounts to finding the minimizer $\vect{p}^*$ of the function $G(\vect{x}) - \angles{\vect{x}, \sum_{i = 1}^m w_i \vect{g}(\vect{p}_i)}$ over $\Delta_n$. If $\vect{p}^*$ is in the interior of $\Delta_n$, then this expression is differentiable at $\vect{p}^*$. If $\vect{p}^*$ is on the boundary, then the expression can be extended to a differentiable function in a neighborhood around $\vect{p}^*$.\footnote{This follows e.g.\ from \textcite[Theorem 1.8]{am15}, where we take $C$ in the theorem statement to be a compact subset of $\Delta_n$ containing $\vect{p}^*$ where $G$ is differentiable. Here we use that $\vect{p}^* \in \mathcal{D}$; by continuity of $\vect{g}$, such a subset necessarily exists.} Thus, applying the KKT conditions (see e.g.\ \textcite[\S5.5.3]{bv04}) tells us that
\[\vect{g}(\vect{p}^*) = \sum_{i = 1}^m w_i \vect{g}(\vect{p}_i) + \lambda \vect{1}_n + \pmb{\mu}\]
for some $\lambda \in \RR$ and $\pmb{\mu} \in \RR^n$ such that $\mu_j \ge 0$ and $\mu_j = 0$ if $p^*_j > 0$. We therefore have
\begin{align*}
	u(\vect{p}^*; j) &= s(\vect{p}^*; j) - \sum_i w_i s(\vect{p}_i; j)\\
	&= G(\vect{p}^*) + \angles{\vect{g}(\vect{p}^*), \pmb{\delta}_j - \vect{p}^*} - \sum_i w_i (G(\vect{p}_i) + \angles{\vect{g}(\vect{p}_i), \pmb{\delta}_j - \vect{p}_i})\\
	&= G(\vect{p}^*) - \sum_i w_i G(\vect{p}_i) + \angles{\sum_i w_i \vect{g}(\vect{p}_i) + \lambda \vect{1}_n + \pmb{\mu}, \pmb{\delta}_j - \vect{p}^*} - \sum_i w_i \angles{\vect{g}(\vect{p}_i), \pmb{\delta}_j - \vect{p}_i}\\
	&= G(\vect{p}^*) - \sum_i w_i G(\vect{p}_i) + \sum_i w_i \angles{\vect{g}(\vect{p}_i), \vect{p}_i - \vect{p}^*} + \angles{\lambda \vect{1}_n + \pmb{\mu}, \pmb{\delta}_j - \vect{p}^*}\\
	&= G(\vect{p}^*) - \sum_i w_i G(\vect{p}_i) + \sum_i w_i \angles{\vect{g}(\vect{p}_i), \vect{p}_i - \vect{p}^*} + \mu_j\\
	&= \sum_i w_i D_G(\vect{p}^* \parallel \vect{p}_i) + \mu_j.
\end{align*}
The second-to-last step follows from the fact that $\angles{\lambda \vect{1}_n, \pmb{\delta}_j - \vect{p}^*} = 0$ and $\angles{\pmb{\mu}, \vect{p}^*} = 0$, and the last step follows by the definition of Bregman divergence. Since $\mu_j = 0$ for every $j$ such that $p^*_j > 0$ (of which there is at least one), the minimum of $u(\vect{p}^*; j)$ over $j$ is achieved simultaneously for all $j$ with $p^*_j > 0$, and this minimum is equal to $\sum_i w_i D_G(\vect{p}^* \parallel \vect{p}_i)$. This quantity is non-negative, and positive except when all $\vect{p}_i$'s with positive weights are equal.

Finally we show that $\vect{x} = \vect{p}^*$ maximizes $\min_j u(\vect{x}; j)$. Suppose that for some report $\vect{q}$ we have that $\min_j u(\vect{q}; j) \ge \min_j u(\vect{p}^*; j)$. Then $u(\vect{q}; j) \ge u(\vect{p}^*; j)$ for \emph{every} $j$ such that $p^*_j > 0$. But this means that the expected score (according to $s$) of an expert who believes $\vect{p}^*$ is larger if the expert reports $\vect{q}$ than if the expert reports $\vect{p}^*$. This contradicts the fact that $s$ is proper.
\end{proof}

\begin{remark}
We can reformulate Theorem~\ref{thm:max_min} as follows: suppose that an agent has access to forecasts $\vect{p}_1, \dots, \vect{p}_m$ and needs to issue a forecast, for which the agent will be scored using a proper scoring rule $s$. The agent can improve upon selecting an expert at random according to weights $w_1, \dots, w_m$, no matter the outcome $j$, by reporting $\vect{p}^*$. This improvement is the same no matter the outcome -- so long as the outcome is assigned positive probability by $\vect{p}^*$ -- and is a strict improvement unless all forecasts with positive weights are the same.
\end{remark}

\begin{remark}
	Theorem~\ref{thm:max_min} is closely related to work by \textcite{gd04} that establishes a connection between entropy maximization and worst-case expected loss minimization. Their work studies a generalized notion of entropy functions that, for the case of a proper scoring rule $s$, is equal to the negative of the expected score function $G$. They show that the forecast $\vect{x}$ that maximizes entropy (minimizes $G$) also maximizes an expert's worst-case score (over outcomes $j$). Considering instead the entropy function $\angles{x, \sum_i w_i \vect{g}(\vect{p}_i)} - G(\vect{x})$ yields results that are very similar to our Theorem~\ref{thm:max_min}. In particular, if $\mathcal{D}$ is a closed set or an open set, then the max-min result in Theorem~\ref{thm:max_min} can be derived from \textcite[Theorem 5.2]{gd04} and \textcite[Theorem 6.2]{gd04}, respectively. However, our proof below captures cases that their results do not address. See also \textcite[Lemma 1]{chl19}, from which (upon considering the scoring rule with expected score function $G(\vect{x}) - \angles{x, \sum_i w_i \vect{g}(\vect{p}_i)}$) yields the simultaneity result of Theorem~\ref{thm:max_min} if $\vect{p}^*$ lies in the interior of $\Delta_n$.
\end{remark}

\section{Convex losses and learning expert weights} \label{sec:convex_losses}
Thus far, when discussing QA pooling, we have regarded expert weights as given. Where do these weights come from? As demonstrated by the results in this section, if the proper scoring rule $s$ has convex exposure (see Definition~\ref{def:convex_exposure}), these weights can be learned from experience. This learning property for weights falls out of the following key observation, which states that an agent's score is a concave function of the weights it uses for the experts.

\begin{theorem} \label{thm:ws_concave}
Let $s$ be a proper scoring rule with convex exposure and forecast domain $\mathcal{D}$, and fix any $\vect{p}_1, \dots, \vect{p}_m \in \mathcal{D}$. Given a weight vector $\vect{w} = (w_1, \dots, w_m) \in \Delta_m$, define the \emph{weight-score} of $\vect{w}$ for an outcome $j$ as
\[\ws_j(\vect{w}) := s \parens{\sideset{}{_\vect{g}}\bigoplus_{i = 1}^m (\vect{p}_i, w_i); j}.\]
Then for every $j \in [n]$, $\ws_j(\vect{w})$ a concave function of $\vect{w}$.
\end{theorem}

\begin{proof}
	Let $\vect{v}$ and $\vect{w}$ be two weight vectors. We wish to show that for any $c \in [0, 1]$, we have
	\[\ws_j(c\vect{v} + (1 - c)\vect{w}) - c \ws_j(\vect{v}) - (1 - c) \ws_j(\vect{w}) \ge 0.\]
	Recall the notation $\vect{p}^*_{\vect{w}}$ from Definition~\ref{def:qa_pooling}. Note that
	\begin{equation} \label{eq:g_linear}
		\vect{g}(\vect{p}^*_{c\vect{v} + (1 - c)\vect{w}}) = \sum_{i = 1}^m (cv_i + (1 - c)w_i) \vect{g}(\vect{p}_i) = c \vect{g}(\vect{p}^*_{\vect{v}}) + (1 - c) \vect{g}(\vect{p}^*_{\vect{w}}).
	\end{equation}
	We have
	\begin{align*}
		&\quad \ws_j(c\vect{v} + (1 - c)\vect{w}) - c \ws_j(\vect{v}) - (1 - c) \ws_j(\vect{w})\\
		&= s(\vect{p}^*_{c\vect{v} + (1 - c)\vect{w}}; j) - c s(\vect{p}^*_{\vect{v}}; j) - (1 - c) s(\vect{p}^*_{\vect{w}}; j)\\
		&= G(\vect{p}^*_{c\vect{v} + (1 - c)\vect{w}}) + \angles{\vect{g}(\vect{p}^*_{c\vect{v} + (1 - c)\vect{w}}), \pmb{\delta}_j - \vect{p}^*_{c\vect{v} + (1 - c)\vect{w}}}\\
		&\qquad - c(G(\vect{p}^*_{\vect{v}}) + \angles{\vect{g}(\vect{p}^*_{\vect{v}}), \pmb{\delta}_j - \vect{p}^*_{\vect{v}}}) - (1 - c)(G(\vect{p}^*_{\vect{w}}) + \angles{\vect{g}(\vect{p}^*_{\vect{w}}), \pmb{\delta}_j - \vect{p}^*_{\vect{w}}})\\
		&= G(\vect{p}^*_{c\vect{v} + (1 - c)\vect{w}}) - \angles{\vect{g}(\vect{p}^*_{c\vect{v} + (1 - c)\vect{w}}), \vect{p}^*_{c\vect{v} + (1 - c)\vect{w}}}\\
		&\qquad - c(G(\vect{p}^*_{\vect{v}}) - \angles{\vect{g}(\vect{p}^*_{\vect{v}}), \vect{p}^*_{\vect{v}}}) - (1 - c)(G(\vect{p}^*_{\vect{w}}) - \angles{\vect{g}(\vect{p}^*_{\vect{w}}), \vect{p}^*_{\vect{w}}})
	\end{align*}
	Step 1 follows from the definition of $\ws$. Step 2 follows from Equation~\ref{eq:s_from_g_2}. Step 3 follows from Equation~\ref{eq:g_linear}, and specifically that the inner product of each side with $\pmb{\delta}_j$ is the same (so the $\pmb{\delta}_j$ terms cancel out, leaving a quantity that does not depend on $j$). Continuing where we left off:
	
	\begin{align*}
		&\quad \ws_j(c\vect{v} + (1 - c)\vect{w}) - c \ws_j(\vect{v}) - (1 - c) \ws_j(\vect{w})\\
		&= G(\vect{p}^*_{c\vect{v} + (1 - c)\vect{w}}) - c \angles{\vect{g}(\vect{p}^*_{\vect{v}}), \vect{p}^*_{c\vect{v} + (1 - c)\vect{w}}} - (1 - c) \angles{\vect{g}(\vect{p}^*_{\vect{w}}), \vect{p}^*_{c\vect{v} + (1 - c)\vect{w}}}\\
		&\qquad - c(G(\vect{p}^*_{\vect{v}}) - \angles{\vect{g}(\vect{p}^*_{\vect{v}}), \vect{p}^*_{\vect{v}}}) - (1 - c)(G(\vect{p}^*_{\vect{w}}) - \angles{\vect{g}(\vect{p}^*_{\vect{w}}), \vect{p}^*_{\vect{w}}})\\
		&= c(G(\vect{p}^*_{c\vect{v} + (1 - c)\vect{w}}) - G(\vect{p}^*_{\vect{v}}) - \angles{\vect{g}(\vect{p}^*_{\vect{v}}), \vect{p}^*_{c\vect{v} + (1 - c)\vect{w}} - \vect{p}^*_{\vect{v}}})\\
		&\qquad + (1 - c)(G(\vect{p}^*_{c\vect{v} + (1 - c)\vect{w}}) - G(\vect{p}^*_{\vect{w}}) - \angles{\vect{g}(\vect{p}^*_{\vect{w}}), \vect{p}^*_{c\vect{v} + (1 - c)\vect{w}} - \vect{p}^*_{\vect{w}}})\\
		&= c D_G(\vect{p}^*_{c\vect{v} + (1 - c)\vect{w}} \parallel \vect{p}^*_{\vect{v}}) + (1 - c) D_G(\vect{p}^*_{c\vect{v} + (1 - c)\vect{w}} \parallel \vect{p}^*_{\vect{w}}) \ge 0.
	\end{align*}
	Step 4 again follows from Equation~\ref{eq:g_linear}. Step 5 is a rearrangement of terms. Finally, step 6 follows from the definition of Bregman divergence, and step 7 follows from the fact that Bregman divergence is always non-negative. This completes the proof.
\end{proof}

\begin{remark}
Theorem~\ref{thm:ws_concave} can be stated in more generality: $s$ need not have convex exposure; it suffices to have that for the particular $\vect{p}_1, \dots, \vect{p}_m$, the QA pool of these forecasts exists for every weight vector.
\end{remark}

Beyond Theorem~\ref{thm:ws_concave}'s instrumental use for no-regret online learning of expert weights (Theorem~\ref{thm:chap5_no_regret} below), the result is interesting in its own right. For example, the following fact -- loosely speaking, that QA pooling cannot benefit from weight randomization -- follows as a corollary. (Recall the definition of $\vect{p}_{\vect{w}}^*$ from Definition~\ref{def:qa_pooling}.)

\begin{corollary}
Consider a randomized algorithm $A$ with the following specifications:
\begin{itemize}
	\item Input: a proper scoring rule $s$ with convex exposure, expert forecasts $\vect{p}_1, \dots, \vect{p}_m$.
	\item Output: a weight vector $\vect{w} \in \Delta_m$.
\end{itemize}
For any input $s, \vect{p}_1, \dots, \vect{p}_m$ and for every $j$, we have
\[s(\vect{p}^*_{\EE[A]{\vect{w}}}; j) \ge \EE[A]{s(\vect{p}^*_{\vect{w}}; j)},\]
where $\vect{p}^*_{\vect{x}}$ denotes the QA pool of $\vect{p}_1, \dots, \vect{p}_m$ with weight vector $\vect{x}$.
\end{corollary}

\begin{remark}
Theorem~\ref{thm:ws_concave} would \emph{not} hold if the pooling operator in the definition of $\ws$ were replaced by linear pooling or by logarithmic pooling.\footnote{For a counterexample to logarithmic pooling, consider $n = 2$, let $s$ be the quadratic scoring rule, $p_1 = (0.1, 0.9)$, $p_2 = (0.5, 0.5)$, $j = 1$, $\vect{v} = (1, 0)$, $\vect{w} = (0, 1)$, and $c = \frac{1}{2}$. For a counterexample to linear pooling, consider $n = 2$, let $s$ be given by $G(p_1, p_2) = \sqrt{p_1^2 + p_2^2}$ (this is known as the \emph{spherical} scoring rule), $p_1 = (0, 1)$, $p_2 = (0.2, 0.8)$, $j = 1$, $\vect{v} = (1, 0)$, $\vect{w} = (0, 1)$, and $c = \frac{1}{2}$.} This is an advantage of QA pooling over using the linear or logarithmic method irrespective of the scoring rule.
\end{remark}

We now state the no-regret result that we have alluded to. This result is quite strong in that it does not merely achieve low regret compared to the best \emph{expert}, but in fact compared to the best possible \emph{weighted pool of experts} in hindsight. This is a substantial distinction, as it is possible for a mixture of experts to substantially outperform any individual expert.

\begin{restatable}{theorem}{noregret} \label{thm:chap5_no_regret}
Let $s$ be a bounded proper scoring rule with convex exposure and forecast domain $\mathcal{D}$. For time steps $t = 1 \dots T$, an agent chooses a weight vector $\vect{w}^t \in \Delta_m$. The agent then receives a score of
\[s \parens{\sideset{}{_\vect{g}}\bigoplus_{i = 1}^m (\vect{p}_i^t, w_i^t); j^t},\]
where $\vect{p}_1^t, \dots, \vect{p}_m^t \in \mathcal{D}$ and $j^t \in [n]$ are chosen adversarially. By choosing $\vect{w}^t$ according to Algorithm~\ref{alg:ogd} (online gradient descent on the experts' weights), the agent achieves $O(\sqrt{T})$ regret in comparison with the best weight vector in hindsight. In particular, if $M$ is an upper bound\footnote{This bound $M$ exists because $s$ is bounded by assumption, and so $\vect{g}$ is also bounded (this follows from Equation~\ref{eq:s_from_g_2}).} on $\norm{\vect{g}}_2$, then for every $\vect{w}^* \in \Delta_m$ we have
\[\sum_{t = 1}^T s \parens{\sideset{}{_\vect{g}}\bigoplus_{i = 1}^m (\vect{p}_i^t, w_i^*); j^t} - s \parens{\sideset{}{_\vect{g}}\bigoplus_{i = 1}^m (\vect{p}_i^t, w_i^t); j^t} \le 3\sqrt{m}M\sqrt{T}.\]
\end{restatable}

\setcounter{algocf}{\value{theorem}}
\begin{algorithm}[ht]
    \caption{Online gradient descent algorithm for Theorem~\ref{thm:chap5_no_regret}} \label{alg:ogd}
    $\vect{w}^1 \gets (1/m, \dots, 1/m)$\;
    \For{$t = 1$ to $T$}{
        $\eta_t \gets \frac{1}{M\sqrt{mt}}$\;
        Play $\vect{w}^t$ and observe loss $L^t(\vect{w}^t)$\;
        $\tilde{\vect{w}}^{t + 1} \gets \vect{w}^t - \eta_t \nabla L^t(\vect{w}^t)$\;
        $\vect{w}^{t + 1} \gets$ projection of $\tilde{\vect{w}}^{t + 1}$ onto $\Delta_m$\;
    }
\end{algorithm}
\setcounter{theorem}{\value{algocf}}

Algorithm~\ref{alg:ogd}, referenced in the theorem statement, is an application of the standard online gradient descent algorithm (see e.g.\ \textcite[Theorem 3.1]{hazan23}) to our particular setting. In our context, the loss function is $L^t(\vect{w}) = -\ws_{j^t}(\vect{w})$, where $\ws$ is as in Theorem~\ref{thm:ws_concave}, relative to forecasts $\vect{p}_1^t, \dots, \vect{p}_m^t$. We defer the proof of Theorem~\ref{thm:chap5_no_regret} to Appendix~\ref{appx:convex_losses}. The proof amounts to applying the standard bounds for online gradient descent, though with an extra step: we use the bound $M$ on $\norm{\vect{g}}$ to bound the gradient of the loss as a function of expert weights.

\section{QA pooling connects two notions of overconfidence} \label{sec:overconfidence}
Overconfidence is a well-studied phenomenon in which forecasters are systematically biased toward reporting probabilities that are too extreme. For example, suppose that an expert assigns probabilities to a large number of events, and that, among those events to which the expert assigns a 1\% probability, 10\% end up happening. This is a sign of overconfidence: the expert's forecasts would be better if all of their forecasts of 1\% were systematically increased.

What, formally, is meant by overconfidence? Arguably, a definition of overconfidence ought to depend on the proper scoring rule used to elicit or grade the expert's forecast. This is because -- as we have already discussed -- a proper scoring rule expresses a subjective opinion about which probabilistic distinctions are important. If an expert assigns a 0.01\% probability to an outcome that has a 1\% chance of occurring, is this a larger mistake than if the expert assigns a 40\% probability to an outcome that has a 41\% chance of occurring? The logarithmic scoring rule says yes; the quadratic scoring rule says no.\footnote{Formally, recall that the Bregman divergence from the true probability to the reported forecast with respect to the expected score function $G$ measures the score that the expert loses out on due to their inaccuracy. We have $D_{G_{\text{log}}}(1\% \parallel 0.01\%) \gg D_{G_{\text{log}}}(41\% \parallel 40\%)$, whereas $D_{G_{\text{quad}}}(1\% \parallel 0.01\%) \approx D_{G_{\text{quad}}}(41\% \parallel 40\%)$.}

In this section, we present two definitions of overconfidence with respect to a proper scoring rule. The first definition is natural and well-motivated, but (to our knowledge) has not previously appeared in the literature. The second definition adapts a standard notion of overconfidence to respect the intuition of the previous paragraph by making use of QA pooling. We then prove that the two definitions are equivalent.

For both definitions, we consider the following setting: an expert forecasts $\vect{p}^1, \dots, \vect{p}^N \in \Delta_n$ for $N$ events. Outcomes $j^1, \dots, j^N \in [n]$ are realized. We are defining what it means for the expert to be overconfident on this sample of $N$ events.\footnote{We define overconfidence with respect to a particular realization of the outcomes (as opposed to on average over realizations) so that our definitions do not require a prior over the probabilities of the outcomes.}

The first definition uses a quite simple principle: \emph{an overconfident expert expects a larger score than they will receive.}

\begin{defin}[Expected score definition of overconfidence] \label{def:exp_score_over}
    Suppose that an expert forecasts $\vect{p}^1, \dots, \vect{p}^N \in \Delta_n$, and outcomes $j^1, \dots, j^N \in [n]$ are realized. The expert is \emph{expected-score overconfident} with respect to a proper scoring rule $s$ (with expected score function $G$) if their expected score exceeds their actual score:
    \begin{equation} \label{eq:exp_score_over}
        \sum_{k = 1}^N G(\vect{p}^k) > \sum_{k = 1}^N s(\vect{p}^k; j^k).
    \end{equation}
\end{defin}

On a basic level, the intuition behind Definition~\ref{def:exp_score_over} is straightforward: an overconfident expert will overestimate their score. But to gain some further intuition, recall Figure~\ref{fig:savage} (repeated here). It shows the score of an expert who reports a probability of $x = 40\%$, under the No outcome (the red dot on the left) and under the Yes outcome (the red dot on the right). Suppose that, in the setting of Definition~\ref{def:exp_score_over}, an expert reports 40\% for all $N$ events. Then the expert expects a score of $G(0.4)$ per event on average: more than that under a No outcome (assuming $G$ is symmetric about $0.5$, as shown) and less than that under a Yes outcome, but $G(0.4)$ on average.

\begin{figure}[ht]
    \centering
    \includegraphics[scale=0.8]{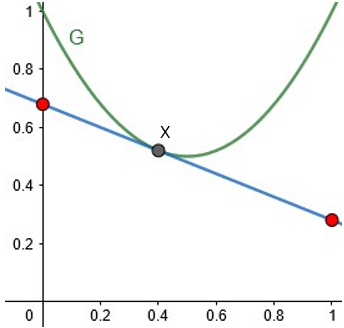}
\end{figure}

If in fact 40\% of the events happen, then the expert is correctly calibrated: neither over- nor underconfident. Their total score will be $N \cdot G(0.4)$ and so the left-hand side of Equation~\ref{eq:exp_score_over} will equal the right-hand side. If only 20\% of the events happen, then the expert is underconfident (should have assigned more extreme probabilities) -- and indeed, the expert's total score will be \emph{higher} than the expert expects. And if more than 40\% of the events happen, then the expert is overconfident, and indeed, that is the case in which Equation~\ref{eq:exp_score_over} holds.\\

Our second definition also uses a simple principle: \emph{an overconfident expert would be better off assigning less extreme probabilities.} This basic notion has previously appeared in the literature. For example, \textcite{schan14} uses the following definition: an expert is overconfident with respect to a proper scoring rule if shrinking all of the expert's forecasts toward the uniform distribution by some positive constant factor will improve the expert's score. Formally, an expert is overconfident by this definition if there exists $w \in (0, 1)$ such that \[\sum_k s \parens{w \vect{p}^k + (1 - w) (1/n, \dots, 1/n); j^k} > \sum_k s(\vect{p}^k; j^k).\]

Our definition is similar in spirit, but importantly different: why shrink the expert's report toward the uniform distribution by a constant factor? Again, it seems that the correct form of shrinkage ought to depend $s$. Perhaps the quadratic score treats 26\% as being equally far from 2\% as it is from 50\% (and so shrinking 2\% halfway toward 50\% would produce a result of 26\%), but the log score certainly does not: shrinking 2\% halfway toward 50\% ``with respect to'' the log scoring rule should give something much less than 26\%. And so our definition instead uses the exposure function $\vect{g}$ of $s$. Much as QA pooling averages the exposures of the experts' forecasts, our method of shrinking a forecast is to average its exposure with the zero vector. Intuitively this means uniformly making the forecast ``more indifferent'' to the eventual outcome.

\begin{defin}
    For a proper scoring rule $s$, we define $\shrink_s(\vect{p}, w)$ to be the forecast $\vect{p}^*$ such that $w\vect{g}(\vect{p})$ is a subgradient of $G$ at $\vect{p}^*$, where $\vect{g}$ is the exposure function of $s$.
\end{defin}

Note that $\shrink_s(\vect{p}, 1) = \vect{p}$, i.e.\ $w = 1$ means no shrinkage (whereas $w = 0$ means complete shrinkage).

\begin{remark} \label{rem:exists_q}
    In the case that there exists $\vect{q}$ such that $\vect{g}(\vect{q}) = 0$ -- i.e.\ a forecast $\vect{q}$ such that the score of $\vect{q}$ does not depend on the outcome -- we have
    \[\shrink_s(\vect{p}, w) = (\vect{p}, w) \oplus (\vect{q}, 1 - w).\]
    For this reason, we will refer to $\shrink_s$ as \emph{QA shrinkage} with respect to $s$. Such a $\vect{q}$ typically exists for proper scoring rules that are used in practice. Indeed, for any proper scoring rule that treats all outcomes symmetrically, $\vect{q}$ is simply the uniform distribution $(1/n, \dots, 1/n)$. In particular, QA shrinkage with respect to the quadratic scoring rule is equivalent to averaging with the uniform distribution. That is, in the case of $s_{\text{quad}}$, our definition coincides with that of \textcite{schan14}. On the other hand, QA shrinkage with respect to $s_{\text{log}}$ is equivalent to taking a logarithmic pool of the forecast with the uniform distribution.
\end{remark}

\begin{defin}[QA shrinkage definition of overconfidence] \label{def:qa_shrinkage_over}
    Suppose that an expert forecasts $\vect{p}^1, \dots, \vect{p}^N \in \Delta_n$, and outcomes $j^1, \dots, j^N \in [n]$ are realized. The expert is \emph{QA-shrinkage overconfident} with respect to a proper scoring rule $s$ if for some $w \in (0, 1)$, we have
    \begin{equation} \label{eq:qa_shrinkage_over}
        \sum_{k = 1}^N s(\shrink_s(\vect{p}^k, w); j^k) > \sum_{k = 1}^N s(\vect{p}^k; j^k).
    \end{equation}
\end{defin}

We can now state our main theorem, which is that (under some weak assumptions about $s$) these two definitions are the same.

\begin{theorem} \label{thm:overconfidence}
    Let $s$ be a proper scoring rule with convex exposure, for which there is a $\vect{q} \in \Delta_n$ such that $\vect{g}(\vect{q}) = \vect{0}$. Suppose that an expert forecasts $\vect{p}^1, \dots, \vect{p}^N \in \Delta_n$, and outcomes $j^1, \dots, j^N \in [n]$ are realized. The expert is expected-score overconfident with respect to $s$ if and only if they are QA-shrinkage overconfident with respect to $s$.
\end{theorem}

\begin{proof}
    First, observe that by using the Savage representation of $s$, we may rewrite the condition for expected-score overconfidence (Equation~\ref{eq:exp_score_over}) as
    \[\sum_{k = 1}^N \angles{\vect{g}(\vect{p}^k), \pmb{\delta}_{j^k} - \vect{p}^k} < 0.\]
    We now show that Definition~\ref{def:qa_shrinkage_over} is equivalent. By Remark~\ref{rem:exists_q}, we may write $\shrink_s(\vect{p}^k, w) = (\vect{p}^k, w) \oplus (\vect{q}, 1 - w)$, where $\vect{q}$ is as in the theorem statement. Furthermore, because $s$ has convex exposure, Equation~\ref{eq:g_convex_exposure} tells us that $\vect{g}(\shrink_s(\vect{p}^k, w)) = w \vect{g}(\vect{p}^k)$.
    
    We now use Theorem~\ref{thm:ws_concave}, which tells us that for all $k$, the quantity
    \[s \parens{(\vect{p}^k, w) \oplus_s (\vect{q}, 1 - w); j^k}\]
    is concave in $w$. This means that the left-hand side of Equation~\ref{eq:qa_shrinkage_over} is a concave function (call it $f$) of $w$. On the other hand, the right-hand side of Equation~\ref{eq:qa_shrinkage_over} is $f(1)$. The question of whether there exists $w \in (0, 1)$ such that $f(w) > f(1)$ is therefore equivalent to the question of whether the derivative of $f$ at $w = 1$ is negative.\footnote{Or more precisely, the left derivative, since $f$ is only defined on $[0, 1]$.} To evaluate this derivative, we note that
    \begin{align*}
        s(\shrink_s(\vect{p}^k, w); j^k) &= G(\shrink_s(\vect{p}^k, w)) + \angles{\vect{g}(\shrink_s(\vect{p}^k, w)), \pmb{\delta}_{j^k} - \shrink_s(\vect{p}^k, w)}\\
        &= G(\shrink_s(\vect{p}^k, w)) + w \angles{\vect{g}(\vect{p}^k), \pmb{\delta}_{j^k} - \shrink_s(\vect{p}^k, w)}.
    \end{align*}
    Therefore, we have
    \[\partl{s(\shrink_s(\vect{p}^k, w); j^k)}{w} = \partl{G(\shrink_s(\vect{p}^k, w))}{w} + \partl{\parens{w \angles{\vect{g}(\vect{p}^k), \pmb{\delta}_{j^k} - \shrink_s(\vect{p}^k, w)}}}{w}.\]
    Considering $\shrink_s(\vect{p}^k, w)$ as a function from $[0, 1]$ to $\RR^n$, let $\vect{J}(w)$ be its Jacobian, i.e.\ the vector of partial derivatives of the $n$ coordinates with respect to $w$. Applying the rules of differentiation, we have that
    \begin{align*}
        \partl{s(\shrink_s(\vect{p}^k, w); j^k)}{w} &= \angles{\vect{g}(\shrink_s(\vect{p}^k, w)), \vect{J}(w)} + \angles{\vect{g}(\vect{p}^k), \pmb{\delta}_{j^k} - \shrink_s(\vect{p}^k, w)}\\
        &\quad - w \angles{\vect{g}(\vect{p}^k), \vect{J}(w)}.
    \end{align*}
    Now, we are specifically interested in the derivative at $w = 1$. Plugging in $w = 1$, we have
    \[\partl{s(\shrink_s(\vect{p}^k, w); j^k)}{w} \mid_{w = 1} = \angles{\vect{g}(\vect{p}^k), \vect{J}(1)} + \angles{\vect{g}(\vect{p}^k), \pmb{\delta}_{j^k} - \vect{p}^k} - \angles{\vect{g}(\vect{p}^k), \vect{J}(1)} = \angles{\vect{g}(\vect{p}^k), \pmb{\delta}_{j^k} - \vect{p}^k}.\]
    Recalling that the expert is QA-shrinkage overconfident if and only if the derivative of $f$ at $w = 1$ is negative, this is the case if and only if
    \[\sum_{k = 1}^N \angles{\vect{g}(\vect{p}^k), \pmb{\delta}_{j^k} - \vect{p}^k} < 0.\]
    As we have already showed, this is the case precisely when the expert is expected-score overconfident. This completes the proof.
\end{proof}

\section{Axiomatization of QA pooling} \label{sec:axiomatization}
In this section, we aim to show that the class of all quasi-arithmetic pooling operators is a natural one, by showing that these operators are precisely those which satisfy a natural set of axioms.

\textcite{kol30, nag30} independently considered the class of quasi-arithmetic means. Given an interval $I \subseteq \RR$ and a continuous, injective function $f: I \to \RR$, the \emph{quasi-arithmetic mean with respect to $f$}, or \emph{$f$-mean}, is the function $M_f$ that takes as input $x_1, \dots, x_m \in I$ (for any $m \ge 1$) and outputs
\[M_f(x_1, \dots, x_m) := f^{-1} \parens{\frac{f(x_1) + \dots + f(x_m)}{m}}.\]
For example, the arithmetic mean corresponds to $f(x) = x$; the quadratic to $f(x) = x^2$; the geometric to $f(x) = \log x$; and the harmonic to $f(x) = \frac{-1}{x}$.

Kolmogorov proved that the class of quasi-arithmetic means is precisely the class of functions $M: \bigcup_{m = 1}^\infty I^m \to I$ satisfying the following natural properties:\footnote{Nagumo also provided a characterization, though with slightly different properties.}

\begin{enumerate}[label=(\arabic*)]
\item $M(x_1, \dots, x_m)$ is continuous and strictly increasing in each variable.
\item $M$ is symmetric in its arguments.
\item $M(x, x, \dots, x) = x$.
\item $M(x_1, \dots, x_k, x_{k + 1}, \dots, x_m)$ = $M(y, \dots, y, x_{k + 1}, \dots, x_m)$, where $y := M(x_1, \dots, x_k)$ appears $k$ times on the right-hand side. Informally, a subset of arguments to the mean function can be replaced with their mean.
\end{enumerate}

The four properties listed above can be viewed as an \emph{axiomatization} of quasi-arithmetic means.\\

Our notion of quasi-arithmetic pooling is exactly that of a quasi-arithmetic mean, except that it is more general in two ways. First, it allows for weights to accompany the arguments to the mean. Second, we are considering quasi-arithmetic means with respect to \emph{vector-valued} functions $\vect{g}$. In the $n = 2$ outcome case, $\vect{g}$ can be considered a scalar-valued function since it is defined on a one-dimensional space (see Remark~\ref{remark:n2} for details); but in general we cannot treat $\vect{g}$ as scalar-valued.\footnote{Another difference is that in the $n = 2$ case, we are restricting $\mathcal{D}$ to be an interval from $0$ to $1$, though this is not a fundamental difference for the purposes of this section.}

Our goal is to extend the above axiomatization of quasi-arithmetic means in these two ways: first (in Section~\ref{sec:axioms_n2}) to include weights as arguments, and second (in Section~\ref{sec:axioms_general_n}) to general $n$ (while still allowing arbitrary weights).\\

\subsection{Generalizing to include weights as arguments} \label{sec:axioms_n2}
The objects that we will be studying in this section are ones of the form $(p, w)$, where $w \ge 0$ and $p \in \mathcal{D}$. In this subsection, $\mathcal{D}$ is a two-outcome forecast domain, whose elements we will identify with by the probability of the first outcome (see Remark~\ref{remark:n2}).\footnote{Proper scoring rules for two outcomes can have four possible forecast domains: $[0, 1]$, $[0, 1)$, $(0, 1]$, and $(0, 1)$.} We will fix the set $\mathcal{D}$ for the remainder of the subsection. Our results generalize to any interval of $\RR$ (as in Kolmogorov's work), but we focus on forecast domains since that is our application.

\begin{defin}
A \emph{weighted forecast} is an element of $\mathcal{D} \times \RR_{>0}$: a probability and a positive weight. Given a weighted forecast $\Pi = (p, w)$ we define $\prb(\Pi) := p$ and $\wt(\Pi) := w$.
\end{defin}

We will thinking of the output of pooling operators as weighted forecasts. This is a simple extension of our earlier definition of quasi-arithmetic pooling (Definition~\ref{def:qa_pooling}), which only output a probability.

\begin{defin}[Quasi-arithmetic pooling with arbitrary weights ($n = 2$)] \label{def:qa_ax_n2}
Given a continuous, strictly increasing function $g: \mathcal{D} \to \RR$, and weighted forecasts $\Pi_1 = (p_1, w_1), \dots, \Pi_m = (p_m, w_m)$, define the \emph{quasi-arithmetic pool} of $\Pi_1, \dots, \Pi_m$ with respect to $g$ as
\[\sideset{}{_g}\bigoplus_{i = 1}^m (p_i, w_i) := \parens{g^{-1} \parens{\frac{\sum_i w_i g(p_i)}{\sum_i w_i}}, \sum_i w_i}.\]
\end{defin}

\begin{remark}
The equivalence of this notion to our earlier one uses the fact that all (continuous) proper scoring rules for binary outcomes have convex exposure (Proposition~\ref{prop:n2_convex_exposure}), so the relationship between $g(p^*)$ and $g(p)$ may be written as in Equation~\ref{eq:g_convex_exposure}.

In the case that $\sum_i w_i = 1$, Definition~\ref{def:qa_ax_n2} reduces to Definition~\ref{def:qa_pooling}. In general, by linearly scaling the weights in Definition~\ref{def:qa_ax_n2} to add to $1$, we recover quasi-arithmetic pooling as previously defined.
\end{remark}

We find the following fact useful.

\begin{prop}
Given two continuous, strictly increasing functions $g_1$ and $g_2$, $\oplus_{g_1}$ and $\oplus_{g_2}$ are the same if and only if $g_2 = ag_1 + b$ for some $a > 0$ and $b \in \RR$.
\end{prop}

\begin{proof}
Clearly if $g_2 = ag_1 + b$ for some $a > 0$ and $b \in \RR$ then $\oplus_{g_1}$ and $\oplus_{g_2}$ are the same. For the converse, suppose that no such $a$ and $b$ exist. Let $x < y$ be such that $g_1$ and $g_2$ are not equal (even up to positive affine transformation) on $[x, y]$. Let $g_2'$ be the positive affine transformation of $g_2$ that makes it equal to $g_1$ at $x$ and $y$, and let $z \in (x, y)$ be such that $g_1(z) \neq g_2'(z)$. Let $\alpha$ be such that $g_1(z) = \alpha g_1(x) + (1 - \alpha) g_1(y)$. Then $(x, \alpha) \oplus_{g_1} (y, 1 - \alpha) = (z, 1)$, but $(x, \alpha) \oplus_{g_2} (y, 1 - \alpha) \neq (z, 1)$, so $\oplus_{g_1}$ and $\oplus_{g_2}$ are different.
\end{proof}

We now define properties (i.e.\ axioms) of a pooling operator $\oplus$, such that these properties are satisfied if and only if $\oplus$ is $\oplus_g$ for some $g$. Our axiomatization will look somewhat different from Kolmogorov's, in part because we choose to define $\oplus$ as a binary operator that (if it satisfies the associativity axiom) extends to the $m$-ary case. This is a simpler domain and will simplify notation. In Appendix~\ref{appx:ax} we exhibit an equivalent set of axioms that more closely resembles Kolmogorov's.

\begin{defin}[Axioms for pooling operators ($n = 2$)] \label{def:wop_n2}
For a pooling operator $\oplus$ on $\mathcal{D}$ (i.e.\ a binary operator on weighted forecasts), we define the following axioms.
\begin{enumerate}
\item \textbf{Weight additivity}: $\wt(\Pi_1 \oplus \Pi_2) = \wt(\Pi_1) + \wt(\Pi_2)$ for every $\Pi_1, \Pi_2$.
\item \textbf{Commutativity}: $\Pi_1 \oplus \Pi_2 = \Pi_2 \oplus \Pi_1$ for every $\Pi_1, \Pi_2$.
\item \textbf{Associativity}: $\Pi_1 \oplus (\Pi_2 \oplus \Pi_3) = (\Pi_1 \oplus \Pi_2) \oplus \Pi_3$ for every $\Pi_1, \Pi_2, \Pi_3$.
\item \textbf{Continuity}: For every $p_1, p_2$, the quantity\footnote{We allow one weight to be $0$ by defining $(p, w) \oplus (q, 0) = (q, 0) \oplus (p, w) = (p, w)$.} $\prb((p_1, w_1) \oplus (p_2, w_2))$
is a continuous function of $(w_1, w_2)$ on $\RR_{\ge 0}^2 \setminus \{(0, 0)\}$.
\item \textbf{Idempotence}: For every $\Pi_1, \Pi_2$, if $\prb(\Pi_1) = \prb(\Pi_2)$ then $\prb(\Pi_1 \oplus \Pi_2) = \prb(\Pi_1)$.
\item \textbf{Monotonicity}: Let $w > 0$ and let $p_1 > p_2 \in \mathcal{D}$. Then for $x \in (0, w)$, the quantity $\prb((p_1, x) \oplus (p_2, w - x))$ is a strictly increasing function of $x$.
\end{enumerate}
\end{defin}

The motivation for the weight additivity axiom is that the weight of a weighted forecast can be thought of as the \emph{amount of evidence} for its prediction. When pooling weighted forecasts, the weight of an individual forecast can be thought of as the strength of its vote in the aggregate.

The monotonicity axiom essentially states that if one pools two forecasts with different probabilities and a fixed total weight, then the larger the share of the weight belonging to the larger of the two probabilities, the larger the aggregate probability.\\

We now state and prove this section's main result: these axioms describe the class of QA pooling operators.

\begin{theorem} \label{thm:representation_n2}
A pooling operator is a QA pooling operator (as in Definition~\ref{def:qa_ax_n2}) with respect to some $g$ if and only if it satisfies the axioms in Definition~\ref{def:wop_n2}.\footnote{
As we mentioned, for an associative pooling operator $\oplus$, $\Pi_1 \oplus \Pi_2 \dots \oplus \Pi_m$ is a well-specified quantity, even without indicating parenthesization. This lets us use the notation $\bigoplus_{i = 1}^m \Pi_i$. This is why the statement of Theorem~\ref{thm:representation_n2} makes sense despite pooling operators not being $m$-ary by default.}
\end{theorem}

We will use $\oplus$ (without a $g$ subscript) to denote an arbitrary pooling operator that satisfies the axioms in Definition~\ref{def:wop_n2}. Before presenting the proof of Theorem~\ref{thm:representation_n2}, we will note a few important facts about weighted forecasts and pooling operators. First, we find it natural to define a notion of multiplying a weighted forecast pair by a positive constant.
	
\begin{defin}
	Given a weighted forecast $\Pi = (p, w)$ and $c > 0$, define $c\Pi := (p, cw)$.
\end{defin}
	
Note that $m\Pi = \bigoplus_{i = 1}^m \Pi$ for any positive integer $m$, by idempotence; this definition is a natural extension to all $c > 0$. We note the following (quite obvious) fact.
	
\begin{prop}
	For every weighted forecast $\Pi$ and $c_1, c_2 > 0$, we have $c_1(c_2 \Pi) = (c_1 c_2)\Pi$.
\end{prop}
	
A natural property that is not listed in Definition~\ref{def:wop_n2} is \emph{scale invariance}, i.e.\ that $\prb((p_1, w_1) \oplus (p_2, w_2)) = \prb((p_1, cw_1) \oplus (p_2, cw_2))$ for any positive $c$; or, equivalently, that $c(\Pi_1 \oplus \Pi_2) = c\Pi_1 \oplus c\Pi_2$. This in fact follows from the listed axioms.
	
\begin{prop}[Distributive property/scale invariance] \label{prop:dist}
	For every $\Pi_1, \Pi_2$ and any operator $\oplus$ satisfying the axioms in Definition~\ref{def:wop_n2}, we have $c(\Pi_1 \oplus \Pi_2) = c\Pi_1 \oplus c\Pi_2$.
\end{prop}

\begin{proof}
	First suppose $c$ is an integer. Then
	\[c \Pi_1 \oplus c \Pi_2 = \bigoplus_{i = 1}^c \Pi_1 \oplus \bigoplus_{i = 1}^c \Pi_2 = \bigoplus_{i = 1}^c (\Pi_1 \oplus \Pi_2) = c(\Pi_1 \oplus \Pi_2).\]
	Here, the first and last steps follow by weight additivity and idempotence. Now suppose that $c = \frac{k}{\ell}$ is a rational number. Let $\Pi_1' = \frac{1}{\ell} \Pi_1$ and $\Pi_2' = \frac{1}{\ell} \Pi_2$. We have
	\[\frac{k}{\ell}(\Pi_1 \oplus \Pi_2) = \frac{k}{\ell} (\ell \Pi_1' \oplus \ell \Pi_2') = \frac{k}{\ell} \cdot \ell(\Pi_1' \oplus \Pi_2') = k(\Pi_1' \oplus \Pi_2') = k \Pi_1' \oplus k \Pi_2' = \frac{k}{\ell} \Pi_1 \oplus \frac{k}{\ell} \Pi_2.\]
	Here, the second and second-to-last steps follow from the fact that the distributive property holds for integers.
		
	Finally, make use of the continuity axiom to extend our proof to all positive real numbers $c$. In particular, it suffices to show that $\prb(\Pi_1 \oplus \Pi_2) = \prb(c\Pi_1 \oplus c\Pi_2)$. Let $p$ be the former quantity; note that $\prb(r\Pi_1 \oplus r\Pi_2) = p$ for positive rational numbers $r$. Since the rationals are dense among the reals, it follows that for every $\epsilon > 0$, we have $\abs{\prb(c\Pi_1 \oplus c\Pi_2) - p} \le \epsilon$. Therefore, $\prb(c\Pi_1 \oplus c\Pi_2) = p$. This completes the proof.
\end{proof}

Armed with these facts, we present a proof of Theorem~\ref{thm:representation_n2}.

\begin{proof}[Proof of Theorem~\ref{thm:representation_n2}]
	We first prove that any QA pooling operator $\oplus_g$ satisfies the axioms in Definition~\ref{def:wop_n2}. Weight additivity, commutativity, and idempotence are trivial. Associativity is also clear: given $\Pi_1 = (p_1, w_1)$ and likewise $\Pi_2, \Pi_3$, we have
	\begin{align*}
		g(\prb((\Pi_1 \oplus_g \Pi_2) \oplus_g \Pi_3)) &= \frac{(w_1 + w_2) g(\prb(\Pi_1 \oplus_g \Pi_2)) + w_3 g(p_3)}{(w_1 + w_2) + w_3}\\
		&= \frac{(w_1 + w_2) \frac{w_1 g(p_1) + w_2 (p_2)}{w_1 + w_2} + w_3 g(p_3)}{w_1 + w_2 + w_3} = \frac{w_1 g(p_1) + w_2 g(p_2) + w_3 g(p_3)}{w_1 + w_2 + w_3}
	\end{align*}
	and likewise for $g(\prb(\Pi_1 \oplus_g (\Pi_2 \oplus_g \Pi_3)))$, so $\prb((\Pi_1 \oplus_g \Pi_2) \oplus_g \Pi_3) = \prb(\Pi_1 \oplus_g (\Pi_2 \oplus_g \Pi_3))$ (since $g$ is strictly increasing and therefore injective). The fact that the weights are also the same is trivial. Continuity follows from the fact that
	\[\prb(\Pi_1 \oplus_g \Pi_2) = g^{-1} \parens{\frac{w_1 g(p_1) + w_2 g(p_2)}{w_1 + w_2}}\]
	is continuous in $(w_1, w_2)$ (when $w_1, w_2$ are not both zero). Here we are using the fact that $g$ is \emph{strictly} increasing, which means that $g^{-1}$ is continuous.
	
	Finally, regarding the monotonicity axiom, for any fixed $w$ and $p_1 > p_2$ (as in the axiom statement), we have
	\[g(\prb((p_1, x) \oplus_g (p_2, w - x))) = \frac{x g(p_1) + (w - x) g(p_2)}{x + w - x} = \frac{x g(p_1) + (w - x) g(p_2)}{w}.\]
	Since $p_1 > p_2$, we have $g(p_1) > g(p_2)$, so the right-hand side strictly increases with $x$. Since $g^{-1}$ is also strictly increasing, it follows that $\prb((p_1, x) \oplus_g (p_2, w - x))$ strictly increases with $x$.\\
	
	The converse -- that every pooling operator satisfying the axioms in Definition~\ref{def:wop_n2} is $\oplus_g$ for some $g$ -- works by constructing $g$ by fixing it at two points and constructing $g$ at all other points. Right now we show how to do this when the forecast domain is $[0, 1]$; see the proof of Theorem~\ref{thm:representation} for the argument in full generality.
	
	Let $\oplus$ be a pooling operator that satisfies our axioms. Define $g$ as follows: let $g(0) = 0$ and $g(1) = 1$. For $0 < p < 1$, define $g(p) = w$ where $(1, w) \oplus (0, 1 - w) = (p, 1)$. (This $w$ exists by continuity and the intermediate value theorem; it is unique by the ``strictly'' increasing stipulation of monotonicity.) Note that $g$ is continuous and increasing by monotonicity.\footnote{As a matter of fact, $g$ is strictly increasing because it is impossible for $g(p_1)$ to equal $g(p_2)$ for $p_1 \neq p_2$, as that would mean that $(1, g(p_1)) \oplus (0, 1 - g(p_1)) = (p_1, 1) = (p_2, 1)$. Another way to look at this is that it comes from the fact that $(1, w) \oplus (0, 1 - w)$ is continuous in $w$ by the continuity axiom. In a sense, the \emph{continuity} of $g$ corresponds to the \emph{strictness} of increase in the monotonicity axiom and the \emph{strictness} of increase of $g$ corresponds to the \emph{continuity} axiom.}
	
	We wish to show that for any $\Pi_1 = (p_1, w_1)$ and $\Pi_2 = (p_2, w_2)$, we have that $\Pi_1 \oplus \Pi_2 = \Pi_1 \oplus_g \Pi_2$. Clearly the weight of both sides is $w_1 + w_2$, so we wish to show that the probabilities on each side are the same. We have\footnote{Steps 3 and 7 uses the distributive property (Proposition~\ref{prop:dist}).}
	\begin{align*}
		\prb(\Pi_1 \oplus \Pi_2) &= \prb(w_1 (p_1, 1) \oplus w_2 (p_2, 1))\\
		&= \prb(w_1 ((1, g(p_1)) \oplus (0, 1 - g(p_1))) \oplus w_2 ((1, g(p_2)) \oplus (0, 1 - g(p_2))))\\
		&= \prb(w_1 (1, g(p_1)) \oplus w_1 (0, 1 - g(p_1)) \oplus w_2 (1, g(p_2)) \oplus w_2 (0, 1 - g(p_2)))\\
		&= \prb((1, w_1 g(p_1)) \oplus (0, w_1(1 - g(p_1))) \oplus (1, w_2 g(p_2)) \oplus (0, w_2(1 - g(p_2))))\\
		&= \prb((1, w_1 g(p_1) + w_2 g(p_2)) \oplus (0, w_1(1 - g(p_1)) + w_2(1 - g(p_2))))\\
		&= \prb \parens{\frac{1}{w_1 + w_2}((1, w_1 g(p_1) + w_2 g(p_2)) \oplus (0, w_1(1 - g(p_1)) + w_2(1 - g(p_2))))}\\
		&= \prb \parens{\parens{1, \frac{w_1 g(p_1) + w_2 g(p_2)}{w_1 + w_2}} \oplus \parens{0, \frac{w_1(1 - g(p_1)) + w_2(1 - g(p_2))}{w_1 + w_2}}},
	\end{align*}
	which by definition of $g$ is equal to the probability $p$ such that $g(p) = \frac{g(p_1) w_1 + g(p_2) w_2}{w_1 + w_2}$. That is, $\prb(\Pi_1 \oplus \Pi_2) = \prb(\Pi_1 \oplus_g \Pi_2)$.
	
	Showing that $\oplus$ and $\oplus_g$ are equivalent for more than two arguments is now trivial:
	\[\sideset{}{_\vect{g}}\bigoplus_{i = 1}^m \Pi_i = \Pi_1 \oplus_g \Pi_2 \oplus_g \Pi_3 \dots \oplus_g \Pi_m = \Pi_1 \oplus \Pi_2 \oplus_g \Pi_3 \dots \oplus_g \Pi_m = \dots = \bigoplus_{i = 1}^m \Pi_i.\]
	(Here we are implicitly using the fact that $\oplus_g$ is associative, as we proved earlier.) This completes the proof.
\end{proof}

\subsection{Generalizing to higher dimensions} \label{sec:axioms_general_n}
In Appendix~\ref{appx:extending}, we generalize our axioms from $n = 2$ outcomes to arbitrary values of $n$. An important challenge is extending the monotonicity axiom: in higher dimensions, what is the appropriate generalization of an increasing function? We show that the correct notion is \emph{cyclical monotonicity}, which we define and discuss. We then present our axiomatization (Definition~\ref{def:qa_ax}) and prove that the axioms constitute a characterization of the class of QA pooling operators (Theorem~\ref{thm:representation}). On a high level, the proof is not dissimilar to that of Theorem~\ref{thm:representation_n2}, though the details are fairly different and more technical.\\

In conclusion, in Definition~\ref{def:wop_n2} we made a list of natural properties that a pooling operator may satisfy. Theorem~\ref{thm:representation_n2} shows that the pooling operators satisfying these properties are exactly the QA pooling operators. In Appendix~\ref{appx:extending}, we generalize this theorem to higher dimensions, thus fully axiomatizing QA pooling. This result gives us an additional important reason to believe that QA pooling with respect to a proper scoring rule is a fundamental notion.

\section{Conclusions and future directions}
While in this work we have focused on proper scoring rules for eliciting probability distributions over possible outcomes, scoring rules can be used to elicit various other properties, such as the expectation or the median of a random variable \parencite{savage71, gr07, lps08}. The topic of \emph{property elicitation} studies such scoring rules. The Savage representation of a proper scoring rule generalizes to arbitrary linear properties \parencite[Theorem 11]{fk15}. That is, consider a random variable (or $n$-tuple of random variables) $X$ and a convex function $G$ whose domain is (a superset of) the possible values of $X$. If $\vect{p}$ is an expert's forecast for $\EE{X}$ and $\vect{x}$ is the realized outcome, then the scoring rule $s(\vect{p}; \vect{x}) := G(\vect{p}) + \angles{\vect{g}(\vect{p}), \vect{x} - \vect{p}}$ is proper.\footnote{In the setting of this work, $X$ is the vector of random variables where the $j$-th variable is $1$ if outcome $j$ happens and $0$ otherwise, and $\vect{x} = \pmb{\delta}_j$ where $j$ is the outcome that happens.} Modulo a fairly straightforward generalization, all proper scoring rules for $\EE{X}$ take this form (see Section~\ref{sec:prelim_bregman}).

Our definition of QA pooling extends verbatim to the setting of eliciting linear properties. In this more general setting, for any proper scoring rule that has convex exposure, it remains the case $\vect{g}(\vect{p}^*) = \sum_i w_i \vect{g}(\vect{p}_i)$; as a consequence, Theorems~\ref{thm:max_min}, \ref{thm:ws_concave}, \ref{thm:chap5_no_regret}, and \ref{thm:overconfidence} generalize.\\

We now discuss several promising directions for future work. First: as we discussed in Section~\ref{sec:generalized_pool}, it is natural to generalize QA pooling by dropping the requirement that weights add to $1$. In Section~\ref{sec:bayesian_justifications}, we gave a Bayesian justification for linear pooling with arbitrary weights, and gave a different and novel Bayesian justification for logarithmic pooling with arbitrary weights. This raises a few questions:
\begin{itemize}
    \item Which of our results generalize if the requirement that weights add to $1$ is dropped? Note that dropping this requirement means that the idempotence axiom of Section~\ref{sec:axiomatization} is no longer satisfied. Is there an axiomatization that characterizes the class of all generalized QA pools?
    \item Just as we gave Bayesian justifications for generalized linear and logarithmic pooling, is there a Bayesian justification for generalized QA pooling with respect to an arbitrary proper scoring rule? What does it look like?
\end{itemize}

Second: it is natural to wonder about connections to prediction markets. After all, a prediction market is a mechanism for eliciting forecasts from multiple experts and aggregating them! As mentioned in Section~\ref{sec:chap5_related_work}, QA pooling can be interpreted in terms of cost function markets: if $\vect{q}_i$ denotes the quantity of shares that each expert would buy in order to bring a market into line with the expert's beliefs, then the QA pool of the experts' beliefs is the probability distribution corresponding to the quantity vector $\vect{q}^* := \sum_i w_i \vect{q}_i$. This follows from \textcite[Eq.\ 22]{acv13}; the underlying reason is the convex duality between an expected score function $G$ and the corresponding cost function $C$. Exploring this connection further may yield insights into forecast aggregation through market mechanisms.

Third: some of our results may be able to be extended or generalized. In fact, we do so in Chapter~\ref{chap:learning}, where we relax the assumption made in Theorem~\ref{thm:chap5_no_regret} that the scoring rule is bounded. Additionally, many of our theorem statements depend on the convex exposure property, but it is possible that some of our definitions and theorem statements could be modified so as to not require this assumption.

QA pooling connects forecast elicitation and aggregation in a way that seems quite fundamental. Further research in this area may yield important insights.
\chapter{Learning weights for logarithmic pooling} \label{chap:learning}
\emph{This chapter presents ``No-Regret Learning with Unbounded Losses: The Case of Logarithmic Pooling'' \parencite{nr22_learning}. It assumes knowledge of logarithmic pooling, presented in Section~\ref{sec:lin_log}. Background on proper scoring rules (Section~\ref{sec:prelim_proper}), as well as on information structures (Section~\ref{sec:prelim_info_struct}), is useful for context but not required. Additionally, Chapter~\ref{chap:qa} is useful for context, as this work builds on it directly.}\\

\emph{Summary:} For each of $T$ time steps, $m$ experts report probability distributions over $n$ outcomes; we wish to learn to aggregate these forecasts in a way that attains a no-regret guarantee. We focus on logarithmic pooling, which is in a certain sense the optimal choice of pooling method if one is interested in minimizing log loss (see Theorem~\ref{thm:max_min}). We consider the problem of learning the best set of parameters (i.e.\ expert weights) in an online adversarial setting.  We assume (by necessity) that the adversarial choices of outcomes and forecasts are consistent, in the sense that experts report calibrated forecasts. Imposing this constraint creates a (to our knowledge) novel semi-adversarial setting in which the adversary retains a large amount of flexibility. In this setting, we present an algorithm based on online mirror descent that learns expert weights in a way that attains $O(\sqrt{T} \log T)$ expected regret as compared with the best weights in hindsight.

\section{Introduction}
\subsection{Learning expert weights for logarithmic pooling}
In Section~\ref{sec:prelim_agg}, we motivated logarithmic pooling as a natural method of forecast aggregation that has compelling theoretical properties and works well in practice. To recall, the logarithmic pool of forecasts $\vect{p}^1, \dots, \vect{p}^m \in \Delta_n$ with weight vector $\vect{w} \in \Delta_m$ is defined by
\[p^*_j(\vect{w}) = c(\vect{w}) \prod_{i = 1}^m (p^i_j)^{w_i},\]
for all $j \in [n]$, where $c(\vect{w})$ is a normalizing constant (that depends on the weights).

But where do the expert weights come from? That is, how should an aggregator determine what weight to assign to each expert? It is common for the aggregator to learn the correct vector of weights over time, based on the experts' track record \parencite{tg15}. Learning expert weights over time is a sequential decision making problem: for each of $T$ time steps, experts submit forecasts to the aggregator, and the aggregator aggregates them using logarithmic pooling, with weights that the aggregator chooses. Then an outcome is realized and the aggregator receives a score depending on the accuracy of the aggregate forecast. The aggregator adjusts expert weights for the next time step, perhaps increasing the weights of experts who did well while decreasing the weights of experts who did poorly.

This problem falls into the well-studied and wide-reaching field of online learning -- and more specifically, online prediction with expert advice \parencite{cl06}. Generally, algorithms for online learning problems are assessed based on their \emph{regret} relative to some baseline. In our case, the natural baseline is the best possible weight vector in hindsight. That is, we are looking for a weight selection algorithm whose overall performance over the $T$ time steps is guaranteed to be almost as good as using the best weight vector $\vect{w}^*$ in hindsight.

We previously discussed online learning of expert weights in Chapter~\ref{chap:qa}. In that chapter, we introduced \emph{quasi-arithmetic (QA) pooling} as a way of aggregating forecasts with respect to a given proper scoring rule $s$. Specifically, the QA pool with respect to a proper scoring rule $s$ is the forecast that guarantees the largest possible expected overperformance (as judged by the scoring rule) compared with the strategy of choosing a random expert to trust.\footnote{More formally, the QA pool maximizes the minimum improvement in the score (over possible outcomes).} We saw that QA pooling with respect to the quadratic scoring rule is linear pooling (averaging the experts' forecasts), while QA pooling with respect to the logarithmic scoring rule is logarithmic pooling.

In Section~\ref{sec:convex_losses}, we gave an algorithm for online-learning expert weights for QA pooling with respect to any \textbf{bounded} proper scoring rule $s$. The algorithm had the property that, asymptotically with the number of time steps $T$, the average score of the algorithm (according to $s$) was guaranteed to be almost as large as the average score of an aggregator who chose the best vector of weights in hindsight. In online learning terminology, this means that the algorithm has \emph{vanishing regret} (or, synonymously, \emph{no regret}).

However, the log scoring rule is not bounded, so the aforementioned result does \emph{not} give a no-regret algorithm for learning weights for logarithmic pooling (if regret is defined with respect to the log scoring rule). The purpose of this chapter is to fill that gap.\\

\emph{A note on terminology:} In this chapter, we will use the \emph{log loss} instead of the log score, as is standard in the learning literature. The log loss is simply the negative of the log score: while the log score of the forecast $\vect{p}$ under outcome $j$ is $\ln(p_j)$, the log loss is $-\ln(p_j)$.

\subsection{Choosing the right benchmark}
In Chapter~\ref{chap:qa}, we established a connection between the log loss and logarithmic pooling, arguing that if the log loss is used for elicitation, then it makes sense to use logarithmic pooling for aggregation. The goal of this work is to develop an algorithm for learning weights for logarithmic pooling in a way that achieves vanishing regret as judged by the log loss function. Within the field of online prediction with expert advice, this is a particularly challenging setting. In part, this is because the losses are potentially unbounded. However, that is not the whole story: finding weights for \emph{linear} pooling, by contrast, is a well-studied problem that has been solved even in the context of log loss. On the other hand, because logarithmic pooling behaves more as a geometric than an arithmetic mean, if some expert assigns a very low probability to the eventual outcome (and the other experts do not) then the logarithmic pool will also assign a low probability, incurring a large loss. This makes the combination of logarithmic pooling with log loss particularly difficult.

We require that our algorithm not have access to the experts' forecasts when choosing weights: an algorithm that chooses weights in a way that depends on forecasts can output an essentially arbitrary function of the forecasts, and thus may do something other than learn optimal weights for logarithmic pooling. For example, suppose that $m = n = 2$ and an aggregator wishes to subvert our intentions and take an equally weighted \emph{linear} pool of the experts' forecasts. Without knowing the experts' forecasts, this is impossible; on the other hand, if the aggregator knew that e.g.\ $\vect{p}_1 = (90\%, 10\%)$ and $\vect{p}_2 = (50\%, 50\%)$, they could assign weights for logarithmic pooling so as to produce the post-hoc desired result, i.e.\ $(70\%, 30\%)$. We wish to disallow this.

One might suggest the following setup: at each time step, the algorithm selects weights for each expert. Subsequently, an adversary chooses each expert's forecast and the outcome, after which the algorithm and each expert incur a log loss. Unfortunately -- due to the unboundedness of log loss and the behavior of logarithmic pooling -- vanishing regret guarantees in this setting are impossible.

\begin{example} \label{example:bad_regret}
	Consider the case of $m = n = 2$. Without loss of generality, suppose that the algorithm assigns Expert 1 a weight $w \ge 0.5$ in the first time step. The adversary chooses reports $(e^{-T}, 1 - e^{-T})$ for Expert 1 and $\parens{\frac{1}{2}, \frac{1}{2}}$ for Expert 2, and for Outcome 1 to happen. The logarithmic pool of the forecasts turns out to be approximately $(e^{-wT}, 1 - e^{-wT})$, so the algorithm incurs a log loss of approximately $wT \ge 0.5 T$, compared to $O(1)$ loss for Expert 2. On subsequent time steps, Expert 2 is perfect (assigns probability $1$ to the correct outcome), so the algorithm cannot catch up.
\end{example}

What goes wrong in Example~\ref{example:bad_regret} is that the adversary has full control over experts' forecast \emph{and} the realized outcome, and is not required to couple the two in any way. This unreasonable amount of adversarial power motivates assuming that the experts are \emph{calibrated}: for example, if an expert assigns a 10\% chance to an outcome, there really is a 10\% chance of that outcome (conditional on the expert's information).

We propose the following setting: an adversary chooses a joint probability distribution over the experts' beliefs and the outcome -- subject to the constraint that each expert is \emph{calibrated.} Loosely speaking, this means that each expert correctly assesses the probability each outcome based on the information they have: if an expert reports $\vect{p}$, then each event $j$ really does have a probability $p_j$ of happening, based on the expert's information. (We formally define calibration in Section~\ref{sec:calibration}.) The adversary retains full control over correlations between forecasts and outcomes, subject to this calibration property. Subsequently, nature randomly samples each expert's belief and the eventual outcome from the distribution. In this setting, we seek to prove upper bounds on the expected value of our algorithm's regret.

Why impose this constraint, instead of a different one? Our reasons are twofold: theoretical and empirical. From a theoretical standpoint, the assumption that experts are calibrated is natural because experts who form Bayesian rational beliefs based on evidence will be calibrated, regardless of how much or how little evidence they have. The assumption is also motivated if we model experts as learners rather than Bayesian agents: even if a forecaster starts out completely uninformed, they can quickly become calibrated in a domain simply by observing the frequency of events \parencite{fv97}.

Second, recent work has shown that modern deep neural networks are calibrated when trained on a proper loss function such as log loss. This is true for a variety of tasks, including image classification \parencite{minderer21, hendrycks20} and language modeling \parencite{kadavath22, dd20, openai23}; see \textcite{bghn23} for a review of the literature. We may wish to use an ensemble of off-the-shelf neural networks for some prediction or classification task. If we trust these networks to be calibrated (as suggested by recent work), then we may wish to learn to ensemble these experts (models) in a way that has strong worst-case theoretical guarantees under the calibration assumption.

Logarithmic pooling is particularly sensible in the context of calibrated experts because it takes confident forecasts ``more seriously'' as compared with linear pooling (simple averaging). If Expert 1 reports probability distribution $(0.1\%, 99.9\%)$ over two outcomes and Expert 2 reports $(50\%, 50\%)$, then the logarithmic pool (with equal weights) is approximately $(3\%, 97\%)$, as compared with a linear pool of roughly $(25\%, 75\%)$. If Expert 1 is calibrated (as we are assuming), then the $(0.1\%, 99.9\%)$ forecast entails very strong evidence in favor of Outcome 2 over Outcome 1. Meanwhile, Expert 2's forecast gives no evidence either way. Thus, it is sensible for the aggregate to point to Outcome 2 over Outcome 1 with a fair amount of confidence.

As another example, suppose that Expert 1 reports $(0.04\%, 49.98\%, 49.98\%)$ and Expert 2 reports $(49.98\%, 0.04\%, 49.98\%)$ (a natural interpretation: Expert 1 found strong evidence against Outcome 1 and Expert 2 found strong evidence against Outcome 2). If both experts are calibrated, a sensible aggregate should arguably assign nearly all probability to Outcome 3. Logarithmic pooling returns roughly $(2.7\%, 2.7\%, 94.6\%)$, which (unlike linear pooling) accomplishes this.

Since we are allowing our algorithm to learn the optimal logarithmic pool, perhaps there is hope to compete not just with the best expert in hindsight, but the optimally weighted logarithmic pool of experts in hindsight. We will aim to compete with this stronger benchmark.

This work demonstrates that the ``calibrated experts'' condition allows us to prove regret bounds when no such bounds are possible for an unrestricted adversary. While that is our primary motivation, the relaxation may also be of independent interest. For example, even in settings where vanishing regret is attainable in the presence of an unrestricted adversary, even stronger regret bounds might be achievable if calibration is assumed.

\subsection{Our main result}
Is vanishing regret possible in our setting? Our main result is that the answer is yes. We exhibit an algorithm that attains expected regret that scales as $O(\sqrt{T} \log T)$ with the number of time steps $T$. Our algorithm uses online mirror descent (OMD) with the \emph{Tsallis entropy regularizer} $R(\vect{w}) = \frac{-1}{\alpha}(w_1^\alpha + \dots + w_m^\alpha)$ and step size $\eta \approx \frac{1}{\sqrt{T} \ln T}$, where any choice of $\alpha \in (0, 1/2)$ attains the regret bound.

Our proof has two key ideas. One is to use the calibration property to show that the gradient of loss with respect to the weight vector is likely to be small (Section~\ref{sec:finishing_up}). This is how we leverage the calibration property to turn an intractable setting into one where -- despite the unboundedness of log loss and the behavior of logarithmic pooling -- there is hope for vanishing regret.

The other key idea (Section~\ref{sec:regret_sga}) involves keeping track of a function that, roughly speaking, reflects how much ``regret potential'' the algorithm has. We show that if the aforementioned gradient updates are indeed small, then this potential function decreases in value at each time step. This allows us to upper bound the algorithm's regret by the initial value of the potential function.

This potential argument is an important component of the proof. A na\"{i}ve analysis might seek to use our bounds on the gradient steps to myopically bound the contribution to regret at each time step. Such an analysis, however, does not achieve our $O(\sqrt{T} \log T)$ regret bound. In particular, an adversary can force a large accumulation of regret if some experts' weights are very small (specifically by making the experts with small weights more informed than those with large weights) -- but by doing so, the small weights increase and the adversary ``spends down'' its potential. Tracking this potential allows us to take this nuance into consideration, improving our bound.

We extend our main result by showing that the result holds even if experts are only approximately calibrated: so long as no expert understates the probability of an outcome by more than a constant factor, we still attain the same regret bound (see Corollary~\ref{cor:approx_calib}). We also show in Section~\ref{sec:chap6_lower_bound} that no OMD algorithm with a constant step size can attain expected regret better than $\Omega(\sqrt{T})$.

\section{Related work} \label{sec:related_work}
In the subfield of \emph{prediction with expert advice,} for $T$ time steps, experts report ``predictions'' from a decision space $\mathcal{D}$ (often, as in our case, the space of probability distributions over a set of outcomes). A forecaster must then output their own prediction from $\mathcal{D}$. Then, predictions are assessed according to a loss function. See \textcite{cl06} for an survey of this field.

We are particularly interested in \emph{mixture forecasters}: forecasters who, instead of choosing an expert to trust at each time step, aggregate the expert' reports. Linear mixtures, i.e.\ convex combinations of predictions, have been especially well-studied, generally with the goal of learning weights for the convex combination to compete with the best weights in hindsight. Standard convex optimization algorithms achieve $O(\sqrt{T})$ regret for bounded, convex losses, but it is sometimes possible to do better. For example, if the loss function is bounded and exp-concave, then logarithmic regret in $T$ is attainable \parencite[\S3.3]{cl06}.

Portfolio theory studies optimal stock selection for maximizing return on investment, often in a no-regret setting. \textcite{cover91} introduced the ``universal portfolio'' algorithm, which, for each of $T$ time steps, selects a portfolio (convex combination of stocks). Our setting translates naturally to Cover's: experts play the role of stocks, and the return of a stock corresponds to the probability that and expert assigns to the eventual outcome. The universal portfolio algorithm achieves logarithmic regret compared with the best portfolio in hindsight \parencite{co96}; in our terms, this means that logarithmic regret (for log loss) is attainable for the linear pooling of experts. See \textcite{lh14} for a survey of this area.

To our knowledge, learning weights for \emph{logarithmic} pooling has not been previously studied. As shown in Example~\ref{example:bad_regret}, it is not possible to achieve vanishing regret if the setting is fully adversarial. We relax our setting by insisting that the experts be calibrated (see Section~\ref{sec:calibration}). To our knowledge, online prediction with expert advice has also not previously been studied under this condition.

The calibration condition can be equivalently restated as follows: the experts' forecasts are based on signals drawn from an information structure -- possibly a different one at each time step -- that is unknown to the aggregator, and experts report accurate probabilities conditional on their signals (see Remark~\ref{rem:info_struct_equiv}). \textcite{bg21} investigated online prediction with expert advice under the assumption that the experts' signals are drawn from a particular type of information structure, which they called the ``partial evidence environment.''\footnote{Essentially, the partial evidence environment is an adaptation of the partial information framework (see Section~\ref{sec:prelim_info_struct}) to the case of binary outcomes. There is a set of signals, which are independent conditional on the outcome, and every expert knows a subset of the signals (the same subset on each time step). Further, the authors assume that knowledge of the information structure would allow a perfect aggregator to infer all relevant information about the experts' signals from their forecasts.} Under the paper's assumptions, the Bayesian optimal aggregate is a generalized logarithmic pool (as introduced in Section~\ref{sec:generalized_pool}), and the authors show that an aggregator can often learn this optimal aggregate with vanishing regret. By contrast, we do not make any assumptions about the information structures, and instead aim to compete not with the optimal aggregate but with the best weighted mixture of experts in hindsight.

\section{Preliminaries} \label{sec:chap6_prelims}
\subsection{Calibration property} \label{sec:calibration}
We define calibration as follows. Note that the definition is in the context of our setting, i.e.\ $m$ experts reporting probability distributions $\vect{p}^1, \dots, \vect{p}^m$ over $n$ outcomes. We will use $J$ to denote the random variable corresponding to the outcome, i.e.\ $J$ takes values in $[n]$.

\begin{defin} \label{def:calib}
	Consider a joint probability distribution $\PP$ over experts' reports and the outcome. We say that expert $i$ is \emph{calibrated} if for all $\vect{p} \in \Delta_n$ and $j \in [n]$, we have that
	\[\pr{J = j \mid \vect{p}^i = \vect{p}} = p_j.\]
	That is, expert $i$ is calibrated if the probability distribution of $J$ conditional on their report $\vect{p}^i$ is precisely $\vect{p}^i$. We say that $\PP$ satisfies the \emph{calibration property} if every expert is calibrated.
\end{defin}

The key intuition behind the usefulness of calibration is that if an expert claims that an outcome is very unlikely, this is strong evidence that the outcome is in fact unlikely. In Section~\ref{sec:finishing_up} we will use the calibration property to show that the gradient of the loss with respect to the weight vector is likely to be relatively small at each time step.

\subsection{Our online learning setting} \label{sec:our_setting}
The setting for our online learning problem is as follows. For each time step $t \in [T]$:
\begin{enumerate}[label=(\arabic*)]
	\item Our algorithm reports a weight vector $\vect{w}^t \in \Delta_m$.
	\item An adversary (with knowledge of $\vect{w}^t$) constructs a probability distribution $\PP$, over reports and the outcome, that satisfies the calibration property.
	\item \label{item:sampling} Reports $\vect{p}^{t, 1}, \dots, \vect{p}^{t, m}$ and an outcome $j^t$ are sampled from $\PP$.
	\item The \emph{loss} of a weight vector $\vect{w}$ is defined as $L^t(\vect{w}) := -\ln(p^*_{j^t}(\vect{w}))$, the log loss of the logarithmic pool of $\vect{p}^{t, 1}, \dots, \vect{p}^{t, m}$ with weights $\vect{w}$. Our algorithm incurs loss $L^t(\vect{w}^t)$.
\end{enumerate}

\begin{remark} \label{rem:info_struct_equiv}
    An equivalent, more mechanistic definition of calibration views $\PP$ instead as a joint probability distribution over signals received by each expert and the outcome, i.e.\ as an information structure (see Section~\ref{sec:prelim_info_struct}). Each expert's probability distribution is then the actual probability distribution over the outcome conditioned on their signal. We can reinterpret our online learning setting in light of this view. Specifically, instead of directly selecting a distribution over the experts' forecasts and the outcome, the adversary selects an information structure describing the probability distribution over the outcome and each expert's signal. The signals and outcome are then drawn at random from the information structure, and each expert's report is the true probability distribution over the outcome conditioned on their signal.
    
    Formally, the adversary selects an information structure $\mathcal{I}_t = (\Omega^t, \PP^t, (\sigma_1^t, \dots, \sigma_m^t), Y^t)$, where $Y^t$ is a random variables taking on one of $n$ possible values, namely the standard basis vectors in $\RR^n$. Then, nature randomly samples $\omega^t$ according to $\PP^t$. The outcome $j^t$ is the coordinate in which $Y^t(\omega^t)$ has a $1$, and each expert $i$'s report $\vect{p}^{t, i}$ is $\EE{Y^t \mid \sigma_i^t(\omega^t)}$.
\end{remark}

We define the \emph{regret} of our algorithm as
\[\text{Regret} = \sum_{t = 1}^T L^t(\vect{w}^t) - \min_{\vect{w} \in \Delta_m} \sum_{t = 1}^T L^t(\vect{w}).\]
That is, the benchmark for regret is the best weight vector in hindsight. Since our setting involves randomness, our goal is to provide an algorithm with vanishing \emph{expected} regret against any adversarial strategy, where the expectation is taken over the sampling in Step~\ref{item:sampling}.

Even subject to the calibration property, the adversary has a large amount of flexibility, because the adversary retains control over the correlation between different experts' forecasts. An unrestricted adversary has exponentially many degrees of freedom (as a function of the number of experts), whereas the calibration property imposes a mere linear number of constraints.\footnote{This follows from the perspective of the adversary choosing an information structure from which experts' signals are drawn. The information structure specifies the probability of every possible combination of signals received by the experts, and thus has dimension that is exponential in the number of experts. The calibration property imposes linearly many constraints on this space.}

\subsection{Our algorithm}
We use Algorithm~\ref{alg:omd} to accomplish this goal. The algorithm is online mirror descent (OMD) on the weight vector. Fix any $\alpha \in (0, 1/2)$. We use the regularizer
\[R(\vect{w}) := \frac{-1}{\alpha}(w_1^\alpha + \dots + w_m^\alpha).\]
This is known as the Tsallis entropy regularizer; see e.g.\ \textcite{zs21} for previous use in the online learning literature. We obtain the same result (up to a multiplicative factor that depends on $\alpha$) regardless of the choice of $\alpha$. Because no choice of $\alpha$ stands out, we prove our result for all $\alpha \in (0, 1/2)$ simultaneously.

We will generally use a step size $\eta = \frac{1}{\sqrt{T} \ln T} \cdot \frac{1}{12m^{(1 + \alpha)/2} n}$. However, in the (unlikely, as we show) event that some expert's weight becomes unusually small, we will reduce the step size.

\setcounter{algocf}{\value{theorem}}
\begin{algorithm}[ht]
	\caption{OMD algorithm for learning weights for logarithmic pooling} \label{alg:omd}
	$R(\vect{w}) := \frac{-1}{\alpha}(w_1^\alpha + \dots + w_m^\alpha)$ \tcp*{Any $\alpha \in (0, 1/2)$ will work}
	$\eta \gets \frac{1}{\sqrt{T} \ln T} \cdot \frac{1}{12m^{(1 + \alpha)/2} n}$\;
	$\vect{w}^1 \gets (1/m, \dots, 1/m)$\;
	\For{$t = 1$ to $T$}{
		\eIf{$\eta \le \min_i ((w_i^t)^\alpha)$}
		{$\eta_t \gets \min(\eta_{t - 1}, \eta)$\;}
		{$\eta_t \gets \min(\eta_{t - 1}, \min_i w_i^t)$ \tcp*{Edge case; happens with low probability}}
		Observe loss function $L^t$ \tcp*{$L^t$ is chosen as described in Section~\ref{sec:our_setting}}
		Define $\vect{w}^{t + 1}$ such that $\nabla R(\vect{w}^{t + 1}) = \nabla R(\vect{w}^t) - \eta_t \nabla L^t(\vect{w}^t)$\;}
\end{algorithm}
\setcounter{theorem}{\value{algocf}}

In Appendix~\ref{appx:chap6}, we prove that Algorithm~\ref{alg:omd} is efficient, taking $O(mn)$ time per time step.

Theorem~\ref{thm:chap6_no_regret} formally states our no-regret guarantee.
\begin{theorem} \label{thm:chap6_no_regret}
	For any adversarial strategy, the expected regret\footnote{The given asymptotics assume that $T \gg m, n$, i.e.\ ignore terms that are lower-order in $T$.} of Algorithm~\ref{alg:omd} is at most
	\[O \parens{m^{(3 - \alpha)/2} n \sqrt{T} \log T}.\]
\end{theorem}

\section{Proof of no-regret guarantee} \label{sec:proof}
In this section, we prove Theorem~\ref{thm:chap6_no_regret}.

\subsection{Outline of proof} \label{sec:outline}
We use the following fact, which follows from the fact that the score of a QA pool of forecasts is concave in the forecasts' weights (see Chapter~\ref{chap:qa}).
\begin{prop}[Follows from Theorem~\ref{thm:ws_concave}]
	Let $\vect{p}^1, \dots, \vect{p}^m$ be forecasts over $n$ outcomes, $j \in [n]$ be an outcome, and $\vect{w} \in \Delta_m$ be a weight vector. Let $\vect{p}^*(\vect{w})$ be the logarithmic pool of the forecasts with weight vector $\vect{w}$ and let $L(\vect{w}) := -\ln(p_j^*(\vect{w}))$ be the log loss of $\vect{p}^*(\vect{w})$ if Outcome $j$ is realized. Then $L$ is a convex function.
\end{prop}
In particular, all of our loss functions $L^t$ are convex, which means that standard regret bounds apply. In particular, to bound the expected regret of Algorithm~\ref{alg:omd}, we will use a well-known regret bound for follow the regularized leader (FTRL) with linearized losses \parencite[Lemma 5.3]{hazan23}, which in our case is equivalent to OMD.\footnote{This equivalence is due to our choice of regularizer, as we never need to project $\vect{w}^t$.}

\begin{lemma}[Follows from {\textcite[Lemma 5.3]{hazan23}}] \label{lem:hazan_omd_bound}
	If $\eta_t = \eta$ for all $t$, the regret of Algorithm~\ref{alg:omd} is at most
	\[\frac{1}{\eta}\parens{\max_{\vect{w} \in \Delta_m} R(\vect{w}) - \min_{\vect{w} \in \Delta_m} R(\vect{w})} + \sum_{t = 1}^T \angles{\nabla L^t(\vect{w}^t), \vect{w}^t - \vect{w}^{t + 1}}.\]
\end{lemma}

Informally, this bound means that if the vectors $\nabla L^t(\vect{w}^t)$ are small in magnitude, our regret is also small. Conversely, if some $\nabla L^t(\vect{w}^t)$ is large, this may be bad for our regret bound. We expect the gradient of the loss to be large if some expert is very wrong (assigns a very low probability to the correct outcome), since the loss would then be steeply increasing as a function of that expert's weight. Fortunately, the calibration property guarantees this to be unlikely. Specifically, we define the \emph{small gradient assumption} as follows.

\begin{defin} \label{def:sga}
	Define $\gamma := 12n \ln T$. The \emph{small gradient assumption} holds for a particular run of Algorithm~\ref{alg:omd} if for every $t \in [T]$ and $i \in [m]$, we have
	\[-\frac{\gamma}{w_i^t} \le \partial_i L^t(\vect{w}^t) \le \gamma,\]
	where $\partial_i$ denotes the partial derivative with respect to the $i$-th weight.\footnote{See Equation~\ref{eq:partl_l_expr} for an expression of this quantity in terms of the experts' reports and weights.}
\end{defin}

In Section~\ref{sec:finishing_up}, we prove that the small gradient assumption is very likely to hold. This is a key conceptual step in our proof, as it is where we leverage the calibration property to prove bounds that ultimately let us bound our algorithm's regret. We then use the low likelihood of the small gradient assumption failing in order to bound the contribution to the expected regret from the case where the assumption fails to hold.

In Sections~\ref{sec:sga} and \ref{sec:regret_sga}, we bound regret under the condition that the small gradient assumption holds. We show that under the assumption, for all $i, t$ we have $(w_i^t)^\alpha \ge \eta$. Consequently, $\eta = \frac{1}{\sqrt{T} \ln T} \cdot \frac{1}{12m^{(1 + \alpha)/2} n}$ at all time steps, so we can apply Lemma~\ref{lem:hazan_omd_bound}. The first term in the bound is $O(1/\eta) = O(\sqrt{T} \log T)$. As for the summation term, we upper bound it by keeping track of the following quantity:
\[\varphi(t) := \sum_{s = 1}^t \angles{\nabla L^s(\vect{w}^s), \vect{w}^s - \vect{w}^{s + 1}} + 19m^2 \gamma^2 \eta(T - t) - 4 \gamma \sum_{i = 1}^m \ln w_i^{t + 1}.\]
The first term is exactly the summation in Lemma~\ref{lem:hazan_omd_bound} up through step $t$. The $19m^2 \gamma^2 \eta$ is something akin to an upper bound on the value of $\angles{\nabla L^t(\vect{w}^t), \vect{w}^t - \vect{w}^{t + 1}}$ at a given time step (times $T - t$ remaining time steps). This upper bound is not strict: in particular, large summands are possible when some weights are small (because of the fact that the lower bound in the small gradient assumption is inversely proportional to $w_i^t$). However, attaining a large summand requires these small weights to increase, thus ``spending potential'' for future large summands. The last term keeps track of this potential.

We show that under the small gradient assumption, $\varphi(t)$ necessarily decreases with $t$. This argument, which we give in Section~\ref{sec:regret_sga}, is another key conceptual step, and is arguably the heart of the proof. Since $\varphi(T)$ is equal to the summation term in Lemma~\ref{lem:hazan_omd_bound} (plus a positive number), and $\varphi(0) \ge \varphi(T)$, the summation term is less than or equal to $\varphi(0)$, which is at most $O(m^{(3 - \alpha)/2} \sqrt{T} \log T)$. This completes the proof.

\subsection{Bounds on $\vect{w}^t$ under the small gradient assumption} \label{sec:sga}
In this section, we state bounds on expert weights and how quickly they change from one time step to the next, conditional on the small gradient assumption. We use the following lemma, whose proof we defer to Appendix~\ref{appx:chap6}.

\begin{restatable}{lemma}{omdwibound} \label{lem:omd_wi_bound}
	Consider a particular run of Algorithm~\ref{alg:omd}. Let $\zeta$ be a constant such that $-\frac{\zeta}{w_i^t} \le \partial_i L^t(\vect{w}^t) \le \zeta$ for all $i, t$. Then for every $i, t$, we have
	\[(w_i^t)^{\alpha - 1} - \parens{\frac{1}{w_i^t} + 1} \eta_t \zeta \le (w_i^{t + 1})^{\alpha - 1} \le (w_i^t)^{\alpha - 1} + \parens{\frac{1}{\min_k w_k} + 1} \eta_t\zeta.\]
	Furthermore, if $\eta_t \zeta \le (1 - \alpha)^2 (w_i^t)^\alpha$ for all $i$, then for every $i$ we have
	\[(w_i^{t + 1})^{\alpha - 1} \le (w_i^t)^{\alpha - 1} + (m + 1) \eta_t\zeta.\]
\end{restatable}

Intuitively, this result states that when the gradient update is small, $w_i^{t + 1}$ is not too different from $w_i^t$. Note that the lower bound $-\frac{\zeta}{w_i^t}$ that we place on the gradient is not a simple Lipschitz bound but instead depends on $w_i^t$; this makes the bounds in Lemma~\ref{lem:omd_wi_bound} less straightforward to prove. In particular, we bound each component $w_i^t$ individually, using bounds on the gradient of the loss for all other components and convexity arguments.

Lemma~\ref{lem:omd_wi_bound} can be translated into bounds on each $w_i^t$ and on the change between $w_i^t$ and $w_i^{t + 1}$:

\begin{restatable}{corollary}{boundscor} \label{cor:bounds}
	Under the small gradient assumption, for sufficiently large $T$ we have for all $i \in [m], t \in [T]$ that:
	\begin{enumerate}[label=(\#\arabic*)]
		\item \label{item:eta_wit} $(w_i^t)^\alpha \ge 4\eta \gamma$ and $w_i^t \ge \frac{1}{10\sqrt{m}} T^{1/(2(\alpha - 1))}$.
		\item \label{item:wit_wit1} $-32(w_i^t)^{1 - \alpha} \eta \gamma \le w_i^t - w_i^{t + 1} \le 2(w_i^t)^{2 - \alpha}(m + 1) \eta \gamma$.
	\end{enumerate}
\end{restatable}

We defer the proof of Corollary~\ref{cor:bounds} to Appendix~\ref{appx:chap6}. The key idea for \ref{item:eta_wit} is to proceed by induction on $t$ on the two sub-statements in parallel: so long as $(w_i^t)^\alpha \ge 4\eta \gamma$, we may use the second part of Lemma~\ref{lem:omd_wi_bound} with $\zeta = \gamma$ to bound $(w_i^{t + 1})^{\alpha - 1}$ in terms of $(w_i^t)^{\alpha - 1}$, which we can leverage to prove both sub-statements for $t + 1$. \ref{item:wit_wit1} then follows from \ref{item:eta_wit} by routine (though nontrivial) algebra.

Armed with the bounds of Corollary~\ref{cor:bounds}, we are now able to show that under the small gradient assumption, Algorithm~\ref{alg:omd} attains vanishing regret.

\subsection{Bounding regret under the small gradient assumption} \label{sec:regret_sga}
Assume the small gradient assumption. Note that since $4\gamma \ge 1$, by Corollary~\ref{cor:bounds}~\ref{item:eta_wit} we have that $\eta_t = \eta$ for all $t$. This means that we may apply the bound in Lemma~\ref{lem:hazan_omd_bound}, and in particular we have
\[\frac{1}{\eta}\parens{\max_{\vect{w} \in \Delta_m} R(\vect{w}) - \min_{\vect{w} \in \Delta_m} R(\vect{w})} = \frac{1}{\eta} \cdot \frac{m}{\alpha} \parens{\frac{1}{m}}^{\alpha} = \frac{m^{1 - \alpha}}{\alpha \eta} = \frac{12}{\alpha} m^{(3 - \alpha)/2}n\sqrt{T} \ln T.\]
It remains to bound the summation component of the regret bound in Lemma~\ref{lem:hazan_omd_bound}. To do so, we prove the following lemma, which we alluded to in Section~\ref{sec:outline} as the heart of the proof of Theorem~\ref{thm:chap6_no_regret}.

\begin{lemma} \label{lem:phi_dec}
	For $t \in \{0, 1, \dots, T\}$, let
	\[\varphi(t) := \sum_{s = 1}^t \angles{\nabla L^s(\vect{w}^s), \vect{w}^s - \vect{w}^{s + 1}} + 19m^2 \gamma^2 \eta(T - t) - 4 \gamma \sum_{i = 1}^m \ln w_i^{t + 1}.\]
	Under the small gradient assumption, for sufficiently large $T$, $\varphi(t)$ is a decreasing function of $t$.
\end{lemma}

To prove this claim, consider a particular $t \in [T]$. We may write
\begin{equation} \label{eq:diff}
	\varphi(t) - \varphi(t - 1) = \sum_{i = 1}^m \parens{(w_i^t - w_i^{t + 1}) \partial_i L^t(\vect{w}^t) - 19m \gamma^2 \eta + 4 \gamma (\ln w_i^t - \ln w_i^{t + 1})}
\end{equation}

\noindent and we wish to show that this quantity is negative. In fact, we show that the contribution from every $i \in [m]$ is negative. The key idea is to consider two cases: $w_i^{t + 1} \le w_i^t$ and $w_i^{t + 1} \ge w_i^t$. In each case, Corollary~\ref{cor:bounds} provides an upper bound on the magnitude of the difference between $w_i^t$ and $w_i^{t + 1}$. If $w_i^{t + 1} \le w_i^t$ then the first and third terms in the summation are positive but small, and are dominated by the middle term. If $w_i^{t + 1} \ge w_i^t$ then the first term may be quite large, because of the asymmetric bound in the small gradient assumption (and the consequently asymmetric bound in Corollary~\ref{cor:bounds}). However, in this case the contribution of the third term is very negative, enough to make the overall expression negative. In this sense, the third term keeps track of unspent potential for future regret, which gets ``spent down'' whenever a large amount of regret is realized (as measured by the first term).

We now prove formally that each term of the summation in Equation~\ref{eq:diff} is negative.

\begin{proof}
	First assume that $w_i^t - w_i^{t + 1} \le 0$. Note that by combining \ref{item:eta_wit} and \ref{item:wit_wit1} of Corollary~\ref{cor:bounds}, we have
	\[w_i^{t + 1} - w_i^t \le 32(w_i^t)^{1 - \alpha} \eta \gamma \le 8 w_i^t.\]
	By the small gradient assumption we have that
	\[(w_i^t - w_i^{t + 1}) \partial_i L^t(\vect{w}^t) \le \frac{\gamma(w_i^{t + 1} - w_i^t)}{w_i^t}.\]
	On the other hand, we have
	\[4 \gamma(\ln w_i^t - \ln w_i^{t + 1}) = - 4 \gamma \ln \parens{1 + \frac{w_i^{t + 1} - w_i^t}{w_i^t}} \le \frac{-\gamma(w_i^{t + 1} - w_i^t)}{w_i^t}\]
	for $T$ large enough. (Here we use that $w_i^{t + 1} - w_i^t \le 8w_i^t$ and that $\ln(1 + x) \ge \frac{x}{4}$ for $x \le 8$.) Thus, the first and third terms in Equation~\ref{eq:diff} are net negative; meanwhile, the second term is also negative, so the expression is negative.\\
	
	Now assume that $w_i^t - w_i^{t + 1} \ge 0$. Again by the small gradient assumption, we have that
	\[(w_i^t - w_i^{t + 1}) \partial_i L^t(\vect{w}^t) \le \gamma(w_i^t - w_i^{t + 1}) \le 2(m + 1) \eta \gamma^2 (w_i^t)^{2 - \alpha} \le 3m \eta \gamma^2\]
	and
	\begin{align*}
		4 \gamma(\ln w_i^t - \ln w_i^{t + 1}) &= -4 \gamma \ln \parens{1 - \frac{w_i^t - w_i^{t + 1}}{w_i^t}} \le -4 \gamma \ln(1 - 2(m + 1)\eta \gamma (w_i^t)^{1 - \alpha})\\
		&\le -4\gamma \ln(1 - 2(m + 1)\eta \gamma) \le -4\gamma \ln(1 - 3m \eta \gamma) \le 16 m \eta \gamma^2
	\end{align*}
	for $T$ sufficiently large, where in the last step we use that $\ln(1 - x) \ge -\frac{4}{3}x$ for $x > 0$ sufficiently small (and we have $\lim_{T \to \infty} 3m \eta \gamma = 0$).
	Since $16 + 3 \le 19$, the right-hand side of Equation~\ref{eq:diff} is negative. This concludes the proof.
\end{proof}

\begin{corollary} \label{cor:main_case}
	For sufficiently large $T$, under the small gradient assumption, the regret of Algorithm~\ref{alg:omd} is at most $\parens{240 + \frac{12}{\alpha}} m^{(3 - \alpha)/2} n \sqrt{T} \ln T$.
\end{corollary}

\begin{proof}
	We have already bounded the first term in the regret bound in Lemma~\ref{lem:hazan_omd_bound}. It remains only to bound the second term. This term is exactly equal to $\varphi(T) + 4\gamma \sum_{i = 1}^m \ln w_i^{T + 1} \le \varphi(T)$, and $\varphi(T) \le \varphi(0)$, by Lemma~\ref{lem:phi_dec}. We have
	\[\varphi(0) = 19m^2 \gamma^2 \eta T + 4m \gamma \ln m \le 20m^2 \gamma^2 \eta T\]
	for sufficiently large $T$. Plugging in $\gamma = 12n \ln T$ and $\eta = \frac{1}{\sqrt{T} \ln T} \cdot \frac{1}{12m^{(1 + \alpha)/2} n}$ concludes the proof.
\end{proof}

\subsection{The case where the small gradient assumption fails} \label{sec:finishing_up}
It remains to consider the case in which the small gradient assumption does not hold. This part of the proof consists primarily of technical lemmas, which we defer to Appendix~\ref{appx:chap6}. The key lemma is a bound on the probability that the small gradient assumption fails by a given margin:

\begin{restatable}{lemma}{lnotsmall} \label{lem:l_not_small}
	For any weight vector $\vect{w}$, $i \in [m]$, and $\zeta \ge 0$, we have that
	\begin{equation} \label{eq:partl_l_ub}
		\pr{\partial_i L(\vect{w}) \ge \zeta} \le n e^{-\zeta}
	\end{equation}
	and
	\begin{equation} \label{eq:partl_l_lb}
		\pr{\partial_i L(\vect{w}) \le -\frac{\zeta}{w_i}} \le mn^2 e^{-\zeta/n}.
	\end{equation}
\end{restatable}

Note that plugging in $\zeta = \gamma$ yields a bound of $mT(ne^{-\gamma} + mn^2 e^{-\gamma/n})$ on the probability that the small gradient assumption fails to hold. (Since $\gamma = 12n \ln T$, this quantity is on the order of $T^{-11}$.)

The proof of Lemma~\ref{lem:l_not_small} is the only part of the proof of Theorem~\ref{thm:chap6_no_regret} that uses the calibration property. While we defer the full proof to Appendix~\ref{appx:chap6}, we highlight how the calibration property is used to prove Equation~\ref{eq:partl_l_ub}. In brief, it is straightforward to show that $\partial_i L(\vect{w}) \le -\ln p_J^i$, where  $J$ is the random variable corresponding to the realized outcome.\footnote{Writing out the expression for $L(\vect{w})$ and differentiating leaves us with $-\ln p_J^i$ plus a negative term (see Equation~\ref{eq:partl_l_expr}).} Therefore, we have
	\begin{align*}
	\pr{\partial_i L(\vect{w}) \ge \zeta} &\le \pr{-\ln p_J^i \ge \zeta} = \pr{p_J^i \le e^{-\zeta}} = \sum_{j = 1}^n \pr{J = j \enskip \& \enskip p_j^i \le e^{-\zeta}}\\
	&= \sum_{j = 1}^n \pr{p_j^i \le e^{-\zeta}} \pr{J = j \mid p_j^i \le e^{-\zeta}} \le \sum_{j = 1}^n \pr{J = j \mid p_j^i \le e^{-\zeta}} \le ne^{-\zeta},
	\end{align*}
where the last step follows by the calibration property, thus proving Equation~\ref{eq:partl_l_ub}.

Combining Lemma~\ref{lem:l_not_small} with an analysis of our algorithm using the standard regret bound for online mirror descent \parencite[Theorem 6.8]{ora21} gives us the following result as a corollary.

\begin{restatable}{corollary}{generalcase} \label{cor:general_case}
	The expected total regret of our algorithm conditional on the small gradient assumption \emph{not} holding, times the probability of this event, is at most $\tilde{O}(T^{(5 - \alpha)/(1 - \alpha) - 10})$.
\end{restatable}

It follows that the contribution to expected regret from the case that the small gradient assumption does not hold is $\tilde{O}(T^{-1})$, which is negligible. Together with Corollary~\ref{cor:main_case} (which bounds regret under the small gradient assumption), this proves Theorem~\ref{thm:chap6_no_regret}.

\subsection{Approximate calibration}
Theorem~\ref{thm:chap6_no_regret} holds even if experts are only \emph{approximately} calibrated.

\begin{defin}
	For $\tau \ge 1$, we say that expert $i$ is \emph{$\tau$-calibrated} if for all $\vect{p} \in \Delta_n$ and $j \in [n]$, we have that $\pr{J = j \mid \vect{p}^j = \vect{p}} \le \tau p_j$. We say that $\PP$ satisfies the \emph{$\tau$-approximate calibration property} if every expert is $\tau$-calibrated.
\end{defin}

\begin{corollary} \label{cor:approx_calib}
	For any $\tau$, Theorem~\ref{thm:chap6_no_regret} holds even if the calibration property is replaced with the $\tau$-approximate calibration property.
\end{corollary}

(Note that the $\tau$ is subsumed by the big-$O$ notation in Theorem~\ref{thm:chap6_no_regret}; Corollary~\ref{cor:approx_calib} does not allow experts to be arbitrarily miscalibrated.)

Technically, Corollary~\ref{cor:approx_calib} is a corollary of the \emph{proof} of Theorem~\ref{thm:chap6_no_regret}, rather than a corollary of the theorem itself.\footnote{Fun fact: the technical term for a corollary to a proof is a \emph{porism}.}

\begin{proof}[Proof of Corollary~\ref{cor:approx_calib}]
	We only used the calibration property in the proofs of Equations~\ref{eq:partl_l_ub} and \ref{eq:partl_l_lb}. In the proof of Equation~\ref{eq:partl_l_ub}, we used the fact that $\pr{J = j \mid p_j^i \le e^{-\zeta}} \le e^{-\zeta}$; the right-hand side now becomes $\tau e^{-\zeta}$, and so the right-hand side of Equation~\ref{eq:partl_l_ub} changes to $\tau n e^{-\zeta}$. Similarly, the right-hand side of Equation~\ref{eq:partl_l_lb} changes\footnote{Specifically, in the proof of Equation~\ref{eq:partl_l_lb}, we use the calibration property in the proof of Lemma~\ref{lem:mnq}; the right-hand side of the lemma changes to $\tau mnq$.} to $\tau mn^2 e^{-\zeta/n}$.
	
	Lemma~\ref{lem:l_not_small} is only used in the proof of Corollary~\ref{cor:general_case}, where $2m^2n^2T$ is replaced by $2\tau m^2 n^2 T$. Since $\tau$ is a constant, Corollary~\ref{cor:general_case} holds verbatim.
\end{proof}

\section{Lower bound} \label{sec:chap6_lower_bound}
In this section, we prove a lower bound result for our setting. Specifically, we show that no OMD algorithm with a constant step size\footnote{While Algorithm~\ref{alg:omd} does not always have a constant step size, it does so with high probability. The examples that prove Theorem~\ref{thm:chap6_lower_bound} cause $\Omega(\sqrt{T})$ regret in the typical case, rather than causing unusually large regret in an atypical case. This makes our comparison of Algorithm~\ref{alg:omd} to this class fair.} substantially outperforms Algorithm~\ref{alg:omd}.

\begin{theorem} \label{thm:chap6_lower_bound}
	For every strictly convex function $R: \Delta_m \to \RR$ that is continuously twice differentiable at its minimum, and $\eta \ge 0$, online mirror descent with regularizer $R$ and constant step size $\eta$ incurs $\Omega(\sqrt{T})$ expected regret.
\end{theorem}

\begin{proof}
	Our examples will have $m = n = 2$. The space of weights is one-dimensional; let us call $w$ the weight of the first expert. We may treat $R$ as a (convex) function of $w$, and similarly for the losses at each time step. We assume that $R'(0.5) = 0$; this allows us to assume that $w_1 = 0.5$ and does not affect the proof idea.
	
	It is straightforward to check that if Experts 1 and 2 assign probabilities $p$ and $\frac{1}{2}$, respectively, to the correct outcome, then
	\[L'(w) = \frac{(1 - p)^w}{p^w + (1 - p)^w} \ln \frac{1 - p}{p}.\]
	If roles are reversed (they say $\frac{1}{2}$ and $p$ respectively) then
	\[L'(w) = -\frac{(1 - p)^{1 - w}}{p^{1 - w} + (1 - p)^{1 - w}} \ln \frac{1 - p}{p}.\]
	
	We first prove the regret bound if $\eta$ is small ($\eta \le T^{-1/2}$). Consider the following setting: Expert 1 always reports $(50\%, 50\%)$; Expert 2 always reports $(90\%, 10\%)$; and Outcome 1 happens with probability $90\%$ at each time step. It is a matter of simple computation that:
	\begin{itemize}
		\item $L'(w) \le 2$ no matter the outcome or the value of $w$.
		\item If $w \ge 0.4$, then $p^*_1(w) \le 0.8$.
	\end{itemize}
	The first point implies that $R'(w_t) \ge -2\eta t$ for all $t$. It follows from the second point that the algorithm will output weights that will result in an aggregate probability of less than $80\%$ for values of $t$ such that $-2\eta t \ge R'(0.4)$, i.e.\ for $t \le \frac{-R'(0.4)}{2\eta}$. Each of these time steps accumulates constant regret compared to the optimal weight vector in hindsight (which with high probability will be near $1$). Therefore, the expected total regret accumulated during these time steps is $\Omega(1/\eta) = \Omega(\sqrt{T})$.\\
	
	Now we consider the case in which $\eta$ is large ($\eta \ge \sqrt{T}$). In this case our example is the same as before, except we change which expert is ``ignorant'' (reports $(50\%, 50\%)$ and which is ``informed'' (reports $(90\%, 10\%)$). Specifically the informed expert will be the one with a lower weight (breaking ties arbitrarily).
	
	We will show that our algorithm incurs $\Omega(\eta)$ regret compared to always choosing weight $0.5$. Suppose without loss of generality that at a given time step $t$, Expert 1 is informed (so $w^t \le 0.5$). Observe that
	\begin{align*}
		L(w^t) - L(0.5)& = -(0.5 - w^t) L'(0.5) + O((0.5 - w)^2)\\
		&= -(0.5 - w^t)\frac{\sqrt{1 - p}}{\sqrt{p} + \sqrt{1 - p}} \ln \frac{1 - p}{p} + O((0.5 - w)^2),
	\end{align*}
	where $p$ is the probability that Expert 1 assigns to the event that happens (so $p = 0.9$ with probability $0.9$ and $p = 0.1$ with probability $0.1$). This expression is (up to lower order terms) equal to $c(0.5 - w^t)$ if $p = 0.9$ and $-3c(0.5 - w^t)$ if $p = 0.1$, where $c \approx 0.55$. This means that an expected regret (relative to $w = 0.5$) of $0.6c(0.5 - w^t)$ (up to lower order terms) is incurred.
	
	Let $D$ be such that $R''(w) \le D$ for all $w$ such that $\abs{w - 0.5} \le \frac{\sqrt{T}}{4D}$. (Such a $D$ exists because $R$ is continuously twice differentiable at $0.5$.) If $\abs{w^t - 0.5} \ge \frac{\sqrt{T}}{4D}$, we just showed that an expected regret (relative to $w = 0.5$) of $\Omega \parens{\frac{\sqrt{T}}{4D}}$ is incurred. On the other hand, suppose that $\abs{w^t - 0.5} \le \frac{\sqrt{T}}{4D}$. We show that $\abs{w^{t + 1} - 0.5} \ge \frac{\sqrt{T}}{4D}$.
	
	To see this, note that $\abs{L'(w^t)} \ge 0.5$, we have that $\abs{R'(w^{t + 1}) - R'(w^t)} \ge 0.5 \eta$. We also have that $D \abs{w^{t + 1} - w^t} \ge \abs{R'(w^{t + 1}) - R'(w^t)}$, so $D \abs{w^{t + 1} - w^t} \ge 0.5 \eta$. Therefore, $\abs{w^{t + 1} - w^t} \ge \frac{\eta}{2D} \ge \frac{\sqrt{T}}{2D}$, which means that $\abs{w^{t + 1} - 0.5} \ge \frac{\sqrt{T}}{4D}$.
	
	This means that an expected regret (relative to $w = 0.5$) of $\Omega \parens{\frac{\sqrt{T}}{4D}}$ is incurred on at least half of time steps. Since $D$ is a constant, it follows that a total regret of at least $\Omega(\sqrt{T})$ is incurred, as desired.
\end{proof}

\section{Conclusion} \label{sec:conclusion}
In this work, we have considered the problem of learning optimal weights for the logarithmic pooling of expert forecasts. It quickly became apparent that under the usual fully adversarial setup, attaining vanishing regret is impossible (Example~\ref{example:bad_regret}). We chose to relax the environment by imposing the constraint on the adversary that experts must be calibrated. Put otherwise, the adversary is allowed to choose a joint probability distribution over the experts' reports and the outcome however it wants to, so long as the experts' reports are calibrated, after which the realized reports and outcome are selected at random from this distribution. To our knowledge, this setting is a novel contribution to the literature on prediction with expert advice. The setting may be of independent interest: we have demonstrated that no-regret bounds are possible in this setting when they are otherwise impossible, and it seems plausible that even in settings where no-regret bounds are attainable in a fully adversarial setting, the calibration property allows for stronger results.

Another important direction for future work is learning weights for \emph{generalized} logarithmic pooling: loosely speaking, logarithmic pooling but without the requirement that weights add to $1$. In Chapter~\ref{chap:prelims}, we introduced generalized logarithmic pooling and exhibited a class of information structures for which generalized logarithmic pooling is the Bayesian optimal aggregation method. This raises the natural question of whether our methods can be adapted to this more general setting.

Finally, we are interested in learning weights for other pooling methods. In particular, it is natural to ask which proper loss functions $\ell$ have the property that it is possible to achieve vanishing regret when learning weights for QA pooling with respect to $\ell$, and under what assumptions. While in Chapter~\ref{chap:qa} we showed that no-regret learning is possible for bounded loss functions, extending our techniques to unbounded loss functions beyond the log loss is a promising avenue for future exploration.
\chapter{Robust aggregation of substitutable signals} \label{chap:robust}
\emph{This chapter presents ``Are You Smarter Than a Random Expert? The Robust Aggregation of Substitutable Signals'' \parencite{nr22_smarter}, although the contents of Section~\ref{sec:general_lb} are original to this thesis. It assumes background on forecast aggregation methods (Section~\ref{sec:prelim_agg}) and information structures (Section~\ref{sec:prelim_info_struct}). Although the relevant definitions will be restated, I strongly recommend reading those sections before reading this chapter in order to gain context and intuition.}\\

\emph{Summary:} The problem of aggregating expert forecasts is ubiquitous in fields as wide-ranging as machine learning, economics, climate science, and national security. Despite this, our theoretical understanding of this question is fairly shallow. The work discussed in this chapter initiates the study of forecast aggregation in a context where experts' knowledge is chosen adversarially from a broad class of information structures. While in full generality it is impossible to achieve a nontrivial performance guarantee, we show that doing so is possible under a condition on the experts' information structure that we call \emph{projective substitutes}. The projective substitutes condition is a notion of informational substitutes: that there are diminishing marginal returns to learning the experts' signals. We show that under the projective substitutes condition, taking the average of the experts' forecasts improves substantially upon the strategy of trusting a random expert. We then consider a more permissive setting, in which the aggregator has access to the prior. We show that by averaging the experts' forecasts and then \emph{extremizing} the average by moving it away from the prior by a constant factor, the aggregator's performance guarantee is substantially better than is possible without knowledge of the prior. Our results give a theoretical grounding to past empirical research on extremization and help give guidance on the appropriate amount to extremize.

\section{Introduction}
Suppose that you wish to estimate how much the GDP of the United States will grow next year: perhaps you are making financial decisions and want to know whether to expect a downturn. You don't personally know much about the question -- just that the historical average rate of GDP growth has been 3\% -- but on the internet you find several forecasts made by machine learning models. One model predicts 3.5\% growth next year; another predicts 1.5\%; a third predicts a downturn: -1\% growth. How might you take this information into account and turn it into one number: your best guess, all things considered?\footnote{Note the difference in setting from Chapters~\ref{chap:qa} and \ref{chap:learning}: we are now interested in aggregating real-valued forecasts, as opposed to probability distributions.}

Because of the ubiquity of its applications, forecast aggregation is of critical importance to many fields: economics, climate science, public health, meteorology, ecology, and sociology, to name a few \parencite{mwgr21}. Despite this, the theoretical tools we have for understanding this problem are fairly limited.

What should we ask of a framework for comparing competing aggregation methods?
First, for each fixed setup, it should allow us to quantitatively assess an aggregation method based on its performance relative to a natural
benchmark (analogous to, for example, assessing an online learning
algorithm via its regret with respect to the best fixed action in
hindsight, as we saw in Chapter~\ref{chap:learning}).
Second, the framework should be general: rather than evaluating an aggregation method based on its performance under a particular assumption about the experts' information sets, it should assess the method based on its performance over a broad range of possible setups.

We can model each expert as having partial information over the state of the world, and thus the quantity being estimated (which we denote $Y$). The experts' information sets may overlap in essentially arbitrary ways, which we formalize using \emph{information structures} (introduced in Section~\ref{sec:prelim_info_struct}).

No aggregation method is simultaneously optimal for every information structure. As such, it is natural to ask which aggregation method optimizes worst-case performance over a broad class of information structures. This is the approach we take, because it has the aforementioned advantages: it assesses aggregation methods based on their performance, but does so broadly rather than under specific assumptions.

\subsection{Our results}
Without any conditions on the experts' information structure, no aggregation strategy can achieve a nontrivial performance guarantee.\footnote{For example, consider the ``XOR information structure'' in which two experts receive independent, random bits, and $Y$ is their XOR. See Section~\ref{sec:improving} for further discussion.} In this work, we optimize for worst-case performance over all information structures that satisfy a condition that we call \emph{projective informational substitutes}. In Section~\ref{sec:prelim_subs}, we introduced the notion of \emph{informational substitutes:} that the value of learning an additional signal has diminishing marginal returns. The projective substitutes condition is a particular formalization of this concept that builds on the notion of ``weak information substitutes'' (Definition~\ref{def:weak_subs_quad}), as we will show that the weak substitutes condition is insufficient for our purposes.\footnote{In Section~\ref{sec:random_expert} we introduce a ``secret sharing'' information structure that shows that with no further assumptions beyond the weak substitutes condition, no aggregation strategy achieves a better performance guarantee than the strategy of choosing a random expert to trust.} Intuitively, substitutable signals allow for effective aggregation because signal interactions are more predictable, so it is possible to infer more from forecasts alone without knowing the information structure.

We consider two settings: the \emph{prior-free setting} and the \emph{known prior setting}. In the prior-free setting, an aggregator receives only the experts' forecasts as input; in the known prior setting, the aggregator additionally knows the prior, i.e.\ the overall expected value of $Y$ (3\% in our leading example). In both settings, the expert must then output an aggregate forecast.

One simple strategy is to pick an expert at random and ``aggregate'' by outputing that expert's forecast. In expectation, this aggregate performs at least as well as the prior; and under the weak substitutes condition, the strategy does at least $1/m$ times as well as someone who knew every expert's signal and the information structure, where $m$ is the number of experts.\footnote{We judge the performance of an aggregation strategy based on its improvement over the prior.} That is, choosing a random expert attains an \emph{approximation ratio} of $1/m$. Unfortunately, we exhibit an information structure that satisfies weak substitutes but on which no aggregation strategy can outperform a $1/m$-approximation (even in the known prior setting).

\begin{figure}[ht]
	\centering
	\includegraphics[width=\textwidth]{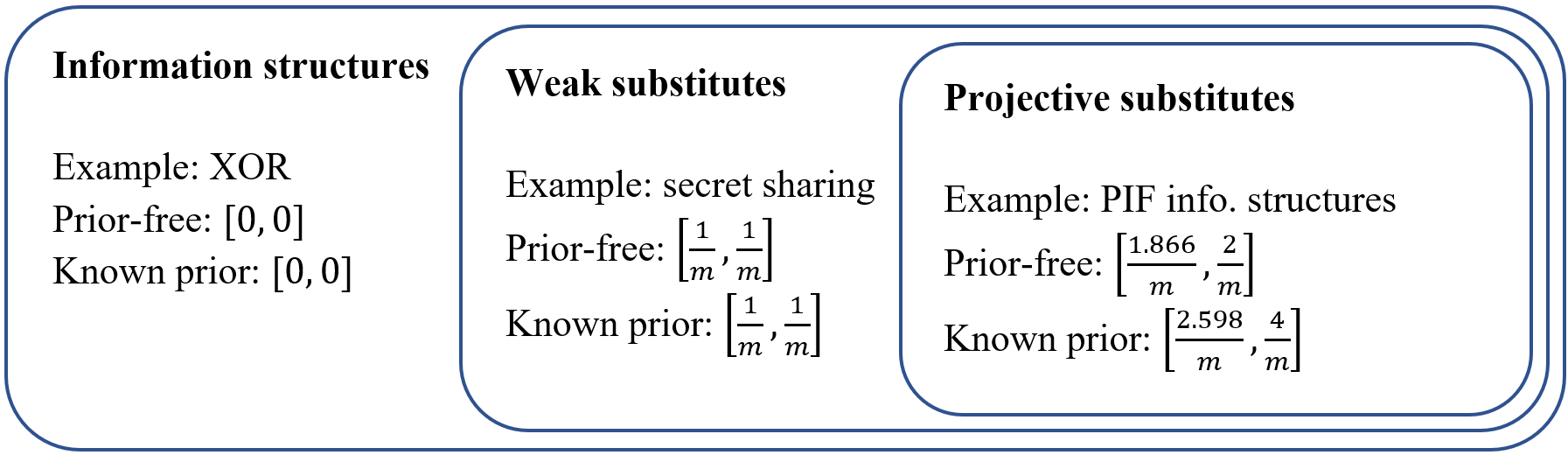}
	\caption[Robust aggregation: results summary]{We are interested in the innermost setting, where nontrivial positive results are possible. Intervals given for each setting indicate positive and negative results, respectively (and are stated as asymptotic approximations in $m$, the number of experts, for the innermost setting).}
	\label{fig:chap7_results}
\end{figure}

However, under our slightly stronger assumption of projective substitutes, it is possible to improve upon this $1/m$ baseline. Thus, while one can ask about robust aggregation in many different settings, the projective substitutes condition appears to be a sweet spot: it allows for a broad array of possible information structures while still allowing at nontrivial performance guarantees in both the prior-free and known prior settings. These results are summarized in Figure~\ref{fig:chap7_results}.\\

In Section~\ref{sec:general_lb}, we begin our investigation with a surprisingly strong negative result. Let $m$ be the number of experts. We show that no aggregation strategy can hope to perform even $\frac{4}{m}$ times as well as someone who knew every expert's signal, \emph{even if the aggregation strategy knows the precise information structure.} We do this by exhibiting a single information structure (one that satisfies the projective substitutes condition) in which the optimal aggregate of the experts' forecasts performs less than $\frac{4}{m}$ times as well as the optimal aggregate given all experts' signals. The information structure is quite natural (it is a PIF information structure -- see Definition~\ref{def:pif}, also repeated later in this chapter), and so our $\frac{4}{m}$ bound serves as a basis of comparison for our other results. How close to this $\frac{4}{m}$ bound can we get with positive results?

In Section~\ref{sec:prior_free}, we investigate the prior-free setting. In this setting, we show that under the projective substitutes condition, the aggregation strategy that averages all experts' forecasts improves upon the random expert strategy, attaining an approximation guarantee of roughly $1.866/m$. We also show that our bound is tight for two experts and close to tight for any number of experts.

In Section~\ref{sec:known_prior}, we investigate the known prior setting. We prove that it is possible to improve upon the aforementioned guarantee of the prior-free setting by \emph{extremizing} the average of the experts' beliefs, i.e.\ moving it away from the prior. Additionally, our results suggest a particular amount by which to extremize. Specifically, we show that by \emph{linearly} extremizing -- moving the average of the experts' forecasts away from the prior by a particular constant factor\footnote{This factor approaches $\sqrt{3}$ as $m$ approaches infinity.} (that depends on $m$) -- it is possible to attain an approximation ratio of roughly $2.598/m$. We show that our positive result is tight for two experts.

\begin{figure}[ht]
	\centering
	\includegraphics[scale=0.8]{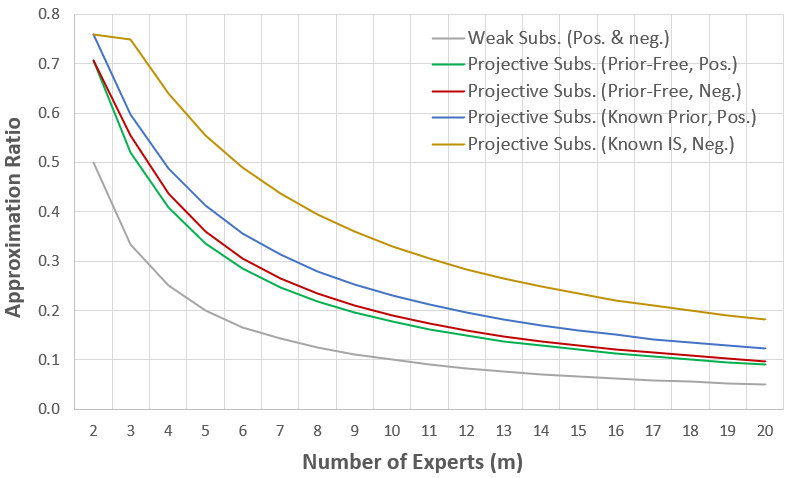}
	\caption[Approximation guarantees as a function of the number of experts]{For $m = 2 \dots 20$: positive and negative results for approximation ratios under various settings (similar to Table~\ref{table:results_small_m}). Grey: weak substitutes (matching positive and negative results). Green: projective substitutes in the prior-free setting (positive). Red: projective substitutes in the prior-free setting (negative). Blue: projective substitutes in the known prior setting (positive). Yellow: projective substitutes in the known prior setting, and in fact the known information structure setting (negative).}
	\label{fig:results_small_m}
\end{figure}

The aforementioned results are stated asymptotically in $m$ for convenience; however, these asymptotics are not our focus. Instead, our goal is to understand which \emph{methods} of aggregation work well in which settings. When is the tried and true method of averaging forecasts about optimal, and when is it possible to attain a substantial improvement? Table~\ref{table:results_small_m} and Figure~\ref{fig:results_small_m} summarize our findings for small values of $m$ (which are plausible for many of the applications that motivate this work). The high-level takeaways are:
\begin{itemize}
	\item Under the projective substitutes condition, it is possible to improve substantially upon selecting a random expert simply by averaging the experts' forecasts.
	\item When only the forecasts are known, no technique can substantially improve upon averaging.
	\item But when the prior is known, extremizing appropriately is substantially better than averaging, and in fact better than any possible aggregation strategy that does not use the prior.
    \item While there is potentially room for improvement on our extremization technique, one cannot hope for a very substantial improvement in the projective substitutes setting -- even for an aggregator who knows the entire information structure.
\end{itemize}

\begin{table}[ht]
	\centering
	\begin{tabular}{c|c|cc|cc}\
		&Weak subs.&\multicolumn{2}{c}{Proj. subs. (prior-free)} &\multicolumn{2}{c}{Proj. subs. (known prior)}\\
		m & Pos.\ \& neg.\ & Averaging (positive) & Negative & Extremizing (positive) & Negative\\
		\hline
		2&0.500&0.706&0.706&0.760&0.760\\
		3&0.333&0.520&0.556&0.596&0.750\\
		4&0.250&0.409&0.438&0.488&0.640\\
		5&0.200&0.336&0.360&0.412&0.556\\
		6&0.167&0.285&0.306&0.356&0.490\\
		7&0.143&0.248&0.265&0.314&0.438
	\end{tabular}
	\caption[Approximation guarantees for small numbers of experts]{For $m = 2 \dots 7$: the approximation ratio guaranteed by choosing a random expert, which is the best one can do under weak substitutes, followed by some of our positive and negative results under the projective substitutes condition. Note that the negative results for the known prior setting hold even if the aggregator knows the entire information structure, not just the prior.}
	\label{table:results_small_m}
\end{table}

\subsection{Related work}
\paragraph{Robust aggregation} In Chapter~\ref{chap:intro}, we discussed the two most common approaches to theoretical questions about forecast aggregation: axiomatic approaches and Bayesian ones. The axiomatic approach seeks to define desirable properties of aggregation methods, and asks which methods satisfy these properties. By contrast, Bayesian approaches take a \emph{parametric} view of aggregation: experts are modeled as Bayesians whose signals are drawn from a particular parameterized family of distributions, and an aggregation method is chosen to optimize an objective function within the model.

Our approach can be thought of as a hybrid of the axiomatic and Bayesian approaches, blending what we believe to be the most appealing parts of each. We draw from the Bayesian approach in using information structures as a formalism for the experts' knowledge, whereas the goal of producing a single role that satisfies some global property (in our case, worst-case optimality) is reminiscent of the axiomatic approach. Our model is non-parametric: rather than assuming a parameterized family of distributions, we seek to optimize our aggregation method against a broad class of information structures.

Our work is most similar to \textcite{abs18}, which likewise seeks to optimize an aggregation method against an adversarially selected information structure. However, the class of information structures that we consider is broader: while they consider the case of two Blackwell-ordered experts (i.e.\ two experts, an unknown one of whom knows strictly more than the other) and two conditionally independent experts, we consider an arbitrary number of experts from any information structure that satisfies the projective substitutes condition. \textcite{lr20} have a similar model, but are also quite restrictive in terms of the information structures they consider. \textcite{dil21} use a similar model, but more distantly related: they consider arbitrary decision problems but restrict the aggregator to a finite number of decisions (just two decisions for many of their results) -- in our setting this would mean forcing the aggregator to choose among finitely many output choices. Our notion of robustness is also similar to that of prior-independent algorithm design (see e.g.\ \textcite[\S3]{hr08}), though in a quite different setting.

Another important difference is that most of the previously mentioned work specifically considers the aggregation of \emph{probabilistic} forecasts, whereas we are interested in aggregating expected value forecasts for arbitrary real-valued quantities.

\paragraph{Extremization} In Section~\ref{sec:prelim_agg}, we discussed the merits of extremization: pushing the average of the experts' forecasts away from the prior. Past empirical work has demonstrated that extremizing the average of the experts' forecasts often improves the aggregate forecast \parencite{sbfmtu14, bmtsu14, su15}. \textcite{bmtsu14} explain this by noting that any individual forecaster should incorporate the fact that they may be missing useful information available to other forecasters, and that simply averaging forecasts would fail to incorporate the full wisdom of the crowd. Studying aggregation in the context of information structures as well, \textcite{su15} note that the forecast average \emph{lacks resolution}, meaning that its variance is provably too low, and finds that extremization helps to solve this issue. \textcite{spu16} note that the more disjoint the experts' information, the more it makes sense to extremize. \textcite{satopaa21} refines this approach, suggesting that the variance of forecasts be taken into consideration.

However, none of the aforementioned works aim to show that extremization produces better results across a broad class of information structures; instead, the authors consider particular structures, such as experts with correlated Gaussian signals. \textcite{su15} note that the information structure framework is in full generality ``too abstract to be applied in practice.'' On the other hand, our approach of robust aggregation is able not only to provide a theoretical justification for extremization, but also to suggest a particular factor of extremization (Theorem~\ref{thm:known_prior_positive}), thus giving rigorous backing to what had previously been justified either by empirical heuristics or by optimization over a quite narrow class of information structures.

\section{Key definitions and preliminaries} \label{sec:chap7_prelims}
\subsection{Information structures}
We introduced information structures in Section~\ref{sec:prelim_info_struct}. Information structures will be essential to this chapter, so we recommend reviewing that discussion. For convenience, we restate the definition below.

\infostruct*

In this chapter, $Y$ will specifically be a real-valued random variable. We will also assume that $Y$ has finite variance. As usual, we will interpret each signal $\sigma_i$ as belonging to expert $i$.

We also recall the XOR information structure:

\xoris*

Just as in Section~\ref{sec:prelim_info_struct}, given a subset $A \subseteq [m]$, we define $Y_A := \EE{Y \mid \sigma_i: i \in A}$. That is, $Y_A$ is the random variable whose value is the expectation of $Y$ given the signals of the experts in $A$. If $A = \{i\}$, we write $Y_i$ in place of $Y_{\{i\}}$.

\subsection{The Pythagorean theorem} \label{sec:chap7_pythag_prelim}
Recall from Section~\ref{sec:prelim_pythag} the Pythagorean theorem:

\pythagsquared*

As discussed in Section~\ref{sec:prelim_pythag}, random variables can be thought of as vectors in a Hilbert space with inner product $\angles{X, Y} := \EE{XY}$. In this space, uncorrelated random variables correspond to orthogonal vectors, and conditional expectations correspond to orthogonal projections. Proposition~\ref{prop:pythag_squared} is simply an application of the usual Pythagorean theorem in the context of this space.

\subsection{Improving on the prior} \label{sec:improving}
In this work, we will be taking the perspective of an aggregator who receives estimates of $Y$ from each expert.\footnote{These estimates are each expert's expectation of $Y$ conditioned on their signal. The aggregator does not receive signals from experts.} The aggregator then produces an estimate $Z$ of $Y$ which is as accurate as possible. In particular, we care about the \emph{robust} estimation of $Y$: a single estimate that is simultaneously as accurate as possible across all possible information structures (satisfying the projective substitutes condition, which we discuss below).

We assess an aggregator's performance by the squared distance between their estimate $Z$ and the true value $Y$. That is, the aggregator wishes to minimize $\EE{(Y - Z)^2}$. We define the function $v(Z)$ as follows to reflect the \emph{quality of $Z$ as an estimate of $Y$}.

\begin{defin}
	Given an information structure $\mathcal{I} = (\Omega, \PP, \pmb{\sigma}, Y)$ and a random variable $Z$, we define
	\[v(Z) := \EE{(Y - \EE{Y})^2} - \EE{(Y - Z)^2}.\]
\end{defin}

Thus, $v(\cdot)$ is the improvement in loss provided by $Z$ over an uninformed estimate. For example, $v(Y_\emptyset) = 0$ and $v(Y) = \EE{(Y - \EE{Y})^2}$ is the variance of $Y$. We cannot possibly hope for any $Z$ such that $v(Z) > Y_{[m]}$, since $Y_{[m]}$ is the estimate produced by knowing all information that exists. This motivates comparing $v(Z)$ against the benchmark $v(Y_{[m]})$.

However, the aggregator does not know the underlying information structure -- only the experts' estimates. Specifically, we will consider two settings:
\begin{enumerate}[label=(\arabic*)]
	\item The \emph{prior-free setting}: the aggregator's estimate is only based on the experts' estimates. That is, $Z$ is a function of $Y_1, \dots, Y_m$.
	\item The \emph{known prior setting}: the aggregator knows the experts' estimates and the prior. That is, $Z$ is a function of $Y_1, \dots, Y_m$ and $\EE{Y}$.
\end{enumerate}
That is, $Z$ is a function of $m$ real numbers (or $m + 1$, in the known prior setting); we call this function the aggregator's \emph{aggregation strategy}. The aggregator's goal is to come up with an aggregation strategy that performs well across information structures (we formalize this below).

In the known prior setting, the aggregator can report $Z = \EE{Y}$; then $v(Z) = 0$ (we call this the \emph{trivial aggregation strategy}). In both settings it is possible to do at least as well as the trivial aggregation strategy by reporting e.g.\ $Z = Y_1$. On the other hand, without any conditions on the information structure, it is not always possible to do strictly better: in the XOR information structure, the aggregator is guaranteed to receive $Y_1 = Y_2 = \frac{1}{2}$, and it is impossible for the aggregator to improve upon simply reporting the prior of $\frac{1}{2}$.

\subsection{Informational complements and substitutes}
Intuitively, in the XOR information structure, the aggregator is impeded by the fact that the experts' signals are informational complements: each signal (and the estimate it produces) is not valuable by itself, but the two signals are valuable when taken together. Perhaps if we assume that the experts' signals are instead informational substitutes, then we will be able to prove nontrivial guarantees about some aggregation strategies. And so, we recall the notion of weak informational substitutes from Section~\ref{sec:prelim_subs}.

\weaksubsquad*

\begin{remark}
    Equivalently, we can say that $\mathcal{I}$ satisfies weak informational substitutes if for all $B \subseteq A \subseteq [m]$ and $i \not \in A$, we have
    \begin{equation} \label{eq:weak_subs_ch7}
        v(Y_{A \cup \{i\}}) - v(Y_A) \le v(Y_{B \cup \{i\}}) - v(Y_B).
    \end{equation}
\end{remark}

\subsection{Random expert strategy under weak substitutes} \label{sec:random_expert}
It is not surprising that with no knowledge of the information structure, it is impossible to outperform the trivial strategy. Perhaps it would be possible to do better with only a coarse constraint on the information structure. It is not \emph{a priori} obvious that this should be possible. However, if $\mathcal{I}$ satisfies weak substitutes, then it is possible to outperform the trivial strategy by reporting a random expert's belief:

\begin{prop} \label{prop:random_expert}
	Suppose that $\mathcal{I} = (\Omega, \PP, \pmb{\sigma}, Y)$ satisfies weak substitutes, and let $Z$ be equal to $Y_i$ for a uniformly random $i \in [m]$ (we call this the \emph{random expert strategy}). Then $v(Z) \ge \frac{1}{m} v(Y_{[m]})$.
\end{prop}

\begin{proof}
	For $j \in [m]$, plug $A = [j - 1], B = \emptyset, i = j$ into Equation~\ref{eq:weak_subs_ch7}. Adding these $m$ inequalities (and noting that $v(\emptyset) = 0$), we find that $\sum_{j = 1}^m v(Y_j) \ge v(Y_{[m]})$. Therefore, for $Z$ as in the proposition statement, we have
	\[v(Z) = \frac{1}{m} \sum_{i = 1}^m v(Y_i)  \ge \frac{1}{m} v(Y_{[m]}),\]
	as desired.
\end{proof}

Put otherwise, the random expert strategy attains an \emph{approximation ratio} of $1/m$.

\begin{defin}
	Given an information structure $\mathcal{I} = (\Omega, \PP, \pmb{\sigma}, Y)$ with $m$ experts, the \emph{approximation ratio} of a random variable $Z$ is given by the quantity $v(Z)/v \parens{Y_{[m]}}$.
\end{defin}

The Pythagorean theorem lets us rewrite the approximation ratio in a more convenient form.
\begin{claim} \label{claim:improvement_rewrite}
	If a random variable $Z$ depends only on $\sigma_1, \dots, \sigma_m$, the approximation ratio of $Z$ (i.e.\ $v(Z)/v(Y_{[m]})$) is equal to
	\[1 - \frac{\EE{(Y_{[m]} - Z)^2}}{\EE{(Y_{[m]} - \EE{Y})^2}}.\]
\end{claim}

\begin{proof}
	We have
	\begin{align*}
		\frac{v(Z)}{v(Y_{[m]})} &= \frac{\EE{(Y - \EE{Y})^2} - \EE{(Y - Z)^2}}{\EE{(Y - \EE{Y})^2} - \EE{(Y - Y_{[m]})^2}}\\
		&= \frac{\parens{\EE{(Y - Y_{[m]})^2} + \EE{(Y_{[m]} - \EE{Y})^2}} - \parens{\EE{(Y - Y_{[m]})^2} + \EE{(Y_{[m]} - Z)^2}}}{\EE{(Y - \EE{Y})^2} - \EE{(Y - Y_{[m]})^2}}\\
		&= \frac{\EE{(Y_{[m]} - \EE{Y})^2} - \EE{(Y_{[m]} - Z)^2}}{\EE{(Y_{[m]} - \EE{Y})^2}} = 1 - \frac{\EE{(Y_{[m]} - Z)^2}}{\EE{(Y_{[m]} - \EE{Y})^2}}.
	\end{align*}
	In the second step, we use the Pythagorean theorem twice: one time we plug in $A = Y$, $B = Y_{[m]}$, $C = \EE{Y}$, and the other time, $A = Y$, $B = Y_{[m]}$, $C = Z$. In the third step, we again use the Pythagorean theorem, plugging in $A = Y$, $B = Y_{[m]}$, $C = \EE{Y}$.
\end{proof}

Unfortunately, the following result shows that with no further assumptions, it is not possible to guarantee an approximation ratio larger than $1/m$:

\begin{prop} \label{prop:weak_bad}
	For every $m$, there is an information structure that satisfies the weak substitutes condition, such that in both the prior-free and known prior settings, no aggregation strategy attains an approximation ratio greater than $1/m$ on the information structure.
\end{prop}

The key idea is to use Shamir secret sharing \parencite{shamir79} to create an $(m, r)$-threshold scheme for a uniformly random $r \in [m]$. Then $v(\cdot)$ is additive (and thus submodular) on the subsets of $[m]$, but an aggregator who only knows the experts' reports will only be able to recover the secret if $r = 1$.

\begin{proof}
	Let $p > m$ be a prime. Consider the following information structure (the \emph{secret sharing information structure}).
	\begin{itemize}
		\item An integer $r \in [m]$ is selected uniformly at random and announced.
		\item A random $(r - 1)$-th degree polynomial $P(x) = a_0 + a_1x + \dots + a_{r - 1}x^{r - 1}$ over $\FF_p$ is selected, with coefficients chosen uniformly at random from $\FF_p$, except that $a_0$ is either $-1$ or $1$ (also uniformly). For each $i \in [m]$, expert $i$ is told $P(i)$.
		\item The quantity $Y$ is equal to $1$ if $a_0 = 1$ and $-1$ if $a_0 = -1$.
	\end{itemize}
	Note that for a fixed choice of $r$ and for any $A \subseteq [m]$, we have $Y_A = 0$ if $\abs{A} < r$ and $Y_A = \pm 1$ if $\abs{A} \ge r$. Therefore, for any $A$ we have that $Y_A = \pm 1$ with probability $\frac{\abs{A}}{m}$ and $0$ otherwise. Therefore, we have that $v(Y_A) = \frac{\abs{A}}{m}$, so $v(\cdot)$ is additive (and thus submodular). Thus, this information structure satisfies weak substitutes.
	
	On the other hand, note that with probability $\frac{m - 1}{m}$, all experts report $0$ to the aggregator, in which case the aggregator cannot do better in expectation than also reporting $0$. Thus, it is impossible for the aggregator to report an estimate $Z$ with $v(Z) > \frac{1}{m}$, and so an approximation ratio larger than $\frac{1}{m}$ is not attainable.
\end{proof}

The secret sharing information structure is a lottery over $m$ different information structures, for each of which $v(\cdot)$ is a threshold function: any $r - 1$ experts know nothing ($v(A) = 0$ if $\abs{A} < r$), while any $r$ experts know everything ($v(A) = 1$ if $\abs{A} \ge r$). Except for $r = 1$, these information structures have experts that should be intuitively regarded as \emph{complementary}. Indeed, these structures generalize the XOR information structure (which is the case of $m = r = 2$). This suggests that properties of $v(\cdot)$ as a set function on $[m]$ are insufficient to capture what we intuitively mean by substitutable signals. This motivates us to seek a natural but stronger notion of informational substitutes -- one that is well-motivated and not too restrictive, but which rules out information structures such as this one and allows an aggregator to outperform the guarantee of the random expert strategy.

\subsection{Projective substitutes} \label{sec:proj_subs}
The following fact helps to motivate our stronger notion of informational substitutes.

\begin{prop} \label{prop:weak_subs_rewrite}
	The weak substitutes condition may be rewritten as: for any $i$ and $B \subseteq A$, we have
	\[\EE{(Y_A - Y_B)^2} \ge \EE{(Y_{A \cup \{i\}} - Y_{B \cup \{i\}})^2}.\]
\end{prop}

\begin{proof}
    By the Pythagorean theorem, we have that
    \[\EE{(Y_A - Y_B)^2} = \EE{(Y - Y_A)^2} - \EE{(Y - Y_B)^2},\]
    and similarly for the right-hand side. Rearranging terms gives us Equation~\ref{eq:weak_subs_quad}.
\end{proof}

Intuitively, this interpretation of substitutes says: \textbf{For any expert $i$, a set $B$ of experts becomes better at predicting the belief of a superset of experts $A$ if $i$'s signal is announced.} Here, by the \emph{belief} of a set $T$ of experts we mean the expected value of $Y$ conditioned on all experts' signals, i.e.\ $Y_T$.

This matches the intuition of substitutes as diminishing marginal returns: if signal $i$ becomes known, the ``information gap'' between $A$ and $B$ decreases.

A more general notion of substitutes would require this to hold \emph{even when $A$ is not a supserset of $B$}. That is: for all $i, A, B$, the experts in $B \cup \{i\}$ can collectively predict the belief of $A \cup \{i\}$ better than the experts in $B$ can collectively predict the belief of $A$. This captures the spirit of diminishing marginal returns in a somewhat broader context.

Let us formalize the notion of $B$'s prediction of $A$'s belief. By this we mean the expected value of $Y_A$ given the signal outcomes of the experts in $B$, i.e.\ $\EE{Y_A \mid \{\sigma_i: i \in B\}}$.

\begin{defin}
	Given an information structure $\mathcal{I} = (\Omega, \PP, \pmb{\sigma}, Y)$ for $m$ experts and subsets $A, B \subseteq [m]$, \emph{$B$'s prediction of $A$'s belief} is defined as the expected value of $Y_A$ given the signal outcomes of the experts in $B$, i.e.
	\[Y_{A \to B} := \EE{Y_A \mid \{\sigma_i: i \in B\}}.\]
\end{defin}

We now state our substitutes definition, which strengthens the weak substitutes condition.
\begin{defin} \label{def:projective_subs}
	An information structure $\mathcal{I} = (\Omega, \PP, \pmb{\sigma}, Y)$ for $m$ experts satisfies \emph{projective substitutes} if for all $A, B \subseteq [m]$ and $i \in [m]$, we have
	\begin{equation} \label{eq:proj_subs}
		\EE{(Y_A - Y_{A \to B})^2} \ge \EE{(Y_{A \cup \{i\}} - Y_{A \cup \{i\} \to B \cup \{i\}})^2}.
	\end{equation}
\end{defin}

The secret sharing information structure does not satisfy projective substitutes: take $A, B, i$ with $\abs{A} \ge \abs{B}$ (but $A \not \supseteq B$) and $i \in B \setminus A$. On the other hand, the example below does satisfy projective substitutes.

\begin{example}
	Consider the following information structure, in which the value of $Y$ is determined by the values of $\sigma_1$ and $\sigma_2$. The table on the left specifies the value of $Y$ depending on the pair of signals, and the table on the right specifies the probability of each pair of signals. (Note that the signal values $w, x, y, z$ are arbitrary labels.)

    \singlespacing
	\[\left\{Y = \begin{tabular}{c|cc}
		&$\sigma_2 = y$&$\sigma_2 = z$\\
		\hline
		$\sigma_1 = w$&0&1\\
		$\sigma_1 = x$&1&2
	\end{tabular}
	\qquad \PP =
	\begin{tabular}{c|cc}
		&$\sigma_2 = y$&$\sigma_2 = z$\\
		\hline
		$\sigma_1 = w$&0.3&0.2\\
		$\sigma_1 = x$&0.2&0.3
	\end{tabular}\right\}
	\]
	Let $A = \{1\}$ and $B = \{2\}$. It is not difficult to compute that
	\[Y_A = \begin{tabular}{c|cc}
		&$\sigma_2 = y$&$\sigma_2 = z$\\
		\hline
		$\sigma_1 = w$&0.4&0.4\\
		$\sigma_1 = x$&1.6&1.6
	\end{tabular} \qquad \qquad
	Y_{A \to B} = \begin{tabular}{c|cc}
		&$\sigma_2 = y$&$\sigma_2 = z$\\
		\hline
		$\sigma_1 = w$&0.88&1.12\\
		$\sigma_1 = x$&0.88&1.12
	\end{tabular}\]
    \doublespacing
 
	For example, in the $(\sigma_1, \sigma_2) = (w, y)$ case we have that $Y_A = \frac{0.3 \cdot 0 + 0.2 \cdot 1}{0.3 + 0.2} = 0.4$. $Y_A$ is then used to compute $Y_{A \to B}$: for example, in the $(\sigma_1, \sigma_2) = (w, y)$ case we have that $Y_{A \to B} = \frac{0.3 \cdot 0.4 + 0.2 \cdot 1.6}{0.3 + 0.2} = 0.88$, as this is the expected value of $Y_A$ conditioned on $\sigma_2 = y$.
	
	It can be computed that $\EE{(Y_A - Y_{A \to B})^2} = 0.3456$, whereas $\EE{(Y_{A \cup \{2\}} - Y_{A \cup \{2\} \to B \cup \{2\}})^2} = \EE{(Y - Y_B)^2} = 0.24$, so Equation~\ref{eq:proj_subs} is satisfied for $A = \{1\}, B = \{2\}, i = 2$. It can be verified that Equation~\ref{eq:proj_subs} is in fact satisfied for \emph{all} $A, B, i$, so this information structure satisfies projective substitutes.
\end{example}

The projective substitutes definition can be interpreted as describing the class of information structures in which full information revelation is a dominant strategy. While in general we are interested in aggregation, not elicitation, we present the following thought experiment in order to motivate the projective substitutes condition.

Consider a central party (call them the \emph{elicitor}) who knows the information structure but does not know the experts' signals. Experts are truthful, but may be strategic: they will not lie about their signal, but may decide not to reveal it. The elicitor wishes to structure incentives that will encourage each expert to reveal their signal. The elicitor puts experts on teams (but does not immediately announce the teams). Then:
\begin{enumerate}
	\item Each expert either reveals their signal to the elicitor, or does not.
	\item The elicitor announces which experts revealed their signals and announces the teams.
	\item Each team makes a prediction about the elicitor's posterior belief (after learning the signals of all experts who decided to reveal) and is scored using a quadratic scoring rule (i.e.\ penalized by the squared distance between their prediction of the elicitor's belief and the elicitor's actual belief).\footnote{Why not elicit $Y$ directly? Eliciting each team's best guess about the \emph{elicitor's} belief is particularly compelling in situations in which the true value of $Y$ will never be known, or will be learned in the far future. Under these circumstances, the elicitor's belief serves as an approximation for $Y$ given the available information.}
\end{enumerate}

This mechanism incentivizes experts to reveal their signals if and only if the information structure satisfies projective substitutes. Formally:
\begin{prop}
	An information structure satisfies projective substitutes if and only if in the above mechanism, revealing one's signal is a dominant strategy for every expert, regardless of who is on their team.
\end{prop}

\begin{proof}
	First suppose that the information structure satisfies projective substitutes. Consider any expert $i$, let $B$ be $i$'s team, and let $A$ be the set of all other experts who reveal their signals. If $i$ does not reveal their signal, then the elicitor's belief will be $Y_A$ and $B$'s prediction of the elicitor's belief will be $Y_{A \to B}$. If $i$ reveals their signal, then the elicitor's belief will be $Y_{A \cup \{i\}}$ and $B$'s prediction of the elicitor's belief will be $Y_{A \cup \{i\} \to B} = Y_{A \cup \{i\} \to B \cup \{i\}}$. Therefore, by the projective substitutes, condition, $B$'s expected prediction error is smaller if $i$ reveals their signal to the expert.
	
	Conversely, suppose that the information structure does not satisfy projective substitutes. Then there are sets $A, B \subseteq [m]$ and $i \in B$ (see Remark~\ref{rem:proj_subs_facts}~\ref{item:proj_subs_equiv}) such that
	\[\EE{(Y_A - Y_{A \to B})^2} < \EE{(Y_{A \cup \{i\}} - Y_{A \cup \{i\} \to B \cup \{i\}})^2}.\]
	Consider expert $i$, suppose their team is $B$, and suppose that the set of experts excluding $i$ who reveal their signal is $A$. Then $i$ is incentivized not to reveal their signal to the elicitor, as revealing their signal will increase $B$'s expected prediction error.
\end{proof}

\begin{remark}[Facts about projective substitutes] \label{rem:proj_subs_facts} \phantom{}
	\begin{enumerate}[label=(\roman*)]
		\item The weak substitutes condition is equivalent to Equation~\ref{eq:proj_subs} holding for all $B \subseteq A$, so the projective substitutes condition is stronger. In fact it is strictly stronger, as it excludes the secret sharing information structure.
		
		\item \label{item:proj_subs_equiv} There are several equivalent formulations of projective substitutes. One definition replaces $\{i\}$ with an arbitrary set $X$. Another modifies the definition by only requiring Equation~\ref{eq:proj_subs} to hold if $i \in B$.\footnote{To see that this is equivalent, for any $B, i$ with $i \not \in B$ define $B' := B \cup \{i\}$. Then Equation~\ref{eq:proj_subs} for $A, B', i$ has the same right-hand side but a smaller or equal left-hand side, and is thus more difficult to satisfy.}
		
		\item The notation $Y_{A \to B}$ comes from the fact that $Y_{A \to B}$ is the orthogonal projection of $Y_A$ onto the space of random variables that depend only on the signals of the experts in $B$. As a consequence of this alternative formulation, $Y_{A \to B}$ is the closest random variable (by expected squared distance) to $Y_A$ among all random variables that depend only on the values $(\sigma_i)_{i \in B}$.
	\end{enumerate}
\end{remark}

\subsection{PIF information structures}
Recall the definition of a PIF information structure from Section~\ref{sec:prelim_info_struct}:

\pifdef*

PIF information structures are natural because each $X_S$ can be thought of as an (additive) piece of evidence about $Y$ that is known by the experts in $S$. They are also quite versatile, as they capture the idea that different experts can have overlapping pieces of evidence in a variety of ways. As we are about to show, all PIF information structures satisfy the projective substitutes condition. This fact further motivates the notion of projective substitutes: it means that the class of projective substitutes information structures is broad enough to include all PIF information structures.

\begin{prop} \label{prop:pif_projective}
    Every PIF information structure satisfies projective substitutes.
\end{prop}

\begin{proof}
    Consider a PIF information structure, and without loss of generality, assume that every $X_S$ is zero-mean. Note that given $A, B \subseteq [m]$, we have
    \[Y_A = \sum_{S: S \cap A \neq \emptyset} X_S.\]
    Also,
    \[Y_{A \to B} = \sum_{S: S \cap A \neq \emptyset, S \cap B \neq \emptyset} X_S,\]
    because $X_S$ counts toward $Y_{A \to B}$ if both someone in $B$ knows $X_S$ and someone in $A$ knows $X_S$. Therefore,
    \[\EE{(Y_A - Y_{A \to B})^2} = \sum_{S: S \cap A \neq \emptyset, S \cap B = \emptyset} \EE{X_S^2}.\]
    Meanwhile,
    \[\EE{(Y_{A \cup \{i\}} - Y_{A \cup \{i\} \to B \cup \{i\}})^2} = \sum_{\substack{S: S \cap (A \cup \{i\}) \neq \emptyset,\\S \cap (B \cup \{i\}) = \emptyset}} \EE{X_S^2}.\]
    Note that if $S \cap (B \cup \{i\}) = \emptyset$, then $S \cap B = \emptyset$ and also $i \not \in S$. If, furthermore, $S \cap (A \cup \{i\}) \neq \emptyset$, it follows that $S \cap A \neq \emptyset$. Thus, any such $S$ also satisfies $S \cap A \neq \emptyset, S \cap B = \emptyset$. Therefore, we have
    \[\EE{(Y_A - Y_{A \to B})^2} \ge \EE{(Y_{A \cup \{i\}} - Y_{A \cup \{i\} \to B \cup \{i\}})^2},\]
    so the information structure indeed satisfies projective substitutes.
\end{proof}

We will use PIF information structures for two of our negative results.

\section{A strong impossibility result for forecast aggregation} \label{sec:general_lb}
In this section, we exhibit a single PIF information structure on which no aggregation strategy can achieve an approximation ratio of $\frac{4}{m}$. This result thus shows that achieving an approximation ratio of $\frac{4}{m}$ is impossible even by an aggregator who knows the entire information structure. \emph{(Thanks to Mark Xu of the Alignment Research Center for the discussion that led to this result.)}

\begin{theorem} \label{thm:tight_lb}
    Consider the PIF information structure $\mathcal{I}$ defined with $X_{\{i\}} \sim \mathcal{N}(0, 1)$, $X_{[m]} \sim \mathcal{N}(0, 1)$, and all other $X_S$'s uniformly zero. No aggregation strategy achieves an approximation ratio of more than $\frac{4m}{(m + 1)^2}$ on $\mathcal{I}$.
\end{theorem}

\begin{proof}
    For convenience, we will write $X_i$ in place of $X_{\{i\}}$. We have $Y_i = X_i + X_{[m]}$ for every $i$.
    
    Let $Z(Y_1, \dots, Y_m)$ be the output of some aggregation strategy. Let $\hat{Y} := \EE{Y \mid Y_1, \dots, Y_m}$. By the Pythagorean theorem, we have that $\EE{(Y - Z)^2} = \EE{(Y - \hat{Y})^2} + \EE{(\hat{Y} - Z)^2} \ge \EE{(Y - \hat{Y})^2}$, with equality if $Z$ is uniformly equal to $\hat{Y}$.

    We claim that $\hat{Y} = \frac{2}{m + 1} \sum_i Y_i$. To see this, observe that for all $j$, $Y_j$ is uncorrelated with $Y - \frac{2}{m + 1} \sum_i Y_i$:
    \begin{align*}
        \EE{Y_j \parens{Y - \frac{2}{m + 1} \sum_i Y_i}} &= \EE{(X_j + X_{[m]}) \parens{\frac{m - 1}{m + 1} \sum_i X_i - \frac{m - 1}{m + 1} X_{[m]}}}\\
        &= \frac{m - 1}{m + 1} \parens{\EE{X_j^2} - \EE{X_{[m]}^2}} = 0.
    \end{align*}
    Because uncorrelated, jointly multinormal vectors are independent \parencite{mkb79}, we in fact have that $Y - \frac{2}{m + 1} \sum_i Y_i$ is \emph{independent} of every $Y_j$. Therefore, the expected value of $Y - \frac{2}{m + 1} \sum_i Y_i$ conditioned on $Y_1, \dots, Y_m$ is uniformly zero, and so $\hat{Y} = \frac{2}{m + 1} \sum_i Y_i$.
    
    Thus, let $Z = \frac{2}{m + 1} \sum_i Y_i$. This is the optimal $Z$, and the approximation ratio achieved by $Z$ is
    \[1 - \frac{\EE{(Y - Z)^2}}{\EE{Y^2}} = 1 - \frac{\parens{\frac{m - 1}{m + 1}}^2(m + 1)}{m + 1} = \frac{4m}{(m + 1)^2},\]
    as desired.
\end{proof}

\section{The prior-free setting} \label{sec:prior_free}
Now that we have established that no aggregator can hope to achieve an approximation ratio of $\frac{4}{m}$, even under the projective substitutes condition and even if the aggregator knows the information structure, we ask: how close to $\frac{4}{m}$ can we get? We begin our investigation with the prior-free setting: that is, we will assume that the aggregator knows \emph{nothing} about the information structure, except that it satisfies the projective substitutes condition. In this setting, what is the largest approximation ratio that the aggregator can guarantee?

In this section we give a positive result and a negative result. The positive result is that \textbf{averaging the experts' reports attains an approximation ratio of at least $(1 + \sqrt{3}/2)/m - O(1/m^2) \approx 1.866/m$}. The negative result is that for all $m$, \textbf{no aggregation strategy attains an approximation ratio of more than $2/m - 1/m^2$}. Thus, projective substitutes enables a significant improvement over the $1/m$-approximation guarantee of the random expert strategy, but no more than by a factor of two.

\subsection{Positive result for the prior-free setting}
\begin{theorem} \label{thm:prior_free_positive}
	Let $\mathcal{I} = (\Omega, \PP, S, Y)$ be an information structure for $m$ experts that satisfies projective substitutes, and let $Z = \frac{1}{m} \sum_{i = 1}^m Y_i$. Then $Z$ attains an approximation ratio of at least
	\[\frac{2}{m} - \frac{m - 1}{2m(2m - 1 + \sqrt{3m^2 - 3m + 1})} - \frac{1}{m^2} \ge \parens{1 + \frac{\sqrt{3}}{2}} \cdot \frac{1}{m} - O \parens{\frac{1}{m^2}}.\]
\end{theorem}

In other words, knowing nothing about an information structure other than the fact that it satisfies the projective substitutes condition, one can significantly improve upon the $1/m$-approximation guarantee of choosing a random expert.

To prove this result, we use the projective substitutes condition to show that one of two things must be true: either (a) the experts' forecasts are (in expectation) fairly different from each other, or (b) the forecasts are somewhat accurate, meaning that they improve substantially upon the prior. (This is a reinterpretation of Lemma~\ref{lem:ab} below.) In case (a), averaging the experts' forecasts guarantees substantial improvement upon a random forecast; in case (b), even though averaging the forecasts does not improve substantially upon a random forecast, a random forecast already substantially outperforms the prior. (Our proof does not rely on casework, instead showing that these two cases quantitatively trade off against each other.)

\begin{proof}
	Let $Z = \frac{1}{m} \sum_{i = 1}^m Y_i$. By Claim~\ref{claim:improvement_rewrite}, showing that $Z$ achieves an approximation ratio of $\alpha$ is equivalent to showing that
	\begin{equation} \label{eq:alpha_rewrite}
		\EE{(Y_{[m]} - Z)^2} \le (1 - \alpha)\EE{(Y_{[m]} - \EE{Y})^2}.
	\end{equation}
	The first step in our proof uses the following fact: for any numbers $y_1, \dots, y_m$ and $y$, we have
	\[\parens{y - \frac{y_1 + \dots + y_m}{m}}^2 = \frac{1}{m} \sum_{i = 1}^m (y - y_i)^2 - \frac{1}{m^2} \sum_{1 \le i < j \le m} (y_i - y_j)^2.\]
	This equality follows from rearranging terms, and applying it in expectation for $y = Y_{[m]}$ and $y_i = Y_i$ gives us the following equality.
	\begin{equation} \label{eq:formula_one}
		\EE{(Y_{[m]} - Z)^2} = \frac{1}{m} \sum_{i = 1}^m \EE{(Y_{[m]} - Y_i)^2} - \frac{1}{m^2} \sum_{1 \le i < j \le m} \EE{(Y_i - Y_j)^2}.
	\end{equation}
	The left-hand side here is the same as in Equation~\ref{eq:alpha_rewrite}; meanwhile, the right-hand side has a term representing the average error of a random expert and another term representing the average expected distance between the experts' forecasts. The following lemma allows us to get a handle on this last term.
	
	\begin{lemma} \label{lem:ab}
		For all $i, j$, and for all $a, b \ge 0$ such that $b \ge \frac{(2a - 1)^2}{4a}$, we have
		\begin{equation} \label{eq:ab}
			\EE{(Y_i - Y_j)^2} \ge a \parens{\EE{(Y_{\{i,j\}} - Y_i)^2} + \EE{(Y_{\{i,j\}} - Y_j)^2}} - b \parens{\EE{(Y_i - \EE{Y})^2} + \EE{(Y_j - \EE{Y})^2}}.
		\end{equation}
	\end{lemma}

	The proof of Lemma~\ref{lem:ab} relies on the projective substitutes assumption. The lemma lets us flexibly lower bound the expected distance between $Y_i$ and $Y_j$ in terms of the average expected distance from $Y_{\{i, j\}}$ to $Y_i$ and $Y_j$. Intuitively, the projective substitutes condition guarantees such a bound because expert $i$ must be able to forecast $Y_{\{i, j\}}$ better than they can forecast $Y_j$. For now we assume the truth of Lemma~\ref{lem:ab} and return to the proof of Theorem~\ref{thm:prior_free_positive}. We note that we may rewrite
	
	\begin{equation} \label{eq:pythag_app}
		\EE{(Y_i - \EE{Y})^2} = \EE{(Y_{[m]} - \EE{Y})^2} - \EE{(Y_{[m]} - Y_i)^2},
	\end{equation}
	by the Pythagorean theorem. Additionally, we note that by weak substitutes (which follows from projective substitutes), for all $i$ we have
	\begin{equation} \label{eq:weak_subs_use}
		\sum_{j \neq i} \EE{(Y_{\{i, j\}} - Y_i)^2} \ge \EE{(Y_{[m]} - Y_i)^2}.
	\end{equation}
	To see this, consider for example $i = 1$. By weak substitutes, $\EE{(Y_{\{1, j\}} - Y_1)^2} \ge \EE{(Y_{[j]} - Y_{[j - 1]})^2}$, so the left-hand side of Equation~\ref{eq:weak_subs_use} is greater than or equal to $\sum_{j > 1} \EE{(Y_{[j]} - Y_{[j - 1]})^2}$, which (by $m - 2$ applications of the Pythagorean theorem) is equal to $\EE{(Y_{[m]} - Y_1)^2}$. Now, combining Equations~\ref{eq:formula_one}, \ref{eq:ab}, \ref{eq:pythag_app}, and \ref{eq:weak_subs_use} gives us that
	\begin{equation} \label{eq:almost_done}
		\EE{(Y_{[m]} - Z)^2} \le \frac{b(m - 1)}{m} \EE{(Y_{[m]} - \EE{Y})^2} + \parens{\frac{1}{m} - \frac{a}{m^2} - \frac{b(m - 1)}{m^2}} \sum_{i = 1}^m \EE{(Y_{[m]} - Y_i)^2}.
	\end{equation}
	for any $a, b$ satisfying Lemma~\ref{lem:ab}. Now, note that by weak substitutes we have
	\begin{equation} \label{eq:conditional_step}
		\sum_{i = 1}^m \EE{(Y_{[m]} - Y_i)^2} = m \EE{(Y_{[m]} - \EE{Y})^2} - \sum_{i = 1}^m \EE{(Y_i - \EE{Y})^2} \le (m - 1) \EE{(Y_{[m]} - \EE{Y})^2},
	\end{equation}
	where the first step uses the Pythagorean theorem and the second step follows from Proposition~\ref{prop:random_expert}. Therefore, if $\frac{1}{m} - \frac{a}{m^2} - \frac{b(m - 1)}{m^2} \ge 0$, we may write Equation~\ref{eq:almost_done} as
	\[\EE{(Y_{[m]} - Z)^2} \le \frac{m - 1}{m} \parens{1 - \frac{a - b}{m}} \EE{(Y_{[m]} - \EE{Y})^2}.\]
	To make this inequality as tight as possible, we wish to make $a - b$ as large as possible; our constraints are that $b \ge \frac{(2a - 1)^2}{4a}$ and $\frac{1}{m} - \frac{a}{m^2} - \frac{b(m - 1)}{m^2} \ge 0$. The optimal values are
	\[a = \frac{2m - 1 + \sqrt{3m^2 - 3m + 1}}{2m} \text{ and } b = \frac{(2a - 1)^2}{4a}.\]
	This gives us
	\[\alpha = 1 - \frac{m - 1}{m} \parens{1 - \frac{a - b}{m}} = \frac{2}{m} - \frac{m - 1}{2m(2m - 1 + \sqrt{3m^2 - 3m + 1})} - \frac{1}{m^2},\]
	as desired.
\end{proof}

\begin{proof}[Proof of Lemma~\ref{lem:ab}]
	Thinking of random variables as vectors (as in Section~\ref{sec:chap7_pythag_prelim}), let $T_i$ be the projection of $Y_j$ onto the space of all affine combinations of $Y_i$ and $Y_{\emptyset}$, i.e.\ $\{\beta Y_i + (1 - \beta) Y_{\emptyset}: \beta \in \RR\}$. (Recall that $Y_{\emptyset}$ is the random variable that is always equal to $\EE{Y}$.) Define $T_j$ analogously. Note that $\EE{(Y_j - Y_{j \to i})^2} \le \EE{(Y_j - T_j)^2}$, since $Y_{j \to i}$ is the closest point to $Y_j$ of the subspace of random variables that depend only on $\sigma_i$, and the aforementioned affine space is a subset of that subspace. Additionally, by projective substitutes (with $A = \{i\}, B = \{j\}, i = i$ in Definition~\ref{def:projective_subs}), we have that $\EE{(Y_{\{i, j\}} - Y_i)^2} \le \EE{(Y_j - Y_{j \to i})^2}$. Therefore, we have that $\EE{(Y_{\{i, j\}} - Y_i)^2} \le \EE{(Y_j - T_j)^2}$, and similarly that $\EE{(Y_{\{i, j\}} - Y_j)^2} \le \EE{(Y_i - T_i)^2}$. It therefore suffices to show that
	\begin{equation} \label{eq:wish_to_show_1}
		\EE{(Y_i - Y_j)^2} \ge a \parens{\EE{(Y_i - T_i)^2} + \EE{(Y_j - T_j)^2}} - b \parens{\EE{(Y_i - Y_{\emptyset})^2} + \EE{(Y_j - Y_{\emptyset})^2}}.
	\end{equation}
	
	By the Pythagorean theorem,\footnote{While in most cases by ``Pythagorean theorem'' we mean Proposition~\ref{prop:pythag_squared}, in this case we are referring to the fact that for orthogonal vectors $\vect{x}$ and $\vect{y}$, we have $\norm{\vect{x}}^2 + \norm{\vect{y}}^2 = \norm{\vect{x} + \vect{y}}^2$ (and applying this fact to e.g.\ $\vect{x} = Y_i - T_i$, $\vect{y} = Y_j - T_i$).} we know the following four facts.
	\begin{align*}
		\EE{(Y_i - Y_j)^2} &= \EE{(Y_i - T_i)^2} + \EE{(Y_j - T_i)^2}; & \EE{(Y_i - Y_j)^2} &= \EE{(Y_j - T_j)^2} + \EE{(Y_i - T_j)^2}\\
		\EE{(Y_i - Y_{\emptyset})^2} &= \EE{(Y_i - T_i)^2} + \EE{(T_i - Y_{\emptyset})^2}; & \EE{(Y_j - Y_{\emptyset})^2} &= \EE{(Y_j - T_j)^2} + \EE{(T_j - Y_{\emptyset})^2}
	\end{align*}
	These let us rewrite Equation~\ref{eq:wish_to_show_1} as such:
	\begin{align} \label{eq:wish_to_show_2}
		&\frac{1}{2} \parens{\EE{(Y_j - T_i)^2} + \EE{(Y_i - T_j)^2}} + \parens{a - \frac{1}{2}} \parens{\EE{(T_i - Y_{\emptyset})^2} + \EE{(T_j - Y_{\emptyset})^2}} \nonumber\\
		&\ge \parens{a - b - \frac{1}{2}} \parens{\EE{(Y_i - Y_{\emptyset})^2} + \EE{(Y_j - Y_{\emptyset})^2}}.
	\end{align}
	We wish to show that this inequality holds so long as $b \ge \frac{(2a - 1)^2}{4a}$. To do so, we note the following fact: for any random variables $Q, R$ and non-negative reals $c_1, c_2$, we have that
	\[c_1 \EE{Q^2} + c_2 \EE{R^2} \ge \frac{c_1 c_2}{c_1 + c_2} \EE{(Q + R)^2}.\]
	This follows (after multiplying through by $c_1 + c_2$ and cancelling terms) from the fact that for all $q, r$ we have $(c_1 q)^2 + (c_2 r)^2 \ge 2 (c_1 q)(c_2 r)$. Now, we apply this identity to $Q := Y_j - T_i$ and $R := T_i - Y_{\emptyset}$, with $c_1 = \frac{1}{2}$ and $c_2 = a - \frac{1}{2}$, and also to $Q := Y_i - T_j$ and $R := T_j - Y_{\emptyset}$. This tells us that
	\begin{align*}
		&\frac{1}{2} \parens{\EE{(Y_j - T_i)^2} + \EE{(Y_i - T_j)^2}} + \parens{a - \frac{1}{2}} \parens{\EE{(T_i - Y_{\emptyset})^2} + \EE{(T_j - Y_{\emptyset})^2}}\\
		&\ge \frac{2a - 1}{4a} \parens{\EE{(Y_i - Y_{\emptyset})^2} + \EE{(Y_j - Y_{\emptyset})^2}}.
	\end{align*}
	Therefore, Equation~\ref{eq:wish_to_show_2} holds so long as $\frac{2a - 1}{4a} \ge a - b - \frac{1}{2}$, which is equivalent to $b \ge \frac{(2a - 1)^2}{4a}$.
\end{proof}

\subsection{Negative results for the prior-free setting}
\begin{theorem} \label{thm:prior_free_negative}
	Fix any $m \ge 1$. For $\mu \in \RR$, let $\mathcal{I}_{\mu}$ be the PIF information structure defined with $X_{\{i\}} \sim \mathcal{N}(0, 1)$, $X_{[m]} = m \mu$ deterministically, and all other $X_S$'s uniformly zero. No aggregation strategy achieves an approximation ratio of more than $\frac{2}{m} - \frac{1}{m^2}$ on every $\mathcal{I}_\mu$.
\end{theorem}

Note that by Proposition~\ref{prop:pif_projective}, $\mathcal{I}_{\mu}$ satisfies the projective substitutes condition, and so Theorem~\ref{thm:prior_free_negative} serves as a negative result for the prior-free setting. The result would still apply if we replaced the projective substitutes assumption with any assumption that permits the class of all PIF information structures (or indeed, any assumption that permits the set of all information structures $\mathcal{I}_{\mu}$).

\begin{proof}
    For simplicity, we write $X_i$ in place of $X_{\{i\}}$. We prove the theorem in two steps. First, we show that taking the average of the experts' reports yields an approximation ratio of exactly $\frac{2}{m} - \frac{1}{m^2}$ for all $\mu$. Second, we show that no aggregation strategy beats taking the average for every $\mu$.
	
	For the first step, assume without loss of generality that $\mu = 0$. Then $Y_{[m]} = \sum_i X_i$, and the average of the $Y_i$'s (which we will denote $Z$) is equal to $\frac{1}{m} \sum_i X_i$. Therefore we have
	\[\EE{(Y_{[m]} - Z)^2} = \EE{\parens{\frac{m - 1}{m} \sum_i X_i}^2} = \parens{\frac{m - 1}{m}}^2 \EE{\parens{\sum_i X_i}^2}.\]
	On the other hand, we have that $\EE{(Y_{[m]} - \EE{Y})^2} = \EE{(\sum_i X_i)^2}$, so (using Claim~\ref{claim:improvement_rewrite}) we have that the approximation ratio is
	\[1 - \parens{\frac{m - 1}{m}}^2 = \frac{2}{m} - \frac{1}{m^2}.\]
	This completes the first step. To complete the second step, we use the following well-known result from statistical theory.\footnote{This fact does not generalize to more than two dimensions, meaning that if the $x_i$ are vectors in three or more dimensions drawn independently from a normal distribution with unknown mean and known covariance matrix, then there is an estimator for the mean that Pareto dominates the sample mean according to expected squared vector distance. One such estimator is the \emph{James-Stein estimator} \parencite{stein56}.}
	
	\begin{prop}[\cite{blyth51, gs51, hl51}] \label{prop:admissible}
		Let $x_1, \dots, x_m$ be drawn independently from a normal distribution with unknown mean $\mu$ and standard deviation $1$. Let $\hat{\mu} := \frac{1}{m} \sum_{i = 1}^m x_i$. Then for every function $Z$ of $x_1, \dots, x_m$, there exists $\mu$ such that $\EE{(Z - \mu)^2} \ge \EE{(\hat{\mu} - \mu)^2}$.
	\end{prop}
	
	Since $\EE{(\hat{\mu} - \mu)^2} = \frac{1}{m}$ for every $\mu$, we have the following fact as a corollary.
	
	\begin{corollary} \label{cor:mean_estimator}
		Let $x_1, \dots, x_m$ be drawn independently from a normal distribution with unknown mean $\mu$ and standard deviation $1$. Then for every aggregation strategy $Z$ that takes as input $x_1, \dots, x_m$, we have $\max_\mu \EE{(Z - \mu)^2} \ge \frac{1}{m}$.
	\end{corollary}
	
	(The only subtlety is that aggregation strategies are not required to be deterministic; however, replacing a randomized aggregation strategy with the deterministic strategy that outputs the expected value of the randomized strategy given the inputs can only reduce expected squared error.)\\
	
	Returning to our proof, observe that each $Y_i$ is equal to $m\mu + X_i$, which is an independent draw from the normally distribution with mean $m\mu$ and standard deviation $1$. Let $Z$ be any aggregation strategy on inputs $Y_1, \dots, Y_m$. We define a new aggregation strategy: $\tilde{Z} := \frac{1}{m - 1}(\sum_i Y_i - Z)$. We claim that if $Z$ achieves an approximation ratio of more than $\frac{2}{m} - \frac{1}{m^2}$ for every $\mu$, then $\tilde{Z}$ violates Corollary~\ref{cor:mean_estimator}. Consider $\tilde{Z}$ as an estimator for $m\mu$. We have
	
	\begin{align*}
		\EE{(m\mu - \tilde{Z})^2} &= \EE{\parens{m\mu - \frac{1}{m - 1}\parens{\sum_i Y_i - Z}}^2}\\
		&= \frac{1}{(m - 1)^2} \EE{\parens{\sum_i Y_i - m(m - 1)\mu - Z}^2} = \frac{1}{(m - 1)^2} \EE{(Y - Z)^2},
	\end{align*}
	where in the last step we use the fact that $Y = m\mu + \sum_i X_i = \sum_i Y_i - m(m - 1)\mu$. Now, suppose for contradiction that $Z$ achieves an approximation ratio of more than $\frac{2}{m} - \frac{1}{m^2}$ on every $\mu$. Then for all $\mu$ we have
	\[\parens{\frac{m - 1}{m}}^2 = 1 - \parens{\frac{2}{m} - \frac{1}{m^2}} > \frac{\EE{(Y - Z)^2}}{\EE{(Y - \EE{Y})^2}} = \frac{\EE{(Y - Z)^2}}{m} = \frac{(m - 1)^2}{m} \EE{(m\mu - \tilde{Z})^2},\]
	so $\EE{(m\mu - \tilde{Z})^2} < \frac{1}{m}$ for every value of $m\mu$ (and thus for every $\mu$). This contradicts Corollary~\ref{cor:mean_estimator} and completes the proof.
\end{proof}

Theorems~\ref{thm:prior_free_positive} and \ref{thm:prior_free_negative} give us non-matching lower and upper bounds on the optimal approximation ratio under the projective substitutes condition. In particular, for $m = 2$ experts, Theorem~\ref{thm:prior_free_positive} tells us that averaging achieves an approximation ratio of $\frac{3 + \sqrt{7}}{8} \approx 0.706$, while Theorem~\ref{thm:prior_free_positive} tells us that no aggregation strategy can achieve an approximation ratio larger than $0.75$. We now show that for two experts, our positive result is tight.

\begin{theorem} \label{thm:prior_free_negative_n2}
	In the prior-free setting, no aggregation strategy achieves an approximation ratio larger than $\frac{3 + \sqrt{7}}{8}$ on every two-expert information structure that satisfies projective substitutes.
\end{theorem}

\begin{proof}
	Let $\mathcal{I}_+$ be the following information structure, where $p = 1 - \frac{\sqrt{7}}{4}$ and $x = \sqrt{14} - 2\sqrt{2}$. We label the signals $-1$ and $1$ because these are the expected values conditional on the respective signals.

    \singlespacing
	\[\mathcal{I}_+ := \left\{Y = \begin{tabular}{c|cc}
		&$\sigma_2 = 1$&$\sigma_2 = -1$\\
		\hline
		$\sigma_1 = 1$&$\frac{1 - (1 - 2p)x}{2p}$&$x$\\
		$\sigma_1 = -1$&$x$&$\frac{-1 - (1 - 2p)x}{2p}$
	\end{tabular}
	\qquad \PP = \begin{tabular}{c|cc}
		&$\sigma_2 = 1$&$\sigma_2 = -1$\\
		\hline
		$\sigma_1 = 1$&$p$&$\frac{1}{2} - p$\\
		$\sigma_1 = -1$&$\frac{1}{2} - p$&$p$
	\end{tabular}\right\}
	\]
    \doublespacing
 
	Let $\mathcal{I}_-$ be the same information structure, but with $x = 2\sqrt{2} - \sqrt{14}$. It is a matter of calculation to verify that these information structures satisfy projective substitutes.\footnote{These information structures were found by finding values that would make the inequalities in the proofs of Theorem~\ref{thm:prior_free_positive} and Lemma~\ref{lem:ab} hold with equality.}
	
	Note that any aggregation strategy that outputs a number other than $1$ on input $(Y_1, Y_2) = (1, 1)$ has an approximation ratio of negative infinity on an information structure where $Y = 1$ deterministically. This is likewise true for $-1$ in place of $1$. Thus, if Theorem~\ref{thm:prior_free_negative_n2} were false, it would be disproved by an information structure that outputs $1$ on $\mathcal{I}_+$ and $\mathcal{I}_-$ if $(\sigma_1, \sigma_2) = (1, 1)$ and $-1$ if $(\sigma_1, \sigma_2) = (-1, -1)$. Conditional on this, the aggregation strategy that minimizes the maximum expected squared distance to $Y_{[m]}$ on $\mathcal{I}_+$ and $\mathcal{I}_-$ is the one that returns $0$ when $(\sigma_1, \sigma_2) = (1, -1)$ or $(\sigma_1, \sigma_2) = (-1, 1)$. It is a matter of calculation to verify that this aggregation strategy achieves an approximation ratio of exactly $\frac{3 + \sqrt{7}}{8}$.
\end{proof}

\section{The known prior setting} \label{sec:known_prior}
Let us now expand the information available to the aggregator by allowing them knowledge of the prior $\EE{Y}$. How might this change the optimal aggregation strategy?

In Section~\ref{sec:prelim_agg}, we saw that it often makes sense to extremize an aggregate forecast -- that is, to push it away from the prior. We gave the following information structure as a motivating example: a coin comes up heads $Y$ fraction of the time, where $Y$ is selected uniformly from $[0, 1]$; each of two experts sees an independent flip of the coin. It can be calculated that an expert who sees heads has a posterior of $\frac{2}{3}$. However, consider the situation in which both experts report heads: collectively they have seen two heads and zero tails, conditional on which the expected value of $Y$ is $\frac{3}{4}$, rather than $\frac{2}{3}$.

More generally, extremization is useful when experts have private information. For example, if many experts update upward from the prior as a result of each of their pieces of evidence, then it stands to reason that observing \emph{all of the evidence} would cause an update that is larger than the average of the experts' individual updates.

\subsection{The extremization factor}
Consider the following aggregation strategy, parameterized by a constant $d$ which we will call the \emph{extremization factor}.
\begin{equation} \label{eq:ext_factor}
	Z := \frac{1}{m} \sum_i Y_i + (d - 1) \parens{\frac{1}{m} \sum_i Y_i - \EE{Y}}.
\end{equation}
Setting $d = 1$ recovers the average of the reports; setting $d = 0$ simply returns the prior. In general, setting $d > 1$ extremizes the average (i.e.\ pushes it away from the prior) by a factor of $d$. As an example, consider the class of information structures in Theorem~\ref{thm:prior_free_negative}, where averaging achieved an approximation ratio of $\frac{2}{m} - \frac{1}{m^2}$. On the other hand, extremizing by a factor of $m$ (i.e.\ setting $d = m$ above) recovers $Y$ exactly (thus achieving an approximation ratio of $1$). This approach, which is a special case of generalized linear pooling (introduced in Section~\ref{sec:generalized_pool}), is known as \emph{linear extremization} \parencite{su15}.

We now prove that by extremizing, we can achieve an approximation ratio that is higher than what we could hope to attain without knowledge of the prior. In particular, we find that \textbf{by applying an appropriate amount of linear extremization, it is possible to achieve an approximation ratio of at least $\frac{3\sqrt{3}}{2m} - O(1/m^2) \approx 2.598/m$.} This is a substantial improvement not only over our positive result in the prior-free setting, but also over our negative result in that setting.

\subsection{Positive result for the known prior setting}
\begin{theorem} \label{thm:known_prior_positive}
	Let $\mathcal{I} = (\Omega, \PP, S, Y)$ be an information structure for $m$ experts that satisfies projective substitutes, and let $Z = \frac{1}{m} \sum_{i = 1}^m Y_i + (d - 1) \parens{\frac{1}{m} \sum_{i = 1}^n Y_i - \EE{Y}}$, where $d = \frac{m(\sqrt{3m^2 - 3m + 1} - 2)}{m^2 - m - 1}$. Then $Z$ attains an approximation ratio of at least
	\[\frac{(3m^2 - 3m + 1)^{3/2} - 9m^2 + 9m + 1}{2(m^2 - m - 1)^2} \ge \frac{3\sqrt{3}}{2m} - O \parens{\frac{1}{m^2}}.\]
\end{theorem}

\begin{figure}[ht]
	\centering
	\includegraphics[scale=0.85]{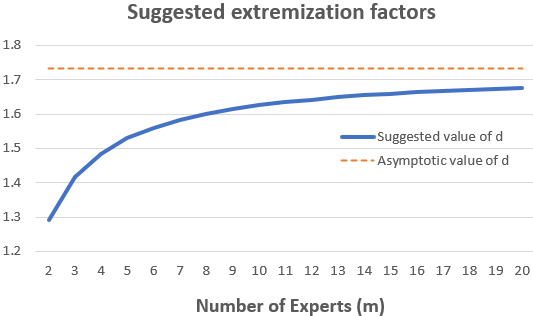}
	\caption[Suggested extremization factor as a function of the number of experts]{For $m = 2 \dots 20$, the value $d = \frac{m(\sqrt{3m^2 - 3m + 1} - 2)}{m^2 - m - 1}$ suggested by Theorem~\ref{thm:known_prior_positive}, together with the asymptotic (with $m$) value $\sqrt{3}$.}
	\label{fig:d}
\end{figure}

In Figure~\ref{fig:d}, we plot the values of $d$ suggested by Theorem~\ref{thm:known_prior_positive}. While $d$ increases with $m$, it reaches a limit -- namely, $\sqrt{3} \approx 1.732$. By contrast, the optimal response to the information structures in our negative result for the prior-free setting (Theorem~\ref{thm:prior_free_negative}) was to extremize by a factor of $m$ (i.e.\ to add up the experts' updates from the prior). This is a consequence of the fact that the signals received by each expert were independent. By contrast, Theorem~\ref{thm:known_prior_positive} suggests a smaller amount of extremization, because it is concerned with the optimal strategy in the worst case over information structures. This means that it must compromise between doing well in settings with independent signals (where a large extremization factor makes sense) and settings in which experts' signals are highly dependent (where little or no extremization is optimal). The extremization factor suggested by Theorem~\ref{thm:known_prior_positive} is also consistent with the Bayesian modeling results in \textcite{satopaa22} and the empirical findings in \textcite{su15}.

The proof of Theorem~\ref{thm:known_prior_positive} is similar to the proof of Theorem~\ref{thm:prior_free_positive}, though with an additional degree of freedom ($d$) to optimize over.

\begin{proof}[Proof of Theorem~\ref{thm:known_prior_positive}]
Let $d > 0$ and let $Z := \frac{1}{m} \sum_i Y_i + (d - 1) \parens{\frac{1}{m} \sum_i Y_i - \EE{Y}}$. As with the proof of Theorem~\ref{thm:prior_free_positive}, we start by upper bounding $\EE{(Y_{[m]} - Z)^2}$. We have
\begin{align*}
	\EE{(Y_{[m]} - Z)^2} &= \EE{\parens{d \parens{Y_{[m]} - \frac{1}{m} \sum_i Y_i} - (d - 1) \parens{Y_{[m]} - \EE{Y}}}^2}\\
	&= d^2 \EE{\parens{Y_{[m]} - \frac{1}{m} \sum_i Y_i}^2} + (d - 1)^2 \EE{(Y_{[m]} - \EE{Y})^2}\\
	&\qquad - \frac{2d(d - 1)}{m} \EE{\parens{m Y_{[m]} - \sum_i Y_i}\parens{Y_{[m]} - \EE{Y}}}\\
	&= d^2 \parens{\frac{1}{m} \sum_i \EE{(Y_{[m]} - Y_i)^2} - \frac{1}{m^2} \sum_{1 \le i < j \le m} \EE{(Y_i - Y_j)^2}}\\
	&\qquad + (d - 1)^2 \EE{(Y_{[m]} - \EE{Y})^2} - \frac{2d(d - 1)}{m} \sum_i \EE{(Y_{[m]} - Y_i)^2}.
\end{align*}
In the last step, we adapt the first term using Equation~\ref{eq:formula_one} and adapt the last term by observing that
\begin{align*}
	\EE{(Y_{[m]} - Y_i)(Y_{[m]} - \EE{Y})} &= \EE{Y_{[m]}(Y_{[m]} - Y_i)} - \EE{Y}\EE{Y_{[m]} - Y_i}\\
	&= \EE{Y_{[m]}(Y_{[m]} - Y_i)} = \EE{(Y_{[m]} - Y_i)^2}
\end{align*}
(where the last step holds because for any given $Y_i$, $\EE{Y_{[m]} \mid Y_i} = Y_i$, so $\EE{Y_i(Y_{[m]} - Y_i)} = 0$). Grouping like terms, we have
\[\EE{(Y_{[m]} - Z)^2} = (d - 1)^2 \EE{(Y_{[m]} - \EE{Y})^2} - \frac{d(d - 2)}{m} \sum_i \EE{(Y_{[m]} - Y_i)^2} - \frac{d^2}{m^2} \sum_{1 \le i < j \le m} \EE{(Y_i - Y_j)^2}.\]
Now, recall Lemma~\ref{lem:ab}. Consider any $a, b \ge 0$ satisfying $b \ge \frac{(2a - 1)^2}{4a}$; then for all $i, j$ we have
\[\EE{(Y_i - Y_j)^2} \ge a \parens{\EE{(Y_{\{i,j\}} - Y_i)^2} + \EE{(Y_{\{i,j\}} - Y_j)^2}} - b \parens{\EE{(Y_i - \EE{Y})^2} + \EE{(Y_j - \EE{Y})^2}}.\]
Therefore we have
\begin{align*}
	&\EE{(Y_{[m]} - Z)^2} \le (d - 1)^2 \EE{(Y_{[m]} - \EE{Y})^2} - \frac{d(d - 2)}{m} \sum_i \EE{(Y_{[m]} - Y_i)^2}\\
	&\qquad - \frac{d^2}{m^2} \sum_{1 \le i < j \le m} \parens{a \parens{\EE{(Y_{\{i,j\}} - Y_i)^2} + \EE{(Y_{\{i,j\}} - Y_j)^2}} - b \parens{\EE{(Y_i - \EE{Y})^2} + \EE{(Y_j - \EE{Y})^2}}}\\
	&= \parens{(d - 1)^2 + \frac{bd^2(m - 1)}{m}} \EE{(Y_{[m]} - \EE{Y})^2} - \parens{\frac{d(d - 2)}{m} + \frac{bd^2(m - 1)}{m^2}} \sum_i \EE{(Y_{[m]} - Y_i)^2}\\
	&\qquad - \frac{ad^2}{m^2} \sum_{1 \le i < j \le m} \parens{\EE{(Y_{\{i,j\}} - Y_i)^2} + \EE{(Y_{\{i,j\}} - Y_j)^2}},
\end{align*}
where in the last step we use the Pythagorean theorem to write $\EE{(Y_i - \EE{Y})^2}$ as $\EE{(Y - \EE{Y})^2} - \EE{(Y - Y_i)^2}$. Now we use Equation~\ref{eq:weak_subs_use}:
\begin{align*}
	\EE{(Y_{[m]} - Z)^2} &\le \parens{(d - 1)^2 + \frac{bd^2(m - 1)}{m}} \EE{(Y_{[m]} - \EE{Y})^2}\\
	&\qquad - \parens{\frac{d(d - 2)}{m} + \frac{bd^2(m - 1)}{m^2} + \frac{ad^2}{m^2}} \sum_i \EE{(Y_{[m]} - Y_i)^2}.
\end{align*}
Now, supposing that $\frac{d(d - 2)}{m} + \frac{bd^2(m - 1)}{m^2} + \frac{ad^2}{m^2}$ is not positive, we may use Equation~\ref{eq:conditional_step} to obtain:
\begin{align*}
	\EE{(Y_{[m]} - Z)^2} &\le \parens{(d - 1)^2 + \frac{bd^2(m - 1)}{m} - \frac{m - 1}{m} \parens{d(d - 2) + \frac{bd^2(m - 1)}{m} + \frac{ad^2}{m}}} \EE{(Y_{[m]} - \EE{Y})^2}\\
	&= \parens{1 + \frac{d^2 - 2d}{m} - (a - b) \frac{m - 1}{m^2} d^2} \EE{(Y_{[m]} - \EE{Y})^2}.
\end{align*}
With $d$ held fixed, our goal is to maximize $a - b$, just as in the proof of Theorem~\ref{thm:prior_free_positive}. This time, our constraints are $b \ge \frac{(2a - 1)^2}{4a}$ (as before) and $\frac{d(d - 2)}{m} + \frac{bd^2(m - 1)}{m^2} + \frac{ad^2}{m^2} \le 0$, which can be rewritten as $a + b(m - 1) \le \frac{2 - d}{d} m$. The optimal values are
\[a = \frac{\frac{2}{d}m - 1 + \sqrt{\parens{\frac{2}{d}m - 1}^2 - m(m - 1)}}{2m} \text{ and } b = \frac{(2a - 1)^2}{4a}.\]
Now, let $a$ and $b$ be as above. We may select $d$ as we please and seek to minimize the expression
\[1 + \frac{d^2 - 2d}{m} - (a - b) \frac{m - 1}{m^2} d^2.\]
We choose the value of $d$ in the theorem statement (which one can verify is optimal using a computer algebra system). This yields the desired approximation ratio.
\end{proof}

\subsection{Negative result for the known prior setting for $m = 2$}
We already have a negative result for the known prior setting: namely, Theorem~\ref{thm:tight_lb}. In the special case of $m = 2$, Theorem~\ref{thm:tight_lb} shows that an approximation ratio larger than $\frac{8}{9} \approx 0.889$ cannot be achieved. By contrast, Theorem~\ref{thm:known_prior_positive} tells us that averaging and extremizing by a factor of $2(\sqrt{7} - 2) \approx 1.292$ achieves an approximation ratio of $\frac{7\sqrt{7} - 17}{2} \approx 0.760$. We prove that this positive result is in fact tight.

\begin{theorem} \label{thm:known_prior_negative_n2}
	In the known prior setting, no aggregation strategy achieves an approximation ratio larger than $\frac{7\sqrt{7} - 17}{2}$ on every two-expert information structure that satisfies projective substitutes.
\end{theorem}

\begin{proof}
Let $\mathcal{I}_+$ be the following information structure, where $p = \frac{2 + \sqrt{7}}{12}$ and $x = \frac{\sqrt{2 + \sqrt{7}}}{3}$. We label the signals $-1$ and $1$ because these are the expected values conditional on the respective signals.

\singlespacing
\[\mathcal{I}_+ := \left\{Y = \begin{tabular}{c|cc}
	&$\sigma_2 = 1$&$\sigma_2 = -1$\\
	\hline
	$\sigma_1 = 1$&$\frac{1 - (1 - 2p)x}{2p}$&$x$\\
	$\sigma_1 = -1$&$x$&$\frac{-1 - (1 - 2p)x}{2p}$
\end{tabular}
\qquad \PP = \begin{tabular}{c|cc}
	&$\sigma_2 = 1$&$\sigma_2 = -1$\\
	\hline
	$\sigma_1 = 1$&$p$&$\frac{1}{2} - p$\\
	$\sigma_1 = -1$&$\frac{1}{2} - p$&$p$
\end{tabular}\right\}
\]
\doublespacing

Let $\mathcal{I}_-$ be the same information structure, but with $x = -\frac{\sqrt{2 + \sqrt{7}}}{3}$. It is a matter of calculation to verify that these information structures satisfy projective substitutes. The quantity $\EE{(Y - \EE{Y})^2}$ is the same for $\mathcal{I}_+$ and $\mathcal{I}_-$, so the aggregation strategy $Z$ that guarantees the largest possible approximation ratio when the information structure is one of $\mathcal{I}_+$ and $\mathcal{I}_-$ is the one that minimizes the maximum value of $\EE{(Y - Z)^2}$ over these two information structures. This is achieved by outputting $0$ when $(Y_1, Y_2)$ is $(1, -1)$ or $(-1, 1)$, $\frac{1}{2p}$ when $(Y_1, Y_2) = (1, 1)$, and $\frac{-1}{2p}$ when $(Y_1, Y_2) = (-1, -1)$. It is a matter of calculation to verify that this aggregation strategy achieves an approximation ratio of exactly $\frac{7\sqrt{7} - 17}{2}$.
\end{proof}

\section{Future directions in robust aggregation} \label{sec:chap7_future_work}
Robust forecast aggregation is a new area. To our knowledge, this work is the first to demonstrate nontrivial guarantees about forecast aggregation under a truly broad class of information structures. But the space of questions to be asked about robust forecast aggregation is much broader than the space of questions that we have considered.

To highlight one example: we found that averaging followed by linear extremization is a robust aggregation method, if the aggregator's error is their squared distance to the truth. However, this aggregation method does not in general make sense for probabilistic forecasts: in particular, aggregating probabilities in this way may result in aggregates outside of $[0, 1]$. A natural question to ask, then, is: what is a robust and sensible way to aggregate \emph{probabilistic} forecasts?

If we change our error measure from squared distance to KL divergence, then answers outside of $[0, 1]$ become unacceptable, as they accrue infinite error for the aggregator. And so we can ask: what aggregation methods achieve a high worst-case approximation ratio, if we use KL divergence as our error measure? A natural first guess might be generalized logarithmic pooling, as introduced in Section~\ref{sec:generalized_pool}. Is this aggregation method in fact robust?\\

To step back from this particular suggestion for future work, let us consider the question that we answered in this chapter. The question was: what \emph{approximation ratio} can be achieved by an aggregator who learns \emph{expected value estimates} of a \emph{real-valued} quantity $Y$ from $m$ \emph{truthful} experts whose signals are drawn from an information structure that satisfies \emph{projective substitutes,} if the aggregator's loss is their \emph{squared error} and the aggregator \emph{knows nothing about the information structure or only knows the prior?}\footnote{We also showed a negative result in the setting where the aggregator knows the entire information structure.}

All of the emphasized phrases in this question can be varied! For example:
\begin{itemize}
    \item What if the aggregator is judged based on their KL divergence from the true value of $Y$, or a different Bregman divergence? What if the aggregator is judged based on an altogether different distance function (perhaps absolute error), such that the aggregator no longer wants to guess about the expectation of $Y$, but a different property (perhaps as the median)?
    \item What if we make a different set of assumptions about the information structure? We could restrict attention to PIF information structures. Or we could use a different notion of informational substitutes. Or -- following the lead of \textcite{abs18} -- we could assume that the experts' signals are independent conditioned on the value of $Y$, or that they are Blackwell ordered.
    \item What if the aggregator knows more about the information structure, such as the covariance matrix of the experts' estimates, or perhaps the entire joint probability distribution of their estimates?\footnote{While our $\frac{4}{m}$ negative result applies in our setting regardless of how much the aggregator knows, in other settings there may be strong approximation guarantees if the expert knows more information than just the prior.}
    \item What if we choose a different benchmark? The approximation ratio as we have defined it seems like a natural choice, but there may be others as well.
    \item What if the aggregator learns information other than the experts' expected values? Perhaps the aggregator instead learns the median of each expert's probability distribution over $Y$. Or, if the aggregator has detailed knowledge of the information structure, then it may make sense to ask what happens if the aggregator learns partial information about the experts' \emph{signals.}
    \item What if the experts are strategic? Perhaps they want to influence the aggregate forecast. What if the experts are biased in some way?
    \item What if $Y$ is vector-valued, rather than real-valued?
\end{itemize}

Many of these settings (and the vast majority of \emph{combinations} of these settings) have yet to be considered. The field of robust forecast aggregation has great potential to grow, and to produce results that are both theoretically interesting and practically useful.
\chapter{When does agreement imply accuracy?} \label{chap:agreement}

\emph{This chapter presents ``Agreement Implies Accuracy for Substitutable Signals'' \parencite{fnw23}. It assumes background on Bregman divergence (Section~\ref{sec:prelim_bregman}) and information structures (Section~\ref{sec:prelim_info_struct}).}\\

\emph{Summary:} Inspired by Aumann's agreement theorem, \textcite{aar05} studied the amount of communication necessary for two Bayesian experts to approximately agree on the expectation of a random variable.
Aaronson showed that, remarkably, the number of bits does not depend on the amount of information available to each expert.
However, in general the agreed-upon estimate may be inaccurate: far from the estimate they would settle on if they were to share all of their information.
We show that if the experts' signals satisfy a particular notion of informational substitutes, then it is the case that if the experts are close to agreement then they are close to the truth.
We prove this result for a broad class of agreement and accuracy measures that includes squared distance and KL divergence.
Additionally, we show that although these measures capture fundamentally different kinds of agreement, Aaronson's agreement result generalizes to them as well.

\section{Introduction} \label{sec:chap8_intro}
Suppose that Alice and Bob are honest, rational Bayesians who wish to estimate some quantity -- say, the unemployment rate one year from now.
Alice is an expert on historical macroeconomic trends, while Bob is an expert on contemporary monetary policy.
They convene to discuss and share their knowledge with each other until they reach an agreement about the expected value of the future unemployment rate.
Alice and Bob could reach agreement by sharing everything they had ever learned, at which point they would have the same information,
but the process would take years.
How, then, should they proceed?

In the seminal work ``Agreeing to Disagree,'' \textcite{aumann76} observed that Alice and Bob can reach agreement simply by taking turns sharing their current expected value for the quantity.
In addition to modeling communication between Bayesian agents, protocols similar to this one model financial markets: each trader shares partial information about their expected value on their turn (discussed in Section~\ref{sec:markets}). 
A remarkable result by \textcite{aar05} shows that if Alice and Bob follow certain protocols of this form, they will agree to within $\epsilon$ with probability $1-\delta$ by communicating $O \parens{\frac{1}{\delta \epsilon^2}}$ bits.\footnote{To ensure that each message is short, Alice and Bob share discretized versions of their estimates; we discuss this in Section~\ref{sec:chap8_prelims}.}
Notably, this bound only depends on the error Alice and Bob are willing to tolerate, and not on the amount of information available to them.

Absent from Aaronson's results, however, is \emph{what estimate Alice and Bob end up agreeing on}.
In particular, there is no guarantee that Alice and Bob will be \emph{accurate}, meaning their agreed-upon estimate will be close (in e.g.\ expected squared distance) to what they would believe if they shared all of their information.
In fact, they might agree on an estimate that is highly inaccurate:
suppose that Alice and Bob have independent, uniformly random bits $b_A,b_B$, and wish to estimate their XOR $b_A \oplus b_B$.\footnote{In Section~\ref{sec:prelim_subs} we introduced this example and called it the ``XOR information structure.''}
Alice and Bob agree from the onset, as from each of their perspectives, the expected value of $b_A \oplus b_B$ is $\frac{1}{2}$.
Yet this expectation is far from the best estimate given their collective knowledge, which is either $0$ or $1$.
So while agreement is fundamental to understanding communication between Bayesians -- in Aumann's terms, they cannot ``agree to disagree'' -- agreement is far from the whole story.
An important open problem, therefore, is what assumptions guarantee that Alice and Bob are accurate once they agree.

We address this open problem by introducing a natural condition, called \emph{rectangle substitutes}, under which agreement implies accuracy.
Rectangle substitutes is a notion of \emph{informational substitutes} (discussed in Section~\ref{sec:prelim_subs}): the property that additional information has diminishing marginal returns.
The notion of substitutes is ubiquitous in optimization problems, and informational substitutes conditions have recently been used to analyze equilibria in markets \parencite{cw16}.
In that context, \textcite{kong2023information} showed for conditionally independent signals convergence of the popular LMSR market implies full information aggregation, i.e.\ accuracy.
We show that under the rectangle substitutes condition, \emph{any} protocol leading to agreement will also lead to accuracy.
We then extend these results beyond the case of squared error, to a broad family of measures of agreement and accuracy including KL divergence.\footnote{Specifically, agreement and accuracy with respect to (almost) arbitrary Bregman divergences.}

\subsection{Overview of approach and results}
In \textcite{aar05}, Alice and Bob are said to \emph{agree} if the squared difference between their estimates is small. Likewise, we can say that Alice and Bob are \emph{accurate} if the squared distance between each of their estimates and the truth is small.
In Section~\ref{sec:quadratic} we present our first main result: under these definitions, \textbf{if the information structure describing Alice and Bob's signals satisfies the rectangle substitutes condition, then agreement implies accuracy}.
In other words, under this assumption, when two Bayesians agree -- regardless of how little information they have shared -- they necessarily agree \emph{on the truth}.

The proof involves carefully partitioning the space of posterior beliefs induced by the protocol.
Agreement is used to show that Alice and Bob usually fall into the same partition element, which means that Bob would not learn much from learning the partition element of Alice's expectation.
Then, the rectangle substitutes condition is used to show that if Bob were to learn Alice's partition element, then he would be very close to knowing the truth.

Aaronson measures agreement in terms of squared error, yet other measurements like KL divergence may be better suited for some settings.
For example, if Alice and Bob estimate the probability of a catastrophic event as $10^{-10}$ and $10^{-2}$, respectively, then under squared error they are said to agree closely, but arguably they disagree strongly, as reflected by their large KL divergence.
Motivated by these different ways to measure agreement, we next ask:
\begin{enumerate}[label=(\arabic*)]
    \item Can Aaronson's protocols be generalized to other notions of agreement, such that the number of bits communicated is independent of the amount of information available to Alice and Bob?
    \item Do other notions of agreement necessarily imply accuracy under rectangle substitutes?
\end{enumerate}
In Section \ref{sec:chap8_bregman}, we give our second and third main results: \textbf{the answer to both questions is yes.}
Specifically, the positive results apply when when measuring agreement and accuracy using Bregman divergences, a class of error measures that includes both squared distance and KL divergence.\footnote{The third result holds under an ``approximate triangle inequality'' condition on the Bregman divergence, which is satisfied by most or all natural choices; indeed, it is nontrivial to construct a Bregman divergence that does not satisfy this property.}

Aaronson's proof of his agreement theorem turns out to be specific to squared distance.
Our agreement theorem (Theorem~\ref{thm:agree_bregman}) modifies Aaronson's protocol to depend on the particular Bregman divergence, i.e.\ the relevant error measure.
It then proceeds in a manner inspired by Aaronson but using several new ideas.
Our proof that agreement implies accuracy under rectangle substitutes for general Bregman divergences also involves some nontrivial changes to our proof for squared distance.
In particular, the fact that the length of an interval cannot be inferred from the Bregman divergence between its endpoints necessitates a closer analysis of the partition of Alice's and Bob's beliefs.

We conclude in Section~\ref{sec:markets} with a discussion of connections between agreement protocols and information revelation in financial markets, and discuss an interesting potential avenue for future work.

\subsection{Related work} \label{sec:chap8_related_work}
\textcite{gp82} discussed the distinction between agreement and full information revelation.
One result shown is that under a natural probability measure on information structures, full agreement and information revelation occur in a single round of communication with probability one.
However, conditions for accuracy and the concept of substitutes are not discussed.

Our setting is related to but distinct from communication complexity.
In that field (e.g.\ \textcite{rao2020communication}), the goal is for Alice and Bob to correctly compute a function of their inputs while communicating as few bits as possible and using any protocol necessary.
By contrast, \textcite{aar05} considered a goal of agreement, not correctness, and focused on specific natural protocols, which he showed achieve this goal in a constant number of bits.
Our work focuses on Aaronson's setting.
We discuss how our results might be framed in terms of communication complexity in Appendix~\ref{appx:comm}.

Our introduction of the substitutes condition is inspired by its usefulness in  prediction markets~\parencite{cw16}.
The ``expectation-sharing'' agreement protocols we study bear a strong similarity to dynamics of market prices.
\textcite{ost12} introduced a condition under which convergence of prices in a market implies that all information is aggregated.
This can be viewed as an ``agreement implies accuracy'' condition.
Similarly, \textcite{kong2023information} presented a result that, for the logarithmic market scoring rule (LMSR) and conditionally independent signals, convergence of the market implies full information revelation.
Our results are conceptually similar, although they are technically quite different as we rely on the novel condition of \emph{rectangle substitutes}.
In the context of the LMSR, the rectangle substitutes notion includes conditionally independent signals as a special case (see discussion in Section~\ref{sec:bregman_prelims}).
We discuss the connection of our work to markets in Section~\ref{sec:markets}.

\section{Preliminaries} \label{sec:chap8_prelims}
\subsection{Information structures}
In Section~\ref{sec:prelim_info_struct}, we introduced the concept of an information structure as a tuple consisting of a probability distribution $\PP$ over a set $\Omega$ of states of the world, a quantity $Y$, and a tuple of $m$ signals that give partial information about the state of the world $\omega$ (and thus about $Y$).

In this chapter, we specifically consider the case of $m = 2$ signals and choose notation accordingly. We will say that there are two experts, Alice and Bob. Alice's signal is $\sigma: \Omega \to \mathcal{S}$ and Bob's signal is $\tau: \Omega \to \mathcal{T}$. Thus, we will think of information structures as 5-tuples $(\Omega, \PP, \sigma, \tau, Y)$. We additionally assume (following Aaronson) that $Y$ takes values in $[0, 1]$.

We denote by $\mu_{\sigma \tau} := \EE{Y \mid \sigma, \tau}$
the random variable that is equal to the expected value of $Y$ conditioned on both Alice's signal $\sigma$ and Bob's signal $\tau$.\footnote{The value of $Y$ need not be determined by $\sigma$ and $\tau$, although for our purposes the case in which it is determined is essentially equivalent.}
We also define $\mu_\sigma := \EE{Y \mid \sigma}$ and $\mu_\tau := \EE{Y \mid \tau}$.
For a measurable set $S \subseteq \mathcal{S}$, we define $\mu_S := \EE{Y \mid \sigma \in S}$; we define $\mu_T$ analogously for $T \subseteq \mathcal{T}$.
Additionally, for $T \subseteq \mathcal{T}$, we define $\mu_{\sigma T} := \EE{Y \mid \tau \in T, \sigma}$, i.e.\ the expected value of $Y$ conditioned on the particular value of $\sigma$ and the knowledge that $\tau \in T$.
If Alice knows that Bob's signal belongs to $T$ (and nothing else about his signal), then the expected value of $Y$ conditional on her information is $\mu_{\sigma T}$; we refer to this as Alice's \emph{expectation}.
Likewise, for $S \subseteq \mathcal{S}$, we define $\mu_{S \tau} := \EE{Y \mid \sigma \in S, \tau}$.
Finally, we define $\mu_{ST} := \EE{Y \mid \sigma \in S, \tau \in T}$. This is the expectation of a third party who only knows that $\sigma \in S$ and $\tau \in T$.

In general we often wish to take expectations conditioned on $\sigma \in S, \tau \in T$ (for some $S \subseteq \mathcal{S}, T \subseteq \mathcal{T}$). We will use the shorthand $\EE{\cdot \mid S, T}$ for $\EE{\cdot \mid \sigma \in S, \tau \in T}$ in such cases.

\subsection{Agreement protocols} \label{subsec:agreement-protocols}
The notion of \emph{agreement} between Alice and Bob is central to our work.
We first define agreement in terms of squared error, and generalize to other error measures in Section~\ref{sec:chap8_bregman}.

\begin{defin}[$\epsilon$-agreement]
Let $a$ and $b$ be Alice's and Bob's expectations, respectively ($a$ and $b$ are random variables defined on $\Omega$). Alice and Bob \emph{$\epsilon$-agree} if $\frac{1}{4} \EE{(a - b)^2} \le \epsilon$.
\end{defin}
\noindent
The constant $\frac{1}{4}$ makes the left-hand side represent Alice's and Bob's squared distance to the average of their expectations.

Our setting follows \textcite{aar05}, which examined communication protocols that cause Alice and Bob to agree.
In a \emph{(deterministic) communication protocol}, Alice and Bob take turns sending each other messages (strings of bits).
On Alice's turns, Alice communicates a message that is a deterministic function of her input (i.e.\ her signal $\sigma$) and all previous communication, and likewise for Bob on his turns.
A \emph{rectangle} is a set of the form $S \times T$ where $S \subseteq \mathcal{S}$ and $T \subseteq \mathcal{T}$.

The \emph{communication transcript} is the ordered tuple of all messages that have been sent. The transcript \emph{at time step $t$} refers to the tuple consisting of the first $t$ messages. The transcript at time step $t$ partitions $\Omega$ into rectangles: for any given sequence of $t$ messages, there are subsets $S_t \subseteq \mathcal{S}, T_t \subseteq \mathcal{T}$ such that the protocol transcript at time $t$ is equal to this sequence if and only if $(\sigma, \tau) \in S_t \times T_t$.\footnote{We can see this inductively: suppose the transcript at time step $t - 1$ partitions $\Omega$ into rectangles, and (without loss of generality) that the $t$-th turn is Alice's. Consider one of these rectangles. Alice's message can only depend on her input and the transcript so far, which means that her message can only partition this rectangle into sub-rectangles.}

For a given communication protocol, we may think of $S_t$ and $T_t$ as random variables.
Alice's expectation at time $t$ (i.e.\ \emph{after} the $t$-th message has been sent) is $\mu_{\sigma T_t}$ and Bob's expectation at time $t$ is $\mu_{S_t \tau}$.
Finally, the protocol terminates at a certain time (which need not be known in advance of the protocol).
While typically in communication complexity a protocol is associated with a final output, in this case we are interested in Alice's and Bob's expectations, so we do not require an output.

It will be convenient to hypothesize a third party observer, whom we call Charlie, who observes the protocol but has no other information.
At time $t$, Charlie has expectation $\mu_{S_t T_t}$.
Charlie's expectation can also be interpreted as the expectation of $Y$ according to Alice and Bob's common knowledge. Note that Alice and Bob each know Charlie's expectation at any given time.

The following definition formalizes the relationship between communication protocols and agreement.

\begin{defin}[$\epsilon$-agreement protocol]
Given an information structure $\mathcal{I}$, a communication protocol \emph{causes Alice and Bob to $\epsilon$-agree} on $\mathcal{I}$ if Alice and Bob $\epsilon$-agree at the end of the protocol, i.e., if $\frac{1}{4} \EE{(\mu_{\sigma T_t} - \mu_{S_t \tau})^2} \le \epsilon$, where the expected value is over Alice's and Bob's inputs. We say that a communication protocol is an \emph{$\epsilon$-agreement protocol} if the protocol causes Alice and Bob to $\epsilon$-agree on every information structure.
\end{defin}

Aaronson defines and analyzes two $\epsilon$-agreement protocols.\footnote{A minor difference to our framing is that \textcite{aar05} focuses on \emph{probable approximate agreement}: protocols that cause the absolute difference between Alice and Bob to be at most $\epsilon$ with probability all but $\delta$. While the results as presented in this section are stronger than those in \textcite{aar05} (the original results follow from these as a consequence of Markov's inequality), these results follow from a straightforward modification of his proofs.} The first of these is the \emph{standard protocol}, in which Alice and Bob take turns stating their expectations for a number of time steps that can be computed by Alice and Bob independently in advance of the protocol, and which is guaranteed to be at most $O(1/\epsilon)$.

The fact that exchanging their expectations for $O(1/\epsilon)$ time steps results in $\epsilon$-agreement is profound and compelling. However, the standard protocol may require an unbounded number of bits of communication, since Alice and Bob are exchanging real numbers. To address this, Aaronson defines another agreement protocol that is truly polynomial-communication (which we slightly modify for our purposes):

\begin{defin}[Discretized protocol, \textcite{aar05}]
Choose $\epsilon > 0$. In the \emph{discretized protocol} with parameter $\epsilon$, on her turn (at time $t$), Alice sends ``low'' if her expectation is smaller than Charlie's by more than $\epsilon/4$, i.e.\ if $\mu_{S_{t - 1} \tau} < \mu_{S_{t - 1} T_{t - 1}} - \epsilon/4$; ``high'' if her expectation is larger than Charlie's by more than $\epsilon/4$; and ``medium'' otherwise. Bob acts analogously on his turn. At the start of the protocol, Alice and Bob use the information structure to independently compute the time $t_{\text{end}} \le \frac{1000}{\epsilon}$ that minimizes $\EE{(\mu_{\sigma T_{t_{\text{end}}}} - \mu_{S_{t_{\text{end}}} \tau})^2}$. The protocol ends at this time.
\end{defin}

\begin{theorem}[{\cite[Theorem 4]{aar05}}] \label{thm:aaronson-thm4}
The discretized protocol with parameter $\epsilon$ is an $\epsilon$-agreement protocol with transcript length $O(1/\epsilon)$ bits.
\end{theorem}

In general, we refer to Aaronson's standard and discretized protocols as examples of \emph{expectation-sharing} protocols.
We will define other examples in Section \ref{sec:chap8_bregman}, similar to Aaronson's discretized protocol but with different cutoffs for low, medium, and high.
We also interpret expectation-sharing protocols in the context of markets in Section~\ref{sec:markets}.

\subsection{Accuracy and informational substitutes}
Most of our main results give conditions such that if Alice and Bob $\epsilon$-agree, then Alice's and Bob's estimates are accurate.
By \emph{accurate}, we mean that Alice's and Bob's expectations are close to $\mu_{\sigma \tau}$, i.e., what they would believe if they knew each other's signals.
(After all, they cannot hope to have a better estimate of $Y$ than $\mu_{\sigma \tau}$; for this reason we sometimes refer to $\mu_{\sigma \tau}$ as the ``truth.'') Formally:

\begin{defin}[$\epsilon$-accuracy]
Let $a$ be Alice's expectation. Alice is \emph{$\epsilon$-accurate} if $\EE{(\mu_{\sigma \tau} - a)^2} \le \epsilon$. We define $\epsilon$-accuracy analogously for Bob.
\end{defin}

One cannot hope for an unconditional result stating that if Alice and Bob agree, then they are accurate.
Consider for instance the \emph{XOR information structure} from the introduction: Alice and Bob each receive independent random bits as input, and $Y$ is the XOR of these bits.
Then from the start Alice and Bob agree that the expected value of $Y$ is exactly $\frac{1}{2}$, but this value is far from $\mu_{\sigma \tau}$, which is either $0$ or $1$.

Intuitively, this situation arises because Alice's and Bob's signals are \emph{informational complements}: each signal is not informative by itself, but they are informative when taken together.
On the other hand, we say that signals are \emph{informational substitutes} if learning one signal is less valuable if you already know the other signal.
We introduced and motivated the concept of informational substitutes in Section~\ref{sec:prelim_subs}. In particular, we defined \emph{weak substitutes} (first introduced by \textcite{cw16}) as a formalization of the notion of diminishing marginal returns to learning an extra signal.
An extreme example of informational substitutes is if $\sigma$ and $\tau$ both specify the value of $Y$ exactly.
In that case, $\sigma$ becomes useless upon learning $\tau$ and vice versa.
Our definition is inspired by the definition of weak substitutes, but we require a stronger notion for our results to hold.
For the following definition, recall that we write $\mid S, T$ as shorthand for $\mid \sigma \in S, \tau \in T$.

\begin{defin} \label{def:rect_subs_quad}
An information structure $\mathcal{I} = (\Omega, \PP, \sigma, \tau, Y)$ satisfies \emph{rectangle substitutes} if for every $S \subseteq \mathcal{S}, T \subseteq \mathcal{T}$ such that $\PP[\sigma \in S, \tau \in T] > 0$, we have
\begin{equation} \label{eq:original_rect_subs}
\EE{(Y - \mu_{S \tau})^2 \mid S, T} - \EE{(Y - \mu_{\sigma \tau})^2 \mid S, T} \le \EE{(Y - \mu_{ST})^2 \mid S, T} - \EE{(Y - \mu_{\sigma T})^2 \mid S, T}.
\end{equation}
\end{defin}

This definition is a strengthening of weak substitutes for two agents: an information structure satisfies \emph{weak substitutes} if Equation~\ref{eq:original_rect_subs} holds specifically for $S = \mathcal{S}$ and $T = \mathcal{T}$. We will show that under the rectangle substitutes condition, if Alice and Bob approximately agree, then they are approximately accurate.

\paragraph{Interpreting substitutes}
Both sides of Equation \ref{eq:original_rect_subs} represent the ``value'' of learning $\sigma$ as measured by a decrease in error.
The left-hand side gives the decrease if one already knows $\tau$ and that $\sigma \in S$; the right-hand side gives the decrease if one only knows that $\sigma \in S, \tau \in T$.
Substitutes thus says: the marginal value of learning $\sigma$ is smaller if one already knows $\tau$ than if one does not.
This statement should hold for every sub-rectangle $S,T$.
Note that the inequality can be rearranged to focus instead on the marginal value of $\tau$ rather than $\sigma$.
Note also that in the XOR information structure, the left-hand side of the inequality is $\frac{1}{4}$ while the right-hand side is zero: a large violation of the substitutes condition.
In the example where $\sigma$ and $\tau$ both specify the value of $Y$, the left side is always zero.

\textcite{cw16} discusses three interpretations of substitutes, which motivate it as a natural condition.
(1) Each side of the inequality measures an improvement in \emph{prediction error}, here the squared loss, due to learning $\sigma$.
Under substitutes, the improvement is smaller if one already knows $\tau$.
(2) Each side measures a \emph{decrease in uncertainty} (here, measured roughly by \emph{variance}) due to learning $\sigma$.
Under substitutes, $\sigma$ provides less information about $Y$ if one already knows $\tau$.%
\footnote{Here, uncertainty is measured by variance of one's belief. Under the KL divergence analogue covered in Section \ref{sec:bregman_prelims}, uncertainty is measured in bits via Shannon entropy.}
(3) Each side measures the \emph{decrease in distance} of a posterior expectation from the truth when learning $\sigma$.
The distance to $Y$ changes less if one already knows $\tau$.

\paragraph{Restrictiveness of substitutes} It is natural to ask about the strength of the rectangle substitutes assumption.
In the case that $\abs{\mathcal{S}} = \abs{\mathcal{T}} = 2$, the condition reduces to the aforementioned and well-established weak substitutes condition.
For larger signal sets, the set of information structures satisfying rectangle substitutes remains nontrivial.
For example, it is satisfied by a positive fraction of information structures (for a natural choice of measure). We show this fact in Appendix~\ref{appx:prelims_omitted} by exhibiting an information structure in which Equation~\ref{eq:original_rect_subs} holds \emph{strictly} for all $S, T$ with $\abs{S}, \abs{T} \ge 2$ (and thus, that all nearby information structures also satisfy rectangle substitutes). Finally, we note that although the rectangle substitutes condition is strong due to the quantification over sub-rectangles, in Section~\ref{sec:graceful_decay} we prove that our main results decay gracefully for information structures that are close to but do not quite satisfy the rectangle substitutes condition.

\subsection{The Pythagorean theorem}
We recall the Pythagorean theorem from Section~\ref{sec:prelim_pythag}:
\pythagsquared*

One application of the Pythagorean theorem in our context takes $A = Y$, $B = \mu_{\sigma \tau}$ (the expected value of $Y$ conditioned on the experts' signals), and $C = \mu_{\sigma T}$ (Alice's expected value, which only depends on her signal and thus on the signal pair).
This particular application, along with the symmetric one taking $C = \mu_{S \tau}$, allows us to rewrite the rectangle substitutes condition in a form that we will find more convenient:

\begin{remark}
An information structure $\mathcal{I}$ satisfies rectangle substitutes if and only if
\begin{equation} \label{eq:rec_sub_quad}
\EE{(\mu_{\sigma \tau} - \mu_{S \tau})^2 \mid S, T} \le \EE{(\mu_{\sigma T} - \mu_{ST})^2 \mid S, T}
\end{equation}
for all $S, T$ such that $\PP[\sigma \in S, \tau \in T] > 0$.
\end{remark}

\section{Results for squared distance} \label{sec:quadratic}
Our main results show that, under the rectangle substitutes condition, any communication protocol that causes Alice and Bob to agree also causes them to be accurate.
We now show the first of these results, which is specific to the squared distance error measure that we have been discussing.

\subsection{Agreement implies accuracy}
\begin{theorem} \label{thm:agreement_accurate_quad}
Let $\mathcal{I} = (\Omega, \PP, \sigma, \tau, Y)$ be an information structure that satisfies rectangle substitutes. For any communication protocol that causes Alice and Bob to $\epsilon$-agree on $\mathcal{I}$, Alice and Bob are $10 \epsilon^{1/3}$-accurate after the protocol terminates.
\end{theorem}

The crux of the argument is the following lemma.

\begin{lemma} \label{lem:bob_close_quad}
Let $\mathcal{I} = (\Omega, \PP, \sigma, \tau, Y)$ be an information structure that satisfies rectangle substitutes. Let $\epsilon = \EE{(\mu_{\sigma} - \mu_{\tau})^2}$. Then
\[\EE{(\mu_{\sigma \tau} - \mu_{\tau})^2} \le 6\epsilon^{1/3}.\]
\end{lemma}

\noindent Let us first prove Theorem~\ref{thm:agreement_accurate_quad} assuming Lemma~\ref{lem:bob_close_quad} is true.

\begin{proof}[Proof of Theorem~\ref{thm:agreement_accurate_quad}]
Consider any protocol that causes Alice and Bob to $\epsilon$-agree on $\mathcal{I}$. Let $S$ be the set of possible signals of Alice at the end of the protocol which are consistent with the protocol transcript, and define $T$ likewise for Bob. Intuitively, $S \times T$ is the set of plausible signal pairs $(\sigma, \tau)$ according to an external observer of the protocol. Observe that $S$ and $T$ are random variables, each a function of both $\sigma$ and $\tau$.
We have
\begin{align*}
\EE{(\mu_{\sigma \tau} - \mu_{S \tau})^2} &= \EE[S, T]{\EE{(\mu_{\sigma \tau} - \mu_{S \tau})^2 \mid S, T}}\\
&\le \EE[S, T]{6 \parens{\EE{(\mu_{\sigma T} - \mu_{S \tau})^2 \mid S, T}}^{1/3}}\\
&\le 6 \EE[S, T]{\EE{(\mu_{\sigma T} - \mu_{S \tau})^2 \mid S, T}}^{1/3}\\
&= 6 \EE{(\mu_{\sigma T} - \mu_{S \tau})^2}^{1/3} \le 6(4\epsilon)^{1/3} \le 10 \epsilon^{1/3}.
\end{align*}
In the second step, we apply Lemma~\ref{lem:bob_close_quad} to the information structure $\mathcal{I}$ restricted to $S \times T$ -- that is, to $\mathcal{I}' = (\Omega', \PP', S, T, Y)$, where $\Omega' = \{\omega \in \Omega: \sigma \in S, \tau \in T\}$ and $\PP'[\omega] = \PP[\omega \mid \sigma \in S, \tau \in T]$. (Note that we use the fact that if $\mathcal{I}$ satisfies rectangle substitutes, then so does $\mathcal{I}'$; this is because a rectangle of $\mathcal{I}'$ is also a rectangle of $\mathcal{I}$.) The third step follows by the concavity of $x^{1/3}$. Therefore, Bob is $10\epsilon^{1/3}$ accurate (and Alice is likewise by symmetry).
\end{proof}

The proof of Lemma~\ref{lem:bob_close_quad} relies on the following claim.

\begin{claim} \label{claim:n_sigma_tau}
In the setting of Lemma~\ref{lem:bob_close_quad}, for any $N \ge 1$, it is possible to partition $[0, 1]$ into $N$ intervals $[0, x_1), [x_1, x_2), \dots,$ $[x_{N - 1}, 1]$ in a way so that each interval has length at most $\frac{2}{N}$, and
\[\PP[k(\sigma) \neq k(\tau)] \le \sqrt{\epsilon} N,\]
where $k(\sigma)$ denotes the $k \in [N]$ such that $x_{k - 1} \le \mu_{\sigma} < x_k$, and $k(\tau)$ is defined analogously.\footnote{For convenience we define $x_0 = 0$ and $x_N$ to be some number greater than $1$.}
\end{claim}

Intuitively, Claim~\ref{claim:n_sigma_tau} is true because if $\EE{(\mu_{\sigma} - \mu_{\tau})^2}$ is small, then $\mu_{\sigma}$ and $\mu_{\tau}$ are likely to fall into the same interval.

\begin{proof}
We claim that in fact we can choose the $x_i$'s so that each $x_i$ is in $\brackets{\frac{i}{N} - \frac{1}{2N}, \frac{i}{N} + \frac{1}{2N}}$. This ensures that each interval has length at most $\frac{2}{N}$.

For $x \in [0, 1]$, let $\rho(x)$ be the probability that $x$ is between $\mu_{\sigma}$ and $\mu_{\tau}$, inclusive. Note that $\PP[k(\sigma) \neq k(\tau)] \le \sum_{i = 1}^{N - 1} \rho(x_i)$.

Observe that if $x$ is selected uniformly from $[0, 1]$, the expected value of $\rho(x)$ is equal to $\abs{\mu_{\sigma} - \mu_{\tau}}$, because both quantities are equal to the probability that $x$ is between $\mu_{\sigma}$ and $\mu_{\tau}$.
Therefore, if $(\sigma, \tau)$ is additionally chosen according to $\PP$, we have
\[\EE[{x \leftarrow [0, 1]}]{\rho(x)} = \EE{\abs{\mu_{\sigma} - \mu_{\tau}}} \le \sqrt{\EE{(\mu_{\sigma} - \mu_{\tau})^2}} = \sqrt{\epsilon}.\]
This means that
\[\sum_{i = 1}^{N - 1} \EE[{x \leftarrow \brackets{\frac{i}{N} - \frac{1}{2N}, \frac{i}{N} + \frac{1}{2N}}}]{\rho(x)} = (N - 1) \EE[{x \leftarrow \brackets{\frac{1}{2N}, 1 - \frac{1}{2N}}}]{\rho(x)} \le \sqrt{\epsilon} N.\]
Thus, if each $x_i$ is selected uniformly at random from $\brackets{\frac{i}{N} - \frac{1}{2N}, \frac{i}{N} + \frac{1}{2N}}$, the expected value of $\PP[k(\sigma) \neq k(\tau)]$ would be at most $\sqrt{\epsilon} N$. In particular this means that there exist choices of the $x_i$'s such that $\PP[k(\sigma) \neq k(\tau)] \le \sqrt{\epsilon} N$.
\end{proof}

We now prove Lemma~\ref{lem:bob_close_quad}.

\begin{proof}[Proof of Lemma~\ref{lem:bob_close_quad}]
Fix a large positive integer $N$ (we will later find it optimal to set $N = \epsilon^{-1/6}$). Consider a partition of $[0, 1]$ into $N$ intervals $[0, x_1), [x_1, x_2), \dots, [x_{N - 1}, 1]$ satisfying the conditions of Claim~\ref{claim:n_sigma_tau}. Let $S^{(k)} := \{\sigma \in \mathcal{S}: x_{k - 1} \le \mu_{\sigma} < x_k\}$. Additionally, let $k(\sigma)$ and $k(\tau)$ be as defined in Claim~\ref{claim:n_sigma_tau}.

Our goal is to upper bound the expectation of $(\mu_{\sigma \tau} - \mu_{\tau})^2$. In pursuit of this goal, we observe that by the Pythagorean theorem, we have
\[\EE{(\mu_{\sigma \tau} - \mu_{\tau})^2} = \EE{(\mu_{\sigma \tau} - \mu_{S^{(k(\sigma))} \tau})^2} + \EE{(\mu_{S^{(k(\sigma))} \tau} - \mu_{\tau})^2}.\]
We now use the rectangle substitutes assumption: for any $k$, by applying Equation~\ref{eq:rec_sub_quad} to $S = S^{(k)}$ and $T = \mathcal{T}$, we know that
\[\EE{(\mu_{\sigma} - \mu_{S^{(k)}})^2 \mid \sigma \in S^{(k)}} \ge \EE{(\mu_{\sigma \tau} - \mu_{S^{(k)} \tau})^2 \mid \sigma \in S^{(k)}}.\]
Taking the expectation over $k$ (i.e.\ choosing each $k$ with probability equal to $\PP[\sigma \in S^{(k)}]$), we have that
\begin{equation} \label{eq:delta_change}
\EE{(\mu_{\sigma} - \mu_{S^{(k(\sigma))}})^2} \ge \EE{(\mu_{\sigma \tau} - \mu_{S^{(k(\sigma))} \tau})^2}.
\end{equation}
Therefore, we have
\begin{equation}
\label{eq:delta_change_2}
\EE{(\mu_{\sigma \tau} - \mu_{\tau})^2} \le \EE{(\mu_{\sigma} - \mu_{S^{(k(\sigma))}})^2} + \EE{(\mu_{S^{(k(\sigma))} \tau} - \mu_{\tau})^2}.
\end{equation}
We will use Claim~\ref{claim:n_sigma_tau} to argue that each of these two summands is small. 
The argument regarding the first summand is straightforward: for any $\sigma$, we have that $x_{k(\sigma)} \le \mu_{\sigma}, \mu_{S^{(k(\sigma))}} < x_{k(\sigma) + 1} \le x_{k(\sigma)} + \frac{2}{N}$, which means that $\EE{(\mu_{\sigma} - \mu_{S^{(k(\sigma))}})^2} \le \frac{4}{N^2}$.

We now upper bound the second summand.%
\footnote{The proof below takes sums over $\hat{\tau} \in \mathcal{T}$ and thus implicitly assumes that $\mathcal{T}$ is finite, but the proof extends to infinite $\mathcal{T}$, with sums over $\tau$ replaced by integrals with respect to the probability measure over $\mathcal{T}$.}
For any $\hat{\tau} \in \mathcal{T}$, let $p(\hat{\tau}) = \PP[\tau = \hat{\tau}]$ and $q(\hat{\tau}) = \PP[\tau = \hat{\tau}, k(\sigma) \neq k(\tau)]$. Then $\sum_{\hat{\tau} \in \mathcal{T}} p(\hat{\tau}) = 1$ and $\sum_{\hat{\tau} \in \mathcal{T}} q(\hat{\tau}) \le \sqrt{\epsilon} N$.
Observe that
\begin{align} \label{eq:two_expectations}
\EE{(\mu_{S^{(k(\sigma))} \tau} - \mu_{\tau})^2} &= \sum_{\hat{\tau}} p(\hat{\tau}) \EE{(\mu_{S^{(k(\sigma))} \hat{\tau}} - \mu_{\hat{\tau}})^2 \mid \tau = \hat{\tau}} \nonumber\\
&= \sum_{\hat{\tau}} (p(\hat{\tau}) - q(\hat{\tau})) \EE{(\mu_{S^{(k(\sigma))} \hat{\tau}} - \mu_{\hat{\tau}})^2 \mid \tau = \hat{\tau}, k(\sigma) = k(\hat{\tau})} \nonumber\\
&\qquad + q(\hat{\tau}) \EE{(\mu_{S^{(k(\sigma))} \hat{\tau}} - \mu_{\hat{\tau}})^2 \mid \tau = \hat{\tau}, k(\sigma) \neq k(\hat{\tau})}.
\end{align}

To handle the first expectation, we note that if $k(\sigma) = k(\hat{\tau})$, then $\abs{\mu_{S^{(k(\sigma))} \hat{\tau}} - \mu_{\hat{\tau}}} \le \frac{q(\hat{\tau})}{p(\hat{\tau})}$.
To see this, observe
\[p(\hat{\tau}) \mu_{\hat{\tau}} = (p(\hat{\tau}) - q(\hat{\tau})) \mu_{S^{(k(\sigma))} \hat{\tau}} + q(\hat{\tau}) \mu_{\mathcal{S} \setminus S^{(k(\sigma))} \hat{\tau}}~.\]
Rearranging and taking absolute values, we conclude
\[p(\hat{\tau}) \abs{\mu_{S^{(k(\sigma))} \hat{\tau}} - \mu_{\hat{\tau}}} = q(\hat{\tau}) \abs{\mu_{S^{(k(\sigma))} \hat{\tau}} - \mu_{\mathcal{S} \setminus S^{(k(\sigma))}}} \le q(\hat{\tau}).\]
Therefore, recalling $q(\hat\tau) \leq p(\hat\tau)$, we have
\begin{align*}
&(p(\hat{\tau}) - q(\hat{\tau})) \EE{(\mu_{S^{(k(\sigma))} \hat{\tau}} - \mu_{\hat{\tau}})^2 \mid \tau = \hat{\tau}, k(\sigma) = k(\hat{\tau})} 
\le (p(\hat{\tau}) - q(\hat{\tau})) \parens{\frac{q(\hat{\tau})}{p(\hat{\tau})}}^2
\le \frac{q(\hat{\tau})^2}{p(\hat{\tau})} 
\le q(\hat{\tau}).
\end{align*}

On the other hand, we can bound the second expectation in Equation~\ref{eq:two_expectations} by $1$. Therefore we have
\[\EE{(\mu_{S^{(k(\sigma))} \tau} - \mu_{\tau})^2} \le \sum_{\hat{\tau}} (q(\hat{\tau}) + q(\hat{\tau})) = 2 \sum_{\hat{\tau}} q(\hat{\tau}) \le 2 \sqrt{\epsilon} N.\]
To conclude, we now know that
\[\EE{(\mu_{\sigma \tau} - \mu_{\tau})^2} \le \frac{4}{N^2} + 2 \sqrt{\epsilon} N.\]
Setting $N = \epsilon^{-1/6}$ makes the right-hand side equal to $6\epsilon^{1/3}$, completing the proof.
\end{proof}

Now that we have proven Lemma~\ref{lem:bob_close_quad}, our proof of Theorem~\ref{thm:agreement_accurate_quad} is complete.

\subsection{Consequences of Theorem~\ref{thm:agreement_accurate_quad}}
Theorem~\ref{thm:agreement_accurate_quad} is a general result about agreement protocols. Applying the result to Aaronson's discretized protocol gives us the following result.

\begin{corollary} \label{cor:aaronson_exp}
Let $\mathcal{I}$ be any information structure that satisfies universal rectangle substitutes. For any $\epsilon > 0$, Alice and Bob will be $\epsilon$-accurate after running Aaronson's discretized protocol with parameter $\epsilon^3/1000$ (and this takes $O(1/\epsilon^3)$ bits of communication).
\end{corollary}

\begin{remark}
The discretized protocol is not always the most efficient agreement protocol.
For example, Proposition~\ref{prop:fast_rect} shows that if the rectangle substitutes condition holds, agreement (and therefore accuracy) can be reached with just $O(\log(1/\epsilon))$ bits, an improvement on Corollary \ref{cor:aaronson_exp}.
We discuss communication complexity further in Appendix~\ref{appx:comm}.
Even if more efficient protocols are sometimes possible, expectation-sharing protocols are of interest because they model naturally-occurring communication processes.
For example, they capture the dynamics of prices in markets, which we also discuss in Section~\ref{sec:markets}.
More generally, we find it remarkable that Alice and Bob become accurate by running Aaronson's agreement protocol (or indeed \emph{any} agreement protocol), despite such protocols being designed with only agreement in mind.
\end{remark}

Finally, we observe the following important consequence of Theorem~\ref{thm:agreement_accurate_quad}: once Alice and Bob agree, they continue to agree.

\begin{corollary}
\label{cor:continue-agreeing-quad}
Let $\mathcal{I} = (\Omega, \PP, \sigma, \tau, Y)$ be an information structure that satisfies rectangle substitutes. Consider a communication protocol with the property that Alice and Bob $\epsilon$-agree after round $t$. Then Alice and Bob $10 \epsilon^{1/3}$-agree on all subsequent time steps.
\end{corollary}

\begin{proof}
If Alice and Bob $\epsilon$-agree then they are $10\epsilon^{1/3}$-accurate, so in particular $\EE{(\mu_{\sigma \tau} - \mu_{\sigma T_t})^2} \le 10\epsilon^{1/3}$. Note that $\EE{(\mu_{\sigma \tau} - \mu_{\sigma T_s})^2}$ is a decreasing function of $s$, since for any $s_1 \le s_2$ we have
\[\EE{(\mu_{\sigma \tau} - \mu_{\sigma T_{s_1}})^2} = \EE{(\mu_{\sigma \tau} - \mu_{\sigma T_{s_2}})^2} + \EE{(\mu_{\sigma T_{s_2}} - \mu_{\sigma T_{s_1}})^2}\]
by the Pythagorean theorem. Therefore, for any $t' > t$, we have $\EE{(\mu_{\sigma \tau} - \mu_{\sigma T_{t'}})^2} \le 10\epsilon^{1/3}$. Symmetrically, we have $\EE{(\mu_{\sigma \tau} - \mu_{S_{t'} \tau})^2} \le 10\epsilon^{1/3}$. Therefore, $\EE{(\mu_{\sigma T_{t'}} - \mu_{S_{t'} \tau})^2} \le 40\epsilon^{1/3}$, which means that after round $t'$, Alice and Bob $10 \epsilon^{1/3}$-agree.
\end{proof}

Corollary~\ref{cor:continue-agreeing-quad} stands in contrast to the more general case, in which it is possible that Alice and Bob ``nearly agree for the first $t - 1$ time steps, then disagree violently at the $t$-th step'' \parencite[\S2.2]{aar05}.
Thus, while the main purpose of Theorem~\ref{thm:agreement_accurate_quad} is a property about \emph{accuracy}, an \emph{agreement} property falls out naturally: under the rectangle substitutes condition, once Alice and Bob are close to agreement, they will remain in relatively close agreement into the future.

\subsection{Graceful decay under closeness to rectangle substitutes} \label{sec:graceful_decay}
In a sense, the rectangle substitutes condition is quite strong: it requires that the weak substitutes condition be satisfied on \emph{every} sub-rectangle. One might hope for a result that generalizes Theorem~\ref{thm:agreement_accurate_quad} to information structures that almost-but-not-quite satisfy the rectangle substitutes.
Let us formally define a notion of closeness to rectangle substitutes.

\begin{defin}
An information structure $\mathcal{I} = (\Omega, \PP, \sigma, \tau, Y)$ satisfies \emph{$\delta$-approximate rectangle substitutes} if for every partition of $\mathcal{S} \times \mathcal{T}$ into rectangles,\footnote{There are partitions into rectangles that cannot arise from a communication protocol. Our results would apply equally if this condition were instead defined for every partition that could arise from a communication protocol, but we state this condition more generally so that it could be applicable in a broader context than the analysis of communication protocols.} the rectangle substitutes condition holds in expectation over the partition, up to an additive constant of $\delta$, i.e., if we have
\begin{equation} \label{eq:approx_rect}
\EE[\sigma, \tau]{(\mu_{\sigma \tau} - \mu_{S_{\sigma, \tau} \tau})^2} \le \EE[\sigma, \tau]{(\mu_{\sigma T_{\sigma, \tau}} - \mu_{S_{\sigma, \tau} T_{\sigma, \tau}})^2} + \delta,
\end{equation}
where $S_{\sigma, \tau} \times T_{\sigma, \tau}$ is the rectangle containing $(\sigma, \tau)$.
\end{defin}

\begin{remark}
The $\delta$-approximate rectangle substitutes property is a relaxation of the rectangle substitutes property, in the sense that the two are equivalent if $\delta = 0$. To see this, first observe that if $\mathcal{I}$ satisfies rectangle substitutes, then it satisfies Equation~\ref{eq:approx_rect} with $\delta = 0$ pointwise across all $S_{\sigma, \tau}, T_{\sigma, \tau}$, and thus in expectation. In the other direction, suppose that $\mathcal{I}$ satisfies $0$-approximate rectangle substitutes. Let $S \subseteq \mathcal{S}, T \subseteq \mathcal{T}$ and consider the partition of $\mathcal{I}$ into rectangles that contains $S \times T$ and, separately, every other signal pair $(\sigma, \tau)$ in its own rectangle. For this partition, Equation~\ref{eq:approx_rect} reduces precisely to Equation~\ref{eq:rec_sub_quad} (the rectangle substitutes condition for $S$ and $T$).
\end{remark}

\noindent Theorem~\ref{thm:agreement_accurate_quad} generalizes to approximate rectangle substitutes as follows.

\begin{theorem} \label{thm:agreement_accurate_graceful_decay}
Let $\mathcal{I} = (\Omega, \PP, \sigma, \tau, Y)$ be an information structure that satisfies $\delta$-approximate rectangle substitutes. For any communication protocol that causes Alice and Bob to $\epsilon$-agree on $\mathcal{I}$, Alice and Bob are $(10 \epsilon^{1/3} + \delta)$-accurate after the protocol terminates.
\end{theorem}

\begin{proof}
We first observe that Lemma~\ref{lem:bob_close_quad} can be modified as follows.

\begin{lemma} \label{lem:bob_close_graceful_decay}
Let $\mathcal{I} = (\Omega, \PP, \sigma, \tau, Y)$ be an information structure that satisfies $\delta$-approximate rectangle substitutes. Let $\epsilon = \EE{(\mu_{\sigma} - \mu_{\tau})^2}$. Then
\[\EE{(\mu_{\sigma \tau} - \mu_{\tau})^2} \le 6\epsilon^{1/3} + \delta.\]
\end{lemma}

\noindent
The proof of Lemma~\ref{lem:bob_close_graceful_decay} is exactly the same as that of Lemma~\ref{lem:bob_close_quad}, except that Equation~\ref{eq:delta_change} includes an additive $\delta$ term on the left-hand side:
\[\EE{(\mu_{\sigma} - \mu_{S^{(k(\sigma))}})^2} + \delta \ge \EE{(\mu_{\sigma \tau} - \mu_{S^{(k(\sigma))} \tau})^2}.\]
This modified inequality follows immediately from the $\delta$-approximate rectangle substitutes condition, noting that one partition of $\mathcal{S} \times \mathcal{T}$ into rectangles is $\{S_1 \times \mathcal{T}, \dots, S_N \times \mathcal{T}\}$.
The extra $\delta$ term produces the $\delta$ term in the lemma statement.

To prove the theorem, let $S$ be the set of possible signals of Alice at the end of the protocol which are consistent with the protocol transcript, and define $T$ likewise for Bob. Let $\delta_{ST}$ be the minimum $\delta$ such that $S \times T$ satisfies $\delta$-approximate rectangle substitutes. Note that $\EE[S, T]{\delta_{ST}} \le \delta$: otherwise, by taking the union over the worst-case partitions for each $S, T$ we would exhibit a partition of $\mathcal{S} \times \mathcal{T}$ into rectangles that would violate the $\delta$-approximate rectangle substitutes property. Therefore we have
\begin{align*}
\EE{(\mu_{\sigma \tau} - \mu_{S \tau})^2} &= \EE[S, T]{\EE{(\mu_{\sigma \tau} - \mu_{S \tau})^2 \mid S, T}}\\
&\le \EE[S, T]{6 \parens{\EE{(\mu_{\sigma T} - \mu_{S \tau})^2 \mid S, T}}^{1/3} + \delta_{ST}}\\
&\le 6 \EE[S, T]{\EE{(\mu_{\sigma T} - \mu_{S \tau})^2 \mid S, T}}^{1/3} + \delta\\
&= 6 \EE{(\mu_{\sigma T} - \mu_{S \tau})^2}^{1/3} + \delta = 6(4\epsilon)^{1/3} + \delta \le 10 \epsilon^{1/3} + \delta.
\end{align*}
As in the proof of Theorem~\ref{thm:agreement_accurate_quad}, the second step follows by applying Lemma~\ref{lem:bob_close_quad} to the information structure $\mathcal{I}$ restricted to $S \times T$.
\end{proof}

\section{Results for other divergence measures} \label{sec:chap8_bregman}
Squared distance is a compelling error measure because it elicits the mean. That is, if you wish to estimate a random variable $Y$ and will be penalized according to the squared distance between $Y$ and your estimate, the strategy that minimizes your expected penalty is to report the expected value of $Y$ (conditional on the information you have). This is in contrast to e.g.\ absolute distance as an error measure, which would instead elicit the median of your distribution. The class of error measures that elicit the mean is precisely the class of Bregman divergences, which we introduced in Section~\ref{sec:prelim_bregman}. (See Proposition~\ref{prop:bregman_max_ev}.)

In this section, our main result is a \textbf{generalization of Theorem~\ref{thm:agreement_accurate_quad} to (almost) arbitrary Bregman divergences} (see e.g.\ Theorem~\ref{thm:agreement_accurate_beta}). Additionally, we provide a \textbf{generalization of Aaronson's discretized protocol to arbitrary Bregman divergences} (Theorem~\ref{thm:agree_bregman}).

\subsection{Preliminaries on Bregman divergences} \label{sec:bregman_prelims}
Recall from Section~\ref{sec:prelim_bregman} the definition of Bregman divergence:
\bregmandef*

In this chapter, we are dealing with scalar quantities, so $n = 1$ and in particular $\mathcal{D} = [0, 1]$. So for us, the \emph{Bregman divergence} from $y$ to $x$ is
\[D_G(y \parallel x) := G(y) - G(x) - (y - x)G'(x).\]

Recall that the Bregman divergence with respect to $G(x) = x^2$ is precisely the squared distance. Another common Bregman divergence is the KL divergence, which corresponds to $G(x) = x \log x + (1 - x) \log(1 - x)$, the negative of the binary entropy function.

We generalize relevant notions such as agreement and accuracy to arbitrary Bregman divergences as follows. In the definitions below, $G: [0, 1] \to \RR$ is a differentiable, strictly convex function.

\begin{defin}
Let $a$ be Alice's expectation. Alice is \emph{$\epsilon$-accurate} if $\EE{D_G(\mu_{\sigma \tau} \parallel a)} \le \epsilon$, and likewise for Bob.\footnote{We discuss our choice of the order of these two arguments (i.e.\ why we do not instead consider the expectation of $D_G(a \parallel \mu_{\sigma \tau})$) in Appendix~\ref{appx:alternative_defs}.}
\end{defin}

We now define $\epsilon$-agreement, and to do so we first define the \emph{Jensen-Bregman divergence}.

\begin{defin}
For $a, b \in [0, 1]$, the \emph{Jensen-Bregman divergence} between $a$ and $b$ with respect to $G$ is
\[\JB_G(a, b) := \frac{1}{2} \parens{D_G \parens{a \parallel \frac{a + b}{2}} + D_G \parens{b \parallel \frac{a + b}{2}}} = \frac{G(a) + G(b)}{2} - G \parens{\frac{a + b}{2}}.\]
\end{defin}

The validity of the second equality can be easily derived from the definition of Bregman divergence. Note that the Jensen-Bregman divergence, unlike the Bregman divergence, is symmetric in its arguments. The Jensen-Bregman divergence is a lower bound on the average Bregman divergence from Alice and Bob to any other point (see Proposition~\ref{prop:chap8_bregman_facts}~\ref{item:min_bregman}).

\begin{defin}
Let $a$ and $b$ be Alice's and Bob's expectations, respectively. Alice and Bob \emph{$\epsilon$-agree} with respect to $G$ if $\JB_G(a, b) \le \epsilon$.
\end{defin}

In Appendix~\ref{appx:alternative_defs} we discuss alternative definitions of agreement and accuracy. The upshot of this discussion is that our definition of agreement is the \emph{weakest} reasonable one, and our definition of accuracy is the \emph{strongest} reasonable one. This means that the main result of this section -- that under a wide class of Bregman divergence, agreement implies accuracy -- is quite powerful: it starts with a weak premise and proves a strong conclusion.

\begin{defin}
Given an information structure $\mathcal{I}$, a communication protocol \emph{causes Alice and Bob to $\epsilon$-agree} on $\mathcal{I}$ with respect to $G$ if Alice and Bob $\epsilon$-agree with respect to $G$ at the end of the protocol. A communication protocol is an \emph{$\epsilon$-agreement protocol} with respect to $G$ if the protocol causes Alice and Bob to $\epsilon$-agree with respect to $G$ on every information structure.
\end{defin}

We also generalize the notion of rectangle substitutes to this domain, following \textcite{cw16}, which explored notions of substitutes for arbitrary Bregman divergences.

\begin{defin} \label{def:rect_subs_bregman}
Let $G: [0, 1] \to \RR$ be a differentiable, strictly convex function. An information structure $\mathcal{I} = (\Omega, \PP, \sigma, \tau, Y)$ satisfies rectangle substitutes with respect to $G$ if for every $S \subseteq \mathcal{S}, T \subseteq \mathcal{T}$, we have
\[\EE{D_G(Y \parallel \mu_{S \tau}) \mid S, T} - \EE{D_G(Y \parallel \mu_{\sigma \tau}) \mid S, T} \le \EE{D_G(Y \parallel \mu_{ST}) \mid S, T} - \EE{D_G(Y \parallel \mu_{\sigma T}) \mid S, T}.\]
\end{defin}

\textcite{cw16} explore the notion of weak substitutes with respect to arbitrary convex functions $G$ as well; just as before, $\mathcal{I}$ is said to satisfy the weak substitutes condition if the above inequality holds for $S = \mathcal{S}$ and $T = \mathcal{T}$.
The authors additionally explore in detail the weak substitutes condition with respect to negative entropy, i.e.\ for $D_G$ equal to the KL divergence. They show that if Alice and Bob have independent signals conditioned on $Y$, then the information structure satisfies weak substitutes with respect to this $G$. In fact, any such information structure also satisfies rectangle substitutes, because an information structure with conditionally independent signals retains the conditional independence when restricted to any sub-rectangle. The rectangle substitutes condition thus covers the specific case of conditionally independent signals under which \textcite{kong2023information} prove their accuracy result. On the other hand, the greater generality of our setting necessitates a different proof strategy.

Recall from Section~\ref{sec:prelim_pythag} that the Pythagorean theorem generalizes to arbitrary Bregman divergences:
\pythagbregman*
(In this chapter we are specifically interested in real-valued random variables, i.e.\ we take $n = 1$.) Just as we did with squared error, this general Pythagorean theorem allows us to rewrite the rectangle substitutes condition for Bregman divergences.

\begin{remark}
An information structure $\mathcal{I}$ satisfies rectangle substitutes with respect to $G$ if and only if for all $S \subseteq \mathcal{S}, T \subseteq \mathcal{T}$ we have
\begin{equation} \label{eq:rec_sub_bregman}
\EE{D_G(\mu_{\sigma \tau} \parallel \mu_{S \tau}) \mid S, T} \le \EE{D_G(\mu_{\sigma T} \parallel \mu_{ST}) \mid S, T}.
\end{equation}
Given the interpretation of Bregman divergences as measures of error, we can interpret the left side as Bob's expected error in predicting the truth while the right side is Charlie's expected error when predicting Alice's expectation (with Charlie as defined in Section~\ref{subsec:agreement-protocols}).
Both sides measure a prediction error due to not having Alice's signal, but from different starting points.
\end{remark}

\subsection{Generalizing the discretized protocol}
In Section~\ref{sec:general_accuracy}, we will show that under some weak conditions, protocols that cause Alice and Bob to agree with respect to $G$ also cause Alice and Bob to be accurate with respect to $G$. However, this raises an interesting question: are there protocols that cause Alice and Bob to agree with respect to $G$?
In particular, we are interested in natural expectation-sharing protocols.
Aaronson's discretized protocol is specific to $G(x) = x^2$, and it is not immediately obvious how to generalize it. We present the following generalization.

\begin{defin}
Let $G$ be a differentiable, strictly convex function, and let $M := \max_x G(x) - \min_x G(x)$. Choose $\epsilon > 0$. In the \emph{discretized protocol with respect to $G$ with parameter $\epsilon$,} on her turn (at time $t$), Alice sends ``medium'' if $D_G(\mu_{\sigma T_{t - 1}} \parallel \mu_{S_{t - 1} T_{t - 1}}) < \frac{\epsilon}{2}$, and otherwise either ``low'' or ``high'', depending on whether $\mu_{\sigma T_t}$ is smaller or larger (respectively) than $\mu_{S_t T_t}$. Bob acts analogously on his turn. At the start of the protocol, Alice and Bob use the information structure to independently compute the time $t_{\text{end}} \le \frac{24M(4M + \epsilon)}{\epsilon^2}$ that minimizes $\EE{D_G(\mu_{\sigma T_{t_{\text{end}}}} \parallel \mu_{S_{t_{\text{end}}} \tau})}$. The protocol ends at this time.
\end{defin}

\begin{restatable}{theorem}{agreebregman} \label{thm:agree_bregman}
The discretized protocol with respect to $G$ with parameter $\epsilon$ is an $\epsilon$-agreement protocol with respect to $G$ that requires $O \parens{\frac{M(M + \epsilon)}{\epsilon^2}}$ bits of communication.
\end{restatable}

Our proof draws inspiration from Aaronson's proof of the discretized protocol, but has significant differences. The key idea is to keep track of the monovariant $\EE{D_G(Y \parallel \mu_{S_t T_t})}$. This is Charlie's expected error (as measured by the Bregman divergence from the correct answer $Y$) after time step $t$ -- recall that Charlie is our name for a third-party observer of the protocol. Note that this quantity is at most $M$ and at least $0$. Hence, if we show that the quantity decreases by at least some value $\beta$ every time Alice and Bob do not $\epsilon$-agree, then we will have shown that Alice and Bob must $\epsilon$-agree within $\frac{\beta}{M}$ time steps. We defer the proof to Appendix~\ref{appx:bregman_omitted}.

\subsection{Approximate triangle inequality}
Our accuracy results in Section~\ref{sec:general_accuracy} will hold for a class of Jensen-Bregman divergences that satisfy an approximate version of the triangle inequality.
Specifically, we will require $\JB_G$ to satisfy the following \emph{$c$-approximate triangle inequality} for some $c>0$.

\begin{defin}
Given a differentiable, strictly convex function $G: [0, 1] \to \RR$ and a positive number $c$, we say that $\JB_G(\cdot, \cdot)$ satisfies the \emph{$c$-approximate triangle inequality} if for all $a, b, x \in [0, 1]$ we have
\[\JB_G(a, x) + \JB_G(x, b) \ge c \JB_G(a, b).\]
\end{defin}

It is possible to construct functions $G$ such that there is no positive $c$ for which $\JB_G$ satisfies the $c$-approximate triangle inequality. However, $\JB_G$ satisfies the $c$-approximate triangle inequality for some positive $c$ for essentially all natural choices of $G$.

\begin{prop} \label{prop:triangle} Let $G: [0, 1] \to \RR$ be a differentiable, strictly convex function.
\begin{enumerate}[label=(\roman*)]
\item \label{item:capprox_1} If $\sqrt{\JB_G(\cdot, \cdot)}$ satisfies the triangle inequality, then $\JB_G$ satisfies the $\frac{1}{2}$-approximate triangle inequality.
\item \label{item:capprox_2} If $G(x) = x^2$ (i.e.\ $D_G$ is squared distance) or if $G(x) = x \log x + (1 - x) \log(1 - x)$ (i.e.\ $D_G$ is KL divergence), then $\sqrt{\JB_G}$ satisfies the triangle inequality (and so $\JB_G$ satisfies the $\frac{1}{2}$-approximate triangle inequality).
\end{enumerate}
\end{prop}

\begin{proof}
Regarding Fact~\ref{item:capprox_1}, suppose that $\sqrt{\JB_G}$ satisfies the triangle inequality. Then for all $a, b, x$ we have $\sqrt{\JB_G(a, x)} + \sqrt{\JB_G(x, b)} \ge \sqrt{\JB_G(a, b)}$. Squaring both sides and observing that $\JB_G(a, x) + \JB_G(x, b) \ge 2\sqrt{\JB_G(a, x) \JB_G(x, b)}$ completes the proof.

Fact~\ref{item:capprox_2} is trivial for $G(x) = x^2$, since $\sqrt{\JB_G}$ is the absolute distance metric (times a constant factor). As for $G(x) = x \log x + (1 - x) \log(1 - x)$, see \textcite{es03}.
\end{proof}

The question of which convex functions $G$ have the property that $\sqrt{\JB_G}$ satisfies the triangle inequality has been explored in previous work \parencite{abb13, ccr08}.

\subsection{Generalized agreement implies generalized accuracy} \label{sec:general_accuracy}
In all of the results in this subsection, we consider the following setting: $G$ is a differentiable convex function; $c$ is a positive real number such that $\JB_G$ satisfies the $c$-approximate triangle inequality; and $\mathcal{I} = (\Omega, \PP, \sigma, \tau, Y)$ is an information structure that satisfies rectangle substitutes with respect to $G$.

We prove generalizations of Theorem~\ref{thm:agreement_accurate_quad}, showing that under the rectangle substitutes condition, if a protocol ends with Alice and Bob in approximate agreement, then Alice and Bob are approximately accurate.

\begin{theorem} \label{thm:agreement_accurate_beta}
Assume that $G$ is symmetric about the line $x = \frac{1}{2}$. For any communication protocol that causes Alice and Bob to $\epsilon$-agree on $\mathcal{I}$, and for any $\beta \ge \frac{2}{c} \epsilon$, Alice and Bob are
\[\parens{\frac{8}{c^2} \beta + 16 \parens{G(0) - G\parens{\parens{\frac{\epsilon}{\beta}}^{1/(1 - \log_2 c)}}}}\text{-accurate}\]
with respect to $G$ after the protocol terminates.
\end{theorem}

This result is not our most general, as it assumes that $G$ is symmetric, but this assumption likely holds for most use cases. To apply the result optimally, one must first optimize $\beta$ as a function of $G$. For example, setting $\beta = \epsilon^{r/(r + 1 - \log_2 c)}$ (with $r$ defined below) gives us the following corollary:\footnote{Corollary~\ref{cor:agreement_accurate_bregman_r} as stated (without the symmetry assumption) is actually a corollary of Theorem~\ref{thm:agreement_accurate_bregman}.}

\begin{corollary} \label{cor:agreement_accurate_bregman_r}
Assume that $G(0) - G(x), G(1) - G(1 - x) \le O(x^r)$. For any communication protocol that causes Alice and Bob to $\epsilon$-agree on $\mathcal{I}$, Alice and Bob are $O \parens{\epsilon^{r/(r + 1 - \log_2 c)}}$-accurate after the protocol terminates, where the constant hidden by $O(\cdot)$ depends on $G$.
\end{corollary}

\begin{remark}
Concretely, if $G'$ is bounded then we can choose $r = 1$, in which case our bound simplifies to $O \parens{\epsilon^{1/(2 - \log_2 c)}}$. If instead we assume that $c = \frac{1}{2}$ (as is the case if $\sqrt{\JB_G(\cdot, \cdot)}$ is a metric), then the bound is $O \parens{\epsilon^{r/(r + 2)}}$. If both of these are true, as is the case for $G(x) = x^2$, then the bound is $O(\epsilon^{1/3})$, which recovers our result in Theorem~\ref{thm:agreement_accurate_quad}.
\end{remark}

For $G$ equal to the negative of Shannon entropy (i.e.\ the $G$ for which $D_G$ is KL divergence), setting $\beta = \epsilon^{1/3} (\log 1/\epsilon)^{2/3}$ in Theorem~\ref{thm:agreement_accurate_beta} gives us the following corollary.

\begin{corollary} \label{cor:agreement_accurate_bregman_log}
If $G(x) = x \log x + (1 - x) \log(1 - x)$, then for any communication protocol that causes Alice and Bob to $\epsilon$-agree on $\mathcal{I}$, Alice and Bob are $O(\epsilon^{1/3} (\log 1/\epsilon)^{2/3})$-accurate after the protocol terminates.
\end{corollary}

Theorem~\ref{thm:agreement_accurate_beta} follows from our most general result about agreement implying accuracy:

\begin{theorem} \label{thm:agreement_accurate_bregman}
Let $\tilde{G}(x) := \max_{a, b: \abs{a - b} \le x}(G(a) - G(b))$ be the maximum possible difference in $G$-values of two points that differ by at most $x$, and let $\tilde{G}^*(x)$ be the concave envelope of $\tilde{G}$, i.e.
\[\tilde{G}^*(x) := \max_{0 \le a, b, w \le 1: wa + (1 - w)b = x} w \tilde{G}(a) + (1 - w) \tilde{G}(b).\]
For any communication protocol that causes Alice and Bob to $\epsilon$-agree on $\mathcal{I}$, and for any $\beta > 0$, Alice and Bob are
\[\parens{\frac{8}{c^2} \beta + 16 \tilde{G}^* \parens{\parens{\frac{\epsilon}{\beta}}^{1/(1 - \log_2 c)}}} \text{-accurate}\]
after the protocol terminates.
\end{theorem}

\begin{proof}
To prove Theorem~\ref{thm:agreement_accurate_bregman}, it suffices to prove the following lemma.
\begin{restatable}{lemma}{bobclosebregman} \label{lem:bob_close_bregman}
Let $G$ be a differentiable convex function on $[0, 1]$ and $c \in (0, 1)$ be such that $\JB_G$ satisfies the $c$-approximate triangle inequality. Let $\mathcal{I} = (\Omega, \PP, \sigma, \tau, Y)$ be an information structure that satisfies rectangle substitutes with respect to $G$. Let $\epsilon = \EE{\JB_G(\mu_{\sigma} \parallel \mu_{\tau})}$. Then for any $\beta > 0$, we have
\[\EE{D_G(\mu_{\sigma \tau} \parallel \mu_{\tau})} \le \frac{8}{c^2} \beta + 16 \tilde{G}^* \parens{\parens{\frac{\epsilon}{\beta}}^{1/(1 - \log_2 c)}}.\]
\end{restatable}

\noindent Let us first prove Theorem~\ref{thm:agreement_accurate_bregman} assuming Lemma~\ref{lem:bob_close_bregman} is true.\\

Consider any protocol that causes Alice and Bob to $\epsilon$-agree on $\mathcal{I}$. Let $S$ be the set of possible signals of Alice at the end of the protocol which are consistent with the protocol transcript, and define $T$ likewise for Bob.

Let $\epsilon_{ST} = \EE{\JB_G(\mu_{\sigma T}, \mu_{S \tau}) \mid S, T}$. Note that
\[\EE[S, T]{\epsilon_{ST}} = \EE[S, T]{\EE{\JB_G(\mu_{\sigma T}, \mu_{S \tau}) \mid S, T}} = \EE{\JB_G(\mu_{\sigma T}, \mu_{S \tau})} \le \epsilon.\]
Therefore, for any $\beta > 0$ we have
\begin{align*}
\EE{D_G(\mu_{\sigma \tau} \parallel \mu_{S \tau})} &\le \EE[S, T]{\frac{8}{c^2} \beta + 16 \tilde{G}^* \parens{\parens{\frac{\epsilon_{ST}}{\beta}}^{1/(1 - \log_2 c)}}}\\
&\le \frac{8}{c^2} \beta + 16 \tilde{G}^* \parens{\EE[S, T]{\parens{\frac{\epsilon_{ST}}{\beta}}^{1/(1 - \log_2 c)}}}\\
&\le \frac{8}{c^2} \beta + 16 \tilde{G}^* \parens{{\parens{\frac{\EE[S, T]{\epsilon_{ST}}}{\beta}}^{1/(1 - \log_2 c)}}}\\
&\le \frac{8}{c^2} \beta + 16 \tilde{G}^* \parens{\parens{\frac{\epsilon}{\beta}}^{1/(1 - \log_2 c)}}.
\end{align*}
In the first step, we apply Lemma~\ref{lem:bob_close_bregman} to the information structure $\mathcal{I}$ restricted to $S \times T$ -- that is, to $\mathcal{I}' = (\Omega', \PP', S, T, Y)$, where $\Omega' = \{\omega \in \Omega: \sigma \in S, \tau \in T\}$ and $\PP'[\omega] = \PP[\omega \mid \sigma \in S, \tau \in T]$. The next two steps follow by the convexity of $\tilde{G}^*$ and $x^{1/(1 - \log_2 c)}$, respectively.
\end{proof}

The basic outline of the proof of Lemma~\ref{lem:bob_close_bregman} is similar to that of Lemma~\ref{lem:bob_close_quad}. Once again, we partition $[0, 1]$ into $N$ intervals. Analogously to Equation~\ref{eq:delta_change_2}, and with $S^{(k(\sigma))}$ defined analogously, we find that
\[\EE{D_G(\mu_{\sigma \tau} \parallel \mu_{\tau})} \le \EE{D_G(\mu_{\sigma} \parallel \mu_{S^{(k(\sigma))}})} + \EE{D_G(\mu_{S^{(k(\sigma))} \tau} \parallel \mu_{\tau})}.\]
As before, we wish to upper bound each summand. However, the fact that the Bregman divergence is now arbitrary introduces complications. First, it is no longer the case that we can directly relate the length of an interval to the Bregman divergence between its endpoints. Second, we consider functions $G$ that become infinitely steep near $0$ and $1$ (such as the negative of Shannon entropy), which makes matters more challenging. This means that we need to be more careful when partitioning $[0, 1]$ into $N$ intervals: see Algorithm~\ref{alg:intervals} for our new approach. Additionally, bounding the second summand involves reasoning carefully about the behavior of the function $G$, which is responsible for the introduction of $\tilde{G}^*$ into the lemma statement. We defer the full proof of Lemma~\ref{lem:bob_close_bregman} to Appendix~\ref{appx:bregman_omitted}.

\section{Connections to markets} \label{sec:markets}

In this work, we established a natural condition on information structures, \emph{rectangle substitutes}, under which any agreement protocol results in accurate beliefs.
\emph{Expectation-sharing} protocols, where Alice and Bob take turns stating their current expected value (or discretizations thereof) are a particularly natural class of agreement protocols.

Expectation-sharing protocols have close connections to financial markets.
In markets, the actions of traders reveal partial information about their believed value for an asset, i.e., their expectation.
Specifically, a trader's decision about whether to buy or sell, and how much, can be viewed as revealing a discretization of this expectation.
In many theoretical models of markets (e.g.\ \textcite{ost12}) traders eventually reach agreement.
The intuition behind this phenomenon is that a trader who disagrees with the price leaves money on the table by refusing to trade.
Our work thus provides a lens into a well-studied question:\footnote{This is related to the \emph{efficient market hypothesis}, the problem of when market prices reflect all available information, which traces back at least to \textcite{fama1970efficient} and \textcite{hayek1945use}. Modern models of financial markets are often based on \textcite{kyle1985continuous}; see e.g.\ \textcite{ost12} and references therein for further information.} when are market prices accurate?
Our results can be viewed as generalizing and conceptually supporting the result presented in \textcite{kong2023information}, under which convergence in a popular prediction market design implies full information revelation in the prices.

An important caveat, however, is that traders behave strategically, and may not disclose their true expected value.
For example, a trader may choose to withhold information until a later point when doing so would be more profitable.
Therefore, to interpret the actions of traders as revealing discretized versions of their expected value, one first has to understand the Bayes-Nash equilibria of the market.
\textcite{cw16} study conditions under which traders are incentivized to reveal all of their information on their first trading opportunity.
They call a market equilibrium \emph{all-rush} if every trader is incentivized to reveal their information immediately.
Their main result, roughly speaking, is that there is an all-rush equilibrium if and only if the information structure satisfies \emph{strong substitutes} -- another strengthening of their weak substitutes condition.
This result is specific to settings in which traders have the option to reveal all of their information on their turn -- a setting that would be considered trivial from the standpoint of communication theory.

An exciting question for further study is therefore:
under what information structure conditions and market settings is it a Bayes-Nash equilibrium to follow an agreement protocol that leads to accurate beliefs?
In other words, what conditions give not only that agreement implies accuracy, but also that the market incentivizes participants to \emph{follow} the protocol?
Together with \textcite{cw16}, our work suggests that certain substitutes-like conditions could suffice.
\chapter{Deductive circuit estimation} \label{chap:elk}
\emph{This chapter presents work done at the Alignment Research Center, in collaboration with Paul Christiano, Jacob Hilton, V\'{a}clav Rozho\v{n}, and Mark Xu. It builds on the ideas of ``Formalizing the Presumption of Independence'' \parencite{cnx22}. The chapter assumes basic familiarity with theoretical computer science. The idea of forecast aggregation in the context of experts holding partial information -- especially as in Section~\ref{sec:bayesian_justifications} -- may be useful for context.}\\

\emph{Summary:} In this chapter, we turn our attention to estimating formal mathematical expressions -- such as the acceptance probability of a boolean circuit -- using deductive arguments. Much as it is possible to be uncertain about whether it will rain next week due to missing information or bounded computational resources, one can also be uncertain about the fraction of inputs on which a boolean circuit will output $1$. While there are simple \emph{inductive} methods for estimating this number, such as by running the circuit on a random sample of inputs, in this chapter we are interested in \emph{deductive} estimation. In other words, we are interested in designing an algorithm that takes as input formal arguments and observations about the structure of a circuit, and uses those observations in order to estimate the circuit's acceptance probability. Deductive estimates have the advantage that they may give insight about the \emph{reasons why} a circuit tends to accept certain classes of inputs.

Deductive estimates of mathematical quantities abound in fields such as number theory, theoretical computer science, and discrete mathematics, but there is little understanding of the rules that govern such estimation. Much as there is a formal definition of a mathematical proof and an algorithm that checks proofs for validity, there may be a formal definition of a valid deductive argument and an algorithm that uses such arguments to estimate mathematical quantities.

In this chapter, we search for such a formalization, focusing our attention on deductively estimating the acceptance probabilities of boolean circuits. We define two properties -- \emph{linearity} and \emph{respect for proofs} -- that a deductive estimation algorithm ought to satisfy, and then provide an efficient algorithm that satisfies those properties. We then show a negative result: that no efficient algorithm can satisfy linearity, respect for proofs, and another property that we call \emph{0-1 boundedness,} assuming that $P \neq PP$. We discuss additional desirable properties for a deductive estimation algorithm and then conclude with a discussion of potential applications to detecting anomalous neural network behavior.

More so than any previous chapter, this one shows a work in progress. It raises more questions than it answers, and some of the raised questions are not formally stated. However, the question of whether deductive estimation can be formalized seems like a fundamental one, and further progress in this direction may have important applications for building advanced AI systems safely.

\section{Introduction} \label{sec:chap9_intro}
Suppose we wish to estimate the acceptance probability $p(C)$ of a boolean circuit $C: \{0, 1\}^n \to \{0, 1\}$ on a uniformly random input -- that is, the fraction of inputs $x$ for which $C(x) = 1$. We can contrast two different types of approaches to such an estimation problem: \emph{inductive} approaches and \emph{deductive} approaches.

By an \emph{inductive} approach, we mean an approach based on observing the input-output behavior of $C$ on some inputs. For example, one inductive approach is to sample inputs to $C$ uniformly at random and observe the fraction of inputs that $C$ accepts (outputs $1$ on).\footnote{We consider pseudorandom sampling to also be inductive. The use of randomness is not required for an argument to be considered inductive.} Such an estimation procedure is very effective, at least in terms of estimating $p(C)$ with a small  additive error. However, it does not provide any insight about \emph{why} $C$'s acceptance probability is what it is.

By a \emph{deductive} approach, we mean an approach that uses structural observations about $C$ in order to estimate $p(C)$. We call such structural observations \emph{deductive arguments.}\footnote{We use the term ``deductive argument" instead of the term ``heuristic argument" (as used in e.g.\ \textcite{cnx22}) to stress the difference with inductive arguments (which we do not consider in this chapter) and to emphasize that a proof is a special type of deductive argument.} While an inductive approaches would generally treat a circuit as a black box and estimate its acceptance probability using input-output behavior, deductive arguments treat circuits as white boxes. Let us illustrate with a few examples.

\begin{example} \label{ex:abc}
    Suppose that $C$ takes as input a triple $(a, b, c)$ of positive integers, computes $\max(a, b)$ and $\max(b, c)$, and accepts if they are equal. A deductive argument about $p(C)$ might point out that if $b$ is the largest of the three integers, then $\max(a, b) = b = \max(b, c)$ and so $C$ will accept, and that this happens with probability roughly $\frac{1}{3}$.
\end{example}

\begin{example} \label{ex:sha}
    Suppose that $C(x)$ computes $\text{SHA-256}(x)$ (the output of SHA-256 is a 256-bit string) and accepts if the first 128 bits (interpreted as an integer) are larger than the last 128 bits. One can make a deductive argument about $p(C)$ by making repeated use of the \emph{presumption of independence} \parencite{cnx22}. In particular, the SHA-256 circuit consists of components (such as the majority circuit, or addition modulo $2^{32}$) that produce uniformly random outputs on independent, uniformly random inputs. Thus, a deductive argument that repeatedly presumes that the inputs to each component are independent concludes that the output of SHA-256 consists of independent, uniformly random bits. It would then follow that the probability that the first 128 bits of the output are larger than the last 128 bits is $\frac{1}{2}$.
\end{example}

\begin{example} \label{ex:cnf}
    Suppose that $C$ is a particular CNF with $k$ clauses of three literals each. A deductive argument about $p(C)$ might point out that $C$ is a CNF with $k$ clauses of three literals each, and that on average, CNFs with this structure accept $\parens{\frac{7}{8}}^k$ fraction of inputs. A more sophisticated argument might point out particular structural patterns in $C$ (e.g.\ that the variable $x_1$ appears with positive sign in every clause) and reason about the average acceptance probability of CNFs with that structure.
\end{example}

\begin{example} \label{ex:primes}
    Suppose that $C$ takes as input an integer $k$ between $e^{100}$ and $e^{101}$ and accepts if $k$ and $k + 2$ are both prime. A deductive argument about $p(C)$ might point out that the density of primes in this range is roughly $1\%$, so if we presume that the event that $k$ is prime and the event that $k + 2$ is prime are independent, then we get an estimate of $0.01\%$. A more sophisticated argument might take this one step further by pointing out that if $k$ is prime, then $k$ is odd, so $k + 2$ is odd, which makes $k + 2$ more likely to be prime (by a factor of $2$), suggesting an estimate of $0.02\%$. A yet more sophisticated argument might point out that additionally, if $k$ is prime, then $k$ is not divisible by $3$, which makes $k + 2$ \emph{more} likely to be divisible by $3$, which reduces the chance that $k + 2$ is prime, and would make similar arguments for divisibility by $5$, $7$, and so on.
\end{example}

In a sense, deductive arguments generalize proofs. A formal proof about the value of $p(C)$ is a type of deductive argument, but in many cases it is infeasible to prove even basic facts about a circuit's acceptance probability. In Example~\ref{ex:sha}, for instance, there is likely no short proof that $p(C)$ is between $\frac{1}{3}$ and $\frac{2}{3}$, but there is a very simple deductive argument. Deductive arguments let us draw conclusions based on circuits' structural properties in many cases where proofs do not.\\

Inductive approaches to circuit estimation hold some obvious advantages over deductive ones. First, they are fast and straightforward. Second, they have strong accuracy guarantees: by sampling random inputs, one can efficiently estimate $p(C)$ with a small additive error with high probability. By comparison, a deductive estimate can be incorrect (for example, the na\"{i}ve argument that gave the $0.01\%$ estimate in Example~\ref{ex:primes} was substantially wrong). So why bother with deductive estimates?

One answer is that deductive estimation may be helpful in situations where sampling-based approaches will fail. For example, suppose we wish to estimate the acceptance probability of a CNF $C$ that has a low acceptance probability, with a small multiplicative error. Sampling inputs will likely fail to find any satisfying assignments. On the other hand, one could hope to be able to reason deductively about $C$'s acceptance probability if given deductive arguments based on $C$'s structure.

More importantly, however, deductive arguments can give us a \emph{mechanistic understanding} of a circuit and provide insight about \emph{why} the circuit has a certain acceptance probability. Such understanding allows us to answer questions like: does $C$ accept inputs $\vect{x}$ and $\vect{y}$ ``for the same reason'' or ``for different reasons''? In Example~\ref{ex:abc}, $C$ might accept an input because $b$ is larger than both $a$ and $c$, or because $a = c$. Deductive arguments can allow us to draw such distinctions; inductive arguments do not. Gaining such a mechanistic understanding is an important motivation for exploring deductive circuit estimation.\\

The purpose of this line of work is to create a deductive estimation algorithm. A \emph{deductive estimation algorithm} takes as input a boolean circuit,\footnote{We are also interested in the more general problem of estimating the expected outputs of arithmetic circuits and of computer programs more generally, but in this chapter we focus on boolean circuits. Boolean circuits are a special case that may capture the core difficulties of the more general problem. Further, boolean circuits have a canonical choice for the distribution of inputs (uniformly random bit strings), and some of our discussion (e.g.\ our definition of linearity in Section~\ref{sec:lin_resp}) is simplified in the case of boolean cirucits.} together with a list of deductive arguments about the circuit, and outputs an estimate of the circuit's acceptance probability based on the provided arguments. Creating such an algorithm requires surmounting several challenges. First, it is necessary to specify a formal language in which deductive arguments may be stated; deductive arguments given as input to the estimation algorithm will be written in this formal language. Second, desiderata for the algorithm should be defined: what properties should a deductive estimation algorithm satisfy? Finally, the third challenge is to create an algorithm that satisfies those properties.

An analogy to formal proof verification may be instructive. The central task of proof verification is the (solved) problem of finding a verification algorithm that, given a mathematical statement and an alleged proof of the statement, verifies the proof. In our setting, the circuit (whose acceptance probability we want to estimate) is analogous to the mathematical statement, and the list of deductive arguments is analogous to the proof. Table~\ref{table:dce_fpv_analogy} describes the analogy in more detail.\\

\begin{table}[ht]
    \centering
    \begin{tabular}{c||c}
    \textbf{Deductive circuit estimation} & \textbf{Formal proof verification}\\
    \hline
    Deductive estimation algorithm & Proof verifier\\
    \hline
    Boolean circuit & Formal mathematical statement\\
    \hline
    List of deductive arguments & Alleged proof of statement\\
    \hline
    Formal language for deductive arguments & Formal language for proofs\\
    \hline
    Desiderata for estimation algorithm & Soundness and completeness\\
    \hline
    Algorithm's estimate of circuit & Proof verifier's output (accept or reject)\\
    \end{tabular}
	\caption[Analogy between deductive circuit estimation and formal proof verification]{We are interested in developing a deductive estimation algorithm for boolean circuits. There are similarities between this task and the (solved) task of developing an algorithm for verifying formal proofs of mathematical statements. This table illustrates the analogy. Importantly, the purpose of a deductive estimation algorithm is to incorporate the deductive arguments that it has been given as input, rather than to generate its own arguments. The output of a deductive estimation algorithm is only as sophisticated as the arguments that it has been given.}
	\label{table:dce_fpv_analogy}
\end{table}

In Section~\ref{sec:linreg}, we begin with a discussion of deductive circuit estimation via linear regression. That is, we will be interested in producing a linear estimator for a class of circuits in terms of a list of features. We illustrate this approach with an example of CNF estimation. The purpose of this section is to provide an example of what deductive arguments for circuit estimation might look like, and how a deductive estimation algorithm might incorporate those arguments into an estimate of the acceptance probability of a circuit.

In Section~\ref{sec:lin_resp}, we examine the problem of deductive circuit estimation in more generality, asking the question: what properties should a deductive estimation algorithm satisfy? We introduce two properties -- which we call \emph{linearity} and \emph{respect for proofs} -- and describe an algorithm that satisfies both properties. Then, in Section~\ref{sec:further_desiderata}, we discuss additional properties. We introduce a basic property called \emph{0-1 boundedness} and show that it is impossible to satisfy in addition to linearity and respect for proofs, unless $P = PP$. We discuss two other properties -- \emph{iterated estimation} and \emph{pulling out known factors} -- that it would be desirable to satisfy, and leave open the question of whether they can be satisfied.

Finally, in Section~\ref{sec:mad}, we conclude with a discussion of potential applications to detecting anomalous behavior in neural networks.

\subsection{Related work}
The work most closely related to ours is \textcite{cnx22}, which posed the question of whether it is possible to formalize the process of deductively estimating the value of a mathematical expression, such as the expectation of a random variable. The authors posited the \emph{presumption of independence} as an underlying principle of deductive estimation: two random variables should be presumed independent until an argument to the contrary is presented. Thus, a deductive estimation algorithm might start by assuming that all sub-expressions are independent, and then update its estimate after incorporating knowledge about various dependencies. This chapter attacks the problem of deductive estimation from a different perspective, but is a continuation of that work.

There has not been much other prior work on formalizing deductive estimation. Perhaps the closest is \textcite{tao12}, which also posits that two quantities should be presumed independent unless there is a good reason to the contrary, and uses this principle to give a heuristic justification for the ABC conjecture from number theory. Heuristic arguments for mathematical claims are common in fields such as number theory, theoretical computer science, and discrete mathematics. For example, \textcite{eu71} give a heuristic argument for the truth of Fermat's last theorem; \textcite{mz02} use heuristic methods to analyze random $k$-SAT instances; and \textcite{ch19} heuristically estimate the frequencies of certain patterns in Conway's game of life. However, little prior work attempts to formalize the rules governing such arguments.

\textcite{barak15, barak16} are also closely related to our work. Barak asks whether it is possible to construct an estimation algorithm for a given quantity that appears reasonable to a broad class of observers. By contrast, in our framing, ``observers" are deductive arguments that are given as input. Our perspective is in some ways less ambitious and in some ways more: on the one hand, we want our estimate to appear reasonable to an observer (i.e.\ incorporate a deductive argument) only if the observer is explicitly given as input; on the other hand, this may allow us to consider a much wider range of possible observers.

More recently, \textcite{gowers23} posited a \emph{no-coincidence principle:} ``If an apparently outrageous coincidence happens in mathematics, then there is a reason for it.'' In the language of this chapter, we might instead write: ``If a circuit exhibits a surprising behavior, then there is a short deductive argument that explains the behavior.'' (An example of a surprising behavior might be that the acceptance probability of a circuit is much larger than a na\"{i}ve estimate would suggest.\footnote{Informally speaking: for a circuit that always outputs $1$, there should be a deductive argument that makes it seem \emph{plausible} that the circuit always outputs $1$. On the other hand, for most circuits that do not always output $1$, no such deductive argument should exist. See \textcite[\S C]{cnx22} for further discussion.})

There is a significant literature on the problem of circuit estimation. Inductive estimates are common, often under the name \emph{approximate model counting.} Beyond straightforward uniform sampling, there are more algorithms based on more sophisticated methods such as Markov chain Monte Carlo \parencite{ws05}. If given access to an NP oracle, hashing-based approaches can give strong guarantees \parencite{ym23}. See \textcite{cmv21} for a survey of this area. There has also been work on derandomization-based deterministic algorithms for approximate DNF counting \parencite{gmr13}.

There has also been extensive research into heuristic methods for the satisfiability problem (see e.g.\ \textcite{gpfw96}). This work is similar in spirit to ours, but our goal is different. Rather than finding particular heuristic algorithms and validating them on particular instances, we are seeking a unified theoretical framework for making estimates based on a very large class of arguments.

\section{Circuit estimation via linear regression} \label{sec:linreg}
Often, families of circuits are easier to reason about than the individual circuits they contain. For example, it may be difficult to compute the acceptance probability of a given 3CNF with $k$ clauses, but it is easy to compute the average acceptance probability of \emph{all} 3CNFs with $k$ clauses. For this reason, estimating the acceptance probability of a circuit $C$ is made easier by instead estimating the acceptance probabilities of \emph{all} circuits in a family that contains $C$.

Let us denote the average acceptance probability of a (finite) family of boolean circuits $\mathscr{C}$ as $p(\mathscr{C})$. That is, $p(\mathscr{C}) := \EE[C \sim \mathscr{C}]{p(C)}$. Consider a circuit $C: \{0, 1\}^n \to \{0, 1\}$ for which $p(C)$ is difficult to compute. Let $\mathscr{C}$ be a family of circuits containing $C$. In a sense, $p(\mathscr{C})$ is a reasonable (if uninformative) estimate for $p(C)$, because this estimate is correct on average over $\mathscr{C}$.

We can ask for more informative estimates: ones that use more information about a circuit than just the fact that it belongs to $\mathscr{C}$. In general, we can hope to estimate the acceptance probabilities of the circuits in $\mathscr{C}$ by using features of those circuits.

To be more precise, suppose that we have $m$ \emph{features}: efficiently computable functions $\varphi_1, \dots, \varphi_m: \mathscr{C} \to \RR$. Suppose further that we can compute the average value of each feature on $\mathscr{C}$ (that is, $\EE[C \sim \mathscr{C}]{\varphi_i(C)}$ for all $i \in m$, covariances between the features (that is, $\text{Cov}_{C \sim \mathscr{C}}(\varphi_i(C), \varphi_j(C))$ for $i, j \in m$), and the covariance of each feature with the acceptance probability (that is, $\text{Cov}_{C \sim \mathscr{C}}(p(C), \varphi_i(C))$ for all $i \in m$). Then the best linear estimator of $p(C)$ in terms of the features, as measured by average squared error over $\mathscr{C}$, is given by the least squares regression formula:

\singlespacing
\begin{equation} \label{eq:linreg}
    \hat{p}(C) = p(\mathscr{C}) + \begin{pmatrix}\text{Cov}_{C' \sim \mathscr{C}}(p(C'), \varphi_1(C')) \\ \vdots \\ \text{Cov}_{C' \sim \mathscr{C}}(p(C'), \varphi_m(C'))\end{pmatrix}^\top \Sigma_{\pmb{\varphi}}^{+} \begin{pmatrix}\varphi_1(C) - \EE[C' \sim \mathscr{C}]{\varphi_1(C')} \\ \vdots \\ \varphi_n(C) - \EE[C' \sim \mathscr{C'}]{\varphi_n(C')}\end{pmatrix},
\end{equation}
\doublespacing

where $\Sigma_{\pmb{\varphi}}$ denotes the covariance matrix of the features and $\Sigma_{\pmb{\varphi}}^+$ is its Moore-Penrose pseudoinverse.

\subsection{Example: CNF estimation} \label{sec:cnf_est}
In this section, we work with a family $\mathscr{C}$ consisting of $2^{3k}$ 3CNFs on $n$ variables with $k$ clauses: specifically, these CNFs will be identical except for the sign of each literal. More formally, for every $j \in [k]$ and $\ell \in [3]$, fix an index $i_{j, \ell}$ in $[n]$. Then, for every $\vect{b} \in \{0, 1\}^{k \times 3}$, we define
\[C_\vect{b}(\vect{x}) := (x_{i_{1, 1}} = b_{1, 1} \vee x_{i_{1, 2}} = b_{1, 2} \vee x_{i_{1, 3}} = b_{1, 3}) \wedge \dots \wedge (x_{i_{k, 1}} = b_{k, 1} \vee x_{i_{k, 2}} = b_{k, 2} \vee x_{i_{k, 3}} = b_{k, 3}).\]
That is, the indices $i_{j, \ell}$ specify which variable appears at each position in the CNF, and $b_{j, \ell}$ specifies the sign of the $\ell$-th literal in clause $j$. Then, for a particular choice of indices $i_{1, 1}, \dots, i_{k, 3}$ (which is implicit in the notation), $\mathscr{C}$ is the set of CNFs $C_\vect{b}$ for all $\vect{b} \in \{0, 1\}^{k \times 3}$.

\begin{claim} \label{claim:78k}
    $p(\mathscr{C}) = (7/8)^k$.
\end{claim}

\begin{proof}
    The quantity $p(\mathscr{C})$ is equal to the fraction of all choices of $(\vect{x}, \vect{b})$ such that $C_\vect{b}(\vect{x}) = 1$. Choose any given $\vect{x}$; we ask: if we select a random $\vect{b}$, what is the probability that $C_\vect{b}(\vect{x}) = 1$? For each of the $k$ clauses, there is a $\frac{7}{8}$ probability that the clause will be satisfied by the chosen $\vect{b}$, and these probabilities are independent.
\end{proof}

Thus, we could estimate $p(C_\vect{b})$ as $(7/8)^k$ for all $\vect{b}$. As we have discussed, this is a reasonable estimate, but not a very informative one. To refine this estimate, we consider $n$ features -- one for each variable -- which we call $\varphi_1, \dots, \varphi_n$:

\begin{defin}
    Given a circuit $C \in \mathscr{C}$, let $\mathscr{C}_i(C)$ (for $i \in [n]$) be the set of circuits in $\mathscr{C}$ that agree define with $C$ on the signs of all instances of the variable $x_i$. Define $\varphi_i(C) := p(\mathscr{C}_i(C))$, i.e.\ the average acceptance probability of these circuits.
\end{defin}

\begin{example}
Suppose that $C(x) = (x_1 = 1 \vee x_2 = 1 \vee x_3 = 1) \wedge (x_1 = 0 \vee x_2 = 1 \vee x_4 = 1)$. There are $2^4 = 16$ circuits in $\mathscr{C}_1(C)$: specifically all circuits of the form
\[(x_1 = 1 \vee x_2 = b_{1, 2} \vee x_3 = b_{1, 3}) \wedge (x_1 = 0 \vee x_2 = b_{2, 2} \vee x_4 = b_{2, 3}),\]
for some choice of bits $b_{1, 2}, b_{1, 3}, b_{2, 2}, b_{2, 3}$. We can compute $\varphi_1(C)$ by conditioning on the value of $x_1$:
\[\varphi_1(C) = \frac{1}{2} \pr[\vect{x}, \vect{b}]{x_2 = b_{1, 2} \vee x_3 = b_{1, 3}} + \frac{1}{2} \pr[\vect{x}, \vect{b}]{x_2 = b_{2, 2} \vee x_4 = b_{2, 3}} = \frac{3}{4}.\]
\\
\end{example}

We can think of each $\varphi_i(C)$ as a more informative estimate of $p(C)$ than $p(\mathscr{C})$ is: it is still correct on average, but now uses more information about $C$.\footnote{In general, features do not need to be reasonable estimates for acceptance probability -- they just happen to be in this case.} An analogy to forecast aggregation may be instructive. Suppose that a circuit $C$ is drawn at random from $\mathscr{C}$, and that there are $n$ experts: Expert $i$ knows the signs of all instances of $x_i$ and no other signs. Then $p(\mathscr{C})$ is the prior for the value of $p(C)$, and $\varphi_i(C)$ is Expert $i$'s estimate of the value of $p(C)$.

Note that $\varphi_i(C)$ is easy to compute. In particular, we have
\[\varphi_i(C) = \pr[\vect{x} \sim \{0, 1\}^n, C' \sim \mathscr{C}_i(C)]{C'(\vect{x}) = 1} = \frac{1}{2} \parens{\pr[\vect{x}: x_i = 1, C' \sim \mathscr{C}_i(C)]{C'(\vect{x}) = 1} + \pr[\vect{x}: x_i = 0, C' \sim \mathscr{C}_i(C)]{C'(\vect{x}) = 1}}.\]
Now, let $C[x_i = 1]$ be the CNF obtained by plugging $x_i = 1$ into $C$ and simplifying. Note that $\pr[\vect{x}: x_i = 1, C' \sim \mathscr{C}_i(C)]{C'(\vect{x}) = 1}$ is equal to the average acceptance probability of all circuits that are identical to $C[x_i = 1]$ except for the signs of the literals -- and finding this average probability is easy (see Claim~\ref{claim:78k}). Computing $\pr[\vect{x}: x_i = 0, C' \sim \mathscr{C}_i(C)]{C'(\vect{x}) = 1}$ is similarly easy.

Thus, we can obtain $n$ estimates $\varphi_1(C), \dots, \varphi_n(C)$, each of which is more informative than $p(\mathscr{C})$. We can combine them into a single estimate of $p(C)$ with linear regression, as per Equation~\ref{eq:linreg}. In particular:
\begin{itemize}
    \item $\EE[C' \sim \mathscr{C}]{\varphi_i(C')} = p(\mathscr{C})$ for all $i$.
    \item The $n$ features are uncorrelated. This is because the $\mathscr{C}_i(C)$ depends only on the signs of the variable $x_i$ in $C$, and these bits are independent for each value of $i$. Thus, $\Sigma_{\pmb{\varphi}}$ is a diagonal matrix containing the feature variances.
    \item $\varphi_i(C')$ is a calibrated estimate of $p(C')$, meaning that $\EE[C' \sim \mathscr{C}]{p(C') \mid \varphi_i(C')} = \varphi_i(C')$. This means that $\text{Cov}_{C' \sim \mathscr{C}}(p(C'), \varphi_i(C')) = \text{Var}_{C' \sim \mathscr{C}}(\varphi_i(C'))$. Therefore, we have

    \singlespacing
    \[\begin{pmatrix}\text{Cov}_{C' \sim \mathscr{C}}(p(C'), \varphi_1(C')) \\ \vdots \\ \text{Cov}_{C' \sim \mathscr{C}}(p(C'), \varphi_m(C'))\end{pmatrix}^\top \Sigma_{\pmb{\varphi}}^{-1} = \begin{pmatrix}1 \\ \vdots \\ 1\end{pmatrix}.\]
    \doublespacing
\end{itemize}
Putting these facts together tells us that the linear regression estimator for $p(C)$ is given by
\begin{equation} \label{eq:cnf_linreg}
    \hat{p}(C) = p(\mathscr{C}) + \sum_{i = 1}^n (\varphi_i(C) - p(\mathscr{C})) = \sum_{i = 1}^n \varphi_i(C) - (n - 1) p(\mathscr{C}).
\end{equation}
Intuitively, the estimator $\hat{p}$ treats $p(\mathscr{C})$ as a prior and treats the quantities $\varphi_i(C) - p(\mathscr{C})$ as independent updates from the prior, which it combines additively.\footnote{We can think of this additive estimate in the context of Section~\ref{sec:bayesian_justifications}. There, we saw that if experts report estimates $Y_1, \dots, Y_m$ for a mean-zero quantity $Y$, then the best linear estimate of $Y$ in terms of $Y_1, \dots, Y_m$ is given by $\text{diag}(\Sigma)^\top \Sigma^{-1} \vect{Y}$, where $\vect{Y} = (Y_1, \dots, Y_m)$ and $\Sigma$ is the covariance matrix of the experts' estimates. If we consider $Y = p(C) - p(\mathscr{C})$, then $Y_i = \varphi_i(C) - p(\mathscr{C})$, and $\Sigma$ is a diagonal matrix (since the $Y_i$'s are independent). This perspective also yields the estimate given by Equation~\ref{eq:cnf_linreg}.}

\begin{figure}[ht]
    \centering
    \includegraphics[scale=0.8]{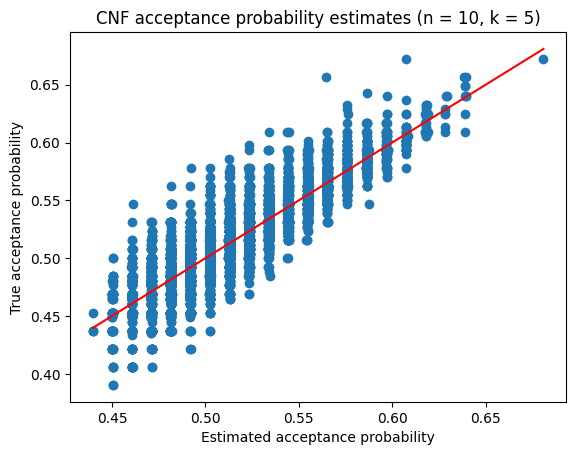}
    \caption[Scatter plot of CNF acceptance probability estimates]{Ten thousand 3CNFs on $n = 10$ variables with $k = 5$ clauses were randomly generated. This scatter plot shows each CNF's estimated acceptance probability $\hat{p}(C)$ on the $x$-axis and its true acceptance probability $p(C)$ on the $y$-axis. In this sample, $\hat{p}(C)$ explains 78\% of the variance in $p(C)$ (that is, $r^2 \approx 0.78$).}
    \label{fig:cnf_estimation}
\end{figure}

It turns out that $\hat{p}(C)$ is a substantial improvement upon the prior $p(\mathscr{C})$. Figure~\ref{fig:cnf_estimation} plots $\hat{p}(C)$ versus $p(C)$ for ten thousand randomly generated 3CNFs on ten variables with five clauses. In this sample, $\hat{p}(C)$ explains 78\% of the variance in $p(C)$.

By adding more features, it would be possible to further refine the estimate, so that it explains even more variance in $p(C)$. For example, one could add features based on the signs of \emph{pairs} of variables, or based on other structural properties of the CNFs in $\mathscr{C}$.

\subsection{Challenges for linear regression} \label{sec:esitmating_features}
One challenge faced by the linear regression estimator given in Equation~\ref{eq:linreg} is that it may output estimates $\hat{p}(C)$ that are not in $[0, 1]$. This makes sense, given that we sought an estimator that minimizes average squared error over $\mathscr{C}$: if $p(C) = 0.1$, the estimate $\hat{p}(C) = -0.2$ is no worse than the estimate $\hat{p}(C) = 0.4$. On the other hand, such errors mean that the linear regression estimator may output estimates that are obviously incorrect. These errors can be corrected for individual circuits by introducing additional features. However, in Section~\ref{sec:01_boundedness}, we show that requiring all estimates to lie between $0$ and $1$ must come at the expense of a different desirable property.

A more significant challenge to using Equation~\ref{eq:linreg} for deductive circuit estimation is that one may want to use features whose average values cannot be easily computed, or whose covariances with $p(C)$ or with each other cannot be easily computed. In such situations, it makes sense to deductively estimate those quantities. How can we do this?

We can think of each feature $\varphi_i$ as being computed by some arithmetic circuit.\footnote{The input to $\varphi_i$ is ordinarily a circuit $C$; when thinking of $\varphi_i$ as a circuit, we can think of its input as a string of bits that describes $C$.} Equation~\ref{eq:linreg} applies just as well for estimating expected values of arithmetic circuits, so we can hope to use linear regression in turn to estimate the quantities we need. However, this is difficult for several reasons. First, it is unclear what family of circuits $\varphi_i$ should be thought of as belonging to (that is, over what family of circuits to do the recursive linear regression step). Second, if the features are almost as complex as $C$ itself, then estimating the covariances between them means estimating quantities that are more complex than $C$. Third, this recursive estimation scheme would need a base case: what if no features are supplied? That is, we want a reasonable ``default guess'' for the expressions in Equation~\ref{eq:linreg}. These challenges are priorities for future work, because fully solving them might yield a much more generic deductive estimation algorithm.

\section{Linearity and respect for proofs} \label{sec:lin_resp}
So far we have discussed how a deductive estimation algorithm might behave when given certain kinds of arguments (features for linear regression). In this section, we take a different perspective: what properties should a deductive estimation algorithm satisfy? We define two properties, which we call \emph{linearity} and \emph{respect for proofs,} and give a deductive estimation algorithm that satisfies both properties.

Let us introduce notation (mostly following \textcite{cnx22}). We will use the letter $G$ to denote a deductive estimation algorithm. $G$ will take two inputs: first, a circuit $C$ (whose acceptance probability it needs to estimate), and second, a set of deductive arguments $\{\pi_1, \dots, \pi_m\}$. We will write $G(C \mid \pi_1, \dots, \pi_m)$ to denote the output of $G$ on these inputs.\footnote{We choose to use the $\mid$ symbol because $G$ needs to estimate $p(C)$ \emph{given} -- i.e.\ in light of -- the arguments $\pi_1, \dots, \pi_m$. As we will discuss later, we may want $G$ to behave akin to a conditional expectation.} For convenience, we will use $\Pi$ to denote a set of arguments and write $G(C \mid \Pi)$.

How do features for linear regression, as in Section~\ref{sec:linreg}, translate into arguments? Since we have not specified a formal language for deductive arguments, we cannot give an exact answer; however, the arguments should specify $\mathscr{C}$ and the features $\varphi_i(C)$. The arguments should also verify the values of the quantities in Equation~\ref{eq:linreg} (namely, $p(\mathscr{C})$, $\text{Cov}(p(C'), \varphi_i(C'))$, $\Sigma_{\pmb{\varphi}}$, $\varphi_i(C)$, and $\EE{\varphi_i(C')}$), so that $G$ can compute $\hat{p}(C)$. For example, there should be a proof of the value of $p(\mathscr{C})$ among the arguments given to $G$.\footnote{More generally, $G$ should be able to accept deductive arguments about the values of these quantities, instead of exact proofs, as briefly discussed in Section~\ref{sec:esitmating_features}.}\\

We now define the following notation for circuit substitutions, and then introduce the aforementioned linearity and respect for proofs properties.

\begin{defin}
    Let $C: \{0, 1\}^n \to \{0, 1\}$ be a circuit with input variables $x_1, \dots, x_n$. For $b \in \{0, 1\}$, we define $C[x_i = b]$ to be the circuit on $n - 1$ variables $x_1, \dots, x_{i - 1}, x_{i + 1}, \dots, x_n$ obtained by setting $x_i = b$ in $C$. More generally, given a partial assignment $A$ to some $\ell$ of the $n$ variables, we define $C[A]$ to be the circuit on the remaining $n - \ell$ variables that is obtained from $C$ by setting the variables in $\ell$ to their corresponding values in $A$. A particular input $\vect{x}$ can be thought of as a full assignment; thus, $C[\vect{x}]$ is a circuit that takes no inputs and outputs $C(\vect{x})$.
\end{defin}

\begin{defin} \label{def:linearity}
    A deductive estimation algorithm $G$ satisfies \emph{linearity} if for all $n$, $C: \{0, 1\}^n \to \{0, 1\}$, $i \in [n]$, and $\Pi$, we have
    \[G(C \mid \Pi) = \frac{1}{2}(G(C[x_i = 0] \mid \Pi) + G(C[x_i = 1] \mid \Pi)).\]
\end{defin}

In other words: a basic fact about circuits is that $p(C) = \frac{1}{2}(p(C[x_i = 0]) + p(C[x_i = 1]))$, and $G$'s estimates should respect this fact. Note that linearity entails the following quite strong property: if $C$ is a circuit on $n$ inputs, then
\[G(C \mid \Pi) = \frac{1}{2^n} \sum_{\vect{x} \sim \{0, 1\}^n} G(C[\vect{x}] \mid \Pi).\]
We note, however, that $G(C[\vect{x}] \mid \Pi)$ need not equal $C(\vect{x})$. Although calculating $C(\vect{x})$ is easy, $G$ does not necessarily do so: $G$'s task is not to estimate a circuit as well as it can, but rather to fully incorporate the arguments in $\Pi$ into its estimate. Unless $\Pi$ contains an explicit computation of $C(\vect{x})$, $G(C[\vect{x}] \mid \Pi)$ may be different from $C(\vect{x})$.

We now introduce the ``respect for proofs'' property. Recall from the introduction that deductive arguments can be thought of as generalizing proofs. The respect for proofs property essentially states that proofs constitute valid deductive arguments that $G$ must respect.

\begin{defin} \label{def:respect_for_proofs}
    A deductive estimation algorithm $G$ satisfies \emph{respect for proofs} if there is a polynomial-time algorithm that, given boolean circuits $C_1, \dots, C_k$ and a proof that $\sum_{i = 1}^k a_i p(C_i) \le b$ for some rational constants $a_1, \dots, a_k, b$, outputs an argument $\pi$ such that for every set of arguments $\Pi$ containing $\pi$, it is the case that $\sum_{i = 1}^k a_i G(C_i \mid \Pi) \le b$.
\end{defin}

In other words, if $G$ is given a proof of a linear constraint on the acceptance probabilities of some circuits, then $G$'s estimates of the acceptance probabilities must also obey that linear constraint. For example:
\begin{itemize}
    \item For a circuit $C$, a proof that $p(C) \le \frac{1}{2}$ can be translated into an argument $\pi$ such that $G(C \mid \pi) \le \frac{1}{2}$. In fact, Definition~\ref{def:respect_for_proofs} gives a stronger guarantee: for all further arguments $\pi_1, \pi_2, \dots, \pi_m$, we must still have $G(C \mid \pi, \pi_1, \dots, \pi_m) \le \frac{1}{2}$.
    \item Similarly, a proof that $p(C) \ge \frac{1}{2}$ can be translated into an argument $\pi'$. If $G$ is given \emph{both} $\pi$ (as in the previous bullet point) and $\pi'$, then $G(C \mid \pi, \pi', \pi_1, \dots, \pi_m)$ must \emph{equal} $\frac{1}{2}$ for any further arguments $\pi_1, \dots, \pi_m$.
    \item For circuits $C_1$ and $C_2$, the property $p(C_1) = p(C_2)$ can likewise be stated as an intersection of two linear constraints: $p(C_1) - p(C_2) \le 0$ and $p(C_2) - p(C_1) \le 0$. A proof of this property can be translated into arguments $\pi$ and $\pi'$ such that $G(C_1 \mid \pi, \pi', \pi_1, \dots, \pi_m) = G(C_2 \mid \pi, \pi', \pi_1, \dots, \pi_m)$ for all $\pi_1, \dots, \pi_m$.
\end{itemize}

Definition~\ref{def:respect_for_proofs} requires the constraints to be linear. Why not allow \emph{all} constraints? Suppose we have a proof that $p(C)$ is either less than $\frac{1}{3}$ or greater than $\frac{2}{3}$. What should $G$'s estimate of $p(C)$ be, in light of such a proof? Answers between $\frac{1}{3}$ and $\frac{2}{3}$ are quite reasonable: while it is known that $p(C) < \frac{1}{3}$ or $p(C) > \frac{2}{3}$, it is not known which of these is true, and so it would be reasonable for $G$ to ``split the difference'' and say $\frac{1}{2}$.

Meanwhile, if it is known that $\sum_i a_i p(C_i) \le b$, then any reasonable tuple of estimates for $p(C_1), \dots, p(C_k)$ ought to satisfy this constraint as well. One formal justification for this is that for any probability distribution over tuples $(p(C_1), \dots, p(C_k))$ that satisfy the constraint, the mean of the distribution will also satisfy the constraint. In fact, this property holds for all convex constraints, not just linear ones, and it would be reasonable to state a stronger version of Definition~\ref{def:respect_for_proofs} that requires $G$ to respect arbitrary \emph{convex} constraints. The results of this section generalize to this stronger definition, although with some caveats; see Remark~\ref{remark:convex_constraints} for further discussion.\\

Essentially, linearity is a basic consistency condition on $G$'s estimates for different circuits derived from the same \emph{base circuit} -- that is, circuits that are derived from the same circuit via different variable substitutions (i.e.\ $C[A_1]$ and $C[A_2]$ for partial assignments $A_1$ and $A_2$). By contrast, respect for proofs is a more general consistency condition on $G$'s estimates, but it only applies if $G$ is supplied a proof. For example, if $C_1$ and $C_2$ provably implement the same function, then $G$ does not need to output the same estimate for $C_1$ and $C_2$ \emph{by default} -- but if it is given a proof that $p(C_1) = p(C_2)$, then its estimates must respect that proof. Taken together, linearity and respect for proofs are fairly powerful.\footnote{For example, given a network of circuits that relate to each other by elementary equivalence transformations (such as distributing AND over OR) and variable substitution, one can turn this network into a set of arguments $\Pi$ that forces $G$ to give consistent estimates for all circuits in the network.}

On the other hand, the respect for proofs property is importantly constrained, because it \emph{cannot} force $G$ to obey a universal quantification of proofs. For example, a proof that $p(C[\vect{x}]) \ge 0$ for all $\vect{x}$ \emph{cannot} be turned into an argument $\pi$ such that $G(C[\vect{x}] \mid \pi) \ge 0$ for all $\vect{x}$. The respect-for-proofs property merely guarantees that for all $\vect{x}$, the proof that $p(C[\vect{x}]) \ge 0$ can be turned into an argument $\pi_{\vect{x}}$ such that $G(C[\vect{x}] \mid \pi_{\vect{x}}) \ge 0$.\footnote{In theory, $G$ can be given an argument $\pi_{\vect{x}}$ for every $\vect{x}$, but in that case the runtime of $G$ may be exponential in the size of $C$, even if $G$ is a polynomial-time algorithm.} In Section~\ref{sec:01_boundedness}, we will show that one cannot hope to satisfy a stronger version of respect for proofs that allows for universal quantification (in conjunction with linearity).

\subsection{An efficient algorithm that satisfies linearity and respect for proofs}
In this section, we give a polynomial-time deductive estimation algorithm that satisfies linearity and respect for proofs. Before we do so, we clarify some conventions and definitions that will become relevant:
\begin{itemize}
    \item Circuits will always have labeled inputs and gates. For example, if $C'$ is an exact gate-for-gate copy of $C$, we still consider $C$ and $C'$ to be different circuits. A linear constraint involving $C$ does not force $G$'s estimate of $p(C')$ to satisfy the same constraint. However, a proof that $p(C) = p(C')$ will force $G$ to output the same estimate for $C$ as for $C'$.
    \item We can formally define a circuit as a DAG of labeled inputs and gates with one sink (output) -- we will call this DAG the \emph{base circuit} -- together with a (possibly empty) partial assignment to the input variables. For example, if $C' = C[x_1 = 0]$ and $C'' = C'[x_2 = 1]$, then we formally define $C''$ to be the base circuit $C$ together with the partial assignment $(x_1 = 0, x_2 = 1)$. That is, $C'' := C[x_1 = 0, x_2 = 1]$. If a circuit $C$ is defined as $\tilde{C}[A]$, then we call $\tilde{C}$ the base circuit of $C$. Every circuit has exactly one base circuit.
    \item The \emph{size} of a partial assignment $A$, denoted $\abs{A}$, is the number of variables that it defines. The \emph{measure} of a partial assignment $A$, denoted $\mu(A)$, is defined as $2^{-\abs{A}}$ (this is the fraction of full assignments that satisfy $A$). Given partial assignments $A_1$ and $A_2$, we define $\mu(A_1 \cup A_2)$ to be the fraction of full assignments satisfying \emph{both} $A_1$ and $A_2$; we have $\mu(A_1 \cup A_2) \le \mu(A_1), \mu(A_2)$, and in particular $\mu(A_1 \cup A_2) = 0$ if $A_1$ and $A_2$ assign opposite values to the same variable.
\end{itemize}

Our deductive estimation algorithm, which we call $G_{\text{lin}}$, is defined in Algorithm~\ref{alg:lin_rfp}. We call this algorithm $G_{\text{lin}}$ because it creates a linear model of a base circuit's behavior on individual inputs, and then computes acceptance probabilities based on that model. (See the discussion immediately after the proof of Theorem~\ref{thm:lin_resp} for details of this interpretation.) The algorithm's implementation details are somewhat complex, but the basic idea behind it is fairly simple. Before proving the algorithm's correctness and efficiency, we describe the intuition behind it.

\setcounter{algocf}{\value{theorem}}
\begin{algorithm}[ht]
    \caption{Deductive estimation algorithm $G_{\text{lin}}$} \label{alg:lin_rfp}
    \SetKwInOut{Input}{input}\SetKwInOut{Output}{output}
    \Input{Boolean circuit $C$, set of deductive arguments $\Pi$}
    \Output{$G_{\text{lin}}(C \mid \Pi)$, a deductive estimate of $p(C)$}
    \begin{enumerate}[leftmargin=*]
        \item \label{step:add_constraints} Initialize an empty list $\ell$ of linear constraints. For each $\pi \in \Pi$, check whether $\pi$ is a valid proof of a rational linear constraint $\sum_{i = 1}^k a_i p(C_i) \le b$. If so, add the constraint to $\ell$.
        \item \label{step:lin_constraints} For every circuit $C'$ that is the base circuit of some circuit that appears in some constraint in $\ell$: let $(A_1, \dots, A_m)$ be the tuple of all partial assignments $A$ such that $C'[A]$ appears in some constraint in $\ell$. Let $M$ be the $m \times m$ matrix defined by $M_{i, j} = \mu(A_i \cup A_j)$. Compute vectors $\vect{v}_1, \dots, \vect{v}_k$ that span the orthogonal complement of the column space of $M$. For each $i \in [k]$, add the constraint $\angles{\vect{v}_i, (\mu(A_1) p(C'[A_1]), \dots, \mu(A_m) p(C'[A_m]))} = 0$ to $\ell$.
        \item \label{step:ellipsoid} Use the ellipsoid method to find a point $\hat{\vect{p}}$ satisfying all constraints in $\ell$ -- that is, an estimate $\hat{p}(C')$ of $p(C')$ for every $C'$ appearing in at least one constraint in $\ell$ \emph{(not just base circuits)}, such that $\hat{\vect{p}}$ simultaneously satisfies all of the constraints in $\ell$.
        \item \label{step:linear_eq} Let $C = \tilde{C}[A]$, where $\tilde{C}$ is $C$'s base circuit. Let $(A_1, \dots, A_m)$ be the (possibly empty) tuple of all partial assignments $A'$ such that $\tilde{C}[A']$ appears in some constraint in $\ell$. Let $M$ be the $m \times m$ matrix defined by $M_{i, j} = \mu(A_i \cup A_j)$. Let $\pmb{\beta}$ be the lowest-norm solution to the equation $M \pmb{\beta} = (\mu(A_1) \hat{p}(\tilde{C}[A_1]), \dots, \mu(A_m) \hat{p}(\tilde{C}[A_m]))$. \emph{(Below we show that this equation has a solution.)} Return $\sum_{j = 1}^m \frac{\mu(A \cup A_j)}{\mu(A)} \beta_j$.
    \end{enumerate}
\end{algorithm}
\setcounter{theorem}{\value{algocf}}

In Step~\ref{step:add_constraints} and \ref{step:lin_constraints}, the algorithm creates a list $\ell$ of constraints on its estimates, which it will then work to satisfy. In Step~\ref{step:add_constraints}, it searches through all linear constraints that have been proven in $\Pi$ and adds them to $\ell$. Its job will be to satisfy all of those constraints. Then, in Step~\ref{step:lin_constraints}, the algorithm adds some additional constraints: ones that it can infer must hold simply from how the acceptance probabilities of circuits derived from the same base circuit must relate to each other. For example, we know that $p(C[x_1 = 0]) + p(C[x_1 = 1]) = 2p(C)$, and that $p(C[x_1 = 0, x_2 = 0]) + p(C[x_1 = 0, x_2 = 1]) + 2p(C[x_1 = 1]) = 4p(C)$, and so on. Adding these constraints to $\ell$ is necessary to make sure that the linear system in Step~\ref{step:linear_eq} is consistent.

Then, in Step~\ref{step:ellipsoid}, the algorithm solves all of the constraints that it found in $\Pi$ or subsequently appended to $\ell$. The result is a point $\hat{\vect{p}}$ whose coordinates correspond to estimated acceptance probabilities of all circuits that appeared in some constraint in $\ell$. For a circuit $C'$, we use $\hat{p}(C')$ to denote the coordinate of $\hat{\vect{p}}$ corresponding to the circuit $C'$.\footnote{We use the ellipsoid algorithm for concreteness, but any worst-case polynomial-time algorithm for solving linear programs would work.}

Finally, in Step~\ref{step:linear_eq}, the algorithm looks at the circuit $C$ that it was provided as input -- it is only at this final step that the algorithm's behavior depends on the input circuit $C$. Let $C = \tilde{C}[A]$, where $\tilde{C}$ is $C$'s base circuit.\footnote{Note that $C$ did not necessarily appear as a constraint in $\ell$, so $\hat{p}(C)$ may not be defined.} While the algorithm will only output an estimate of $C$, it must satisfy linearity, which means that its estimate for $p(C)$ must be consistent with the counterfactual estimates that it would have given, had it been asked to estimate the acceptance probability of a different circuit with the same base circuit $\tilde{C}$. In order to do this, it finds coefficients $\beta_i$ -- one for every circuit $\tilde{C}[A_i]$ that appeared in $\ell$ -- with the following property: if for every $\vect{x}$, $\tilde{C}(\vect{x})$ were equal to the sum of $\beta_j$ over all $j$ such that $\vect{x}$ is consistent with $A_j$, then for every $A_i$, the average output of $\tilde{C}[A_i]$ would be equal to $\hat{p}(\tilde{C}[A_i])$. That is, the algorithm creates a linear model of $\tilde{C}'s$ behavior on individual inputs that is consistent with the estimates given by $\hat{\vect{p}}$. From there, it is straightforward to compute the average output of $C$ based on this linear model -- and this is the number that the algorithm outputs.

\begin{example} \label{ex:cxb}
    Recall from Section~\ref{sec:cnf_est} our class of $2^{3k}$ 3CNFs on $n$ variables with $k$ clauses, parameterized by $\vect{b} \in \{0, 1\}^{k \times 3}$:
    \[C_\vect{b}(\vect{x}) = (x_{i_{1, 1}} = b_{1, 1} \vee x_{i_{1, 2}} = b_{1, 2} \vee x_{i_{1, 3}} = b_{1, 3}) \wedge \dots \wedge (x_{i_{k, 1}} = b_{k, 1} \vee x_{i_{k, 2}} = b_{k, 2} \vee x_{i_{k, 3}} = b_{k, 3}).\]
    (Here, the indices $i_{j, \ell} \in [n]$ are fixed.) Define $\tilde{C}(\vect{x}, \vect{b}) := C_\vect{b}(\vect{x})$.

    Now, suppose we want to estimate $p(C)$, where $C = C_\vect{z}$ for some particular $\vect{z}$. We can recover the estimate from Section~\ref{sec:cnf_est} as a simple example of Algorithm~\ref{alg:lin_rfp}. Note that $C = \tilde{C}[\vect{b} = \vect{z}]$. For $i \in [n]$, define $A_i$ to be the partial assignment that sets all bits in $\vect{b}$ corresponding to appearances of the variable $x_i$ in the CNF formula the same way that they are set in $\vect{z}$. Note that $C = \tilde{C}[A_1, \dots, A_n]$. For convenience, let $\varphi_i := p(\tilde{C}[A_i])$.

    Let $\pi$ be the proof that $p(\tilde{C}) = (7/8)^k$ ($\pi$ might look essentially like the proof of Claim~\ref{claim:78k}, though in a formal language).\footnote{More precisely, we should have two arguments: $\pi^{\le}$, which proves that $p(\tilde{C}) \le (7/8)^k$, and $\pi^{\ge}$, which proves that $-p(\tilde{C}) \le (7/8)^k$, and similarly for $\pi_1, \dots, \pi_n$.} Let $\pi_1, \dots, \pi_n$ be proofs of the values of $\varphi_i$ (these proofs are straightforward, as discussed in Section~\ref{sec:cnf_est}). Let $\Pi = \{\pi, \pi_1, \dots, \pi_n\}$.

    What is $G_{\text{lin}}(C \mid \Pi)$? In Step~\ref{step:add_constraints}, the algorithm adds constraints that enforce the values of $\hat{p}(\tilde{C}), \hat{p}(\tilde{C}[A_1]), \dots, \hat{p}(\tilde{C}[A_n])$. Assuming that every variable appears at least once in the CNF formula, no further constraints are added in Step~\ref{step:lin_constraints}, because the matrix $M$ defined in that step has full rank. Step~\ref{step:ellipsoid} is trivial, because the constraints in Step~\ref{step:add_constraints} define a polytope consisting of a single point -- at this point we have found $\hat{\vect{p}}$, and we have $\hat{p}(\tilde{C}) = p(\tilde{C})$ and $\hat{p}(\tilde{C}[A_i]) = \varphi_i$ for $i \in [n]$.

    What happens in Step~\ref{step:linear_eq}? Call $\beta_0, \beta_1, \dots, \beta_n$ the $\beta$-values corresponding to $\tilde{C}, \tilde{C}[A_1], \dots, \tilde{C}[A_n]$, respectively. Because $C = \tilde{C}[A_1, \dots, A_n]$, the algorithm will return $\beta_0 + \beta_1 + \dots + \beta_n$. Now, let's compute $\pmb{\beta}$. Note that $\mu(A_i \cup A_j) = \mu(A_i) \mu(A_j)$ for $i \neq j$. This fact lets us simplify the matrix equation, giving us the following $n + 1$ linear equations:
    \begin{align*}
        \beta_0 + \sum_{j = 1}^n \mu(A_j) \beta_j &= p(\tilde{C})\\
        \beta_0 + \beta_i + \sum_{j \neq i} \mu(A_j) \beta_j &= \varphi_i \enskip \forall i \in [n].
    \end{align*}
    Subtracting the first equation from the last $n$ equations, we find that any solution must have $(1 - \mu(A_i)) \beta_i = \varphi_i - p(\tilde{C})$ for all $i$, and so $\beta_i = \frac{\varphi_i - p(\tilde{C})}{1 - \mu(A_i)}$. This means that
    \[\beta_0 = p(\tilde{C}) - \sum_{j = 1}^n \mu(A_j) \beta_j = p(\tilde{C}) - \sum_{j = 1}^n \mu(A_j) \cdot \frac{\varphi_j - p(\tilde{C})}{1 - \mu(A_j)}.\]
    Therefore, the algorithm returns
    \[\beta_0 + \beta_1 + \dots + \beta_n = p(\tilde{C}) + \sum_{j = 1}^n (1 - \mu(A_j)) \cdot \frac{\varphi_j - p(\tilde{C})}{1 - \mu(A_j)} = \sum_{j = 1}^n \varphi_j - (n - 1) p(\tilde{C}).\]
    This is exactly the same as Equation~\ref{eq:cnf_linreg}, our linear regression estimate from Section~\ref{sec:cnf_est}!\\
\end{example}

We now prove that Algorithm~\ref{alg:lin_rfp} is efficient, and that it satisfies linearity and respect for proofs.

\begin{theorem} \label{thm:lin_resp}
    The deductive estimation algorithm $G_{\text{lin}}$, as defined in Algorithm~\ref{alg:lin_rfp}, runs in polynomial time in the length of its input, and satisfies linearity and respect for proofs.
\end{theorem}

\begin{proof}
    We begin by proving that Algorithm~\ref{alg:lin_rfp} runs in polynomial time. Step~\ref{step:add_constraints} is straightforwardly fast, because checking a proof for validity takes linear time. Step~\ref{step:lin_constraints} is also fast: first, computing $M$ is straightforward. To compute $\vect{v}_1, \dots, \vect{v}_k$, we first (greedily) find a subset of $M$'s column space that spans the column space (which will have size $m - k$) and use the Gram-Schmidt process to find $m - k$ orthogonal vectors that have the same span. Then we continue the Gram-Schmidt process in order to find $k$ vectors in $\RR^m$ that are orthogonal to these $m - k$ orthogonal vectors. These are our $\vect{v}_1, \dots, \vect{v}_k$.

    Step~\ref{step:ellipsoid} is an application of the ellipsoid method, which -- given rational linear constraints that define a nonempty region -- can find a point in the region in polynomial time \parencite{gls93}. The only detail that we must check is that the intersection of the constraints in $\ell$ is nonempty. We show this by showing that the vector $\vect{p}$ of true acceptance probabilities satisfies all constraints.
    
    Clearly, $\vect{p}$ satisfies the constraints that are added to $\ell$ in Step~\ref{step:add_constraints} (after all, $\Pi$ contains proofs that $\vect{p}$ satisfies the constraints). What about the constraints added in Step~\ref{step:lin_constraints}? For $C'$, $(A_1, \dots, A_m)$, and $M$ as defined in Step~\ref{step:lin_constraints}, we must check that if $\vect{v}$ is orthogonal to the column space of $M$, then $\angles{\vect{v}, (\mu(A_1) p(C'[A_1]), \dots, \mu(A_m) p(C'[A_m]))} = 0$. In order to do so, it suffices to show that $(\mu(A_1) p(C'[A_1]), \dots, \mu(A_m) p(C'[A_m]))$ lies in the column space of $M$.

    \begin{claim}
        For $C'$, $(A_1, \dots, A_m)$, and $M$ as in Step~\ref{step:lin_constraints}, $(\mu(A_1) p(C'[A_1]), \dots, \mu(A_m) p(C'[A_m]))$ lies in the column space of $M$.
    \end{claim}

    \begin{proof}
        Let $n$ be the number of inputs to $C'$. Associate to each coordinate of $\RR^{2^n}$ a unique element of $\{0, 1\}^n$. Let $\Phi$ be the $2^n \times m$ matrix whose $i$-th column is the indicator vector for the partial assignment $A_i$ -- that is, $\Phi_i$ contains a $1$ in the coordinate corresponding to $\vect{x} \in \{0, 1\}^n$ if $\vect{x}$ is consistent with $A_i$. Let $\vect{y} \in \RR^{2^n}$ be the indicator vector for $C'$ -- that is, $\vect{y}$ contains a $1$ in the coordinate corresponding to $\vect{x}$ if $C'(\vect{x}) = 1$.

        Let $\pmb{\beta} \in \RR^m$ be such that $\Phi \pmb{\beta} = \proj_\Phi(\vect{y})$ -- that is, $\Phi \pmb{\beta}$ is the projection of $\vect{y}$ onto the column space of $\Phi$. We claim that $M \pmb{\beta} = (\mu(A_1) p(C'[A_1]), \dots, \mu(A_m) p(C'[A_m]))$. To see this, we first observe that $M = \frac{1}{2^n} \Phi^T \Phi$. This is because $\Phi^T \Phi$ is an $m \times m$ matrix whose $(i, j)$-entry is the number of assignments satisfying both $A_i$ and $A_j$, which is $2^n \mu(A_i \cup A_j)$. Therefore, we have
        \[M \pmb{\beta} = \frac{1}{2^n} \Phi^T \Phi \pmb{\beta} = \frac{1}{2^n} \Phi^T \proj_\Phi(\vect{y}) = \frac{1}{2^n} \Phi^T \vect{y},\]
        where the last step follows from the fact that $\Phi^T(\vect{y} - \proj_\Phi(\vect{y})) = \vect{0}$, as $\vect{y} - \proj_\Phi(\vect{y})$ is orthogonal to the column space of $\Phi$. Finally, observe that $\Phi^T \vect{y}$ is a vector in $\RR^m$ whose $i$-th coordinate counts the number of inputs $\vect{x} \in \{0, 1\}^n$ that satisfy $C$ \emph{and} are consistent with partial assignment $A_i$. This means that $\frac{1}{2^n} \Phi^T \vect{y}$ is the \emph{fraction} of inputs with this property, so the $i$-th coordinate of $\frac{1}{2^n} \Phi^T \vect{y}$ is equal to $\mu(A_i) p(C'[A_i])$. This completes the proof.
    \end{proof}

    Finally, we consider Step~\ref{step:linear_eq}. The equation $M \pmb{\beta} = (\mu(A_1) \hat{p}(\tilde{C}[A_1]), \dots, \mu(A_m) \hat{p}(\tilde{C}[A_m]))$ has a solution. After all, we enforced this in Step~\ref{step:lin_constraints}! Assuming that $\tilde{C}$ is the base circuit of some constraint in $\ell$,\footnote{And if not, then the empty vector is a solution to the linear system (which has zero equations and zero variables).} when we considered $C' = \tilde{C}$ in Step~\ref{step:lin_constraints}, we added linear constraints that enforced that $(\mu(A_1) \hat{p}(\tilde{C}[A_1]), \dots, \mu(A_m) \hat{p}(\tilde{C}[A_m]))$ is orthogonal to every vector in the orthogonal complement of the column space of $M$. This means that $(\mu(A_1) \hat{p}(\tilde{C}[A_1]), \dots, \mu(A_m) \hat{p}(\tilde{C}[A_m]))$ lies in $M$'s column space.
    
    Now, since a solution exists, the space of solutions is an affine subspace of $\RR^m$. The lowest-norm solution is just the projection of $\vect{0}$ onto the space, which can be found in polynomial time. Computing the output is also straightforward. This completes the proof that Algorithm~\ref{alg:lin_rfp} runs in polynomial time.\\

    We now prove that $G_{\text{lin}}$ satisfies linearity and respect for proofs.

    \begin{claim}
        $G_{\text{lin}}$ satisfies respect for proofs.
    \end{claim}

    \begin{proof}
        We begin by noting the following fact: if $C$ appears in a constraint in $\ell$, then $G_{\text{lin}}(C \mid \Pi) = \hat{p}(C)$. To see this, let $\tilde{C}, A_1, \dots, A_m, M, \pmb{\beta}$ be as in Step~\ref{step:linear_eq} of the algorithm, and without loss of generality assume that $C = \tilde{C}[A_1]$. The matrix equation defining $\pmb{\beta}$ gives us $m$ linear constraints, the first of which says that
        \[\sum_{j = 1}^m \mu(A_1 \cup A_j) \beta_j = \mu(A_1) \hat{p}(C),\]
        and so indeed we have
        \[G_{\text{lin}}(C \mid \Pi) = \sum_{j = 1}^m \frac{\mu(A_1 \cup A_j)}{\mu(A_1)} \beta_j = \hat{p}(C).\]

        Now, to prove that $G_{\text{lin}}$ satisfies respect for proofs, let us consider boolean circuits $C_1, \dots, C_k$ and a proof that $\sum_{i = 1}^k a_i p(C_i) \le b$. The polynomial-time algorithm in Definition~\ref{def:respect_for_proofs} that converts the proof into an argument $\pi$ is the trivial algorithm that just outputs the proof that it receives as input. Now, consider any $\Pi$ containing $\pi$. Let us consider the behavior of Algorithm~\ref{alg:lin_rfp} if given $\Pi$ as input.
        
        Note that the vector $\hat{\vect{p}}$ computed in Step~\ref{step:ellipsoid} only depends on $\Pi$, and not at all on $C$: after all the algorithm first looks at $C$ in Step~\ref{step:linear_eq}. Note also that $C_1, \dots, C_k$ all appear in some constraint in $\ell$ (because the algorithm receives $\pi$ as part of $\Pi$), and so the quantities $\hat{p}(C_1), \dots, \hat{p}(C_k)$ are well-defined and do not depend on the input $C$ given to Algorithm~\ref{alg:lin_rfp}.

        Because the constraint $\sum_{i = 1}^k a_i p(C_i) \le b$ appeared in $\ell$, we have $\sum_{i = 1}^k a_i \hat{p}(C_i) \le b$. Therefore, we have that $\sum_{i = 1}^k a_i G_{\text{lin}}(C_i \mid \Pi) = \sum_{i = 1}^k a_i \hat{p}(C_i) \le b$, as desired.
    \end{proof}

    \begin{claim}
        $G_{\text{lin}}$ satisfies linearity.
    \end{claim}

    \begin{proof}
        Let $C$, $\Pi$, and $i$ be as in Definition~\ref{def:linearity}. We are interested in the algorithm's outputs when run on $(C, \Pi)$, $(C[x_i = 0], \Pi)$, and $C[x_i = 1], \Pi)$. Let $\tilde{C}$ be the base circuit of $C$, which is also the base circuit of $C[x_i = 0]$ and $C[x_i = 1]$. Observe that $\pmb{\beta}$ as defined in Step~\ref{step:linear_eq} is the same in all three cases. This is because the steps before Step~\ref{step:linear_eq} only depend on $\Pi$ and not on the input circuit, and $\pmb{\beta}$ depends only $\Pi$ and the base circuit defined in Step~\ref{step:linear_eq}, which is $\tilde{C}$ in all three cases.

        Let $A_1, \dots, A_m$ be as in Step~\ref{step:linear_eq}, which again are the same in all three cases. Let $A$ be such that $C = \tilde{C}[A]$; then $C[x_i = 0] = \tilde{C}[A \cup (x_i = 0)]$ and $C[x_i = 1] = \tilde{C}[A \cup (x_i = 1)]$. We want to show that
        \begin{equation} \label{eq:lin_beta}
            \sum_{j = 1}^m \frac{\mu(A \cup A_j)}{\mu(A)} \beta_j = \frac{1}{2} \parens{\sum_{j = 1}^m \frac{\mu(A \cup (x_i = 0) \cup A_j)}{\mu(A \cup (x_i = 0))} \beta_j + \sum_{j = 1}^m \frac{\mu(A \cup (x_i = 1) \cup A_j)}{\mu(A \cup (x_i = 1))} \beta_j}.
        \end{equation}
        We have that $\mu(A \cup (x_i = 0)) = \mu(A \cup (x_i = 1)) = \frac{1}{2} \mu(A)$. We also have that $\mu(A \cup (x_i = 0) \cup A_j) + \mu(A \cup (x_i = 1) \cup A_j) = \mu(A \cup A_j)$. Equation~\ref{eq:lin_beta} follows straightforwardly from these two facts.
    \end{proof}

    We have now proven that Algorithm~\ref{alg:lin_rfp} runs in polynomial time and that it satisfies linearity and respect for proofs, and so we have completed the proof of Theorem~\ref{thm:lin_resp}.
\end{proof}

Observe that inputs to a circuit $C$ are in correspondence with full assignments to the variables in $C$. That is, we have $C(\vect{x}) = p(C[\vect{x}])$, where on the right-hand side we think of $\vect{x}$ as a partial assignment that happens to be a \emph{full} assignment. This lets us define $G_{\text{lin}}$'s estimate of $C(\vect{x})$ \emph{on a particular input $\vect{x}$}: namely, it is the quantity $G_{\text{lin}}(C[\vect{x}] \mid \Pi)$. So, what is this quantity? For $A_1, \dots, A_m, \pmb{\beta}$ as in Step~\ref{step:linear_eq} of Algorithm~\ref{alg:lin_rfp}, we have
\[G_{\text{lin}}(C[\vect{x}] \mid \Pi) = \sum_{j = 1}^m \frac{\mu(\vect{x} \cup A_j)}{\mu(\vect{x})} \beta_j = \sum_{j \text{: } \vect{x} \text{ is consistent with } A_j} \beta_j.\]
This gives us a different perspective on $G_{\text{lin}}$: instead of thinking of $G_{\text{lin}}(C[\vect{x}] \mid \Pi)$ as a special case in which we substitute values for all variables, we can instead think of $G_{\text{lin}}$ as \emph{constructing a linear model} for $\tilde{C}$'s behavior on individual inputs: namely, the sum of all $\beta_j$ such that the input is consistent with $A_j$. Then, for a given circuit $C = \tilde{C}[A]$, $G_{\text{lin}}(C \mid \Pi)$ is simply the average value of $G_{\text{lin}}(\tilde{C}[\vect{x}] \mid \Pi)$ over all $\vect{x}$ that are consistent with $A$.

\begin{remark} \label{remark:convex_constraints}
    As discussed earlier, it makes sense to extend the definition of respect for proofs to allow for \emph{convex} constraints. Can we can extend Algorithm~\ref{alg:lin_rfp} to deal with convex constraints? The answer is essentially yes, but with a few caveats:
    \begin{itemize}
        \item In order to run the ellipsoid method, we must have access to a separation oracle for the convex constraints. It is easy for $G$ to implement a separation oracle under the following assumptions: first, the convex constraints are expressed as $f(C(x_1), \dots, C(x_k)) \le b$ for a convex function $f$. Second, there is a fast algorithm for evaluating $f$ and the gradient of $f$ at a point (or a subgradient, if $f$ is not differentiable). Third, $G$ is given a proof that $f$ is convex, as well as algorithms for evaluating $f$ and its gradient and proofs that the algorithms are fast and correct.
        \item Even so, for general convex constraints, the ellipsoid method may not be able to find an \emph{exact} solution to all constraints. However, the ellipsoid method can quickly find a point that that satisfies all constraints to within $\varepsilon$ tolerance. Thus, we can satisfy ``near''-respect-for-proofs, but not necessarily exact respect for proofs.
    \end{itemize}
\end{remark}

\section{Desiderata beyond linearity and respect for proofs} \label{sec:further_desiderata}
Although $G_{\text{lin}}$ satisfies linearity and respect for proofs, its estimates are in some sense arbitrary. The ellipsoid method finds \emph{some} $\hat{\vect{p}}$ satisfying the constraints in $\Pi$, but this $\hat{\vect{p}}$ is by no means canonical.

As a more canonical choice, we could instead find the centroid of the feasible region.\footnote{In order to make the centroid well-defined, we can bound the feasible region by adding the constraints $0 \le p(C') \le 1$ for every $C'$ appearing in some constraint in $\ell$.} While we cannot efficiently compute the exact centroid, we can compute it approximately by sampling from the feasible region, which can be done efficiently (see e.g.\ \textcite{cdwy18}).

However, choosing the centroid of the feasible region has some undesirable properties, even if it can be computed exactly. Let $G_{\text{cent}}$ be the modified version of $G_{\text{lin}}$ that defines $\hat{\vect{p}}$ to be the centroid of the feasible region. Consider three circuits $C_1, C_2, C_3$, about which no information is supplied except that $0 \le p(C_1), p(C_2), p(C_3) \le 1$. Then $G_{\text{cent}}$ calculates $\hat{p}(C_1) = \hat{p}(C_2) = \hat{p}(C_3) = \frac{1}{2}$ -- so far so good.

Now, let $C = C_1 \wedge C_2 \wedge C_3$. Note that $0 \le p(C) \le p(C_1), p(C_2), p(C_3)$ and also that $p(C) \ge p(C_1) + p(C_2) + p(C_3) - 2$. This second fact follows from the fact that
\[1 - p(C) = p(\neg C_1 \vee \neg C_2 \vee \neg C_3) \le p(\neg C_1) + p(\neg C_2) + p(\neg C_3) = 3 - p(C_1) - p(C_2) - p(C_3).\]
These are the only linear constraints on $p(C)$ in terms of $p(C_1)$, $p(C_2)$, and $p(C_3)$ that always hold, no matter what $C_1$, $C_2$, and $C_3$ are. Now, if $G$ is supplied this additional information, it turns out $\hat{p}(C_1) = \frac{14}{25}$ (as can be shown by taking an integral). In other words, giving $G_{\text{cent}}$ irrelevant information -- information about some auxiliary circuit that was guaranteed to be true no matter what the circuits $C_1$, $C_2$, and $C_3$ are -- changed its estimate of $p(C_1)$.

This source of potential dissatisfaction with $G_{\text{lin}}$ and $G_{\text{cent}}$ raises the question: what other properties should a deductive estimation algorithm $G$ have, beyond linearity and respect for proofs? \textcite{cnx22} informally discuss an ``independence of irrelevant arguments'' property as a potential desideratum: that $G$'s estimates ought not to depend ``irrelevant'' information. However, formally defining what it means for information to be irrelevant appears to be difficult. In this section we will consider other desiderata for $G$.

\subsection{0-1 boundedness} \label{sec:01_boundedness}
In this section, we propose a simple property that we call 0-1 boundedness.

\begin{defin}
    A deductive estimation algorithm $G$ satisfies \emph{0-1 boundedness} if for every boolean circuit $C$ and set of arguments $\Pi$, we have $0 \le G(C \mid \Pi) \le 1$.
\end{defin}

As we discussed in Section~\ref{sec:lin_resp}, the respect for proofs property does not force $G$ to respect \emph{universal} proofs. For example, a proof that $0 \le p(C) \le 1$ for all boolean circuits $C$ cannot be turned into an argument $\pi$ such that $0 \le G(C \mid \pi) \le 1$ for all $C$. The 0-1 boundedness property is weaker than ``respect for universal proofs'' -- and yet, it cannot be satisfied together with linearity and respect for proofs.

\begin{theorem}
    Unless $P = PP$, there is no polynomial-time deductive estimation algorithm that satisfies linearity, respect for proofs, and 0-1 boundedness.
\end{theorem}

\begin{proof}
    Let $G$ be a deductive estimation algorithm that satisfies linearity, respect for proofs, and 0-1 boundedness. We will show that a polynomial-time algorithm with oracle access to $G$ can solve $\#3CNF$ in polynomial time. This means that either $P = PP$ or $G$ is not a polynomial-time algorithm.
    
    Recall the circuit $\tilde{C}(\vect{x}, \vect{b})$ from Example~\ref{ex:cxb}:
    \[\tilde{C}(\vect{x}, \vect{b}) = (x_{i_{1, 1}} = b_{1, 1} \vee x_{i_{1, 2}} = b_{1, 2} \vee x_{i_{1, 3}} = b_{1, 3}) \wedge \dots \wedge (x_{i_{k, 1}} = b_{k, 1} \vee x_{i_{k, 2}} = b_{k, 2} \vee x_{i_{k, 3}} = b_{k, 3}),\]
    for some fixed indices $i_{1, 1,}, \dots, i_{k, 3}$ that are implicit in the notation.

    Consider the proof that $p(\tilde{C}) = (7/8)^k$. Because $G$ satisfies respect for proofs, this proof can be turned into an argument $\pi_0$ such that $G(\tilde{C} \mid \Pi) = (7/8)^k$ for all $\Pi$ containing $\pi_0$.

    We now consider $8k$ additional arguments: eight arguments for every clause. Let $j \in [k]$ and $\vect{y} = (y_1, y_2, y_3) \in \{0, 1\}^3$, and define $A_{j, \vect{y}}$ to be the partial assignment
    \[A_{j, \vect{y}} = (b_{j, 1} = y_1, b_{j, 2} = y_2, b_{j, 3} = y_3, x_{i_{j, 1}} = \neg y_1, x_{i_{j, 2}} = \neg y_2, x_{i_{j, 3}} = \neg y_3).\]
    Then $p(\tilde{C}[A_{j, \vect{y}}]) = 0$. This is because $A$ makes clause $j$ false, and thus -- no matter how the rest of the variables are set -- the entire formula is false. The proof of this fact can be turned into an argument $\pi_{j, \vect{y}}$ such that $G(\tilde{C}[A_{j, \vect{y}}] \mid \Pi) = 0$ for all $\Pi$ containing $\pi_{j, \vect{y}}$.

    Let $\tilde{\Pi}$ consist of $\pi_0$ together with $\pi_{j, \vect{y}}$ for all $k$ choices of $j$ and eight choices of $\vect{y}$.

    \begin{claim} \label{claim:G_correct}
        For all assignments $(\vect{x}, \vect{b})$ to $\tilde{C}$, we have
        \[G(\tilde{C}[\vect{x}, \vect{b}] \mid \tilde{\Pi}) = \tilde{C}(\vect{x}, \vect{b}).\]
    \end{claim}

    \begin{proof}
        We first show that $G(\tilde{C}[\vect{x}, \vect{b}] \mid \tilde{\Pi}) \le \tilde{C}(\vect{x}, \vect{b})$ for all $(\vect{x}, \vect{b})$. First, if $\tilde{C}(\vect{x}, \vect{b}) = 1$, then $G(\tilde{C}[\vect{x}, \vect{b}] \mid \tilde{\Pi}) \le 1 = \tilde{C}(\vect{x}, \vect{b})$ by 0-1 boundedness.
        
        Now, let $(\vect{x}, \vect{b})$ be such that $\tilde{C}(\vect{x}, \vect{b}) = 0$, and suppose for contradiction that $G(\tilde{C}[\vect{x}, \vect{b}] \mid \tilde{\Pi}) > 0$. $(\vect{x}, \vect{b})$ violates some clause; call it $j$. Let $\vect{y} = (b_{j, 1}, b_{j, 2}, b_{j, 3})$. By respect for proofs, we have that $G(\tilde{C}[A_{j, \vect{y}}] \mid \tilde{\Pi}) = 0$. By (repeated applications of) linearity, this means that
        \[\EE[\vect{x}', \vect{b}' \text{ consistent with } A_{j, \vect{y}}]{G(\tilde{C}[\vect{x}', \vect{b}'] \mid \tilde{\Pi})} = G(\tilde{C}[A_{j, \vect{y}}] \mid \tilde{\Pi}) = 0.\]
        But $(\vect{x}, \vect{b})$ is consistent with $A_{j, \vect{y}}$ and $G(\tilde{C}[\vect{x}, \vect{b}] \mid \tilde{\Pi}) > 0$ by assumption. It follows that for \emph{some} $(\vect{x}', \vect{b}')$, $G(\tilde{C}[\vect{x}', \vect{b}'] \mid \tilde{\Pi}) < 0$. But this contradicts 0-1 boundedness.

        We now know that $G(\tilde{C}[\vect{x}, \vect{b}] \mid \tilde{\Pi}) \le \tilde{C}(\vect{x}, \vect{b})$ for all $(\vect{x}, \vect{b})$. Now, suppose for contradiction that $G(\tilde{C}[\vect{x}, \vect{b}] \mid \tilde{\Pi}) < \tilde{C}(\vect{x}, \vect{b})$ for some $(\vect{x}, \vect{b})$. By (repeated applications of) linearity, it follows that
        \[G(\tilde{C} \mid \tilde{\Pi}) = \EE[\vect{x}, \vect{b}]{G(\tilde{C}[\vect{x}, \vect{b}] \mid \tilde{\Pi})} < \EE[\vect{x}, \vect{b}]{\tilde{C}(\vect{x}, \vect{b})} = p(\tilde{C}) = (7/8)^k.\]
        But this contradicts respect for proofs, because $\pi_0 \in \tilde{\Pi}$.
    \end{proof}
    
    By linearity, it follows from Claim~\ref{claim:G_correct} that for every $\vect{b}$, we have
    \[G(\tilde{C}[\vect{b}] \mid \tilde{\Pi}) = \EE[\vect{x}]{\tilde{C}(\vect{x}, \vect{b})} = p(\tilde{C}[\vect{b}]).\]
    We can now give an algorithm that, with oracle access to $G$, solves $\#3CNF$ in polynomial time. Consider any 3CNF $C$ on $n$ variables. We have that $C = \tilde{C}[\vect{b}]$ for a particular $\tilde{C}$ (i.e.\ for particular choices of indices $i_{1, 1}, \dots, i_{k, 3}$) and $\vect{b}$. Create $\tilde{\Pi}$ as above. Return $2^n G(\tilde{C}[\vect{b}] \mid \tilde{\Pi})$. The resulting output will be $2^n p(\tilde{C}[\vect{b}])$, which is equal to the number of satisfying assignments to $C$.
\end{proof}

\subsection{Desiderata inspired by conditional expectation} \label{sec:cond_exp_desiderata}
We have now shown that we cannot hope for $G$ to satisfy linearity and respect for proofs while requiring $G$ to output estimates in $[0, 1]$. What other properties might we hope for $G$ to satisfy? In this section, we informally discuss two such properties that are inspired by properties of conditional expectations.

We call the first of these properties \emph{iterated estimation,} because it is inspired by the law of iterated expectations from probability theory. Roughly speaking, the law of iterated expectations states that for a probability space $(\Omega, \mathcal{F}, \PP)$, a random variable $X$ defined on the space, and $\sigma$-algebras\footnote{For readers unfamiliar with $\sigma$-algebras, we suggest thinking of $\mathcal{H}_1$ as partial information about $X$ and $\mathcal{H}_2$ as more fine-grained partial information about $X$.} $\mathcal{H}_1 \subseteq \mathcal{H}_2 \subseteq \mathcal{F}$, we have
\[\EE{\EE{X \mid \mathcal{H}_2} \mid \mathcal{H}_1} = \EE{X \mid \mathcal{H}_1}.\]
Why is this true? We can think of $\mathcal{H}_1$ as revealing partial information about the state of the world $\omega \in \Omega$, and $\mathcal{H}_2$ as revealing more fine-grained information than $\mathcal{H}_1$. The right-hand side is a random variable that is the expectation of $X$ conditioned on knowing the information in $\mathcal{H}_1$. The left-hand side is the expected value of what you \emph{will} think the expectation of $X$ is after learning the information in $\mathcal{H}_2$, if you only know the information in $\mathcal{H}_1$. These are the same because, if you knew that in expectation you would update your guess about $X$ in a particular direction once you learned $\mathcal{H}_2$, it would make sense to update your guess about $X$ before learning $\mathcal{H}_2$.

\begin{defin}[Informal]
    A deductive estimation algorithm $G$ satisfies \emph{iterated estimation} if for all boolean circuits $C$ and for all sets of arguments $\Pi$ and $\Pi'$, we have
    \[G(G(C \mid \Pi, \Pi') \mid \Pi) = G(C \mid \Pi).\]
\end{defin}

In other words: if $G$ only knows the arguments in $\Pi$, then its estimate for what its estimate of $p(C)$ \emph{would} be if it also considered $\Pi'$, is just its current estimate of $p(C)$.

The statement of iterated estimation does not \emph{quite} make sense, because we said that the first argument to $G$ is a boolean circuit -- whereas $G(C \mid \Pi, \Pi')$ is not quite a boolean circuit.

However, we can make sense of this property by expanding the scope of the first argument to $G$. Suppose we allow $G$ to accept not just boolean circuits, but arbitrary computer programs (Turing machines) that output a real number. Then $G(C \mid \Pi, \Pi')$ is one such computer program -- namely, it takes no inputs, runs $G$ on the input $(C, (\Pi, \Pi'))$, and outputs the output of $G$.\footnote{The fact that this program takes no inputs is nothing new: we have already considered expressions such as $G(C[\vect{x}] \mid \Pi)$, where $\vect{x}$ is a full assignment to $C$, and so $C[\vect{x}]$ is a circuit with no inputs that outputs $C(\vect{x})$.} While in this chapter we have dealt with boolean circuits for simplicity, we are interested in the deductive estimation of the outputs of computer programs in general, and do not see a fundamental obstacle to enriching the space of inputs to $G$ to allow for arbitrary programs.

Returning to the iterated estimation property, it makes sense to ask: does the parallel to the law of iterated estimation make sense, or should we want $G$ to behave differently from an iterated expectation? One reason for skepticism is that it requires the outer $G$ in the expression $G(G(C \mid \Pi, \Pi') \mid \Pi)$ to be ``blinded'' to $\Pi'$ when estimating $G(C \mid \Pi, \Pi')$. That is, the outer $G$ sees the expression $G(C \mid \Pi, \Pi')$, but does not ``process'' it, beyond the extent allowed by $\Pi$.

But in a sense, this is the behavior we desire from $G$. We do not expect $G$ to be able to deduce properties of the circuit (or program) that it is estimating, except for the properties that are given to $G$ as arguments. In that sense, we should not expect the outer $G$ to be able to draw conclusions from $\Pi'$, except to the extent that it can do so based on the arguments that it has been given (i.e.\ $\Pi$).\\

The second property is inspired by a different property of conditional expectations, called ``pulling out known factors.'' The pulling out known factors property of conditional expectations states roughly that for an $\mathcal{H}$-measurable random variable $X$ and a random variable $Y$, we have $\EE{XY \mid \mathcal{H}} = X \EE{Y \mid \mathcal{H}}$. In other words, if the information supplied by $\mathcal{H}$ is sufficient to determine the value of $X$, then $X$ can be pulled out of the conditional expectation.

\begin{defin}[Informal]
    A deductive estimation algorithm $G$ satisfies the \emph{pulling out known factors} property if for boolean circuits $C_1, C_2$ and sets of arguments $\Pi, \Pi'$, we have
    \[G(G(C_1 \mid \Pi) p(C_2) \mid \Pi, \Pi') = G(C_1 \mid \Pi) G(C_2 \mid \Pi, \Pi').\]
\end{defin}

Again, we require $G$ to accept arbitrary programs as input, and interpret $G(C_1 \mid \Pi) p(C_2)$ as the program that computes $G(C_1 \mid \Pi)$, computes $p(C_2)$, and returns their product.\footnote{This program may not run in polynomial time, but that is fine: $G$ can estimate the outputs of programs that may take a long time to run.} The intuition is the same as for the corresponding property of conditional expectations: in the expression $G(G(C_1 \mid \Pi) p(C_2) \mid \Pi, \Pi')$, the outer $G$ knows $\Pi$, so it knows what its estimate of $p(C_1)$ would be if it only knew $\Pi$. To the outer $G$, the value $G(C_1 \mid \Pi)$ is a particular constant, rather than an unknown, so it can be ``pulled out.''

\section{Mechanistic anomaly detection for neural networks} \label{sec:mad}
Let $G_{\text{good}}$ be a hypothetical deductive estimation algorithm that we would find satisfying: perhaps it satisfies linearity, respect for proofs, and some other important desiderata. One potential use case for $G_{\text{good}}$ is to tell apart different reasons for why a circuit might output $1$. Recall Example~\ref{ex:abc} from the introduction: $C$ takes as input a triple $(a, b, c)$ of positive integers and accepts if $\max(a, b) = \max(b, c)$. Note that $C$ can accept an input for one of two ``reasons'': either $b \ge a, c$ (which happens with probability $\frac{1}{3}$ or so), or $a = c$ (which happens extremely rarely).

Let $\pi$ be the proof that if $b \ge a, c$ then $C(a, b, c) = 1$. We might expect the following behavior from $G_{\text{good}}$: for all $(a, b, c)$ such that $b \ge a, c$, we have $G_{\text{good}}(C[a, b, c] \mid \pi) = 1$. Meanwhile, if $b < a$ or $b < c$, then $G_{\text{good}}(C[a, b, c] \mid \pi)$ is close to zero -- even if in fact $a = c$ -- because $\pi$ does not point out the fact that if $a = c$, then $C(a, b, c) = 1$. Thus, $G_{\text{good}}$ allows us to distinguish inputs on which $C$ outputs $1$ for the ``usual'' reason ($b \ge a, c$) from inputs on which $C$ outputs $1$ for a different reason ($a = c$).

Of course, this approach is only useful insofar as it can be applied to circuits that we don't already understand. For such circuits, there is an obvious barrier to using $G_{\text{good}}$ in this way: how do we find the deductive arguments to give to $G_{\text{good}}$? In general, finding deductive arguments is as hard as finding proofs of mathematical statements. However, \emph{in the particular case of machine learning,} we may be able to find deductive arguments that allow us to understand the neural networks that we have trained.

The key fact about neural networks is that they are trained from random initialization via gradient descent. We can hope to learn deductive arguments for the behavior of a neural network \emph{also} via gradient descent, in parallel with training the network. These arguments might be features in some continuous parameterized space, somewhat akin to the features $\varphi_i$ in Section~\ref{sec:linreg}. One possibility is that these features will be simple functions of the neural network's weights and other features. The features would be learned to maximize predictive accuracy of the neural network's behavior: for example, if the neural network (call it $N$) is trained to get a high reward according to a reward model (call it $R$), and it learns to get high reward on the input distribution $D$, then we would hope to learn features $\varphi_1, \dots, \varphi_m$ such that $G_{\text{good}}(\EE[\vect{x} \sim D]{R(\vect{x}, N(\vect{x}))} \mid \varphi_1, \dots, \varphi_m)$ is (correctly) high.\footnote{How does respect for proofs figure into this? The respect for proofs property is meant to capture that the space of possible features must be extremely expressive: enough that features can encode formal proofs about $N$ and $R$. In practice, the features that we learn are unlikely to be proofs; however, respect for proofs is an easily formalizable property that is necessary (even if not sufficient) for the features to be arbitrarily expressive. It is thus a convenient property to work with when taking a theoretical approach to finding a satisfying deductive estimation algorithm.}\\

We may then be able to use these features for \emph{mechanistic anomaly detection} \parencite{christiano22}. The mechanistic anomaly detection problem can be stated informally as follows: suppose that a neural network $N$ gets high average reward from a reward model $R$ over an input distribution $D$ (that is, $\EE[\vect{x} \sim D]{R(\vect{x}, N(\vect{x}))}$ is large). Let $\vect{x}^* \not \in D$ be a new input on which $N$ again gets high reward (that is, $R(\vect{x}^*, N(\vect{x}^*))$ is also large). Did $N$ get high reward on $\vect{x}^*$ \emph{for the same reasons} that it got high average reward on $D$, or \emph{for a different reason?}

For example, suppose that $N$ is an financial assistant that takes actions such as buying and selling stocks, transferring money between bank accounts, and paying taxes. Suppose that $R$ sanity-checks $N$'s actions for legal compliance, and then rewards $N$ if it has turned a profit. Then $N$ may have two ways of getting high reward: \emph{successful task completion} (making good trades on the stock market) and \emph{successful deception} (for example, making a profit via money laundering and then covering its tracks). Suppose that $D$ is a ``clean'' distribution of input scenarios: on each scenario in $D$, $N$ achieves high reward via the ``successful task completion'' route. If $\vect{x}^*$ is a new input scenario on which $N$ again achieves high reward, we want to be able to determine whether $N$ successfully completed the task (got high reward \emph{for the same reasons} as it did on $D$) or if it instead successfully deceived $R$ (got high reward \emph{for a different reason}). In the latter case, we would like to classify $N$'s behavior on $\vect{x}^*$ as ``anomalous'' (and perhaps refer the input for more extensive oversight).

Mechanistic anomaly detection is a particularly challenging problem, because it cannot in general be solved by looking at $N$'s input-output behavior. After all, a sufficiently advanced reward model $R$ would be able to integrate observations of input-output behavior into its oversight process. To catch successful deception, it could be necessary to analyze to analyze $N$'s internals. This is why sophisticated machinery such as deductive estimation may be necessary.

How can deductive estimation be used to solve the mechanistic anomaly detection problem? Suppose that we successfully learn features $\varphi_1, \dots, \varphi_m$ that explain $N$'s high average reward to $G_{\text{good}}$ -- that is, $G_{\text{good}}(\EE[\vect{x} \sim D]{R(\vect{x}, N(\vect{x}))} \mid \varphi_1, \dots, \varphi_m)$ is appropriately large. Then consider the quantity
\[G_{\text{good}}(R(\vect{x}^*, N(\vect{x}^*)) \mid \varphi_1, \dots, \varphi_m).\]
If $N$ gets high reward on $\vect{x}^*$ for the same reasons than it got high reward on inputs in $D$, then $\varphi_1, \dots, \varphi_m$ should be sufficient to explain $N$'s high reward on $\vect{x}^*$ as well, and so we expect $G_{\text{good}}(R(\vect{x}^*, N(\vect{x}^*)) \mid \varphi_1, \dots, \varphi_m)$ to be large as well. On the other hand, suppose that $N$ gets high reward on $\vect{x}^*$ for a \emph{different} reason than it got high reward on $D$. Then $G_{\text{good}}(R(\vect{x}^*, N(\vect{x}^*)) \mid \varphi_1, \dots, \varphi_m)$ will probably be small, because the reasons why $N$ did well on $D$ (which are represented as features $\varphi_1, \dots, \varphi_m$) are \emph{insufficient} to explain $N$'s high reward on $\vect{x}^*$. So if $R(\vect{x}^*, N(\vect{x}^*))$ is large, but $G_{\text{good}}(R(\vect{x}^*, N(\vect{x}^*)) \mid \varphi_1, \dots, \varphi_m)$ is small, then we classify $N$'s behavior on $\vect{x}^*$ as anomalous.\footnote{In fact, given the input $\vect{x}^*$, we may further optimize our features $\varphi_1, \dots, \varphi_m$, with the goal of finding features that still cause $G$ to estimate that $N$'s average reward on $D$ is large, while causing $G$ to estimate that $N$'s reward on $\vect{x}^*$ is small. If we find any such features, we label $N$'s behavior on $\vect{x}^*$ anomalous. Of course, care must be taken to not over-optimize, as this runs the danger of labeling every input ``anomalous.''}

\section{Conclusion} \label{sec:chap9_conclusion}
In my view, the problem of deductive circuit estimation -- finding an efficient deductive estimation algorithm that satisfies important properties such as linearity, respect for proofs, and more -- has two compelling motivations.

The first is theoretical elegance and importance. In the introduction, we gave several examples of deductive arguments about a circuit. Those arguments were not proofs, but in an important sense, they were sound. Barring further arguments, it is reasonable to guess that the first 128 bits of the output of SHA-256 are as likely to be larger than the last 128 bits as they are to be smaller. Barring further observations about the structure of a 3CNF with $k$ clauses, it is reasonable to guess that its acceptance probability is $(7/8)^k$. Such deductive reasoning is commonplace in computer science and mathematics.

\textcite{cnx22} asked whether it is possible to formalize such reasoning: that is, to design an algorithm that estimates quantities given formal arguments similar to the ones in the introduction. This question seems fundamental: much as mathematicians have found a formalization of a mathematical proof, one could hope for a formalization of a much broader class of deductive arguments.

In this chapter, we continued this formalization project by designing an algorithm for boolean circuit estimation that satisfies linearity and respect for proofs, and by suggesting further properties that deductive estimation algorithms ought to satisfy. However, we have not satisfied those further properties, nor am I convinced that we have found the right set of properties to satisfy. We have raised more questions than we have answered. But in my opinion, the questions that we have raised are fundamental and exciting.\\

The second motivation for this work is practical relevance to the AI alignment problem.\footnote{See \textcite{ji_alignment_23} for a survey of the AI alignment problem.} AI labs such as OpenAI and Anthropic are putting substantial effort into \emph{scalable oversight:} methods of overseeing training that will continue to be reliable even as AI capabilities advance \parencite{ls23, anthropic23}. However, a sufficiently advanced AI could find ways to exploit loopholes in its oversight process. While oversight approaches that rely on observations of input-output behavior may work, this is not guaranteed. Approaches that use model internals as part of the oversight process may be necessary to build safe advanced AI.

In Section~\ref{sec:mad}, we saw how deductive estimation algorithms could be used to oversee an AI by flagging anomalous behavior. There are many challenges ahead for this approach, and this path forward is more complex and speculative than most approaches to scalable oversight. However, this approach has the potential to succeed even under somewhat pessimistic assumptions about the difficulty of the AI alignment problem.\footnote{See \href{https://www.alignment.org/blog/}{\texttt{https://www.alignment.org/blog/}} for further discussion.}\\

For both of these reasons -- the theoretical and the practical -- I find deductive estimation to be a compelling research direction. I am excited to see what the future holds for this area.


\clearpage
\phantomsection
\addcontentsline{toc}{chapter}{Epilogue}
\begin{center}
\pagebreak
\vspace*{5\baselineskip}
\textbf{\large Epilogue}
\end{center}

\setcounter{footnote}{0}

Each technical chapter of this thesis explored a different facet of algorithmic Bayesian epistemology. Chapters~\ref{chap:precision} and \ref{chap:arbitrage} explored the elicitation of knowledge under strategic constraints. Chapter~\ref{chap:qa} bridged the problem of knowledge elicitation with the problem knowledge aggregation. Chapters~\ref{chap:learning} and \ref{chap:robust} focused on the knowledge aggregation problem, now under informational constraints. Chapter~\ref{chap:agreement} also explored aggregation,\footnote{After all, the question of Chapter~\ref{chap:agreement} can be framed as: under what circumstances can Alice and Bob aggregate their knowledge by sharing a small amount of their information?} but now under communication constraints. Finally, Chapter~\ref{chap:elk} explored belief formation under both computational and informational constraints.

We closed every chapter with a discussion of future directions. Some chapters had a few suggestions, while others had many. Depending on how you count, Chapter~\ref{chap:robust} had as many as 1,152 suggestions! Some closed with fully formal problem statements (Chapter~\ref{chap:arbitrage}), others with problem statements so informal that formalizing them would in itself be a breakthrough (Chapter~\ref{chap:elk}). Some suggestions were intellectual curiosities, while others had significant potential for application. Some were merely interesting, while others were (in my opinion) utterly fascinating.

In light of all these differences, I'd like to close by highlighting the directions in ABE that I find \emph{most} exciting. These directions are listed in order of the chapter(s) to which they are relevant.

\paragraph{Bayesian justifications for generalized QA pooling} In Section~\ref{sec:bayesian_justifications}, we gave Bayesian justifications for generalized linear and logarithmic pooling (i.e.\ linear and logarithmic pooling but with weights that do not necessarily add to $1$). That is, we presented information structures in which a generalized linear pool is the \emph{exactly correct} aggregation method, and similarly for generalized logarithmic pooling. Can we do this for \emph{every} QA pooling method -- that is, generalized QA pooling with respect to every proper scoring rule?

\paragraph{Directions in robust forecast aggregation} As discussed in Section~\ref{sec:chap7_future_work}, robust forecast aggregation seems like a \emph{particularly} exciting area for future work. Highlights include:
\begin{itemize}
    \item \emph{Generalization to KL divergence.} In our exploration of the known prior setting, we found that averaging experts' forecasts and then moving the average away from the prior by a constant factor results in a robustly high-quality forecast. However, if the value $Y$ being forecast is a probability, then this procedure can result in aggregate forecasts outside of $[0, 1]$. If the error measure were changed from squared error to KL divergence, then such forecasts would naturally be disallowed, as they would incur an infinite penalty. So, what is a robust way to aggregate probabilistic forecasts if the error measure is KL divergence? The answer to this question may have important implications for the aggregation of probabilistic forecasts!
    \item \emph{Giving the aggregator additional information.} In our problem, the aggregator learns each expert's expectation of $Y$ and (sometimes) the prior. How much better can the aggregator do if they learn additional information, either about the experts' beliefs or about the information structure itself? Note that the answer might vary considerably depending on which assumptions are made about the information structure.
\end{itemize}

\paragraph{Sophisticated Bayesian models for forecast aggregation} While this thesis focused primarily on robust forecast aggregation, there is significant low-hanging fruit in Bayesian aggregation: creating a Bayesian model of experts' information and finding the optimal aggregate (or a good aggregate) under that model. While much more well-studied than robust aggregation, most work on Bayesian aggregation focuses on specific application domains (see \textcite{mwgr21} for a survey).

One particularly compelling direction is to build on the Gaussian partial information framework introduced by \textcite{spu16} (see our exposition in Section~\ref{sec:prelim_info_struct}). In this framework, the optimal aggregate is a linear combination of the experts' forecasts, with weights determined by the overlap between the experts' information sets (see Section~\ref{sec:bayesian_justifications}). \textcite{sjpu17} explore how this overlap may be inferred when multiple quantities are being forecast. But how might one aggregate forecasts within this framework if only one quantity is being forecast?

Here is one possible Bayesian model, inspired by this question. Let us say that the experts are forecasting a quantity $Y = X_1 + X_2 + \dots$, where each $X_j$ is independently and normally distributed, with mean $0$ and standard deviation $\frac{1}{j}$. (Essentially, this means that the factors contributing to $Y$ have ``importances'' that follow a Zipfian distribution. Note that $Y$ is almost surely finite.) There are $m$ experts, each of which knows a subset of the $X_j$'s. In particular, we can model each expert $i$ as having an \emph{expertise score} $e_i$ (over which we have some prior), and each $X_j$ as having an \emph{obviousness score} $o_j$ (over which we have some prior). The odds that Expert $i$ knows the value of $X_j$ are $e_i \cdot o_j$ to $1$ (or in other words, the probability that expert $i$ knows $X_j$ is $\frac{e_i o_j}{e_i o_j + 1}$). Each expert tells the aggregator their estimate of $Y$, which is the sum of all $X_j$'s whose value the expert knows. The aggregator can then do (approximate) Bayesian inference to find the optimal way to combine the experts' estimates into a single estimate of $Y$.

Note that it is straightforward to adapt this model to the context of forecasting a binary event. Just as before, we have $Y = X_1 + X_2 + \dots$, with each expert knowing a subset of the $X_j$'s, but now the experts are estimating the probability that $Y \ge 0$, conditioned on their information. The aggregator's task is now to combine these probabilities into an overall probability that $Y \ge 0$.

\paragraph{Finding a good deductive estimator} In Chapter~\ref{chap:elk}, we found a deductive estimation algorithm that satisfies linearity and respect for proofs. We then exhibited an informal property (independence of irrelevant information) that the algorithm does not satisfy, thus prompting the question of whether a ``better'' deductive estimation algorithm exists. This question decomposes into two sub-questions. First, what does it mean for a deductive estimation algorithm to be good? A formalization of ``independence of irrelevant information'' would help answer this question, as would formulating other desirable properties of an estimation algorithm (perhaps ``iterated estimation" and ``pulling out known factors," as discussed in Section~\ref{sec:cond_exp_desiderata}). Second, how do we create a deductive estimation algorithm that is good (as defined by the answer to the first question)? As discussed in Section~\ref{sec:chap9_conclusion}, these questions seem quite fundamental.

\paragraph{Wagering mechanisms that produce good aggregate forecasts} In Chapter~\ref{chap:intro}, we briefly discussed wagering mechanisms, which are a type of mechanism for eliciting probabilistic forecasts from multiple experts. Each expert submits a forecast and a \emph{wager,} and the wagers get redistributed according to each expert's performance.

Wagering mechanisms are similar to prediction markets, in that they elicit forecasts by giving experts opportunities to place bets. Prediction markets also aggregate experts' forecasts into a single forecast, which can be inferred from market prices. It is natural to wonder whether wagering mechanisms can also be used for aggregation.

\textcite[\S6]{llwcrsp15} ask the natural question of what Bayesian experts' wagers look like in equilibrium.\footnote{To circumvent the no-trade theorem, \textcite{llwcrsp15} posit that the experts get utility from gambling. Alternatively, we could circumvent the no-trade theorem by subsidizing the wagering mechanism.} For a given information structure, we could say that a wagering mechanism produces good aggregate forecasts if pooling the experts' forecasts according to their equilibrium wagers outperforms pooling the forecasts with equal weights.\footnote{Intuitively, a wagering mechanism should produce good aggregate forecasts for an information structure if each expert's equilibrium wager reflects their confidence. By \emph{confidence,} we (informally) mean the extent to which the expert expects their forecast to remain unchanged upon learning other experts' signals. If experts' wagers reflect their confidence, then an expert who knows almost all of the available information will make a large wager, while an expert who knows that other experts have important private information will make a small wager. In that case, using wagers as weights for aggregation should produce a sensible result.} Perhaps it is possible to create a wagering mechanism that produces good aggregate forecasts for a large, natural class of information structures.

It seems to me that wagering mechanisms have the potential to be a powerful tool for eliciting and aggregating forecasts, alongside prediction markets. However, such wagering mechanisms would need to be designed so that optimal wagers are reflective of confidence.\\

In summary, algorithmic Bayesian epistemology presents a wide range of fascinating questions and possible directions for further research. I hope that this thesis has conveyed my enthusiasm for ABE, and that it has sparked a similar curiosity in you as well.

\clearpage
\phantomsection 
\titleformat{\chapter}[display]
{\normalfont\bfseries\filcenter}{}{0pt}{\large\bfseries\filcenter{#1}}  
\titlespacing*{\chapter}
  {0pt}{0pt}{30pt}

\begin{singlespace}  
	\setlength\bibitemsep{\baselineskip}  
	\addcontentsline{toc}{chapter}{References}  
	\printbibliography[title={References}]
\end{singlespace}


\titleformat{\chapter}[display]
{\normalfont\bfseries\filcenter}{}{0pt}{\Large\chaptertitlename\ \Large\thechapter : \Large\bfseries\filcenter{#1}}  
\titlespacing*{\chapter}
  {0pt}{0pt}{30pt}	
  



\addtocontents{toc}{\protect\setcounter{tocdepth}{0}}
\begin{appendices}

\addtocontents{toc}{\protect\renewcommand{\protect\cftchappresnum}{\appendixname\space}}
\addtocontents{toc}{\protect\renewcommand{\protect\cftchapnumwidth}{6em}}


\chapter{Details omitted from Chapter~\ref{chap:precision}}
\section{Details omitted from Section~\ref{sec:chap3_prelims}}\label{app:prelim}
\begin{fact} \label{fact:beta}
    If a coin's bias is chosen uniformly from $[0, 1]$, then for all $n$ and $0 \le k \le n$, the probability that exactly $k$ of $n$ flips come up heads is $\frac{1}{n + 1}$.
\end{fact}

\begin{proof}
    Consider the following method of choosing the $n$ flip outcomes: first, choose a number $r$ in $[0, 1]$ (the bias); then, choose $n$ numbers in $[0, 1]$, each of which corresponds to a heads flip if it is less than $r$. The probability of $k$ heads is the probability that the first of the $n + 1$ numbers is the $(k + 1)$-th smallest, which is $\frac{1}{n + 1}$.
\end{proof}

\weaklyproper*

\begin{proof}
Fix a scoring rule $s$ and let $r_p(x)$ be the expected score of an expert with belief $p$ who reports $x$ (so $r_p(x) = p s(x) + (1 - p) s(1 - x)$).

We first prove that if $s$ is weakly proper then it satisfies the two stated constraints. Suppose that $s$ is weakly proper. It is clear that $s$ satisfies the first equation: for all $p$, in order for $r_p(x)$ to have a maximum at $x = p$, its derivative $ps'(x) - (1 - p)s'(1 - x)$ must be $0$ at $x = p$. So we first conclude that we must have $xs'(x) = (1-x)s'(1-x)$ for all $x \in (0,1)$. Next, observe that
\begin{equation} \label{rpxeq}
r_p'(x) = p s'(x) - (1-p)s'(1-x) = ps'(x) - (1 - p) \frac{xs'(x)}{1 - x} = s'(x) \parens{p - (1 - p) \frac{x}{1 - x}}.
\end{equation}
Suppose for contradiction that for some $p \in (0, 1)$, we have $s'(p) < 0$. Since $s'$ is continuous, $s'(x) < 0$ on some open interval containing $p$. On that open interval, then, the sign of $r_p'(x)$ is the opposite of the sign of $p - (1 - p) \frac{x}{1 - x}$ -- that is, negative when $x < p$ and positive when $x > p$. But then $r_p(x)$ is strictly \emph{minimized}, rather than maximized, at $x = p$ on this interval, contradicting that $s$ is weakly proper.

To prove the stronger claim when $s$ is proper, assume for contradiction that $s'(p)$ is not strictly positive almost everywhere. Then because $s'$ is continuous, there is an interval of non-zero length in which $s'(x) = 0$ on the entire interval. Let $p$ lie on the interior of this interval. Equation~\eqref{rpxeq} then establishes that $r_p'(x)$ is $0$ in an interval around $x = p$, meaning that $p$ is not the unique maximizer of $r_p$, contradicting that $s$ is proper.

Conversely, suppose that $s$ satisfies the two stated constraints. We show that $s$ is weakly proper by showing a stronger statement: that for all $p$, $r_p(x)$ weakly increases on $(0, p]$ and weakly decreases on $[p, 1)$. By the first constraint, (\ref{rpxeq}) holds. By the second constraint, for all $x$, $r_p'(x)$ is either $0$ or has the sign of $p - (1 - p)\frac{x}{1 - x}$, i.e.\ positive if $x < p$ and negative if $x > p$. This means that $r_p(x)$ is weakly increasing on $(0, p]$ and weakly decreasing on $[p, 1)$, and so attains a weak global maximum at $x = p$, as desired.

To prove the stronger claim when $s'(x) > 0$ almost everywhere, we show that $r_p(x)$ strictly increases almost everywhere on $(0, p]$ and strictly decreases almost everywhere on $[p, 1)$. Again, (\ref{rpxeq}) holds, so by the second constraint we have that $r_p'(x)$ has the sign of $p - (1 - p)\frac{x}{1 - x}$ almost everywhere, i.e.\ positive if $x < p$ and negative if $x > p$. Thus, $r_p(x)$ is strictly increasing almost everywhere on $(0, p]$ and strictly decreasing almost everywhere on $[p, 1)$, and so attains a strict global maximum at $x = p$, as desired.
\end{proof}

\strictlyproper*

\begin{proof}
As $\frac{1 - p}{p}$ is strictly positive on $(\half, 1)$, and $s'(1-p)$ is strictly (resp., weakly) positive almost everywhere on $(\half,1)$, we immediately conclude that $s'(p)$ is also strictly (resp., weakly) positive almost everywhere on $(\half, 1)$. Therefore, $s$ is proper (resp., weakly proper) by Lemma~\ref{lem:weaklyproper}.
\end{proof}

\begin{prop} \label{prop:cost}
Let $s$ be a proper scoring rule and define $\cost_c(s)$ to be the expected score of a globally-adaptive expert who is scored with $s$. Then $\lim_{c \rightarrow 0} \cost_c(s) = \int_0^1 G_s(x) dx$. That is, $\cost_c(s) = \int_0^1 G_s(x) dx \pm o(1)$.
\end{prop}

\begin{proof}
First, we wish to argue that as $c \rightarrow 0$, \emph{no matter the true bias,} the number of flips any expert will choose to make approaches $\infty$. To see this, observe that after $n$ flips, the expert's current belief will always be an integer multiple of $\frac{1}{n+1}$. So if
\[c(m):= \min_{k<n \leq m, \text{ and } k,n \in \mathbb{N}}\left\{\frac{k+1}{n+2}G_s \parens{\frac{k+2}{n+3}} + \frac{n-k+1}{n+2}G_s \parens{\frac{k+1}{n+3}} - G_s \parens{\frac{k+1}{n+2}}\right\}\]
then the expert will always flip the coin at least $m$ times as long as $c \leq c(m)$. Observe also that the minimum is taken over finitely many terms, all of which are strictly positive, so $c(m)$ is strictly positive. Therefore, for all $m$, there is a sufficiently small $c(m) > 0$ such that whenever the cost is at most $c(m)$, the expert flips at least $m$ times no matter the true bias. Note that while these calculations are done for an adaptive expert, they hold for a non-adaptive expert as well because the non-adaptive expert would want to flip at least $m$ coins \emph{no matter the outcomes}. 

Now, let's consider the expected score of an expert who makes exactly $m$ flips no matter what. For all $k$, such an expert sees $k$ total heads with probability $\frac{1}{m + 1}$ (Fact~\ref{fact:beta}). And conditioned on seeing $k$ heads, the expert's expected score tomorrow is $G_s \parens{\frac{k+1}{m+2}}$. Therefore, we can conclude that for all $m$, the expert's expected score after $m$ coin flips is $\frac{1}{m + 1} \sum_{k=0}^m G_s \parens{\frac{k+1}{m+2}}$.

Now, we want to understand the limit of this sum as $m \rightarrow \infty$. Observe that for each $m$, the sum is a Riemann sum for the function $G_s(x)$ on $[0,1]$ (i.e.\ each $\frac{k+1}{m+2}$ lies inside the interval $\left[\frac{k}{m+1}, \frac{k+1}{m+1}\right]$). Therefore, the limit as $m \rightarrow \infty$ is just the Riemann integral, and we get:

$$\lim_{m \rightarrow \infty} \frac{1}{m + 1} \sum_{k=0}^m G_s \parens{\frac{k+1}{m+2}} = \int_0^1 G_s(x)dx.$$

So now we can conclude that the non-adaptive expert gets expected payoff $\int_0^1 G_s(x)dx + o(1)$. As $c \rightarrow 0$, the number of flips $m \rightarrow \infty$, and the expected payoff as $m \rightarrow \infty$ approaches $\int_0^1 G_s(x) dx$. For an adaptive expert, observe that by Lemma~\ref{lem:increward}, as long as they flip the coin at least $m$ times with probability $1$, their expected score is at least as large as if they flipped it exactly $m$ times with probability $1$. As their expected score can certainly not exceed $\int_0^1 G_s(x) dx$ (as this is the score of a perfect expert who knows exactly the bias), their expected score must also approach $\int_0^1 G_s(x) dx$ as $m \rightarrow \infty$ (and therefore as $c \rightarrow \infty$ as well).
\end{proof}

\normalizedequiv*
\begin{proof}
We have
\[\int_0^1 G_s(x) dx = \int_0^1 (xs(x) + (1 - x)s(1 - x)) dx = \int_0^1 2x s(x) dx\]
where the last step follows by separating $(1 - x)s(1 - x)$ into its own integral and substituting $u = 1 - x$. Now we integrate by parts, letting $u = s(x)$ and $dv = 2x dx$, so that $du = s'(x)$ and $v = x^2$, to get
\begin{align*}
    \int_0^1 G_s(x) dx &= x^2 s(x) \mid_0^1 - \int_0^1 x^2 s'(x) dx = s(1) - \int_0^1 x^2 s'(x) dx\\
    &= s(1) - \int_0^\half x^2 s'(x) dx - \int_\half^1 x^2 s'(x) dx\\
    &= s(1) - \int_0^\half x(1 - x)s'(1 - x) dx - \int_\half^1 x^2 s'(x) dx\\
    &= s(1) - \int_\half^1 (x(1 - x) + x^2)s'(x) dx = s(1) - \int_\half^1 xs'(x) dx.
\end{align*}
Expressing $s(1) - s \parens{\half}$ as $\int_\half^1 s'(x) dx$, we obtain the desired equality.
\end{proof}

\section{Properties of respectful scoring rules}\label{app:respect}
We state several sufficient conditions for a scoring rule to be respectful, confirm that typical scoring rules are respectful, and provide a brief discussion.

\begin{claim} \label{claim:easy}
If Conditions 1 and 2 of Definition~\ref{def:respectful} hold and $\abs{G_s'''(x)}$ is bounded on $(0, 1)$ then $s$ is respectful.
\end{claim}

This should be clear: take $c$ to be small enough such that $c^{-.16}$ times the lower bound on $G_s''$ exceeds the upper bound on $\abs{G_s'''}$. Scoring rules such as the quadratic scoring rule and the spherical scoring rule satisfy the hypotheses of Claim~\ref{claim:easy}. One well-known generalization of the quadratic scoring rule is the Tsallis rule~\parencite{tsa88}. This rule, parametrized by $\gamma > 1$, is defined to be the scoring rule $s$ for which $G_s(x) = x^\gamma + (1 - x)^\gamma$. When $\gamma = 2$, this yields the quadratic scoring rule.

For $\gamma \ge 3$, it is evident that the Tsallis rule satisfies the hypotheses of Claim~\ref{claim:easy}. However, this is not so for $\gamma < 3$ (except for $\gamma = 2$). Perhaps more importantly, the logarithmic scoring rule does not satisfy Claim~\ref{claim:easy} either. This motivates the following result (the proof appears at the end of this section).

\begin{claim} \label{claim:interesting}
Suppose that Conditions 1 and 2 of Definition~\ref{def:respectful} are satisfied. Suppose further that $G_s'''$ is bounded on any closed sub-interval of $(0, 1)$, and that there exist constants $k \neq 0$ and $r$ such that $\lim_{x \to 0} x^r G_s'''(x) = k$. Then $s$ is respectful.
\end{claim}

The logarithmic scoring rule satisfies the hypotheses of Claim~\ref{claim:interesting} ($r = 2$ and $k = -1$). The Tsallis rule with $\gamma < 3$ (and $\gamma \neq 2$) also satisfies these hypotheses ($r = 3 - \gamma$ and $k = \gamma(\gamma - 1)(\gamma - 2)$). The proof of Claim~\ref{claim:interesting} is more involved, and deferred to Appendix~\ref{app:index}.


We briefly discuss ways in which proper scoring rules can fail to be respectful. One way a scoring rule can be disrespectful is if $G_s''$ grows extremely quickly near zero (e.g.\ $G_s''(x) = e^{1/x}$). Such functions, however, are outside the scope of this entire exercise because they are not normalizable. That is, such $G_s$ have $\int_0^1 G_s(x) dx = \infty$, and provide infinite expected payment to the expert. So this ``limitation'' of respectfulness is more of a restatement of normalizability.

Another way a scoring rule could be disrespectful is if $G_s''$ is not bounded away from zero. For example: $G_s''(x) = \parens{x - \frac{1}{2}}^2$ or $G_s''(x) = x(1 - x)$. If $G_s''$ remains ``very flat'' near $0$ for a ``large interval'', then $\ind^\ell(s)$ is infinite anyway. This implies that we should expect the error to be a \emph{larger} order of magnitude than $c^{-\ell/4}$, and for such functions to not incentivize precision well at all (although we do not explicitly prove this). This makes sense: if $G_s''(x) \approx 0$, then the expert gains $\approx 0$ by flipping the coin to refine their current belief (Lemma~\ref{lem:increward}). It is also possible that $G_s''$ is not bounded away from zero, but also not ``very flat''. We conjecture that Theorem~\ref{thm:global} does hold for such functions, but that our approach does not establish this. While it is possible to come up with such functions (e.g., the two above) which elude Theorem~\ref{thm:global}'s precise statement, this does not affect commonly-studied scoring rules, nor the scoring rules designed in this paper (sometimes leaning on Theorem~\ref{thm:approx}).

Finally, as with any exercise in analysis, there are continuous functions that behave erratically near zero, such as $G_s''(x) = \sin \parens{\frac{1}{x}} + \sin \parens{\frac{1}{1 - x}} + 3$. While it may or may not be the case that Theorem~\ref{thm:global} extends to such functions, this does not seem particularly relevant.

\begin{proof}[Proof of Claim~\ref{claim:interesting}]
Let $r, k$ be as in Claim~\ref{claim:interesting}. If $r \le 0$ then the claim is uninteresting: $G_s'''$ is bounded on $(0, 1)$ and so the statement is subsumed by Claim~\ref{claim:easy}. The interesting case is when $r > 0$.

We first consider the case when $r > 1$. Note that $\lim_{x \to 0} G_s''(x) = \infty$. To see this, suppose for contradiction that this limit is finite. We may write
\[\lim_{x \to 0} G_s''(x) = \lim_{x \to 0} \frac{xG_s''(x)}{x} = \lim_{x \to 0} xG_s'''(x) + G_s''(x)\]
by L'H\^{o}pital's rule, so $\lim_{x \to 0} xG_s'''(x) = 0$, contradicting that $\lim_{x \to 0} x^r G_s'''(x) \neq 0$.

Now, the fact that $\lim_{x \to 0} G_s''(x) = \infty$ lets us apply L'H\^{o}pital's rule:
\[\lim_{x \to 0} x^{r - 1} G_s''(x) = \lim_{x \to 0} \frac{G_s''(x)}{x^{1 - r}} = \lim_{x \to 0} \frac{G_s'''(x)}{(1 - r)x^{-r}} = \frac{1}{1 - r} \lim_{x \to 0} x^r G_s'''(x) = \frac{k}{1 - r}.\]
This means that
\[\lim_{x \to 0} x \frac{G_s'''(x)}{G_s''(x)} = 1 - r\]
so in particular, there exists $\delta > 0$ such that for all $x \le \delta$ we have $x \frac{G_s'''(x)}{G_s''(x)} \in [-r, 2 - r]$ and so $\frac{\abs{G_s'''(x)}}{G_s''(x)} \le \frac{r}{x}$. On the other hand, $G_s'''$ is bounded on $[\delta, 1 - \delta]$ by assumption.

To finish, let $t$ as in Definition~\ref{def:respectful} equal $0.3$. Assume $c$ is small enough that the following conditions are satisfied:
\begin{itemize}
    \item $\frac{\abs{G_s'''(x)}}{G_s''(x)} \le c^{-.16}$ on $[\delta, 1 - \delta]$.
    \item $c^{-.01} \ge r$.
\end{itemize}
Then the last condition of Definition~\ref{def:respectful} will be satisfied on $[\delta, 1 - \delta]$; it will also be satisfied on $[c^{.3}, \delta]$ for any $c$ satisfying the second condition above, because on that interval we have

\[\frac{\abs{G_s'''(x)}}{G_s''(x)} \le \frac{r}{x} \le \frac{1}{c^{.16} \sqrt{x(1 - x)}} \cdot \frac{c^{.16}r}{\sqrt{x}} \le \frac{1}{c^{.16} \sqrt{x(1 - x)}} \cdot \frac{c^{.16}r}{c^{.15}} \le \frac{1}{c^{.16} \sqrt{x(1 - x)}}.\]

By symmetry of $G_s''$ about $\frac{1}{2}$ (and antisymmetry of $G_s'''$) we have that the condition also holds on $[1 - \delta, 1 - c^t]$, as desired.

Now we consider the case that $r = 1$. As above, we have $\lim_{x \to 0} G_s''(x) = \infty$. Proceeding similarly, we have
\[\lim_{x \to 0} \frac{G_s''(x)}{\ln x} = \lim_{x \to 0} \frac{G_s'''(x)}{\frac{1}{x}} = \lim_{x \to 0} xG_s'''(x) = k.\]
This means that there exists $\delta > 0$ such that for all $x \le \delta$ we have $x \ln x \frac{G_s'''(x)}{G_s''(x)} \in [0, 2]$ and so $\frac{\abs{G_s'''(x)}}{G_s''(x)} \le \frac{-2}{x \ln x}$. We finish as before.

Finally, consider the case that $0 < r < 1$. Let $a$ be a lower bound on $G_s''$, as in the statement of Claim~\ref{claim:interesting}. It suffices to show that for for $c$ small enough, we have $\abs{G_s'''(x)} \le \frac{a}{c^{.16}\sqrt{x}}$ on $[c^{.3}, 1 - c^{.3}]$. Let $\delta$ be such that $x^r G_s'''(x) \in [k - 1, k + 1]$ for all $x \le \delta$. On $[c^{.3}, \delta]$ we have
\[\abs{G_s'''(x)} \le \frac{\abs{k} + 1}{x^r} \le \frac{\abs{k} + 1}{x} = \frac{a}{c^{.16} \sqrt{x}} \cdot \frac{\abs{k} + 1}{ac^{-.16} \sqrt{x}} \le \frac{a}{c^{.16} \sqrt{x}} \cdot \frac{\abs{k} + 1}{ac^{-.16} \cdot c^{.15}}\]
if $c$ is small enough that $c^{-.01} \ge \frac{\abs{k} + 1}{a}$. (As before, we also need to make sure that $c$ is small enough that the condition is satisfied on $[\delta, 1 - \delta]$.) This concludes the proof.
\end{proof}

\section{Details omitted from Section~\ref{sec:index}}\label{app:index}
\subsection{Details omitted from Section~\ref{sec:one}}
\rdoubleprime*

\begin{proof}
By Lemma~\ref{lem:increward}, we have
\[\Delta_{n + 1} = \frac{h + 1}{n + 2} G_s \parens{\frac{h + 2}{n + 3}} + \frac{n - h + 1}{n + 2} G_s \parens{\frac{h + 1}{n + 3}} - G_s \parens{\frac{h + 1}{n + 2}}.\]
Since $G_s$ is twice differentiable, we may use Taylor's approximation theorem to write
\[G_s \parens{\frac{h + 1}{n + 3}} = G_s \parens{\frac{h + 1}{n + 2}} + \parens{\frac{h + 1}{n + 3} - \frac{h + 1}{n + 2}} G_s' \parens{\frac{h + 1}{n + 2}} + \frac{1}{2} \parens{\frac{h + 1}{n + 3} - \frac{h + 1}{n + 2}}^2 G_s''(c_1)\]
for some $c_1 \in \brackets{\frac{h + 1}{n + 3}, \frac{h + 1}{n + 2}}$. Similarly we have
\[G_s \parens{\frac{h + 2}{n + 3}} = G_s \parens{\frac{h + 1}{n + 2}} + \parens{\frac{h + 2}{n + 3} - \frac{h + 1}{n + 2}} G_s' \parens{\frac{h + 1}{n + 2}} + \frac{1}{2} \parens{\frac{h + 2}{n + 3} - \frac{h + 1}{n + 2}}^2 G_s''(c_1)\]
for some $c_2 \in \brackets{\frac{h + 1}{n + 2}, \frac{h + 2}{n + 3}}$. When we plug these expressions into the formula for $\Delta_{n + 1}$ above, the zeroth- and first-order terms cancel. We are left with
\begin{align*}
\Delta_{n + 1} &= \frac{n - h + 1}{n + 2} \cdot \frac{1}{2} \parens{\frac{h + 1}{n + 3} - \frac{h + 1}{n + 2}}^2 G_s''(c_1) + \frac{h + 1}{n + 2} \cdot \frac{1}{2} \parens{\frac{h + 2}{n + 3} - \frac{h + 1}{n + 2}}^2 G_s''(c_2)\\
&= \frac{(h + 1)^2(n - h + 1)}{2(n + 2)^3(n + 3)^2} G_s''(c_1) + \frac{(h + 1)(n - h + 1)^2}{2(n + 2)^3(n + 3)^2} G_s''(c_2)\\
&= \frac{q(1 - q)}{2(n + 3)^2}(qG_s''(c_1) + (1 - q)G_s''(c_2)).
\end{align*}
Note that $\abs{c_1 - q} \le \frac{h + 1}{n + 2} - \frac{h + 1}{n + 3} \le \frac{1}{n}$, so $c_1 \in [q - \frac{1}{n}, q + \frac{1}{n}]$, and similarly for $c_2$. This completes the proof.
\end{proof}

\onethirdbound*

\begin{proof}
Suppose that $G_s''(x) \ge a$ for all $x \in (0, 1)$. By Claim~\ref{claim:rdoubleprime} we have
\[\Delta_{n + 1} \ge \frac{q(1 - q)a}{2(n + 3)^2}.\]
Now, we have that $\frac{1}{n + 2} \le q \le \frac{n - 1}{n + 2}$, and $q(1 - q)$ decreases as $q$ gets farther from $\frac{1}{2}$. This means that
\[\Delta_{n + 1} \ge \frac{\frac{n + 1}{(n + 2)^2} a}{2(n + 3)^2} \ge \frac{a}{72(n + 1)^3}.\]
Therefore, if $\Delta_{n + 1} < c$ then $n + 1 > \frac{1}{\parens{\frac{72}{a}}^{1/3}c^{1/3}}$, so $n > \frac{1}{\alpha c^{1/3}}$ for some $\alpha$ (not to be confused with $a$), if $c$ is small enough.
\end{proof}

\subsection{Details omitted from Section~\ref{sec:two}}
\omegap*

\begin{proof}
Let $1 \le j_p \le n - 2$ be such that $\frac{j_p}{n} \le p \le \frac{j_p + 1}{n}$. We have
\[\frac{j_p}{n} - \frac{\sqrt{j_p(n - j_p)}}{2n^{1.49}} \le Q_{j_p/n}(n) \le Q_p(n) \le Q_{(j_p + 1)/n}(n) \le \frac{j_p + 1}{n} + \frac{\sqrt{(j_p + 1)(n - 1 - j_p)}}{2n^{1.49}}\]
so
\begin{align*}
\abs{Q_p(n) - p} &\le \max \parens{p - \frac{j_p}{n} + \frac{\sqrt{j_p(n - j_p)}}{2n^{1.49}}, \frac{j_p + 1}{n} - p + \frac{\sqrt{(j_p + 1)(n - 1 - j_p)}}{2n^{1.49}}}\\
&\le \frac{1}{n} + \frac{1}{2n^{1.49}} \max \parens{\sqrt{j_p(n - j_p)}, \sqrt{(j_p + 1)(n - 1 - j_p)}}.
\end{align*}
For fixed $n$ and for $\frac{1}{n} \le p \le 1 - \frac{1}{n}$, this maximum divided by $n\sqrt{p(1 - p)}$ is maximized when $p = \frac{1}{n}$ (or $p = 1 - \frac{1}{n}$), in which case the ratio is $\sqrt{\frac{2(n - 2)}{n - 1}} \le \sqrt{2}$. Therefore we have
\[\frac{\abs{Q_p(n) - p}}{\sqrt{p(1 - p)}} \le \frac{1}{n\sqrt{p(1 - p)}} + \frac{n\sqrt{2}}{2n^{1.49}} \le \frac{1}{n^{.49}}\]
for $n$ large enough. (Here we again use that $p \ge \frac{1}{n}$, so $\sqrt{p(1 - p)}$ is minimized at $p = \frac{1}{n}$.)
\end{proof}

\omegaunlikely*

\begin{proof}
We have
\[\pr{\overline{\Omega_N}} \le \sum_{n = N}^\infty \sum_{j = 1}^{n - 1} \pr{\abs{Q_{j/n}(n) - \frac{j}{n}} > \frac{\sqrt{j(n - j)}}{2n^{1.49}}}.\]
Now, let $\mathcal{H}_{j/n}(n)$ be the fraction of the first $n$ coin flips that were heads (so $\mathcal{H}_{j/n}(n)$ is an average of $n$ i.i.d. Bernoulli random variables that are $1$ with probability $\frac{j}{n}$). Note that $Q_{j/n}(n)$ is within $\frac{1}{n}$ of $\mathcal{H}_{j/n}(n)$, and for large $n$ we have $\frac{1}{n} \le \frac{\sqrt{j(n - j)}}{2n^{1.49}}$ for all $j$. This means that for large $n$, by the triangle inequality we have that if $\abs{\mathcal{H}_{j/n}(n) - \frac{j}{n}} > \frac{\sqrt{j(n - j)}}{n^{1.49}}$ then $\abs{Q_{j/n}(n) - \frac{j}{n}} > \frac{\sqrt{j(n - j)}}{2n^{1.49}}$. Therefore, for large $N$ we have
\[\pr{\overline{\Omega_N}} \le \sum_{n = N}^\infty \sum_{j = 1}^{n - 1} \pr{\abs{\mathcal{H}_{j/n}(n) - \frac{j}{n}} > \frac{\sqrt{j(n - j)}}{n^{1.49}}}.\]
We bound each of these probabilities. Recall the following version of the Chernoff bound: for $0 \le \delta \le 1$, if $X = \sum_{i = 1}^n X_i$ is a sum of i.i.d. Bernoulli random variables with $\EE{X} = \mu$, then
\[\pr{\abs{X - \mu} \ge \delta \mu} \le 2e^{-\mu \delta^2/3}.\]
We apply this to our random variables (so $X = n\mathcal{H}_{j/n}(n)$ and $\mu = j$). Assume $j \le \frac{n}{2}$. Let $\delta = n^{-.49} \sqrt{\frac{n - j}{j}}$. Then
\[\pr{\abs{\frac{X}{n} - \frac{j}{n}} \ge \frac{\sqrt{j(n - j)}}{n^{-1.49}}} \le 2e^{-n^{.02}/6}.\]
If $j \ge \frac{n}{2}$, a symmetry argument yields the same result. Therefore, for sufficiently large $N$ we have
\begin{align*}
\pr{\overline{\Omega_N}} &\le \sum_{n = N}^\infty \sum_{j = 1}^{n - 1} 2e^{-n^{.02}/6} \le \sum_{n = N}^\infty 2ne^{-n^{.02}/6} \le \sum_{n = N}^\infty e^{-n^{.02}/7} = e^{-N^{.02}/7} \sum_{n = N}^\infty e^{-(n^{.02} - N^{.02})/7}\\
&\le e^{-N^{.02}/7} \sum_{n = N}^\infty e^{-n^{.02}/14} \le e^{-N^{.02}/7} \sum_{n = N}^\infty \frac{14^{100} \cdot 100!}{n^2}.
\end{align*}
The last step comes from observing that $e^x \ge \frac{x^{100}}{100!}$ for positive $x$ and plugging in $x = \frac{n^{.02}}{14}$. Now, this summand is bounded by a constant, since $\sum_{n = 1}^\infty \frac{1}{n^2}$ converges, and so we have
\[\pr{\overline{\Omega_N}} \le O \parens{e^{-N^{.02}/7}} \le O \parens{e^{-N^{.01}}},\]
as desired.
\end{proof}

\subsection{Details omitted from Section~\ref{sec:three}}
\nstopbound*

\begin{proof}
We assume for convenience that $t < 0.3$ (which is safe, as Definition~\ref{def:respectful} holds for all $t \leq t'$ whenever it holds for $t'$).

Fix $p \in [2c^t, 1 - 2c^t]$. Let $\alpha$ be as in Claim~\ref{claim:one_third_bound}. As before, let $Q(n)$ be the predictor's estimate for the bias of the coin after $n$ flips. Then for $c$ small enough that $\frac{1}{\alpha c^{1/3}} \ge N$ and $c^{1/30} \le \frac{2}{\alpha}$, for $n \ge \frac{1}{\alpha c^{1/3}}$, we have
\[p - \sqrt{p(1 - p)}(\alpha c^{1/3})^{.49} \le Q(n) \le p + \sqrt{p(1 - p)}(\alpha c^{1/3})^{.49}.\]
This follows from Claim~\ref{claim:omega_p}, noting that if $c^{1/30} \le \frac{2}{\alpha}$ and $n \ge \frac{1}{\alpha c^{1/3}}$ then $\frac{1}{n} \le p \le 1 - \frac{1}{n}$.

Now, recall Claim~\ref{claim:rdoubleprime}:
\[\Delta_{n + 1} = \frac{Q(n)(1 - Q(n))}{2(n + 3)^2}(Q(n)G_s''(c_1) + (1 - Q(n))G_s''(c_2))\]
for some $c_1, c_2 \in [Q(n) - \frac{1}{n}, Q(n) + \frac{1}{n}]$. In the remainder of this proof, what we essentially argue is that $G_s''$ on this interval is not too far from $G_s''(p)$, because of our bound on $Q(n)$ as $p$ plus or minus a small quantity.

We ask: for a given (possibly negative) $\epsilon$, how far from $G_s''(p)$ can $G_s''(p + \epsilon)$ be? Well, since $G_s'''$ is integrable, we have
\[\abs{G_s''(p + \epsilon) - G_s''(p)} = \abs{\int_p^{p + \epsilon} G_s'''(x) dx} \le \abs{\int_p^{p + \epsilon} \abs{G_s'''(x)} dx}.\]
Now, since $s$ is respectful we have that for $c$ small enough, if $p, p + \epsilon \in [c^t, 1 - c^t]$ then
\[\abs{G_s'''(x)} \le \frac{1}{c^{.16}\sqrt{x(1 - x)}} G_s''(x) \le \frac{1}{c^{.16}\sqrt{\hat{p}(1 - \hat{p})}} G_s''(x),\]
where $\hat{p}$ is defined to be the number on the interval between $p$ and $p + \epsilon$ minimizing $\sqrt{x(1 - x)}$ (i.e.\ farthest from $\frac{1}{2}$).

Define $r := \frac{1}{c^{.16}\sqrt{\hat{p}(1 - \hat{p})}}$. Then $\abs{G_s'''(x)} \le rG_s''(x)$. Below, we will use this fact to prove the following claim.

\begin{claim} \label{claim:diff_ineq}
$\abs{G_s''(p + \epsilon) - G_s''(p)} \le G_s''(p)(e^{r\abs{\epsilon}} - 1)$.
\end{claim}

Assuming the claim for now: how large of an $\epsilon$ do we care about? The farthest that $c_1$ and $c_2$ can be from $p$ is
\[\sqrt{p(1 - p)}(\alpha c^{1/3})^{.49} + \alpha c^{1/3} \le 2\sqrt{p(1 - p)}(\alpha c^{1/3})^{.49},\]
for small $c$. (This is because we assumed for convenience that $t < 0.3$, which means that $p(1 - p) \ge c^{.3}$, so $\alpha c^{1/3} \le \sqrt{p(1 - p)}(\alpha c^{1/3})^{.49}$.) Therefore, by Claim~\ref{claim:diff_ineq} we have
\begin{align*}
\abs{G_s''(p) - G_s''(c_1)}, \abs{G_s''(p) - G_s''(c_2)} &\le G_s''(p)(e^{2r\sqrt{p(1 - p)}(\alpha c^{1/3})^{.49}} - 1)\\
&= G_s''(p) \exp \parens{\frac{2\sqrt{p(1 - p)}(\alpha c^{1/3})^{.49}}{c^{.16}\sqrt{\hat{p}(1 - \hat{p})}}} - G_s''(p)
\end{align*}
where $\hat{p}$ is either $p \pm 2\sqrt{p(1 - p)}(\alpha c^{1/3})^{.49}$, whichever is farther from $\frac{1}{2}$. It is easy to check\footnote{Without loss of generality assume $p \le \frac{1}{2}$, so $\hat{p} = p - 2\sqrt{p(1 - p)}(\alpha c^{1/3})^{.49}$. Then $\frac{\sqrt{p(1 - p)}}{\sqrt{\hat{p}(1 - \hat{p})}} \le \frac{p(1 - p)}{\hat{p}(1 - \hat{p})} \le \frac{p}{\hat{p}}$, so it suffices to show that $2\sqrt{p(1 - p)}(\alpha c^{1/3})^{.49} \le p$. This is indeed the case, as $p \ge c^{.3} \ge 4(\alpha c^{1/3})^{.98}$ for small $c$, so $2\sqrt{p}(\alpha c^{1/3})^{.49} \le p$.} that for small enough $c$ we have that $\frac{\sqrt{p(1 - p)}}{\sqrt{\hat{p}(1 - \hat{p})}} \le 2$, and so we have
\[\abs{G_s''(p) - G_s''(c_1)}, \abs{G_s''(p) - G_s''(c_2)} \le G_s''(p) \exp \parens{\frac{1}{2} \beta c^{1/300}} - G_s''(p) \le \beta c^{1/300} G_s''(p)\]
for small enough $c$, where $\beta = 8\alpha^{.49}$. (Here we use that $e^x \le 1 + 2x$ for small positive $x$.) It follows, then, by Claim~\ref{claim:rdoubleprime}, that
\[\frac{Q(n)(1 - Q(n))}{2(n + 3)^2} G_s''(p)(1 - \beta c^{1/300}) \le \Delta_{n + 1} \le \frac{Q(n)(1 - Q(n))}{2(n + 3)^2} G_s''(p)(1 + \beta c^{1/300}).\]
Note that since $Q(n) \ge p - \sqrt{p(1 - p)}(\alpha c^{1/3})^{.49}$ and $1 - Q(n) \ge 1 - p - \sqrt{p(1 - p)}(\alpha c^{1/3})^{.49}$, we may write
\begin{align*}
Q(n)(1 - Q(n)) &\ge p(1 - p)(1 + (\alpha c^{1/3})^{.98}) - \sqrt{p(1 - p)}(\alpha c^{1/3})^{.49} = p(1 - p) \parens{1 - \frac{(\alpha c^{1/3})^{.49}}{\sqrt{p(1 - p)}}}\\
&\ge p(1 - p) \parens{1 - \frac{2(\alpha c^{1/3})^{.49}}{c^{t/2}}} \ge p(1 - p)(1 - \alpha^{.49} c^{.01})
\end{align*}
for $c$ small enough that the second-to-last step holds. (In the last step we use that $t < 0.3$.) A similar calculation shows that $Q(n)(1 - Q(n)) \le p(1 - p)(1 + \alpha^{.49} c^{.01})$ for $c$ small enough.\footnote{An extra $\alpha^{.49} c^{.49/3}$ appears, but this term is dominated by $\alpha^{.49} c^{.01}$ for small $c$.} Also note that $n^2 \le (n + 3)^2 \le n^2(1 + 4\alpha c^{1/3})^2$. Putting these approximations all together, we note that the $c^{1/300}$ approximation is the dominant one, which means that there is a constant $\gamma$ such that
\begin{equation} \label{eq:delta_bound}
\frac{p(1 - p)}{2n^2}G_s''(p)(1 - \gamma c^{1/300}) \le \Delta_{n + 1} \le \frac{p(1 - p)}{2n^2}G_s''(p)(1 + \gamma c^{1/300}).
\end{equation}
Therefore, since the expert stops flipping when $\Delta_{n + 1} < c$, we have
\[\sqrt{\frac{p(1 - p)G_s''(p)}{2c}(1 - \gamma c^{1/300})} \le \nstop \le \sqrt{\frac{p(1 - p)G_s''(p)}{2c}(1 + \gamma c^{1/300})}.\]
This holds for any $p$ such that $p \pm 2\sqrt{p(1 - p)}(\alpha c^{1/3})^{.49} \in [c^t, 1 - c^t]$; a sufficient condition is $p \in [2c^t, 1 - 2c^t]$.
\end{proof}

\begin{proof}[Proof of Claim~\ref{claim:diff_ineq}]
We prove this for positive $\epsilon$. The result then follows for negative $\epsilon$ because if some $G_s$ is a counterexample for some negative $\epsilon$, then a function $G_{\tilde{s}}$ defined so that $G_{\tilde{s}}'''(p + x) := -G_s'''(p - x)$ for $x \in [0, -\epsilon]$ serves as a counterexample for $-\epsilon$. (This is because $G_{\tilde{s}}''(p + x) = G_s''(p - x)$ for all $x \in [0, -\epsilon]$, by the fundamental theorem of calculus.) Additionally, we may assume that $G_s''(p) = 1$, because if there is a counterexample function $G_s$ to the claim then $G_{\tilde{s}}(x) := \frac{G_s(x)}{G_s''(p)}$ also serves as a counterexample.

We prove that
\[e^{-r\epsilon} - 1 \le G_s''(p + \epsilon) - 1 \le e^{r\epsilon} - 1.\]
The left inequality suffices because $1 - e^{-x} \le e^x - 1$ for all $x$, so in particular $1 - e^{r\epsilon} \le e^{-r\epsilon} - 1$.

We begin with the right inequality. Suppose for contradiction that $G_s''(p + \epsilon) > e^{r\epsilon}$. Let $S$ be the set of points in $[p, p + \epsilon]$ where $G_s''(x) > e^{r(x - p)}$. Since $S$ contains $p + \epsilon$, it is nonempty; let $p_1 = \inf_S$. Since $G_s''$ is continuous, we have $G_s''(p_1) - e^{r(p_1 - p)} = 0$. Pick $\delta > 0$ small enough that the set $T$ of points $x \in \brackets{p_1, p_1 + \min(\epsilon, \frac{1}{3r})}$ with $G_s''(x) - e^{r(x - p)} > \delta$ is nonempty. Let $p_2 = \inf_T$, so $G_s''(p_2) - e^{r(p_2 - p)} = \delta$. Note that
\[\delta = G_s''(p_2) - e^{r(p_2 - p)} = G_s''(p_1) - e^{r(p_1 - p)} + \int_{p_1}^{p_2} \frac{d}{dx} \parens{G_s''(x) - e^{r(x - p)}} dx = \int_{p_1}^{p_2} \parens{G_s'''(x) - re^{r(x - p)}} dx.\]
It follows that $G_s'''(p_3) - re^{r(p_3 - p)} \ge \frac{\delta}{p_2 - p_1} \ge 2r\delta$ for some $p_3 \in [p_1, p_2]$. (Otherwise the value of the integral would be at most the integral of $2r\delta$ from $p_1$ to $p_2$, which is at most $\frac{2}{3} \delta$, since $p_2 - p_1 \le \frac{1}{3r}$.) Therefore, because $\abs{G_s'''(x)} \le rG_s''(x)$ for all $x \in [p, p + \epsilon]$, we have that
\[rG_s''(p_3) - re^{r(p_3 - p)} \ge 2r\delta\]
so $G_s''(p_3) - e^{r(p_3 - p)} \ge 2\delta$. But then we have that $p_2 < p_3$ and $p_3 \in T$, contradicting the definition of $p_2$ as the infimum of $T$.

The proof of the left inequality above proceeds similarly, but is not exactly analogous. Suppose for contradiction that $t := G_s''(p + \epsilon) - e^{-r\epsilon} < 0$. Define $p_1$ to be the supremum of points in $[p, p + \epsilon]$ where $G_s''(x) \ge e^{-r(x - p)}$ (so $G_s''(p_1) = e^{-r(p_1 - p)}$). Then $G_s''(x) - e^{-r(x - p)}$ is zero at $x = p_1$ and $t$ at $x = p + \epsilon$, so
\[\int_{p_1}^{p + \epsilon} (G_s'''(x) + re^{-r(x - p)}) dx = t < 0,\]
which means that for some $p_2 \in [p_1, p + \epsilon]$ we have that $G_s'''(p_2) + re^{-r(p_2 - p)} \le \frac{t}{2(p + \epsilon - p_1)}$ (otherwise the value of the integral would be at least $\frac{t}{2}$). Since $\abs{G_s'''(x)} \le rG_s''(x)$ for all $x \in [p, p + \epsilon]$, we have that
\[-rG_s''(p_2) + re^{-r(p_2 - p)} \le \frac{t}{2(p + \epsilon - p_1)} < 0,\]
so $G_s''(p_2) > e^{-r(p_2 - p)}$. This is a contradiction, since on the one hand we have $p_2 > p_1$, but on the other hand $p_1$ was defined as the supremum of points where $G_s''(x) \ge e^{-r(x - p)}$. This completes the proof.
\end{proof}

\subsection{Details omitted from Section~\ref{sec:four}}
\close*

\emph{A note on terminology:} We will sometimes speak of limits that hold ``uniformly over $p$.'' If we say that a function $g(c)$ is $o(h(c))$ \emph{uniformly} over $p$, we mean that $g$ and $h$ are implicitly functions of $p$ as well, and that $\frac{h(x)}{g(x)}$ approaches zero uniformly in $p$ (i.e.\ for all $\epsilon$ there exists $c_\epsilon$ such that for all $c \le c_\epsilon$, we have $\frac{h(x)}{g(x)} < \epsilon$ for all (relevant) values of $p$). So for instance, the $o(1)$ in the statement of Lemma~\ref{lem:close} is uniform in $p$.

\begin{proof}
Fix any $c$ and $p \in [2c^t, 1 - 2c^t]$. Let
\[n_0 := \sqrt{\frac{p(1 - p)G_s''(p)}{2c}(1 - \gamma c^{1/300})}.\]
Let $d_1$ be the expert's error after $n_0$ flips, i.e.\ $\abs{Q(n_0) - p}$. Let $d_2$ be the distance from their guess after $n_0$ flips to their guess after $\nstop$ flips. Then the expert's error after $\nstop$ flips lies between $d_1 - d_2$ and $d_1 + d_2$ by the triangle inequality. That is, we have
\[\max(0, d_1 - d_2) \le \err_c(p) \le d_1 + d_2,\]
so
\[\EE{\max(0, d_1 - d_2)^\ell \mid \Omega_N} \le \EE{(\err_c(p))^\ell \mid \Omega_N} \le \EE{(d_1 + d_2)^\ell \mid \Omega_N}\]
which means that
\[\frac{\EE{\max(0, d_1 - d_2)^\ell \mid \Omega_N}}{\EE{d_1^\ell \mid \Omega_N}} \le \frac{\EE{(\err_c(p))^\ell \mid \Omega_N}}{\EE{d_1^\ell \mid \Omega_N}} \le \frac{\EE{(d_1 + d_2)^\ell \mid \Omega_N}}{\EE{d_1^\ell \mid \Omega_N}}.\]
Below, we will prove the following facts:

\begin{claim} \label{claim:d2_o_d1}
We have
\[\lim_{c \to 0} \frac{\EE{d_2^\ell \mid \Omega_N}}{\EE{d_1^\ell \mid \Omega_N}} = 0\]
uniformly over $p \in [2c^t, 1 - 2c^t]$. That is, for all $\epsilon$ there exists $c_\epsilon$ such that for all $c < c_\epsilon$, the fraction above is less than $\epsilon$ for all $p \in [2c^t, 1 - 2c^t]$.
\end{claim}

\begin{prop} \label{prop:exp_l}
Let $X_{c,p}$ and $Y_{c,p}$ be random variables taking values in $[0, 1]$ for each real number $c > 0$ and $p \in \mathcal{P}_c$ (some arbitrary set that depends on $c$). Let $\ell > 0$. If $\lim_{c \to 0} \frac{\EE{Y_{c,p}^\ell}}{\EE{X_{c,p}^\ell}} = 0$ uniformly over $p \in \mathcal{P}_c$, then $\lim_{c \to 0} \frac{\EE{(X_{c,p} + Y_{c,p})^\ell}}{\EE{X_{c,p}^\ell}} = 1$ uniformly over $p \in \mathcal{P}_c$. Separately, if $\lim_{c \to 0} \frac{\EE{Y_{c,p}^\ell}}{\EE{(X_{c,p} + Y_{c,p})^\ell}} = 0$ uniformly over $p \in \mathcal{P}_c$, then $\lim_{c \to 0} \frac{\EE{X_{c,p}^\ell}}{\EE{(X_{c,p} + Y_{c,p})^\ell}} = 1$ uniformly over $p \in \mathcal{P}_c$.
\end{prop}

Now, by combining Claim~\ref{claim:d2_o_d1} with the first statement of Proposition~\ref{prop:exp_l} (with $X_{c,p} = d_1$ and $Y_{c,p} = d_2$), we have
\[\lim_{c \to 0} \frac{\EE{(d_1 + d_2)^\ell \mid \Omega_N}}{\EE{d_1^\ell \mid \Omega_N}} = 1.\]
By the second statement of Proposition~\ref{prop:exp_l} (with $X_{c,p} = \max(0, d_1 - d_2)$ and $Y_{c,p} = d_1 - \max(0, d_1 - d_2)$), we have
\[\lim_{c \to 0} \frac{\EE{\max(0, d_1 - d_2)^\ell \mid \Omega_N}}{\EE{d_1^\ell \mid \Omega_N}} = 1.\]
Note that the premise of the second statement of Proposition~\ref{prop:exp_l} holds for these $X_{c,p}$ and $Y_{c,p}$, because $Y_{c,p} = d_1 - \max(0, d_1 - d_2) \le d_2$ and so certainly if $\lim_{c \to 0} \frac{\EE{d_2^\ell \mid \Omega_N}}{\EE{d_1^\ell \mid \Omega_N}} = 0$ then $\lim_{c \to 0} \frac{\EE{(d_1 - \max(0, d_1 - d_2))^\ell \mid \Omega_N}}{\EE{d_1^\ell \mid \Omega_N}} = 0$.

By the squeeze theorem, it follows that
\[\lim_{c \to 0} \frac{\EE{(\err_c(p))^\ell \mid \Omega_N}}{\EE{d_1^\ell \mid \Omega_N}} = 1.\]

Note that this limit holds uniformly over $p \in [2c^t, 1 - 2c^t]$.

To complete the proof of Lemma~\ref{lem:close}, we use the following fact, which we will also prove later.

\begin{claim} \label{claim:be}
\[\EE{d_1^\ell \mid \Omega_N} = \mu_\ell \parens{\frac{p(1 - p)}{n_0}}^{\ell/2}(1 + o(1))\]
where the $o(1)$ term is a function of $c$ (but not $p$) that approaches zero as $c$ approaches zero.
\end{claim}
We have
\[n_0 = \sqrt{\frac{p(1 - p)G_s''(p)}{2c}(1 - \gamma c^{1/300})} = \sqrt{\frac{p(1 - p)G_s''(p)}{2c}}(1 + o(1)).\]
Therefore we have
\[\EE{d_1^\ell \mid \Omega_N} = \mu_\ell \parens{\frac{2c}{p(1 - p)G_s''(p)}}^{\ell/4}(1 + o(1)),\]
as desired.
\end{proof}

We now return to the proofs that we deferred.

\begin{proof}[Proof of Proposition~\ref{prop:exp_l}]
The condition that $X_{c,p}, Y_{c,p} \in [0, 1]$ is simply a convenient one to guarantee that all relevant expectations are finite. Now, for any $a \in (0, 1)$, we have
\begin{align} \label{eq:exp_l}
\EE{(X_{c,p}+Y_{c,p})^\ell} &= \int_0^\infty \pr{(X_{c,p}+Y_{c,p})^\ell > z}dz = \int_0^\infty \pr{X_{c,p}+Y_{c,p} > z^{1/\ell}}dz \nonumber\\
&\leq \int_0^\infty \parens{\pr{X_{c,p} > (1-a)z^{1/\ell}} + \pr{Y_{c,p} > az^{1/\ell}}}dz \nonumber\\
&= \int_0^\infty \parens{\pr{\frac{X_{c,p}^\ell}{(1-a)^\ell} > z} + \pr{\frac{Y_{c,p}^\ell}{a^\ell} > z}} dz = \frac{\EE{X_{c,p}^\ell}}{(1-a)^\ell} + \frac{\EE{Y_{c,p}^\ell}}{a^\ell},
\end{align}
where the inequality follows by a union bound.

We start with the first statement. Dividing Equation~\ref{eq:exp_l} by $\EE{X_{c,p}^\ell}$, we have
\[1 \le \frac{\EE{(X_{c,p} + Y_{c,p})^\ell}}{\EE{X_{c,p}^\ell}} \le \frac{1}{(1 - a)^\ell} + \frac{\EE{Y_{c,p}^\ell}}{\EE{X_{c,p}^\ell}} \cdot \frac{1}{a^\ell}.\]
The limit of $\frac{\EE{Y_{c,p}^\ell}}{\EE{X_{c,p}^\ell}}$ as $c$ approaches zero is $0$ by assumption, so for $c$ small enough we have that $\frac{\EE{Y_{c,p}^\ell}}{\EE{X_{c,p}^\ell}} \cdot \frac{1}{a^\ell} \le a$ for all $p \in \mathcal{P}_c$. In other words, for every $a$ there exists $c_a$ such that for all $c \le c_a$ we have that $1 \le \frac{\EE{(X_{c,p} + Y_{c,p})^\ell}}{\EE{X_{c,p}^\ell}} \le \frac{1}{(1 - a)^\ell} + a$. Since $\lim_{a \to 0} \frac{1}{(1 - a)^\ell} + a = 1$, we have that for all $\epsilon > 0$ there exists $c_\epsilon$ such that for all $c \le c_\epsilon$ and $p \in \mathcal{P}_c$ we have that $1 \le \frac{\EE{(X_{c,p} + Y_{c,p})^\ell}}{\EE{X_{c,p}^\ell}} \le 1 + \epsilon$. This proves the first statement.

As for the second statement, we divide Equation~\ref{eq:exp_l} by $\EE{(X_{c,p} + Y_{c,p})^\ell}$ to obtain
\[1 \le \frac{1}{(1 - a)^\ell} \cdot \frac{\EE{X_{c,p}^\ell}}{\EE{(X_{c,p} + Y_{c,p})^\ell}} + \frac{1}{a^\ell} \cdot \frac{\EE{Y_{c,p}^\ell}}{\EE{(X_{c,p} + Y_{c,p})^\ell}}\]
so
\[(1 - a)^\ell\parens{1 - \frac{1}{a^\ell} \cdot \frac{\EE{Y_{c,p}^\ell}}{\EE{(X_{c,p} + Y_{c,p})^\ell}}} \le \frac{\EE{X_{c,p}^\ell}}{\EE{(X_{c,p} + Y_{c,p})^\ell}} \le 1.\]
The limit of $\frac{\EE{Y_{c,p}^\ell}}{\EE{(X_{c,p} + Y_{c,p})^\ell}}$ as $c$ approaches zero is $0$ by assumption, so for $c$ small enough we have that $\frac{1}{a^\ell} \cdot \frac{\EE{Y_{c,p}^\ell}}{\EE{(X_{c,p} + Y_{c,p})^\ell}} \le a$ for all $p \in \mathcal{P}_c$. In other words, for every $a$ there exists $c_a$ such that for all $c \le c_a$ we have that $(1 - a)^{\ell + 1} \le \frac{\EE{X_{c,p}^\ell}}{\EE{(X_{c,p} + Y_{c,p})^\ell}} \le 1$. Since $\lim_{a \to 0} (1 - a)^{\ell + 1} = 1$, we have that for all $\epsilon > 0$ there exists $c_\epsilon$ such that for all $c \le c_\epsilon$ and $p \in \mathcal{P}_c$ we have that $1 - \epsilon \le \frac{\EE{X_{c,p}^\ell}}{\EE{(X_{c,p} + Y_{c,p})^\ell}} \le 1$. This proves the second statement.
\end{proof}

\begin{proof}[Proof of Claim~\ref{claim:be}]
We use the Berry-Esseen theorem, a result about the speed of convergence of a sum of i.i.d. random variables to a normal distribution.

\begin{theorem}[Berry-Esseen theorem]
Let $X_1, \dots, X_n$ be i.i.d. random variables with $\EE{X_1} = 0$, $\EE{X_1^2} \equiv \sigma^2 > 0$, and $\EE{\abs{X_1}^3} = \rho < \infty$. Let $Y = \frac{1}{n} \sum_i X_i$ and let $F$ be the CDF of $\frac{Y\sqrt{n}}{\sigma}$. Let $\varphi(x)$ be the standard normal distribution. Then for all $x$ we have
\[\abs{F(x) - \varphi(x)} \le \frac{C \rho}{\sigma^3 \sqrt{n}},\]
for some universal constant $C$ independent of $n$ and the distribution of the $X_i$.
\end{theorem}

Define $X_i$ to be $1 - p$ if the expert flips heads (which happens with probability $p$) and $-p$ if the expert flips tails (which happens with probability $1 - p$). Let $Y = \sum_i X_i$. Then $\sigma = \sqrt{p(1 - p)}$ and $\rho = p(1 - p)(p^2 + (1 - p)^2) \le p(1 - p)$. Plugging in these $X_i$ and $n = n_0$ into the Berry-Esseen theorem, we have
\[\abs{F(x) - \varphi(x)} \le \frac{C p(1 - p)}{(p(1 - p))^{3/2} \sqrt{n_0}} = \frac{C}{\sqrt{p(1 - p)n_0}}.\]
Now, we want to approximate $\EE{d_1^\ell} = \EE{\abs{Q(n_0) - p}^\ell}$. Note that $Q(n_0)$ is within $\frac{1}{n_0}$ of $Y + p$, the number of heads flipped divided by $n_0$. This means that $Y - \frac{1}{n_0} \le Q(n_0) - p \le Y + \frac{1}{n_0}$. For this reason, we focus on computing $\EE{\abs{Y}^\ell}$ and subsequently correct for this small difference.

Observe that
\begin{align*}
\EE{\abs{Y}^\ell} &= \int_0^\infty \pr{\abs{Y}^\ell > x} dx = \int_0^\infty \pr{\abs{Y}^\ell > u^\ell} \cdot \ell u^{\ell - 1} du\\
&= \int_0^\infty (\pr{Y > u} + \pr{Y < -u}) \cdot \ell u^{\ell - 1} du = \int_0^\infty\\
&= \int_0^\infty \parens{\pr{\frac{Y\sqrt{n_0}}{\sigma} > \frac{u\sqrt{n_0}}{\sigma}} + \pr{\frac{Y\sqrt{n_0}}{\sigma} < \frac{-u\sqrt{n_0}}{\sigma}}} \cdot \ell u^{\ell - 1} du\\
&= \int_0^\infty \parens{1 - F \parens{\frac{u \sqrt{n_0}}{\sigma}} + F \parens{\frac{-u \sqrt{n_0}}{\sigma}}} \cdot \ell u^{\ell - 1} du.
\end{align*}
Now, observe on the other hand that
\[\int_0^\infty \parens{1 - \varphi \parens{\frac{u \sqrt{n_0}}{\sigma}} + \varphi \parens{\frac{-u \sqrt{n_0}}{\sigma}}} \cdot \ell u^{\ell - 1} du = \EE{\abs{Z}^\ell},\] 
where $Z$ is a random variable drawn from a normal distribution with mean zero and variance $\frac{\sigma^2}{n_0}$. Now we ask: how different is this second quantity (the integral involving $\varphi$) from the first one (the integral involving $F$)?

The answer is: not that different. Indeed, as we derived, $F \parens{\frac{u \sqrt{n_0}}{\sigma}}$ is within $\frac{C}{\sqrt{p(1 - p)n_0}}$ of $\varphi \parens{\frac{u \sqrt{n_0}}{\sigma}}$ for all $u$. Furthermore, for any $r < \frac{1}{2}$, if $u \le -\frac{\sqrt{p(1 - p)}}{n_0^r}$ then $F \parens{\frac{u\sqrt{n_0}}{\sigma}}$ and $\varphi \parens{\frac{u\sqrt{n_0}}{\sigma}}$ are both $e^{-n_0^{\Omega(1)}}$. (For $\varphi$ this follows by concentration of normal distributions; the claim for $F$ follows from Claim~\ref{claim:omega_unlikely}, realizing the fact that there is nothing special about the $0.49$ in the exponent except that it is less than $\frac{1}{2}$.) Similarly, if $u \ge \frac{\sqrt{p(1 - p)}}{n_0^r}$ then both $F \parens{\frac{u\sqrt{n_0}}{\sigma}}$ and $\varphi \parens{\frac{u\sqrt{n_0}}{\sigma}}$ are exponentially close to $1$. Finally, for $u \le -1$ we have $F \parens{\frac{u\sqrt{n_0}}{\sigma}} = 0$ and $\varphi \parens{\frac{u\sqrt{n_0}}{\sigma}} = O(e^{-u^2 n_0})$. This means that $\EE{\abs{Y}^\ell}$ is within
\[\int_0^{\frac{\sqrt{p(1 - p)}}{n_0^{r}}} \frac{2C}{\sqrt{p(1 - p)n_0}} \cdot \ell u^{\ell - 1} du + e^{-n_0^{\Omega(1)}} = 2C(p(1 - p))^{(\ell - 1)/2} n_0^{-r\ell - 1/2} + e^{-n_0^{\Omega(1)}}\]
of $\EE{\abs{Z}^\ell}$. 
Now, since $p(1 - p) \ge c^t \ge c^{.3}$ and $n_0 \ge \frac{1}{\alpha c^{1/3}}$, we have that $p(1 - p) \ge (\alpha n_0)^{-.9}$. It is easy to check that setting any $r > \frac{1}{2} - \frac{1}{20\ell}$ shows that
\[\EE{\abs{Y}^\ell} = \EE{\abs{Z}^\ell}(1 + o(1))\]
where the $o(1)$ depends only on $c$, not on $p$. Note that $\EE{\abs{Z}^\ell} = \Theta \parens{\frac{p(1 - p)}{n_0}}^{\ell/2} = \omega \parens{\frac{1}{n_0^\ell}}$ uniformly over $p$. This means that $\frac{1}{n_0^\ell} = o \parens{\EE{\abs{Y}}^\ell}$ uniformly over $p$.

Now, recall that $Y - \frac{1}{n_0} \le Q(n_0) - p \le Y + \frac{1}{n_0}$, so
\[\max \parens{0, \abs{Y} - \frac{1}{n_0}} \le \abs{Q(n_0) - p} \le \abs{Y} + \frac{1}{n_0},\]
and thus
\[\EE{\max \parens{0, \abs{Y} - \frac{1}{n_0}}^\ell} \le \EE{\abs{Q(n_0) - p}^\ell} = \EE{d_1} \le \EE{\parens{\abs{Y} + \frac{1}{n_0}}^\ell}.\]
By the first statement of Proposition~\ref{prop:exp_l} (with $X_{c,p} = \abs{Y}$ and $Y_{c,p} = \frac{1}{n_0}$), we have that
\[\lim_{c \to 0} \frac{\EE{\parens{\abs{Y} + \frac{1}{n_0}}^\ell}}{\EE{\abs{Y}^\ell}} = 1\]
uniformly over $p$. By the second statement (with $X_{c,p} = \max \parens{0, \abs{Y} - \frac{1}{n_0}}$ and $Y_{c,p} = \abs{Y} - \max \parens{0, \abs{Y} - \frac{1}{n_0}}$), we have that
\[\lim_{c \to 0} \frac{\EE{\max \parens{0, \abs{Y} - \frac{1}{n_0}}^\ell}}{\EE{\abs{Y}^\ell}} = 1\]
uniformly over $p$. (The premise of the second statement is satisfied because $\abs{Y} - \max \parens{0, \abs{Y} - \frac{1}{n_0}} \le \frac{1}{n_0}$.) Therefore, the squeeze theorem tells us that
\[\lim_{c \to 0} \frac{\EE{d_1}}{\EE{\abs{Y}^\ell}} = 1\]
uniformly over $p$. Therefore, we have $\EE{d_1} = \EE{\abs{Z}^\ell}(1 + o(1))$ where the $o(1)$ term only depends on $c$.

Finally, note that
\[\EE{d_1} = \EE{d_1 \mid \Omega_N} \pr{\Omega_N} + \EE{d_1 \mid \overline{\Omega_N}} \pr{\overline{\Omega_N}}.\]
Since $0 \le \EE{d_1 \mid \overline{\Omega_N}} \le 1$ and $\pr{\overline{\Omega_N}} = e^{-n_0^{\Omega(1)}}$ (where the $\Omega(1)$ does not depend on $p$), we have that $\EE{d_1 \mid \Omega_N}$ is within $e^{-n_0^{\Omega(1)}}$ of $\EE{d_1}$. Applying Proposition~\ref{prop:exp_l} in the same way as earlier, we find that $\EE{d_1^\ell \mid \Omega_N} = \EE{d_1^\ell}(1 + o(1)) = \EE{\abs{Z}^\ell}(1 + o(1))$. We know that $\EE{\abs{Z}^\ell} = \mu_\ell \parens{\frac{p(1 - p)}{n_0}}^{\ell/2}$. This completes the proof.
\end{proof}


\begin{proof}[Proof of Claim~\ref{claim:d2_o_d1}]
Let $k = n_{\text{stop}} - n_0$. Define $\{Y_i\}_{i = 0}^k$ as follows: $Y_0 = 0$ and for $i > 0$, $Y_i$ is either $Y_{i - 1} + 1 - p$ (if the $n_0 + i$-th flip is heads, i.e.\ with probability $p$) or $Y_{i - 1} - p$ (if the $n_0 + i$-th flip is tails, i.e.\ with probability $1 - p$. Note that $\{Y_i\}$ is a martingale.

Now, observe that for any $0 \le i \le k$, we have
\[Q(n_0 + i) = \frac{(n_0 + 2)Q(n_0) + Y_i + pi}{n_0 + i + 2}.\]
This is because $(n_0 + 2)Q(n_0)$ is one more than the number of heads in the first $n_0$ flips and $(n_0 + i + 2)Q(n_0 + i)$ is one more than the number of heads in the first $n_0 + i$ flips. Thus,
\[\abs{Q(n_0 + i) - Q(n_0)} = \abs{\frac{Y_i + i(p - Q(n_0))}{n_0 + i + 2}} \le \frac{\max_i \abs{Y_i} + k \abs{p - Q(n_0)}}{n_0} = \frac{\max_i \abs{Y_i} + kd_1}{n_0}.\]
Therefore we have
\begin{align*}
\EE{d_2^\ell} &\le \EE{\max_{i \le k} \parens{\abs{Q(n_0 + i) - Q(n_0)}^\ell}} \le \frac{\EE{(\max_i \abs{Y_i} + kd_1)^\ell}}{n_0^\ell} = \frac{2^\ell \EE{\parens{\frac{\max_i \abs{Y_i} + kd_1}{2}}^\ell}}{n_0^\ell}\\
&\le \frac{2^\ell \EE{\frac{(\max_i \abs{Y_i})^\ell + (kd_1)^\ell}{2}}}{n_0^\ell} = \frac{2^{\ell - 1}}{n_0^\ell} \parens{\EE{\parens{\max_i \abs{Y_i}}^\ell} + k^\ell \EE{d_1^\ell}}.
\end{align*}
Here, the last inequality follows from the fact that the arithmetic mean of $\max_i \abs{Y_i}$ and $kd_1$ is less than or equal to the $\ell$-power mean (since $\ell \ge 1$).

Now, it is clear that $\frac{2^{\ell - 1}k^\ell}{n^\ell} \EE{d_1^\ell} = o(\EE{d_1^\ell})$, since $k = o(n_0)$ by Proposition~\ref{prop:n_stop_bound}. We now show that $\frac{2^{\ell - 1}}{n_0^\ell}\EE{\parens{\max_i \abs{Y_i}}^\ell} = o(\EE{d_1^\ell})$. We make use of a tool called the Burkholder-Davis-Gundy inequality.

\begin{defin}
Let $Y = \{Y_i\}_{i = 0}^k$ be a martingale. The \emph{quadratic variation} of $Y$, denoted $[Y]$, is equal to
\[[Y] = \sum_{i = 1}^k (Y_i - Y_{i - 1})^2.\]
Note that $[Y]$ is a random variable, not a number.
\end{defin}
\begin{theorem}[Burkholder-Davis-Gundy inequality]
Let $\ell \ge 1$. There is a constant $C_\ell$ such that for every martingale $Y = \{Y_i\}_{i = 0}^k$ with $Y_0 = 0$, we have
\[\EE{\parens{\max_{i = 0}^k \abs{Y_i}}^\ell} \le C_\ell \EE{[Y]^{\ell/2}}.\]
\end{theorem}

We wish to bound $\EE{\parens{\max_{i = 0}^k \abs{Y_i}}^\ell}$ above. To do so, we bound $\EE{[Y]^{\ell/2}}$ above. Observe that $[Y]$ is a sum of $k$ independent random variables that are each either $p^2$ (with probability $1 - p$) or $(1 - p)^2$ (with probability $p$). Thus, $\EE{[Y]} = kp(1 - p) \equiv \mu$. Observe that
\begin{align*}
\EE{[Y]^{\ell/2}} &= \int_0^\infty \pr{[Y]^{\ell/2} \ge x} dx = \int_0^\infty \pr{[Y] \ge x^{2/\ell}} dx \le (3\mu)^{\ell/2} + \int_{(3\mu)^{\ell/2}}^\infty \pr{[Y] \ge x^{2/\ell}} dx\\
&\le (3\mu)^{\ell/2} + \int_{(3\mu)^{\ell/2}}^\infty e^{\frac{\mu - x^{2/\ell}}{2}} dx.
\end{align*}
The last line comes from a Chernoff bound. In particular, we have that $\pr{[Y] \ge \mu(1 + \delta)} \le e^{-\delta^2 \mu/(2 + \delta)} \le e^{-\delta \mu/2}$ for $\delta \ge 2$. Setting $\delta = \frac{x^{2/\ell}}{\mu} - 1$ gives us the expression above. Now, we can bound the integral as follows:
\[\int_{(3\mu)^{\ell/2}}^\infty e^{\frac{\mu - x^{2/\ell}}{2}} dx \le ((5\mu)^{\ell/2} - (3\mu)^{\ell/2}) e^{-\mu} + ((7\mu)^{\ell/2} - (5\mu)^{\ell/2}) e^{-2\mu} + \dots \equiv B(\mu).\]
Note that $B(\mu)$ is continuous, converges on $[0, \infty)$, and approaches zero as $\mu \to \infty$. It follows that $B(\mu)$ is bounded on $[0, \infty)$; in other words, our integral is $O(1)$ (i.e.\ possibly depends on $\ell$ but is at most a constant for fixed $\ell$). Therefore, we have
\[\EE{\parens{\max_i \abs{Y_i}}^\ell} \le \EE{[Y]^{\ell/2}} \le 3^{\ell/2}(kp(1 - p))^{\ell/2} + O(1).\]
Therefore we have
\[\frac{2^{\ell - 1}}{n_0^\ell}\EE{\parens{\max_i \abs{Y_i}}^\ell} = O \parens{\parens{\frac{p(1 - p)}{n_0}}^{\ell/2} \parens{\frac{k}{n_0}}^{\ell/2}} = O \parens{\parens{\frac{k}{n_0}}^{\ell/2} \EE{d_1^\ell}} = o(\EE{d_1^\ell}).\]
Therefore, we have that $\EE{d_2^\ell} = o(\EE{d_1^\ell})$. By the same reasoning as in the proof of Claim~\ref{claim:be} below, it follows that $\EE{d_2 \mid \Omega_N} = o(\EE{d_1^\ell})$. We previously showed that $\EE{d_1^\ell} = \Theta(\EE{d_1^\ell \mid \Omega_N})$. This completes the proof.
\end{proof}

\subsection{Details omitted from Section~\ref{sec:five} and proof of Theorem~\ref{thm:global}}
In this section, we complete the proof of Theorem~\ref{thm:global}. We begin by proving the analog of Theorem~\ref{thm:global} for a locally adaptive expert.

\begin{theorem} \label{thm:local}
If $s$ is a respectful, normalizable, continuously differentiable proper scoring rule, and $\err_c(p)$ is the expected error of a locally adaptive expert scored by $s$ when the coin has bias $p$ and the cost of a flip is $c$, then
\[\lim_{c \to 0} c^{-\ell/4} \EE[p \leftarrow U_{[0, 1]}]{\err_c(p)^\ell} = \mu_\ell \int_0^1 \parens{\frac{2x(1 - x)}{G_s''(x)}}^{\ell/4} dx.\]
\end{theorem}

\begin{proof}
Let $N = \frac{1}{\alpha c^{1/3}}$, i.e.\ a large enough function of $c$ that it is guaranteed that the expert flips the coin at least $N$ times. We have
\[c^{-\ell/4} \EE[p \leftarrow U_{[0, 1]}]{\err_c(p)^\ell} = c^{-\ell/4}\parens{\EE[p \leftarrow U_{[0, 1]}]{\err_c(p)^\ell \mid \Omega_N} \pr{\Omega_N} + \EE[p \leftarrow U_{[0, 1]}]{\err_c(p)^\ell \mid \overline{\Omega_N}} \pr{\overline{\Omega_N}}}.\]
We wish to compute the limit of this quantity as $c$ approaches zero. Note that
\[\lim_{c \to 0} c^{-\ell/4} \EE[p \leftarrow U_{[0, 1]}]{\err_c(p)^\ell \mid \overline{\Omega_N}} \pr{\overline{\Omega_N}} = 0.\]
This is because $\err_c(p)$ is bounded between $0$ and $1$ and $\pr{\overline{\Omega_N}} = O(e^{-N^{.01}}) = O(e^{-\Omega(c^{-1/300})})$, which goes to zero faster than $c^{-\ell/4}$ goes to infinity. Therefore we have
\[\lim_{c \to 0} c^{-\ell/4} \EE[p \leftarrow U_{[0, 1]}]{\err_c(p)^\ell} = \lim_{c \to 0} c^{-\ell/4} \EE[p \leftarrow U_{[0, 1]}]{\err_c(p)^\ell \mid \Omega_N}.\]
(We may ignore the $\pr{\Omega_N}$ term above because it approaches $1$ in the limit.) We may write this quantity as
\[\lim_{c \to 0} c^{-\ell/4} \parens{(1 - 4c^t) \EE[p \leftarrow U_{[2c^t, 1 - 2c^t]}]{\err_c(p)^\ell \mid \Omega_N} + 4c^t \EE[p \leftarrow U_{[0, 2c^t] \cup [1 - 2c^t, 1]}]{\err_c(p)^\ell \mid \Omega_N}}.\]
Let us focus on the second summand. Let $p \in [0, 2c^t] \cup [1 - 2c^t, 1]$. We assume $p \in [0, 2c^t]$; the other case is analogous.

We consider two sub-cases: $p \in [0, \alpha c^{1/3}]$ and $p \in [\alpha c^{1/3}, 2c^t]$. First suppose that $p \in [c^{1/3}, 2c^t]$. Note that since $\Omega_N$ holds, we have for all $n \ge N$ that
\[\abs{Q(n) - p} \le \frac{\sqrt{p(1 - p)}}{n^{.49}} \le \frac{\sqrt{2c^t}}{N^{.49}} = \sqrt{2}\alpha^{.49} c^{t/2 + .49/3}.\]
This in particular is true of $n = \nstop$, so
\[\err_c(p)^\ell \le (\sqrt{2}\alpha^{.49})^\ell c^{\ell(t/2 + .49/3)} = o(c^{\ell/4})\]
since $t > \frac{1}{4}$ and so $\frac{t}{2} + \frac{.49}{3} > \frac{1}{4}$.

Now suppose that $p \in [0, \alpha c^{1/3}]$. Recall the notation $Q_p(n)$ from the discussion preceding the definition of $\Omega_N$. For any $n \ge N$, we have
\begin{align*}
\abs{Q(n) - p} &\le \abs{Q(n) - Q_{\alpha c^{1/3}}(n)} + \abs{Q_{\alpha c^{1/3}}(n) - \alpha c^{1/3}} + \abs{\alpha c^{1/3} - p}\\
&\le Q_{\alpha c^{1/3}}(n) + \abs{Q_{\alpha c^{1/3}}(n) - \alpha c^{1/3}} + \alpha c^{1/3} \le 2\alpha c^{1/3} + 2\abs{Q_{\alpha c^{1/3}}(n) - \alpha c^{1/3}}\\
&\le 2\alpha c^{1/3} + \frac{2\sqrt{\alpha c^{1/3}}}{n^{.49}} \le 2\alpha c^{1/3} + 2\sqrt{\alpha} c^{1/6 + .49/3} = o(c^{1/4})
\end{align*}
so $\err_c(p)^\ell = o(c^{\ell/4})$.

This means that
\[\lim_{c \to 0} c^{-\ell/4} \cdot 4c^t \EE[p \leftarrow U_{[0, 2c^t] \cup [1 - 2c^t, 1]}]{\err_c(p)^\ell \mid \Omega_N} = 0\]
so we can ignore this summand. Therefore, we have
\[\lim_{c \to 0} c^{-\ell/4} \EE[p \leftarrow U_{[0, 1]}]{\err_c(p)^\ell} = \lim_{c \to 0} c^{-\ell/4} (1 - 4c^t) \EE[p \leftarrow U_{[2c^t, 1 - 2c^t]}]{\err_c(p)^\ell \mid \Omega_N}.\]
From Lemma~\ref{lem:close}, we have that
\begin{align*}
(1 - o(1)) \int_{2c^t}^{1 - 2c^t} \mu_\ell \parens{\frac{2x(1 - x)}{G_s''(x)}}^{\ell/4} dx &\le c^{-\ell/4} (1 - 4c^t) \EE[p \leftarrow U_{[2c^t, 1 - 2c^t]}]{\err_c(p)^\ell \mid \Omega_N}\\
&\le (1 + o(1)) \int_{2c^t}^{1 - 2c^t} \mu_\ell \parens{\frac{2x(1 - x)}{G_s''(x)}}^{\ell/4} dx.
\end{align*}
By the squeeze theorem, we conclude that.
\[\lim_{c \to 0} c^{-\ell/4} \EE[p \leftarrow U_{[0, 1]}]{\err_c(p)^\ell} = \lim_{c \to 0} \int_{2c^t}^{1 - 2c^t} \mu_\ell \parens{\frac{2x(1 - x)}{G_s''(x)}}^{\ell/4} dx = \mu_\ell \int_0^1 \parens{\frac{2x(1 - x)}{G_s''(x)}}^{\ell/4} dx.\]
\end{proof}

\globalhelper*

\begin{proof}
Our approach will be to compare the behavior of a locally adaptive expert to that of a globally adaptive one. We will assume that the experts observe the same stream of coin flips (each heads with probability $p$ unknown to the experts) but that they may decide to stop at different times. As before, we will let $Q(n) = \frac{h + 1}{n + 2}$ where $h$ is the number of the first $n$ flips to have come up heads; since the experts see the same coin flips, we do not need to distinguish between $Q(n)$ for the locally adaptive expert and for the globally adaptive expert. We will let $n_l$ and $n_g$ be the number of times the locally and globally adaptive experts flip the coin, respectively (so $n_g \ge n_l$). (We used the notation $\nstop$ in place of $n_l$ in Proposition~\ref{prop:n_stop_bound}.) Let $t$ be as in the definition of respectful scoring rules, and in particular we will assume that $t < 0.3$ as before (for any $t$ that witnesses that a scoring rule is respectful, any smaller $t > \frac{1}{4}$ also works).

Suppose the globally adaptive expert flips the coin $N := (1 + 6\gamma c^{1/300}) n_l$ times. We show that they do not flip the coin another time.

By definition of $\Omega_{n_l}$, we have that
\[p - \frac{\sqrt{p(1 - p)}}{n_l^{.49}} \le Q(n_l) \le p + \frac{\sqrt{p(1 - p)}}{n_l^{.49}}.\]
It is easy to check that because $n_l = \Omega(c^{-1/3})$ (by Claim~\ref{claim:one_third_bound}) and $4c^t \le Q(n_l) \le 1 - 4c^t$ (so $Q(n_l), 1 - Q(n_l) = \Omega(n_l^{-.9})$), the above relationship between $Q(n_l)$ and $p$ implies that $2c^t \le p \le 1 - 2c^t$. This allows us to use some results from our analysis of locally adaptive experts. In particular, by Equation~\ref{eq:delta_bound} in the proof of Proposition~\ref{prop:n_stop_bound}, we have that
\[\frac{p(1 - p)}{2n_l^2}G_s''(p)(1 - \gamma c^{1/300}) \le \Delta_{n_l + 1} \le \frac{p(1 - p)}{2n_l^2}G_s''(p)(1 + \gamma c^{1/300}).\]
Conditional on $\Omega_N$ (and by definition $\Omega_{n_l}$ implies $\Omega_N$), we also have
\[\frac{p(1 - p)}{2N^2}G_s''(p)(1 - \gamma c^{1/300}) \le \Delta_{N + 1} \le \frac{p(1 - p)}{2N^2}G_s''(p)(1 + \gamma c^{1/300}).\]
In particular this means that
\[\Delta_{N + 1} \le \frac{n_l^2}{N^2} \cdot \frac{1 + \gamma c^{1/300}}{1 - \gamma c^{1/300}} \Delta_{n_l + 1} \le \parens{\frac{n_l}{N}}^2(1 + 3\gamma c^{1/300}) \Delta_{n_l + 1}\]
for $c$ small enough. This means that if $\parens{\frac{n_l}{N}}^2 \le \frac{1}{1 + 6\gamma c^{1/300}}$ then $\Delta_{N + 1} \le (1 - 2\gamma c^{1/300}) \Delta_{n_l + 1} < c(1 - 2\gamma c^{1/300})$. Furthermore, for any $n \ge N$ we will have $\Delta_{n + 1} < c(1 - 2\gamma c^{1/300})$.

However, this does not mean that the globally adaptive expert won't flip the coin for the $N + 1$-th time, because they don't know that $\Omega_{n_l}$ is true. From the expert's perspective, \emph{if} they knew $\Omega_{n_l}$ (or even $\Omega_N$) to be true, they would stop flipping the coin; but perhaps they should keep flipping the coin because of the outside chance that $\Omega_N$ is false.

This turns out not to be the case, because the probability that $\Omega_N$ is false is so small. In particular, from the expert's perspective, if $\Omega_N$ is false, they cannot achieve a score that is better than the expectation of $G_s(p)$ conditional on the coins they've flipped and on $\Omega_N$ being false. We show that if the scoring rule $s$ is normalizable (i.e.\ $\int_0^1 G_s(x) dx$ is finite), then this quantity isn't too large. In particular, we show the following:

\begin{claim}
Let $H_N$ be the random variable corresponding to the number of heads flipped in the first $N$ flips. Then for any $3c^tN \le h \le (1 - 3c^t)N$, we have
\[\EE{G_s(p) \mid \overline{\Omega_N}, H_N = h} \le \frac{2}{c^t} \int_0^1 G_s(x) dx.\]
\end{claim}

\begin{proof}
We have
\[\EE{G_s(p) \mid \overline{\Omega_N}, H_N = h} \le \EE{G_s(p) \mid \overline{\Omega_N}, H_N = h, p < \frac{1}{2}} + \EE{G_s(p) \mid \overline{\Omega_N}, H_N = h, p > \frac{1}{2}}.\]
Let us consider the expectation conditioned on $p < \frac{1}{2}$. Consider the distribution $D$ of $p$ conditioned on $\overline{\Omega_N}$, $H_N = h$, and $p < \frac{1}{2}$. Consider also the uniform distribution $D'$ on $[0, c^t]$.

We claim that $D$ stochastically dominates $D'$, i.e.\ $\pr[x \leftarrow D]{x \le y} \le \pr[x \leftarrow D']{x \le y}$ for all $y$. To see this, observe that the PDF of $D$ is an increasing function on $[0, c^t]$. This is because $D$ on $[0, c^t]$ is a constant multiple of the distribution $D''$ of $p$ conditioned on $\overline{\Omega_N}$, $H_N = h$, and $p \le c^t$; but in this case the condition $\overline{\Omega_N}$ is redundant because if $p \le c^t$ then $\overline{\Omega_N}$ holds. So $D''$ is the distribution of $p$ conditioned on $H_N = h$ and $p \le c^t$. Clearly the PDF of $D''$ increases on $[0, c^t]$ (because the expert starts with uniform priors and updates more strongly on against values of $p$ farther from $\frac{h + 1}{N + 2}$, which is greater than $2.9c^t$ for $c$ small enough).

Now, the expectation of $G_s(p)$ if $p$ were drawn from $D'$ instead of $D$ is equal to $\frac{1}{c^t} \int_0^{c^t} G_s(x) dx \le \frac{1}{c^t} \int_0^1 G_s(x) dx$. On the other hand, the actual expectation of $G_s(p)$ (i.e.\ with $p$ drawn from $D$) is necessarily smaller. This is because $G_s$ is convex and symmetric about $\frac{1}{2}$, meaning that $G_s$ is decreasing on $(0, \frac{1}{2})$. Since $D$ stochastically dominates $D'$, we conclude that
\[\EE{G_s(p) \mid \overline{\Omega_N}, H_N = h, p < \frac{1}{2}} \le \frac{1}{c^t} \int_0^1 G_s(x) dx.\]
The same inequality holds conditional instead on $p > \frac{1}{2}$, which concludes the proof.
\end{proof}

From the expert's perspective, this means that if they flip the coin for the $N + 1$-th time, then:
\begin{itemize}
\item In the case that $\Omega_N$ is true, the best case is that they never flip the coin again, in which case they will pay a total cost of $c$ and get expected score at most $c(1 - 2\gamma c^{1/300})$.
\item In the case that $\Omega_N$ is false, the best case is that they get a score of $\frac{2}{c^t} \int_0^1 G_s(x) dx$.
\end{itemize}
In other words, the expert's expected score if they flip the coin for the $N + 1$-th time and pursue the optimal strategy from there is at most
\[\frac{2}{c^t} \int_0^1 G_s(x) dx \cdot \pr{\overline{\Omega_N}} - 2\gamma c^{1/300} \pr{\Omega_N} \le O \parens{\frac{e^{-N^{.01}}}{c^t}} - \gamma c^{1/300} = O \parens{\frac{e^{-\Omega(c^{1/300})}}{c^t}} - \gamma c^{1/300}.\]
The first step is nontrivial: it uses the fact that the probability that the expert assigns to $\overline{\Omega_N}$ after the first $N$ flips is $O(e^{-N^{.01}})$. This doesn't immediately follow from Claim~\ref{claim:omega_unlikely} because the claim only states that the prior probability of $\overline{\Omega_N}$, i.e.\ before any flips, is $O(e^{-N^{.01}})$. To see that the posterior probability (after the first $N$ flips) is also of this order, we first observe that the posterior probability cannot depend on the order of the flip outcomes; this is apparent from the definition of $\Omega_N$. However, perhaps the \emph{number} of heads, i.e.\ the value of $H_N$, affects the posterior probability of $\overline{\Omega_N}$. This may be so, but it cannot increase the probability by more than a factor of $N + 1$. That is because the prior for $H_N$ is uniform over $\{0, \dots, N\}$ (Fact~\ref{fact:beta}).

Now, the quantity on the right is negative for $c$ small enough, so the expert will not flip the $N + 1$-th coin. This proves the claim.
\end{proof}

The following corollary is essentially identical to Proposition~\ref{prop:n_stop_bound} but for globally adaptive experts.
\begin{corollary} \label{cor:global_close}
Assume that $\Omega_N$ holds for some $N$. For sufficiently small $c$, for all $p \in [8c^t, 1 - 8c^t]$, we have
\[\sqrt{\frac{p(1 - p)G_s''(p)}{2c}(1 - 14\gamma c^{1/300})} \le n_g \le \sqrt{\frac{p(1 - p)G_s''(p)}{2c}(1 + 14\gamma c^{1/300})}.\]
\end{corollary}

\begin{proof}
Because $\Omega_N$ holds for some $N$, for sufficiently small $c$ the fact that $p \in [8c^t, 1 - 8c^t]$ implies that $Q(n_l) \in [4c^t, 1 - 4c^t]$. This means that we may apply Lemma~\ref{lem:global_helper} to say that $n_g \le N$. Consequently we have that
\begin{align*}
&\sqrt{\frac{p(1 - p)G_s''(p)}{2c}(1 - \gamma c^{1/300})} \le n_l \le n_g \le N = (1 + 6\gamma c^{1/300}) n_l\\
&\le (1 + 6\gamma c^{1/300}) \sqrt{\frac{p(1 - p)G_s''(p)}{2c}(1 + \gamma c^{1/300})} \le \sqrt{\frac{p(1 - p)G_s''(p)}{2c}(1 + 14\gamma c^{1/300})}
\end{align*}
for $c$ small enough. Therefore we have
\[\sqrt{\frac{p(1 - p)G_s''(p)}{2c}(1 - 14\gamma c^{1/300})} \le n_g \le \sqrt{\frac{p(1 - p)G_s''(p)}{2c}(1 + 14\gamma c^{1/300})}.\]
\end{proof}

Theorem~\ref{thm:global} follows as a simple corollary.

\globalthm*

\begin{proof}
The lemma analogous to Lemma~\ref{lem:close} but for expected globally adaptive error, and for $p \in [8c^t, 1 - 8c^t]$, follows immediately from Corollary~\ref{cor:global_close}. This is because the proof of Lemma~\ref{lem:close} makes no assumptions about the specific value of $\gamma$ (other than that it is positive), which means that the proof goes through just as well for $14\gamma$ in place of $\gamma$. Theorem~\ref{thm:global} follows from this fact exactly in the same way that Theorem~\ref{thm:local} followed from Lemma~\ref{lem:close}.
\end{proof}

\section{Details omitted from Section~\ref{sec:optimal}}\label{app:optimal}
\besth*

\begin{proof}
Let $\lambda_\ell = \frac{\ell}{4 \kappa_\ell^{\ell/4 + 1}}$. Consider the functional
\begin{align*}
\chi(h) &:= \int_{\half}^1 \parens{\frac{x(1 - x)^2}{h(x)}}^{\ell/4} dx + \lambda_\ell \parens{\int_\half^1 (1 - x)h(x) dx - 1}\\
&= \int_\half^1 \parens{\parens{\frac{x(1 - x)^2}{h(x)}}^{\ell/4} + \lambda_\ell (1 - x)h(x)}dx - \lambda_\ell.
\end{align*}
It suffices to show that among all $h$ satisfying $\int_\half^1 (1 - x)h(x) dx = 1$, $\tilde{h}$ minimizes $\chi(h)$. This is because among such $h$, the second summand in the definition of $\chi$ is always zero. In fact, we prove something stronger: $\tilde{h}$ minimizes $\chi$, among all functions from $[\half, 1)$ to $\RR$. To show this, it suffices to show that for every $x \in [\half, 1)$, the value $y$ that minimizes
\[\parens{\frac{x(1 - x)^2}{y}}^{\ell/4} + \lambda_\ell(1 - x)y\]
is $y = \tilde{h}(x)$. The derivative with respect to $y$ of this expression is
\[\frac{-\ell}{4} (x(1 - x)^2)^{\ell/4} y^{-(\ell/4 + 1)} + \lambda_\ell(1 - x),\]
which is an increasing function of $y$ (since $y^{-(\ell/4 + 1)}$ is a decreasing function of $y$ and $\frac{-\ell}{4} (x(1 - x)^2)^{\ell/4}$ is negative). It is equal to $0$ precisely when
\[y = \parens{\frac{\ell}{4\lambda_\ell}}^{4/(\ell+4)} (x^\ell (1 - x)^{2\ell - 4})^{1/(\ell + 4)} = \kappa_\ell (x^\ell (1 - x)^{2\ell - 4})^{1/(\ell + 4)} = \tilde{h}(x).\]
It remains only to note that $\int_\half^1 (1 - x)\tilde{h}(x) dx = 1$, and this follows immediately from the definition of $\tilde{h}$ and $\kappa_\ell$.
\end{proof}

\hunique*

\begin{proof}
Suppose for contradiction that there is another continuous function $\hat{h}$ satisfying the above properties that achieves the minimum. Then $\chi(\hat{h}) = \chi(\tilde{h})$, with $\chi$ as in the proof of Lemma~\ref{lem:besth}, since $\tilde{h}$ minimizes $\chi$ and the second summand in the definition of $\chi$ is zero for both $\hat{h}$ and $\tilde{h}$. In particular, we have that $\chi(\hat{h}) - \chi(\tilde{h}) = 0$, i.e.
\[\int_\half^1 \parens{\parens{\parens{\frac{x(1 - x)^2}{\hat{h}(x)}}^{\ell/4} + \lambda_\ell (1 - x)\hat{h}(x)} - \parens{\parens{\frac{x(1 - x)^2}{\tilde{h}(x)}}^{\ell/4} + \lambda_\ell (1 - x)\tilde{h}(x)}} dx = 0.\]
Let $\Delta(x)$ be the integrand. Note that $\Delta$ is always nonnegative, and is zero precisely for those values of $x$ where $\hat{h}(x) = \tilde{h}(x)$ (since, as we showed earlier, $y = \tilde{h}(x)$ is the unique value minimizing $\parens{\frac{x(1 - x)^2}{y}}^{\ell/4} + \lambda_\ell (1 - x)y$). Since $\hat{h} \neq \tilde{h}$, $\Delta$ is positive at some $x_0$; say $\Delta(x_0) = y_0$. Also, note that $\Delta$ is continuous because $\hat{h}$ and $\tilde{h}$ are continuous. This means that for some $\delta > 0$, $\abs{\Delta(x) - y_0} < \frac{y_0}{2}$ for all $x$ such that $x_0 \le x \le x_0 + \delta$. But this means that the integral of $\Delta$ on $[x_0, x_0 + \delta]$ is at least $\frac{\delta \cdot y_0}{2} > 0$, so $\int_\half^1 \Delta(x) > 0$, a contradiction. Therefore, $\tilde{h}$ is indeed the unique continuous function satisfying the stated constraints.
\end{proof}

\optimal*

\begin{proof}
We have reasoned that $s_{\ell,\opt}$ is the antiderivative of $\tilde{h}$ on $[\half, 1)$, which gives us $s_{\ell,\opt}$ for $x \ge \half$. For $x < \half$, we have
\[s_{\ell,\opt}'(x) = \frac{1 - x}{x} s_{\ell,\opt}'(1 - x) = \kappa_\ell \frac{1 - x}{x} ((1 - x)^\ell x^{2\ell - 4})^{1/(\ell + 4)} = \kappa_\ell (x^{\ell - 8} (1 - x)^{2\ell + 4})^{1/(\ell + 4)},\]
which extends to the stated function $s_{\ell,\opt}$ by the fundamental theorem of calculus. The constant we want to add (upon taking the antiderivative) is zero so that $s_{\ell,\opt} \parens{\half} = 0$. We need to check that $s_{\ell,\opt}$ is continuously differentiable at $\half$, which means checking that $\tilde{h}$ is continuous at $\half$ when extended to $(0, 1)$. This is indeed the case because
\[\lim_{x \to \half^-} s_{\ell, \opt}'(x) = \lim_{x \to \half^-} \frac{1 - x}{x} s_{\ell, \opt}'(1 - x) = \lim_{x \to \half^-} s_{\ell, \opt}'(1 - x) = \lim_{x \to \half^+} s_{\ell, \opt}'(x).\]
Finally, $s_{\ell,\opt}$ is the unique continuous normalized minimizer because its derivative is unique, by Corollary~\ref{cor:hunique}. Note that $s_{\ell,\opt}$ is in fact strictly proper since $\tilde{h}(x)$ is positive on $[\half, 1)$.
\end{proof}

\respect*

\begin{proof}[{Proof of Remark~\ref{rem:respect} for $1 \le \ell \le 8$}]
This proof handles the case of $\ell \in [1,8]$. Let us write $G_{\ell, \opt}$ to mean $G_{s_{\ell, \opt}}$. Then
\[G''_{\ell, \opt}(x) = \frac{s'_{\ell, \opt}(x)}{1 - x} = \begin{cases}\kappa_\ell (x^{\ell - 8}(1 - x)^\ell)^{1/(\ell + 4)} & x \le \half \\
\kappa_\ell (x^\ell(1 - x)^{\ell - 8})^{1/(\ell + 4)} & x \ge \half.\end{cases}\]
First note that $G_{\ell, \opt}$ is strongly convex. Since $G''_{\ell, \opt}$ is symmetric, it suffices to show this on $(0, \half]$. We have $1 - x \ge \half$ on this interval, and $x^{(\ell - 8)/(\ell + 4)}$ is bounded away from zero when $x \le 8$. Next, the fact that $G'''_{\ell, \opt}$ is Riemann integral on any closed sub-interval of $(0, 1)$ is evident. Finally, there are constants $k \neq 0$ and $r$ such that $\lim_{x \to 0} x^r G'''_{\ell, \opt}(x) = k$: in particular, $r = \frac{12}{\ell + 4}$ and $k = \frac{(\ell - 8)\kappa_\ell}{\ell + 4}$.

(Note that for $\ell > 8$, $s_{\ell, \opt}$ is not respectful, since $\lim_{x \to 0} G''_{\ell, \opt}(x) = 0$.)
\end{proof}

\section{Details omitted from Section~\ref{sec:weierstrass}}\label{app:weierstrass}
\analytic*

\begin{proof}
Suppose that $s$ is a proper scoring rule. Then $s$ is nonconstant, $s'(x) \ge 0$ everywhere by Lemma~\ref{lem:weaklyproper}, and by the same lemma we have that
\[xs'(x) = (1 - x)s'(1 - x).\]
Taking successive derivatives of both sides, we have
\begin{align*}
s'(x) + xs''(x) &= -s'(1 - x) - (1 - x)s''(1 - x)\\
2s''(x) + xs'''(x) &= 2s''(1 - x) + (1 - x)s'''(1 - x)\\
3s'''(x) + xs^{(4)}(x) &= -3s'''(1 - x) - (1 - x)s^{(4)}(1 - x)
\end{align*}
and so on. Plugging in $x = \half$, we have
\begin{align*}
s' \parens{1/2} + 1/2 s'' \parens{1/2} &= -s' \parens{1/2} - 1/2 s'' \parens{1/2}\\
2s'' \parens{1/2} + 1/2 s''' \parens{1/2} &= 2s'' \parens{1/2} + 1/2 s''' \parens{1/2}\\
3s''' \parens{1/2} + 1/2 s^{(4)} \parens{1/2} &= -3s''' \parens{1/2} - 1/2 s^{(4)} \parens{1/2}
\end{align*}
and so on. These equations alternate between giving us tautologies and simplifying to the following identities:
\[s'' \parens{1/2} = -2s' \parens{1/2}; \quad s^{(4)} \parens{1/2} = -6s''' \parens{1/2}; \quad s^{(6)} \parens{1/2} = -10f^{(5)} \parens{1/2};\]
and so on, the general form of the identities being that for $k$ odd, we have
\[s^{(k + 1)}\parens{1/2} = -2k s^{(k)}\parens{1/2}.\]
Since $s$ is analytic, we have
\begin{align*}
s(x) &= \sum_{n = 0}^\infty \frac{1}{n!} s^{(n)} \parens{\half} \parens{x - \half}^n\\
&= s \parens{\half} + \sum_{k > 0 \text{ odd}} \frac{1}{(k + 1)!} \parens{(k + 1) s^{(k)}\parens{\half} \parens{x - \half}^k + s^{(k + 1)} \parens{\half} \parens{x - \half}^{k + 1}}.
\end{align*}
Letting $c_k = \frac{1}{(k + 1)!} s^{(k)} \parens{\half}$ for $k = 0, 1, 3, 5, \dots$, we have
\begin{align*}
s(x) &= c_0 + \sum_{k > 0 \text{ odd}} c_k \parens{(k + 1)\parens{x - \half}^k - 2k \parens{x - \half}^{k + 1}}\\
&= c_0 + \sum_{k > 0 \text{ odd}} c_k(2k + 1 - 2kx)\parens{x - \half}^k.
\end{align*}
This proves the forward direction. Conversely, we claim that if $s$ is nonconstant, $s'(x) \ge 0$ everywhere, and $s$ can be written in the stated form for some $c_0, c_1, c_3, \dots$, then $s$ is a proper scoring rule. We only have to verify that $xs'(x) = (1 - x)s'(1 - x)$ everywhere (by Lemma~\ref{lem:weaklyproper} and Lemma~\ref{lem:constant} (stated and proven below)). Taking the derivative of $(1 - x)s'(1 - x)$ term by term, we have
\begin{align*}
(1 - x)s'(1 - x) &= (x - 1)(s(1 - x))' = (x - 1) \frac{d}{dx} \parens{c_0 + \sum_{k > 0 \text{ odd}} c_k(2kx + 1)\parens{\half - x}^k}\\
&= (x - 1) \sum_{k > 0 \text{ odd}} c_k \parens{2k \parens{\half - x}^k - k(2kx  + 1) \parens{\half - x}^{k - 1}}\\
&= \sum_{k > 0 \text{ odd}} (x - 1)kc_k \parens{\half - x}^{k - 1}(1 - 2x - 2kx - 1)\\
&= \sum_{k > 0 \text{ odd}} 2k(k + 1)c_kx(1 - x)\parens{\half - x}^{k - 1}.
\end{align*}

Similarly, we have
\begin{align*}
xs'(x) &= x \frac{d}{dx} \parens{c_0 + \sum_{k > 0 \text{ odd}} c_k(2k + 1 - 2kx)\parens{x - \half}^k}\\
&= x \sum_{k > 0 \text{ odd}} c_k \parens{-2k \parens{x - \half}^k + k(2k + 1 - 2kx) \parens{x - \half}^{k - 1}}\\
&= \sum_{k > 0 \text{ odd}} xkc_k \parens{x - \half}^{k - 1}(1 - 2x + 2k + 1 - 2kx)\\
&= \sum_{k > 0 \text{ odd}} 2k(k + 1)c_kx(1 - x)\parens{x - \half}^{k - 1} = (1 - x)s'(1 - x),
\end{align*}
as desired.
\end{proof}

\begin{lemma} \label{lem:constant}
The only infinitely differentiable scoring rules that are weakly proper but not proper are constant functions.
\end{lemma}

\begin{proof}
Let $s$ be an infinitely differentiable scoring rule that is weakly proper but not proper. Recall that in the proof of Lemma~\ref{lem:weaklyproper}, we showed that for all $p$, the function $r_p(x)$ weakly increases on $(0, p]$ and weakly decreases on $[p, 1)$. Since $s$ is not proper, there is some $p$ such that $r_p(x)$ does not strictly increase on $(0, p]$ or does not strictly decrease on $[p, 1)$. But this means that $r_p(x)$ is constant on some open interval, which means that $r_p(x)$ is constant (because $s$ is infinitely differentiable, which means that $r_p(x)$ is also infinitely differentiable). Thus, for some $c \in \RR$ we have that $ps(x) + (1 - p)s(1 - x) = c$ for all $x$. Taking the derivative, we have that $ps'(x) = (1 - p)s'(1 - x)$ for all $x$. But we also have that $xs'(x) = (1 - x)s'(1 - x)$ for all $x$. The only way for both of these equations to hold is for $s'(x)$ to be uniformly zero, so $s$ is indeed constant.
\end{proof}

\taylorphil*

\begin{proof}
By Theorem~\ref{thm:analytic}, $s$ is a proper scoring rule if and only if $s$ is nonconstant, $s'(x) \ge 0$ everywhere, and
\[s(x) = c_0 + \sum_{k > 0 \text{\emph{ odd}}} c_k(2k + 1 - 2kx)\parens{x - \half}^k\]
for some $c_0, c_1, c_3, c_5, \dots \in \RR$. Equivalently,
\begin{align*}
s'(x) &= \sum_{k > 0 \text{\emph{ odd}}} kc_k(2k + 1 - 2kx)\parens{x - \half}^{k - 1} - 2kc_k\parens{x - \half}^k\\
&= \sum_{k > 0 \text{\emph{ odd}}} kc_k\parens{2k + 1 - 2kx - 2 \parens{x - \half}}\parens{x - \half}^{k - 1}\\
&= \sum_{k > 0 \text{\emph{ odd}}} 2kc_k(k + 1)(1 - x)\parens{x - \half}^{k - 1}\\
&= \sum_{k \ge 0 \text{\emph{ even}}} 2(k + 1)c_{k + 1}(k + 2)(1 - x)\parens{x - \half}^k = (1 - x) \sum_{k \ge 0 \text{\emph{ even}}} d_k \parens{x - \half}^k,
\end{align*}
where $d_k = 2(k + 1)(k + 2)c_{k + 1}$. Noting that $s$ is constant if and only if $G_s''$ is uniformly zero, and that $s'(x) \ge 0$ if and only if $G_s''(x) \ge 0$, this completes the proof.
\end{proof}

In proving Theorem~\ref{thm:approx}, we will make substantial use of $G_s''$, the second derivative of the expected score function of $s$. For convenience, we will write $\varphi$ instead of $G_s''$.

Note that if $s$ is a continuously differentiable (but not necessarily infinitely differentiable) proper scoring rule, then we have
\[\ind^\ell(s) = 2 \int_\half^1 \parens{\frac{x(1 - x)}{\varphi(x)}}^{\ell/4} dx.\]

Also, note that $s$ is normalized if and only if $s \parens{\half} = 0$ and (by Claim~\ref{claim:normalized_equiv} and Remark~\ref{remark:basic})
\[\int_\half^1 (1 - x)s'(x) dx = \int_\half^1 (1 - x)^2\varphi(x) dx = 1.\]

\approxthm*

\begin{proof}
The Weierstrass approximation theorem says that any continuous function can be uniformly approximated by polynomials on a closed interval. A constructive proof of this theorem (for the interval $[0, 1]$) is given by the Bernstein polynomials: $b_{i, n}(x) = \binom{n}{i}x^i(1 - x)^{n - i}$. Given a continuous function $\psi: [0, 1] \to \RR$, define
\[B_n(\psi)(x) = \sum_{i = 0}^n \psi \parens{\frac{i}{n}} b_{i, n}(x).\]
Then the polynomials $B_n(\psi)$ converge uniformly to $\psi$ \parencite[\S36]{Estep}. Suppose that $\psi$ also satisfies $\psi(x) = \psi(1 - x)$. Then $\psi \parens{\frac{i}{n}} = \psi \parens{\frac{n - i}{n}}$, which means $B_n(\psi)$ can be written as a linear combination of polynomials $(b_{i, n} + b_{n - i, n})(x)$. These polynomials are equal at $x$ and $1 - x$, and thus $B_n(\psi)(x) = B_n(\psi)(1 - x)$. From this we conclude that $\psi$ can be uniformly approximated on $[0, 1]$ by a sequence polynomials $p_i$ that satisfy $p_i(x) = p_i(1 - x)$.

Let $\varphi_\ell$ be the $\varphi$ corresponding to $s_{\ell, \opt}$. Recall that
\begin{equation} \label{eq:phi_ell}
\varphi_\ell(x) = \begin{cases}\kappa_\ell (x^{\ell - 8}(1 - x)^\ell)^{1/(\ell + 4)} & x \le \half \\
\kappa_\ell (x^\ell(1 - x)^{\ell - 8})^{1/(\ell + 4)} & x \ge \half.\end{cases}
\end{equation}

Let $0 < \epsilon < \min(\half, \varphi_\ell(\half))$. Consider the following function $\varphi_{\ell, \epsilon}: [0, 1] \to \RR$.

\singlespacing
\[\varphi_{\ell, \epsilon}(x) = \begin{cases}\varphi_\ell(\epsilon) & x \le \epsilon \\ \varphi_\ell(x) & \epsilon \le x \le 1 - \epsilon \\ \varphi_\ell(1 - \epsilon) & x \ge 1 - \epsilon.\end{cases}\]
\doublespacing

Observe that $\varphi_{\ell, \epsilon}(x) = \varphi_{\ell, \epsilon}(1 - x)$ for all $x \in [0, 1]$; this is a straightforward consequence of the fact that $\varphi_\ell$ is symmetric about $\half$. Per our discussion above, there exists a polynomial $p_\epsilon$ satisfying $p_\epsilon(x) = p_\epsilon(1 - x)$ such that for all $x \in [0, 1]$, $\abs{p_\epsilon(x) - \varphi_{\ell, \epsilon}(x)} \le \epsilon$. In particular, we take $p_\epsilon = B_{n(\epsilon)}(\varphi_{\ell, \epsilon})$, where $n(\epsilon)$ is any $n$ large enough that $p_\epsilon$ is uniformly within $\epsilon$ of $\varphi_{\ell, \epsilon}$.

Observe that such a polynomial, when written as a sum of powers of $x - \half$, must only contain even powers of $x - \half$, since $p_\epsilon(x) - p_\epsilon(1 - x)$ must be the zero polynomial. Consequently, by Lemma~\ref{lem:taylorphil}, $(1 - x)p_\epsilon(x)$ is the derivative of a proper scoring rule.\footnote{The fact that $p_\epsilon$ is nonnegative everywhere follows from the fact that it is a uniform $\epsilon$-approximation of $\varphi_{\ell, \epsilon}$, which is greater than $1$ on $[0, 1]$.} To find the associated normalized proper scoring rule (call it $s_\epsilon$), we take the antiderivative (taking the constant coefficient in the $\parens{x - \half}$-expansion to be zero), and divide by $\int_\half^1 (1 - x)^2p_\epsilon(x) dx$. Thus, corresponding to each $\epsilon$ we have a normalized polynomial scoring rule $s_\epsilon$ with incentivization index
\[\ind^\ell(s_\epsilon) = 2 \int_\half^1 \parens{\frac{x(1 - x)}{p_\epsilon(x) \cdot \frac{1}{\int_\half^1 (1 - x)^2p_\epsilon(x) dx}}}^{\ell/4} dx = 2 \parens{\int_\half^1 (1 - x)^2p_\epsilon(x) dx}^{\ell/4} \int_\half^1 \parens{\frac{x(1 - x)}{p_\epsilon(x)}}^{\ell/4} dx.\]

\begin{claim}
$s_\epsilon$ is respectful.
\end{claim}

\begin{proof}
Since $s_\epsilon$ is polynomial (and thus bounded and infinitely differentiable), it suffices to show that the second derivative of its expected score function is bounded away from zero. The second derivative of $s_\epsilon$'s expected score function is a positive multiple of $p_\epsilon$, so it suffices to show that $p_\epsilon$ is bounded away from zero. This is indeed the case. To see this, note that $\varphi_{\ell, \epsilon}$ is bounded away from zero (as $\varphi_\ell$ is bounded away from zero on $[\epsilon, 1 - \epsilon]$); let $L > 0$ be such that $\varphi_{\ell, \epsilon}(x) \ge L$ on $[0, 1]$. Then
\[p_\epsilon(x) = \sum_{i = 0}^{n(\epsilon)} \varphi_{\ell, \epsilon} \parens{\frac{i}{n(\epsilon)}} \binom{n(\epsilon)}{i} x^i (1 - x)^{n(\epsilon) - i} \ge \sum_{i = 0}^{n(\epsilon)} L \binom{n(\epsilon)}{i} x^i (1 - x)^{n(\epsilon) - i} = L.\]
\end{proof}

Our goal is to upper bound $\ind^\ell(s_\epsilon)$ in a way that shows that $\lim_{\epsilon \to 0} \ind^\ell(s_\epsilon) = \ind^\ell(s_{\ell,\opt})$. To do this, it suffices to show that the first of the two integrals in our formula for $s_\epsilon$ converges to $1$ as $\epsilon \to 0$ and that twice the second integral converges to $\ind^\ell(s_\ell, \opt)$ as $\epsilon \to 0$. We begin by working with the first of the two integrals.

\begin{claim}
\[\limsup_{\epsilon \to 0} \int_\half^1 (1 - x)^2p_\epsilon(x) dx \le 1.\]
\end{claim}

\begin{proof}
First observe that, since $s_{\ell,\opt}$ is normalized, we have $\int_\half^1 (1 - x)^2\varphi_\ell(x) dx = 1$. Next, note that for $1 \le \ell \le 8$, $\varphi_\ell$ is increasing on $[\half, 1)$ (as is evident from Equation~\ref{eq:phi_ell}), which means that
\[\int_\half^1 (1 - x)^2\varphi_{\ell, \epsilon}(x) dx \le \int_\half^1 (1 - x)^2 \varphi_\ell(x) dx = 1. \qquad (1 \le \ell \le 8)\]
On the other hand, for $\ell > 8$, observe that $\varphi_\ell$ is bounded above on $[\half, 1)$, say by a constant $M_\ell$, which means that in this case
\[\int_\half^1 (1 - x)^2\varphi_{\ell, \epsilon}(x) dx \le \int_\half^1 (1 - x)^2 \varphi_\ell(x) dx + \epsilon M_\ell = 1 + \epsilon M_\ell. \qquad (\ell > 8)\]

Now, we have
\[\parens{\int_\half^1 (1 - x)^2p_\epsilon(x) dx}^{\ell/4} \le \parens{\int_\half^1 (1 - x)^2(\varphi_{\ell, \epsilon}(x) + \epsilon) dx}^{\ell/4} \le \parens{\epsilon + \int_\half^1 (1 - x)^2\varphi_{\ell, \epsilon}(x) dx}^{\ell/4}\]
which is at most $(1 + \epsilon)^{\ell/4}$ (for $\ell \le 8$) and at most $(1 + (M_\ell + 1)\epsilon)^{\ell/4}$ (for $\ell > 8$).
\end{proof}

Next we work with the second integral.

\begin{claim}
\[\limsup_{\epsilon \to 0} 2 \int_\half^1 \parens{\frac{x(1 - x)}{p_\epsilon(x)}}^{\ell/4} dx \le I(s_\ell, \opt).\]
\end{claim}

\begin{proof}
We have
\begin{align*}
\int_\half^1 \parens{\frac{x(1 - x)}{p_\epsilon(x)}}^{\ell/4} dx &\le \int_\half^1 \parens{\frac{x(1 - x)}{\varphi_{\ell, \epsilon}(x) - \epsilon}}^{\ell/4} dx = \int_\half^1 \parens{\frac{x(1 - x)}{\varphi_{\ell, \epsilon}(x)\parens{1 - \frac{\epsilon}{\varphi_{\ell, \epsilon}(x)}}}}^{\ell/4} dx\\
&\le \parens{1 - \max_{x \in [\half, 1]} \frac{\epsilon}{\varphi_{\ell, \epsilon}(x)}}^{\ell/4} \int_\half^1 \parens{\frac{x(1 - x)}{\varphi_{\ell, \epsilon}(x)}}^{\ell/4} dx.
\end{align*}

Let us consider $\max_{x \in [\half, 1]} \frac{\epsilon}{\varphi_{\ell, \epsilon}(x)}$ as a function of $\epsilon$. For $\ell \le 8$, since $\varphi_\ell$ is increasing on $[\half, 1)$, this is just $\frac{\epsilon}{\varphi_\ell(\half)}$, a quantity that approaches zero as $\epsilon$ approaches zero.

For $\ell > 8$, as $\epsilon$ approaches zero we have that $\min_{x \in [\half, 1]} \varphi_{\ell, \epsilon}(x)$ approaches $\kappa_\ell \epsilon^{(\ell - 8)/(\ell + 4)} = \omega(\epsilon)$. This means that $\max_{x \in [\half, 1]} \frac{\epsilon}{\varphi_{\ell, \epsilon}(x)}$ approaches zero as $\epsilon$ approaches zero.

Therefore, we have
\[\limsup_{\epsilon \to 0} 2 \int_\half^1 \parens{\frac{x(1 - x)}{p_\epsilon(x)}}^{\ell/4} dx \le \limsup_{\epsilon \to 0} 2 \int_\half^1 \parens{\frac{x(1 - x)}{\varphi_{\ell, \epsilon}(x)}}^{\ell/4}.\]

Next, note that for $x \in [1 - \epsilon, 1]$, we have
\begin{align*}
\parens{\frac{x(1 - x)}{\varphi_{\ell, \epsilon}(x)}}^{\ell/4} &= \parens{\frac{x(1 - x)}{\varphi_\ell(1 - \epsilon)}}^{\ell/4} \le \parens{\frac{\epsilon(1 - \epsilon)}{\varphi_\ell(1 - \epsilon)}}^{\ell/4} = \parens{\frac{\epsilon(1 - \epsilon)}{\kappa_\ell ((1 - \epsilon)^\ell \epsilon^{\ell - 8})^{1/(\ell + 4)}}}^{\ell/4}\\
&= \frac{(\epsilon^3(1 - \epsilon))^{\ell/(\ell + 4)}}{\kappa_\ell^{\ell/4}} \le \frac{1}{\kappa_\ell^{\ell/4}}.
\end{align*}
Therefore we have
\[\int_\half^1 \parens{\frac{x(1 - x)}{\varphi_{\ell, \epsilon}(x)}}^{\ell/4} dx \le \int_\half^1 \parens{\frac{x(1 - x)}{\varphi_\ell(x)}}^{\ell/4} dx + \frac{\epsilon}{\kappa_\ell^{\ell/4}} = \half \ind^\ell(s_{\ell,\opt}) + \frac{\epsilon}{\kappa_\ell^{\ell/4}},\]
so
\[\limsup_{\epsilon \to 0} 2 \int_\half^1 \parens{\frac{x(1 - x)}{\varphi_{\ell, \epsilon}(x)}}^{\ell/4} \le \ind^\ell(s_{\ell, \opt}).\]
\end{proof}

It therefore follows that $\limsup_{\epsilon \to 0} \ind^\ell(s_\epsilon) \le \ind^\ell(s_{\ell,\opt})$. But in fact, the inequality is an equality; this is because no continuously differentiable function has incentivization index less than that of $s_{\ell,\opt}$. This completes the proof of Theorem~\ref{thm:approx}.
\end{proof}

\newpage
\section{Simulation results} \label{app:simulation}
\small
\singlespacing

\begin{table}[ht]
\begin{center}
\begin{tabular}{|c c||c|c|c|c|c|}
\hline
Cost & Rule & Avg. Error & Predicted Avg. Error & Ratio & Avg. \# Flips & Max. \# Flips\\
\hline \hline
0.1 & $s_\text{quad}$ & 0.1616 & 0.1490 & 1.0845 & 2.3341 & 3\\
& $s_\text{log}$ & 0.1609 & 0.1389 & 1.1582 & 2.3317 & 3\\
& $s_{1,\opt}$ & 0.1553 & 0.1348 & 1.1522 & 2.6658 & 3\\
\hline
0.03 & $s_\text{quad}$ & 0.1136 & 0.1103 & 1.0298 & 6.0933 & 7\\
& $s_\text{log}$ & 0.1093 & 0.1028 & 1.0636 & 6.7133 & 7\\
& $s_{1,\opt}$ & 0.1110 & 0.0997 & 1.1126 & 7.1547 & 10\\
\hline
0.01 & $s_\text{quad}$ & 0.0850 & 0.0838 & 1.0147 & 11.9745 & 15\\
& $s_\text{log}$ & 0.0816 & 0.0781 & 1.0444 & 13.2780 & 14\\
& $s_{1,\opt}$ & 0.0802 & 0.0758 & 1.0577 & 15.3399 & 23\\
\hline
0.003 & $s_\text{quad}$ & 0.0626 & 0.0620 & 1.0096 & 23.2076 & 29\\
& $s_\text{log}$ & 0.0590 & 0.0578 & 1.0199 & 26.3918 & 27\\
& $s_{1,\opt}$ & 0.0580 & 0.0561 & 1.0349 & 31.2093 & 52\\
\hline
0.001 & $s_\text{quad}$ & 0.0472 & 0.0471 & 1.0014 & 41.5592 & 52\\
& $s_\text{log}$ & 0.0448 & 0.0439 & 1.0193 & 47.4845 & 48\\
& $s_{1,\opt}$ & 0.0434 & 0.0426 & 1.0179 & 57.5323 & 107\\
\hline
0.0003 & $s_\text{quad}$ & 0.0349 & 0.0349 & 1.0016 & 77.2931 & 97\\
& $s_\text{log}$ & 0.0329 & 0.0325 & 1.0113 & 89.2927 & 90\\
& $s_{1,\opt}$ & 0.0320 & 0.0315 & 1.0130 & 108.8362 & 230\\
\hline
0.0001 & $s_\text{quad}$ & 0.0265 & 0.0265 & 0.9999 & 134.6477 & 171\\
& $s_\text{log}$ & 0.0248 & 0.0247 & 1.0047 & 157.1403 & 158\\
& $s_{1,\opt}$ & 0.0241 & 0.0240 & 1.0043 & 192.5842 & 460\\
\hline
0.00003 & $s_\text{quad}$ & 0.0196 & 0.0196 & 1.0013 & 247.0952 & 314\\
& $s_\text{log}$ & 0.0184 & 0.0183 & 1.0056 & 289.8925 & 291\\
& $s_{1,\opt}$ & 0.0177 & 0.0177 & 0.9980 & 356.4421 & 979\\
\hline
\end{tabular}
\end{center}\caption[Simulation results for incentivization index accuracy]{Results of a simulation showing the behavior of a locally adaptive expert for small-to-medium costs $c$. ``Predicted average error'' means the error that Theorem~\ref{thm:global} predicts in the limit as $c$ approaches $0$ (but multiplied by $c^{-1/4}$ in the stated value of $c$). ``Ratio'' refers to the ratio between the average error and the predicted average error. ``Maximum number of flips'' refers to the maximum number of flips for that cost and rule in the $100,000$ simulations. (Note that because of the large number of simulations and comparatively small number of flips, these are likely to be universal upper bounds on the number of flips given that cost and rule.)}\label{tab:simulation}
\end{table}

\normalsize
\doublespacing
\chapter{Details omitted from Chapter~\ref{chap:qa}}
\section{Details omitted from Section~\ref{sec:convex_losses}} \label{appx:convex_losses}
\noregret*

\begin{proof}
	We apply Theorem~3.1 of \textcite{hazan23}; this theorem tells us that in order to prove the stated bound, it suffices to show that for all $t$ and $\vect{w}$, $\norm{\nabla L^t(\vect{w})}_2 \le \sqrt{2m}M$.
	
	Let $L$ be an arbitrary loss function, i.e.\ $L(\vect{w}) = -\ws_j(\vect{w})$ for some $j, \vect{p}_1, \dots, \vect{p}_m$. Let $\vect{p}^*(\vect{w}) = \sideset{}{_\vect{g}}\bigoplus\limits_{i = 1}^m (\vect{p}_i, w_i)$. We claim that

    \singlespacing
	\begin{equation} \label{eq:delLw}
		\nabla L(\vect{w}) = \begin{pmatrix} \text{--- } \vect{g}(\vect{p}_1) \text{ ---} \\ \vdots \\ \text{--- } \vect{g}(\vect{p}_m) \text{ ---} \end{pmatrix} (\vect{p}^*(\vect{w}) - \pmb{\delta}_j),
	\end{equation}
    \doublespacing
 
	where this $m$-dimensional vector should be interpreted modulo translation by $\vect{1}_m$ (see Remark~\ref{remark:mod_T1n}). To see this, observe that
	\[\nabla L(\vect{w}) = - \nabla WS_j(\vect{w}) = - \nabla_{\vect{w}} s(\vect{p}^*(\vect{w}); j) = - \nabla_{\vect{w}} \parens{G(\vect{p}^*(\vect{w})) + \angles{\vect{g}(\vect{p}^*(\vect{w})), \pmb{\delta}_j - \vect{p}^*(\vect{w})}},\]
	where $\nabla_{\vect{w}}$ denotes the gradient with respect to change in the weight vector $\vect{w}$ (as opposed to change in the probability vector). Now, by the chain rule for gradients, we have
	\[\nabla_{\vect{w}} G(\vect{p}^*(\vect{w})) = (J_{\vect{p}^*}(\vect{w}))^{\top} \vect{g}(\vect{p}^*(\vect{w})),\]
	where $J_{\vect{p}^*}$ denotes the Jacobian matrix of the function $\vect{p}^*(\vect{w})$. Also, we have
	\[\vect{g}(\vect{p}^*(\vect{w})) = \sum_{i = 1}^m w_i \vect{g}(\vect{p}_i),\]
	so (again by the chain rule) we have

    \singlespacing
	\[\nabla_{\vect{w}}(\angles{\vect{g}(\vect{p}^*(\vect{w})), \pmb{\delta}_j - \vect{p}^*(\vect{w})}) = \begin{pmatrix} \vect{g}(\vect{p}_1) \\ \vdots \\ \vect{g}(\vect{p}_m) \end{pmatrix} (\pmb{\delta}_j - \vect{p}^*(\vect{w})) - (J_{\vect{p}^*}(\vect{w}))^{\top} \vect{g}(\vect{p}^*(\vect{w})).\]
    \doublespacing
 
	This gives us Equation~\ref{eq:delLw}.\footnote{Note that the cancellation of the Jacobian terms stems not from the specific relationship between $\vect{p}^*$ and $\vect{w}$ but from the nature of proper scoring rules. We obtain the same cancellation if we consider $\nabla s(\vect{p}; j)$, where after differentiating $G(\vect{p}) + \angles{\vect{g}(\vect{p}), \pmb{\delta}_j - \vect{p}}$ we find that the $\vect{g}(\vect{p})$ terms cancel.} Now, for any $i$, we have
	\[\abs{\angles{\vect{g}(\vect{p}_i), \vect{p}^*(\vect{w}) - \pmb{\delta}_j}} \le \norm{\vect{g}(\vect{p}_i)}_2 \norm{\vect{p}^*(\vect{w}) - \pmb{\delta}_j}_2 \le \sqrt{2}M.\]
	Therefore,
	\[\norm{\nabla L(\vect{w})}_2 \le \sqrt{m \cdot \parens{\sqrt{2}M}^2} = \sqrt{2m}M,\]
	completing the proof.
\end{proof}

\section{Details omitted from Section~\ref{sec:axiomatization}} \label{appx:ax}
	We claim that our axioms in Definition~\ref{def:qa_ax_n2} can be restated equivalently in a form similar to that of Kolmogorov introduced at the top of Section~\ref{sec:axiomatization} (though with weights.)
	\begin{claim}
		Given a pooling operator $\oplus$ on $\mathcal{D}$ satisfying Definition~\ref{def:qa_ax_n2}, the function $M$ defined on arbitrary tuples of weighted forecasts defined by $M(\Pi_1, \dots, \Pi_m) := \bigoplus_{i = 1}^m \Pi_i$ satisfies the following axioms:
		\begin{enumerate}[label=(\arabic*)]
			\item $M(\Pi_1, \dots, \Pi_m)$ is strictly increasing in each $\prb(\Pi_i)$ and continuous in its inputs.\footnote{That is, it is a continuous function of its input in $\mathcal{D}^m \times (\RR_{\ge 0}^m \setminus \vect{0})$, where weighted forecasts with weight $0$ are ignored when computing $M$.}
			\item $M$ is symmetric in its arguments.
			\item $M((p, w_1), \dots, (p, w_m)) = (p, \sum_i w_i)$.
			\item $M(\Pi_1, \dots, \Pi_k, \Pi_{k + 1}, \dots, \Pi_m) = M(\Pi', \Pi_{k + 1}, \dots, \Pi_m)$, where $y := M(\Pi_1, \dots, \Pi_k)$.
			\item $M((p_1, w_1), \dots, (p_m, w_m))$ has weight $w_1 + \dots + w_m$.
		\end{enumerate}
		Additionally, given any $M$ defined on arbitrary tuples of weighted forecasts, the operator $\oplus$ defined by $\Pi_1 \oplus \Pi_2 := M(\Pi_1, \Pi_2)$ satisfies Definition~\ref{def:qa_ax_n2}.
	\end{claim}

	\begin{proof}
		We first prove that given $\oplus$ satisfying Definition~\ref{def:qa_ax_n2}, $M$ satisfies the stated axioms. The last four axioms are clear, so we prove the first one. The fact that $M$ is strictly increasing in each probability follows immediately by considering the continuous, strictly increasing function $g$ such that $\oplus = \oplus_g$, which exists by Theorem~\ref{thm:representation_n2}. Continuity likewise follows, since the quantity in Definition~\ref{def:qa_ax_n2} is continuous.
		
		We now prove that given $M$ satisfying the stated axioms, $\oplus$ satisfies Definition~\ref{def:qa_ax_n2}. Weight additivity, commutativity, continuity, and idempotence are clear. To prove associativity, note that
		\[\Pi_1 \oplus (\Pi_2 \oplus \Pi_3) = M(\Pi_1, M(\Pi_2, \Pi_3)) = M(\Pi_1, \Pi_2, \Pi_3) = M(M(\Pi_1, \Pi_2), \Pi_3) = (\Pi_1 \oplus \Pi_2) \oplus \Pi_3.\]
		To prove monotonicity, let $p_1 > p_2$ and $w > x > y$. We wish to prove that $\prb((p_1, x) \oplus (p_2, w - x)) > \prb((p_1, y) \oplus (p_2, w - y))$. We have
		\begin{align*}
			\prb((p_1, x) \oplus (p_2, w - x)) &= \prb(M((p_1, x), (p_2, w - x))) = \prb(M((p_1, y), (p_1, x - y), (p_2, w - x)))\\
			&> \prb(M((p_1, y), (p_2, x - y), (p_2, w - x))) = \prb(M((p_1, y), (p_2, w - y)))\\
			&= \prb((p_1, y) \oplus (p_2, w - y)).
		\end{align*}
	\end{proof}

\subsection{Extending the results of Section~\ref{sec:axiomatization} to $n > 2$ outcomes} \label{appx:extending}
We now discuss extending our axiomatization to arbitrary values of $n$ in a way that, again, describes the class of QA pooling operators. Just as we fixed a two-outcome forecast domain $\mathcal{D}$ in Section~\ref{sec:axiomatization}, we now fix an $n$-outcome forecast domain $\mathcal{D}$ for any $n \ge 2$. Our definition of weighted forecasts remains the same (except that now $\prb(\Pi)$ is a vector). Our definition of quasi-arithmetic pooling, however, needs to change to make $\vect{g}$ vector-valued. This raises the question: what is the analogue of ``increasing'' for vector-valued functions? It turns out that the relevant notion for us is \emph{cyclical monotonicity}, introduced by \textcite{roc70_paper} (see also \textcite[\S27]{roc70_book}). We will define this notion shortly, but first we give the definition of quasi-arithmetic pooling with arbitrary weights (analogous to Definition~\ref{def:qa_ax_n2}) for this setting. Throughout this section, we will use the notation $H_n(c) := \{\vect{x} \in \RR^n: \sum_i x_i = c\}$. Recall from Remark~\ref{remark:mod_T1n} that the range of the gradient of a function defined on $\mathcal{D}$ is a subset of $H_n(0)$.

\begin{defin}[Quasi-arithmetic pooling with arbitrary weights] \label{def:qa_ax}
	Given a continuous, strictly cyclically monotone vector-valued function $\vect{g}: \mathcal{D} \to H_n(0)$ whose range is a convex set, and weighted forecasts $\Pi_1 = (\vect{p}_1, w_1), \dots, \Pi_m = (\vect{p}_m, w_m)$, define the \emph{quasi-arithmetic pool} of $\Pi_1, \dots, \Pi_m$ with respect to $\vect{g}$ as
	\[\sideset{}{_{\vect{g}}}\bigoplus_{i = 1}^m (\vect{p}_i, w_i) := \parens{\vect{g}^{-1} \parens{\frac{\sum_i w_i \vect{g}(\vect{p}_i)}{\sum_i w_i}}, \sum_i w_i}.\]
\end{defin}

Note that QA pooling as defined in Definition~\ref{def:qa_pooling} can be written in the form of Definition~\ref{def:qa_ax} if and only if the scoring rule has convex exposure; if it does not, then for some choices of parameters, $\sum_i w_i \vect{g}(\vect{p}_i)$ will be equal to a subgradient -- but not the gradient -- of $G$ at some point.

\begin{defin}[Cyclical monotonicity] \label{def:cyc_mon}
	A function $\vect{g}: U \subseteq \RR^n \to \RR^n$ is \emph{cyclically monotone} if for every list of points $\vect{x}_0, \vect{x}_1, \dots, \vect{x}_{k - 1}, \vect{x}_k = \vect{x}_0 \in U$, we have
	\[\sum_{i = 1}^k \angles{\vect{g}(\vect{x}_i), \vect{x}_i - \vect{x}_{i - 1}} \ge 0.\]
	We also say that $\vect{g}$ is \emph{strictly cyclically monotone} if the inequality is strict except when $\vect{x}_0 = \dots = \vect{x}_{k - 1}$.
\end{defin}

To gain an intuition for this notion, consider the case of $k = 2$; then this condition says that $\angles{\vect{g}(\vect{x}_1) - \vect{g}(\vect{x}_0), \vect{x}_1 - \vect{x}_0} \ge 0$. In other words, the change in $\vect{g}$ from $\vect{x}_0$ to $\vect{x}_1$ is in the same general direction as the direction from $\vect{x}_0$ to $\vect{x}_1$. This property is called \emph{2-cycle} (or \emph{weak}) \emph{monotonicity}.

Cyclical monotonicity is a stronger notion, and has applications to mechanism design and revealed preference theory (see e.g.\ \textcite{ls07, abhm10, fk14, vohra07}). In such settings, it is usually the case that two-cycle and cyclical monotonicity are equivalent. Indeed, Saks and Yu showed that these conditions are equivalent in settings where the set of outcomes (i.e.\ the range of $\vect{g}$) is finite \parencite{sy05}. However, cyclical monotonicity is substantially stronger than two-cycle monotonicity when the range of $\vect{g}$ is infinitely large, as in our setting. In fact, the difference between these two conditions is that a two-cycle monotone function is cyclically monotone if and only if it is also \emph{vortex-free} \parencite[Theorem 3.9]{ak14}. \emph{Vortex-freeness} means that the path integral of $\vect{g}$ along any triangle vanishes. See \textcite{ak14} for a detailed comparison of these two notions.\\

The immediately relevant fact for us is that cyclically monotone functions are gradients of convex functions (and vice versa). Speaking more precisely:

\begin{theorem} \label{thm:cyc_mon_convex}
	A vector-valued function $\vect{g}$ is continuous and strictly cyclically monotone if and only if it is the gradient of a differentiable, strictly convex function $G$.
\end{theorem}

\begin{proof}
	Per a theorem of Rockafellar (\textcite{roc70_paper}, see also \textcite[Theorem 24.8]{roc70_book}), a function $\vect{g}$ is cyclically monotone if and only if it is a subgradient of a convex function $G$. The proof of this fact shows just as easily that a function is strictly cyclically monotone if and only if it is a subgradient of a strictly convex function.
	
	Consider a differentiable, strictly convex function $G$. Its gradient is continuous (see \textcite[Theorem 25.5]{roc70_book}). Conversely, consider a continuous, strictly cyclically monotone vector-valued function $\vect{g}$. As we just discussed, it is a subgradient of some strictly convex function $G$. A convex function with a continuous subgradient is differentiable \parencite[Proposition 17.41]{bc11}.
\end{proof}

This means that the conditions on $\vect{g}$ in Definition~\ref{def:qa_ax} are precisely those necessary to let $\vect{g}$ be any function that it could be in our original definition of quasi-arithmetic pooling (Definition~\ref{def:qa_pooling}). Our new definition is thus equivalent to the old one (after normalizing weights to add to $1$).\\

We now discuss our axioms for pooling operators that will again capture the class of QA pooling operators. We will keep the weight additivity, commutativity, associativity, and idempotence verbatim from our discussion of the $n = 2$ case. We will slightly strengthen the continuity argument (see below).

We will also add a new axiom, \emph{subtraction}, which states that if $\Pi_1 \oplus \Pi_2 = \Pi_1 \oplus \Pi_3$ then $\Pi_2 = \Pi_3$. Subtraction in the $n = 2$ case follows from monotonicity; in this case, however, we the subtraction axiom will help us state the monotonicity axiom. In particular, it allows us to make the following definition, which essentially extends the notion of pooling to allow for negative weights.

\begin{defin} \label{def:p}
	Let $\oplus$ be a pooling operator satisfying weight additivity, commutativity, associativity, and subtraction. Fix $\vect{p}_1, \dots, \vect{p}_k \in \mathcal{D}$. Define a function $\vect{p}: \Delta_k \to \mathcal{D}$ (with $\vect{p}_1, \dots, \vect{p}_k$ serving as implicit arguments) defined by
	\[\vect{p}(w_1, \dots, w_k) = \prb \parens{\bigoplus_{i = 1}^k (\vect{p}_i, w_i)}.\]
	We extend the definition of $\vect{p}$ to a partial function on $H_k(1)$, as follows: given input $(w_1, \dots, w_k)$, let $S \subseteq [k]$ be the set of indices $i$ such that $w_i < 0$ and $T \subseteq [k]$ be the set of indices $i$ such that $w_i > 0$. We define $\vect{p}(w_1, \dots, w_k)$ to be the $\vect{q} \in \mathcal{D}$ such that
	\[(\vect{q}, 1) \oplus \parens{\bigoplus_{i \in S} (\vect{p}_i, -w_i)} = \bigoplus_{i \in T} (\vect{p}_i, w_i).\]
	Note that $\vect{q}$ is not guaranteed to exist, which is why we call $\vect{p}$ a partial function. However, if $\vect{q}$ exists then it is unique, by the subtraction axiom.
\end{defin}

We can now state the full axiomatization, including the monotonicity axiom.

\begin{defin}[Axioms for pooling operators] \label{def:wop}
	For a pooling operator $\oplus$ on $\mathcal{D}$, we define the following axioms.
	\begin{enumerate}
		\item \textbf{Weight additivity}: $\wt(\Pi_1 \oplus \Pi_2) = \wt(\Pi_1) + \wt(\Pi_2)$ for every $\Pi_1, \Pi_2$.
		\item \textbf{Commutativity}: $\Pi_1 \oplus \Pi_2 = \Pi_2 \oplus \Pi_1$ for every $\Pi_1, \Pi_2$.
		\item \textbf{Associativity}: $\Pi_1 \oplus (\Pi_2 \oplus \Pi_3) = (\Pi_1 \oplus \Pi_2) \oplus \Pi_3$ for every $\Pi_1, \Pi_2, \Pi_3$.
		\item \textbf{Continuity}: For every positive integer $k$ and $\vect{p}_1, \dots, \vect{p}_k$, the quantity\footnote{The continuity axiom is only well-defined conditioned on $\oplus$ being associative, which is fine for our purposes. We allow a proper subset of weights to be zero by defining the aggregate to ignore forecasts with weight zero.}
		\[\prb \parens{\bigoplus_{i = 1}^k (\vect{p}_i, w_i)}\]
		is a continuous function of $(w_1, \dots, w_k)$ on $\RR_{\ge 0}^k \setminus \{\vect{0}\}$.
		\item \textbf{Idempotence}: For every $\Pi_1$ and $\Pi_2$, if $\prb(\Pi_1) = \prb(\Pi_2)$ then $\prb(\Pi_1 \oplus \Pi_2) = \prb(\Pi_1)$.
		\item \textbf{Subtraction}: If $\Pi_1 \oplus \Pi_2 = \Pi_1 \oplus \Pi_3$ then $\Pi_2 = \Pi_3$.
		\item \textbf{Monotonicity}: There exist vectors $\vect{p}_1, \dots, \vect{p}_n \in \mathcal{D}$ such that $\vect{p}$ (as in Definition~\ref{def:p}) is a strictly cyclically monotone function from its domain to $\RR^n$.
	\end{enumerate}
\end{defin}

This monotonicity axiom essentially extends our previous monotonicity axiom (in Definition~\ref{def:wop_n2}) to a multi-dimensional setting. It states that there are $n$ ``anchor points'' in $\mathcal{D}$ such that the function $\vect{p}$ from weight vectors to $\mathcal{D}$ that pools the anchor points with the weights given as input obeys a notion of monotonicity (namely cyclical monotonicity). Informally, this means that the vector of weights that one would need to give to the anchor points in order to arrive at a forecast $\vect{p}$ ``correlates'' with the forecast $\vect{p}$ itself.\\

We now state the main theorem of our axiomatization.

\begin{theorem} \label{thm:representation}
	A pooling operator satisfies the axioms in Definition~\ref{def:wop} if and only if it is a QA pooling operator as in Definition~\ref{def:qa_ax}.\footnote{Recall that Definition~\ref{def:qa_ax} is narrower than Definition~\ref{def:qa_pooling}, since it excludes QA pools with respect to scoring rules that do not have convex exposure. Without convex exposure, the associativity, subtraction, and monotonicity axioms may be violated. For example, consider $\Pi_i = (\pmb{\delta}_i, 1)$ for $i = 1, 2, 3$, for the scoring rule with expected score function $G(\vect{x}) = x_1^4 + x_2^4 + x_3^4$. We have that $(\Pi_1 \oplus_\vect{g} \Pi_2) \oplus_\vect{g} \Pi_3 = ((1/2, 1/2, 0), 2) \oplus_\vect{g} (\pmb{\delta}_3, 1) \approx ((0.182, 0.182, 0.635), 3)$, whereas $\Pi_1 \oplus_\vect{g} (\Pi_2 \oplus_\vect{g} \Pi_3) = ((0.635, 0.182, 0.182), 3)$}.
\end{theorem}

\begin{proof}
	We begin by noting the following fact, which follows from results in \textcite[\S26]{roc70_book}.
	
	\begin{prop} \label{prop:g_inverse}
		A strictly cyclically monotone function $\vect{g}: \mathcal{D} \to \RR^n$ is injective, and its inverse $\vect{g}^{-1}$ is strictly cyclically monotone and continuous.\footnote{Why can't we apply this result again to $\vect{g}^{-1}$ to conclude that $\vect{g}$ is continuous, even though we did not assume it to be? The reason is that the proof of continuity relies on the convexity of $\mathcal{D}$; if $\vect{g}$ is discontinuous then the domain of $\vect{g}^{-1}$ may not be convex (or even connected), so we cannot apply the result to $\vect{g}^{-1}$.}
	\end{prop}
	
	We provide a partial proof below; it relies on the following observation.
	
	\begin{remark} \label{remark:cyc_mon_equiv}
		We can instead write the condition as
		\[\sum_{i = 1}^k \angles{\vect{g}(\vect{x}_i) - \vect{g}(\vect{x}_{i - 1}), \vect{x}_i} \ge 0.\]
		This is equivalent to the condition in Definition~\ref{def:cyc_mon}, because it is the same statement (with rearranged terms) when the $\vect{x}_i$'s are listed in reverse order.
	\end{remark}
	
	\begin{proof}
		First, suppose that $\vect{g}(\vect{x}) = \vect{g}(\vect{y})$. Then
		\[\vect{g}(\vect{x})(\vect{x} - \vect{y}) + \vect{g}(\vect{y})(\vect{y} - \vect{x}) = 0.\]
		Since $\vect{g}$ is \emph{strictly} cyclically monotone, this implies that $\vect{x} = \vect{y}$. (Note that we only use two-cycle monotonicity.)
		
		We now show that $\vect{g}^{-1}$ is strictly cyclically monotone. That is, we wish to show that
		\[\sum_{i = 1}^k \angles{\vect{x}_i, \vect{g}^{-1}(\vect{x}_i) - \vect{g}^{-1}(\vect{x}_{i - 1})} > 0\]
		for any $\vect{x}_1, \dots, \vect{x}_k = \vect{x}_0$ that are not all the same. (See Remark~\ref{remark:cyc_mon_equiv}.) By the cyclical monotonicity of $\vect{g}$, we have that
		\[\sum_{i = 1}^k \angles{\vect{g}(\vect{p}_i), \vect{p}_i - \vect{p}_{i - 1}} > 0\]
		(the strictness of the inequality follows by the injectivity of $\vect{g}$: if $\vect{x}_i \neq \vect{x}_j$ then $\vect{p}_i \neq \vect{p}_j$). This means that
		\[\sum_{i = 1}^k \angles{\vect{x}_i, \vect{g}^{-1}(\vect{x}_i) - \vect{g}^{-1}(\vect{x}_{i - 1})} > 0,\]
		as desired. As for continuity, we defer to \textcite[Theorem 26.5]{roc70_book}.
	\end{proof}
	
	Back to the proof of Theorem~\ref{thm:representation}, we first prove that any such $\oplus_{\vect{g}}$ satisfies the stated axioms. Weight additivity, commutativity, associativity, and idempotence are clear. Continuity follows from the formula
	\[\prb((\vect{p}_1, w_1) \oplus_{\vect{g}} (\vect{p}_2, w_2)) = \vect{g}^{-1} \parens{\frac{w_1 \vect{g}(\vect{p}_1) + w_2 \vect{g}(\vect{p}_2)}{w_1 + w_2}},\]
	noting that $\vect{g}^{-1}$ is continuous by Proposition~\ref{prop:g_inverse}. Likewise, subtraction follows from the fact that $\vect{g}$ is injective (by Proposition~\ref{prop:g_inverse}), as is $\vect{g}^{-1}$ (likewise). Monotonicity remains.
	
	The range of $\vect{g}$ contains an open subset\footnote{This follows from the invariance of domain theorem, which states that the image of an open subset of a manifold under an injective continuous map is open.} of $H_n(0)$, so in particular it contains the vertices of some translated and dilated copy of the standard simplex. That is, there are $n$ points $\vect{x}_1, \dots, \vect{x}_n$ in the range of $\vect{g}$ for which there is a positive scalar $a$ and vector $\vect{b}$ such that $a \pmb{\delta}_i + \vect{b} = \vect{x}_i$ for every $i$. (Here $\pmb{\delta}_i$ is the $i$-th standard basis vector in $\RR^n$.) We will let $\vect{p}_i$ be the pre-image of $\vect{x}_i$ under $\vect{g}$, so that $\vect{g}(\vect{p}_i) = a \pmb{\delta}_i + \vect{b}$.
	
	Observe that for any $\vect{w}$ in the domain of $\vect{p}$, we have
	\[\vect{g}(\vect{p}(\vect{w})) = \sum_{i = 1}^n w_i \vect{g}(\vect{p}_i) = \sum_{i = 1}^n w_i(a \pmb{\delta}_i + \vect{b}) = a\vect{w} + \vect{b},\]
	so
	\[\vect{p}(\vect{w}) = \vect{g}^{-1}(a\vect{w} + \vect{b}).\]
	We have that $\vect{g}^{-1}$ is strictly cyclically monotone (by Proposition~\ref{prop:g_inverse}), and it is easy to verify that for any strictly cyclically monotone function $\vect{f}$ and any $a > 0$ and $\vect{b}$, $\vect{f}(a\vect{x} + \vect{b})$ is a strictly cyclically monotone function of $\vect{x}$. Therefore, $\vect{p}(\vect{w}) = \vect{g}^{-1}(a\vect{w} + \vect{b})$ is strictly cyclically monotone, as desired.\\
	
	Now we prove the converse. Assume that we have a pooling operator $\oplus$ satisfying the axioms in Definition~\ref{def:wop}. We wish to show that $\oplus$ is $\oplus_{\vect{g}}$ for some $\vect{g}: \mathcal{D} \to H_n(0)$.
	
	For the remainder of this proof, let $\vect{p}_1, \dots, \vect{p}_n$ be vectors certifying the monotonicity of $\oplus$, and let $\vect{p}(\cdot)$ be as in Definition~\ref{def:p}.
	
	For any $\vect{q} \in \mathcal{D}$, let $\vect{g}(\vect{q}) := \vect{w} - \frac{1}{n} \vect{1}_n$, where $\vect{w} \in H_n(1)$ is such that $\vect{p}(\vect{w}) = \vect{q}$ and $\vect{1}_n$ is the all-ones vector. This raises the question of well-definedness: does this $\vect{w}$ necessarily exist, and if so, is it unique? The following claim shows that this is indeed the case.
	
	\begin{claim} \label{claim:p_bijective}
		The function $\vect{p}$, from the subset of $H_n(1)$ where it is defined to $\mathcal{D}$, is bijective.
	\end{claim}
	
	\begin{proof}
		The fact that $\vect{p}$ is injective follows from the fact that it is strictly cyclically monotone (see Proposition~\ref{prop:g_inverse}). We now show that $\vect{p}$ is surjective.
		
		Let $\vect{q} \in \mathcal{D}$. Define the function $\tilde{\vect{p}}: \Delta_{n + 1} \to \mathcal{D}$ by
		\[\tilde{\vect{p}}(w_1, \dots, w_{n + 1}) := \prb \parens{\parens{\bigoplus_{i = 1}^n (\vect{p}_i, w_i)} \oplus (\vect{q}, w_{n + 1})}.\]
		Since $\tilde{\vect{p}}$ is a continuous map\footnote{By the continuity axiom; here we use the more generalized form we stated earlier.} from $\Delta_{n + 1}$ (an $n$-dimensional manifold) to $\mathcal{D}$ (an $(n - 1)$-dimensional manifold), $\tilde{\vect{p}}$ is not injective.\footnote{This follows e.g.\ from the Borsuk-Ulam theorem.} So in particular, let $\vect{w}_1 \neq \vect{w}_2 \in \Delta_{n + 1}$ be such that $\vect{\tilde{p}}(\vect{w}_1) = \vect{\tilde{p}}(\vect{w}_2)$. That is, we have
		
		\begin{equation} \label{eq:ptilde_not_injective}
			\parens{\bigoplus_{i = 1}^n (\vect{p}_i, w_{1, i})} \oplus (\vect{q}, w_{1, n + 1}) = \parens{\bigoplus_{i = 1}^n (\vect{p}_i, w_{2, i})} \oplus (\vect{q}, w_{2, n + 1}).
		\end{equation}
		
		Observe that $w_{1, n + 1} \neq w_{2, n + 2}$; for otherwise it would follows from the subtraction axiom that two different combinations of the $\vect{p}_i$'s would give the same probability, contradicting the fact that $\vect{p}$ is injective. Without loss of generality, assume that $w_{1, n + 1} > w_{2, n + 1}$. We can rearrange the terms in Equation~\ref{eq:ptilde_not_injective} to look as follows.
		\[(\vect{q}, w_{1, n + 1} - w_{2, n + 1}) \oplus \parens{\bigoplus_{i \in S} (\vect{p}_i, v_i)} = \bigoplus_{i \in T \subseteq [n] \setminus S} (\vect{p}_i, v_i)\]
		for some positive $v_1, \dots, v_n$. By the distributive property, we may multiply all weights by $\frac{1}{w_{1, n + 1} - w_{2, n + 1}}$. The result will be an equation as in Definition~\ref{def:p}, certifying that $\vect{q}$ is in the range of the function $\vect{p}$, as desired.
	\end{proof}
	
	We return to our main proof, now that we have shown that our function $\vect{g}(\vect{q}) := \vect{w} - \frac{1}{n} \vect{1}$, where $\vect{w} \in H_n(1)$ is such that $\vect{p}(\vect{w}) = \vect{q}$, is well-defined. In fact, we can simply write $\vect{g}(\vect{q}) = \vect{p}^{-1}(\vect{q}) - \frac{1}{n} \vect{1}$. (The vector $\frac{1}{n} \vect{1}$ is fairly arbitrary; it only serves the purpose of forcing the range of $\vect{g}$ to lie in $H_n(0)$ instead of $H_n(1)$.)
	
	We first show that the equation that defines $\oplus_{\vect{g}}$ holds -- that is, that if $(\vect{q}_1, v_1) \oplus (\vect{q}_2, v_2) = (\vect{q}, v_1 + v_2)$ (with $v_1, v_2 \ge 0$, not both zero), then
	\[\vect{g}(\vect{q}) = \frac{v_1 \vect{g}(\vect{q}_1) + v_2 \vect{g}(\vect{q}_2)}{v_1 + v_2}.\]
	Let $\vect{w}_1, \vect{w}_2 \in H_n(1)$ be such that $\vect{q}_1 = \vect{p}(\vect{w}_1)$ and $\vect{q}_2 = \vect{p}(\vect{w}_2)$. It is intuitive that $\vect{q} = \vect{p} \parens{\frac{v_1 \vect{w_1} + v_2 \vect{w_2}}{v_1 + v_2}}$, but we show this formally.
	
	\begin{claim} \label{claim:p_linear}
		Given $\vect{q}_1, \vect{q}_2 \in \mathcal{D}$ with $\vect{q}_1 = \vect{p}(\vect{w}_1), \vect{q}_2 = \vect{p}(\vect{w}_2)$, and $0 \le \alpha \le 1$, we have
		\[\vect{p}(\alpha \vect{w}_1 + (1 - \alpha) \vect{w}_2) = (\vect{q}_1, \alpha) \oplus (\vect{q}_2, 1 - \alpha).\]
	\end{claim}
	
	\begin{proof}
		Note that
		\begin{align*}
			(\vect{q}_1, 1) \oplus \parens{\bigoplus_{i: w_{1, i} < 0} (\vect{p}_i, w_{1, i})} &= \bigoplus_{i: w_{1, i} > 0} (\vect{p}_i, w_{1, i})\\
			(\vect{q}_2, 1) \oplus \parens{\bigoplus_{i: w_{2, i} < 0} (\vect{p}_i, w_{2, i})} &= \bigoplus_{i: w_{2, i} > 0} (\vect{p}_i, w_{2, i}).
		\end{align*}
		Applying the distributive property to the two above equations with constants $\alpha$ and $1 - \alpha$, respectively, and adding them, we get that
		\begin{align*}
			&\parens{\vect{q}_1, \alpha} \oplus \parens{\vect{q}_2, 1 - \alpha} \oplus \parens{\bigoplus_{i: w_{1, i} < 0} (\vect{p}_i, \alpha w_{1, i})} \oplus \parens{\bigoplus_{i: w_{2, i} < 0} (\vect{p}_i, (1 - \alpha) w_{2, i})}
			\\&= \parens{\bigoplus_{i: w_{1, i} > 0} (\vect{p}_i, \alpha w_{1, i})} \oplus \parens{\bigoplus_{i: w_{2, i} > 0} (\vect{p}_i, (1 - \alpha) w_{2, i})}.
		\end{align*}
		We have that $\parens{\vect{q}_1, \alpha} \oplus \parens{\vect{q}_2, 1 - \alpha} = (\vect{q}, 1)$. It follows (after rearranging terms, from Definition~\ref{def:p}) that $\vect{q} = \vect{p} \parens{\frac{v_1 \vect{w_1} + v_2 \vect{w_2}}{v_1 + v_2}}$.
	\end{proof}
	
	Applying Claim~\ref{claim:p_linear} with $\alpha = \frac{v_1}{v_1 + v_2}$, we find that
	\[\vect{g}(\vect{q}) = \frac{v_1 \vect{w_1} + v_2 \vect{w_2}}{v_1 + v_2} - \frac{1}{n} \vect{1} = \frac{v_1 \parens{\vect{g}(\vect{q}_1) + \frac{1}{n} \vect{1}} + v_2\parens{\vect{g}(\vect{q}_2) + \frac{1}{n} \vect{1}}}{v_1 + v_2} - \frac{1}{n} \vect{1} = \frac{v_1 \vect{g}(\vect{q}_1) + v_2 \vect{g}(\vect{q}_2)}{v_1 + v_2},\]
	as desired.\\
	
	It remains to show that $\vect{g}$ is continuous, strictly cyclically monotone, and has convex range. By the monotonicity axiom, $\vect{p}$ is strictly cyclically monotone. It follows by Proposition~\ref{prop:g_inverse} that its inverse its continuous and strictly cyclically monotone. Therefore, $\vect{g}$ is continuous and cyclically monotone (as it is simply a translation of $\vect{p}^{-1}(\vect{q})$ by $\frac{1}{n} \vect{1}$).
	
	Finally, to show that $\vect{g}$ has convex range, we wish to show that $\vect{p}^{-1}$ has convex range; or, in other words, that the domain on which $\vect{p}$ is defined is convex. And indeed, this follows straightforwardly from Claim~\ref{claim:p_linear}. Let $\vect{w}_1, \vect{w}_2$ be in the domain of $\vect{p}$, with $\vect{p}(\vect{w}_1) = \vect{q}_1, \vect{p}(\vect{w}_2) = \vect{q}_2$. Then for any $0 \le \alpha \le 1$, we have that
	\[\vect{p}(\alpha \vect{w}_1 + (1 - \alpha) \vect{w}_2) = (\vect{q}_1, \alpha) \oplus (\vect{q}_2, 1 - \alpha),\]
	so in particular $\alpha \vect{w}_1 + (1 - \alpha) \vect{w}_2$ is in the domain of $\vect{p}$. This concludes the proof.
\end{proof}

\section{The convex exposure property} \label{appx:convex_exposure}
Several of our results have been contingent on the convex exposure property. In this e-companion, we consider when the convex exposure property holds. Our first result is that it always holds in the case of a binary outcome (i.e.\ $n = 2$).

\begin{prop}\leavevmode \label{prop:n2_convex_exposure}
	If $n = 2$, every (continuous) proper scoring rule has convex exposure.
\end{prop}

\begin{proof}
	Consider a proper scoring rule $s$ with forecast domain $\mathcal{D}$. Since $\mathcal{D}$ is connected and $\vect{g}$ is continuous on $\mathcal{D}$, the range of $\vect{g}$ over $\mathcal{D}$ is connected. In the $n = 2$ outcome case, the range of $\vect{g}$ lies on the line $\{(x_1, x_2): x_1 + x_2 = 0\}$, and a connected subset of a line is convex.
\end{proof}

As we shall see, the convex exposure property holds for nearly all of the most commonly used scoring rules even in higher dimensions.

(A note on notation: in this section we use $p_j$ instead of $p(j)$ to refer to the $j$-th coordinate of a probability distribution $\vect{p}$.)\\

We now show that scoring rules that -- like the logarithmic scoring rule -- ``go off to infinity'' have convex exposure.

\begin{prop} \label{prop:steep_convex}
Let $s$ be a proper scoring rule whose forecast domain is the interior of $\Delta_n$, such that for any point $\vect{x}$ on the boundary of $\Delta_n$, and for any sequence $\vect{x}_1, \vect{x}_2, \dots$ converging to $\vect{x}$, $\lim_{k \to \infty} \norm{\vect{g}(\vect{x}_k)}_2 = \infty$.\footnote{\textcite{acv13} say that $G$ is a \emph{pseudo-barrier function} if this condition is satisfied.} Then $s$ has convex exposure.
\end{prop}

This is a statement of convex analysis -- namely that if $\norm{\vect{g}}_2$ approaches $\infty$ on the boundary of a convex set, then the range of $\vect{g}$ is convex (assuming $\vect{g}$ is the gradient of a differentiable convex function). See \textcite[Theorem 26.5]{roc70_book} for the proof. In non-pathological cases, the basic intuition is that \emph{every} $\vect{v} \in \{\vect{x}: \sum_i x_i = 0\}$ is the gradient of $G$ at some point. In these cases, $\nabla G(\vect{x}) = \vect{v}$ where $\vect{x}$ minimizes $G(\vect{v}) - \angles{\vect{v}, \vect{x}}$; the $\lim_{k \to \infty} \norm{g(\vect{x}_k)}_2 = \infty$ condition means that this minimum does not occur on the boundary of $\Delta_n$.

\begin{corollary} \label{cor:steep}
The following scoring rules have convex exposure:
\begin{itemize}
\item The logarithmic scoring rule.
\item The scoring rule given by $G(\vect{p}) = -\sum_j p_j^{\gamma}$ for $\gamma \in (0, 1)$.
\item The scoring rule given by $G(\vect{p}) = -\sum_j \ln p_j$, which can be thought of as the limit of the $G$ in the previous bullet point as $\gamma \to 0$.\footnote{This is a natural way to think of this scoring rule because $\nabla G(\vect{p}) = -(p_1^{-1}, \dots, p_n^{-1})$.}
\item The scoring rule $hs$ given by $G_{hs}(\vect{p}) = -\prod_j p_j^{1/n}$.
\end{itemize}
\end{corollary}

\paragraph{\textbf{The $hs$ scoring rule}} The last of these scoring rules is a generalization of the scoring rule $hs(q) = 1 - \sqrt{\frac{1 - q}{q}}$ used by \textcite{bb20} as part of proving their minimax theorem for randomized algorithms.\footnote{Here we are using the shorthand notation for the $n = 2$ outcome case discussed in Remark~3.5 of the main article.} The authors used this scoring rule as a key ingredient in their minimax theorem for randomized algorithms. The key property of the scoring rule was a result about its \emph{amplification} \parencite[Lemma 3.10]{bb20}. The authors define a \emph{forecasting algorithm} to be a generalization of a randomized algorithm that outputs an estimated \emph{probability} that an output should be accepted. Then, roughly speaking, the authors show that given a forecasting algorithm $R$, it is possible to create a forecasting algorithm $R'$ that has a much larger expected score from the scoring rule $hs$ by combining running $R$ a small number of times and combining the outputs. This is an important new result in theoretical computer science and suggests that $hs$ deserves more attention.

Since additive and multiplicative constants are irrelevant, we may treat $hs(q) = -\frac{1}{2} \sqrt{\frac{1 - q}{q}}$. Observe that (in the case of two outcomes), the expected score $G_{hs}$ on a report of $q$ is
\[G_{hs}(q) = q \parens{-\frac{1}{2} \sqrt{\frac{1 - q}{q}}} + (1 - q) \parens{-\frac{1}{2} \sqrt{\frac{1 - q}{q}}} = -\sqrt{q(1 - q)}.\]
That is, $G_{hs}$ is precisely negative the geometric mean of $q$ and $1 - q$. This motivates us to generalize $hs$ to a setting with $n$ outcomes by setting
\[G_{hs}(\vect{p}) := -\prod_{i = 1}^n p_i^{1/n}.\]
It should not be obvious that this function is convex, but it turns out to be; this is the precise statement of an inequality known as Mahler's inequality \parencite{wiki:mahler}.\\

Next we note that the quadratic scoring rule has convex exposure, since its exposure function $\vect{g}(\vect{p}) = 2\vect{p}$ (modulo $\vect{1}_n$ as discussed in Remark~3.12 of the main article) maps any convex set to a convex set.
\begin{prop}
The quadratic scoring rule has convex exposure.
\end{prop}

\emph{Spherical} scoring rules -- the third most studied proper scoring rules, after the quadratic and logarithmic rules -- also have convex exposure.
\begin{defin}[Spherical scoring rules]
\parencite[Example 2]{gr07}
For any $\alpha > 1$, define the \emph{spherical scoring rule with parameter $\alpha$} to be the scoring rule given by
\[G_{\text{sph}, \alpha}(\vect{p}) := \parens{\sum_{i = 1}^n p_i^{\alpha}}^{1/\alpha}.\]
If the ``spherical scoring rule'' is referenced with no parameter $\alpha$ given, $\alpha$ is presumed to equal $2$.
\end{defin}

\begin{prop} \label{prop:sph_convex}
For any $\alpha > 1$, the spherical scoring rule with parameter $\alpha$ has convex exposure.
\end{prop}

\begin{proof}
Fix $\alpha > 1$. We will write $G$ in place of $G_{\text{sph}, \alpha}$. We have\footnote{As discussed in Remark~3.12 of the main article, the range of $\vect{g}$ should be thought of as modulo $T(\vect{1}_n)$. However, we find it convenient for this proof to think of it as lying in $\RR^n$ and project later.}
\begin{equation} \label{eq:g_sph}
\vect{g}(\vect{p}) = \parens{\sum_{j = 1}^n p_j^{\alpha}}^{(1/\alpha) - 1} (p_1^{\alpha - 1}, \dots, p_n^{\alpha - 1}).
\end{equation}
Now, define the \emph{$n$-dimensional unit $\beta$-sphere}, i.e.\ $\{\vect{x}: \sum_j x_j^{\beta} = 1\}$, and define the \emph{$n$-dimensional unit $\beta$-ball} correspondingly (i.e.\ with $\le$ in place of $=$). The range of $\vect{g}$ is precisely the part of the $n$-dimensional unit $\frac{\alpha}{\alpha - 1}$-sphere with all non-negative coordinates. Indeed, on the one hand, for any $\vect{p}$ we have
\[\sum_j g_j(\vect{p})^{\alpha/(\alpha - 1)} = \parens{\sum_j p_j^\alpha}^{-1} \cdot \sum_j p_j^\alpha = 1\]
(where $g_j(\vect{p})$ denotes the $j$-th coordinate of $\vect{g}(\vect{p})$ as in Equation~\ref{eq:g_sph}). On the other hand, given a point $\vect{x}$ on the unit $\frac{\alpha}{\alpha - 1}$-sphere with all non-negative coordinates,
\[\vect{p} = \parens{\sum_j x_j^{1/(\alpha - 1)}}^{-1} \parens{x_1^{1/(\alpha - 1)}, \dots, x_n^{1/(\alpha - 1)}}\]
lies in $\Delta_n$ and satisfies $\vect{g}(\vect{p}) = \vect{x}$.

The crucial point for us is that for $\beta > 1$, the unit $\beta$-ball is convex. This means that for any such $\beta$, the convex combination of any number of points on the unit $\beta$-sphere will lie in the unit $\beta$-ball. Since $\frac{\alpha}{\alpha - 1} > 1$ for $\alpha > 1$, we have that for arbitrary $\vect{p}, \vect{q} \in \Delta_n$ and $w \in [0, 1]$, $w \vect{g}(\vect{p}) + (1 - w) \vect{g}(\vect{q})$ lies in the unit $\beta$-ball -- in fact, in the part with all non-negative coordinates. Now, consider casting a ray from this convex combination point in the positive $\vect{1}_n$ direction. All points on this ray are equivalent to this point modulo $T(\vect{1}_n)$, and this ray will intersect the unit $\beta$-sphere at some point $\vect{x}$ with all non-negative coordinates. The point $\vect{p} \in \Delta_n$ with $\vect{g}(\vect{p}) = \vect{x}$ satisfies
\[\vect{g}(\vect{p}) = \vect{g}(\vect{p}) + (1 - w) \vect{g}(\vect{q}).\]
This completes the proof.
\end{proof}

\begin{remark}
The above proof gives a geometric interpretation of the QA pooling with respect to the spherical scoring rule, particularly for $\alpha = 2$. In the $\alpha = 2$ case, pooling amounts to taking the following steps:
\begin{enumerate}[label=(\arabic*)]
\item Scale each forecast so it lies on the unit sphere.
\item Take the weighted average of the resulting points in $\RR^n$.
\item Shift the resulting point in the positive $\vect{1}_n$ direction to the unique point in that direction that lies on the unit sphere.
\item Scale this point so that its coordinates add to $1$.
\end{enumerate}
\end{remark}

Finally we consider the parametrized family known as \emph{Tsallis scoring rules} \parencite{tsa88}.
\begin{defin}[Tsallis scoring rules]
For $\gamma > 1$, the \emph{Tsallis scoring rule with parameter $\gamma$} is the rule given by
\[G_{\text{Tsa},\gamma}(\vect{p}) = \sum_{j = 1}^m p_j^\gamma.\]
\end{defin}
Setting $\gamma = 2$ above yields the quadratic scoring rule. Note also that we have already addressed the scoring rule given by $G(\vect{p}) = \pm \sum_j p_j^\gamma$ for $\gamma \le 1$ (except $\gamma = 0, 1$, which are degenerate), with the sign chosen to make $G$ convex: these scoring rules have convex exposure by Proposition~\ref{prop:steep_convex}. The following proposition completes our analysis for this natural class of scoring rules.

\begin{prop} \label{prop:tsallis}
For $\gamma \le 2$, the Tsallis scoring rule with parameter $\gamma$ has convex exposure. For $\gamma > 2$, this is not the case if $n > 2$.
\end{prop}

\begin{proof}
Fix $\gamma > 1$. We will write $G$ in place of $G_{\text{Tsa},\gamma}$. Up to a multiplicative factor of $\gamma$ that we are free to ignore, we have
\[\vect{g}(\vect{p}) = (p_1^{\gamma - 1}, \dots, p_n^{\gamma - 1}).\]
Let $\vect{p}, \vect{q} \in \Delta_n$ and $w \in [0, 1]$. We wish to find an $\vect{x} \in \Delta_n$ such that $\vect{g}(\vect{x}) = w \vect{g}(\vect{p}) + (1 - w) \vect{g}(\vect{q})$, i.e.
\[w p_j^{\gamma - 1} + (1 - w) q_j^{\gamma - 1} + c = x_j^{\gamma - 1},\]
for all $j \in [n]$, for some $c$. Since $\sum_j x_j = 1$, this $c$ must satisfy
\begin{equation} \label{eq:tsallis_requirement}
\sum_j (w p_j^{\gamma - 1} + (1 - w) q_j^{\gamma - 1} + c)^{1/(\gamma - 1)} = 1.
\end{equation}
Let $h(x) := \sum_j (w p_j^{\gamma - 1} + (1 - w) q_j^{\gamma - 1} + x)^{1/(\gamma - 1)}$. Note that $h$ is increasing in $x$.

First consider the case that $\gamma \le 2$. By concavity, we have that $w p_j^{\gamma - 1} + (1 - w) q_j^{\gamma - 1} \le (wp_j + (1 - w)q_j)^{\gamma - 1}$. This means that
\[h(0) = \sum_j (w p_j^{\gamma - 1} + (1 - w) q_j^{\gamma - 1})^{1/(\gamma - 1)} \le \sum_j (wp_j + (1 - w)q_j) = 1.\]
On the other hand, $\lim_{x \to \infty} h(x) = \infty$. Since $h$ is continuous, there must be some $x \in [0, \infty)$ such that $h(x) = 1$; call this value $c$. Then let
\[x_j = (w p_j^{\gamma - 1} + (1 - w) q_j^{\gamma - 1} + c)^{1/(\gamma - 1)}.\]
Then every $x_j$ is nonnegative and $\sum_j x_j = 1$, so we have succeeded.

Now consider the case that $\gamma > 2$, and consider as a counterexample $\vect{p} = (1, 0, \dots, 0)$, $\vect{q} = (0, 1, 0, \dots, 0)$, and $w = \frac{1}{2}$. To satisfy Equation~\ref{eq:tsallis_requirement}, we are looking for $c$ such that
\[h(c) = 2 \parens{\frac{1}{2} + c}^{1/(\gamma - 1)} + (n - 2)c^{1/(\gamma - 1)} = 1.\]
Note that $h(0) = 2 \cdot 2^{-1/(\gamma - 1)} = 2^{(\gamma - 2)/(\gamma - 1)} > 1$, so $c < 0$ (as $h$ is increasing). But in that case $x_j^{\gamma - 1} < 0$ for any $j \ge 3$, a contradiction (assuming $n > 2$).
\end{proof}

Note that because $\nabla G_{\text{Tsa}, \gamma}(\vect{p}) = (p_1^{\gamma - 1}, \dots, p_n^{\gamma - 1})$ (up to a constant factor), QA pooling with respect to the Tsallis scoring rule  can be thought of as an appropriately scaled coordinate-wise $(\gamma - 1)$-power mean. For $\gamma = 2$ it is the coordinate-wise arithmetic average. For $\gamma = 3$ it is the coordinate-wise root mean square, but with the average of the squares scaled by an appropriate additive constant so that, upon taking the square roots, the probabilities add to $1$. (However, as the Tsallis score with parameter $3$ does not have convex exposure, this is not always well-defined.)

In Corollary~\ref{cor:steep} we mentioned that the scoring rule given by $G(\vect{p}) = -\sum_j \ln p_j$ can be thought of as an extension to $\gamma = 0$ of (what we are now calling) the Tsallis score, because the derivative of $\ln x$ is $x^{-1}$. QA pooling with respect to this scoring rule is, correspondingly, the $-1$-power mean, i.e.\ the harmonic mean. This pooling method is appropriately referred to as \emph{harmonic pooling}, see e.g.\ \textcite[\S4.2]{ddmc95}.

Finally, we note that the logarithmic scoring rule can likewise be thought of as an extension of the Tsallis score to $\gamma = 1$, in that the second derivative of $x \ln x$ is $x^{-1}$. It is likewise natural to call the geometric mean the $0$-power mean; notice that logarithmic pooling is precisely an appropriately scaled coordinate-wise geometric mean.
\chapter{Details omitted from Chapter~\ref{chap:learning}} \label{appx:chap6}
\section{Efficiency of Algorithm~\ref{alg:omd}}
The only nontrivial step of the algorithm is finding the weight vector satisfying the equation on the last line of the algorithm. To do so, it is first necessary to compute the gradient of the loss. This gradient, given by Equation~\ref{eq:partl_l_expr} below, can clearly be computed in time $O(mn)$. After that, it is necessary to find the weight vector $\vect{w}^{t + 1}$ that satisfies the equation on the last line. This can be done efficiently through local search: the goal amounts to find weights $(w_1, \dots, w_m)$ such that the vector $(w_1^{\alpha - 1}, \dots, w_m^{\alpha - 1})$ is equal to a target vector (call it $\vect{v}$) plus a constant $c$ times the all-ones vector. That is, we need to simultaneously solve the equation $w_i^{\alpha - 1} = v_i + c$ for all $i$, with weights that add to $1$. (Here, the $v_i$ are knowns and the $w_i$ and $c$ are unknowns.)

We start by finding $c$, by solving the equation $\sum_i (v_i + c)^{1/(\alpha - 1)} = 1$. Such a $c$ exists because the left-hand side of this equation is continuous and monotone decreasing, going from infinity to zero as $c$ ranges from $-\min_i v_i$ to infinity. We can solve for $c$ very efficiently, e.g.\ with Newton's method. Once we know $c$, we know each $w_i$: we have $w_i = (v_i + c)^{1/(\alpha - 1)}$. Thus, Algorithm 1 takes $O(mn)$ time.

\section{Details omitted from Section~\ref{sec:proof}}
\omdwibound*

\begin{proof}
	Fix any $t$. Note that since the space of possible weights is $\Delta_m$, it is most natural to think of $\nabla R$ as a function from $\Delta_m$ to $\RR^m/T(\vect{1}_m)$, i.e.\ $\RR^m$ modulo translation by the all-ones vector (which is orthogonal to $\Delta_m$ in $\RR^m$). That is, $\nabla R(\vect{w}) = -((w_1)^{\alpha - 1}, \dots, (w_m)^{\alpha - 1})$, where this vector may be thought of as modulo translation by the all-ones vector. Nevertheless, we find it convenient to define $\partial_i R(\vect{w}) := -(w_i)^{\alpha - 1}$. We define $\partial_i L^t(\vect{w})$ similarly (see Section~\ref{sec:finishing_up}).
	
	Define $\vect{h} \in \RR^m$ to have coordinates $h_i := \partial_i R(\vect{w}^t) - \eta_t \partial_i L^t(\vect{w}^t)$. Per the update rule, we have that $h_i \equiv R(\vect{w}^{t + 1}) \mod T(\vect{1}_m)$. We have
	
	\begin{equation} \label{eq:hi_lb}
		-(w_i^t)^{\alpha - 1} - \eta_t \zeta = \partial_i R(\vect{w}^t) - \eta_t \zeta \le h_i \le \partial_i R(\vect{w}^t) + \frac{\eta_t \zeta}{w_i^t} = -(w_i^t)^{\alpha - 1} + \frac{\eta_t \zeta}{w_i^t}
	\end{equation}
	Applying the first and last claims of Lemma~\ref{lem:renormalize_bound} (below) with $a = \alpha - 1$, $\vect{v} = \vect{w}^t$, $\kappa = \eta_t \zeta$, and $\vect{g} = -\vect{h}$, we have that there exists a unique $c \in \RR$ such that
	\[\sum_{i = 1}^m (-h_i + c)^{1/(\alpha - 1)} = 1,\]
	and in fact that $-\eta_t \zeta \le c \le m \eta_t \zeta$. (Equation~\ref{eq:hi_lb} is relevant here because it is equivalent to the $v_i^a - \frac{\kappa}{w_i^t} \le g_i \le v_i^a + \kappa$ conditions in Lemma~\ref{lem:renormalize_bound}. This is also where we use that $\eta_t \zeta \le (1 - \alpha)^2 (w_i^t)^\alpha$, which is equivalent to $\kappa \le a^2 v_i^{a + 1}$.) The significance of this fact is that $(-h_i + c)^{1/(\alpha - 1)}$ is precisely $w_i^{t + 1}$, since (in $\RR^m$) we have that $\parens{\partial_i R(\vect{w}^{t + 1}), \dots, \partial_i R(\vect{w}^{t + 1})} = \vect{h} - c \cdot \vect{1}$ for \emph{some} $c$, and in particular this $c$ must be such that $\sum_i w_i^{t + 1} = 1$. In particular, this means that for all $i$, we have
	\[(w_i^{t + 1})^{\alpha - 1} = -h_i + c \le (w_i^t)^{\alpha - 1} + \eta_t \zeta + \frac{1}{\min_k w_k} \eta_t \zeta = (w_i^t)^{\alpha - 1} + \parens{\frac{1}{\min_k w_k} + 1} \eta_t\zeta.\]
	Here, the inequality comes from the left inequality of Equation~\ref{eq:hi_lb} and the fact that $c \le \frac{1}{\min_k w_k} \eta_t \zeta$. If we also have that $\eta_t \zeta \le (1 - \alpha)^2 (w_i^t)^\alpha$, then the last claim of Lemma~\ref{lem:renormalize_bound} gives us that
	\[(w_i^{t + 1})^{\alpha - 1} = -h_i + c \le (w_i^t)^{\alpha - 1} + \eta_t \zeta + m \eta_t \zeta = (w_i^t)^{\alpha - 1} + (m + 1)\eta_t \zeta.\]
	Similarly, we have
	\[(w_i^{t + 1})^{\alpha - 1} = -h_i + c \ge (w_i^t)^{\alpha - 1} - \frac{\eta_t \zeta}{w_i^t} - \eta_t \zeta = (w_i^t)^{\alpha - 1} - \parens{\frac{1}{w_i^t} + 1} \eta_t \zeta.\]
\end{proof}

\begin{lemma} \label{lem:renormalize_bound}
	Let $-1 < a < 0$ and $\vect{g} \in \RR^m$. There is a unique $c \in \RR$ such that $\sum_i (g_i + c)^{1/a} = 1$. Furthermore, let $\vect{v} \in \Delta_m$ and $\kappa \ge 0$. Then:
	\begin{itemize}
		\item If $g_i \le v_i^a + \kappa$ for all $i$, then $c \ge -\kappa$.
		\item If $g_i \ge v_i^a - \frac{\kappa}{v_i}$ for all $i$, then $c \le \frac{\kappa}{\min_i v_i}$.
		\begin{itemize}
			\item And if, furthermore, $\kappa \le a^2 v_i^{a + 1}$ for all $i$, then $c \le m\kappa$.
		\end{itemize}
	\end{itemize}
\end{lemma}

\begin{proof}
	Observe that $\sum_i (g_i + c)^{1/a}$ is a continuous, monotone decreasing function on $c \in (-\min_i g_i, \infty)$; the range of the function on this interval is $(0, \infty)$. Therefore, there is a unique $c \in (-\min_i g_i, \infty)$ such that the sum equals $1$.
	
	We now prove the first bullet. Since $x^{1/a}$ decreases in $x$ and $g_i \le v_i^a + \kappa$, we have that
	\[1 = \sum_i (g_i + c)^{1/a} \ge \sum_i (v_i^a + \kappa + c)^{1/a}.\]
	Suppose for contradiction that $c < -\kappa$. Then $v_i^a + \kappa + c < v_i^a$ for all $i$, so
	\[\sum_i (v_i^a + \kappa + c)^{1/a} > \sum_i (v_i^a)^{1/a} = \sum_i v_i = 1.\]
	This is a contradiction, so in fact $c \ge -\kappa$.
	
	The first claim of the second bullet is analogous. Since $x^{1/a}$ decreases in $x$ and $g_i \ge v_i^a - {\kappa}{v_i}$, we have that
	\begin{equation} \label{eq:first_ub_ineq}
		1 = \sum_i (g_i + c)^{1/a} \le \sum_i \parens{v_i^a - \frac{\kappa}{v_i} + c}^{1/a}.
	\end{equation}
	Suppose for contradiction that $c > \frac{\kappa}{v_i}$ for every $i$. Then $v_i^a - \frac{\kappa}{v_i} + c > v_i^a$ for all $i$, so
	\[\sum_i \parens{v_i^a - \frac{\kappa}{v_i} + c}^{1/a} < \sum_i (v_i^a)^{1/a} = \sum_i v_i = 1.\]
	This is a contradiction, so in fact $c \le \frac{\kappa}{\min_i v_i}$.
	
	We now prove the second claim of the second bullet. To do so, we note the following technical lemma (proof below).
	\begin{lemma} \label{lemma:renormalize_technical}
		For $-1 < a < 0$ and $\kappa, c \ge 0$, the function $f(x) = \parens{x^a - \frac{\kappa}{x} + c}^{1/a}$ is defined and concave at any value of $x > 0$ such that $a^2 x^{a + 1} \ge \kappa$.
	\end{lemma}
	
	Since for a general concave function $f$ it holds that $\frac{1}{m} \sum_{i = 1}^m f(x_i) \le f \parens{\frac{1}{m} \sum_{i = 1}^m x_i}$, the following inequality follows from Lemma~\ref{lemma:renormalize_technical}:
	\[\sum_i \parens{v_i^a - \frac{\kappa}{v_i} + c}^{1/a} \le m \parens{\parens{\frac{1}{m}}^a - \kappa m + c}^{1/a}.\]
	(Here we are using the fact that $\sum_i v_i = 1$.) Now, combining this fact with Equation~\ref{eq:first_ub_ineq}, we have that
	\begin{align*}
		m \parens{\parens{\frac{1}{m}}^a - \kappa m + c}^{1/a} &\ge 1\\
		\parens{\frac{1}{m}}^a - \kappa m + c &\le \parens{\frac{1}{m}}^a
	\end{align*}
	so $c \le m\kappa$, as desired.
\end{proof}

\begin{proof}[Proof of Lemma~\ref{lemma:renormalize_technical}]
	To show that $f$ is defined for any $x$ such that $a^2 x^{a + 1} \ge \kappa$, we need to show that $x^a - \frac{\kappa}{x} + c > 0$ for such values of $x$. This is indeed the case:
	\[x^a - \frac{\kappa}{x} + c \ge x^a - a^2 x^a + c = (1 - a^2) x^a + c > c \ge 0.\]
	
	Now we show concavity. We have
	
	\footnotesize
	\[f''(x) = \frac{1}{-a} \parens{\parens{1 + \frac{1}{-a}} \parens{x^a - \frac{\kappa}{x} + c}^{1/a - 2} \parens{ax^{a - 1} + \frac{\kappa}{x^2}}^2 - \parens{x^a - \frac{\kappa}{x} + c}^{1/a - 1} \parens{a(a - 1) x^{a - 2} - \frac{2\kappa}{x^3}}}\]
	
	\normalsize
	so we wish to show that
	\[\parens{1 + \frac{1}{-a}} \parens{x^a - \frac{\kappa}{x} + c}^{1/a - 2} \parens{ax^{a - 1} + \frac{\kappa}{x^2}}^2 \le \parens{x^a - \frac{\kappa}{x} + c}^{1/a - 1} \parens{a(a - 1) x^{a - 2} - \frac{2\kappa}{x^3}}\]
	for every $x$ such that $a^2 x^{a + 1} \ge \kappa$. Fix any such $x$, and let $d = \frac{\kappa}{x^{a + 1}}$ (so $0 \le d \le a^2$). We have
	\begin{align*}
		d &\le a^2\\
		(1 + a)(a^2 - d)d &\ge 0\\
		(1 - a)(a + d)^2 &\le -a(1 - d)(a(a - 1) - 2d) & \text{(rearrange terms)}\\
		\parens{1 - \frac{1}{a}}(a + d)^2 x^a &\le ((1 - d) x^a)(a(a - 1) - 2d) & \text{(multiply by $\frac{x^a}{-a}$)}\\
		\parens{1 - \frac{1}{a}}(a + d)^2 x^a &\le ((1 - d) x^a + c)(a(a - 1) - 2d) & \text{($c(a(a - 1) - 2d) \ge 0$)}\\
		\parens{1 - \frac{1}{a}}((a + d) x^{a - 1})^2 &\le ((1 - d) x^a + c)(a(a - 1) - 2d) x^{a - 2} &\text{(multiply by $x^{a - 2}$)}\\
		\parens{1 - \frac{1}{a}} \parens{ax^{a - 1} + \frac{\kappa}{x^2}}^2 &\le \parens{x^a - \frac{\kappa}{x} + c} \parens{a(a - 1)x^{a - 2} - \frac{2\kappa}{x^3}}. & \text{(substitute $d = \kappa x^{-a - 1}$)}
	\end{align*}
	Note that the fifth line is justified by the fact that $c \ge 0$ and $a(a - 1) \ge 2d$ (because $a^2 \ge d$ and $-a > a^2 \ge d$). Now, multiplying both sides by $\parens{x^a - \frac{\kappa}{x} + c}^{1/a - 2}$ completes the proof.
\end{proof}

\boundscor*

\begin{proof}
	Note that $\eta \gamma = \frac{1}{\sqrt{T} m^{(1 + \alpha)/2}}$ and also that $\eta_t \le \eta$ for all $t$; we will be using these facts.
	
	To prove (\#1), we proceed by induction on $t$. In the case of $t = 1$, all weights are $1/m$, so the claim holds for sufficiently large $T$. Now assume that the claim holds for a generic $t < T$; we show it for $t + 1$.
	
	By the small gradient assumption, we may use Lemma~\ref{lem:omd_wi_bound} with $\zeta = \gamma$. By the inductive hypothesis (and the fact that $\eta_t \le \eta$), we may apply the second part of Lemma~\ref{lem:omd_wi_bound}:
	\begin{align*}
		(w_i^{t + 1})^{\alpha - 1} &\le (w_i^t)^{\alpha - 1} + (m + 1) \eta \gamma \le \dots \le (1/m)^{\alpha - 1} + t(m + 1) \eta \gamma.\\
		&\le (1/m)^{\alpha - 1} + \frac{(T - 1)(m + 1)}{m^{(1 + \alpha)/2} \sqrt{T}} \le 3m^{(1 - \alpha)/2} \sqrt{T}.
	\end{align*}
	Since $\frac{-1}{2} < \alpha - 1 < 0$, this means that $w_i^t \ge \frac{1}{10\sqrt{m}} T^{1/(2(\alpha - 1))}$.
	
	We also have that
	\[(w_i^{t + 1})^\alpha \ge \frac{1}{(10\sqrt{m})^\alpha} T^{\alpha/(2(\alpha - 1))} \ge \frac{4}{m^{(1 + \alpha)/2}} T^{-1/2} = 4\eta \gamma\]
	for $T$ sufficiently large, since $\frac{\alpha}{2(\alpha - 1)} > \frac{-1}{2}$. This completes the inductive step, and thus the proof of (\#1).\\
	
	To prove (\#2), we use the following technical lemma (see below for the proof).
	\begin{lemma} \label{lemma:fy}
		Fix $x > 0$ and $-1 < a < 0$. Let $f(y) = (x^a + y)^{1/a}$. Then for all $y > -x^a$, we have
		\begin{equation} \label{eq:fy_1}
			x - f(y) \le \frac{-1}{a} x^{1 - a}y
		\end{equation}
		and for all $-1 < c \le 0$, for all $c x^a \le y \le 0$, we have
		\begin{equation} \label{eq:fy_2}
			f(y) - x \le \frac{1}{a} (1 + c)^{1/a - 1} x^{1 - a}y.
		\end{equation}
	\end{lemma}
	
	We apply Equation~\ref{eq:fy_1} to $x = w_i^t$, $y = (m + 1) \eta \gamma$, and $a = \alpha - 1$. This tells us that
	\[w_i^t - w_i^{t + 1} \le w_i^t - ((w_i^t)^{\alpha - 1} + (m + 1) \eta \gamma)^{1/(\alpha - 1)} \le 2 (w_i^t)^{2 - \alpha} (m + 1) \eta \gamma.\]
	The first step follows by the second part of Lemma~\ref{lem:omd_wi_bound} and the fact that $\eta_t \le \eta$. The second step follows from Equation~\ref{eq:fy_1} and uses the fact that $\frac{1}{1 - \alpha} > 2$.
	
	For the other side of (\#2), we observe that since by (\#1) we have $(w_i^t)^\alpha \ge 4\eta \gamma$, it follows that $\frac{1}{2} (w_i^t)^\alpha \ge (w_i^t + 1) \eta \gamma$, and so $\parens{\frac{1}{w_i^t} + 1} \eta \gamma \le \frac{1}{2} (w_i^t)^{\alpha - 1}$. Therefore, we can apply Equation~\ref{eq:fy_2} to $x = w_i^t$, $y = -\parens{\frac{1}{w_i^t} + 1} \eta \gamma$, $a = \alpha - 1$, and $c = -\frac{1}{2}$. This tells us that
	\begin{align*}
		w_i^{t + 1} - w_i^t &\le \parens{(w_i^t)^{\alpha - 1} - \parens{\frac{1}{w_i^t} + 1} \eta \gamma}^{1/(\alpha - 1)} - w_i^t \le 16 (w_i^t)^{2 - \alpha} \parens{\frac{1}{w_i^t} + 1} \eta \gamma\\
		&\le 32 (w_i^t)^{1 - \alpha} \eta \gamma.
	\end{align*}
	This completes the proof.
\end{proof}

\begin{proof}[Proof of Lemma~\ref{lemma:fy}]
	For all $y > -x^a$, we have
	\[f'(y) = \frac{1}{a} (x^a + y)^{1/a - 1}\]
	and
	\[f''(y) = \frac{1}{a} \parens{\frac{1}{a} - 1} (x^a + y)^{1/a - 2} > 0,\]
	so $f'$ is increasing. Thus, for positive values of $y$ we have
	\[f'(0) \le \frac{f(y) - f(0)}{y} = \frac{f(y) - x}{y} \le f'(y)\]
	and for negative values of $y$ we have
	\[f'(y) \le \frac{f(y) - f(0)}{y} = \frac{f(y) - x}{y} \le f'(0).\]
	Regardless of whether $y$ is positive or negative, this means that $x - f(y) \le -yf'(0) = \frac{-1}{a} x^{1 - a}y$.
	
	Now, let $-1 < c \le 0$ and suppose that $c x^a \le y \le 0$. Since $f'$ is increasing, we have that
	\[f'(y) \ge f'(c x^a) = \frac{1}{a} ((1 + c)x^a)^{1/a - 1} = \frac{1}{a} (1 + c)^{1/a - 1} x^{1 - a},\]
	so
	\[f(y) - x \le y f'(y) \le \frac{1}{a} (1 + c)^{1/a - 1} x^{1 - a}y.\]
\end{proof}

\lnotsmall*

\begin{proof}
	We first derive an expression for $\partial_i L(\vect{w})$ given expert reports $\vect{p}^1, \dots, \vect{p}^m$, where $L(\vect{w})$ is the log loss of the logarithmic pool $\vect{p}^*(\vect{w})$ of $\vect{p}^1, \dots, \vect{p}^m$ with weights $\vect{w}$, and $j$ is the realized outcome. We have\footnote{It should be noted that $\nabla L(\vect{w})$ is most naturally thought of as living in $\RR^m / T(\vect{1}_m)$, i.e.\ $m$-dimensional space modulo translation by the all-ones vector, since $\vect{w}$ lives in a place that is orthogonal to the all-ones vector. As an arbitrary but convenient convention, we define $\partial_i L(\vect{w})$ to be the specific value derived below, and define the small gradient assumption accordingly.}
	\begin{align} \label{eq:partl_l_expr}
		\partial_i L(\vect{w}) &= -\partial_i \ln \frac{\prod_{k = 1}^m (p^k_j)^{w_k}}{\sum_{\ell = 1}^n \prod_{k = 1}^m (p^k_\ell)^{w_k}} = \partial_i \ln \parens{\sum_{\ell = 1}^n \prod_{k = 1}^m (p^k_\ell)^{w_k}} - \partial_i \ln \parens{\prod_{k = 1}^m (p^k_j)^{w_k}} \nonumber\\
		&= \frac{\sum_{\ell = 1}^n \ln p^i_\ell \cdot \prod_{k = 1}^m (p^k_\ell)^{w_k}}{\sum_{\ell = 1}^n \prod_{k = 1}^m (p^k_\ell)^{w_k}} - \ln p^i_j = \sum_{\ell = 1}^n p^*_\ell(\vect{w}) \ln p^i_\ell - \ln p^i_j.
	\end{align}
	
	Equation~\ref{eq:partl_l_ub} now follows fairly straightforwardly. Equation~\ref{eq:partl_l_expr} tells us that $\partial_i L(\vect{w}) \le -\ln p^i_J$, where $J$ is the random variable corresponding to the realized outcome. Therefore, we have
	\begin{align*}
		\pr{\partial_i L(\vect{w}) \ge \zeta} &\le \pr{-\ln p_J^i \ge \zeta} = \pr{p_J^i \le e^{-\zeta}} = \sum_{j = 1}^n \pr{J = j \enskip \& \enskip p_j^i \le e^{-\zeta}}\\
		&= \sum_{j = 1}^n \pr{p_j^i \le e^{-\zeta}} \pr{J = j \mid p_j^i \le e^{-\zeta}} \le \sum_{j = 1}^n \pr{J = j \mid p_j^i \le e^{-\zeta}} \le ne^{-\zeta},
	\end{align*}
	where the last step follows by the calibration property. This proves Equation~\ref{eq:partl_l_ub}.
	
	We now prove Equation~\ref{eq:partl_l_lb}. The proof has a similar idea, but is somewhat more technical. We begin by proving the following lemma; we again use the calibration property in the proof.
	
	\begin{lemma} \label{lem:mnq}
		For all $q$, we have
		\[\pr{\forall j \exists i: p^i_j \le q} \le mnq.\]
	\end{lemma}
	
	\begin{proof}
		Let $J$ be the random variable corresponding to the index of the outcome that ends up happening. We have
		\begin{align*}
			\pr{\forall j \exists i: p^i_j \le q} &\le \pr{\exists i: p^i_J \le q} = \sum_{j \in [n]} \pr{J = j \enskip \& \enskip \exists i: p^i_j \le q}\\
			&\le \sum_{j \in [n]} \sum_{i \in [m]} \pr{J = j \enskip \&  \enskip p^i_j \le q}\\
			&= \sum_{j \in [n]} \sum_{i \in [m]} \pr{p^i_j \le q} \pr{J = j \mid p^i_j \le q} \le \sum_{j \in [n]} \sum_{i \in [m]} 1 \cdot q = mnq,
		\end{align*}
		where the fact that $\pr{J = j \mid p^i_j \le q} \le q$ follows by the calibration property.
	\end{proof}
	
	\begin{corollary} \label{cor:mnq}
		For any reports $\vect{p}^1, \dots, \vect{p}^m$, weight vector $\vect{w}$, $i \in [m]$, and $j \in [n]$, we have
		\[\pr{p^*_j(\vect{w}) \ge \frac{(p^i_j)^{w_i}}{q}} \le mnq.\]
	\end{corollary}
	
	\begin{proof}
		We have
		\[p^*_j(\vect{w}) = \frac{\prod_{k = 1}^m (p_j^k)^{w_k}}{\sum_{\ell = 1}^n \prod_{k = 1}^m (p_{\ell}^{k})^{w_{k}}} \le \frac{(p_j^i)^{w_i}}{\sum_{\ell = 1}^n \prod_{k = 1}^m (p_{\ell}^{k})^{w_{k}}}.\]
		Now, assuming that there is an $\ell$ such that for every $k$ we have $p^{k}_{\ell} > q$, the denominator is greater than $q$, in which case we have $p^*_j(\vect{w}) < \frac{(p^i_j)^{w_i}}{q}$. Therefore, if $p^*_j(\vect{w}) \ge \frac{(p^i_j)^{w_i}}{q}$, it follows that for every $\ell$ there is a $k$ such that $p^k_\ell \le q$. By Lemma~\ref{lem:mnq}, this happens with probability at most $mnq$.
	\end{proof}
	
	We now use Corollary~\ref{cor:mnq} to prove Equation~\ref{eq:partl_l_lb}. Note that the equation is trivial for $\zeta < n$, so we assume that $\zeta \ge n$. By setting $q := e^{\frac{-\zeta}{n}}$, we may restate Equation~\ref{eq:partl_l_lb} as follows: for any $q \le \frac{1}{e}$, any $i \in [m]$, and any weight vector $\vect{w}$,
	\[\pr{\partial_i L(\vect{w}) \le -\frac{n \ln 1/q}{w_i}} \le mn^2q.\]
	(Note that the condition $q \le \frac{1}{e}$ is equivalent to $\zeta \ge n$.) We prove this result.
	
	From Equation~\ref{eq:partl_l_expr}, we have
	\[\partial_i L(\vect{w}) = \sum_{j = 1}^n p^*_j(\vect{w}) \ln p^i_j - \ln p^i_j \ge \sum_{j = 1}^n p_j(\vect{w}) \ln p^i_j.\]
	Now, it suffices to show that for each $j \in [n]$, the probability that $p_j(\vect{w}) \ln p^i_j \le - \frac{\ln 1/q}{w_i} = \frac{\ln q}{w_i}$ is at most $mnq$; the desired result will then follow by the union bound. By Corollary~\ref{cor:mnq}, for each $j$ we have that
	\[\pr{p_j(\vect{w}) \ln p_j^i \le \frac{(p_j^i)^{w_i}}{q} \ln p_j^i} \le mnq.\]
	Additionally, we know for a fact that $p_j(\vect{w}) \ln p^i_j \ge \ln p^i_j$ (since $p_j(\vect{w}) \le 1$), so in fact
	\[\pr{p_j(\vect{w}) \ln p_j^i \le \max \parens{\frac{(p_j^i)^{w_i}}{q} \ln p_j^i, \ln p_j^i}} \le mnq.\]
	It remains only to show that $\max \parens{\frac{(p_j^i)^{w_i}}{q} \ln p_j^i, \ln p_j^i} \ge \frac{\ln q}{w_i}$. If $p_j^i \ge q^{1/w_i}$ then this is clearly true, since in that case $\ln p_j^i \ge \frac{\ln q}{w_i}$. Now suppose that $p_j^i < q^{1/w_i}$. Observe that $\frac{x^{w_i}}{q} \ln x$ decreases on $(0, e^{-1/w_i})$, and that (since $q \le \frac{1}{e}$) we have $q^{1/w_i} \le e^{-1/w_i}$. Therefore,
	\[\frac{(p_j^i)^{w_i}}{q} \ln p_j^i \le \frac{(q^{1/w_i})^{w_i}}{q} \ln q^{1/w_i} = \frac{\ln q}{w_i}.\]
	This completes the proof of Equation~\ref{eq:partl_l_lb}, and thus of Lemma~\ref{lem:l_not_small}.
\end{proof}

The following lemma lower bounds the regret of Algorithm~\ref{alg:omd} as a function of $\zeta$.

\begin{lemma} \label{lem:bad_case}
	Consider a run of Algorithm~\ref{alg:omd}. Let $\zeta$ be such that $-\frac{\zeta}{w_i^t} \le \partial_i L^t(\vect{w}^t) \le \zeta$ for all $i, t$. The total regret is at most
	\[O \parens{\zeta^{2(2 - \alpha)/(1 - \alpha)} T^{(5 - \alpha)/(1 - \alpha)}}.\]
\end{lemma}

\begin{proof}
	We first bound $w_i^t$ for all $i, t$. From Lemma~\ref{lem:omd_wi_bound}, we have that
	\[(w_i^{t + 1})^{\alpha - 1} \le (w_i^t)^{\alpha - 1} + \parens{\frac{1}{\min_i w_i^t} + 1} \eta_t \zeta \le (w_i^t)^{\alpha - 1} + 2\zeta.\]
	Here we use that $\frac{1}{\min_i w_i} + 1 \le \frac{2}{\min_i w_i}$ and that $\eta_t \le \min_i w_i$. Therefore, we have that
	\[(w_i^t)^{\alpha - 1} \le (w_i^{t - 1})^{\alpha - 1} + 2\zeta \le \dots \le m^{1 - \alpha} + 2\zeta(t - 1) \le m^{1 - \alpha} + 2\zeta T.\]
	Thus, $w_i^t \ge (m^{1 - \alpha} + 2 \zeta T)^{1/(\alpha - 1)} \ge \Omega((\zeta T)^{1/(\alpha - 1)})$ for all $i, t$.
	
	We now use the standard regret bound for online mirror descent, see e.g.\ \textcite[Theorem 6.8]{ora21}:
	
	\begin{equation} \label{eq:orabona_regret_bound}
		\text{Regret} \le \max_t \frac{B_R(\vect{u}; \vect{w}^t)}{\eta_T} + \frac{1}{2\lambda} \sum_{t = 1}^T \eta_t \norm{\nabla L^t(\vect{w}^t)}_*^2
	\end{equation}
	where $B_R(\cdot ; \cdot)$ is the Bregman divergence of with respect to $R$, $\vect{u}$ is the optimal (overall loss-minimizing) point, $\lambda$ is a constant such that $R$ is $\lambda$-strongly convex with respect to a norm of our choice over $\Delta_m$, and $\norm{\cdot}_*$ is the dual norm of the aforementioned norm.
	
	Note that for any $\vect{x} \in \Delta_m$,  we have
	\[\max_{\vect{v} \in \Delta_m} B_R(\vect{v}; \vect{x}) = \max_{\vect{v} \in \Delta_m} R(\vect{v}) - R(\vect{x}) - \angles{\nabla R(\vect{x}), \vect{v} - \vect{x}} \le \frac{m^{1 - \alpha}}{\alpha} + (\min_i x_i)^{\alpha - 1}.\]
	In the last step, we use the fact that $-\nabla R(\vect{x}) = (x_1^{\alpha - 1}, \dots, x_m^{\alpha - 1})$ (all of these coordinates are positive), so $-\angles{\nabla R(\vect{x}), \vect{v} - \vect{x}} \le \angles{(x_1^{\alpha - 1}, \dots, x_m^{\alpha - 1}), \vect{v}}$, and that all coordinates of $\vect{v}$ are non-negative and add to $1$.
	
	Therefore, given our bound on $w_i^t$, this first component of our regret bound (\ref{eq:orabona_regret_bound}) is at most
	\[\frac{1}{\eta_T} \parens{\frac{m^{1 - \alpha}}{\alpha} + m^{1 - \alpha} + 2 \zeta T} \le O \parens{\frac{\zeta T}{\eta_T}} \le O \parens{\frac{\zeta T}{(\zeta T)^{1/(\alpha - 1)}}} = O \parens{(\zeta T)^{(2 - \alpha)/(1 - \alpha)}}.\]
	
	To bound the second term, we choose to work with the $\ell_1$ norm. To show that $R$ is $\lambda$-convex it suffices to show that for all $\vect{x}, \vect{y} \in \Delta_m$ we have $\angles{(\nabla^2 R)(\vect{x}) \vect{y}, \vect{y}} \ge \lambda \norm{\vect{y}}^2$, where $\nabla^2 R$ is the Hessian matrix of $R$ (\textcite[Lemma 14]{ss07}; see also \textcite[Theorem 4.3]{ora21}). Equivalently, we wish to find a $\lambda$ such that
	\[(1 - \alpha) \sum_i x_i^{\alpha - 2} y_i^2 \ge \lambda.\]
	Since $x_i^{\alpha - 2} \ge 1$ for all $i$, the left-hand side is at least $(1 - \alpha) \sum_i y_i^2 \ge \frac{1 - \alpha}{m}$, so $\lambda = \frac{1 - \alpha}{m}$ suffices.
	
	Now, given $\vect{\theta} \in \RR^m$, we have $\norm{\vect{\theta}}_* = \max_{\vect{x}: \norm{\vect{x}} \le 1} \angles{\vect{\theta}, \vect{x}}$. In the case of the $\ell_1$ primal norm, the dual norm is the largest absolute component of $\vect{\theta}$. Thus, we have
	\[\norm{\nabla L^t(\vect{x}^t)}_* \le \frac{\zeta}{w_i^t} \le O \parens{\zeta(\zeta T)^{1/(1 - \alpha)}} = O \parens{\zeta^{(2 - \alpha)/(1 - \alpha)} T^{1/(1 - \alpha)}}.\]
	Since $\eta_t \le O(T^{-1/2})$, we have that the second component of our regret bound (\ref{eq:orabona_regret_bound}) is at most
	\[O \parens{T \cdot T^{-1/2} \cdot \zeta^{2(2 - \alpha)/(1 - \alpha)} T^{2/(1 - \alpha)}} \le O \parens{\zeta^{2(2 - \alpha)/(1 - \alpha)} T^{(5 - \alpha)/(1 - \alpha)}}.\]
	This component dominates our bound on the regret of the first component, in both $\zeta$ and $T$. This concludes the proof.
\end{proof}

\generalcase*
\begin{proof}
	Let $Z$ be the minimum value of $\zeta$ such that $-\frac{\zeta}{w_i^t} \le \partial_i L^t(\vect{w}^t) \le \zeta$ for all $i, t$. Note that by Lemma~\ref{lem:l_not_small}, we have that
	\[\pr{Z \ge x} \le \sum_{i = 1}^m \sum_{t = 1}^T (mn^2 e^{-\frac{x}{n}} + ne^{-x}) \le 2m^2 n^2 T e^{-\frac{x}{n}}.\]
	Let $\mu$ be the constant hidden in the big-O of Lemma~\ref{lem:bad_case}, i.e.\ a constant (dependent on $m$, $n$, and $\alpha$) such that
	\[\text{Regret} \le \mu Z^{2(2 - \alpha)/(1 - \alpha)} T^{(5 - \alpha)/(1 - \alpha)}.\]
	Let $r(Z, T)$ be the expression on the right-hand side. The small gradient assumption not holding is equivalent to $Z > 12n \ln T$, or equivalently, $r(Z, T) > r(12n \ln T, T)$. The expected regret of our algorithm conditional on the small gradient assumption \emph{not} holding, times the probability of this event, is therefore at most the expected value of $r(Z, T)$ conditional on the value being greater than $r(12n \ln T, T)$, times this probability. This is equal to
	\begin{align*}
		&r(12n \ln T, T) \cdot \pr{Z > 12n \ln T} + \int_{x = r(12n \ln T, T)}^\infty \pr{r(Z, T) \ge x} dx\\
		&\le \sum_{k = 11}^\infty r((k + 1)n \ln T, T) \cdot \pr{Z \ge kn \ln T}\\
		&\le \sum_{k = 11}^\infty \mu \cdot ((k + 1) n \ln T)^{2(2 - \alpha)/(1 - \alpha)} T^{(5 - \alpha)/(1 - \alpha)} \cdot 2m^2 n^2 T \cdot T^{-k}\\
		&\le \sum_{k = 11}^\infty \tilde{O}(T^{1 + (5 - \alpha)/(1 - \alpha) - k}) = \tilde{O}(T^{(5 - \alpha)/(1 - \alpha) - 10}),
	\end{align*}
	as desired. (The first inequality follows by matching the first term with the $k = 11$ summand and upper-bounding the integral with subsequent summands, noting that $r((k + 1)n \ln T, T) \ge 1$.)
\end{proof}

Note that $\frac{5 - \alpha}{1 - \alpha} - 10 \le \frac{5 - 1/2}{1 - 1/2} - 10 = -1$. Therefore, the contribution to expected regret from the case that the small gradient assumption does not hold is $\tilde{O}(T^{-1})$, which is negligible. Together with Corollary~\ref{cor:main_case} (which bounds regret under the small gradient assumption), this proves Theorem~\ref{thm:chap6_no_regret}.
\chapter{Details omitted from Chapter~\ref{chap:agreement}}
\section{Details omitted from Section~\ref{sec:chap8_prelims}} \label{appx:prelims_omitted}
In Section~\ref{sec:chap8_prelims}, we claimed that a positive measure of $n \times n$ information structures satisfy rectangle substitutes.
To formalize this claim, we choose a natural measure over $n \times n$ information structures, specified via the following probability distribution over the values of $Y$ and $\PP[\sigma,\tau]$:
\begin{itemize}
    \item Alice has signals labeled $\sigma_0, \dots, \sigma_{n - 1}$; Bob has signals labeled $\tau_0, \dots, \tau_{n - 1}$. Correspondingly, there are $n^2$ states which we identify with the pair $(i, j)$. For each $i, j$, whenever $\sigma = \sigma_i$ and $\tau = \tau_j$, $Y = y(i, j)$ where $y(i, j)$ is uniformly random in $[0, 1]$.
    \item The probability distribution over states $(i, j)$ is selected uniformly from the space of probability distributions over $n^2$ states.
\end{itemize}

\begin{theorem}
For every $n$, a positive measure of $n \times n$ information structures (per the above measure) satisfy the rectangle substitutes condition.
\end{theorem}

\begin{proof}
The proof is conceptually quite simple. It suffices to exhibit an information structure in which the weak substitutes condition (i.e.\ Equation~\ref{eq:original_rect_subs}) holds \emph{strictly} for every $S, T$ such that $\abs{S}, \abs{T} \ge 2$. It then follows that for a sufficiently small $\delta$, every information structure in the $\delta$-ball around this one\footnote{We can for example define the distance between information structures $\mathcal{I}$ and $\mathcal{I}'$ as $\sum_{i, j} (y(i, j) - y'(i, j))^2 + (\PP[(i, j)] - \PP'[(i, j)])^2$.} also satisfies rectangle substitutes, completing the proof.\footnote{This uses the continuity of the terms in Equation~\ref{eq:original_rect_subs}. Note that the continuity of conditional expectations relies on the conditioning events having positive probability, as is the case in the information structure that we exhibit. Note also that we need not concern ourselves with cases in which $\abs{S} = 1$ or $\abs{T} = 1$, since in those cases the equation is necessarily an equality.}

The information structure $\mathcal{I}$ that we exhibit is as follows: choose any increasing, strictly concave function $f: [0, 2(n - 1)] \to \RR$ (for example, $f(x) = \sqrt{x}$). Let $y(i, j) = \frac{i + j}{2n}$, and let $\PP[(i, j)]$ be proportional to $\epsilon^{f(i + j)}$.

For convenience, define the \emph{substitutes slack} of an information structure to be the additive margin by which the information structure satisfies weak substitutes, i.e.\ the right-hand side of Equation~\ref{eq:original_rect_subs} minus the left-hand side for $S = \mathcal{S}$ and $T = \mathcal{T}$.

Fix a particular $S$ and $T$ such that $\abs{S}, \abs{T} \ge 2$. We wish to show that for sufficiently small positive values of $\epsilon$, Equation~\ref{eq:original_rect_subs} holds strictly. We will show that the substitutes slack of $\mathcal{I} \mid_{S, T}$, i.e.\ $\mathcal{I}$ restricted to $S \times T$, is positive when $\epsilon$ is sufficiently small.

In order to prove this, we first consider the following (different) information structure for values $v, a, a', b, b', c, x, x', y, y'$ (obeying comparisons that we specify below). Each row corresponds to a possible signal value $\sigma$ for Alice, and each column a possible signal value $\tau$ for Bob.
\[
Y = \begin{bmatrix}
v&v + b&b'\\
v + a&v + a + b&-\\
a'&-&-
\end{bmatrix}
\qquad \text{with probability proportional to} \qquad
\begin{bmatrix}
1&y&y'\\
x&cxy&0\\
x'&0&0
\end{bmatrix}
\]
In this information structure, suppose that $x' \le x \ll 1$; $y' \le y \ll 1$; and $1 \ll c \le \frac{1}{x}, \frac{1}{y}$ (so $xy \ll cxy \le x, y$). It can be verified (e.g.\ with a computer algebra system) that the substitutes slack of this information structure is $2abcxy + O(xy)$.

We will transform this information structure into $\mathcal{I}_{S, T}$ while (approximately) preserving substitutes slack. To foreshadow the correspondence, define $i_S$ and $i'_S$ be the smallest and second smallest values of $i$ such that $\sigma_i \in S$, and define $j_T$ and $j'_T$ analogously. The rows of the information structure above will correspond to $\sigma = \sigma_{i_S}, \sigma_{i'_S}$, and all other values of $\sigma \in S$, in that order; the columns will correspond to $\tau = \tau_{j_T}, \tau_{j'_T}$, and all other values of $\tau \in T$, in that order.

Set $x := \epsilon^{f(i_S' + j_T) - f(i_S + j_T)}$, $y := \epsilon^{f(i_S + j'_T) - f(i_S + j_T)}$, and $c := \epsilon^{f(i_S + j_T) + f(i'_S + j'_T) - f(i_S + j'_T) - f(i'_S + j_T)}$, so that $cxy = \epsilon^{f(i'_S + j'_T) - f(i_S + j_T)}$. Note that these values satisfy the aforementioned inequalities involving $x$, $y$, and $c$. (The fact that $1 \ll c$ follows from the strict concavity of $f$.) Set $v := \frac{i_S}{2n}$, $a := \frac{i'_S - i_S}{2n}$, and $b := \frac{j'_T - j_T}{2n}$. Set $x'$ so that $\PP[i > i'_S \mid i \in S, j = j_T] = \frac{x'}{1 + x + x'}$ and $y'$ so that $\PP[j > j'_T \mid i = i_S, j \in T] = \frac{y'}{1 + y + y'}$. We set $a' := \EE{y(i, j_T) \mid i > i'_S, i \in S}$ and set $b' := \EE{y(i_S, j) \mid j > j'_T, j \in T}$.

We now make the following transformation to this information structure: we replace the third row with $\abs{S} - 2$ rows, each corresponding to a different $i > i'_S$. As before, each signal will only be possible in conjunction with Bob's first signal; the value of $Y$ for the signal corresponding to $\sigma_i$ in $\mathcal{I}$ will be $\frac{i + j_T}{2n}$, and the probability will be $\PP[(i, j_T)]$. Note that this simply ``splits'' Alice's third signal into multiple (more informative) signals while preserving the total probability and expectation (this is due to how we picked $a', b', x', y'$ above). This does not affect the substitutes slack of the information structure, because the value of Bob's signal does not change as a result of the transformation (regardless of whether Alice's signal is known).

We make the same transformation but this time to Bob, replacing the third column with $\abs{T} - 2$ columns. The transformation is otherwise analogous, and the substitutes slack again does not change.

Finally, in our last transformation we make this information structure match $\mathcal{I}$ exactly. Note that the information structures already match in the first row $(i = i_S)$, and in the first column $(j = j_T)$, and in the (second row, second column) entry ($(i, j) = (i'_S, j'_T)$). All other entries in $\mathcal{I}$ have probabilities that are $o(cxy)$ (recall that $cxy = \epsilon^{f(i'_S + j'_T) - f(i_S + j_T)}$). As a consequence, adding these entries to the information structure that we are transforming only changes the substitutes slack by $o(cxy)$.

Therefore, $\mathcal{I}$ has substitutes slack $2abcxy + o(cxy) \ge \frac{2}{n^2} \epsilon^{f(i'_S + j'_T) - f(i_S + j_T)} (1 + o(1))$. This is positive for $\epsilon$ sufficiently small, as desired.

We complete the proof by setting $\epsilon$ to be such that it is sufficiently small (in the above argument) for all $S, T$ such that $\abs{S}, \abs{T} \ge 2$.
\end{proof}

\section{Details omitted from Section~\ref{sec:quadratic}} \label{appx:quad_omitted}
\begin{prop} \label{prop:fast_rect}
Consider the following protocol, parametrized by $\epsilon > 0$. Alice and Bob send their initial expectations to each other, rounding to the nearest multiple of $\epsilon$. This protocol entails communicating $O(\log 1/\epsilon)$ bits. At the end of the protocol, Alice and Bob $\epsilon^2/2$-agree and are $\epsilon^2$-accurate (with respect to $G(x) = x^2$).
\end{prop}

\begin{proof}
Let $S$ be the set of possible signals of Alice at the end of the protocol which are consistent with the protocol transcript, and define $T$ likewise for Bob. Recall that we use $\mathcal{S}$ and $\mathcal{T}$ to denote the sets of all of Alice's and Bob's possible signals, respectively. We have
\[\EE{(\mu_{\sigma \tau} - \mu_{S \tau})^2} \le \EE{(\mu_{\sigma} - \mu_{S})^2} \le \epsilon^2,\]
since $\mu_{\sigma}$ and $\mu_{S})$ are guaranteed to be within $\epsilon$ of each other by construction. Thus, Bob is $\epsilon^2$-accurate, and likewise for Alice. By the $\frac{1}{2}$-approximate triangle inequality for $G(x) = x^2$ (see Section~\ref{sec:chap8_bregman}), it follows that Alice and Bob $\epsilon^2/2$-agree.
\end{proof}

\section{Details omitted from Section~\ref{sec:chap8_bregman}} \label{appx:bregman_omitted}
\begin{prop} \label{prop:chap8_bregman_facts}
Let $G$ be a differentiable convex function on the interval $[0, 1]$. For all $0 \le a \le b \le 1$, we have
\begin{enumerate}[label=(\roman*)]
    \item \label{item:min_bregman} $\frac{1}{2}(D_G(a \parallel x) + D_G(b \parallel x)) \ge \JB_G(a, b)$ for every $x \in [0, 1]$.
    \item \label{item:triangle} $\JB_G$ satisfies the reverse triangle inequality: for every $x \in [a, b]$, we have $\JB_G(a, x) + \JB_G(x, b) \le \JB_G(a, b)$.
    \item \label{item:shortening} For all $a \le a' \le b' \le b$, we have $\JB_G(a', b') \le \JB_G(a, b)$.
    \item \label{item:fact1} For a random variable $X$ supported on $[a, b]$, we have
    \[\EE{D_G(X \parallel \EE{X})} = \EE{G(X)} - G(\EE{X}) \le 2 \JB_G(a, b).\]
\end{enumerate}
\end{prop}

\begin{proof}
Fact~\ref{item:min_bregman} follows from Proposition~\ref{prop:bregman_max_ev}. Regarding Fact~\ref{item:triangle}, without loss of generality assume that $x \le \frac{a + b}{2}$ and that $G(x) = G \parens{\frac{a + b}{2}}$ (uniformly adding a constant to the derivative of $G$ does not change any Jensen-Bregman divergence, hence the second assumption). Then $G \parens{\frac{a + x}{2}} \ge G(x)$, so $\JB_G(a, x) \le \frac{G(a) - G(x)}{2}$. Since $\frac{b + x}{2} \ge \frac{a + b}{2}$, we also have that $G \parens{\frac{b + x}{2}} \ge G(x)$, so $\JB_G(b, x) \le \frac{G(b) - G(x)}{2}$. Thus, we have
\[\JB_G(a, x) + \JB_G(b, x) \le \frac{G(a) + G(b)}{2} - G(x) = \frac{G(a) + G(b)}{2} - G \parens{\frac{a + b}{2}} = \JB_G(a, b).\]

\noindent Fact~\ref{item:shortening} follows from Fact~\ref{item:triangle}: we have
\[\JB_G(a, b) = \JB_G(a, a') + \JB_G(a', b') + \JB_G(b', b) \ge \JB_G(a', b').\]

\noindent Regarding the equality in Fact~\ref{item:fact1}, we have
\begin{align*}
\EE{D_G(X \parallel \EE{X})} &= \EE{G(X) - G(\EE{X}) - (X - \EE{X})G'(\EE{X})}\\
&= \EE{G(X) - G(\EE{X})} = \EE{G(X)} - G(\EE{X}),
\end{align*}
where the first step follows from the fact that $\EE{(X - \EE{X}) G'(\EE{X})} = G'(\EE{X}) \EE{X - \EE{X}}$, and $\EE{X - \EE{X}} = 0$.

Regarding the inequality in Fact~\ref{item:fact1}, without loss of generality assume that $\EE{X} \le \frac{a + b}{2}$. By convexity we have that
\[G \parens{\frac{a + b}{2}} \le \frac{\frac{b - a}{2}}{b - \EE{X}} G(\EE{X}) + \frac{\frac{a + b}{2} - \EE{X}}{b - \EE{X}} G(b),\]
so
\begin{align*}
\JB_G(a, b) &= G(a) + G(b) - 2G \parens{\frac{a + b}{2}}\\
&\ge G(a) + G(b) - \frac{b - a}{b - \EE{X}} G(\EE{X}) - \frac{a + b - 2\EE{X}}{b - \EE{X}} G(b)\\
&= G(a) + \frac{\EE{X} - a}{b - \EE{X}} G(b) - \frac{b - a}{b - \EE{X}} G(\EE{X})\\
&= \frac{b - a}{b - \EE{X}} \parens{\frac{b - \EE{X}}{b - a} G(a) + \frac{\EE{X} - a}{b - a} G(b) - G(\EE{X})}\\
&\ge \frac{b - \EE{X}}{b - a} G(a) + \frac{\EE{X} - a}{b - a} G(b) - G(\EE{X}) \ge \EE{G(X)} - G(\EE{X}).
\end{align*}
In the last step we use the fact that for a convex function $f$ and a random variable $X$ defined on an interval $[a, b]$ with mean $\mu$, the maximum possible value of $\EE{f(X)}$ is attained if $X$ is either $a$ or $b$ with the appropriate probabilities.
\end{proof}

\agreebregman*
\begin{proof}
Suppose that Alice and Bob do not $\epsilon$-agree at time step $t$, and without loss of generality assume that the next turn (number $t + 1$) is Alice's. We begin by observing that, by Proposition~\ref{prop:chap8_bregman_facts}~\ref{item:min_bregman}, we have
\[\EE{D_G(\mu_{\sigma T_t} \parallel \mu_{S_t T_t}) + D_G(\mu_{S_t \tau} \parallel \mu_{S_t T_t})} \ge 2 \EE{\JB_G(\mu_{\sigma T_t}, \mu_{S_t \tau})} > 2\epsilon.\]
Therefore, either $\EE{D_G(\mu_{\sigma T_t} \parallel \mu_{S_t T_t})} \ge \frac{2 \epsilon}{3}$ or $\EE{D_G(\mu_{S_t \tau} \parallel \mu_{S_t T_t})} \ge \frac{4 \epsilon}{3}$.

\paragraph{Case 1:} $\EE{D_G(\mu_{\sigma T_t} \parallel \mu_{S_t T_t})} \ge \frac{2 \epsilon}{3}$. Let us use ``hi,'' ``lo,'' and ``md'' to denote the events that Alice says ``high,'' Alice says ``low,'' and Alice says ``medium,'' respectively. We have
\begin{align*}
\frac{2 \epsilon}{3} &\le \EE{D_G(\mu_{\sigma T_t} \parallel \mu_{S_t T_t})} = \EE{\EE{D_G(\mu_{\sigma T_t} \parallel \mu_{S_t T_t}) \mid S_t, T_t}}\\
&= \EE{\EE{D_G(\mu_{\sigma T_t} \parallel \mu_{S_t T_t}) \cdot \mathbbm{1}_{\text{hi or lo}} \mid S_t, T_t}} + \EE{\EE{D_G(\mu_{\sigma T_t} \parallel \mu_{S_t T_t}) \cdot \mathbbm{1}_{\text{md}} \mid S_t, T_t}}\\
&\le \EE{\EE{D_G(\mu_{\sigma T_t} \parallel \mu_{S_t T_t}) \cdot \mathbbm{1}_{\text{hi or lo}} \mid S_t, T_t}} + \frac{\epsilon}{2},
\end{align*}
where ``$\mid S_t, T_t$'' is short for ``$\mid \sigma \in S_t, \tau \in T_t$,'' a notation we use throughout the proof. We thus have
\begin{equation}
\label{eq:raf-too-lazy-to-name-this-properly}
\EE{\EE{D_G(\mu_{\sigma T_t} \parallel \mu_{S_t T_t}) \cdot \mathbbm{1}_{\text{hi}} \mid S_t, T_t}} + \EE{\EE{D_G(\mu_{\sigma T_t} \parallel \mu_{S_t T_t}) \cdot \mathbbm{1}_{\text{lo}} \mid S_t, T_t}} \ge \frac{\epsilon}{6}.
\end{equation}

\noindent We now make use of the following lemma.

\begin{lemma} \label{lem:charlie_learns}
Suppose that turn $t + 1$ is Alice's. Let ``hi'' denote the event that Alice says ``high.'' Let $\alpha := \EE{D_G(\mu_{\sigma T_t} \parallel \mu_{S_t T_t}) \cdot \mathbbm{1}_{\text{hi}} \mid S_t, T_t}$. Then
\[\EE{D_G(\mu_{S_{t + 1} T_{t + 1}} \parallel \mu_{S_t T_t}) \cdot \mathbbm{1}_{\text{hi}} \mid S_t, T_t} \ge \frac{\alpha \epsilon}{8M + 2\epsilon}.\]
The analogous statement is true if Alice says ``low,'' and likewise if it is instead Bob's turn.
\end{lemma}

We assume Lemma~\ref{lem:charlie_learns} and return to prove it afterward.
This lemma translates Equation~\ref{eq:raf-too-lazy-to-name-this-properly} into a statement about how much Charlie learns.
Specifically, we have that
\begin{align*}
&\EE{D_G(\mu_{S_{t + 1} T_{t + 1}} \parallel \mu_{S_t T_t})} = \EE{\EE{D_G(\mu_{S_{t + 1} T_{t + 1}} \parallel \mu_{S_t T_t}) \mid S_t, T_t}}\\
&\ge \EE{\EE{D_G(\mu_{S_{t + 1} T_{t + 1}} \parallel \mu_{S_t T_t}) \cdot \mathbbm{1}_{\text{hi}} \mid S_t, T_t}} + \EE{\EE{D_G(\mu_{S_{t + 1} T_{t + 1}} \parallel \mu_{S_t T_t}) \cdot \mathbbm{1}_{\text{lo}} \mid S_t, T_t}}\\
&\ge \frac{\epsilon}{8M + 2\epsilon} (\EE{\EE{D_G(\mu_{\sigma T_t} \parallel \mu_{S_t T_t}) \cdot \mathbbm{1}_{\text{hi}} \mid S_t, T_t}} + \EE{\EE{D_G(\mu_{\sigma T_t} \parallel \mu_{S_t T_t}) \cdot \mathbbm{1}_{\text{lo}} \mid S_t, T_t}})\\
&\ge \frac{\epsilon^2}{6(8M + 2\epsilon)}.
\end{align*}

\paragraph{Case 2:} $\EE{D_G(\mu_{S_t \tau} \parallel \mu_{S_t T_t})} \ge \frac{4\epsilon}{3}$. Using the Pythagorean theorem to write the same Bregman divergence in two ways, we have that
\begin{align*}
&\EE{D_G(\mu_{S_{t + 1} \tau} \parallel \mu_{S_{t + 1} T_{t + 1}})} + \EE{D_G(\mu_{S_{t + 1} T_{t + 1}} \parallel \mu_{S_t T_t})} = \EE{D_G(\mu_{S_{t + 1} \tau} \parallel \mu_{S_t T_t})}\\
&= \EE{D_G(\mu_{S_{t + 1} \tau} \parallel \mu_{S_t \tau})} + \EE{D_G(\mu_{S_t \tau} \parallel \mu_{S_t T_t})} \ge \EE{D_G(\mu_{S_t \tau} \parallel \mu_{S_t T_t})} \ge \frac{4\epsilon}{3}.
\end{align*}
This means that one of the two summands on the left-hand side is at least $\frac{2\epsilon}{3}$.\\

\textbf{Case 2a:} $\EE{D_G(\mu_{S_{t + 1} \tau} \parallel \mu_{S_{t + 1} T_{t + 1}})} \ge \frac{2\epsilon}{3}$. In that case we have that
\[\EE{D_G(\mu_{S_{t + 2} T_{t + 2}} \parallel \mu_{S_{t + 1} T_{t + 1}})} \ge \frac{\epsilon^2}{6(8M + 2\epsilon)}\]
by the same logic as in Case 1.\\

\textbf{Case 2b:} $\EE{D_G(\mu_{S_{t + 1} T_{t + 1}} \parallel \mu_{S_t T_t})} \ge \frac{2\epsilon}{3} \ge \frac{\epsilon^2}{12\epsilon} \ge \frac{\epsilon^2}{6(8M + 2\epsilon)}$.\\

\noindent In each of our cases, we have that
\begin{align*}
&\EE{D_G(Y \parallel \mu_{S_{t} T_{t}}) - D_G(Y \parallel \mu_{S_{t + 2} T_{t + 2}})} = \EE{D_G(\mu_{S_{t + 2} T_{t + 2}} \parallel \mu_{S_t T_t})}\\
&= \EE{D_G(\mu_{S_{t + 2} T_{t + 2}} \parallel \mu_{S_{t + 1} T_{t + 1}})} + \EE{D_G(\mu_{S_{t + 1} T_{t + 1}} \parallel \mu_{S_t T_t})} \ge \frac{\epsilon^2}{6(8M + 2\epsilon)}.
\end{align*}

\noindent Therefore, the total number of steps until agreement is first reached cannot be more than
\[2 \cdot \frac{M}{\frac{\epsilon^2}{6(8M + 2\epsilon)}} = \frac{24M(4M + \epsilon)}{\epsilon^2}.\]
This completes the proof.
\end{proof}

We now prove Lemma~\ref{lem:charlie_learns}.

\begin{proof}[Proof of Lemma~\ref{lem:charlie_learns}]
We will restrict our probability space to outcomes where Charlie knows $S_t, T_t$ at time $t$ (and thus omit ``$\mid S_t, T_t$'' from here on). For convenience, we will let $A := \mu_{\sigma T_t}$ be Alice's expectation (a random variable) and $c := \mu_{S_t T_t}$ be Charlie's expectation (which is a particular number in $[0, 1]$). We will let $\epsilon' := \frac{\epsilon}{2}$, so that if Alice says ``high'' then Charlie knows that $A > c$ and that $D_G(A \parallel c) \ge \epsilon'$.

Let $D(x) := D_G(x \parallel c) = G(x) - G(c) - G'(c)(x - c)$, and let $\hat{a}_h := \EE{A \mid \text{hi}}$. Note that if Alice says ``high'' then $\mu_{S_{t + 1} T_{t + 1}} = \hat{a}_h$. In our new notation, we may write $\alpha = \EE{D(A) \mid \text{hi}} \cdot \PP[\text{hi}]$, and we wish to show that $D(\hat{a}_h) \cdot \PP[\text{hi}] \ge \frac{\alpha \epsilon'}{2(M + \epsilon')}$. Put otherwise, our goal is to show that
\[\frac{D(\hat{a}_h)}{\EE{D(A) \mid \text{hi}}} \ge \frac{\epsilon'}{2(M + \epsilon')}.\]
For convenience we will let $B$ denote the quantity on the left-hand side.

Let $a_{\text{hmin}}$ be the number larger than $c$ such that $D(a) = \epsilon'$, so that $A \ge a_{\text{hmin}}$ whenever Alice says ``high.''\footnote{If $D(a) < \epsilon'$ for all $a > c$ then Alice never says ``high'' and the lemma statement is trivial.} Observe that since $D$ is convex (Bregman divergences are convex in their first argument), for a fixed value of $\hat{a}_h$, the value of $\EE{D(A) \mid \text{hi}}$ is maximized when $A$ is either $a_{\text{hmin}}$ or $1$ (with probabilities $\frac{1 - \hat{a}_h}{1 - a_{\text{hmin}}}$ and $\frac{\hat{a}_h - a_{\text{hmin}}}{1 - a_{\text{hmin}}}$, respectively). Therefore we have
\begin{equation} \label{eq:first_b_bound}
B = \frac{D(\hat{a}_h)}{\EE{D(A) \mid \text{hi}}} \ge \frac{D(\hat{a}_h)(1 - a_{\text{hmin}})}{(1 - \hat{a}_h) \epsilon' + (\hat{a}_h - a_{\text{hmin}}) D(1)}.
\end{equation}

\paragraph{Case 1:} $(1 - \hat{a}_h) \epsilon' \ge (\hat{a}_h - a_{\text{hmin}}) D(1)$. In that case we have
\[B \ge \frac{D(\hat{a}_h)(1 - a_{\text{hmin}})}{2(1 - \hat{a}_h) \epsilon'} \ge \frac{\epsilon'(1 - a_{\text{hmin}})}{2(1 - \hat{a}_h) \epsilon'} \ge \frac{1}{2} \ge \frac{\epsilon'}{2(M + \epsilon')}.\]

\paragraph{Case 2:} $(1 - \hat{a}_h) \epsilon' \le (\hat{a}_h - a_{\text{hmin}}) D(1)$. In that case we have

\begin{equation} \label{eq:b_case2}
B \ge \frac{D(\hat{a}_h)(1 - a_{\text{hmin}})}{2(\hat{a}_h - a_{\text{hmin}}) D(1)}.
\end{equation}

\textbf{Case 2a:} $D(1) \le \frac{1 - c}{\hat{a}_h - c}(M + \epsilon')$. Then we have
\[B \ge \frac{D(\hat{a}_h)(1 - a_{\text{hmin}})}{2(\hat{a}_h - a_{\text{hmin}}) \cdot \frac{1 - c}{\hat{a}_h - c}(M + \epsilon')} \ge \frac{\epsilon'}{2(M + \epsilon')} \cdot \frac{(1 - a_{\text{hmin}})(\hat{a}_h - c)}{(\hat{a}_h - a_{\text{hmin}})(1 - c)}.\]
(In the last step we again use that $D(\hat{a}_h) \ge \epsilon$.) Now, it is easy to verify that the second fraction is at least $1$ (this comes down to the fact that $a_{\text{hmin}} \ge c$), so we indeed have that $B \ge \frac{\epsilon'}{2(M + \epsilon')}$.\\

\textbf{Case 2b}: $D(1) \ge \frac{1 - c}{\hat{a}_h - c}(M + \epsilon')$. We claim that for all $x \ge c$, we have that
\begin{equation} \label{eq:m}
D(x) \ge \frac{x - c}{1 - c} D(1) - M.
\end{equation}
To see this, suppose for contradiction that for some $x$ we have $D(x) < \frac{x - c}{1 - c} D(1) - M$. Then
\begin{align*}
G(x) - G(c) - G'(c)(x - c) &< \frac{x - c}{1 - c}(G(1) - G(c) - G'(c)(1 - c)) - M\\
(1 - c)G(x) - (1 - c)G(c) &< (x - c)G(1) - (x - c) G(c) - (1 - c)M\\
G(x) + M &< \frac{(1 - x)G(c) + (x - c)G(1)}{1 - c}.
\end{align*}
On the other hand, we have that both $G(c)$ and $G(1)$ are less than or equal to $G(x) + M$, by definition of $M$. This means that
\[G(1), G(c) < \frac{(1 - x)G(c) + (x - c)G(1)}{1 - c}\]
but this implies that $G(1) < G(c)$ and that $G(c) < G(1)$, a contradiction.\\

Plugging in $x = \hat{a}_h$ into Equation~\ref{eq:m}, we find that
\[D(\hat{a}_h) \ge \frac{\hat{a}_h - c}{1 - c} D(1) - M.\]
Plugging this bound into Equation~\ref{eq:b_case2}, we get that
\begin{align*}
B &\ge \frac{\parens{\frac{\hat{a}_h - c}{1 - c} D(1) - M}(1 - a_{\text{hmin}})}{2(\hat{a}_h - a_{\text{hmin}}) D(1)} = \frac{1 - a_{\text{hmin}}}{2(\hat{a}_h - a_{\text{hmin}})} \cdot \frac{\hat{a}_h - c}{1 - c} \parens{1 - \frac{M}{\frac{\hat{a}_h - c}{1 - c}D(1)}}\\
&\ge \frac{1 - a_{\text{hmin}}}{2(\hat{a}_h - a_{\text{hmin}})} \cdot \frac{\hat{a}_h - c}{1 - c} \parens{1 - \frac{M}{M + \epsilon'}} \ge \frac{\epsilon'}{2(M + \epsilon')},
\end{align*}
where in the second-to-last step we use that $D(1) \ge \frac{1 - c}{\hat{a}_h - c}(M + \epsilon')$ and in the last step we again use the fact that $\frac{(1 - a_{\text{hmin}})(\hat{a}_h - c)}{(\hat{a}_h - a_{\text{hmin}})(1 - c)} \ge 1$.
\end{proof}

\bobclosebregman*

\begin{proof}
We will partition $[0, 1]$ into a number $N$ of small intervals $I_1 = [x_0 = 0, x_1)$, $I_2 = [x_1, x_2)$, $I_3 = [x_2, x_3)$, \dots, $I_N = [x_{N - 1}, x_N = 1]$ with certain desirable properties (which we will describe below). For $k \in [N]$, we will let $S^{(k)} := \{\sigma \in \mathcal{S}: \mu_{\sigma} \in I_k\}$. For a given $\sigma \in \mathcal{S}$, we will let $k(\sigma)$ be the $k$ such that $\sigma \in S^{(k)}$.

Our goal is to upper bound the expectation of $D_G(\mu_{\sigma \tau} \parallel \mu_{\tau})$. In pursuit of this goal, we observe that by Proposition~\ref{prop:pythag_bregman} we have
\begin{equation} \label{eq:sum_of_parts}
\EE{D_G(\mu_{\sigma \tau} \parallel \mu_{\tau})} = \EE{D_G(\mu_{\sigma \tau} \parallel \mu_{S^{(k(\sigma))} \tau})} + \EE{D_G(\mu_{S^{(k(\sigma))} \tau} \parallel \mu_{\tau})}.
\end{equation}
Now, for any $k$, by applying Equation~\ref{eq:rec_sub_bregman} to $S = S^{(k)}$ and $T = \mathcal{T}$, we know that
\[\EE{D_G(\mu_{\sigma} \parallel \mu_{S^{(k)}}) \mid S^{(k)}} \ge \EE{D_G(\mu_{\sigma \tau} \parallel \mu_{S^{(k)} \tau}) \mid S^{(k)}}.\]
(Here, ``$\mid S^{(k)}$'' is short for ``$\mid \sigma \in S^{(k)}$.'') This is our only use of the rectangle substitutes assumption. Now, taking the expectation over $k$ (i.e.\ choosing each $k$ with probability equal to $\PP[\sigma \in S^{(k)}]$), we have that
\[\EE{D_G(\mu_{\sigma} \parallel \mu_{S^{(k(\sigma))}})} \ge \EE{D_G(\mu_{\sigma \tau} \parallel \mu_{S^{(k(\sigma))} \tau})}.\]
Together with Equation~\ref{eq:sum_of_parts}, this tells us that
\begin{equation} \label{eq:sum_of_parts_main}
\EE{D_G(\mu_{\sigma \tau} \parallel \mu_{\tau})} \le \EE{D_G(\mu_{\sigma} \parallel \mu_{S^{(k(\sigma))}})} + \EE{D_G(\mu_{S^{(k(\sigma))} \tau} \parallel \mu_{\tau})}.
\end{equation}
Our goal will be to bound the two summands in Equation~\ref{eq:sum_of_parts_main}. We will specify the boundaries of the intervals $I_1, \dots, I_N$ with this goal in mind.\\

On an intuitive level, we are hoping for two things to be true:
\begin{itemize}
\item In order for the first summand to be small, we want $\mu_{\sigma}$ and $\mu_{S^{(k(\sigma))}}$ to be similar in value. In other words, we want each interval is ``short'' (for a notion of shortness with respect to $G$ that we are about to discuss).
\item In order for the second summand to be small, we want $\mu_{S^{(k(\sigma))} \tau}$ and $\mu_{\tau}$ to be similar in value. In other words, the estimate of a third party who knows $\tau$ shouldn't change much upon learning $k(\sigma)$. One way to ensure this is by creating the intervals in a way that makes the third party very confident about the value of $k(\sigma)$ before learning it. Intuitively this should be true because Alice and Bob approximately agree, so Alice's estimate is likely to be close to Bob's. However, we must be careful to strategically choose the boundaries of our intervals $x_1, \dots, x_{N - 1}$ so that Alice's and Bob's estimates are unlikely to be on opposite sides of a boundary.\footnote{This limits how many intervals we can reasonably use, which is why we cannot make our intervals arbitrarily short to satisfy the first of our two criteria.}
\end{itemize}

What, formally, do we need for the first summand to be small? For any $k$, we have $\mu_{S^{(k(\sigma))}} = \EE{\mu_{\sigma} \mid \sigma \in S^{(k)}}$. We can apply Proposition~\ref{prop:chap8_bregman_facts}~\ref{item:fact1} to the random variable $X = \mu_{\sigma}$ on the probability subspace given by $\sigma \in S^{(k)}$. Since $X$ takes on values in $I_k$, we have that
\begin{equation} \label{eq:2jb}
\EE{D_G(\mu_{\sigma} \parallel \mu_{S^{(k)}}) \mid S^{(k)}} \le 2\JB_G(I_k),
\end{equation}
where $\JB_G(I_k)$ is shorthand for the Jensen-Bregman divergence between the endpoints of $I_k$. Therefore, if $\JB_G(I_k)$ is small for all $k$, then the first summand (which is an expected value of $\EE{D_G(\mu_{\sigma} \parallel \mu_{S^{(k)}}) \mid S^{(k)}}$ over $k \in [N]$) is also small.\\

What about the second summand? As per the intuition above, we wish to choose our boundary points $x_1, \dots, x_{N - 1}$ so that Alice's and Bob's estimates are unlikely to be on opposite sides of any boundary. Let $\mu_- = \min(\mu_{\sigma}, \mu_{\tau})$ be the smaller of the two estimates and $\mu_+ = \max(\mu_{\sigma}, \mu_{\tau})$ be the larger one. We say that $\mu_-, \mu_+$ \emph{thwart} a point $x \in (0, 1)$ if $\mu_- \le x \le \mu_+$ and $\mu_- \neq \mu_+$. We define the \emph{thwart density} of $x$ to be
\[\rho(x) := \PP[\mu_-, \mu_+ \text{ thwart } x].\]
Roughly speaking, we will choose $x_1, \dots, x_{N - 1}$ such that $\rho(x_k)$ is small on average.\\

We will approach this problem by first creating intervals to satisfy the first criterion (short intervals), without regard to the second, and then modifying them to satisfy the second without compromising the first. Formally, we choose our intervals according to the following algorithm.

\setcounter{algocf}{\value{theorem}}
\begin{algorithm}[ht]
\caption{Partitioning $[0, 1]$ into intervals $I_1, \dots, I_N$} \label{alg:intervals} \phantom{}
\begin{enumerate}
    \item Choose points $0 < x_1' < x_2' < \dots < x_{N - 2}' < 1$ such that the $N - 1$ intervals thus created all have Jensen-Bregman divergence between $\beta$ and $\frac{2\beta}{c}$, inclusive, where $\beta$ and $c$ are as in the statement of Lemma~\ref{lem:bob_close_bregman}. ($N$ is not pre-determined; it is defined as one more than the number of intervals created.) (See footnote for why this is possible.\footnotemark)
    
    \item Let $x_0' := 0, x_{N - 1}' := 1$ for convenience. Define $I_k' := [x_{k - 1}', x_k']$. For $k \in [N - 1]$, let $\alpha_k := \inf_{x \in I_k'} \rho(x)$. Let $x_k \in I_k'$ be such\footnotemark \phantom{ }that $\rho(x_k) \le 2\alpha_k$.
    
    \item Return the intervals $I_1 = [0, x_1), I_2 = [x_1, x_2), \dots, I_N = [x_{N - 1}, 1]$.
\end{enumerate}
\end{algorithm}
\setcounter{theorem}{\value{algocf}}
\singlespacing
\addtocounter{footnote}{-1}
\footnotetext{Define $x_1'$ so that $\JB_G(0, x_1') = \frac{2\beta}{c}$ (this is possible because $\JB_G$ is continuous in its arguments). Define $x_2'$ so that $\JB_G(x_1', x_2') = \frac{2\beta}{c}$. Keep going until an endpoint $x_{N - 3}'$ is defined such that adding $x_{N - 2}'$ as before would leave an interval $(x_{N - 2}', 1)$ with Jensen-Bregman divergence less than $\frac{2\beta}{c}$. Now, instead of defining $x_{N - 2}'$ in this way, define it so that $\JB_G(x_{N - 3}', x_{N - 2}') = \JB_G(x_{N - 2}', 1)$. Since $\JB_G(x_{N - 3}', 1) \ge \frac{2\beta}{c}$, the $c$-approximate triangle inequality that we have by assumption tells us that $\JB_G(x_{N - 3}', x_{N - 2}') = \JB_G(x_{N - 2}', 1) \ge \beta$.}
\stepcounter{footnote}
\footnotetext{If the infimum is achieved (e.g.\ if the space of signals to Alice and Bob is finite), then we can set $x_k := \arg \min_x \rho(x)$. Our algorithm works in more generality, at the expense of a factor of $2$ in our final bound. Note that by replacing $2$ with a smaller constant can arbitrarily reduce this factor.}
\doublespacing

\noindent We begin by observing that for any $k \in [N]$, we have
\[\JB_G(I_k) = \JB_G(x_{k - 1}, x_k) \le \JB_G(x_{k - 2}', x_k') \le \frac{1}{c}(\JB_G(x_{k - 2}', x_{k - 1}') + \JB_G(x_{k - 1}', x_k')) \le \frac{4\beta}{c^2}\]
where for convenience we define $x_{-1}' := 0, x_N' := 1$.
Therefore, by Equation~\ref{eq:2jb}, we have
\begin{equation} \label{eq:bound_1}
\EE{D_G(\mu_{\sigma} \parallel \mu_{S^{(k(\sigma))}})} \le \frac{8\beta}{c^2}.
\end{equation}

It remains to bound the second summand of Equation~\ref{eq:sum_of_parts_main}, $\EE{D_G(\mu_{S^{(k(\sigma))} \tau} \parallel \mu_{\tau})}$, which is the bulk of the proof. We proceed in two steps:
\begin{enumerate}[label=(\arabic*)]
    \item (Lemma~\ref{lem:jbgi_small}) We show that $\sum_{k = 1}^N \alpha_k$ is small. This means that Alice's and Bob's estimates are unlikely to lie on opposite sides of some boundary point $x_k$. As a consequence, Bob is highly likely to know $k(\sigma)$ with a lot of confidence
    \item (Lemma~\ref{lem:second_summand}) We bound the second summand as a function of $\sum_{k = 1}^N \alpha_k$. The intuition is that if $\sum_k \alpha_k$ is small, then Bob is highly likely to know $k(\sigma)$ with a lot of confidence, which means that he does not learn too much from learning $k(\sigma)$.
\end{enumerate}

\noindent We begin with the first step; recall our notation $\mu^- := \min(\mu_{\sigma}, \mu_{\tau})$ and $\mu^+ := \max(\mu_{\sigma}, \mu_{\tau})$.

\begin{lemma} \label{lem:jbgi_small}
\[2 \sum_{k = 1}^N \alpha_k \le 4 \parens{\frac{\epsilon}{\beta c}}^{1/(1 - \log_2 c)}.\]
\end{lemma}

\begin{proof}
We use the following claim, whose proof we provide afterward.

\begin{claim} \label{claim:alpha_2}
Let $I = [x^-, x^+]$ be any sub-interval of $[0, 1]$ and let $\alpha = \inf_{x \in I} \rho(x)$. Then there is an increasing sequence of points $z_0 := x^-, z_1, z_2, \dots, z_{L - 1}, z_L := x^+$, such that for every $\ell \in [L]$, $\PP[\mu_- \le z_{\ell - 1}, \mu_+ \ge z_\ell] \ge \frac{\alpha}{2}$, and where
\[L \le \frac{2}{\alpha} \sum_{\ell \in [L]} \PP[\mu_- \le z_{\ell - 1} < \mu_+ \le z_\ell].\]
\end{claim}

We apply Claim~\ref{claim:alpha_2} to the intervals $I_1', \dots, I_{N - 1}'$, with $\alpha = \alpha_k$. Let $z_{k, 0}, \dots, z_{k, L_k}$ be the points whose existence the claim proves, and let $r_k := \sum_{\ell \in [L_k]} \PP[\mu_- \le z_{k, \ell - 1} < \mu_+ \le z_{k, \ell}]$, so that $L_k \le \frac{2}{\alpha_k} r_k$. Observe that $\sum_k r_k \le 1$, because the intervals $(z_{k, \ell - 1}, z_{k, \ell}]$ are disjoint for all $k, \ell$. We make the following claim (we provide the proof afterward).

\begin{claim} \label{claim:alphak_bound}
\begin{equation} \label{eq:alphak_bound}
\sum_{k \in [N - 1]} r_k \parens{\frac{\alpha_k}{2r_k}}^{1 - \log_2 c} \le \frac{\epsilon}{\beta c}.
\end{equation}
\end{claim}

\noindent We may rewrite Equation~\ref{eq:alphak_bound} as
\[\parens{\sum_{k \in [N - 1]} r_k \parens{\frac{\alpha_k}{2r_k}}^{1 - \log_2 c}}^{1/(1 - \log_2 c)} \le \parens{\frac{\epsilon}{\beta c}}^{1/(1 - \log_2 c)}.\]

Recall that $\sum_k r_k \le 1$. Scaling the $r_k$'s to add to $1$ decreases the left-hand side above, so we may assume that $\sum_k r_k = 1$. Note that $x^{1 - \log_2 c}$ is convex. Thus, by using a weighted Jensen inequality on the left-hand side with weights $r_k$, we find that
\[\frac{1}{2} \sum_k \alpha_k = \sum_k r_k \cdot \frac{\alpha_k}{2r_k} \le \parens{\sum_{k \in [N - 1]} r_k \parens{\frac{\alpha_k}{2r_k}}^{1 - \log_2 c}}^{1/(1 - \log_2 c)} \le \parens{\frac{\epsilon}{\beta c}}^{1/(1 - \log_2 c)}.\]
This completes the proof of Lemma~\ref{lem:jbgi_small}.
\end{proof}

\begin{proof}[Proof of Claim~\ref{claim:alpha_2}]
Let $z_1 = \inf \{z: \PP[\mu_- \le z_0 < \mu_+ \le z] \ge \frac{\alpha}{2}\}$, or $x^+$ if this number does not exist or is larger than $x^+$.
Note that $\PP[\mu_- \le z_0 < \mu_+] \ge \alpha$, as we have $\rho(z_0) = \PP[\mu_- \le z_0 < \mu_+] + \PP[\mu_- < z_0 = \mu_+] \ge \alpha$, so if the first term were less than $\alpha$ we would have some $z' > z_0$ with $\rho(z') < \alpha$.
On the other hand, $\PP[\mu_- \le z_0 < \mu_+ < z_1] \le \frac{\alpha}{2}$, since
\[\PP[\mu_- \le z_0 < \mu_+ < z_1] = \lim_{z \to z_1 \text{ from below}} \PP[\mu_- \le z_0 < \mu_+ \le z]\]
and if the right-hand side were more than $\frac{\alpha}{2}$ then that would contradict the definition of $z_1$ as an infimum.
Therefore, $\PP[\mu_- \le z_0, \mu_+ \ge z_1] \ge \frac{\alpha}{2}$.

If $z_1 = x^+$, we are done. Otherwise, let $z_2 = \inf \{z: \PP[\mu_- \le z_1 < \mu_+ \le z] \ge \frac{\alpha}{2}\}$. Then $\PP[\mu_- \le z_1, \mu_+ \ge z_2] \ge \frac{\alpha}{2}$. Define $z_3$ analogously, and so forth.

All that remains to show is the upper bound on $L$. This is where we use the fact that (by construction) $\PP[\mu_- \le z_{\ell - 1} < \mu_+ \le z_\ell] \ge \frac{\alpha}{2}$. Summing over all $\ell$, we have
\[\sum_{\ell \in [L]} \PP[\mu_- \le z_{\ell - 1} < \mu_+ \le z_\ell] \ge \frac{\alpha}{2} L,\]
which (after rearranging) completes the proof.
\end{proof}

\begin{proof}[Proof of Claim~\ref{claim:alphak_bound}]
First note that by construction, $\JB_G(I_k') \ge \beta$ for all $k$. By repeated use of the $c$-approximate triangle inequality,\footnote{We sub-divide $I_k'$ into $[z_{k, 0}, z_{k, L_k/2}]$ and $[z_{k, L_k/2}, z_{k, L}]$, then subdivide each of these, and so on.} we find that
\[\sum_{\ell \in [L_k]} \JB_G(z_{k, \ell - 1}, z_{k, \ell}) \ge c^{\ceil{\log_2 L_k}} \JB_G(I_k') \ge c^{1 + \log_2 \frac{2r_k}{\alpha_k}} \JB_G(I_k') \ge c^{1 + \log_2 \frac{2r_k}{\alpha_k}} \beta = c \parens{\frac{2r_k}{\alpha_k}}^{\log_2 c} \beta.\]
On the other hand, we have
\begin{align*}
\epsilon &\ge \EE{\JB_G(\mu_{\sigma}, \mu_{\tau})} = \sum_{\sigma, \tau} \PP[\sigma, \tau] \JB_G(\mu_{\sigma}, \mu_{\tau}) \ge \sum_{\sigma, \tau} \PP[\sigma, \tau] \sum_{\substack{k, \ell: \mu_- \le z_{k, \ell - 1} \\ \mu_+ \ge z_{k, \ell}}} \JB_G(z_{k, \ell - 1}, z_{k, \ell})\\
&= \sum_{k, \ell} \PP[\mu_- \le z_{k, \ell - 1}, \mu_+ \ge z_{k, \ell}] \JB_G(z_{k, \ell - 1}, z_{k, \ell}) \ge \sum_{k, \ell} \frac{\alpha_k}{2} \JB_G(z_{k, \ell - 1}, z_{k, \ell}).
\end{align*}
Here, the third step follows by the reverse triangle inequality (Fact~\ref{item:triangle} of Proposition~\ref{prop:chap8_bregman_facts}) and the fourth step follows by rearranging the order of summation.\footnote{The case that the space of signals is infinite is identical except that the summation is replaced by an integral over the probability space.} Combining the last two facts gives us that
\[\epsilon \ge \sum_k \frac{\alpha_k}{2} \cdot c \parens{\frac{2r_k}{\alpha_k}}^{\log_2 c} \beta = \sum_k r_k  \parens{\frac{2r_k}{\alpha_k}}^{\log_2 c - 1} \beta c,\]
which rearranges to the desired identity.
\end{proof}

We are now ready to bound the second summand, i.e.\ $\EE{D_G(\mu_{S^{(k(\sigma))} \tau} \parallel \mu_{\tau})}$, where $k(\sigma)$ is the $k$ such that Alice's estimate $\mu_{\sigma}$ lies in $I_k$. For convenience we will define $k(\tau)$ for Bob by analogy as the $k$ such that $\mu_{\tau}$ lies in $I_k$. By Lemma~\ref{lem:jbgi_small} and the preceding discussion, we know that
\[\PP[k(\sigma) \neq k(\tau)] \le 4 \parens{\frac{\epsilon}{\beta c}}^{1/(1 - \log_2 c)}.\]

\begin{lemma} \label{lem:second_summand}
Let $Q = \PP[k(\sigma) \neq k(\tau)]$. Then
\[\EE{D_G(\mu_{S^{(k(\sigma))} \tau} \parallel \mu_{\tau})} \le 2\tilde{G}^*(Q).\]
\end{lemma}

The key idea is that because $k(\sigma) = k(\tau)$ with probability near $1$, learning $k(\sigma)$ is unlikely to make Bob update his estimate much.

\begin{proof}
Consider any signal $\hat{\tau} \in \mathcal{T}$ and let $p(\hat{\tau}) = \PP[\tau = \hat{\tau}]$. We have\footnote{This proof takes sums over $\hat{\tau} \in \mathcal{T}$ and thus implicitly assumes that $\mathcal{T}$ is finite, but the proof extends to infinite $\mathcal{T}$, with sums over $\tau$ replaced by integrals with respect to the probability measure over $\mathcal{T}$.}
\[\EE{D_G(\mu_{S^{(k(\sigma))} \tau} \parallel \mu_{\tau})} = \sum_{\hat{\tau} \in \mathcal{T}} p(\hat{\tau}) \EE{D_G(\mu_{S^{(k(\sigma))} \hat{\tau}} \parallel \mu_{\hat{\tau}}) \mid \tau = \hat{\tau}}.\]
Note that $\mu_{\hat{\tau}} = \EE{\mu_{S^{(k(\sigma))} \hat{\tau}} \mid \tau = \hat{\tau}}$, so by Proposition~\ref{prop:chap8_bregman_facts} we have that
\[\EE{D_G(\mu_{S^{(k(\sigma))} \tau} \parallel \mu_{\tau})} = \sum_{\hat{\tau} \in \mathcal{T}} p(\hat{\tau}) \parens{\EE{G(\mu_{S^{(k(\sigma))}\hat{\tau}}) \mid \tau = \hat{\tau}} - G(\mu_{\hat{\tau}})}.\]
Let $q(\hat{\tau}) = \PP[\tau = \hat{\tau}, k(\sigma) \neq k(\hat{\tau})]$, so $\sum_{\hat{\tau} \in \mathcal{T}} q(\hat{\tau}) = Q$. Then
\begin{align*}
&\EE{G(\mu_{S^{(k(\sigma))}\hat{\tau}}) \mid \tau = \hat{\tau}} - G(\mu_{\hat{\tau}}) =\\
&\frac{p(\hat{\tau}) - q(\hat{\tau})}{p(\hat{\tau})} \parens{\EE{G(\mu_{S^{(k(\hat{\tau}))} \hat{\tau}}) - G(\mu_{\hat{\tau}})}} + \frac{q(\hat{\tau})}{p(\hat{\tau})} \parens{\EE{G(\mu_{S^{(k(\sigma))}\hat{\tau}}) \mid \tau = \hat{\tau}, k(\sigma) \neq k(\hat{\tau})} - G(\mu_{\hat{\tau}})}.
\end{align*}
The second term is at most $\frac{q(\hat{\tau})}{p(\hat{\tau})} M$, since $M$ is the range of $G$. To bound the first term, we note that $\mu_{S^{(k(\hat{\tau}))}\hat{\tau}}$ cannot differ from $\mu_{\hat{\tau}}$ by more than $\frac{q(\hat{\tau})}{p(\hat{\tau}) - q(\hat{\tau})}$, as otherwise the average value of $\mu_{S^{(k(\sigma))} \hat{\tau}}$ could not be $\mu_{\hat{\tau}}$. Therefore, $\EE{G(\mu_{S^{(k(\hat{\tau}))} \hat{\tau}}) - G(\mu_{\hat{\tau}})}$ is bounded by the largest possible difference in $G$-values of two points that differ by at most $\frac{q(\hat{\tau})}{p(\hat{\tau}) - q(\hat{\tau})}$. Therefore, we have
\begin{align*}
\EE{D_G(\mu_{S^{(k(\sigma))} \tau} \parallel \mu_{\tau})} &\le \sum_{\hat{\tau} \in \mathcal{T}} p(\hat{\tau}) \parens{\frac{p(\hat{\tau}) - q(\hat{\tau})}{p(\hat{\tau})} \tilde{G} \parens{\frac{q(\hat{\tau})}{p(\hat{\tau}) - q(\hat{\tau})}} + \frac{q(\hat{\tau})}{p(\hat{\tau})} M}\\
&\le QM + \sum_{\hat{\tau} \in \mathcal{T}} (p(\hat{\tau}) - q(\hat{\tau})) \tilde{G} \parens{\frac{q(\hat{\tau})}{p(\hat{\tau}) - q(\hat{\tau})}},
\end{align*}
where $\tilde{G}$ is defined as in the statement of Lemma~\ref{lem:second_summand}. If $G$ is symmetric on $[0, 1]$, then $\tilde{G}(x) = G(0) - G(x)$ for $x \le \frac{1}{2}$ and $M$ otherwise. This is a concave function, but $\tilde{G}$ is not in general concave. However, consider $\tilde{G}^*$ as defined in the lemma statement, so $\tilde{G}(x) \le \tilde{G}^*(x)$ for all $x$. Then
\begin{align*}
\EE{D_G(\mu_{S^{(k(\sigma))} \tau} \parallel \mu_{\tau})} &\le QM + \sum_{\hat{\tau} \in \mathcal{T}} (p(\hat{\tau}) - q(\hat{\tau})) \tilde{G}^* \parens{\frac{q(\hat{\tau})}{p(\hat{\tau}) - q(\hat{\tau})}}\\
&\le QM + \parens{\sum_{\hat{\tau} \in \mathcal{T}} (p(\hat{\tau}) - q(\hat{\tau}))} \cdot \tilde{G}^* \parens{\frac{\sum_{\hat{\tau} \in \mathcal{T}} q(\hat{\tau})}{\sum_{\hat{\tau} \in \mathcal{T}} (p(\hat{\tau}) - q(\hat{\tau}))}}\\
&= QM + (1 - Q) \tilde{G}^* \parens{\frac{Q}{1 - Q}} \le Q M + \tilde{G}^*(Q) \le 2\tilde{G}^*(Q).
\end{align*}
Here, the second step follows by Jensen's inequality with terms $\frac{q(\hat{\tau})}{p(\hat{\tau}) - q(\hat{\tau})}$ and weights $p(\hat{\tau}) - q(\hat{\tau})$, the second-to-last step follows from the fact that $\tilde{G}^*$ is convex and $\tilde{G}^*(0) = 0$, and the last step follows from the fact that $\tilde{G}^*$ is convex and $\tilde{G}^*(1) = M$.
\end{proof}

Since $Q \le 4 \parens{\frac{\epsilon}{\beta c}}^{1/(1 - \log_2 c)}$, combining Lemma~\ref{lem:second_summand} with Equation~\ref{eq:bound_1} gives us the following result.
\[\EE{D_G(\mu_{\sigma \tau} \parallel \mu_{\tau})} \le \frac{8\beta}{c^2} + 2 \tilde{G}^* \parens{4 \parens{\frac{\epsilon}{\beta c}}^{1/(1 - \log_2 c)}}.\]
Noting that $\tilde{G}^*$ is concave and $c^{-1/(1 - \log_2 c)} \le 2$ (which is true for all $0 < c < 1$) completes the proof of Lemma~\ref{lem:bob_close_bregman}.
\end{proof}

\section{Alternative definitions of agreement and accuracy} \label{appx:alternative_defs}
For arbitrary Bregman divergences, there are several notions of agreement and accuracy that are worth considering. Before we discuss these, we make a note about the order of arguments in a Bregman divergence. In our context, it makes the most sense to talk of the Bregman divergence \emph{from a more informed estimate to a less informed estimate}. By a ``more informed estimate'' we mean a finer-grained one, i.e.\ one that is informed by more knowledge. For example, in terms of estimating $Y$ in the context of this work explores, $Y$ is more informed than $\mu_{\sigma \tau}$, which is more informed than $\mu_{\sigma}$ and $\mu_{S \tau}$, which are each more informed than $\mu_{ST}$, which is more informed than $\EE{Y}$.

To see that this is the natural order of the arguments, recall that Bregman divergences are motivated by the property that they elicit the mean (see Proposition~\ref{prop:bregman_max_ev}): if an agent who gives an estimate of $x$ for the value of a random variable $Y$ incurs a loss of $D_G(Y \parallel x)$, then the agent minimizes their expected loss by reporting $x = \EE{Y}$. This means that the expert ought to report the expected value of $Y$ given the information that the expert knows.

This means that given two estimates of $Y$, $Z_1$ and $Z_2$, of which $Z_1$ is more informed, the quantity $D_G(Z_1 \parallel Z_2)$ has a natural interpretation: it is the expected amount the expert gains by learning more and refining their estimate from $Z_2$ to $Z_1$. This follows by the Pythagorean theorem:
\[\EE{D_G(Z_1 \parallel Z_2)} = \EE{D_G(Y \parallel Z_2)} - \EE{D_G(Y \parallel Z_1)}.\]

\subsection{Alternative definitions of agreement}
One important motivation for using the Jensen-Bregman divergence to the midpoint as the definition of agreement is that this quantity serves as a lower bound on the expected amount that Charlie disagrees with Alice and Bob. Formally:

\begin{defin} \label{def:agree_charlie}
Let $a$, $b$, and $c$ be Alice's, Bob's, and Charlie's expectations, respectively (these are random variables on $\Omega$). Alice and Bob \emph{$\epsilon$-agree with Charlie} if $\frac{1}{2}(\EE{D_G(a \parallel c) + D_G(b \parallel c)}) \le \epsilon$.
\end{defin}

(This is the order of arguments because Alice and Bob are more informed than Charlie.) By Proposition~\ref{prop:chap8_bregman_facts}~\ref{item:min_bregman}, we know that \textbf{if Alice and Bob $\epsilon$-agree with Charlie then they $\epsilon$-agree}.

As it happens the fact that under this (stronger) definition of agreement implies accuracy under rectangle substitutes follows immediately:

\begin{prop}
Let $\mathcal{I} = (\Omega, \PP, \sigma, \tau, Y)$ be an information structure that satisfies rectangle substitutes. For any communication protocol that causes Alice and Bob to $\epsilon$-agree with Charlie on $\mathcal{I}$, Alice and Bob are $2 \epsilon$-accurate after the protocol terminates.
\end{prop}

\begin{proof}
Let $S$ be the set of possible signals of Alice at the end of the protocol which are consistent with the protocol transcript, and define $T$ likewise for Bob. Recall that Charlie's expectation is $\mu_{ST}$. We have
\[\EE{D_G(\mu_{\sigma \tau} \parallel \mu_{S \tau})} \le \EE{D_G(\mu_{\sigma T} \parallel \mu_{ST})} \le \EE{D_G(\mu_{\sigma T} \parallel \mu_{ST})} + \EE{D_G(\mu_{S \tau} \parallel \mu_{ST})} \le 2 \epsilon,\]
where the first inequality follows by rectangle substitutes and the last inequality follows because Alice and Bob $\epsilon$-agree with Charlie.
\end{proof}

The drawback of Definition~\ref{def:agree_charlie} is that it is not so much a definition of Alice and Bob's agreement with each other, so much as a definition of agreement with respect to the protocol being run (since Charlie only exists within the context of the protocol). Put otherwise, it is impossible to determine whether Alice and Bob $\epsilon$-agree with Charlie simply by knowing Alice and Bob's expectations; one must also know Charlie's expectation, which cannot be determined from Alice's and Bob's expectations. The question ``how far from agreement are Alice and Bob if Alice believes 25\% and Bob believes 30\%?'' makes sense in the context of $\epsilon$-agreement, but not in the context of $\epsilon$-agreement with Charlie.\\

A different notion of agreement, which (like $\epsilon$-agreement) only depends on Alice's and Bob's expectations, uses the \emph{symmetrized Bregman divergence} between these expectations: $\frac{1}{2}(D_G(a \parallel b) + D_G(b \parallel a))$.

\begin{defin}
Let $a$ and $b$ be Alice's and Bob's expectations, respectively (these are random variables on $\Omega$). Alice and Bob satisfy \emph{symmetrized $\epsilon$-agreement} if $\frac{1}{2}(D_G(a \parallel b) + D_G(b \parallel a))$.
\end{defin}

By Proposition~\ref{prop:chap8_bregman_facts}~\ref{item:shortening}, we know that \textbf{if Alice and Bob satisfy symmetrized $\epsilon$-agreement then they $\epsilon$-agree}.

In our context, symmetrized Bregman divergence is less natural than Jensen-Bregman divergence. This is symmetrized Bregman divergence (unlike Jensen-Bregman divergence) does not seem to closely relate to our previous discussion of the Bregman divergence from a more informed to a less informed estimate being most natural.

\subsection{Alternative notions of accuracy}
Our definition of Alice's accuracy as the expected Bregman divergence from the truth $\mu_{\sigma \tau}$ to Alice's expectation seems like the most natural one. However, one may desire a definition of accuracy that takes both Alice's and Bob's expectations into account, judging the pair's accuracy based on their consensus belief, rather than each of their individual beliefs. For instance, one could say that Alice and Bob are \emph{$\epsilon$-midpoint-accurate} if $\EE{D_G \parens{\mu_{\sigma \tau} \parallel \frac{a + b}{2}}} \le \epsilon$. By this definition, Alice's and Bob's expectations could individually be far from the truth, but they are considered accurate because the average of their expectations is close to correct.

\begin{prop} \label{prop:midpoint_accurate}
If Alice and Bob are $\epsilon$-accurate, then they are $2 \epsilon$-midpoint-accurate.
\end{prop}

\begin{proof}
Observe that for all $a, b, y$ it is the case that
\[D_G \parens{y \parallel \frac{a + b}{2}} \le \max(D_G(y \parallel a), D_G(y \parallel b) \le D_G(y \parallel a) + D_G(y \parallel b).\]
The first inequality is true simply because $\frac{a + b}{2}$ lies in between $a$ and $b$. Therefore,
\[\EE{D_G \parens{y \parallel \frac{a + b}{2}}} \le \EE{D_G(y \parallel a) + D_G(y \parallel b)} \le 2\epsilon.\]
\end{proof}

Another natural choice for Alice's and Bob's consensus belief is the QA pool (see \textcite{nr23_qa}). Proposition~\ref{prop:midpoint_accurate} likewise holds for the QA pool in place of the midpoint, and indeed holds for any choice of consensus belief that is guaranteed to lie in between Alice's and Bob's expectations. Thus, any such definition will be weaker than our definition of $\epsilon$-accuracy for Alice and Bob (up to a constant factor).\\

To summarize, among the above definitions of agreement, $\epsilon$-agreement is the weakest; and among the above definitions of accuracy, Alice's and Bob's $\epsilon$-accuracy is the strongest. This is an indication of strength for Theorem~\ref{thm:agreement_accurate_bregman}: it starts from a relatively weak premise and reaches a relatively strong conclusion.

\section{Implications for communication complexity} \label{appx:comm}
Our results can be framed in a communication complexity context, where they imply that ``substitutable'' functions can be computed with probability $1-\delta$ (over the inputs) with a transcript length depending only on $\delta$.
This is a nonstandard and weak notion of computing the function, but sketching the reduction may inspire future work on connections between substitutes and communication complexity.

In a classic deterministic communication complexity setup (e.g.\ \textcite{rao2020communication}), Alice holds $\sigma \in \mathcal{S}$, Bob holds $\tau \in \mathcal{T}$, and the goal is to compute some function $g: \mathcal{S} \times \mathcal{T} \to \{0,1\}$ using a communication protocol (see Section \ref{subsec:agreement-protocols}).
Our setting captures this model when $Y = g(\sigma,\tau)$.
Observe that in this case, $Y = \mu_{\sigma \tau}$, i.e.\ Alice and Bob's information together determine $Y$ completely.
A communication protocol defines its output by a function $h: \Pi \to \{0,1\}$ where $\Pi$ is the space of transcripts.
We can simply let $h(\pi) = \text{round}(\mu_{ST})$, i.e.\ rounding the \emph{ex post} expectation $\EE{Y \mid \pi} = \mu_{ST}$ to either zero or one.
This is equivalent to the belief of ``Charlie'', or the common knowledge of Alice and Bob after the protocol is completed.

\begin{defin}[Rectangle substitutes, $(1-\delta)$-computes]
  Given a function $g$ and a distribution $\mathcal{D}$ over $\mathcal{S} \times \mathcal{T}$, we say $(g,\mathcal{D})$ satisfy \emph{rectangle substitutes} if the corresponding information structure with $Y = g(\sigma,\tau)$ satisfies rectangle substitutes (Definition \ref{def:rect_subs_quad}).
  We say a protocol $(1-\delta)$-computes $g$ over $\mathcal{D}$ if, with probability at least $1-\delta$ over $(\sigma,\tau) \sim \mathcal{D}$, the protocol has $h(\pi) = g(\sigma,\tau)$.
\end{defin}

By our results, under rectangle substitutes $(g,\mathcal{D})$, any agreement protocol approximately computes $g$ over $\mathcal{D}$.
More precisely, using a fast substitutes-agreement protocol similar to Proposition \ref{prop:fast_rect}, we obtain the following.
\begin{corollary}
  Suppose $(g, \mathcal{D})$ satisfy rectangle substitutes.
  Then for every $\delta \in (0,1)$, there is a deterministic communication protocol using $O(\log(1/\delta))$ bits of communication that $(1-\delta)$-computes $g$ over $\mathcal{D}$.
\end{corollary}
\begin{proof}
  In round one, Alice sends her current expectation $\mu_{\sigma}$ rounded to a multiple of $\epsilon$; call this message $A$.
  In round two, Bob sends his updated expectation $\mu_{S \tau}$ rounded to a multiple of $\epsilon$; call this message $B$.
  The protocol then halts, and the output is $B$ rounded to either zero or one.
  It uses $O(\log(1/\epsilon))$ bits.
  Let $S,T$ be the random rectangle associated with the protocol.
  
  By construction, $|\mu_{\sigma} - A| \leq \epsilon$, and $\mu_{S}$ is the expectation of $Y$ conditioned on $A$, so it follows that $|\mu_{\sigma} - \mu_{S}| \leq \epsilon$.
  Using substitutes (just as in Proposition~\ref{prop:fast_rect}),
  \[ \EE{(\mu_{\sigma \tau} - \mu_{S \tau})^2} \leq \EE{(\mu_{\sigma} - \mu_{S})^2} \leq \epsilon^2 . \]
  By construction, $|B - \mu_{S\tau}| \leq \epsilon$.
  Therefore, by the $\tfrac{1}{2}$-approximate triangle inequality for squared distance (e.g.\ Proposition~\ref{prop:triangle})),
  \[ \EE{(\mu_{\sigma \tau} - B)^2}
      ~\leq~ 2\EE{(\mu_{\sigma \tau} - \mu_{S \tau})^2} + 2\EE{(\mu_{S \tau} - B)^2}
      ~\leq~ 2\epsilon^2 . \]
  Now, the protocol is incorrect if $|B - \mu_{\sigma \tau}| \geq \tfrac{1}{2}$.
  Using Markov's inequality,
  \begin{align*}
    \PP[|B - \mu_{\sigma \tau}| \geq \tfrac{1}{2}]
    &=    \PP[(B - \mu_{\sigma \tau})^2 \geq \tfrac{1}{4}]  \\
    &\leq 4\EE{(B - \mu_{\sigma \tau})^2}  \\
    &\leq 8\epsilon^2 .
  \end{align*}
  Therefore, given $\delta \in (0,1)$, we run the protocol with $\epsilon = \sqrt{\delta/8}$.
  The probability of an incorrect output is at most $\delta$, and we use $O(\log(1/\epsilon) = O(\log(1/\delta))$ bits of communication.
\end{proof}
\end{appendices}

\end{document}